\documentclass[a4paper,12pt]{report}
\pdfoutput=1
\usepackage[utf8]{inputenc}
\usepackage{listings}
\usepackage[T1]{fontenc}
\usepackage{fancyhdr}
\usepackage[Sonny]{fncychap}
\usepackage{xcolor,lipsum}
\usepackage{comment}
\usepackage{amsmath}
\usepackage[left=2.5cm,right=2.5cm]{geometry}
\usepackage{amsmath}
\usepackage{amssymb}
\usepackage{graphicx}
\usepackage{float}
\usepackage{slashed}
\usepackage{amsfonts}
\usepackage{hyperref}
\usepackage[toc,page]{appendix}
\usepackage[compat=1.0.0]{tikz-feynman}
\usepackage[makeroom]{cancel}
\usepackage{subcaption}
\usepackage{transparent}
\usepackage{tikz}
\usepackage{multirow}
\usepackage{mathabx}
\usepackage{stackengine}
\usepackage{cleveref}
\newcommand*\circled[1]{\tikz[baseline=(char.base)]{
            \node[shape=circle,draw,inner sep=2pt] (char) {#1};}}
\usepackage{pdfpages}
\usepackage{dsfont}
\newcommand\xrowht[2][0]{\addstackgap[.5\dimexpr#2\relax]{\vphantom{#1}}}

\let\orgautoref\autoref

\renewcommand{\autoref}[1]{\def\equationautorefname{Eq.}\orgautoref{#1}}

\newcommand{\nl}{\newline}
\newcommand{\nn}{\nonumber}

\definecolor{darkgreen}{rgb}{0.0, 0.45, 0.0}

\hypersetup{
    colorlinks=true,
    citecolor=red,
    linkcolor=blue,
    filecolor=magenta,      
    urlcolor=cyan,
    pdftitle={New solutions in supergravity: A phenomenological study at the LHC}
    }

\newcommand{\beq}{\begin{equation}}
\newcommand{\eeq}{\end{equation}}
\newcommand{\bee}{\end{equation}}
\newcommand{\beqa}{\begin{eqnarray}}
\newcommand{\eeqa}{\end{eqnarray}}
\newcommand{\noi}{\noindent}
\newcommand{\bpm}{\begin{pmatrix}}
\newcommand{\epm}{\end{pmatrix}}

\def\ga{\mathrel{\raise.3ex\hbox{$>$\kern-.75em\lower1ex\hbox{$\sim$}}}}
\def\la{\mathrel{\raise.3ex\hbox{$<$\kern-.75em\lower1ex\hbox{$\sim$}}}}

\newtheorem{theorem}{Theorem}[section]
\newtheorem{proposition}[theorem]{Proposition}
\newcommand{\qeed}{\hfill\textrm{$\blacksquare$}\break\null}
\newenvironment{demo}{\noindent\textit{Proof.}~}{\qeed}
\newenvironment{rcases}
  {\left.\begin{aligned}}
  {\end{aligned}\right\rbrace}

\newcommand{\eK}{e^{\big|\langle z \rangle \big|^2}}
\newcommand{\eKs}{e^{  \frac12 \big|\langle z \rangle \big|^2}}
\newcommand{\pp}[1]{(\langle \phi^{#1} \rangle + \phi^{#1})}

\newcommand{\pb}[1]{(\langle \phi^\dag_{#1} \rangle + \phi^\dag_{#1})}
\newcommand{\bS}[1]{(\langle S^{#1} \rangle + S^{#1})}
\newcommand{\bSb}[1]{(\langle S^\dag_{#1} \rangle + S^\dag_{#1})}

\newcommand{\vp}[1]{\langle \phi^{#1} \rangle}

\newcommand{\vpb}[1]{\langle \phi^\dag_{#1} \rangle }
\newcommand{\vS}[1]{\langle S^{#1} \rangle }
\newcommand{\vSb}[1]{\langle S^\dag_{#1} \rangle}
\newcommand{\lag}{\langle}
\newcommand{\rag}{\rangle}
\newcommand{\lb}{\big|}
\newcommand{\tPhi}{\widetilde{\Phi}}

\def\cc#1{\kern .7em\hfill #1 \hfill\kern .7em}

\newcommand{\ba}{\color{black}}

\definecolor{ora}{cmyk}{0,0.61,0.87,0}

\newcommand{\is}{{i^\ast}}

\def\ben{\begin{enumerate}}
\def\een{\end{enumerate}}
\def\bit{\begin{itemize}}
\def\eit{\end{itemize}}
\def\beq{\begin{equation}}
\def\eeq{\end{equation}}
\def\ba{\begin{array}}
\def\ea{\end{array}}
\def\bea{\begin{eqnarray}}
\def\eea{\end{eqnarray}}
\def\nn{\nonumber}
\def\noi{\noindent}
\def\nl{\newline}
\def\k{\kappa}
\def\l{\lambda}
\def\b{\beta}

\def\e{\varepsilon}

\DeclareMathOperator{\sgn}{sgn}

\title{New Solutions in Supergravity A phenomenological study at the LHC and in Cosmology}
\author{Robin Ducrocq, Groupe Théorie, IPHC, Strasbourg}

\begin{document}

\includepdf[pages={1}]{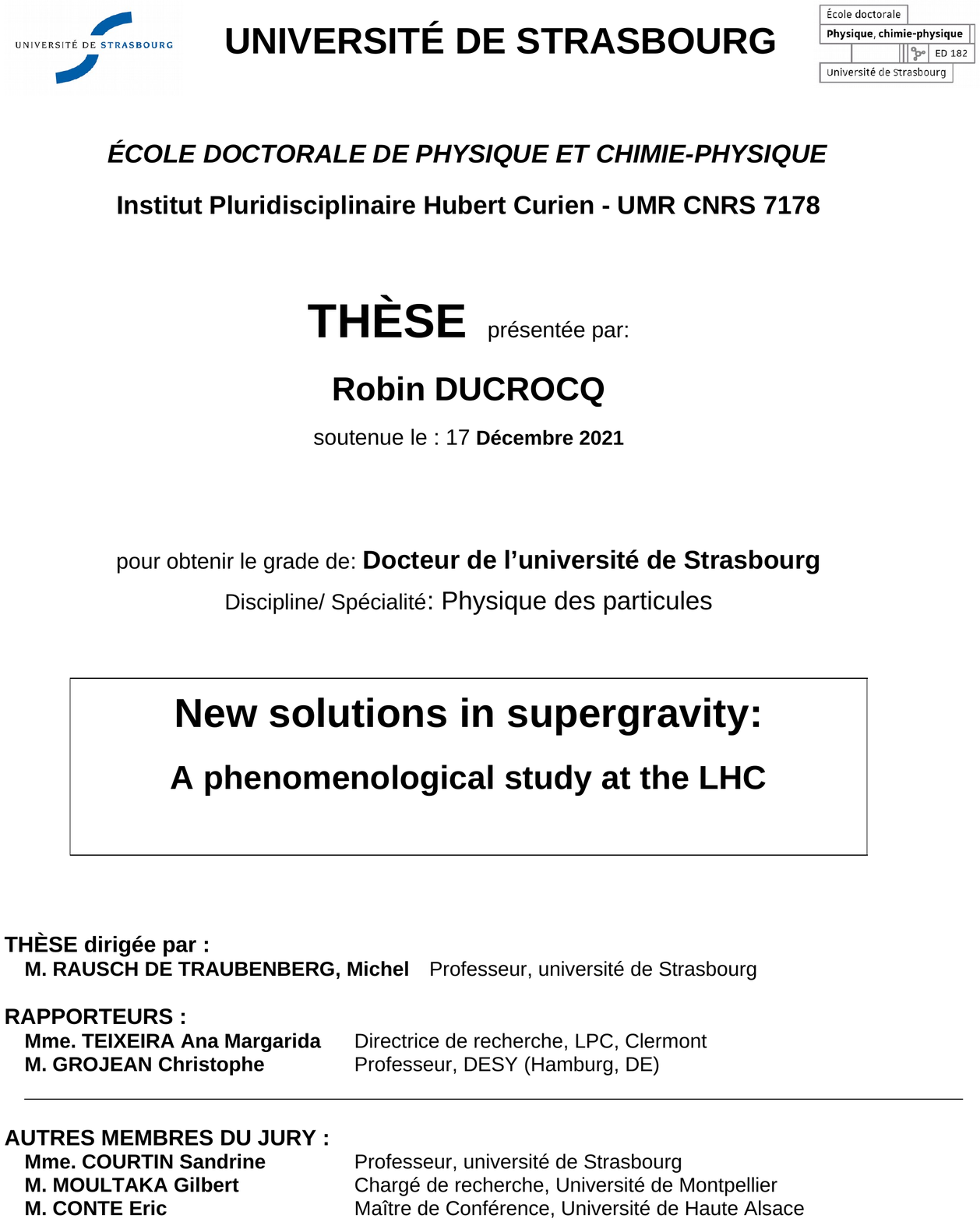}

\chapter*{Remerciements}

Je me permets avec ces quelques phrases de remercier toutes les personnes qui ont contribué de près ou de loin à la réalisation de cette thèse. \medskip

Je tiens à remercier chaleureusement mon directeur de thèse Michel Rausch de Traubenberg. Tu auras été d'un grand soutien durant ces trois ans. Tes compétences en physique ainsi qu'en algèbre n'arrêterons jamais de m'épater ! Un grand merci aussi à Eric Conte. Tu m'auras finalement suivi durant toutes mes années d'études. C'est toujours un plaisir de discuter et travailler avec toi. Merci aussi à toute l'équipe Théorie de m'avoir accueilli pour ces trois ans de thèse. \medskip

Merci aux personnes avec qui j'ai pu travailler dans le cadre du projet SLowSUGRA à savoir Gilbert Moultaka, Cyril Hugonie, Vincent Venin, Julien Lavalle et Gaétan Facchinetti. Grâce à vous j'ai pu enrichir mes connaissances dans divers domaines de la physique des hautes énergies. Un merci tout particulier à toi Gilbert pour ta gentillesse (et parfois tes super blagues !) et à Cyril pour ton aide sur la compréhension des chaînes de Markov.  \medskip

Un grand merci aussi aux membres de l'équipe CMS de l'IPHC, notamment Jérémy Andrea, Daniel Bloch et Emery Nibigira avec qui j'ai eu la chance de collaborer.\medskip

Je souhaiterais aussi remercier les membres de mon jury qui ont acceptés de lire ce manuscrit et de suivre mes travaux. \medskip

Je remercie aussi les autres doctorants du laboratoire : Alex, Clément, Guillaume, Douja, Dylan, Julie et tous les autres ! \medskip

L'expérience d'une thèse me semble infaisable sans soutien en dehors du laboratoire. Je remercie ainsi toutes les personnes qui m'ont soutenu dans les moments compliqués. Merci à ma famille, à mes amis et en particulier Thomas et surtout Adélaïde.

\tableofcontents

\chapter{Résumé détaillé}
\section{Introduction}
Le modèle standard de la physique des particules est le cadre théorique qui décrit les particules élémentaires et leurs interactions (à l'exception de l'interaction gravitationnelle). Il est essentiellement basé sur la théorie quantique des champs (QFT) et sur des principes de symétrie liés aux symétries d'espace-temps (relativité restreinte) et aux symétries internes (invariance de jauge). Les théories de jauges nous permettent d'exprimer ces symétries internes en termes du groupe de jauge $SU(3)_C\times SU(2)_L\times U(1)_Y$ où $SU(3)_C$ correspond à l'interaction forte et $SU(2)_L\times U(1)_Y$ à l'interaction électrofaible, l'unification des interactions faibles et électromagnétiques. La découverte du boson de Higgs par les collaborations CMS \cite{cmsH} et ATLAS \cite{atlasH} en 2012 au CERN (Genève) est souvent considérée comme la dernière preuve expérimentale du modèle standard. Avec le mécanisme de Brout-Englert-Higgs, le boson de Higgs brise spontanément le secteur électrofaible en acquérant une valeur dans le vide (\textit{v.e.v}) $\lag h \rag$ non nulle:
\begin{eqnarray}
SU(2)_L\times U(1)_Y\underset{ \left< h \right>\ne 0}{\rightarrow}  U(1)_{e.m.} \nn
\end{eqnarray}
et génère des termes de masses pour les fermions et les bosons de jauges.\medskip 

Encore aujourd'hui, ce modèle reste l'un des cadres théoriques les plus précis en physique. Malheureusement, plusieurs indices montrent que le modèle standard ne peut pas être la théorie fondamentale pour comprendre la physique à haute énergie. Par exemple, des divergences quadratiques apparaissent lors de l'évaluation des corrections radiatives du boson de Higgs, ce qui conduit au problème de hiérarchie entre la masse du boson de Higgs de $125\ \text{GeV}$ et la masse de Planck $m_p=10^{19}\ \text{GeV}$. Il faut alors appliquer un réglage fin sur les paramètres, ce qui conduit à un problème de naturalité. De plus, les mesures astrophysiques et cosmologiques actuelles ont révélé que les particules du modèle standard ne couvrent que quelques pour cent de la totalité de l'univers, l'autre partie provenant de la matière noire et de l'énergie noire. Les interactions gravitationnelles ne sont, par ailleurs, pas prises en compte. La communauté des physiciens essaye alors d'intégrer le modèle standard dans un modèle plus fondamental de sorte que le modèle standard soit une théorie effective. Il existe plusieurs approches. L'une d'entre elles émerge des principes de symétrie.\medskip

La supersymétrie \cite{martin_supersymmetry_1998}\cite{book_susy} et la supergravité \cite{review}\cite{WessBagger}\cite{book_sugra} sont des extensions naturelles du modèle standard, étendant de manière non triviale l'algèbre de Poincaré. Dans ces théories, une nouvelle symétrie entre les fermions et les bosons est considérée, ce qui conduit à un spectre en particules plus riche et permet de résoudre certains problèmes du modèle standard. Cependant, la supersymétrie impose l'égalité entre les masses des particules dans le même supermultiplet ce qui est en contradiction avec les mesures expérimentales. Ce problème peut être résolu via des mécanismes brisant explicitement cette symétrie fermion-boson et générant des termes de brisure \textit{douce} (c'est-à-dire des termes conduisant à des divergences logarithmiques) dans le Lagrangien. Plusieurs mécanismes existent déjà, introduisant des interactions entre les champs classiques et les champs d'un nouveau secteur (appelé secteur caché) à haute énergie. L'un de ces mécanismes est introduit dans le contexte de la supergravité. La brisure de la supergravité dans le secteur caché est transmise au secteur visible par l'intermédiaire d'effets gravitationnels, ce qui conduit à la brisure de la supersymétrie à basse énergie. Ce mécanisme a été étudié dans les années 80 par Soni \& Weldon \cite{soni_analysis_1983} qui ont classé les formes possibles de deux fonctions fondamentales en supergravité : le superpotentiel et le potentiel de Kähler. Ces solutions sont à la source de toutes les études actuelles de la supersymétrie avec brisure de supersymétrie par médiation gravitationnelle.\medskip

Dans un article récent \cite{moultaka_low_2018}, la classification obtenue par Soni \& Weldon s'est avérée incomplète. De nouvelles solutions, introduisant un nouveau type de superchamp (appelé \textit{hybride}) ayant des caractéristiques à la fois du secteur caché et du secteur visible, ont été trouvées. Une autre spécificité de ces solutions est que les nouvelles interactions génèrent dans le Lagrangien de nouveaux termes de brisure \textit{dure} (c'est-à-dire des termes conduisant à des divergences quadratiques) supprimés par une échelle d'énergie intermédiaire inférieure à la masse de Planck. Ces nouveaux termes contribuent à l'arbre ainsi qu'au niveau des boucles avec le secteur visible et semblent pouvoir expliquer la non-détection des particules supersymétriques dans les expériences des collisionneurs. Par ailleurs, ils semblent aussi permettre la réduction du réglage fin de la masse du boson de Higgs. La structure du superpotentiel impose au moins deux nouveaux superchamps hybrides afin de coupler le nouveau secteur et le secteur visible, contenant les particules du modèle standard. Cela correspond alors à une extension à deux singlets du MSSM avec quelques spécificités.\medskip

Dans le premier chapitre de ce manuscrit, la notion de symétrie en physique des particules est introduite. À partir du théorème de Haag–Łopuszański–Sohnius, la supersymétrie est définie et le Lagrangien de la supergravité $N=1$ en dimension quatre est construit selon des principes géométriques. Le lien entre la supergravité et la supersymétrie à basse énergie est mis en évidence et les solutions de Soni \& Weldon \cite{soni_analysis_1983} pour le mécanisme de brisure de supersymétrie par médiation gravitationnelle sont données. Quelques modèles supersymétriques sont introduits comme le MSSM (pour \textit{Minimal Supersymmetric Standard Model}) et le N2MSSM, nouvelle extension à deux singlets du MSSM. Un résumé de ce chapitre est présenté Section \ref{subsec:resume1} et Section \ref{subsec:resume2}. \medskip

Le chapitre 2 est consacré à l'analyse des nouvelles solutions (appelées NSW pour Non-Soni-Weldon) conduisant à la brisure de supersymétrie par interactions gravitationnelles. Après avoir donné quelques généralités concernant ces solutions, nous montrons, en utilisant le potentiel de Coleman Weinberg, que les nouveaux termes de brisure de la supersymétrie génèrent de nouvelles divergences quadratiques et de nouvelles contributions aux matrices de masse. Un modèle non canonique spécifique, déduit des solutions NSW (le S2MSSM) est introduit. Dans ce modèle, deux superchamps $S^p$ ($p=1,2$) du secteur hybride sont couplés au secteur visible dans le superpotentiel. Le potentiel complet à basse énergie est calculé. Enfin, une étude de l'ordre de grandeur des corrections radiatives sur la masse du boson de Higgs est effectuée dans un modèle simplifié. Nous identifions les régions de l'espace des paramètres qui permettent de reproduire une masse de boson de Higgs proche de $125\ \text{GeV}$. Ce chapitre est résumé dans les Sections \ref{subsec:resume3} et \ref{subsec:resume4}. \medskip

Une analyse préliminaire du N2MSSM est effectuée au chapitre 3. Ce modèle est introduit pour plusieurs raisons. Premièrement, il présente certaines similitudes avec le S2MSSM dans le contexte des solutions classiques de Soni-Weldon car il contient deux superchamps singlets. Une comparaison entre ces deux modèles peut être intéressante pour mettre en évidence les différences phénoménologiques entre les solutions Soni-Weldon et Non-Soni-Weldon. De plus, l'ajout d'un second singlet pourrait résoudre certaines difficultés du MSSM et du NMSSM, comme le réglage fin (\textit{fine-tuning}) des paramètres pour reconstruire une masse cohérente de boson de Higgs ainsi que pour obtenir une échelle électrofaible valide. Dans ce manuscrit, nous analyserons principalement les différences entre le N2MSSM et le NMSSM. Un générateur de spectre pour le N2MSSM (avec les programmes \textsc{SARAH} et \textsc{SPheno}) est mis en place et
un algorithme de Monte-Carlo à chaîne de Markov (ou MCMC pour \textit{Markov-Chain Monte-Carlo}) est implémenté pour un balayage efficace de l'espace des paramètres. Plusieurs contraintes sur la masse du boson de Higgs et sur le gluino sont imposées. La différence de \textit{fine-tuning} de l'échelle électrofaible entre le N2MSSM et le NMSSM est également étudiée. La section \ref{subsec:resume5} résume cette analyse.\medskip

Dans le dernier chapitre (découplé de l'étude des nouvelles solutions NSW et du N2MSSM), nous analysons la pertinence d'éventuelles signatures de quarks top déplacés au LHC. Notre étude comporte deux analyses phénoménologiques basées sur deux modèles supersymétriques simples. Le premier est une désintégration d'un squark stop en un quark top et un neutralino LSP (pour Lightest Supersymmetric Particle) $\tilde{t}\rightarrow t\chi^0_1$ dans le MSSM avec mécanisme de brisure de supersymétrie induite par gravitation. Le second est un squark stop se désintégrant en un quark top et un gravitino LSP $\tilde{t}\rightarrow t\psi_\mu$ dans le MSSM suivant un mécanisme de brisure de supersymétrie par médiation de jauge (ou GMSB pour \textit{Gauge Mediated Supersymmetry Breaking}) où le gravitino est naturellement la LSP. Plusieurs outils sont utilisés tels que \textsc{MadGraph\_aMC@NLO} pour la génération d'événements, \textsc{Pythia} pour la simulation de l'hadronisation ou \textsc{Madanalysis5} pour l'analyse des données. Via cette analyse, différents benchmarks sont définis afin d'étudier les distributions des observables utilisées au sein des expériences CMS et ATLAS. Le résumé de cette étude se trouve Section \ref{subsec:resume6}. \medskip

\section{Reconstruction du lagrangien de la supergravité $N=1$ en $4D$}\label{subsec:resume1}
La supergravité est une théorie locale de la supersymétrie. De part sa construction, elle correspond naturellement à une théorie de la gravitation où l'espace associé à la supersymétrie (appelé le superespace) peut être courbe. La supergravité peut être construite en adoptant une approche équivalente à la relativité générale en se basant sur le principe d'équivalence et en définissant de nouveaux champs : le graviton $e_{\mu}{}^{\tilde{\mu}}$ (champ de spin 2) et le gravitino $\psi_{\tilde{\mu}}$ (champ de spin $3/2$). \newline

Dans ce manuscrit, le Lagrangien de la supergravité $N=1$ en quatre dimensions a été construit en suivant le formalisme du superespace et via l'introduction de deux objets fondamentaux liés à la géométrie du superespace courbe : le supervierbein $E_{\tilde{M}}{}^M$ et la superconnection de spin $\Omega_{\tilde{M}MN}$. Le supervierbein permet de relier les objets du superespace courbe au superespace plat tangent alors que la superconnection de spin permet de définir les dérivées covariantes. Le calcul complet du Lagrangien et un processus long et complexe. Chaque étape est explicitée afin de reconstruire l'action de la supergravité $N=1$ en dimension quatre :
\begin{gather}
\mathcal{S}_{\text{{\tiny SUGRA}}} = \frac38 \int d^4x\, d^2\Theta\mathcal{E}\Big( \bar{\mathcal{D}}\cdot\bar{\mathcal{D}} - 8\mathcal{R} \Big)e^{-\frac13 K(\Phi,\Phi^{\dagger}e^{-2gV})} \nn \\
 + \int d^4x\, d^2\Theta \mathcal{E}W(\Phi) \nn \\
+ \frac{1}{16g^2}\int d^4x\, d^2\Theta \mathcal{E}h_{ab}(\Phi)\mathcal{W}^{\alpha a}\mathcal{W}_{\alpha}{}^b + \mathrm{h.c.}\nn
\end{gather}
où trois fonctions fondamentales sont introduites : le potentiel de Kähler $K(\Phi,\Phi^{\dagger})$ définissant les termes cinétiques des superchamps chiraux, le superpotentiel $W(\Phi)$ contenant les couplages de Yukawas et la fonction cinétique de jauge $h_{ab}$ correspondant aux termes cinétiques des superchamp vectoriels (bosons de jauges et jauginos).\medskip

En supposant une forme canonique au potentiel de Kähler et à la fonction cinétique de jauge
\begin{gather}
K(\Phi,\Phi^{\dagger})=\Phi^i\Phi^\dagger_{i}\ , \quad h_{ab}=\delta_{ab}\ , \nn
\end{gather}
ainsi qu'un superpotentiel cubique
\begin{gather}
W(\Phi)=\alpha_i\Phi^i + \frac12 \beta_{ij}\Phi^i\Phi^j + \frac16\lambda_{ijk}\Phi^i\Phi^j\Phi^k\ ,  \nn
\end{gather}
il est alors possible de montrer que la limite basse énergie du Lagrangien de la supergravité correspond naturellement au Lagrangien de la supersymétrie : 
\begin{gather}
\displaystyle \lim_{m_p\rightarrow \infty} \Big( \mathcal{L}_{SUGRA} -\mathcal{L}_{Pure\ SUGRA} \Big) = \mathcal{L}_{SUSY}^{kin.} +\mathcal{L}_{SUSY}^{int.}\label{resume:1}
\end{gather}
où $\mathcal{L}_{SUSY}^{kin.}$ contient les termes cinétiques et $\mathcal{L}_{SUSY}^{int.}$ les termes d'interactions. 
\section{Brisure de supersymétrie, MSSM et N2MSSM}\label{subsec:resume2}
La supersymétrie impose que les particules au sein d'un même multiplet possèdent la même masse. Or, les mesures expérimentales rejettent la présence de particules supersymétriques à ces échelles d'énergies. Ainsi, afin de rendre compte de ces observations, la supersymétrie doit être brisée. Il existe plusieurs mécanismes de brisure de la supersymétrie. Le travail présenté dans ce manuscrit porte principalement sur les mécanismes de brisure induite par la gravitation.\\
Les solutions classiques de la supergravité (classifiées par Soni et Weldon en 1983 \cite{soni_analysis_1983}) induisent par ce mécanisme des termes de brisure \textit{douce} dans le Lagrangien \autoref{resume:1} ayant de bonnes propriétés de renormalisation (corrections uniquement logarithmiques). Ces termes s'obtiennent via l'introduction d'un secteur de champ caché et le calcul du potentiel scalaire qui est fonction du potentiel de Kähler $K(\Phi , \Phi^\dagger,z, z^\dagger)$ ($\Phi$ et $z$ étant, respectivement, des champs du secteur visible et caché) et du superpotentiel $W(\Phi )$. Le potentiel scalaire est :
\begin{eqnarray}
V = e^{K(\Phi , \Phi^\dagger,z, z^\dagger)/m_p^2}\left( \mathcal{D}_IW\mathcal{D}^{I*}W^{*}K^I{}_{I*}-\frac{3}{m_p^2}|W|^2 \right)
\label{eq:potsugra}
\end{eqnarray}
où $m_p$ est la masse de Planck, $\mathcal{D}_I$ sont des dérivées covariantes et $K^I{}_{I*}$ est la métrique de Kähler.\\

De part ces mécanismes, on peut alors construire des modèles de supersymétrie \sloppy phénoménologiquement viable. Le plus simple d'entre eux est le MSSM (pour \textit{Minimal Supersymmetric Standard Model}), basé sur le groupe de jauge $SU(3)_C\times SU(2)_L\times U(1)_Y$ contenant uniquement les particules du modèle standard (à l'exception de la présence de deux doublets de Higgs) ainsi que leurs partenaires supersymétriques. Le superpotentiel du MSSM peut être écrit comme 
\begin{eqnarray}
W_{MSSM} &=& -\hat{L}\cdot \hat{H}_D\mathbf{Y_E}\hat{E}- \hat{Q}\cdot \hat{H}_D \mathbf{Y_D} \hat{D} +  \hat{Q}\cdot \hat{H}_U \mathbf{Y_U} \hat{U} \nn\\
&&+ \mu \hat{H}_U\cdot \hat{H}_D + W_{N} +W_{RPV} \ ,\nn
\end{eqnarray}
où $\mathbf{Y_I}$ sont des matrices $3\times 3$ contenant les couplages de Yukawa, $W_N$ correspondant au couplage de Yukawa et la masse de Dirac des neutrinos :
\begin{gather}
W_N = L\cdot \hat{H}_D \mathbf{Y_N} \hat{N} + m_N \hat{N}\hat{N} \nn
\end{gather}
et le produit $SU(2)$ "$\cdot$" tel que $\hat{H}_U\cdot\hat{H}_D=\hat{H}_U^{+}\hat{H}_D^{-} - \hat{H}_U^{0}\hat{H}_D^{0}$. Le dernier terme, $W_{RPV}$, 
\begin{gather}
W_{RPV} = \lambda_{ijk}\hat{L}^i\cdot \hat{L}^j\hat{E}^k+\lambda'_{ijk}\hat{L}^i\cdot \hat{Q}^j\hat{D}^k + \kappa_i\hat{L}^i\cdot \hat{H}_U + \lambda''_{ijk}\hat{U}^i\hat{D}^j\hat{D}^k \nn 
\end{gather}
contient des termes violant une symétrie discrète appelé la R-parité \cite{MSSMRPV}.\\

Le spectre du MSSM contient deux Higgs scalaires $h_i$ ($i=1,2$), un pseudoscalaire $A_0$ et un chargé $h^\pm$. Dans le secteur des sfermions (partenaires supersymétriques des fermions du modèle standard), deux états sont associés à chaque fermions, soit deux stops $\tilde{t}_i$, deux sbottoms $\tilde{b}_i$ et deux staus $\tilde{\tau}_i$ ($i=1,2$) (on suppose uniquement la troisième famille de fermion). De plus, quatre nouveaux états appelés neutralinos $\tilde{\chi}^0_i$ ($i=1,...,4$) sont générés via le mélange entre les higgsinos (superpartenaires des bosons de Higgs) et les jauginos neutres (partenaires supersymétriques des bosons de jauges). Le higgsino chargé et les états de winos (partenaire supersymétrique du boson W) génèrent, quant à eux, deux charginos $\tilde{\chi}^\pm_i$ ($i=1,2$) (voir \cite{book_susy}\cite{martin_supersymmetry_1998} pour une description plus complète du MSSM). \medskip 

Le MSSM à cependant des problèmes conceptuels qui peuvent être supprimés par l'ajout de nouveaux superchamps singlet de jauge. L'ajout d'un seul superchamp singlet défini le modèle appelé NMSSM. Nous avons défini une extension à deux superchamps singlet de jauge appelé le N2MSSM. \\

Le N2MSSM est défini comme le MSSM contenant deux singlets de jauge supplémentaires $\hat{S}^p=(\underset{\sim}{1},\underset{\sim}{1},0)$ $(p=1,\ 2)$. En ne considérant ni les couplages violant la R-parité ni les neutrinos, le superpotentiel peut alors être écrit comme:
\begin{eqnarray}
W_{N2MSSM} &=& \mu\hat{H}_U\cdot \hat{H}_D  + \lambda_i \hat{S}^i \hat{H}_U\cdot \hat{H}_D  + \frac13\kappa_{ijk}\hat{S}^i\hat{S}^j\hat{S}^k + \frac12 \mu'_{ij}\hat{S}^i\hat{S}^j + \xi_{F,i}\hat{S}^i \nn\\
&& + \hat{Q}\cdot \hat{H}_U \mathbf{Y_U} \hat{U} - \hat{Q}\cdot \hat{H}_D \mathbf{Y_D} \hat{D} - \hat{L}\cdot \hat{H}_D \mathbf{Y_E} \hat{E} \ .\nn
\end{eqnarray}
On suppose ici que tous les paramètres sont réels afin d'éviter une brisure de la symétrie CP. De plus, uniquement la troisième famille de fermions est prise en compte (voir \autoref{yukawa}). Le NMSSM peut être obtenu comme une limite du N2MSSM en supprimant tous les couplages du second singlet $\hat{S}^2$.\medskip

Le superpotentiel présenté ci-dessus peut être simplifié en supposant une invariance sous $\mathbb{Z}_3$, c'est-à-dire invariant sous la transformation $\hat{\Phi}\rightarrow e^{\frac{2\pi i}{3}}\hat{\Phi}$. Cette symétrie apparaît lorsque tous les paramètres dimensionnés dans $W_{N2MSSM}$ ($\mu'_{ij} = \xi_{F,i} = 0$) sont nul\footnote{On peut supposer une invariance superconforme du modèle, ce qui impose alors la symétrie $\mathbb{Z}_3$ du superpotentiel (voir \cite{book_sugra})}. Cette hypothèse génère cependant des observables cosmologiques appelés mur de domaine (ou \textit{domain wall problem}) \cite{domainwall}. Les \textit{v.e.vs.} non-nulles des champs scalaires brisent cette symétrie discrète et génèrent des régions de l'univers ayant la même énergie du vide mais possédant des phases différentes venant de la transformation $\hat{\Phi}\rightarrow e^{\frac{2\pi i}{3}}\hat{\Phi}$. La surface de séparation entre ces régions (les murs de domaines) doit alors pouvoir être détectée par les mesures expérimentales. Des solutions à ce problème ont déjà été proposées \cite{domwalsol1}\cite{domwalsol2}. \medskip

L'ajout des singlets de jauge étend aussi le spectre en particule du MSSM. Les parties scalaires des superchamps $\hat{S}^p$ vont se mélanger avec le secteur des Higgs scalaires et pseudoscalaires pour donner 4 états scalaires et 3 états pseudoscalaires. Les parties fermioniques vont quant à eux contribuer au secteur des neutralinos et générer 6 états (contrairement à 4 dans le MSSM). Les autres secteurs restent inchangés. 

\section{Solutions Non-Soni-Weldon et S2MSSM}\label{subsec:resume3}
Il a récemment été mis en évidence que la classification des solutions par Soni \& Weldon est incomplète \cite{moultaka_low_2018}. Il existe une nouvelle zoologie de solutions, introduisant un nouveau type de superchamp chiral (appelé \textit{hybride} et noté par la suite $S$) ayant des propriétés à la fois du secteur visible et du secteur caché. On peut alors montrer (après brisure de supersymétrie dans le secteur caché) que le potentiel \eqref{eq:potsugra} lié à ces solutions peut s'écrire sous la forme :
\begin{eqnarray}
V = V_{SUSY} + V_{soft} + V_{hard} + m_p^2\Lambda \nn
\end{eqnarray}
avec $V_{soft}$ des termes de brisure douce (termes présents dans les modèles classiques brisant la supersymétrie), $\Lambda$ la constante cosmologique et $V_{hard}$ la partie du potentiel contenant de nouveaux termes de brisure dite dure ayant des propriétés de renormalisation différentes des termes doux et paramétriquement supprimés. La particularité de ces termes est qu'ils pourraient permettre des particules supersymétriques très massives permettant d'expliquer leurs non-détections au sein des expériences de physique des particules. \\

Ces calculs nous ont permis de faire ressortir deux modèles jamais étudiés auparavant et intéressants pour notre étude. Le premier est le N2MSSM, basé sur les solutions de Soni et Weldon et déjà introduit dans la section précédente. Le second est un modèle ayant une structure inédite provenant des nouvelles solutions baptisé le S2MSSM étant équivalent au N2MSSM en nombre de particules. L'étude de ces deux modèles permettra de mettre en évidence l'apport phénoménologique des termes de brisure dure par rapport aux termes dit doux. Le superpotentiel et le potentiel de Kähler du S2MSSM correspondent à : 
\beqa
W(z, S, \widetilde{\Phi}) &=& 
m_{p\ell}\left(W_{1,0}(z) + 
 S^p W_{1, p} (z)\right)
+ \   W_0(z, \widetilde{\Phi},\mathcal{U}^{12}) +  S^p W_{0, p} (z) \ ,  \nn\ \\
K(z,z^\dag,S,S^\dag,\tilde{\Phi},\tilde{\Phi}^\dag) &=& m_p^2z^iz^\dag_i + S_p^\dag S^p + \tilde{\Phi}_{a*}^\dag\Lambda^{a*}{}_a(z,z^{\dagger})\tilde{\Phi}^a \nn\\
&=& m_p^2\hat{K}(z,z^\dag) + \tilde{K}(z,z^\dag,S,S^\dag,\tilde{\Phi},\tilde{\Phi}^\dag),  \nn 
\eeqa
où :
\begin{gather}
\mathcal{U}^{12} = \mu_2 S^1 - \mu^1S^2 (\equiv \mathcal{U})\nn
\end{gather}
et :
\begin{eqnarray}
W_0(z,\tilde{\Phi},S) &=& \lambda(z) \mathcal{U}\hat{H}_U\cdot\hat{H}_D + \kappa(z)\mathcal{U}^3 + y_u(z) \hat{Q}\cdot\hat{H}_U\hat{U} -y_d(z) \hat{Q}\cdot\hat{H}_D\hat{D} \nn\\
&& -y_e(z) \hat{L}\cdot\hat{H}_D\hat{E} \nn
\end{eqnarray}
(avec $\{\tilde{\Phi}^a\}=\{\hat{H}_U,\hat{H}_D,\hat{Q},\hat{U},\hat{D},\hat{L},\hat{E}\}$, le contenu en champ du MSSM). Notons que les deux champs hybrides $S^1$ et $S^2$ se couplent aux champs de matières $\tilde{\Phi}^a$ via la combinaison linéaire $\mathcal{U}^{12}\equiv \mathcal{U}$. Le potentiel de Kähler est supposé non-canonique dans le secteur de matière $\{\tilde{\Phi}^a\}$ afin de générer des termes de brisures de la supersymétrie non-universels. Afin de simplifier le calcul, nous supposons que le secteur caché n'est composé que d'un seul champ $z$.

\section{Fine-tuning de la masse du Higgs au sein du S2MSSM}\label{subsec:resume4}
L'apport des nouvelles solutions de supergravité semble permettre de résoudre un problème du modèle standard appelé le problème de naturalité. Ce problème correspond au fait qu'une masse de boson de Higgs de $125\ \text{GeV}$ n'est théoriquement pas naturelle. En effet, les nouvelles solutions apportent de nouvelles corrections à boucles provenant des nouveaux champs $S$ et amenant potentiellement à une valeur de masse de Higgs plus proche naturellement des $125\ \text{GeV}$.
Ainsi, Le calcul du potentiel, de ses dérivées et de la matrice de masse du secteur scalaire ont étés effectués dans le cadre d'un modèle générique à n nouveaux champs $\{S^1,\dots,S^n\}$ en supposant des valeurs des champs dans le vide (\textit{v.e.vs}) non-nulles ($\left<S^p\right> \neq 0$) interagissant avec un seul champ du secteur caché $z$.\medskip

L'analyse s'est par la suite restreinte sur un modèle sans interactions entre le secteur observable et les nouveaux champs $S^p$ en supposant des \textit{v.e.vs} nulles ($\left<S^p\right> = 0$). Nous avons alors déterminé l'ordre de grandeur de la correction sur la masse du boson de Higgs $m_h^{(1L)}$ dans le cadre d'un modèle avec un seul nouveau champ avec :
\begin{gather}
m_h^{(1L)} = \sqrt{m_h^2{}^{(TL)} + \frac{1}{16\pi^2m_p^2}e^{|\lag z\rag|^2}\Big( m_{\zeta '}^2\sin^2\theta +  m_{S '}^2\cos^2\theta  \Big)\Big(\sqrt{|{\cal I}|^2} + \frac{1}{m_p^2}\big(4|\xi_{3/2}|^2 - 2\big)|{\cal I}|^2\Big)} \nn
\end{gather}
où $m_h^{(TL)}=115\ \text{GeV}$ est la masse du boson de Higgs contenant les contributions à l'arbre et à boucle des termes doux, $\theta$ est l'angle de mélange dans la base $\{z,S\}$, $m_{\zeta '}^2$ et $m_{S '}^2$ les états de masses liés aux champs $z$ et $S$ et $\xi_{3/2}$, $|\mathcal{I}|^2$ des paramètres de notre modèle.\medskip

\begin{figure}[H]
    \centering
\includegraphics[width=.6\linewidth]{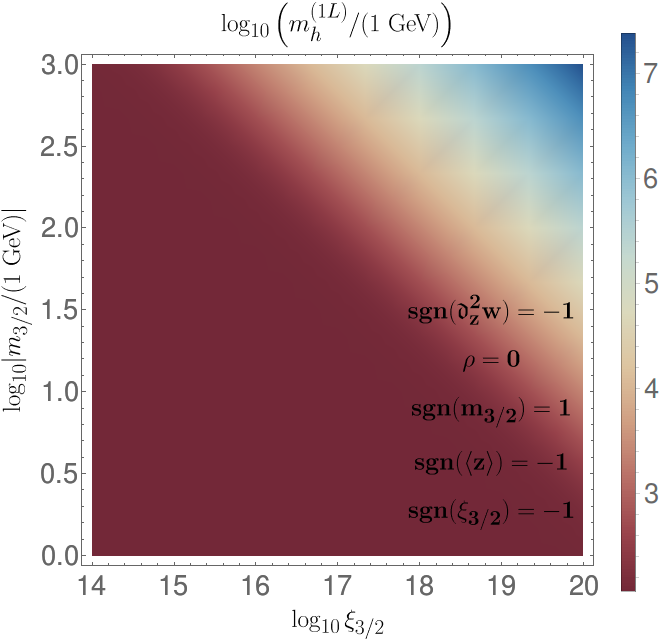} 
\caption{Évolution de la masse du boson de Higgs à une boucle $m_h^{(1L)}$ en fonction de $\log_{10}\big(|m_{3/2}/(1\ \mathrm{GeV})|\big)$ (avec $m_{3/2}$ la masse du gravitino) et $\log_{10}\big(|\xi_{3/2}|\big)$ ($\xi_{3/2}$ étant un certain paramètre du modèle relié à la masse de gravitino $m_{3/2}$) pour une configuration de paramètres. La masse du boson de Higgs à l'arbre avec les contributions des termes doux est supposée à $m_h^{(T.L.)}=115\ \text{GeV}$. Les seules configurations amenant à une correction importante sont obtenues pour de très hautes valeurs de $\xi_{3/2}$ ce qui amène un problème de naturalité.}
\label{fig:rho0mh1L_resume}
\end{figure}

 En supposant une échelle GUT dans le secteur visible, nous avons mis en évidence des zones de l'espace des paramètres amenant des corrections radiatives relativement importantes et permettant une masse du boson de Higgs à $125\ \text{GeV}$ à l'ordre d'une boucle. Bien que \textit{fine-tunées} (voir \autoref{fig:rho0mh1L_resume}), ces configurations semblent pouvoir être obtenues de manière naturelle au sein de modèles plus complexes contenant plusieurs champs dans le secteur caché. \newline

La difficulté de ces calculs provient des nombreuses contraintes sur la nullité de la constante cosmologique ($\left<V\right>=0$) ainsi que le minimum du potentiel ($\left<\frac{\partial V}{\partial X^i}\right>=0$ avec $X^i$ les champs du modèle). L’étude du cas général à n nouveaux champs est naturellement plus complexe. Les premiers résultats montrent que la structure de la matrice de masse du secteur $\{z,S^p\}$ $(p=1,\ \dots,\ n)$ correspond à :
\begin{gather}
\mathcal{M}^2 = \begin{pmatrix} a \mathbb{I}_{n-1}&0&0\\
  0&a + b |\mathcal{I}|^2&c |\mathcal{I}|\\
  0 &\bar c |\mathcal{I}|&d\\
  \end{pmatrix}
  \label{eq:matrixprime}
\end{gather}
avec $a$, $b$, $c$, $d$ et $\mathcal{I}$ des paramètres du modèle. Cette structure montre $n-1$ états dégénérés ayant une masse proche de $a = m_{3/2}$, tandis que le dernier singlet se couple au champ du secteur caché de manière équivalente au cas à un seul nouveau champ. Une analyse numérique complète dans cette configuration mettra en évidence l’ordre de grandeur de la correction radiative.

Après avoir effectué cette analyse dans le cas du modèle simplifié, le but sera de prendre en compte les \textit{v.e.vs} non-nulles des champs ainsi que les termes d'interactions entre les nouveaux champs $S^p$ et les bosons de Higgs. Le calcul des équations du groupe de renormalisation (RGEs) permettra aussi de mettre en place un générateur de spectre complet pour les nouvelles solutions Non-Soni-Weldon.\footnote{Notons que le théorème de non-renormalisation n'est pas forcément applicable dans le cadre de modèles contenant des termes de brisures durs. Cela complexifie le calcul des RGEs.} 

\section{Vers la phénoménologie du N2MSSM}\label{subsec:resume5}
Il existe plusieurs raisons d'étudier le N2MSSM :
\begin{itemize}
\item ce modèle étant l'équivalent du S2MSSM pour les solutions classiques de Soni \& Weldon, cela permettrait d'analyser les différences phénoménologiques entre les nouvelles solutions NSW et les solutions classiques,
\item l'ajout d'un deuxième singlet au NMSSM pourrait faciliter la reconstruction d'une masse de boson de Higgs ainsi que d'une densité relique de matière noire en accord avec les mesures expérimentales,
\item cela permet de définir de nouvelles signatures expérimentales non étudiées au sein du LHC avec des processus contenant potentiellement des scalaires très légers.
\end{itemize}
Pour effectuer l'étude du N2MSSM, plusieurs outils existants ont été utilisés. Le calcul des matrices de masses, des équations du groupe de renormalisation (RGE) à deux boucles (essentielle pour l'analyse phénoménolo\-gique), des équations de minimisations du potentiel ainsi que des corrections radiatives à une boucle sur les masses des particules (deux boucles pour les bosons de Higgs scalaires) sont obtenus via le package \textsc{SARAH} \cite{Staub_sarah} de \textsc{Mathematica}. Le générateur de spectre \textsc{SPheno} \cite{Porod_spheno} est utilisé pour calculer le spectre du modèle ainsi que certaines observables comme le \textit{fine-tuning} $\Delta_{FT}$ sur la masse du boson Z $m_Z$ (donnant une indication sur la naturalité de la masse du boson Z et donc de l'échelle électrofaible):
\begin{gather}
\Delta_{FT} =  max\left( Abs\left[\Delta_i\right]\right) \quad \textrm{avec} \quad \Delta_i=\frac{\partial\textrm{ln} m_Z^2}{\partial \textrm{ln} q_i} \nonumber
\end{gather}
($q_i$ étant les paramètres du modèle) ou le moment magnétique anomal du muon $a_{\mu} = \frac12 \left( g_{\mu} - 2 \right)$. \medskip

Les zones de l'espace des paramètres écartées par les mesures expérimentales étant non-négligeable et la dimension de cet espace importante, il a été choisi d'utiliser un algorithme permettant de scanner efficacement l'espace des paramètres, l'algorithme utilisé étant un algorithme MCMC pour {\it Markov-Chain Monte-Carlo}. Cet algorithme impose la définition et la réduction d'une fonction de pénalité représentant la validité expérimentale du point de l'espace des paramètres étudié (un point amenant à un spectre de masse cohérent aura une pénalité plus faible qu'un point ayant un spectre divergent), permettant alors de scanner rapidement des zones de l'espace des paramètres valides du point de vue expérimental. La dernière étape du scan consiste à réduire le \textit{fine-tuning} $\Delta_{FT}$ jusqu'à obtenir la convergence de la fonction de pénalité (une variation de la fonction de pénalité inférieure à $1\%$ entre deux itérations signifie que le minimum de la fonction de pénalité est atteint).\medskip

\begin{figure}[H]
\centering
\includegraphics[scale=0.6]{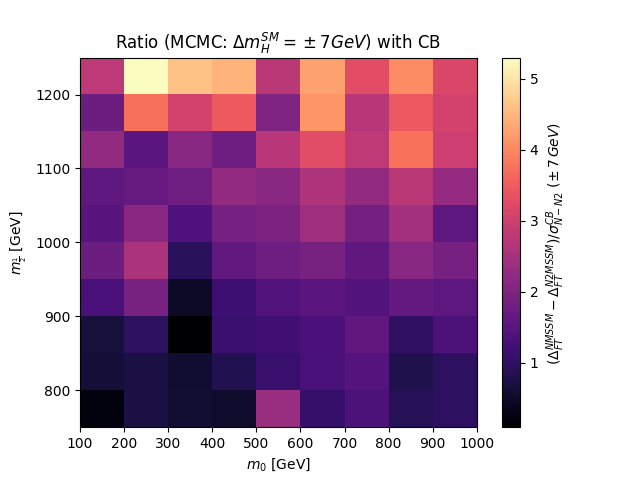}
\caption{Différence de {\it fine-tuning} $\Delta_{FT}$ entre les modèles NMSSM et N2MSSM (Scan en $m_0\ [\text{GeV}]$ et $m_{1/2}\ [\text{GeV}]$ via le MCMC). Les incertitudes $\sigma_{N-N2}$ ont étés obtenues via un fit de Crystal-Ball sur les distributions.}
\label{fig:scan}
\end{figure}

La Figure \ref{fig:scan} représente un résultat de scan sur les paramètres $m_0$ et $m_{1/2}$ (paramètres de masses brisant la supersymétrie) en mettant en place l'algorithme MCMC sur tous les autres paramètres (soit 14 paramètres au total). Les contraintes lors de ce scan sont :
\begin{itemize}
\item un secteur de squarks avec une masse supérieure à $1\ \text{TeV}$,
\item une masse de gluino $m_{\tilde{g}}> 1.6\ \text{TeV}$,
\item  une masse de boson de Higgs correspondant à celui du Modèle Standard de $m_{h} = 125\pm 7\ \text{GeV}$.
\end{itemize}
Le but de ce graphique est de comparer le {\it fine-tuning} du NMSSM (modèle équivalent au N2MSSM mais avec un champ singlet en moins) et du N2MSSM. De part la structure de ces deux modèles, le résultat attendu est un {\it fine-tuning} plus important pour le NMSSM que pour le N2MSSM. Dû à la nature stochastique de la chaîne de Markov, il est important de faire suivre les résultats des scans par une analyse statistique de ces derniers. Pour cela, une incertitude sur $\Delta_{FT}^{NMSSM} - \Delta_{FT}^{N2MSSM}$ est calculée pour chaque cellule du scan. On peut alors calculer la variable
\begin{gather}
X_{\sigma}=\frac{(\Delta_{FT}^{NMSSM} - \Delta_{F}^{N2MSSM})}{\sigma_{N-N2}}\nn
\end{gather}
(avec $\sigma_{N-N2}$ l'incertitude associé à la différence de fine-tuning entre les deux modèles) qui donne une estimation de la \textit{p-value} liée à l'hypothèse : 
\begin{center}
\textit{Les valeurs de fine-tuning pour le NMSSM et le N2MSSM ne sont pas statistiquement significatif}.\end{center}
On peut constater sur la figure  \ref{fig:scan} que les écarts sont pour le moment trop importants (peu de cellules avec $X_{\sigma}>5$ correspondant à un écart à $5\sigma$) et donc que la différence de fine-tuning n'est pour le moment pas statistiquement significatif. Une analyse plus précise (actuellement en cours) doit alors être effectuée en améliorant la précision sur la convergence de la fonction de pénalité.\medskip


Les résultats de ce scan nous a cependant montré que la masse du boson de Higgs $m_{H} = 125\pm 7\ GeV$ est plus facilement reconstruite dans le cas du modèle N2MSSM que dans le NMSSM. Cela est naturellement dû aux nouvelles contributions venant du deuxième champ singlet. Enfin, l'incertitude utilisée sur la masse du boson de Higgs reconstruite a été imposée, comme expliqué précédemment, à $\Delta m_H=\pm 7\ \text{GeV}$. Cette valeur est plus importante que l'incertitude théorique couramment admise qui est de $\Delta m_H=\pm 3\ \text{GeV}$. Cet écart s'explique par le temps de calcul beaucoup plus important pour reconstruire une masse de boson de Higgs avec un écart autorisé de $3\ \text{GeV}$. Un écart sur la masse du boson de Higgs de $\Delta m_H=\pm 3\ \text{GeV}$ ainsi que d'autres observables ($a_{\mu}$, $\Omega_{DM}$, ...) pourraient être utilisées (via l'utilisation d'une grille de calcul) pour une étude plus approfondie. 

\section{Particules à long temps de vie en Supersymétrie}\label{subsec:resume6}
Les particules à long temps de vie existent déjà au sein du modèle standard. Elles sont dues à des symétries légèrement brisées ou à des médiateurs lourds. Il est aussi possible de produire ce type de particules dans des modèles supersymétriques.\medskip

En collaboration avec plusieurs expérimentateurs de l'équipe CMS de l'IPHC, nous avons travaillé sur un projet commun d'étude de quark top déplacé provenant de particules supersymétriques à long temps de vie se désintégrant dans le volume du trajectographe. Le but de cette étude est d'effectuer diverses analyses sur plusieurs modèles supersymétriques en parallèle. Une étude de production de squark stop se désintégrant en quark top et gravitino ($\tilde{t}\rightarrow t\psi_\mu$) a ainsi été effectuée dans le cadre du modèle Supersymétrique MSSM avec un mécanisme de brisure appelé GMSB pour {\it Gauge Mediated Supersymmetric Breaking}. Via ce mécanisme, la masse du gravitino est donné par
\begin{gather}
m_{3/2} \propto \frac{M_{SUSY}^2}{m_p}\nn
\end{gather}
avec $m_p$ la masse de Planck et $M_{SUSY}$ l'échelle de brisure de supersymétrie, plus petite que la masse de Planck. Le gravitino est alors naturellement la LSP (\textit{Lightest Supersymmetric Particle}) dans ces modèles. Il a été supposé que le squark stop correspondait à la seconde particule supersymétrie la plus légère (ou NLSP). \nl

\begin{figure}[H]
\centering
\includegraphics[scale=0.6]{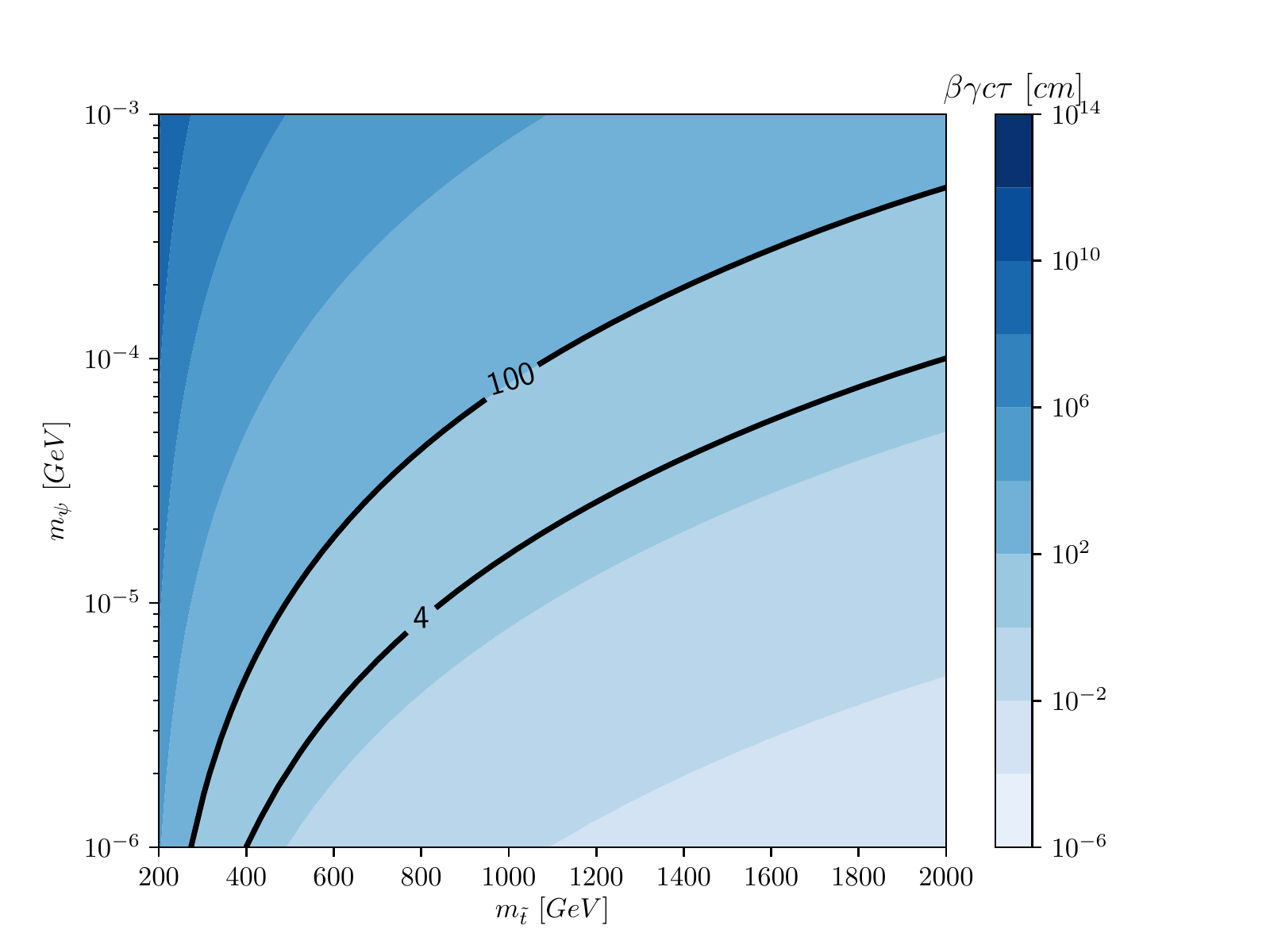}
\caption{Distance de vol corrigée des effets relativistes ($\beta\gamma c\tau$) du squark stop en fonction de la masse du stop et du gravitino. Les deux lignes noires correspondent aux limites géométriques du trajectographe du détecteur CMS.}
\label{fig:flightrésumé}
\end{figure}

Dans un premier temps, l'évolution de la section efficace de production de pairs de squark stop au LHC ($\sqrt{s}=14\ \text{TeV}$) a été générée grâce au programme \sloppy \textsc{MadGraph\_aMC@NLO} afin de déterminer les valeurs possibles de masse de stop. En imposant un nombre suffisamment important d'évènements pour une luminosité intégrée de $\mathcal{L}\approx 300\ \text{fb}^{-1}$, il a été décidé d'étudier la désintégration du stop avec $m_{\tilde{t}}\in[1\ \text{TeV}, 1.4\ \text{TeV}]$. 
La largeur de désintégration du processus $\tilde{t}_i\rightarrow t\psi_{\mu}$ a été calculée afin de déterminer le temps de vie du squark stop : 
\begin{eqnarray}
\Gamma (\tilde{t}_i\rightarrow \psi_{\mu}t ) &=& \frac{1}{48\pi M_p^2m_{3/2}^2m_{\tilde{t}}^3}\Big((-m_{3/2}^2+m_t^2+m_{\tilde{t}}^2)^2-4m_t^2m_{\tilde{t}}^2\Big)^{3/2}\nonumber \\
&& \times (m_{\tilde{t}}^2-m_{3/2}^2-m_t^2 +2\sin 2\theta m_tm_{3/2}) \nn
\end{eqnarray}
avec $m_{3/2}$, la masse de gravitino, $\theta$ l'angle de mélange de la matrice de masse des stops, $M_p=m_p/\sqrt{8\pi}$ et $m_t$ la masse du quark top. La distance de vol du squark stop peut alors être calculée grâce à la relation 
\begin{gather}
c\tau = \frac{\hbar c}{\Gamma}\ . \nn
\end{gather}
L'angle de mélange ne modifiant peu la cinématique du processus, il a été supposé que $\theta=0$. En simulant des évènements de collisions (notamment via le programme \textsc{MadGraph\_aMC@NLO} \cite{madgraph}), on peut alors déterminer le facteur correctif $\beta\gamma$ prenant 	en compte les effets relativistes et ainsi en déterminer la distance de vol du squark stop dans le référentiel du laboratoire $\beta\gamma c\tau$. Nous avons pu alors déterminer l'évolution de $\beta\gamma c\tau$ en fonction de la masse de stop $m_{\tilde{t}}$ et de gravitino $m_{3/2}$ (voir \autoref{fig:flightrésumé}).\medskip
Cette étude à montré qu'il existe des valeurs de masses de squark stop et de gravitino permettant d'obtenir un squark stop se désintégrant dans le trajectographe du détecteur CMS (entre $4\ \text{cm}$ et $100\ \text{cm}$). Des benchmarks $(m_{\tilde{t}}, m_{3/2})$ ont ainsi pu être définis correspondant à plusieurs scénarios possibles : une faible (haute) masse de stop correspondant à une haute (faible) section efficace de production de paire de stop et une distance de vol $\beta\gamma c\tau \in [10\ \text{cm}, 30\ \text{cm}, 50\ \text{cm}, 70\ \text{cm}]$ correspondant à une désintégration du squark stop dans les premières ou dernières couches du trajectographe. Un total de 8 différentes configurations ont été définies avec $m_{\tilde{t}}=\{1\ \text{TeV}, 1.4\ \text{TeV}\}$. \medskip

\begin{figure}[H]
    \centering
      \includegraphics[width=.7\linewidth]{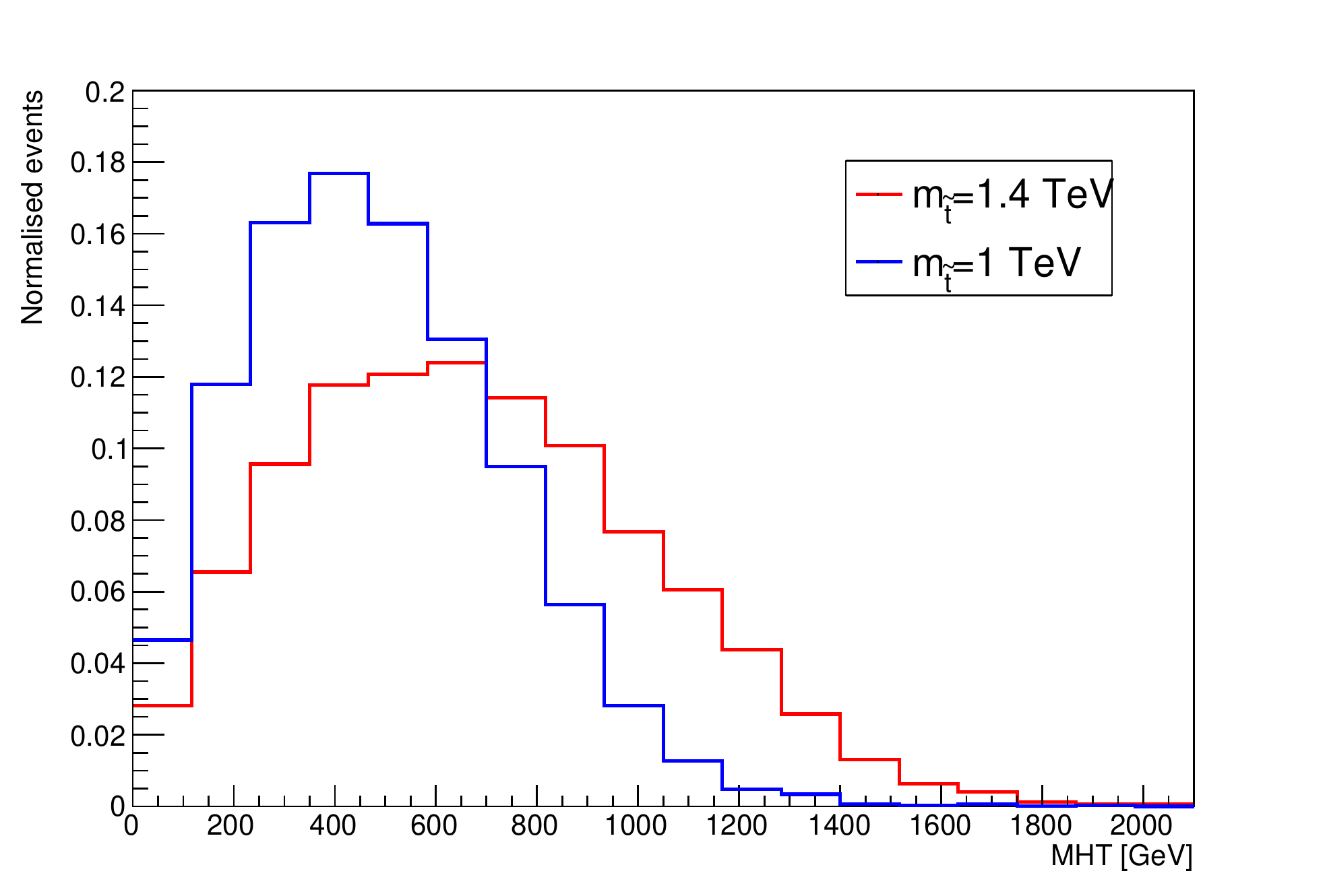}
    \caption{Distributions de $MHT$ pour $m_{\tilde{t}}=1\ \text{TeV}$ et $m_{\tilde{t}}=1.4\ \text{TeV}$ (le choix de $m_{3/2}$ de modifie pas la cinématique).}
    \label{fig:mhtR}
\end{figure}

Les collisions proton-proton ont été simulées grâce aux programmes \textsc{MadGraph\_aMC@NLO} et \textsc{Pythia} en prenant en compte la R-hadronisation. Les distributions des observables liées à la cinématique des évènements ont étés obtenus à l'aide de l'outil \textsc{Madanalysis5}. Le gravitino étant très léger ($m_{3/2}\sim 10^{-5}\ \text{GeV}$), les variations de $m_{3/2}$ au sein des benchmarks ne modifie pas les distributions. On se retrouve alors avec uniquement deux distrbutions correspondant aux deux valeurs possibles de $m_{\tilde{t}}$. Ces distributions ont mis en évidence les potentielles variables à implémenter dans le système de déclenchement des expériences CMS et ATLAS. Par exemple, l'énergie transverse manquante, ou $MET$ (voir Figure \ref{fig:mhtR}), définie comme 
\begin{gather}
MET=||\vec{MET}|| = ||-\sum \vec{p}_T{}_{visible}||\nn
\end{gather}
et la $MHT$ (variable équivalente à $MET$ mais uniquement pour les hadrons, voir Figure \ref{fig:mhtR}) semblent être des variables intéressantes pour le système de déclenchement.
\begin{figure}[H]
    \centering
      \includegraphics[width=.7\linewidth]{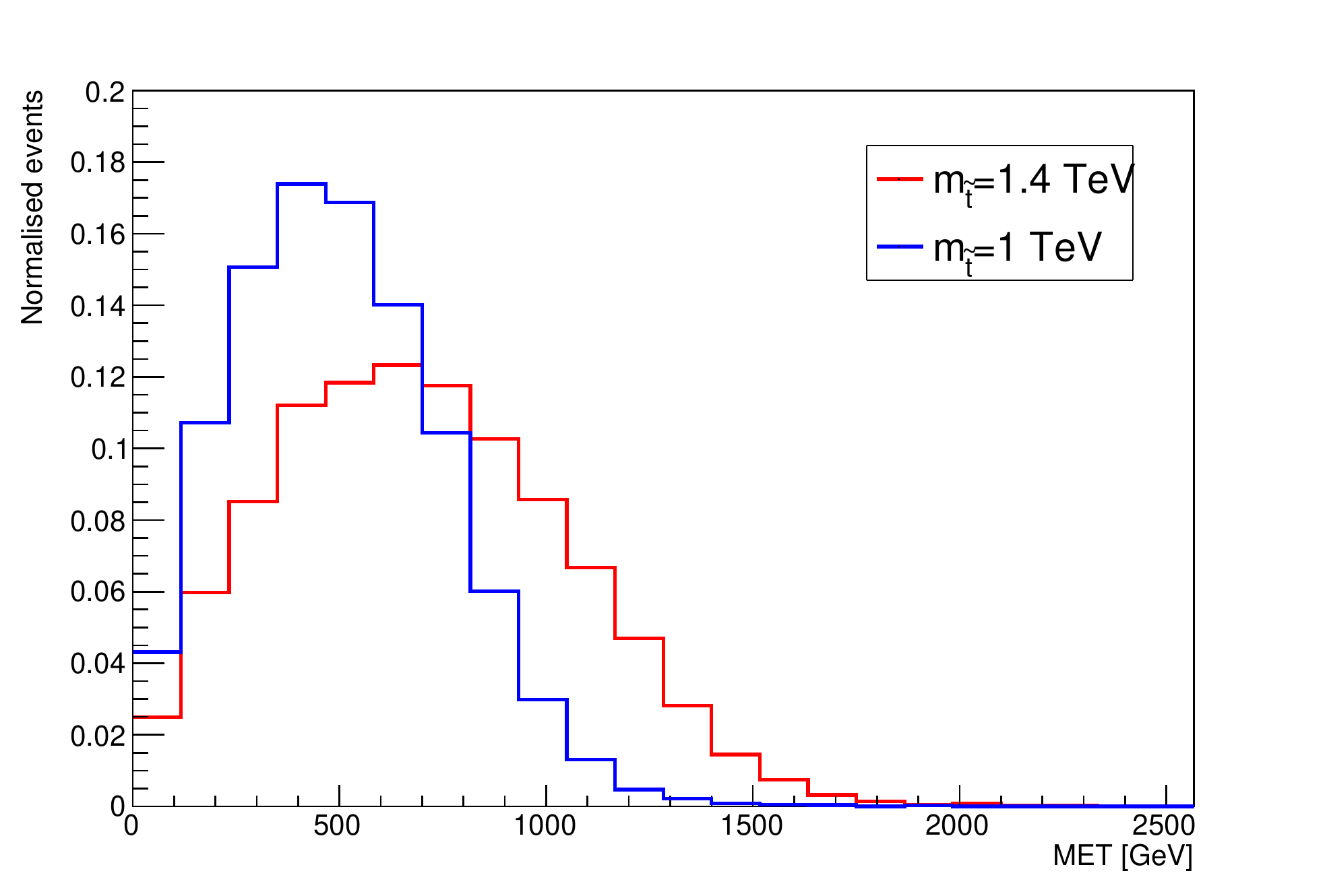}
    \caption{Distributions de $MET$ pour $m_{\tilde{t}}=1\ \text{TeV}$ et $m_{\tilde{t}}=1.4\ \text{TeV}$ (le choix de $m_{3/2}$ de modifie pas la cinématique).}
    \label{fig:metR}
\end{figure}
Cette étude pourra être prise en compte par les expérimentateurs des expériences CMS et ATLAS afin d'étudier la sensibilité de ces expériences à la signature de quarks top déplacés.

\chapter{Introduction}
The Standard Model of particle physics is the theoretical framework describing elementary particles and their interactions (except for gravitational interaction). It is based on quantum field theory (QFT) and symmetry principles linked to space-time symmetries (special relativity) and internal symmetries (gauge invariance). Gauge theories allow us to express those internal symmetries in terms of the $SU(3)_C\times SU(2)_L\times U(1)_Y$ gauge group, i.e. the $SU(3)_C$ group for strong interactions and the $SU(2)_L\times U(1)_Y$ group for electroweak interactions, \textit{i.e.}, the unification of the weak and electromagnetic interactions. The experimental discovery of the Higgs boson by the CMS \cite{cmsH} and ATLAS \cite{atlasH} collaboration in 2012 at CERN (Geneva) is often considered the last experimental evidence of the Standard Model. With the Brout-Englert-Higgs mechanism, the Higgs boson spontaneously breaks the electroweak sector by acquiring a non-zero vacuum expectation value (or \textit{v.e.v}) $\lag h \rag$:
\begin{eqnarray}
SU(2)_L\times U(1)_Y\underset{ \left< h \right>\ne 0}{\rightarrow}  U(1)_{e.m.} \nn
\end{eqnarray}
and generates mass terms for fermions and gauge bosons.\medskip

This model is, still nowadays, one of the most accurate physical frameworks. Unfortunately, several hints show that the Standard Model cannot be the fundamental theory for understanding high energy scale physics. For example, quadratic divergences emerge when evaluating radiative corrections to the Higgs boson, leading to the hierarchy problem between the Higgs boson mass of $125\ \mathrm{GeV}$ and the Planck mass $m_p=10^{19}\ \mathrm{GeV}$. A fine-tuning on the parameters is then applied to fit the experimental measurement, which turn the hierarchy problem in a problem of naturality. Moreover, current astrophysical and cosmological measurements show that the Standard Model only explains a few per cent of the universe, the others coming from dark matter and dark energy. In addition, gravitational interactions are not taken into account in this framework. The physicist community then tries to incorporate the Standard Model into a more fundamental and more accurate model. There are several approaches. One of them emerges from symmetry principles.\medskip

Supersymmetry \cite{martin_supersymmetry_1998}\cite{book_susy} and supergravity \cite{review}\cite{WessBagger}\cite{book_sugra} are natural extensions of the Standard Model, extending in a non-trivial way the Poincaré algebra. In those theories, a new symmetry between fermions and bosons is considered, leading to a richer spectrum of particles and solving several problems of the Standard Model. However, supersymmetry imposes equality between the masses of particles in the same supermultiplet, which is in total contradiction with the experimental measurements. This problem can be solved by introducing mechanisms that explicitly break this symmetry and introduce \textit{soft} supersymmetric breaking terms (\textit{i.e.}, terms generating logarithmic divergences) in the Lagrangian. Several mechanisms already exist introducing interactions between the classical fields and fields from a new sector (called the hidden sector) at high energy. One mechanism is introduced in the context of supergravity. Supergravity breaking in the hidden sector is mediated to the visible sector (through gravitational effects), leading to supersymmetry breaking at low energy. This mechanism was studied in the eighties by Soni \& Weldon \cite{soni_analysis_1983}, who classified the possible forms of the two fundamental functions in supergravity: the superpotential and the Kähler potential. Those solutions are the source of all the current studies of supersymmetry with \textit{Gravity Mediated Supersymmetry Breaking}.\medskip

In a recent article \cite{moultaka_low_2018}, the classification obtained by Soni \& Weldon has been found to be incomplete. New solutions, introducing a new type of superfield (named \textit{hybrid}) with characteristics of the hidden and visible sectors were found. Another specificity of these solutions is that new interactions generate in the Lagrangian new \textit{hard}-breaking terms (\textit{i.e.}, terms introducing quadratic divergences) parametrically suppressed by an intermediate energy scale lower than the Planck mass. These new terms interact at tree-level and loop-levels with the visible sector and may explain the non-detection of supersymmetric particles at colliders experiments and potentially reduce the fine-tuning of the Higgs boson mass. Moreover, the structure of the superpotential imposes at least two new hybrid superfields to couple with the visible sector, corresponding then to a two-singlet extension-like of the MSSM with, however, some specific features. \medskip

In the first chapter of this manuscript, the notion of symmetries in particle physics is discussed. From the Haag–Łopuszański–Sohnius theorem, supersymmetry is introduced and the Lagrangian of the $N=1$ \& $D=4$ supergravity theory is constructed following geometrical principles. The link between supergravity and supersymmetry at low energy is pointed out and the solutions of Soni \& Weldon \cite{soni_analysis_1983} for the \textit{Gravity Mediated Supersymmetry Breaking} mechanism are given. Some supersymmetric models are introduced such as the MSSM (for \textit{Minimal Supersymmetric Standard Model}) and the N2MSSM, new two-singlet-extension of the MSSM.\medskip

Chapter 2 is devoted to the analysis of the new solutions (called NSW for Non-Soni-Weldon) leading to supersymmetry breaking through gravitational interactions. After giving some generalities concerning these solutions, we show, using the Coleman Weinberg potential, that the new supersymmetric breaking terms generate new quadratic divergences and new contribution to the mass matrices. A specific non-canonical model, deduced from the NSW solutions (the S2MSSM) is introduced. In this model, two superfields $S^p$ ($p=1,2$) from the hybrid sector are coupled to the visible sector in the superpotential. The full potential at low energy is calculated. Finally, an investigation of the order of magnitude of the radiative corrections on the Higgs boson mass is done in a simplified model. We identify the regions of the parameters space which help to reproduce a Higgs boson mass near $125\ \text{GeV}$.\medskip

A preliminary analysis of the N2MSSM is done in Chapter 3. This model is introduced for several reasons. Firstly, this model have some similarities with the S2MSSM in the context of classical Soni-Weldon solutions because it contains two singlet superfields. A comparison between those two models can be interesting to highlight the phenomenological differences between Soni-Weldon and Non-Soni-Weldon solutions. Moreover, the addition of a second singlet may potentially resolve some difficulties of the MSSM and the NMSSM, such as the fine-tuning of the parameters to reconstruct an acceptable Higgs boson mass or to get a valid electroweak scale. In this manuscript, we will mainly analyse the differences between the N2MSSM and the NMSSM. The construction of a spectrum generator for the N2MSSM (with the programs \textsc{SARAH} and \textsc{SPheno}) is done and a Markov-Chain Monte-Carlo algorithm is implemented for the scan of the parameters space. Several constraints on the mass of the Higgs boson and on the gluino are imposed. The difference of fine-tuning of the electroweak-scale is also studied. \medskip

In this last chapter (decoupled from the NSW and N2MSSM study), we analyse the relevance of possible displaced-top signatures at the LHC. We start our investigation by two phenomenological analysis based on two simple supersymmetric models. The first one is a decay of a stop squark to a top quark and a LSP neutralino (for \textit{Lightest Supersymmetric Particle}) $\tilde{t}\rightarrow t\chi^0_1$ in the MSSM with \textit{Gravity Mediated Supersymmetry Breaking} mechanism. The second is a stop squark decaying into a top quark and a LSP gravitino $\tilde{t}\rightarrow t\psi_\mu$ in the MSSM following a \textit{Gauge Mediated Supersymmetry Breaking} mechanism (or GMSB) where the gravitino is naturally the LSP. Several tools are used such as \textsc{MadGraph\_aMC@NLO} for the generation of events, \textsc{Pythia} for the simulation of hadronisation or \textsc{Madanalysis5} for the data analysis. This phenomenological analysis will allow to define several benchmarks and so discuss on the possible trigger to apply.\medskip

A conclusion and a discussion of perspective are given.  

\label{intro}

\chapter{Four dimensional $N=1$ Supergravity}\label{sec:sugra}
The Standard Model of particle physics, which describes all the fundamental particles and their interactions, is a huge theoretical and experimental success for the physicist community. Strongly based on space-time and internal symmetries, it has been validated through years thanks to the experimental collaborations. The last building block of this model, the Higgs boson, has been successfully discovered by the CMS and ATLAS experiments at the LHC \cite{cmsH}\cite{atlasH}. \medskip

However, it exists many theoretical and experimental observations that are still incoherent in the Standard Model context. Those issues could be explained if the Standard Model is only a low energy limit of a more fundamental theory. One possible extension of the Standard Model is to add to the known symmetries a new global symmetry (transformation parameters are independent of the space-time coordinates) between fermions and bosons. Those types of models are called supersymmetry. Extending this transformation locally leads naturally to a non-renormalisable theory including gravitation called supergravity.\medskip

In this section, some general features of $N=1$ supergravity in four dimensions will be given. By analysing the possible symmetries of a theory with bosonic and fermionic transformations, a structure called $N=1$ superalgebra will naturally emerge. Following this new algebraic structure, we will construct the $N=1$ supergravity in four dimensions using appropriate geometrical quantities. We will then define fields coming from the study of curved manifolds called the supervierbein $E_{M}{}^{\tilde{M}}$ and the superconnection $\Omega_{\tilde{M}N}{}^{Q}$. All the various steps of calculation of the action will be presented. We will also introduce the three fundamental functions in supersymmetry and supergravity, namely the Kähler potential $K$, the superpotential $W$ and the gauge kinetic function $h_{ab}$. 

Historically, supersymmetry was the first theory studied, while supergravity was introduced as a gauge theory of supersymmetry. In this manuscript, we will not follow this approach. After constructing the $N=1$ supergravity in four dimensions, we will show that supersymmetry will naturally emerge as a low energy limit of supergravity. However, since those theories have theoretical predictions which are non-compatible with the actual measurements, they must then be broken symmetries. The various supersymmetry breaking mechanisms, highlighting the \textit{Gravity Mediated Supersymmetry Breaking} mechanism, will be described. Those mechanisms imply the existence of a new superfield sector, the \textit{hidden} sector, where supergravity is broken. Choosing the Kähler potential $K$, the superpotential $W$ and the gauge kinetic function $h_{ab}$, models of supersymmetry such as the \textit{Minimal Supersymmetric Standard Model} (or MSSM) can be constructed. A new model called N2MSSM, which will be studied in Chapter \ref{chap:N2MSSM}, will also be defined.

\section{Symmetry in particle physics}\label{sec:sym}
It is known that symmetry principles are very powerful tools for classifying all the particles and their interactions. In this section, the possible internal symmetries of the action will be studied. Supposing two types of transformations (bosonic and fermionic), we will obtain two different algebraic structures: Lie algebra and Lie superalgebra. Focusing on the second, we will define $N=1$ supersymmetry algebra, which is central to construct the supergravity action.

\subsection{Lie algebras and the Coleman-Mandula theorem}\label{subsec:CM}
In quantum field theory, the S-matrix (or scattering matrix) is the key to determining the probabilities of interactions between particles. Since a trivial S-matrix means a non-interacting theory, a constraint on this non-triviality must be imposed. This leads to a restriction on the possible symmetries of the theory.\medskip

It exists two types of particles in particle physics, depending on their spin: bosons $\phi^i(x)$ with integer spin and fermions $\psi^a(x)$ with half-integer spin where $\phi^i(x)$ and $\psi^a(x)$ are relativistic fields depending on the space-time coordinates. Their dynamics are imposed through the action $\mathcal{S}$ of the theory:\medskip
\begin{eqnarray}
\mathcal{S} &=& \int dtdx^3 \mathcal{L}(\phi^i,\partial_\mu\phi^i,\psi^a,\partial_\nu\psi^a)\label{eq:actionS}
\end{eqnarray}
(where $\mathcal{L}(\phi^i,\partial_\mu\phi^i,\psi^a,\partial_\nu\psi^a)$ is the Lagrangian) and the principle of least action.\medskip

In the context of quantum field theory, the relativistic fields $\phi^i$ and $\psi^a$ can be quantised using the canonical quantisation\footnote{Note that the gauge bosons must be treated with care.}, introducing then relativistic quantum fields (we will not follow the quantisation using path-integral method). Using the Spin-Statistic theorem \cite{spinstat}, bosonic fields $\phi^i$ are then quantised using commutation relations, whereas fermionic fields $\psi^a$ are quantised using anti-commutation relations. \medskip

This framework is well-suited to analyse the symmetries of the action. A symmetry is a transformation $\mathcal{T}$ of the action \autoref{eq:actionS} such as the action if invariant, \textit{i.e.} $\delta_\mathcal{T} \mathcal{S} = 0$ or can be written as a total derivative. Emmy Noether established a theorem relating symmetry of the action with conserved quantities \cite{Noether}. For any continuous internal symmetry of the action $\mathcal{S}$, which infinitesimally transforms the fields as:
\begin{eqnarray}
\phi^i\rightarrow \phi^i + \delta_S \phi^i \quad, \quad \psi^a \rightarrow \psi^a + \delta_S \psi^a \ ,\nonumber
\end{eqnarray}
we can associate a conserved charge $\mathcal{Q}_S$ that can be written in terms of the two fields $\phi^i$ and $\psi^j$:
\begin{eqnarray}
\mathcal{Q}_S = \int \left( \Pi_i\delta_S\phi^i + \Xi_a\delta_S\psi^a \right)d^3x\nn
\end{eqnarray}
where $\Pi$ and $\Xi$ are the conjugate momenta of respectively $\phi$ and $\psi$. After quantisation and using the Spin-Statistics theorem, the charges $\mathcal{Q}_S$ appear as generators of the symmetry:
\begin{eqnarray}
\left[ \phi^i(x), \mathcal{Q}_S \right] = i\delta_S\phi^i(x),\qquad \left[ \psi^a(x), \mathcal{Q}_S \right] = i\delta_S\psi^a(x)\ .\nn
\end{eqnarray}
Since the charges $\mathcal{Q}_S$ are written in terms of bosonic and fermionic fields $\phi^i$ and $\psi^a$, we can then introduce \textit{a priori} bosonic and fermionic charges. From this statement, two types of symmetry naturally appear: bosonic and fermionic symmetries. \medskip

In the first case, we assume that the action is invariant under bosonic transformations:
\begin{eqnarray}
\delta_\mathcal{A} \phi^i = (B_\mathcal{A}^1)^i{}_j\phi^j,\qquad \delta_\mathcal{A} \psi^a = (B_\mathcal{A}^2)^a{}_b\psi^b\ ,\label{eq:bosonsym}
\end{eqnarray}
(with $B_\mathcal{A}^1$ and $B_\mathcal{A}^2$, the generators of the bosonic symmetry in the considered representation). The conserved charges $\mathcal{B_A}$ generate then a Lie algebra:
\begin{gather}
\left[ \mathcal{B_A}, \mathcal{B_B} \right] = i f_{\mathcal{AB}}{}^\mathcal{C} \mathcal{B_C}.\nn
\end{gather}
Note that the Jacobi identity is automatically valid since we consider a given representation and the Lie bracket is the classical matrix commutator. However, those results do not take into account the requirement of a non-trivial S-matrix. We will see that this requirement constrains the possible algebras $\mathfrak{g}$. Assuming the following hypothesis:
\begin{itemize}
\item[1)] a S-matrix constructed from a quantum field theory in four dimensions is considered,
\item[2)] the theory contains a discrete and finite spectrum with all particles in positive energy representation of Poincaré algebra $\mathfrak{Iso}(1,3)$,
\item[3)] there exists an energy gap between the vacuum state and the one-particle state,
\item[4)] the symmetry group is a Lie group $G$ associated to a Lie-algebra $\mathfrak{g}$,
\end{itemize}   
the Coleman-Mandula theorem\footnote{The points 1)-4) of the Coleman-Mandula and 1)-4') of the Haag–Łopuszański–Sohnius theorem (see \autoref{subsec:supalgebra}) are the validity conditions of these theorems. Some conditions can be relaxed, leading to new sorts of symmetries (leading, for example, to higher-spin or superconformal algebra). This discussion is out of the scope of this thesis.} \cite{ColemanMandula} states that internal and space-time symmetries combine only in a trivial way. The algebra $\mathfrak{g}$ takes the form:
\begin{eqnarray}
\mathfrak{g} = \mathfrak{Iso}(1,3)\times \mathfrak{g}_C \nn
\end{eqnarray}
where $\mathfrak{Iso}(1,3)$ is the Poincaré algebra and $\mathfrak{g}_C$ is a compact Lie algebra associated to the internal symmetries. The Poincaré algebra is defined as $\mathfrak{Iso}(1,3)=\mathrm{Span}\big\{ P^\mu, J^{\mu\nu},\,\mu,\nu=0,\dots ,3 \big\}$, with $J^{\mu\nu}$ the generators of Lorentz transformations and $P^{\mu}$ the generators of translations in space-time. This result is central for the construction of the Standard Model, where $\mathfrak{g}_C =\mathfrak{su}(3)\times\mathfrak{su}(2)\times\mathfrak{u}(1)$.

\subsection{Lie superalgebras and the Haag–Łopuszański–Sohnius theorem}\label{subsec:supalgebra}

In this section, we consider a new type of algebra associated to fermionic transformations. In addition to the bosonic symmetries \autoref{eq:bosonsym}, fermionic symmetries can be added to the action \autoref{eq:actionS} such as:
\begin{equation}
\delta_\mathcal{I} \phi^i = (F_\mathcal{I})^i{}_a\psi^a\quad , \quad \delta_\mathcal{I} \psi^a = (F_\mathcal{I})^a{}_i\phi^i\ .
\label{ferm:trsfo}
\end{equation}
The specific characteristic of such symmetries is that it changes the nature of particles, \textit{i.e.}, a boson is transformed into a fermion and \textit{vice versa}. We see that the introduction of fermionic generators also imposes bosonic generators since the composition of two fermionic transformations leads to a bosonic transformation. \medskip
Denoting fermionic charges as $\mathcal{F_I}$, the following structure appears:
\begin{eqnarray}
\Big[ \mathcal{B_A}, \mathcal{B_B} \Big] = i f_{\mathcal{AB}}{}^\mathcal{C} \mathcal{B_C}\quad , \quad \Big\{ \mathcal{F_I}, \mathcal{F_J} \Big\} = Q_{\mathcal{IJ}}{}^\mathcal{K} \mathcal{F_K}\quad , \quad \Big[ \mathcal{B_A},\mathcal{F_I} \Big] = R_{\mathcal{AI}}{}^\mathcal{J}\mathcal{F_J} \ ,  \label{alg3}
\end{eqnarray}
with the various Bianchi identities (trivially satisfied): 
\begin{eqnarray}
\Big[\big[\mathcal{B_A},\mathcal{B_B}  \big],\mathcal{B_C}  \Big] + \Big[\big[\mathcal{B_B},\mathcal{B_C}  \big],\mathcal{B_A}  \Big] +\Big[\big[\mathcal{B_C},\mathcal{B_A}  \big],\mathcal{B_B}  \Big] = 0 \ ,\label{eq:idbianchi1}\\
\Big[\big[\mathcal{B_A},\mathcal{B_B}  \big],\mathcal{F_I}  \Big] + \Big[\big[\mathcal{B_B},\mathcal{F_I}  \big],\mathcal{B_A}  \Big] +\Big[\big[\mathcal{F_I},\mathcal{B_A}  \big],\mathcal{B_B}  \Big] = 0 \ ,\label{eq:idbianchi2}\\
\Big[\big[\mathcal{B_A},\mathcal{F_I}  \big],\mathcal{F_J}  \Big] + \Big[\big\{\mathcal{F_I},\mathcal{F_J}  \big\},\mathcal{B_A}  \Big] -\Big[\big[\mathcal{F_J},\mathcal{B_A}  \big],\mathcal{F_I}  \Big] = 0 \ ,\label{eq:idbianchi4}\\
\Big\{\big\{\mathcal{F_I},\mathcal{F_J}  \big\},\mathcal{F_K}  \Big\} + \Big\{\big\{\mathcal{F_J},\mathcal{F_K}  \big\},\mathcal{F_I}  \Big\} +\Big\{\big\{\mathcal{F_K},\mathcal{F_I}  \big\},\mathcal{F_J}  \Big\} = 0 \ .\label{eq:idbianchi3}
\end{eqnarray}
From the first relation of \autoref{alg3} and the Bianchi identity \autoref{eq:idbianchi1}, we recover the Lie algebra structure of the previous section (\ref{subsec:CM}). We will denote this algebra as:
\begin{eqnarray}
\mathfrak{g}_0 &=& \mathrm{Span}\Big\{  \mathcal{B_A},\ \mathcal{A}=1,\dots,m\Big\}\ .\nn
\end{eqnarray}
Using the second equality of \autoref{alg3} combined with \autoref{eq:idbianchi2}, the vectorial space $\mathfrak{g}_1$ generated by the fermionic charges $\mathcal{F_I}$:
\begin{eqnarray}
\mathfrak{g}_1 = \mathrm{Span}\Big\{ \mathcal{F_I},\ \mathcal{I}=1,\dots,n \Big\}\nn
\end{eqnarray} 
form a $\mathfrak{g}_0$-representation. All those relations combined with the relations \ref{eq:idbianchi4} and \ref{eq:idbianchi3} generate the $\mathbb{Z}_2$-graded algebra $\mathfrak{g}$ called Lie superalgebra:
\begin{eqnarray}
\mathfrak{g}=\mathfrak{g}_0\oplus\mathfrak{g}_1=\Big\{ \mathcal{B_A},\ \mathcal{A}=1,\dots,m \Big\} \oplus \Big\{ \mathcal{F_I},\ \mathcal{I}=1,\dots,n  \Big\}\ .\nn
\end{eqnarray}
We now take into account the non-triviality of the S-matrix and assume the same hypothesis as in Section \ref{subsec:CM}, with 4) substituted with:
\begin{itemize}
\item[4')] the symmetry group is a Lie supergroup $G$ associated to a Lie superalgebra $\mathfrak{g}$,
\end{itemize}
The Haag–Łopuszański–Sohnius theorem \cite{HLS} states that $\mathfrak{g}$ must be a supersymmetric extension of the Poincaré algebra (called Poincaré superalgebra) where the supersymmetric extension is defined using $N$ supercharges (Majorana spinors) ($0 < N \leq 8$). \medskip

We will focus on the simplest supersymmetric extension describing $N=1$ supersymmetry theory. The superalgebra is generated by:
\begin{eqnarray}
\mathfrak{g}=\mathfrak{g}_0\oplus\mathfrak{g}_1 = \big\{ \mathfrak{Iso}(1,3)\times \mathfrak{g}_C \big\}\oplus \mathrm{Span}\big\{ Q_\alpha , \overline{Q}_{\dot{\alpha}},\alpha,\dot{\alpha}=1,2 \big\} \label{eq:supalg}
\end{eqnarray} 
with $Q_\alpha$, $\overline{Q}_{\dot{\alpha}}$ (using Van der Waerden notation), the left \& right-handed spinors corresponding to a Majorana spinor $(Q_\alpha)^{\dagger} = \bar{Q}_{\dot{\alpha}}$. The different commutation and anticommutation relations leading to the superalgebra \autoref{eq:supalg} are:
\begin{gather}
\label{susy:algebra}
\Big\{ Q_\alpha, \bar{Q}_{\dot{\alpha}} \Big\} = 2\sigma^\mu{}_{\alpha\dot{\alpha}} P_\mu , \quad \Big\{ Q_\alpha, Q_\beta \Big\} = 0 , \quad \Big\{ \bar{Q}^{\dot{\alpha}} ,\bar{Q}^{\dot{\beta}} \Big\} = 0 \ ,\\
\Big[ Q_\alpha , P_\mu \Big] = 0, \qquad \Big[ \overline{Q}^{\dot{\alpha}} , P_\mu \Big] = 0\nn\ ,\\
\Big[ Q_\alpha , J^{\mu\nu} \Big] = i\left( \sigma^{\mu\nu} \right)_\alpha{}^\beta Q_\beta ,\qquad \Big[ \overline{Q}^{\dot{\alpha}} , J^{\mu\nu} \Big] = i\left( \overline{\sigma}^{\mu\nu} \right)^{\dot{\alpha}}{}_{\dot{\beta}} \overline{Q}^{\dot{\beta}}\ , \nn
\end{gather}
with the various $\sigma$ matrices introduced in Appendix \ref{app:conventions}. The two Casimir operators $C_i$ associated to the superalgebra \autoref{eq:supalg} are: 
\begin{gather}
C_1 = P^{\mu}P_{\mu} , \\
C_2 =  W^{\mu\nu}W_{\mu\nu} \quad , \quad W_{\mu\nu}=\big( W_{\mu} - \frac14 \bar{Q}\bar{\sigma}_{\mu}Q\big)P_{\nu}-\big(  W_{\nu} - \frac14 \bar{Q}\bar{\sigma}_{\nu}Q \big)P_{\mu} \nn
\end{gather} 
(with $W_{\mu}=\frac12 \epsilon_{\mu\nu\rho\sigma}P^\nu J^{\rho\sigma}$ the Pauli-Lubanski operator). Using the proper values of $C_i$, it is then possible to construct the irreducible representations of $\mathfrak{g}$ called supermultiplets. For more technical details on superalgebra and symmetries in particle physics, see \cite{book_susy} and \cite{groupmichel} respectively.

\section{Curved Superspace}
In Section \ref{sec:sym}, we have introduced the $N=1$ supersymmetry algebra. We want now to construct a viable physical theory based on local supersymmetry, leading to supergravity. \medskip

We will not adopt the historical approach of supergravity, \textit{i.e.}, constructing supergravity theory from the supersymmetry framework. There exist several approaches for calculations in supergravity (see, for example \cite{freedman_van_proeyen}). The superspace approach will be used in this manuscript (see \cite{superspace}\cite{WessBagger}\cite{book_sugra}\cite{review}), following general relativity as a guideline. \medskip

In general relativity, the equivalence principle allows to associate to a curved frame a local tangent frame where gravitation is cancelled. In the curved frame, the theory is invariant under diffeomorphisms where the coordinates in the tangent space transform with Lorentz transformations. There is, however, an issue for describing spinors with the classical formulation of general relativity. This problem can be solved using the tetrad formalism, introducing two new fields called the vierbein $e_{\mu}{}^{\tilde{\mu}}$ and the spin connection $\omega_{\tilde{\mu}\nu}{}^{\rho}$.\medskip

The construction of supergravity theory can be done by adopting the same approach using an appropriate space called superspace, describing superfields. A superspace is constructed by extending the usual Minkowsky space using fermionic coordinates $(\theta^\alpha, \bar{\theta}_{\dot{\alpha}} )$. A point of the superspace $\mathfrak{z}$ is then defined by the coordinates $\mathfrak{z}^{M} = (x^\mu, \theta^\alpha , \bar{\theta}_{\dot{\alpha}} )$, and functions of the superspace are called superfields $\Phi(x,\theta,\bar{\theta})$. The theory is then invariant by superdiffeomorphisms in the curved frame where coordinates transform under Lorentz transformations in the tangent flat superspace.

\subsection{Key objects in curved superspace: supervierbein \& superconnection}\label{sec:EOmega}
In order to correctly derive the $N=1$ supergravity action in four dimensions, we will follow the tetrad (or Cartan) formalism and present the supervierbein $E_{M}{}^{\tilde{M}}$ and the superconnection $\Omega_{\tilde{M}NP}$.\medskip

We define a point in the flat tangent superspace $\mathfrak{z}^M = (x^\mu , \theta^\alpha , \bar{\theta}_{\dot{\alpha}})$ where there is invariance under local Lorentz transformations:
\begin{eqnarray}
{\mathfrak{z}'}^M = \mathfrak{z}^N\Lambda_N{}^M (\mathfrak{z} ) \Rightarrow
\begin{cases}
{x'}^\mu = x^\nu \Lambda_\nu{}^\mu (\mathfrak{z} )\ ,\\
{\theta '}^\beta = \theta^\alpha \Lambda_\alpha{}^\beta (\mathfrak{z})\ , \\
{\bar{\theta}}_{\dot{\beta}} = {\bar{\theta}}_{\dot{\alpha}} \Lambda^{\dot{\alpha}}{}_{\dot{\beta}} (\mathfrak{z})\ .
  \end{cases} \label{eq:lorentz}
\end{eqnarray} 
The superfields matrix $\Lambda_N{}^M(\mathfrak{z}) = \left( \Lambda_\mu{}^\nu(\mathfrak{z}) , \Lambda_\alpha{}^\beta(\mathfrak{z}) , \Lambda^{\dot{\alpha}}{}_{\dot{\beta}}(\mathfrak{z}) \right)$, corresponding respectively to the vector, left-handed and right-handed representations, have a block diagonal structure. This means that the elements with indices of different natures vanish (since Lorentz transformations do not mix indices of different natures). \medskip

Defining a point $\mathfrak{z}^{\tilde{M}}$ in the curved superspace, we have invariance under superdiffeomorphisms:
\begin{eqnarray}
\mathfrak{z'}^{\tilde{M}} = \mathfrak{z}^{\tilde{M}} + \mathfrak{\xi}^{\tilde{M}}(\mathfrak{z})
\Rightarrow
\begin{cases}
{x'}^{\tilde{\mu}} = x^{\tilde{\mu}} + \xi^{\tilde{\mu}} ( \mathfrak{z} )\ , \\
{\theta'}^{\tilde{\alpha}} = {\theta}^{\tilde{\alpha}} + \xi^{\tilde{\alpha}} ( \mathfrak{z} )\ , \\
{\bar{\theta}'}_{\tilde{\dot{\alpha}}} = {\bar{\theta}}_{\tilde{\dot{\alpha}}} + \bar{\xi}_{\tilde{\dot{\alpha}}} ( \mathfrak{z} )\ , \label{eq:superdiffeo}
  \end{cases}
\end{eqnarray}
where $\xi^{\tilde{M}}$ are differentiable functions. The conjugate variables $\partial_{\tilde{M}}$ are also introduced such as:
\begin{eqnarray}
{\partial_{\tilde{M}}} = ({\partial_{\tilde{\mu}}} , {\partial_{\tilde{\alpha}}} , {\bar{\partial}^{\tilde{\dot{\alpha}}}} ) &\Rightarrow& \Big[ {\partial_{\tilde{M}}}, {\mathfrak{z}^{\tilde{N}}} \Big]{}_{|\tilde{M}||\tilde{N}|} = {\delta_{\tilde{M}}}{}^{\tilde{N}}\nn \nonumber
\end{eqnarray}
with $|\mu|=0$, $|\alpha| = |\dot{\alpha}| = 1$ and where the graded bracket is given by:
\begin{eqnarray}
\Big[ \partial_{\tilde{M}} , \mathfrak{z}^{\tilde{N}} \Big]_{|{\tilde{M}}||{\tilde{N}}|} \equiv \partial_{\tilde{M}} \mathfrak{z}^{\tilde{N}} - ( - )^{|{\tilde{M}}||{\tilde{N}}|}\mathfrak{z}^{\tilde{N}}\partial_{\tilde{M}} \ .\label{eq:gradedcomu}
\end{eqnarray}
The relation between the curved and the flat tangent superspace is obtained by introducing a dynamical superfield called \textit{supervierbein} $E_{\tilde{M}}{}^{M}\equiv \frac{\partial \mathfrak{z}^{M}}{\partial \mathfrak{z}^{\tilde{M}}}$. The relation between a superfield expressed in the curved superspace $V^{\tilde{M}}$ and in the tangent flat superspace $V^{M}$ can then be written as:
\begin{equation}
V_{\tilde{M}} = \frac{\partial \mathfrak{z}^{M}}{\partial \mathfrak{z}^{\tilde{M}}} V^M \equiv E_{\tilde{M}}{}^M V_M , \nn
\end{equation}
The inverse supervierbein $E_M{}^{\tilde{M}}$ is also defined such as:
\begin{eqnarray}
V_M = \frac{\partial \mathfrak{z}^{\tilde{M}}}{\partial \mathfrak{z}^{M}} V^{\tilde{M}} \equiv E_M{}^{\tilde{M}} V_{\tilde{M}} ,\nn
\end{eqnarray}
with $E_{\tilde{M}}{}^M E_M{}^{\tilde{N}} = \delta_{\tilde{M}}{}^{\tilde{N}}$ and $E_M{}^{\tilde{M}} E_{\tilde{M}}{}^N = \delta_M{}^N$. See also that in the local tangent frame we get:
\begin{eqnarray}
\partial_N\xi^{M} = E_{N}{}^{\tilde{M}}\partial_{\tilde{M}}\xi^N = E_{N}{}^{\tilde{M}}E_{\tilde{M}}{}^{M}= \delta_{N}{}^M \ .\nn
\end{eqnarray}
As in general relativity, it is then possible to express all objects of the theory in a frame where gravitation is cancelled, and coordinates transform as Lorentz transformations.\medskip

The second dynamical superfield, \textit{i.e.}, the superconnection $\Omega_{\tilde{M}MN}$ is introduced with the symmetry property:
\begin{equation}
\Omega_{\tilde{M}MN} = - (-)^{|M||N|} \Omega_{\tilde{M}NM}\nn
\end{equation}
coming from the Lorentz algebra structure. Since the Lorentz transformations do not mix indices of different natures, the only non-zero elements of the superconnection are $\Omega_{\tilde{M}\mu\nu}$, $\Omega_{\tilde{M}\alpha\beta}$ and $\Omega_{\tilde{M}}{}^{\dot{\alpha}\dot{\beta}}$. Those new variables aim to construct covariant derivatives with respect to Lorentz transformations:
\begin{eqnarray}
\mathcal{D}_{\tilde{M}}X^M = \partial_{\tilde{M}}X^M + (-)^{|\tilde{M}||N|}X^N\Omega_{\tilde{M}N}{}^M\ , & \,  & \mathcal{D}_{\tilde{M}}X_M = \partial_{\tilde{M}}X_M - \Omega_{\tilde{M}M}{}^N X_N \ ,\label{eq:dercov}\\
\mathcal{D}_M X^M = E_M{}^{\tilde{M}}\mathcal{D}_{\tilde{M}}X^M \ , & \,  & \mathcal{D}_N X_M = E_N{}^{\tilde{M}}\mathcal{D}_{\tilde{M}}X_M \ . \nn
\end{eqnarray}
Note that the sign structure of the first relation is obtained using the commutation:
\begin{eqnarray}
\Omega_{\tilde{M}N}{}^{M}X^{N} = (-)^{|N|(|\tilde{M}| + |N| + |M|)}X^{N}\Omega_{\tilde{M}N}{}^{M} = (-)^{|N||\tilde{M}|}X^{N}\Omega_{\tilde{M}N}{}^{M}
\end{eqnarray}
because $M$ and $N$ are of the same nature. From the invariance of the covariant derivative under a local Lorentz transformation, the superconnection transforms as (with a matrix $\Lambda$):
\begin{equation}
\Omega_{\tilde{M}M}{}^N \rightarrow \Omega'_{\tilde{M}M}{}^N = \left( \Lambda^{-1} \right)_M{}^P\Omega_{\tilde{M}P}{}^Q\Lambda_Q{}^N - \left( \Lambda^{-1} \right)_M{}^P \partial_{\tilde{M}}\Lambda_P{}^N\ .\nn
\end{equation}
The structure of the covariant derivative $\mathcal{D}_{M}$ \autoref{eq:dercov} has a lie algebra structure, leading to the definition of the torsion $\mathcal{T}_{MN}{}^{Q}$ and the curvature tensor $R_{MNPQ}$:
\begin{equation}
\left[ \mathcal{D}_M , \mathcal{D}_N \right]_{|M||N|} \equiv \mathcal{T}_{MN}{}^{Q}\mathcal{D}_{Q} - \frac12 R_{MN\underline{\alpha}\underline{\beta}} J^{\underline{\alpha}\underline{\beta}}\label{eq:commder}
\end{equation}
with $\underline{\alpha}=(\alpha,\dot{\alpha})$. Remark that the last term contains only $J_{\alpha\beta}$ and $J^{\dot{\alpha}\dot{\beta}}$ since the generators obey the relations:
\begin{eqnarray}
J_{\mu\nu} &=& \frac14 \left(\bar{\sigma}_{\nu}\right)^{\dot{\alpha}\alpha}\left(\bar{\sigma}_{\mu}\right)^{\dot{\beta}\beta}\big( \epsilon_{\alpha\beta}J_{\dot{\alpha}\dot{\beta}} - \epsilon_{\dot{\alpha}\dot{\beta}}J_{\alpha\beta} \big)\ ,\nn\\
J_{\alpha}{}^{\beta} &=& \left( \sigma^{\nu\mu} \right)_{\alpha}{}^{\beta}J_{\mu\nu}\ ,\nn \\
J^{\dot{\alpha}}{}_{\dot{\beta}} &=& \left( \bar{\sigma}^{\nu\mu} \right)^{\dot{\alpha}}{}_{\dot{\beta}}J_{\mu\nu}\ ,\nn
\end{eqnarray}
thus we only sum over independent generators. 

From the definition of the graded commutator \autoref{eq:gradedcomu}, we have the following relations:
\begin{eqnarray}
\mathcal{T}_{MNP} &=& -(-)^{|M||N|}\mathcal{T}_{NMP}\ , \nn\\
R_{MNPQ} &=& -(-)^{|M||N|} R_{NMPQ}\ , \nn\\
R_{MNPQ} &=& -(-)^{|P||Q|}R_{MNQP}\ .\label{eq:symR}
\end{eqnarray}
Calculating the graded commutator \autoref{eq:commder} using the relation $\mathcal{D}_{M}=E_{M}{}^{\tilde{N}}\mathcal{D}_{\tilde{N}}$, it is possible to find the general form of the supertorsion and supercurvature tensors:
\begin{eqnarray}
\mathcal{T}_{\tilde{P}\tilde{Q}}{}^Q &=& -\mathcal{D}_{\tilde{P}}E_{\tilde{Q}}{}^Q + (-)^{|\tilde{P}||\tilde{Q}|}\mathcal{D}_{\tilde{Q}}E_{\tilde{P}}{}^Q \ ,\label{eq:torsion}\\
R_{\tilde{M}\tilde{N}S}{}^Q &=& \partial_{\tilde{M}}\Omega_{\tilde{N}S}{}^Q-\Omega_{\tilde{M}S}{}^R\Omega_{\tilde{N}R}{}^Q -(-)^{|\tilde{M}||\tilde{N}|}\left( \partial_{\tilde{N}}\Omega_{\tilde{M}S}{}^Q - \Omega_{\tilde{N}S}{}^R\Omega_{\tilde{M}R}{}^S \right)\ ,\nn
\end{eqnarray}
with 
\begin{eqnarray}
R_{\tilde{M}\tilde{N}S}{}^Q &=& (-)^{|\tilde{M}|\left( |\tilde{N}| + |N| \right)}E_{\tilde{M}}{}^M E_{\tilde{N}}{}^N R_{MNS}{}^Q\ , \nn\\
\mathcal{T}_{\tilde{Q}\tilde{S}}{}^S &=& (-)^{|\tilde{P}|\left( |N| + |\tilde{S}|  \right)}E_{\tilde{S}}{}^N E_{\tilde{P}}{}^M \mathcal{T}_{MN}{}^S\ .\nn
\end{eqnarray}
In supersymmetry, those results simplify since the superconnection vanishes:
\begin{equation}
\Omega_{\tilde{M}N}{}^{Q} = 0 \quad \rightarrow \quad \mathcal{D}_{\tilde{M}} = \partial_{\tilde{M}}\ .\nn
\end{equation}
Using known relations in supersymmetry \cite{martin_supersymmetry_1998}\cite{book_susy}, covariant derivatives take the form:
\begin{gather}
\mathcal{D}_\mu = E_\mu{}^{\tilde{M}}\partial_{\tilde{M}} = \partial_\mu \quad , \quad \mathcal{D}_\alpha = E_\alpha{}^{\tilde{M}}\partial_{\tilde{M}} = \partial_\alpha + i\sigma^\mu{}_{\alpha\dot{}\alpha}\bar{\theta}^{\dot{\alpha}}\partial_\mu\ , \nn\\
\bar{\mathcal{D}}^{\dot{\alpha}} = E^{\dot{\alpha}\tilde{M}}\partial_{\tilde{M}} = \bar{\partial}^{\dot{\alpha}}-i\theta_\alpha\bar{\sigma}^\mu{}^{\alpha\dot{\alpha}}\partial_\mu \quad , \quad \Big\{ \mathcal{D}_{\alpha} ,\mathcal{D}_{\dot{\alpha}}\Big\} = -2i\sigma^{\mu}{}_{\alpha\dot{\alpha}}\mathcal{D}_{\mu} \ ,  \nn
\end{gather}
which leads to the supersymmetry algebra. The supervierbein can then be identified through:
\begin{eqnarray}
E_M{}^{\tilde{M}} = \begin{pmatrix}
\delta_\mu{}^{\tilde{\mu}} & 0 & 0 \\
i\delta_\mu{}^{\tilde{\mu}}(\sigma^\mu\bar{\theta})_\alpha & \delta_\alpha{}^{\tilde{\alpha}} & 0 \\
-i\delta_\mu{}^{\tilde{\mu}}(\theta\sigma^\mu)_{\dot{\beta}}\epsilon^{\dot{\alpha}\dot{\beta}} & 0 & \delta^{\dot{\alpha}}{}_{\tilde{\dot{\alpha}}}
\end{pmatrix}\quad , \quad
E_{\tilde{M}}{}^M = \begin{pmatrix}
\delta_{\tilde{\mu}}{}^\mu & 0 & 0 \\
-i\delta_{\tilde{\mu}}{}^\mu ( \sigma^{\tilde{\mu}}\bar{\theta} )_{\tilde{\alpha}} & \delta_{\tilde{\alpha}}{}^\alpha & 0 \\
i\delta_{\tilde{\mu}}{}^\mu ( \theta\sigma^{\tilde{\mu}} )_{\tilde{\dot{\beta}}}\epsilon^{\tilde{\dot{\alpha}}\tilde{\dot{\beta}}} & 0 & \delta^{\tilde{\dot{\alpha}}}{}_\alpha  
\end{pmatrix}\ .\nn
\end{eqnarray}

\subsection{Bianchi identities: Constraints on the torsion components}\label{sec:bianchi}
As we have seen in \autoref{sec:EOmega} (\autoref{eq:commder}), the graded commutators of covariant derivatives involve supertorsion and supercurvature tensors. Therefore, the components of such tensors are not independent since the covariant derivatives satisfy the Bianchi identities [\autoref{eq:idbianchi1},\autoref{eq:idbianchi2},\autoref{eq:idbianchi3}]:
\begin{eqnarray}
0 &=& (-)^{|M_1||M_3|}\left[ \mathcal{D}_{M_1},\left[ \mathcal{D}_{M_2},\mathcal{D}_{M_3} \right]_{|M_2||M_3|} \right]_{|M_1|\left( |M_2| + |M_3| \right)}\nn\\
&& +  (-)^{|M_2||M_1|}\left[ \mathcal{D}_{M_2},\left[ \mathcal{D}_{M_3},\mathcal{D}_{M_1} \right]_{|M_3||M_1|} \right]_{|M_2|\left( |M_3| + |M_1| \right)}\nn\\
&& +  (-)^{|M_3||M_2|}\left[ \mathcal{D}_{M_3},\left[ \mathcal{D}_{M_1},\mathcal{D}_{M_2} \right]_{|M_1||M_2|} \right]_{|M_3|\left( |M_1| + |M_2| \right)}\ .\nn
\end{eqnarray}
Using \autoref{eq:commder}, it directly follows the two relations:
\begin{eqnarray}
0 &=& (-)^{|M_1||M_3|}\left[ \mathcal{D}_{M_1}\mathcal{T}_{M_2M_3 S}-\mathcal{T}_{M_1M_2}{}^{R}\mathcal{T}_{RM_3 S} + R_{M_1M_2M_3S} \right] \label{bianchiid1}\\
&& + (-)^{|M_2||M_1|}\left[ \mathcal{D}_{M_2}\mathcal{T}_{M_3M_1 S}-\mathcal{T}_{M_2M_3}{}^{R}\mathcal{T}_{RM_1 S} + R_{M_2M_3M_1S} \right] \nonumber\\
&& + (-)^{|M_3||M_2|}\left[ \mathcal{D}_{M_3}\mathcal{T}_{M_1M_2 S}-\mathcal{T}_{M_3M_1}{}^{R}\mathcal{T}_{RM_2 S} + R_{M_3M_1M_2S} \right]\ , \nonumber\\
0 &=& (-)^{|M_1||M_3|}\left[ \mathcal{T}_{M_1M_2}{}^RR_{RM_3PQ} - \mathcal{D}_{M_1}R_{M_2M_3PQ} \label{bianchiid2}\right] \\
&& + (-)^{|M_2||M_1|}\left[ \mathcal{T}_{M_2M_3}{}^RR_{RM_1PQ} - \mathcal{D}_{M_2}R_{M_3M_1PQ} \right] \nonumber\\
&& + (-)^{|M_3||M_2|}\left[ \mathcal{T}_{M_3M_1}{}^RR_{RM_2PQ} - \mathcal{D}_{M_3}R_{M_1M_2PQ} \right] \ .\nonumber
\end{eqnarray}
However, the torsion and curvature tensors contain too many degrees of freedom, and the supersymmetry theory is not generally recovered in the flat superspace limit. The number of degrees of freedom is then too large and must be reduced, taking into account the flat limit of the theory. \medskip

There are several possibles sets of constraints on $\mathcal{T}_{MN}{}^{P}$ that leads to various supergravity models. For this presentation, the following constraints are imposed \cite{WessBagger}\cite{GRIMM_torsion_constraints}\cite{WESS_ZUMINO_torsion_constraints}:
\begin{gather}
\mathcal{T}_{\underline{\alpha}\underline{\beta}}{}^{\underline{\gamma}} = 0 \quad , \quad \mathcal{T}_{\alpha\beta}{}^{\mu} = \mathcal{T}_{\dot{\alpha}\dot{\beta}}{}^{\mu} = 0 \quad , \quad \mathcal{T}_{\mu\nu}{}^{\rho} = 0 \ ,\label{eq:torsionconstraints}\\
\mathcal{T}_{\alpha\dot{\alpha}}{}^\mu = \mathcal{T}_{\dot{\alpha}\alpha}{}^\mu = -2i\sigma^\mu{}_{\alpha\dot{\alpha}} \quad , \quad \mathcal{T}_{\underline{\alpha}\mu}{}^{\nu} = \mathcal{T}_{\mu\underline{\alpha}}{}^{\nu} = 0 \ .\nonumber 
\end{gather}
Remark that those constraints are consistent with the definition of chiral superfields (satisfying $\bar{\mathcal{D}}_{\dot{\alpha}}\Phi=0$, see \autoref{sec:chiralSF}):
\begin{equation}
\Big\{ \bar{\mathcal{D}}_{\dot{\alpha}} , \bar{\mathcal{D}}_{\dot{\beta}} \Big\}\Phi = \mathcal{T}_{\dot{\alpha}\beta}{}^M\mathcal{D}_M \Phi =\mathcal{T}_{\dot{\alpha}\beta}{}^{\mu}\mathcal{D}_{\mu} \Phi+\mathcal{T}_{\dot{\alpha}\beta}{}^{\gamma}\mathcal{D}_{\gamma} \Phi+\mathcal{T}_{\dot{\alpha}\beta}{}{}_{\dot{\gamma}}\bar{\mathcal{D}}^{\dot{\gamma}} \Phi= 0\ ,\nn
\end{equation}
which means that the chiral superfield of supersymmetry naturally extends in supergravity. It can be shown that \autoref{eq:torsionconstraints} taken into account in (\autoref{bianchiid1}-\autoref{bianchiid2}) allow us to rewrite the non-zero elements of the torsion and curvature tensors with only three independent superfields $\mathcal{R}$, $G_{\alpha\dot{\alpha}}$ and $W_{(\alpha\beta\gamma)}$ (symmetric under exchange of indices). \medskip

The calculation of the non-zero elements of $\mathcal{T}_{MN}{}^{P}$ and $R_{MNPQ}$ is time-consuming and so will not be presented here (see \cite{WessBagger}\cite{book_sugra} for a complete derivation). The elements $\mathcal{T}_{MN}{}^{P}$ can be written as:
\begin{eqnarray}
\mathcal{T}_{\alpha\dot{\alpha}}{}^\mu = \mathcal{T}_{\dot{\alpha}\alpha}{}^\mu &=& -2i\sigma^\mu{}_{\alpha\dot{\alpha}}\ ,\nn\\
\mathcal{T}_{\alpha\mu}{}^\beta=-\mathcal{T}_{\mu\alpha}{}^\beta = \frac12 \bar{\sigma}_\mu{}^{\dot{\gamma}\gamma}\mathcal{T}_{\alpha\gamma\dot{\gamma}}{}^\beta \, &,& \, \mathcal{T}_{\alpha\mu}{}^{\dot{\beta}}=-\mathcal{T}_{\mu\alpha}{}^{\dot{\beta}} = \frac12 \bar{\sigma}_\mu{}^{\dot{\gamma}\gamma}\mathcal{T}_{\alpha\gamma\dot{\gamma}}{}^{\dot{\beta}}\ , \nn\\
\mathcal{T}_{\dot{\alpha}\mu}{}^{\beta} = - \mathcal{T}_{\mu\dot{\alpha}}{}^{\beta}=\frac12 \bar{\sigma}^{\dot{\gamma}\gamma}\mathcal{T}_{\dot{\alpha}\gamma\dot{\gamma}}{}^{\beta}\, &,& \, \mathcal{T}_{\dot{\alpha}\mu}{}^{\dot{\beta}} = - \mathcal{T}_{\mu\dot{\alpha}}{}^{\dot{\beta}}=\frac12 \bar{\sigma}^{\dot{\gamma}\gamma}\mathcal{T}_{\dot{\alpha}\gamma\dot{\gamma}}{}^{\dot{\beta}}\ ,\nn \\
\mathcal{T}_{\mu\nu}{}^{\alpha} = - \mathcal{T}_{\nu\mu}{}^{\alpha} = \frac14 \bar{\sigma}_\mu{}^{\dot{\delta}\delta}\bar{\sigma}_\nu{}^{\dot{\gamma}\gamma}\mathcal{T}_{\delta\dot{\delta}\gamma\dot{\gamma}}{}^{\alpha} \, &,& \, \mathcal{T}_{\mu\nu}{}^{\dot{\alpha}} = - \mathcal{T}_{\nu\mu}{}^{\dot{\alpha}} = \frac14 \bar{\sigma}_\mu{}^{\dot{\delta}\delta}\bar{\sigma}_\nu{}^{\dot{\gamma}\gamma}\mathcal{T}_{\delta\dot{\delta}\gamma\dot{\gamma}}{}^{\dot{\alpha}}\ ,\nn
\end{eqnarray}
with
\begin{eqnarray}
\mathcal{T}_{\alpha\gamma\dot{\gamma}\dot{\beta}} &=& 2i\epsilon_{\alpha\gamma}\epsilon_{\dot{\gamma}\dot{\beta}}\mathcal{R}^\dag \, ; \, \mathcal{T}_{\dot{\alpha}\gamma\dot{\gamma}\beta} = 2i\epsilon_{\dot{\alpha}\dot{\gamma}}\epsilon_{\gamma\beta}\mathcal{R}\ , \nn\\
\mathcal{T}_{\alpha\gamma\dot{\gamma}\beta} &=& \frac{i}{4} \left( \epsilon_{\gamma\beta}G_{\alpha\dot{\gamma}} - 3\epsilon_{\alpha\beta}G_{\gamma\dot{\gamma}}-3\epsilon_{\alpha\gamma}G_{\beta\dot{\gamma}} \right)\, ; \, \mathcal{T}_{\dot{\alpha}\gamma\dot{\gamma}\dot{\beta}} = \frac{i}{4} \left( \epsilon_{\dot{\gamma}\dot{\beta}}G_{\gamma\dot{\alpha}} - 3\epsilon_{\dot{\alpha}\dot{\beta}}G_{\gamma\dot{\gamma}}-3\epsilon_{\dot{\alpha}\dot{\gamma}}G_{\gamma\dot{\beta}} \right)\ , \nn\\
\mathcal{T}_{\delta\dot{\delta}\gamma\dot{\gamma}\alpha}&=&-2\epsilon_{\dot{\delta}\dot{\gamma}}W_{(\delta\gamma\alpha)} + \frac12 \epsilon_{\dot{\gamma}\dot{\delta}}\left( \epsilon_{\alpha\gamma}\bar{\mathcal{D}}^{\dot{\epsilon}}G_{\delta\dot{\epsilon}} + \epsilon_{\alpha\delta}\bar{\mathcal{D}}^{\dot{\epsilon}}G_{\gamma\dot{\epsilon}} \right) + \frac12 \epsilon_{\delta\gamma} \left( \bar{\mathcal{D}}_{\dot{\delta}} G_{\alpha\dot{\gamma}}+ \bar{\mathcal{D}}_{\dot{\gamma}}G_{\alpha\dot{\delta}}\right)\ , \nn\\
\mathcal{T}_{\delta\dot{\delta}\gamma\dot{\gamma}\dot{\alpha}}&=&-2\epsilon_{\delta\gamma}\bar{W}_{(\bar{\delta}\bar{\gamma}\bar{\alpha})} + \frac12 \epsilon_{\gamma\delta}\left( \epsilon_{\dot{\alpha}\dot{\gamma}}\mathcal{D}^{\epsilon}G_{\epsilon\dot{\delta}} + \epsilon_{\dot{\alpha}\dot{\delta}}\mathcal{D}^{\epsilon}G_{\epsilon\dot{\gamma}} \right) + \frac12 \epsilon_{\dot{\delta}\dot{\gamma}} \left( \mathcal{D}_{\delta} G_{\gamma\dot{\alpha}} + \mathcal{D}_{\gamma}G_{\delta\dot{\alpha}}\right) \ .\label{eq:torsioncomp}
\end{eqnarray}
The relations between the superfields $\mathcal{R}$, $G_{\alpha\dot{\alpha}}$, $W_{(\alpha\beta\gamma)}$ and the components of the curvature tensor $R_{MNPQ}$ can be found in \cite{WessBagger}\cite{book_sugra}.\medskip

Nevertheless, some remarks can be added for the computation. The Bianchi identities lead to two equalities \autoref{bianchiid1} and \autoref{bianchiid2}. Using the Dragon theorem \cite{Dragon}, it can be shown that the first relations automatically imply the validity of the seconds, meaning that we can only consider \autoref{bianchiid1}. Moreover, the conversion of spin indices to vector indices is also used intensively. For example, the following relation for the curvature tensor is valid:
\begin{equation}
R_{\dot{\alpha}\dot{\beta}\gamma\dot{\gamma}\delta\dot{\delta}} = \sigma^{\mu}_{\gamma\dot{\gamma}}\sigma^{\nu}_{\delta\dot{\delta}}R_{\dot{\alpha}\dot{\beta}\mu\nu} =2\epsilon_{\gamma\delta}R_{\dot{\alpha}\dot{\beta}\dot{\gamma}\dot{\delta}}-2\epsilon_{\dot{\gamma}\dot{\delta}}R_{\dot{\alpha}\dot{\beta}\gamma\delta}\ ,\nonumber
\end{equation}
where the second equality is obtained by decomposing the tensor in irreducible representations of $SL(2,\mathbb{C})=\overline{SO(1,3)}$. To be more precise, we first denote, 
\begin{gather}
R_{MN\alpha}{}^{\beta} = \frac{1}{4}R_{MN\mu\nu}\left(\sigma^{\mu}\bar{\sigma}^{\nu}\right)_{\alpha}{}^{\beta}\ , \ R_{MN}{}^{\dot{\alpha}}{}_{\dot{\beta}} = \frac{1}{4}R_{MN\mu\nu}\left(\bar{\sigma}^{\mu}\sigma^{\nu}\right)^{\dot{\alpha}}{}_{\dot{\beta}}\ .\label{eq:Rspinor}
\end{gather}
Since $R_{MN\mu\nu}$ is skew-symmetric by exchange of the indices $(\mu,\nu)$ (see \autoref{eq:symR}), we can then write:
\begin{eqnarray}
R_{MN\mu\nu} &=& \frac12 \left( R_{MN\mu\nu} - R_{MN\nu\mu} \right) \nn\\
 &=& \frac12 \left( \left( \bar{\sigma}_{\nu}\sigma_{\mu} \right)^{\dot{\beta}}{}_{\dot{\alpha}}R_{MN}{}^{\dot{\alpha}}{}_{\dot{\beta}} - \left( \sigma_{\mu}\bar{\sigma}_{\nu} \right)_{\beta}{}^{\alpha} R_{MN\alpha}{}^{\beta} \right) \ .\nn
\end{eqnarray}
By expanding the elements of the curvature tensor in spinor indices as the first equality of \autoref{eq:Rspinor} we then get
\begin{gather}
R_{MN\gamma\dot{\gamma}\delta\dot{\delta}} = 2\epsilon_{\gamma\delta}R_{MN\dot{\gamma}\dot{\delta}}-2\epsilon_{\dot{\gamma}\dot{\delta}}R_{MN\gamma\delta}\ .\nn
\end{gather}
Decomposing the elements $R_{MN\gamma\dot{\gamma}\delta\dot{\delta}}$ into irreducible representations of $SL(2,\mathbb{C})$:
\begin{gather}
R_{MN\gamma\dot{\gamma}\delta\dot{\delta}} = \epsilon_{\gamma\delta} w_{\dot{\gamma}\dot{\delta}} + \epsilon_{\dot{\gamma}\dot{\delta}}w'_{\gamma\delta} \nn
\end{gather}
where the tensors $w$ and $w'$ can be easily obtained through \autoref{eq:Rspinor}.
\medskip

Finally, we get:
\begin{eqnarray}
\Big\{ \mathcal{D}_\alpha , \mathcal{D}_\beta \Big\} &=& -\frac12 R_{\alpha\beta\gamma\delta} J^{\delta\gamma}\ , \label{eq:alg1}\\
\Big\{ \bar{\mathcal{D}}_{\dot{\alpha}} , \bar{\mathcal{D}}_{\dot{\beta}} \Big\} &=& - \frac12 R_{\dot{\alpha}\dot{\beta}\dot{\gamma}\dot{\delta}}J^{\dot{\delta}\dot{\gamma}}\ , \nn\\
\Big\{ \mathcal{D}_{\alpha} , \bar{\mathcal{D}}_{\dot{\alpha}} \Big\} &=& \mathcal{T}_{\alpha\dot{\alpha}}{}^{\mu}\mathcal{D}_\mu - \frac12 R_{\alpha\dot{\alpha}\beta\gamma}J^{\gamma\beta} - \frac12 R_{\alpha\dot{\alpha}\dot{\beta}\dot{\gamma}}\ , \nn\\
\Big[ \mathcal{D}_\alpha , \mathcal{D}_\mu \Big] &=& \mathcal{T}_{\alpha\mu}{}^{\beta}\mathcal{D}_\beta - \mathcal{T}_{\alpha\mu}{}^{\dot{\beta}}\bar{\mathcal{D}}_{\dot{\beta}} - \frac12 R_{\alpha\mu\beta\gamma}J^{\gamma\beta} - \frac12 R_{\alpha\mu\dot{\beta}\dot{\gamma}} J^{\dot{\gamma}\dot{\beta}}\ ,\nn \\
\Big[ \bar{\mathcal{D}}_{\dot{\alpha}} , \mathcal{D}_\mu \Big] &=& \mathcal{T}_{\dot{\alpha}\mu}{}^{\beta}\mathcal{D}_\beta - \mathcal{T}_{\dot{\alpha}\mu}{}^{\dot{\beta}}\bar{\mathcal{D}}_{\dot{\beta}} - \frac12 R_{\dot{\alpha}\mu\beta\gamma}J {\gamma\beta} - \frac12 R_{\dot{\alpha}\mu\dot{\beta}\dot{\gamma}} J^{\dot{\gamma}\dot{\beta}}\ , \nn\\
\Big[ \mathcal{D}_\mu , \mathcal{D}_\nu \Big] &=& \mathcal{T}_{\mu\nu}{}^{\beta}\mathcal{D}_\beta - \mathcal{T}_{\mu\nu}^{\dot{\beta}}\bar{\mathcal{D}}_{\dot{\beta}} - \frac12 R_{\mu\nu\beta\gamma}J^{\gamma\beta} - \frac12 R_{\mu\nu\dot{\beta}\dot{\gamma}}J^{\dot{\gamma}\dot{\beta}}\ .\nn
\end{eqnarray}

\subsection{Invariance of supergravity under Weyl supergroup}\label{subsec:weyl}
The algebra \autoref{eq:alg1} is invariant under superdiffeomorphisms \autoref{eq:superdiffeo} and local Lorentz transformations \autoref{eq:lorentz}. However, there exists a larger group of transformations of $J^{MN}$ and $\mathcal{D}_{M}$, which leaves the graded commutator \autoref{eq:commder} invariant. Indeed, the automorphism group of \autoref{eq:alg1} is the super-Weyl group, also called superconformal group (which is equivalent to the Weyl group for general relativity) \cite{HoweTucker}\cite{siegelsuperconforme}. The aim of this section is to identify the superconformal variations $\delta_{\Sigma}J^{\underline{\alpha}\underline{\beta}}$ and $\delta_{\Sigma}\mathcal{D}_{M}$. \medskip

Using the well-known commutation relations of the generators $J^{MN}$:
\begin{equation}
\Big[ J^{MN} , J^{PQ} \Big] = i\left(-\eta^{NP} J^{MQ} + \eta^{PM} J^{NQ} - \eta^{NQ} J^{PM} + \eta^{QM} J^{PN}\right)\ ,\nn
\end{equation}
with $\eta^{MN}=\left( \eta^{\mu\nu} , \epsilon^{\alpha\beta} , \epsilon_{\dot{\alpha}\dot{\beta}} \right)$, the equality $\delta J^{\underline{\alpha}\underline{\beta}} = 0$ follows immediately. For the transformation of the covariant derivative, we define:
\begin{equation}
\delta \mathcal{D}_{\alpha} = - \Phi \mathcal{D}_\alpha + \left( \Phi_\alpha \right)^{\beta\gamma}J_{\beta\gamma}\nn
\end{equation}
where $\Phi$ and $(\Phi_\alpha)^{\beta\gamma}$ are arbitrary superfields (with $(\Phi_\alpha)^{\beta\gamma}=(\Phi_\alpha)^{\gamma\beta}$). \medskip

Making a transformation of the anti-commutator $\left\{ \mathcal{D}_{\alpha} , \mathcal{D}_{\beta} \right\} $ gives:
\begin{eqnarray}
\delta\left\{ \mathcal{D}_{\alpha} , \mathcal{D}_{\beta} \right\} &=& -2\Phi \left\{ \mathcal{D}_{\alpha} , \mathcal{D}_{\beta} \right\}  - \mathcal{D}_{\beta}\Phi\mathcal{D}_{\alpha} - \mathcal{D}_{\alpha}\Phi\mathcal{D}_{\beta} - (\Phi_{\alpha})^{\gamma\delta}\left[ \epsilon_{\delta\beta}\mathcal{D}_{\gamma} + \epsilon_{\gamma\beta}\mathcal{D}_{\delta} \right] \nonumber \\
&& -(\Phi_{\beta})^{\gamma\delta}\left[ \epsilon_{\delta\alpha}\mathcal{D}_{\gamma} + \epsilon_{\gamma\alpha}\mathcal{D}_{\delta} \right] + \mathcal{D}_{\beta}(\Phi_{\alpha})^{\gamma\delta}J_{\delta\gamma} + \mathcal{D}_{\alpha}(\Phi_{\beta})^{\gamma\delta}J_{\delta\gamma}\label{eq:var}
\end{eqnarray}
The structure \autoref{eq:alg1} is invariant under super-Weyl transformations. The variation \ref{eq:var} must then be expressed only with Lorentz generators in the left-handed representation. Imposing that the other terms vanish leads to
\begin{eqnarray}
(\Phi_{\alpha})^{\gamma\beta} &=& \frac12 \left[ \delta^{\beta}{}_{\alpha}\mathcal{D}^{\gamma}\Phi + \delta^{\gamma}{}_{\alpha}\mathcal{D}^{\beta}\Phi \right]\ ,\nn
\end{eqnarray}
and thus we obtain
\begin{eqnarray}
\delta \mathcal{D}_{\alpha} = -\Phi\mathcal{D}_{\alpha} + \mathcal{D}^{\gamma}\Phi J_{\alpha\gamma}\ .\nn
\end{eqnarray}
Similarly, we can deduce $\delta\bar{\mathcal{D}}_{\dot{\alpha}}$. The variation $\delta\mathcal{D}_{\mu}$ is obtained from the development of $\delta\left\{ \mathcal{D}_{\alpha} , \bar{\mathcal{D}}_{\dot{\alpha}} \right\}$ with the help of \autoref{eq:alg1}. Defining the chiral superfield $\Sigma$ and introducing $\Phi = 2\Sigma^{\dagger} - \Sigma$, we get:
\begin{eqnarray}
\delta_{\Sigma} \mathcal{D}_\alpha &=& \Big[ \Sigma - 2\Sigma^\dagger \Big]\mathcal{D}_{\alpha} - \mathcal{D}^{\gamma} \Sigma J_{\alpha\gamma} \ ,\nn\\
\delta_{\Sigma} \bar{\mathcal{D}}_{\dot{\alpha}} &=& \Big[ \Sigma^\dagger - 2\Sigma \Big]\bar{\mathcal{D}}_{\dot{\alpha}} - \bar{\mathcal{D}}^{\dot{\gamma}} \Sigma^\dagger J_{\dot{\alpha}\dot{\gamma}} \ ,\nn\\
\delta_{\Sigma} \mathcal{D}_\mu &=&  - \Big( \Sigma + \Sigma^\dagger \Big) \mathcal{D}_{\mu} - \frac{i}{2} \Big[ \bar{\mathcal{D}}\Sigma^\dagger\bar{\sigma}_{\mu}\mathcal{D} + \mathcal{D}\Sigma\bar{\sigma}_{\mu}\bar{\mathcal{D}} \Big] - \mathcal{D}^{\nu}\Big( \Sigma + \Sigma^\dagger \Big) J_{\mu\nu} \ ,\nn\\
\delta_{\Sigma} J_{\alpha\beta} &=& 0 \ ,\nn\\
\delta_{\Sigma} J_{\dot{\alpha}\dot{\beta}} &=& 0\ .\nn
\end{eqnarray}
From the expressions of the non-vanishing torsion components \autoref{eq:torsioncomp} and \autoref{eq:alg1}, the variation of the three superfields $\mathcal{R}$, $G_{\mu}$ and $W_{(\alpha\beta\gamma)}$ can also be calculated. Using the variation \ref{eq:var}:
\begin{gather}
\delta\left\{ \mathcal{D}_{\alpha} , \mathcal{D}_{\beta} \right\} = \left[ \left( 2\Sigma^{\dagger} - \Sigma \right)R_{\alpha\beta\gamma\delta} + \frac12 \epsilon_{\alpha\delta}\epsilon_{\beta\gamma}\mathcal{D}\cdot\mathcal{D} \Sigma + \frac12 \epsilon_{\beta\delta}\epsilon_{\alpha\gamma}\mathcal{D}\cdot\mathcal{D} \Sigma \right] J^{\delta\gamma} = -\frac12 \delta R_{\alpha\beta\gamma\delta}J^{\delta\gamma}\nn
\end{gather}
with:
\begin{eqnarray}
\left\{ \mathcal{D}_{\alpha} , \mathcal{D}_{\beta} \right\}\Sigma = 0 \quad , \quad \left[ \mathcal{D}_{\alpha} , \mathcal{D}_{\beta} \right] = \epsilon_{\alpha\beta}\mathcal{D}\cdot\mathcal{D} \Sigma\ ,\nn
\end{eqnarray}
the transformation of $\mathcal{R}^{\dagger}$ can be computed using the relation between the components of curvature tensor $R_{MNPQ}$ \cite{WessBagger}\cite{book_sugra}:
\begin{gather}
R_{\alpha\beta\gamma\delta} = 4\left( \epsilon_{\alpha\gamma}\epsilon_{\beta\delta} + \epsilon_{\beta\gamma}\epsilon_{\alpha\delta} \right)\mathcal{R}^{\dagger} \ . \label{eq:Rcurv}
\end{gather}
Those calculations give: 
\begin{eqnarray}
\delta_{\Sigma} \mathcal{R} &=& -4\Sigma \mathcal{R} - \frac14 \Big( \bar{\mathcal{D}}\cdot\bar{\mathcal{D}} - 8\mathcal{R} \Big)\Sigma^\dagger\ , \nn\\
\delta_{\Sigma} G_{\mu} &=& - \Big[ \Sigma + \Sigma^\dagger \Big]G_{\mu} + i\mathcal{D}_{\mu}\Big[ \Sigma - \Sigma^\dagger \Big] \ ,\nn\\
\delta_{\Sigma} W_{(\alpha\beta\gamma)} &=& -3\Sigma W_{(\alpha\beta\gamma)}\ .\nn
\end{eqnarray}
Transformation laws of the superconnection can also be deduced by developing the covariant derivatives as \autoref{eq:dercov}. Lengthy calculus give:
\begin{eqnarray}
\delta_{\Sigma} \Omega_{\tilde{M}\beta\gamma} &=& E_{\tilde{M}\beta} \mathcal{D}_{\gamma} \Sigma + E_{\tilde{M}\gamma}\mathcal{D}_{\beta}\Sigma - E_{\tilde{M}\gamma}\epsilon_{\gamma\alpha}(\sigma^{\nu\rho})_{\beta}{}^{\alpha}\mathcal{D}_{\rho}\Big[ \Sigma + \Sigma^\dagger \Big] \ ,\nn\\
\delta_{\Sigma} \Omega_{\tilde{M}\dot{\beta}\dot{\gamma}} &=& E_{\tilde{M}\dot{\beta}}\bar{\mathcal{D}}_{\dot{\gamma}}\Sigma^\dagger + E_{\tilde{M}\dot{\gamma}}\bar{\mathcal{D}}_{\dot{\beta}}\Sigma^\dagger - E_{\tilde{M}\nu}\epsilon_{\dot{\beta}\dot{\alpha}}(\bar{\sigma}^{\nu\rho})^{\dot{\alpha}}{}_{\dot{\gamma}}\mathcal{D}_{\rho}\Big[ \Sigma + \Sigma^\dagger \Big]\ ,\nn \\
\delta_{\Sigma} \Omega_{\tilde{M}\nu\rho} &=& -E_{\tilde{M}\nu}\mathcal{\rho}\Big[ \Sigma + \Sigma^\dagger \Big] + E_{\tilde{M}\rho}\mathcal{D}_{\nu}\Big[ \Sigma + \Sigma^\dagger \Big] -2E_{\tilde{M}}{}^{\alpha}(\sigma_{\nu\rho})_{\alpha}{}^{\beta}\mathcal{D}_{\beta}\Sigma - 2E_{\tilde{M}\dot{\alpha}}(\bar{\sigma}_{\nu\rho})^{\dot{\alpha}}{}_{\dot{\beta}}\bar{\mathcal{D}}^{\dot{\beta}}\Sigma^\dagger\ .\nn
\end{eqnarray}
Those results will be helpful in a future section (see \autoref{subsec:mattergauge}). Indeed, as we will see, the computation of the supergravity action will lead to non-correctly normalised kinetic terms. In order to rescale the Lagrangian in order to get correctly normalised kinetic terms, a Weyl-rescaling followed by a shift of the gravitino field (so-called \textit{gravitino-shift}) will be performed.

\subsection{Supergravity transformations and gauge fixing conditions}
We will now investigate the transformations of the main superfields under local Lorentz transformations (\autoref{eq:lorentz}) and superdiffeomorphisms (\autoref{eq:superdiffeo}). Considering a vector superfield $V^{M}$, the supervierbein $E_{M}{}^{\tilde{M}}$ and the superconnection $\Omega_{\tilde{M}P}{}^{Q}$, their transformations can be written as:
\begin{eqnarray}
V^{M}(\mathfrak{z}) &\rightarrow& V'{}^{M}(\mathfrak{z}' ) = V^{N}(\mathfrak{z})\Lambda_{N}{}^{M}(\mathfrak{z}) \ ,\label{eq:transfoV}\\
\Omega_{\tilde{M}P}{}^{Q}(\mathfrak{z}) &\rightarrow& \Omega'_{\tilde{M}P}{}^{Q}(\mathfrak{z}') = \frac{\partial \mathfrak{z}^{\tilde{N}}}{\partial \mathfrak{z}'{}^{\tilde{M}}}(\Lambda^{-1})_{P}{}^{R}(\mathfrak{z})\Big[ \Omega_{\tilde{N}R}{}^{S}(\mathfrak{z})\Lambda_{S}{}^{Q}(\mathfrak{z}) - \partial_{\tilde{N}}\Lambda_{R}{}^{Q}(\mathfrak{z}) \Big] \ ,\nn\\
E_{\tilde{M}}{}^{M}(\mathfrak{z}) &\rightarrow& E'_{\tilde{M}}{}^{M}(\mathfrak{z}') = \frac{\partial \mathfrak{z}^{\tilde{N}}}{\partial \mathfrak{z}'{}^{\tilde{M}}}E_{\tilde{N}}{}^{N}(\mathfrak{z})\Lambda_{N}{}^{M}(\mathfrak{z})\ .\nn
\end{eqnarray}
Developing the parameters around the identity, the matrix $\Lambda_{M}{}^{N}(\mathfrak{z})$ can be written as:
\begin{equation}
\Lambda_{N}{}^{M} = \delta_{N}{}^{M} +L_{N}{}^{M}\ .  \nn
\end{equation}
Using \autoref{eq:transfoV} we obtain:
\begin{eqnarray}
\delta V^{M} &=& V'{}^{M}(\mathfrak{z}) - V{}^{M}(\mathfrak{z}) \nonumber \\
&=& -\xi^{\tilde{P}}\partial_{\tilde{P}}V^M + V^NL_{N}{}^{M} \nonumber \\
&=& -\xi^{\tilde{P}}\mathcal{D}_{\tilde{P}}V^M + V^N\left( \xi^{\tilde{P}}\Omega_{\tilde{P}N}{}^{M} + L_N{}^{M} \right)\ ,\nn
\end{eqnarray}
where the relation can be simplified by redefining $L_{N}{}^{M} \rightarrow L'{}_N{}^{M} = \xi^{\tilde{P}}\Omega_{\tilde{P}N}{}^{M} + L_{N}{}^{M}$. The variation of $G^{\mu}$ can then be computed. In the same manner, we have:
\begin{eqnarray}
\delta E_{\tilde{M}}{}^{M} &=& -\mathcal{D}_{\tilde{M}}\xi^{M} + \xi^{\tilde{N}}\mathcal{T}_{\tilde{N}\tilde{M}}{}^{M} + E_{\tilde{M}}{}^{P}L_{P}{}^{M} \ ,\label{eq:deltaE}\\
\delta\Omega_{\tilde{M}P}{}^{Q} &=& -\xi^{\tilde{N}}R_{\tilde{N}\tilde{M}P}{}^{Q} + \Omega_{\tilde{M}P}{}^{R}L_{R}{}^{Q}-L_{P}{}^{R}\Omega_{\tilde{M}R}{}^{Q} - \partial_{\tilde{M}}L_{P}{}^{Q}\ .\label{eq:deltaOmega}
\end{eqnarray}
For the variation of the superfields $\mathcal{R}$, we obtain directly:
\begin{eqnarray}
\delta \mathcal{R} &=& -\xi^{\tilde{M}}\mathcal{D}_{\tilde{M}}\mathcal{R}\ . \label{eq:trsfochrial}
\end{eqnarray}

We will see in the next sections that the development of the supergravity action is based on the lowest components in the $\theta$-expansion of the superfields. Recalling that $\xi^{M}(\mathfrak{z})$ and $L_{M}{}^{N}$ are superfields, they can be expanded in terms of Grassmann variables in the Einstein frame. For the superfield $\xi^{M}(\mathfrak{z})$ defining the superdiffeomorphisms:
\begin{eqnarray}
\xi^{M}(\mathfrak{z}) &=& \xi^{(0,0)M}(x) + \theta^{\tilde{\alpha}}\xi_{\tilde{\alpha}}^{(1,0)M}(x) + \bar{\theta}_{\tilde{\dot{\alpha}}}\xi^{(0,1)M}{}^{\tilde{\dot{\alpha}}}(x) + \theta\sigma_{\tilde{\mu}}\bar{\theta}\xi^{(1,1)M}{}^{\tilde{\mu}}(x) \label{eq:expXi}\\
&&+ \theta\cdot\theta\xi^{(2,0)M}(x) + \bar{\theta}\cdot\bar{\theta}\xi^{(0,2)M}(x) + \bar{\theta}\cdot\bar{\theta}\theta^{\tilde{\alpha}}\xi_{\tilde{\alpha}}^{(1,2)M}(x) \nonumber \\
&&+ \theta\cdot\theta \bar{\theta}_{\tilde{\dot{\alpha}}}\xi^{(2,1)M}{}^{\tilde{\dot{\alpha}}}(x) + \theta\cdot\theta\bar{\theta}\cdot\bar{\theta}\xi^{(2,2)M}(x)\ .\nn
\end{eqnarray}

The same type of expansions can also be done for $L_{M}{}^{N}$. Using the large arbitrariness of the parameters, some components can be set to zero. Defining $X\big| = X\big|_{\theta=\bar{\theta} = 0}$, we can fix, after analysing the variation $\delta\Omega_{MN}{}^{P}\big|$ from \autoref{eq:deltaOmega}:
\begin{equation}
\Omega_{\tilde{\mu}M}{}^{N}\big| = \omega_{\tilde{\mu}M}{}^{N}(x) \quad , \quad \Omega_{\tilde{\alpha}M}{}^{N}\big| = 0 \quad , \quad \Omega^{\tilde{\dot{\alpha}}}_{M}{}^{N}\big| = 0 \ ,\label{eq:omegamin}
\end{equation}
where $\omega_{\tilde{\mu}M}{}^{N}=\left(\omega_{\tilde{\mu}\mu}{}^{\nu},\omega_{\tilde{\mu}\alpha}{}^{\beta},\omega_{\tilde{\mu}}{}^{\dot{\alpha}}{}_{\dot{\beta}} \right)$ is the spin-connection of general relativity. By examining the variation of the lowest components of the supervierbein $\delta E_{\tilde{M}}{}^{M}\big|$ from \autoref{eq:deltaE}, and taking into account the invertibility of $E_{\tilde{M}}{}^{M}$, a proper choice for the components of $\xi^{\tilde{M}}$ in \autoref{eq:expXi} and for $L_{M}{}^{N}$ allows to eliminate some components of $E_{\tilde{M}}{}^{M}\big|$:
\begin{eqnarray}
E_{\tilde{M}}{}^{M}(\mathfrak{z})\big| &=& \begin{pmatrix}
e_{\tilde{\mu}}{}^{\mu}(x) & \frac12\psi_{\tilde{\mu}}{}^{\alpha}(x) & \frac12 \bar{\psi}_{\tilde{\mu}\dot{\alpha}}(x) \\
0 & \delta_{\tilde{\alpha}}{}^{\alpha} & 0 \\
0 & 0 & \delta^{\tilde{\dot{\alpha}}}{}_{\dot{\alpha}}
\end{pmatrix} \ ,\label{eq:Emin1}\\
E_{M}{}^{\tilde{M}}(\mathfrak{z})\big| &=& \begin{pmatrix}
e_{\mu}{}^{\tilde{\mu}}(x) & -\frac12 \psi_{\mu}{}^{\tilde{\alpha}}(x) & -\frac12 \bar{\psi}_{\mu\tilde{\dot{\alpha}}}(x) \\
0 & \delta_{\alpha}{}^{\tilde{\alpha}} & 0 \\
0 & 0 & \delta^{\dot{\alpha}}{}_{\tilde{\dot{\alpha}}}
\end{pmatrix}\ ,\label{eq:Emin2}
\end{eqnarray}
where $e_{\tilde{\mu}}{}^{\mu}(x)$ is the helicity-2 graviton field and $(\psi_{\tilde{\mu}}{}^{\alpha},\bar{\psi}_{\tilde{\mu}}{}^{\dot{\alpha}})$ the Weyl components of the spin-$3/2$ gravitino field. Since the relations $E_{\tilde{M}}{}^{M}E_{M}{}^{\tilde{N}} = \delta_{\tilde{M}}{}^{\tilde{N}}$ and $E_{M}{}^{\tilde{M}}E_{\tilde{M}}{}^{N} = \delta_{M}{}^{N}$ hold, we have:
\begin{eqnarray}
e_{\tilde{\mu}}{}^{\mu}e_{\nu}{}^{\tilde{\nu}} = \delta_{\tilde{\mu}}{}^{\tilde{\nu}} \, &,& \, e_{\mu}{}^{\tilde{\mu}}e_{\tilde{\mu}}{}^{\nu} = \delta_{\mu}{}^{\nu} \ ,\nn\\
\psi_{\mu}{}^{\tilde{\alpha}} = e_{\mu}{}^{\tilde{\mu}}\psi_{\tilde{\mu}}{}^{\alpha}\delta_{\alpha}{}^{\tilde{\alpha}} \, &,& \, \bar{\psi}_{\mu\tilde{\dot{\alpha}}} = e_{\mu}{}^{\tilde{\mu}}\bar{\psi}_{\tilde{\mu}\dot{\alpha}}\delta^{\dot{\alpha}}{}_{\tilde{\dot{\alpha}}}\ ,\nn
\end{eqnarray}
It can be shown that the lowest components of the superfields $\mathcal{R}$ and $G_{\mu}$ cannot be eliminated. We then define a complex scalar field $M(x)$ and a real vector field $b_{\mu}$ by:
\begin{equation}
\mathcal{R}(\mathfrak{z})\big| = -\frac16 M(x) \quad , \quad G_{\mu}(\mathfrak{z})\big| = -\frac13 b_{\mu}(x) .\nn
\end{equation}
We thus obtain the degrees of freedom of the supergravity multiplet that contains a helicity-2 field $e_{\mu}{}^{\tilde{\mu}}(x)$ (the graviton), Weyl components $(\psi_{\tilde{\mu}}{}^{\alpha}(x),\bar{\psi}_{\tilde{\mu}}{}^{\dot{\alpha}}(x))$ of the spin-$3/2$ gravitino and two auxiliary fields $M(x)$ and $b_{\mu}(x)$. The superconnection is not a fundamental field (see \autoref{subsec:lowestcompo}), so it doesn't belong to the multiplet. This multiplet is called the \textit{minimal} supergravity multiplet. Other constraints can lead to different numbers of auxiliary fields (see \cite{new-minimal}\cite{non-minimal1}\cite{non-minimal2}). This is out of the scope of this thesis. \medskip

The gauge fixing conditions (Equations \ref{eq:omegamin}, \ref{eq:Emin1} and \ref{eq:Emin1}) are, nevertheless,  not preserved by the general transformations \autoref{eq:transfoV}. We must then specify the transformations which preserve the gauge previously defined. In a first step, we assume
\begin{equation}
\xi^{\mu}(\mathfrak{z})\big| = 0 \quad , \quad \xi^{\alpha}(\mathfrak{z})\big| = \epsilon^{\alpha} \quad , \quad \bar{\xi}_{\dot{\alpha}}(\mathfrak{z})\big| = \bar{\epsilon}_{\dot{\alpha}} \quad , \quad L_{MN}(\mathfrak{z})\big| = 0\ . \nonumber
\end{equation}
Since $\delta E_{\tilde{\alpha}}{}^{\mu}\big| = \delta E^{\tilde{\dot{\alpha}}\mu}\big| = 0$, using \autoref{eq:deltaE}, we obtain:
\begin{eqnarray}
\delta E_{\tilde{\alpha}}{}^{\mu}\big| &=& -\partial_{\tilde{\alpha}}\xi^{\mu}\big| + 2i\delta_{\tilde{\alpha}}{}^{\alpha}\sigma^{\mu}{}_{\alpha\dot{\alpha}}\bar{\epsilon}^{\dot{\alpha}} = 0\ , \nonumber \\
\delta E^{\tilde{\dot{\alpha}}\mu}\big| &=& - \bar{\partial}^{\tilde{\dot{\alpha}}}\xi^{\mu}\big| -2i\delta^{\tilde{\dot{\alpha}}}{}_{\dot{\alpha}}\epsilon^{\dot{\alpha}\dot{\beta}}\epsilon^{\alpha}\sigma^{\mu}{}_{\alpha\dot{\beta}} = 0 \ .\nonumber 
\end{eqnarray} 
It is then possible to express the parameters $\xi^{\mu}(\mathfrak{z})$ as:
\begin{gather}
\xi^{\mu}(\mathfrak{z}) = 2i\left( \theta^{\tilde{\alpha}}\delta_{\tilde{\alpha}}{}^{\alpha}\sigma^{\mu}{}_{\alpha\dot{\alpha}}\bar{\epsilon}^{\dot{\alpha}} + \epsilon^{\alpha}\sigma^{\mu}_{\alpha\dot{\beta}}\epsilon^{\dot{\alpha}\dot{\beta}}\bar{\theta}_{\tilde{\dot{\alpha}}}\delta^{\tilde{\dot{\alpha}}}{}_{\dot{\alpha}}\right) + \cdots\nn
\end{gather}
where the dots correspond to other terms coming from higher-order components. In order to recover a parameter invariant under \ref{eq:transfoV}, we can just make the substitution $\left( \delta_{\tilde{\alpha}}{}^{\alpha} , \delta^{\dot{\tilde{\alpha}}}{}_{\dot{\alpha}} \right)\rightarrow\left( E_{\tilde{\alpha}}{}^{\alpha} , E^{\dot{\tilde{\alpha}}}{}_{\dot{\alpha}} \right)$ leading to:
\begin{gather}
\xi^{\mu}(\mathfrak{z}) = 2i\Big[ \tilde{\theta}\sigma^{\mu}\bar{\epsilon} - \epsilon\sigma^{\mu}\bar{\tilde{\theta}} \Big] \nn
\end{gather}
with $\tilde{\theta}_{\alpha}=\theta^{\tilde{\alpha}}E_{\tilde{\alpha}}{}^{\beta}\epsilon_{\alpha\beta}$. The parameters $L_{MN}(\mathfrak{z})$ associated with the transformations which preserve the gauge fixing constraints can be obtained using $\delta\Omega_{\tilde{\alpha}M}{}^{N}\big|=\delta\Omega^{\tilde{\dot{\alpha}}}{}_{M}{}^{N}=0$. We finally obtain:
\begin{eqnarray}
\xi^{\mu}(\mathfrak{z}) &=& 2i\Big[ \bar{\theta}\sigma^{\mu}\bar{\epsilon} - \epsilon\sigma^{\mu}\bar{\theta} \Big] \ ,\nn\\
\xi^{\alpha}(\mathfrak{z}) &=& \epsilon^{\alpha} \ ,\nn\\
\bar{\xi}_{\dot{\alpha}}(\mathfrak{z}) &=& \bar{\epsilon}_{\dot{\alpha}} \ ,\nn\\
L_{\alpha\beta}(\mathfrak{z}) &=& \frac13 \Big[ \tilde{\theta}_{\alpha}(2\epsilon_{\beta}M^* + b_{\beta\dot{\gamma}}\bar{\epsilon}^{\dot{\gamma}}) + \tilde{\theta}_{\beta}(2\epsilon_{\alpha}M^* + b_{\alpha\dot{\gamma}}\bar{\epsilon}^{\dot{\gamma}}) \Big] \ ,\label{eq:sugratrso}\\
L_{\dot{\alpha}\dot{\beta}}(\mathfrak{z}) &=& \frac13 \Big[ \tilde{\bar{\theta}}_{\dot{\alpha}} ( 2\bar{\epsilon}_{\dot{\beta}}M + \epsilon^{\gamma}b_{\gamma\dot{\beta}} ) + \tilde{\bar{\theta}}_{\dot{\beta}}( 2\bar{\epsilon}_{\dot{\alpha}}M + \epsilon^{\gamma}b_{\gamma\dot{\alpha}} ) \Big] \ ,\nn\\
L_{\mu\nu}(\mathfrak{z}) &=& -\frac12 \Big[ (\bar{\sigma}_{\mu}\sigma_{\nu})^{\dot{\alpha}}{}_{\dot{\beta}}L_{\dot{\alpha}}{}^{\dot{\beta}} + (\sigma_{\mu}\bar{\sigma}_{\nu})_{\alpha}{}^{\beta}L^{\alpha}{}_{\beta} \Big]\ .\nn
\end{eqnarray}

\subsection{Lowest order component fields}\label{subsec:lowestcompo}
All those relations can now be used in order to calculate the lowest order components of the three fundamental superfields $\mathcal{R}$, $G_{\mu}$ and $W_{(\alpha\beta\gamma)}$ as well as their derivatives. This is a very lengthy computation for some derivatives of the fields. Therefore, only the method and the final results will be mentioned. \medskip

Since $\mathcal{R}$, $G_{\mu}$ and $W_{(\alpha\beta\gamma)}$ are related to the torsion tensor $\mathcal{T}_{MN}{}^{Q}$, the components $\mathcal{T}_{MN}{}^{Q}\big|$ must be firstly computed. From \autoref{eq:torsion} and the lowest components of the supervierbein (\ref{eq:Emin1},\ref{eq:Emin2}) and the superconnection \autoref{eq:omegamin}:
\begin{eqnarray}
\mathcal{T}_{\tilde{\mu}\tilde{\nu}}{}^{\mu}\big| &=& -\partial_{\tilde{\mu}}e_{\tilde{\nu}}{}^{\mu} + \partial_{\tilde{\nu}}e_{\tilde{\mu}}{}^{\mu} - \omega_{\tilde{\mu}\tilde{\nu}}{}^{\mu} + \omega_{\tilde{\nu}\tilde{\mu}}{}^{\mu}\ , \label{eq:lowtorsion1}\\
\mathcal{T}_{\tilde{\mu}\tilde{\nu}}{}^{\alpha}\big| &=& - \frac12 \big( \mathcal{D}_{\tilde{\mu}} \psi_{\tilde{\nu}}{}^{\alpha} - \mathcal{D}_{\tilde{\nu}} \psi_{\tilde{\mu}}{}^{\alpha} \big) = -\frac12 \psi_{\tilde{\mu}\tilde{\nu}}{}^{\alpha} \ ,\nn\\
\mathcal{T}_{\tilde{\mu}\tilde{\nu}\dot{\alpha}}\big| &=& - \frac12 \big( \mathcal{D}_{\tilde{\mu}}\bar{\psi}_{\tilde{\nu}\dot{\alpha}} - \mathcal{D}_{\tilde{\nu}}\bar{\psi}_{\tilde{\mu}\dot{\alpha}} \big) = -\frac12 \bar{\psi}_{\tilde{\mu}\tilde{\nu}\dot{\alpha}}\ ,\nn
\end{eqnarray}
where we have introduced:
\begin{eqnarray}
\omega_{\tilde{\mu}\tilde{\nu}}{}^{\mu} &=& e_{\tilde{\nu}}{}^{\nu}\omega_{\tilde{\mu}\nu}{}^{\mu} \ ,\nn\\
\psi_{\tilde{\mu}\tilde{\nu}}{}^{\alpha} = \mathcal{D}_{\tilde{\mu}}\psi_{\tilde{\nu}}{}^{\alpha} - \mathcal{D}_{\tilde{\nu}}\psi_{\tilde{\mu}}{}^{\alpha} \quad &,& \quad \bar{\psi}_{\tilde{\mu}\tilde{\nu}\dot{\alpha}} = \mathcal{D}_{\tilde{\mu}}\bar{\psi}_{\tilde{\nu}\dot{\alpha}} - \mathcal{D}_{\tilde{\nu}}\bar{\psi}_{\tilde{\mu}\dot{\alpha}}\ , \nn\\
\mathcal{D}_{\tilde{\mu}}\psi_{\tilde{\nu}}{}^{\alpha} = \partial_{\tilde{\mu}}\psi_{\tilde{\nu}}{}^{\alpha} + \psi_{\tilde{\nu}}{}^{\beta}\omega_{\tilde{\mu}\beta}{}^{\alpha} \quad &,& \quad \mathcal{D}_{\tilde{\mu}}\bar{\psi}_{\tilde{\nu}\dot{\alpha}} = \partial_{\tilde{\mu}}\bar{\psi}_{\tilde{\nu}\dot{\alpha}} + \bar{\psi}_{\tilde{\nu}\dot{\beta}}\omega_{\tilde{\mu}}{}^{\dot{\beta}}{}_{\dot{\alpha}}\ ,\nn
\end{eqnarray}
The component $\mathcal{T}_{\tilde{\mu}\tilde{\nu}}{}^{\mu}\big|$ can also be developed in flat indices using the supervierbein:
\begin{eqnarray}
\mathcal{T}_{\tilde{\mu}\tilde{\nu}}{}^{\mu}\big| &=& E_{\tilde{\mu}}{}^{M}E_{\tilde{\nu}}{}^{N}\mathcal{T}_{MN}{}^{\mu}\big| \nn\\
&=& \big( E_{\tilde{\mu}}{}^{\gamma}E_{\tilde{\nu}\dot{\delta}} + E_{\tilde{\mu}\dot{\delta}}E_{\tilde{\nu}}{}^{\gamma}  \big)\mathcal{T}_{\gamma}{}^{\dot{\delta}\mu}\big| \nonumber \\
&=& - \frac{i}{2}\big( \psi_{\tilde{\mu}}\sigma^{\mu}\bar{\psi}_{\tilde{\nu}} - \psi_{\tilde{\nu}}\sigma^{\mu}\bar{\psi}_{\tilde{\mu}} \big) \ .\label{eq:lowtorsion2}
\end{eqnarray}
From \autoref{eq:lowtorsion1}, \autoref{eq:lowtorsion2} and symmetry property of the spin-connection, the expression of $\omega_{\tilde{\mu}\tilde{\nu}\tilde{\rho}}=e_{\mu\tilde{\rho}}\omega_{\tilde{\mu}\tilde{\nu}}{}^{\mu}$ can be found:
\begin{eqnarray}
\omega_{\tilde{\mu}\tilde{\nu}\tilde{\rho}} &=& -\frac12 e_{\mu\tilde{\rho}}\big( \partial_{\tilde{\mu}}e_{\tilde{\nu}}{}^{\mu} - \partial_{\tilde{\nu}}e_{\tilde{\mu}}{}^{\mu} \big) +\frac12 e_{\mu\tilde{\mu}}\big( \partial_{\tilde{\nu}}e_{\tilde{\rho}}{}^{\mu} - \partial_{\tilde{\rho}}e_{\tilde{\nu}}{}^{\mu} \big) \nonumber \\
&& -\frac12 e_{\mu\tilde{\nu}}\big( \partial_{\tilde{\rho}}e_{\tilde{\mu}}{}^{\mu} - \partial_{\tilde{\mu}}e_{\tilde{\rho}}{}^{\mu} \big) + \frac{i}{4}e_{\mu\tilde{\rho}}\big( \psi_{\tilde{\mu}}\sigma^{\mu}\bar{\psi}_{\tilde{\nu}} - \psi_{\tilde{\nu}}\sigma^{\mu}\bar{\psi}_{\tilde{\mu}} \big) \nonumber \\
&& - \frac{i}{4}e_{\tilde{\mu}\mu}\big( \psi_{\tilde{\nu}}\sigma^{\mu}\bar{\psi}_{\tilde{\rho}} - \psi_{\tilde{\rho}}\sigma^{\mu}\bar{\psi}_{\tilde{\nu}} \big) + \frac{i}{4}e_{\mu\tilde{\nu}}\big( \psi_{\tilde{\rho}}\sigma^{\mu} \bar{\psi}_{\tilde{\mu}} - \psi_{\tilde{\mu}}\sigma^{\mu}\bar{\psi}_{\tilde{\rho}} \big)\ .\nn
\end{eqnarray}
Thus, the spin-connection $\omega_{\tilde{\mu}\tilde{\nu}\tilde{\rho}}$ is then not a fundamental field and can be written in terms of the graviton and the gravitino. \medskip

Thereafter, using the relations between the torsion components $\mathcal{T}_{MN}{}^{Q}$ and the fundamental superfields $\mathcal{R}$, $G_{\mu}$ and $W_{(\alpha\beta\gamma)}$ \autoref{eq:torsioncomp}, it is then possible to evaluate their lowest components as well as their derivatives. In a similar way, the connection between those superfields and the components of the curvature tensor $R_{MNPQ}$ is also used (see \cite{book_sugra}\cite{WessBagger}).\medskip

The lowest components read:
\begin{eqnarray}
\mathcal{R}\big| &=& -\frac16 M \ ,\nonumber\\
\mathcal{R}^\dagger \big| &=& -\frac16 M^* \ ,\nonumber\\
\mathcal{D}_{\alpha} \mathcal{R} \big| &=& -\frac13 (\sigma^{\mu\nu}\psi_{\mu\nu})_{\alpha} + \frac{i}{6} (\sigma^{\mu}\bar{\psi}_{\mu})_{\alpha}M - \frac{i}{6} \psi_{\mu\alpha}b^{\mu}\ , \nonumber\\
\bar{\mathcal{D}}^{\dot{\alpha}}\mathcal{R}^\dagger \big| &=& -\frac13(\bar{\sigma}^{\mu\nu}\bar{\psi}_{\mu\nu})^{\dot{\alpha}}+ \frac{i}{6} (\bar{\sigma}^{\mu}\psi_{\mu})^{\dot{\alpha}}M^* +\frac{i}{6} \bar{\psi}_{\mu}{}^{\dot{\alpha}}b^\mu \ ,\nonumber\\
\mathcal{D}\cdot\mathcal{D} \mathcal{R} \big| &=& \frac13 e_{\mu}{}^{\tilde{\mu}}e_\nu{}^{\tilde{\nu}}R_{\tilde{\mu}\tilde{\nu}}{}^{\mu\nu}\Big| + \frac49 \left| M \right|^2 - \frac29 b_\mu b^\mu -\frac{2i}{3}e_{\mu}{}^{\tilde{\mu}}\mathcal{D}_{\tilde{\mu}}b^\mu - \frac13\bar{\psi}_{\mu}\cdot\bar{\psi}^{\mu} M  \ ,\nonumber\\
&& - \frac13 \psi_{\nu}\sigma^{\nu}\bar{\psi}_{\mu}b^{\mu} - \frac{2i}{3}\bar{\psi}^{\mu}\bar{\sigma}^{\nu}\psi_{\mu\nu} - \frac{1}{12}\epsilon^{\mu\nu\rho\sigma}\Big[ \psi_{\mu}\sigma_{\sigma}\bar{\psi}_{\nu\rho} + \bar{\psi}_{\mu}\bar{\sigma}_{\sigma}\psi_{\nu\rho} \Big] \ ,\nonumber\\
\bar{\mathcal{D}}\cdot\bar{\mathcal{D}}\mathcal{R}^\dagger \big| &=& \frac13 e_{\mu}{}^{\tilde{\mu}}e_{\nu}{}^{\tilde{\nu}}R_{\tilde{\mu}\tilde{\nu}}{}^{\mu\nu}\Big| + \frac49 \left| M \right|^2 - \frac29 b_\mu b^\mu + \frac{2i}{3}e_{\mu}{}^{\tilde{\mu}}\mathcal{D}_{\tilde{\mu}}b^\mu  - \frac13 \psi_\mu \cdot \psi^{\mu} M^* \ ,\nonumber\\
&& - \frac13 \psi_{\mu}\sigma^{\nu}\bar{\psi}_{\nu} b^{\nu} - \frac{2i}{3}\psi^{\mu}\sigma^{\nu} \bar{\psi}_{\mu\nu} + \frac{1}{12}\epsilon^{\mu\nu\rho\sigma}\Big[ \psi_{\mu}\sigma_{\sigma}\bar{\psi}_{\nu\rho} + \bar{\psi}_{\mu}\bar{\sigma}_{\sigma}\psi_{\nu\rho} \Big] \ ,\nonumber\\
G_{\alpha\dot{\alpha}}\Big| &=& -\frac13 b_{\alpha\dot{\alpha}} \ ,\nonumber\\
\mathcal{D}_\delta G_{\gamma\dot{\alpha}}\big| &=& -\frac14 \bar{\psi}_{\delta}{}^{\dot{\epsilon}}{}_{\gamma\dot{\epsilon\dot{\alpha}}} - \frac{1}{12}\epsilon_{\delta\gamma}\bar{\psi}^{\epsilon\dot{\epsilon}}{}_{\epsilon\dot{\alpha}\dot{\epsilon}} + \frac{i}{6}\psi_{\gamma\dot{\alpha}\delta}M^* + \frac{i}{12}(\bar{\psi}_{\delta\dot{\epsilon}\dot{\alpha}}b_{\gamma}{}^{\dot{\epsilon}} + \bar{\psi}_{\gamma\dot{\epsilon}}{}^{\dot{\epsilon}}b_{\delta\dot{\alpha}} + \bar{\psi}_{\delta\dot{\epsilon}}{}^{\dot{\epsilon}}b_{\gamma\dot{\alpha}}) \ ,\nonumber\\
\bar{\mathcal{D}}_{\dot{\delta}}G_{\alpha\dot{\gamma}} \big| &=& -\frac14 \psi^{\epsilon}{}_{\dot{\delta}\epsilon\dot{\gamma}\alpha} - \frac{1}{12} \epsilon_{\dot{\delta}\dot{\gamma}}\psi_{\alpha}{}^{\dot{\epsilon}\epsilon}{}_{\dot{\epsilon}\epsilon} - \frac{i}{6}\bar{\psi}_{\alpha\dot{\gamma}\dot{\delta}}M - \frac{i}{12}(\psi_{\epsilon\dot{\delta}\alpha}b^{\epsilon}{}_{\dot{\gamma}}+ \psi_{\epsilon\dot{\gamma}}{}^{\epsilon}b_{\alpha\dot{\delta}}+ \psi_{\epsilon\dot{\delta}}{}^{\epsilon}b_{\alpha\dot{\gamma}}) \ ,\nonumber\\
\mathcal{D}_{\delta}G^{\mu} \big| &=& \frac13 (\sigma^{\nu}\bar{\psi}_{\nu}{}^{\mu})_{\delta} - \frac{i}{12}\epsilon^{\mu\nu\rho\sigma}(\sigma_{\sigma}\bar{\psi}_{\nu\rho})_{\delta} + \frac{i}{6}\psi^{\mu}{}_{\delta}M^* + \frac{i}{6}(\sigma^{\nu}\bar{\psi}_{\nu})_{\delta}b^{\mu} + \frac{1}{12} \epsilon^{\mu\nu\rho\sigma}(\sigma_{\sigma}\bar{\psi}_{\nu})_{\delta}b_{\rho}\ , \nonumber\\
\bar{\mathcal{D}}^{\dot{\delta}}G^{\mu}\big| &=& -\frac13 (\bar{\sigma}^{\nu}\psi_{\nu}{}^{\mu})^{\dot{\delta}} - \frac{i}{12}\epsilon^{\mu\nu\rho\sigma}(\bar{\sigma}_{\sigma}\psi_{\nu\rho})^{\dot{\delta}} - \frac{i}{6}\bar{\psi}^{\mu\dot{\delta}}M + \frac{i}{6} (\bar{\sigma}^{\nu\psi_{\nu}})^{\dot{\delta}}b^{\mu} - \frac{1}{12}\epsilon^{\mu\nu\rho\sigma}(\bar{\sigma}_{\sigma}\psi_{\nu})^{\dot{\delta}}b_{\mu} \ ,\nonumber\\
W_{(\alpha_1 \alpha_2 \alpha_3)}\big| &=& -\frac{1}{48} \sum_{r\in S_3}\left( \psi_{\alpha_{\tau(1)}\dot{\delta}\alpha_{\tau (2)}}{}^{\dot{\delta}}{}_{\alpha_{\tau (3)}} - i\psi_{\alpha_{\tau (1)}\dot{\delta}\alpha_{\tau (2)}}b_{\alpha_{\tau (3)}}{}^{\dot{\delta}} \right)\ , \nonumber\\
\bar{W}_{(\dot{\alpha}_1 \dot{\alpha}_2 \dot{\alpha}_3)}\big| &=& -\frac{1}{48} \sum_{r\in S_3}\left( \bar{\psi}_{\delta\dot{\alpha}_{\tau(1)}}{}^{\delta}{}_{\dot{\alpha}_{\tau (2)}\dot{\alpha}_{\tau (3)}} + i\bar{\psi}_{\delta\dot{\alpha}_{\tau (1)}\dot{\alpha}_{\tau (2)}}b^{\delta}{}_{\dot{\alpha}_{\tau (3)}} \right) \ .\nonumber
\end{eqnarray}
with $S^3$ the permutation group with three elements and:
\begin{gather}
G_{\alpha\dot{\alpha}}=\sigma^{\mu}{}_{\alpha\dot{\alpha}}G_{\mu} \ , \ b_{\alpha\dot{\alpha}}=\sigma^{\mu}{}_{\alpha\dot{\alpha}}b_{\mu} \ , \ \psi_{\beta\dot{\beta}\gamma\dot{\gamma}}{}^{\alpha} = \sigma^{\mu}{}_{\beta\dot{\beta}}\sigma^{\nu}{}_{\gamma\dot{\gamma}}e_{\mu}{}^{\tilde{\mu}}e_{\nu}{}^{\tilde{\nu}}\psi_{\tilde{\mu}\tilde{\nu}}{}^{\alpha} \ , \nn\\
\bar{\psi}_{\beta\dot{\beta}\gamma\dot{\gamma}\dot{\alpha}} = \sigma^{\mu}{}_{\beta\dot{\beta}}\sigma^{\nu}{}_{\gamma\dot{\gamma}}e_{\mu}{}^{\tilde{\mu}}e_{\nu}{}^{\tilde{\nu}}\bar{\psi}_{\tilde{\mu}\tilde{\nu}\dot{\alpha}} \ .\nn
\end{gather}
The supergravity transformation of the fields $e_{\tilde{\mu}}{}^{\mu}$, $\psi_{\mu}$, $M$ and $b_{\mu}$ can also be derived from the previous results. For example, the variation of the vierbein can be obtained from:
\begin{eqnarray}
\delta e_{\tilde{\mu}}{}^{\mu} = \delta E_{\tilde{\mu}}{}^{\mu}\big| = -\mathcal{D}_{\tilde{\mu}}\xi^{\mu}\big| + \xi^{M}\mathcal{T}_{M\tilde{\mu}}{}^{\mu}\big| + E_{\tilde{\mu}}{}^{\nu}L_{\nu}{}^{\mu}\big|\ .\nn
\end{eqnarray}
Using \autoref{eq:torsion}, we easily find:
\begin{eqnarray}
\delta e_{\tilde{\mu}}{}^{\mu} = -i\big[ \epsilon(x)\sigma^{\mu}\bar{\psi}_{\tilde{\mu}} - \psi_{\tilde{\mu}}\sigma^{\mu}\bar{\epsilon}(x) \big]\ .\label{eq:sugramltp_trsfo1}
\end{eqnarray}
The transformations of the other fields also follow using the same process:
\begin{eqnarray}
\delta \psi_{\tilde{\mu}}{}^{\alpha} &=& -2\mathcal{D}_{\tilde{\mu}}\epsilon^{\alpha}(x) + \frac{i}{3}e_{\tilde{\mu}}{}^{\nu}\big( \bar{\epsilon}(x)\bar{\sigma}_{\nu} \big)^{\alpha}M + \frac{i}{3}e_{\tilde{\mu}}{}^{\nu}\big[ \big( \epsilon (x)\sigma_{\rho}\bar{\sigma}_{\nu} \big)^{\alpha}b^{\rho} - 3\epsilon^{\alpha} (x) b_{\gamma} \big] \ ,\nn\\
\delta \bar{\psi}_{\tilde{\mu}\dot{\alpha}} &=& -2\mathcal{D}_{\tilde{\mu}}\bar{\epsilon}_{\dot{\alpha}}(x) + \frac{i}{3}e_{\tilde{\mu}}{}^{\nu} \big( \epsilon (x)\sigma_{\nu} \big)_{\dot{\alpha}}M^* - \frac{i}{3}e_{\tilde{\mu}}{}^{\nu} \big[ \big( \bar{\epsilon}(x)\bar{\sigma}_{\rho}\sigma_{\nu} \big)_{\dot{\alpha}}b^{\rho} - 3\bar{\epsilon}_{\dot{\alpha}}(x)b_{\nu} \big] \ ,\nn\\
\delta M &=& -2(\epsilon\sigma^{\mu\nu}\psi_{\mu\nu}) + i(\epsilon\sigma^{\mu}\bar{\psi}_{\mu})M -i(\epsilon\cdot\psi_{\mu})b^{\mu} \ ,\label{eq:sugramltp_trsfo2}\\
\delta M^* &=& -2(\bar{\epsilon}\bar{\sigma}^{\mu\nu}\bar{\psi}_{\mu\nu}) + i(\bar{\epsilon}\bar{\sigma}^{\mu}\psi_{\mu})M^* + i(\bar{\epsilon}\cdot\bar{\psi}_{\mu})b^{\mu} \ ,\nn\\
\delta b^{\mu} &=& \epsilon\sigma^{\nu}\bar{\psi}_{\nu}{}^{\mu} - \bar{\epsilon}\bar{\sigma}^{\nu}\psi_{\nu}{}^{\mu} - \frac{i}{4}\epsilon^{\mu\nu\rho\sigma}\Big[ \epsilon\sigma_{\sigma}\bar{\psi}_{\nu\rho} + \bar{\epsilon}\bar{\sigma}_{\sigma}\psi_{\nu\rho} \Big] + \frac{i}{2}\Big[ \epsilon\cdot\psi^{\mu} M^* - \bar{\epsilon}\cdot\bar{\psi}^{\mu}M \Big] \nn\\
&& + \frac{i}{2}\Big[ \epsilon\sigma^{\nu}\bar{\psi}_{\nu} + \bar{\epsilon}\bar{\sigma}^{\nu}\psi_{\nu} \Big] b^{\mu} + \frac14 \epsilon^{\mu\nu\rho\sigma}\Big[ \epsilon \sigma_{\sigma} \bar{\psi}_{\nu} -\bar{\epsilon}\bar{\sigma}_{\sigma}\psi_{\nu} \Big] b_{\rho}\ .\nn
\end{eqnarray}
We clearly see that the fields $\big( e_{\tilde{\mu}}{}^{\mu}, \psi_{\tilde{\mu}} , M , b_{\mu} \big)$ form a multiplet under supergravity transformations as explained previously.

\section{Superfield approach in curved superspace}
There exist two types of superfields relevant to particle physics in the context of $N=1$ supersymmetry theory: the chiral superfield $\Phi(\mathfrak{z})=\left( \phi,\chi,F  \right)$ containing a scalar field $\phi$, a Weyl spinor $\chi$ and a scalar auxiliary field $F$ and the vector superfield $V(\mathfrak{z})=\left( v_{\mu}, \lambda, D\right)$ with $v_{\mu}$ a vector field, $\lambda$ a Majorana spinor and $D$ an auxiliary field.\medskip

As briefly explained in \autoref{sec:bianchi}, the definition of chiral superfields in supersymmetry does not generally extend in supergravity. However, imposing the proper torsion constraints such as \autoref{eq:torsionconstraints} allows to extend the notion of chiral superfields in supergravity.\medskip

In this section, the chiral superfields $\Phi$ and the vector superfields $V$ will be introduced. The components and derivatives of those superfields and their transformations according to supergravity transformations will be given. The interactions between those two superfields will also be derived.
\subsection{Chiral superfields}
\label{sec:chiralSF}

In supersymmetry, chiral (anti-chiral) superfields $\Phi(\mathfrak{z}^M)$ ($\Phi^{\dagger}(\mathfrak{z}^M)$) are defined through the constraints:
\begin{gather}
\bar{D}_{\dot{\alpha}}\Phi(\mathfrak{z}^M) = 0 \quad , \quad D_{\alpha}\Phi^{\dagger}(\mathfrak{z}^M) = 0 \ ,
\end{gather}
where the $D_{\alpha}$ and $\bar{D}_{\dot{\alpha}}$ are the superderivatives in supersymmetry. Following the torsion constraints \autoref{eq:torsionconstraints}, a similar definition can be extended in the context of supergravity. Modifying the superderivatives to covariant derivatives, we can then write:
\begin{eqnarray}
\bar{\mathcal{D}}_{\dot{\alpha}}\Phi(\mathfrak{z}^{\tilde{M}}) = 0 \quad , \quad \mathcal{D}_{\alpha}\Phi^{\dagger}(\mathfrak{z}^{\tilde{M}}) = 0\ .\nn
\end{eqnarray}
Note that although the superfield $\Phi(\mathfrak{z}^{\tilde{M}})$ depends on curved coordinates $\mathfrak{z}^{\tilde{M}}=\big( x^{\tilde{\mu}},\theta^{\tilde{\alpha}}, \bar{\theta}_{\tilde{\dot{\alpha}}} \big)$, its definition relies on flat indices. \medskip

In the same manner as \autoref{eq:expXi}, the chiral superfield $\Phi$ can be expressed in supersymmetry using an expansion in $\theta$-variables. Even though this method is still possible in supergravity, it is more tedious since the expansion is done in the curved frame, which imposes to know all the components of the supervierbein $E_{\tilde{M}}{}^{M}$ in $\theta$-expansion. It is then more suitable to define the components of the chiral superfield according to the (covariant) derivatives of the superfield: 
\begin{eqnarray}
\phi &=& \Phi\big| \quad , \quad \chi_{\alpha} = \frac{1}{\sqrt{2}}\mathcal{D}_{\alpha}\Phi \big| \quad , \quad F = \frac14 \mathcal{D}_{\alpha}\mathcal{D}^{\alpha}\Phi\big|\ , \label{eq:chiral}\\
\phi^\dagger &=& \Phi^{\dagger}\big| \quad , \quad \bar{\chi}^{\dot{\alpha}} = \frac{1}{\sqrt{2}}\bar{\mathcal{D}}^{\dot{\alpha}}\Phi^\dagger \big| \quad , \quad F^\dagger = \frac14 \bar{\mathcal{D}}^{\dot{\alpha}}\bar{\mathcal{D}}_{\dot{\alpha}}\Phi^\dagger\big| \ .\label{eq:antichiral} 
\end{eqnarray}
The calculation of the other lowest components of the derivatives of $\Phi$ (up to the fourth derivatives) is also mandatory to fully expand the Lagrangian of supergravity. Since covariant derivatives are defined in the Einstein frame (see \autoref{eq:dercov}), the indices must firstly be changed to tilted indices where the gauge fixing conditions can be imposed (\ref{eq:omegamin}\ref{eq:Emin1}\ref{eq:Emin2}). Only some examples of calculations will be shown. A complete presentation can be found in \cite{WessBagger}\cite{book_sugra}.\medskip

Using \autoref{eq:chiral}, \autoref{eq:dercov} and $E_{\tilde{\mu}}{}^{\mu}\big|=e_{\tilde{\mu}}{}^{\mu}$, the covariant derivative $\mathcal{D}_{\mu}\Phi$ can be easily calculated:
\begin{eqnarray}
\mathcal{D}_{\mu}\Phi\big| &=& E_{\mu}{}^{\tilde{M}}\mathcal{D}_{\tilde{M}}\Phi\big| = E_{\mu}{}^{\tilde{\mu}}\mathcal{D}_{\tilde{\mu}}\Phi\big| + E_{\mu}{}^{\tilde{\alpha}}\mathcal{D}_{\tilde{\alpha}}\Phi\big| + E_{\mu\tilde{\dot{\alpha}}}\bar{\mathcal{D}}^{\tilde{\dot{\alpha}}}\Phi\big| \nonumber \\
&=& e_{\mu}{}^{\tilde{\mu}}\big( \partial_{\tilde{\mu}}\phi - \frac{\sqrt{2}}{2}\psi_{\tilde{\mu}}\cdot\chi \big) \equiv \hat{D}_{\mu}\phi\ ,\nn
\end{eqnarray} 
which defines the derivative covariant with respect to supergravity transformations. \medskip

The value of $\mathcal{D}_{\alpha}\mathcal{D}_{\beta}\Phi\big|$ directly follows the definition of the auxiliary field $F$ in \autoref{eq:chiral}. Using the property $A_{\alpha}B_{\beta} - A_{\beta}B_{\alpha}=\epsilon_{\alpha\beta}A\cdot B$ and the graded commutator \autoref{eq:commder} with \autoref{eq:torsionconstraints} we get the two identities:
\begin{gather}
\frac12 \left( \mathcal{D}_{\alpha}\mathcal{D}_{\beta} - \mathcal{D}_{\beta}\mathcal{D}_{\alpha} \right)\Phi\big| = 2\epsilon_{\alpha\beta}F \quad , \quad \left\{ \mathcal{D}_{\alpha} , \mathcal{D}_{\beta} \right\} \Phi\big| = 0 \ ,\nn
\end{gather}
which gives $\mathcal{D}_{\alpha}\mathcal{D}_{\beta}\Phi\big| =2\epsilon_{\alpha\beta}F $.\medskip

The derivation of the third derivative $\mathcal{D}_{\alpha}\mathcal{D}_{\beta}\mathcal{D}_{\gamma}\Phi\big|$ leads to interesting results. Using the property:
\begin{gather}
\mathcal{D}_{\alpha}\mathcal{D}_{\beta}\mathcal{D}_{\gamma}+\mathcal{D}_{\beta}\mathcal{D}_{\gamma}\mathcal{D}_{\alpha}+\mathcal{D}_{\gamma}\mathcal{D}_{\alpha}\mathcal{D}_{\beta}-\mathcal{D}_{\alpha}\mathcal{D}_{\gamma}\mathcal{D}_{\beta}-\mathcal{D}_{\beta}\mathcal{D}_{\alpha}\mathcal{D}_{\gamma}-\mathcal{D}_{\gamma}\mathcal{D}_{\beta}\mathcal{D}_{\alpha}=0\nn
\end{gather}
and the vanishing commutator $\left\{ \mathcal{D}_{\alpha}, \mathcal{D}_{\beta} \right\}\Phi =0$ (see \autoref{eq:commder}), we can write:
\begin{gather}
\mathcal{D}_{\alpha}\mathcal{D}_{\beta}\mathcal{D}_{\gamma} \Phi = \frac13 \left( \mathcal{D}_{\alpha}\mathcal{D}_{\beta}\mathcal{D}_{\gamma} + \mathcal{D}_{\alpha}\mathcal{D}_{\gamma}\mathcal{D}_{\beta} - \mathcal{D}_{\beta}\mathcal{D}_{\gamma}\mathcal{D}_{\alpha} - \mathcal{D}_{\gamma}\mathcal{D}_{\alpha}\mathcal{D}_{\beta} \right)\ . \nn
\end{gather} 
By contracting the last two indices using $\epsilon^{\beta\gamma}$ and with \autoref{eq:Rcurv}:
\begin{gather}
\mathcal{D}_{\alpha}\mathcal{D}\cdot\mathcal{D}\Phi = -\frac23 \left\{ \mathcal{D}_{\alpha} , \mathcal{D}_{\beta} \right\}\mathcal{D}^{\beta}\Phi = 8\mathcal{R}^{\dagger}\mathcal{D}_{\alpha} \Phi
\end{gather}
which can be rewritten as
\begin{gather}
\mathcal{D}_{\alpha}\Big( \mathcal{D}\cdot\mathcal{D} - 8\mathcal{R}^{\dagger} \Big)\Phi = 0 \label{eq:projphi}
\end{gather}
since $\mathcal{R}^{\dagger}$ is an anti-chiral superfield, \textit{i.e.},  $\mathcal{D}_{\alpha} \mathcal{R}^{\dagger}  =0$. This is an important result. The equality \autoref{eq:projphi} means that the operator $\mathcal{D}\cdot\mathcal{D} - 8\mathcal{R}^{\dagger}$ acting on the chiral superfield $\Phi$ generates an anti-chiral superfield. Hence, it is possible to associate a chiral (anti-chiral) superfield $\Xi$ ($\Xi^{\dagger}$) to any anti-chiral (chiral) superfield $\Phi^{\dagger}$ ($\Phi$) through the operator  $\mathcal{D}\cdot\mathcal{D} - 8\mathcal{R}^{\dagger}$ ($\bar{\mathcal{D}}\cdot\bar{\mathcal{D}} - 8\mathcal{R}$):
\begin{equation}
\Xi = \left( \bar{\mathcal{D}}\cdot\bar{\mathcal{D}} - 8\mathcal{R} \right)\Phi^\dagger \quad , \quad \Xi^{\dagger} = \left( \mathcal{D}\cdot\mathcal{D} - 8\mathcal{R}^{\dagger} \right)\Phi\ .\nn
\end{equation}
This operator is the analogue of the operator $D\cdot D$ in supersymmetry (where $D$ is the superderivative in supersymmetry) and will be central to construct invariant action (given in Subsection \ref{sec:actionsugra}).\medskip

We now turn to the transformation of the chiral superfields under supergravity transformations. Using the definition of a supergravity transformation \autoref{eq:sugratrso} and the variation
\begin{gather}
\delta\Phi = -\xi^{\tilde{M}}\mathcal{D}_{\tilde{M}}\Phi \ ,\nn
\end{gather}
the transformations of $\phi$, $\chi_{\alpha}$ and $F$ are:
\begin{eqnarray}
\delta\phi &=& -\sqrt{2}\epsilon\cdot\chi\ , \label{eq:chiraltrso1} \\
\delta \chi_{\alpha} &=& \sqrt{2}\Big( \epsilon_{\alpha}F - i\big( \sigma^{\mu}\bar{\epsilon} \big)_{\alpha}\hat{D}_{\mu}\phi \Big) \ ,\label{eq:chiraltrso2} \\
\delta F &=& \frac{\sqrt{2}}{3}\epsilon\cdot\chi M^* - i\sqrt{2} \hat{D}_{\mu}\chi \sigma^{\mu}\bar{\epsilon} - \frac{\sqrt{2}}{6}b_{\mu}\chi\sigma^{\mu}\bar{\epsilon} \ .\label{eq:chiraltrso3}
\end{eqnarray}
Note that the variation of the auxiliary field $F$ is not a total derivative (compared to what happens in supersymmetry).

\subsection{Vector superfields}
The vector superfield is defined in the same way as in supersymmetry by the reality condition
\begin{equation}
V=V^\dagger\ .\nn
\end{equation}
We suppose a non-abelian gauge group with the Lie algebra $\mathfrak{g}$ and $V=V^aT_a$, $T_a$ being the generators of the gauge group, in an appropriate matrix representation, taken to be hermitian. The vector superfield transforms then as:
\begin{equation}
e^{2gV}\rightarrow e^{-2ig\Lambda} e^{2gV} e^{2ig\Lambda^{\dagger}}\ ,\nn
\end{equation}
with the chiral superfield $\Lambda=\Lambda^aT_a$ (where $a$ corresponds to the index in the adjoint representation of $\mathfrak{g}$). Similarly than in supersymmetry, we adopt the Wess-Zumino gauge in order to eliminate all the non-physical fields \cite{WessZumino}:
\begin{eqnarray}
V\big|= 0 \quad , \quad \mathcal{D}_{\alpha}V\big| &=&  0 \quad ,\quad \mathcal{D}_{\dot{\alpha}}V\big| = 0\ , \label{eq:WZgauge} \\
\mathcal{D}\cdot\mathcal{D}V\big| = 0\quad &,&\quad  \bar{\mathcal{D}}\cdot\bar{\mathcal{D}}V\big| = 0\ .\nn
\end{eqnarray}
We also denote the strength tensor by:
\begin{eqnarray}
\mathcal{W}_{\alpha} &=& - \frac14\left[ \bar{\mathcal{D}}\cdot\bar{\mathcal{D}}-8\mathcal{R} \right]e^{2gV}\mathcal{D}_{\alpha}e^{-2gV} \ ,\nonumber \\
\bar{\mathcal{W}}_{\dot{\alpha}} &=& \frac14 \left[ \mathcal{D}\cdot\mathcal{D} - 8\mathcal{R}^{\dagger} \right]e^{-2gV}\bar{\mathcal{D}}_{\dot{\alpha}}e^{2gV}\ , \label{eq:strenghtensor}
\end{eqnarray}
with $g$, the coupling constant related to the defined gauge group. The various physical components of the vector superfield are defined using the covariant derivative: 
\begin{gather}
\lambda_{\alpha} = \frac{i}{2g}\mathcal{W}_{\alpha}\big| \quad ,\quad \bar{\lambda} = \frac{-i}{2g}\bar{\mathcal{W}}_{\dot{\alpha}}\big| \quad ,\quad D^a = \frac{1}{4g}\mathcal{D}^{\alpha}\mathcal{W}^{a}_{\alpha}\big| = \frac{1}{4g}\bar{\mathcal{D}}_{\dot{\alpha}}\bar{\mathcal{W}}^{a\dot{\alpha}}\big|\ , \\
 v_{\mu} = \frac14 \bar{\sigma}_{\mu}^{\dot{\alpha}\alpha}\big[ \mathcal{D}_{\alpha},\bar{\mathcal{D}}_{\dot{\alpha}} \big]V\big| \ ,\nn
\end{gather}
where $v_{\mu}$ is a gauge boson, $(\lambda , \bar{\lambda})$ a Majorana spinor (called the gaugino) and $D$ an auxiliary field. \medskip

As explained in Section \ref{sec:chiralSF}, the calculus of the various derivatives of $V$ and $\mathcal{W}_{\alpha}$ is required for the computation of the final supergravity action. Because the superfield strength tensor is a function of the vector superfield $V$, the derivatives of $V$ must firstly be computed. The method is similar to the calculation with the chiral superfield. In the Wess-Zumino gauge \autoref{eq:WZgauge} and using the definition of the covariant derivative \autoref{eq:dercov}, $\mathcal{D}_{\tilde{\mu}}V\big|$ vanishes and so:
\begin{gather}
\mathcal{D}_{\mu} V\big| = E_{\mu}{}^{\tilde{M}}\mathcal{D}_{M}V\big| =E_{\mu}{}^{\tilde{\mu}}\mathcal{D}_{\mu}V\big| +E_{\mu}{}^{\tilde{\alpha}}\mathcal{D}_{\alpha}V\big| +E_{\mu\tilde{\dot{\alpha}}}\bar{\mathcal{D}}^{\tilde{\dot{\alpha}}}V\big| = 0 \ .\nn 
\end{gather}
The computation of the other derivatives is fully presented in \cite{WessBagger}\cite{book_sugra}. We also have in the Wess-Zumino gauge (since $V^3=0$):
\begin{eqnarray}
\mathcal{W}_{\alpha}&=&- \frac14 \Big[ \bar{\mathcal{D}}\cdot\bar{\mathcal{D}} - 8\mathcal{R} \Big] e^{2gV}\mathcal{D}_{\alpha}e^{-2gV} \nonumber \\
&=& - \frac14 \Big[ \bar{\mathcal{D}}\cdot\bar{\mathcal{D}} - 8\mathcal{R} \Big] \Big( -2g\mathcal{D}_{\alpha}V + 2g^2\mathcal{D}_{\alpha}V^2 - 4g^2 V \mathcal{D}_{\alpha}V +\cdots \Big) \nonumber \\
&=& \frac{g}{2} \Big[ \bar{\mathcal{D}}\cdot\bar{\mathcal{D}} - 8\mathcal{R} \Big] \Big( \mathcal{D}_{\alpha}V + g\big[ V, \mathcal{D}_{\alpha}V \big] \Big)\ . \label{eq:WZWa}
\end{eqnarray}
The various components of $\mathcal{W}_{\alpha}$ can then be computed starting from \autoref{eq:WZWa}:
\begin{eqnarray}
\mathcal{W}_{\alpha}\Big| &=& -2ig\lambda_{\alpha}\ ,  \label{eq:Walphacompo1} \\
\bar{\mathcal{W}}_{\dot{\alpha}} \Big| &=& 2ig\bar{\lambda}_{\dot{\alpha}} \ ,\nonumber \\
\mathcal{D}_{\alpha}\mathcal{W}_{\beta} \Big| &=& 2g\left[ i(\sigma^{\mu\nu})_{\alpha\beta}(\hat{F}_{\mu\nu}) + \epsilon_{\alpha\beta}D \right] \ ,\nonumber \\
\bar{\mathcal{D}}_{\dot{\alpha}} \bar{\mathcal{W}}_{\dot{\beta}}\Big| &=& -2g\left[ i(\bar{\sigma}^{\mu\nu})_{\dot{\alpha}\dot{\beta}}(\hat{F}_{\mu\nu}) + \epsilon_{\dot{\alpha}\dot{\beta}}D \right]\ , \nonumber \\
\mathcal{D}\cdot\mathcal{D} \mathcal{W}_{\alpha}\Big| &=& -2g\left[ 4\sigma^{\mu}{}_{\alpha\dot{\alpha}}\hat{D}_{\mu}\bar{\lambda}^{\dot{\alpha}} - 2ib_{\mu}(\sigma^{\mu}\bar{\lambda})_{\alpha} - 2iM^*\lambda_{\alpha} \right]\ ,\nonumber \\
\bar{\mathcal{D}}\cdot\bar{\mathcal{D}} \bar{\mathcal{W}}_{\dot{\alpha}}\Big| &=& -2g\left[ 4\sigma^{\mu}{}_{\alpha\dot{\alpha}}\hat{D}_{\mu}\lambda^{\alpha} + 2ib_{\mu}(\lambda\sigma^{\mu})_{\dot{\alpha}} + 2iM\bar{\lambda}_{\dot{\alpha}} \right] \ ,\label{eq:Walphacompo2}
\end{eqnarray}
with:
\begin{eqnarray}
\hat{F}_{\mu\nu} &=& \hat{D}_{\mu}v_{\nu} - \hat{D}_{\nu}v_{\mu} + ig\left[ v_\mu , v_\nu \right] = e_{\mu}{}^{\tilde{\mu}}e_{\nu}{}^{\tilde{\nu}} \hat{F}_{\tilde{\mu}\tilde{\nu}}\ , \nonumber \\
\hat{F}_{\tilde{\mu}\tilde{\nu}} &=& F_{\tilde{\mu}\tilde{\nu}} - \frac{i}{2} \left[ \bar{\lambda}\bar{\sigma}_{\tilde{\nu}}\psi_{\tilde{\mu}} + \lambda\sigma_{\tilde{\nu}}\bar{\psi}_{\tilde{\mu}} - \bar{\lambda}\bar{\sigma}_{\tilde{\mu}}\psi_{\tilde{\nu}} - \lambda\sigma_{\tilde{\mu}}\bar{\psi}_{\tilde{\nu}} \right] \ ,\nonumber \\
\hat{D}_{\mu}v_{\nu} &=& e_{\mu}{}^{\tilde{\mu}}\mathcal{D}_{\tilde{\mu}}v_{\nu} - \frac{i}{2}\left[ \bar{\lambda}\bar{\sigma}_{\nu}\psi_{\mu} - \bar{\psi}_{\mu}\bar{\sigma}_{\nu}\lambda \right] - \frac{i}{2} ( \psi_{\nu}\sigma^{\rho}\bar{\psi}_{\mu} )v_{\rho}\ , \nonumber \\
\hat{D}_{\mu} \lambda_{\alpha} &=& e_{\mu}{}^{\tilde{\mu}}\left( \mathcal{D}_{\tilde{\mu}}\lambda_{\alpha} + ig\left[ v_{\tilde{\mu}},\lambda_{\alpha} \right] \right) - \frac12(\sigma^{\nu\rho}\psi_{\mu})_{\alpha}(\hat{F}_{\nu\rho}) + \frac{i}{2}D\psi_{\mu\alpha}\ ,\nn\\
\hat{D}_{\mu}\bar{\lambda}_{\dot{\alpha}} &=& e_{\mu}{}^{\tilde{\mu}}\left( \mathcal{D}_{\tilde{\mu}}\bar{\lambda}_{\dot{\alpha}} + ig\left[ v_{\tilde{\mu}} , \bar{\lambda}_{\dot{\alpha}} \right] \right) + \frac12(\bar{\psi}_{\mu}\bar{\sigma}^{\nu\rho})_{\dot{\alpha}}(\hat{F}_{\nu\rho}) - \frac{i}{2}D\bar{\psi}_{\mu\dot{\alpha}}\nn
\end{eqnarray}
and $F_{\tilde{\mu}\tilde{\nu}}=e_{\tilde{\mu}}{}^{\mu}e_{\tilde{\nu}}{}^{\nu}F_{\mu\nu}=e_{\tilde{\mu}}{}^{\mu}e_{\tilde{\nu}}{}^{\nu}(\hat{D}_{\mu}v_{\nu} - \hat{D}_{\nu}v_{\mu})$. We now consider an infinitesimal supergravity transformation with parameters $\xi^{M}$. By checking the variation of the non-physical fields we get, for example:
\begin{eqnarray}
\chi_{\alpha} \equiv \mathcal{D}_{\alpha} V \Big| = 0 \quad\rightarrow\quad \delta\mathcal{D}_{\alpha} V \Big| = -\xi^{\tilde{M}}\mathcal{D}_{\tilde{M}}\mathcal{D}_{\alpha} V \Big| = -(\sigma^{\mu}\bar{\epsilon})_{\alpha}v_{\mu} \neq 0\ ,\nn
\end{eqnarray}
meaning that supergravity transformations do not preserve the Wess-Zumino gauge. To remain in this gauge, we have to combine a supergravity transformation with a specific gauge transformation. Identifying the parameters of the gauge transformation which compensate the supergravity transformation, we finally obtain:
\begin{eqnarray}
\delta v_{\mu} &=& i\big( \epsilon\sigma_{\mu}\bar{\lambda} - \lambda\sigma^{\mu}\bar{\epsilon} \big) - \frac{i}{2} \big( \psi_{\mu}\sigma^{\nu} \bar{\epsilon} - \epsilon\sigma^{\nu}\bar{\psi}_{\mu} \big) v_{\nu}\ , \nonumber \\
\delta \lambda_{\alpha} &=& \big( \sigma^{\mu\nu}\epsilon\big)_{\alpha}\hat{F}_{\mu\nu} + i\epsilon_{\alpha}D\ , \nonumber \\
\delta\bar{\lambda}_{\dot{\alpha}} &=& -\big( \bar{\epsilon}\bar{\sigma}^{\mu\nu} \big)_{\dot{\alpha}}\hat{F}_{\mu\nu}-i\bar{\epsilon}_{\dot{\alpha}} D \ ,\nonumber \\
\delta D &=& -\epsilon\sigma^{\mu}\hat{D}\bar{\lambda} - \hat{D}_{\mu}\lambda \sigma^{\mu}\bar{\epsilon} + \frac{i}{2} b_{\mu} \big( \epsilon\sigma^{\mu}\bar{\lambda} - \lambda\sigma^{\mu}\bar{\epsilon} \big) + \frac{i}{2} \big( M^*\epsilon\cdot\lambda - M\bar{\epsilon}\cdot\bar{\lambda} \big)\ .\nn
\end{eqnarray}
Similarly, we also have to specify the variation of the chiral supermultiplets $\big( \phi, \chi_{\alpha}, F \big)$. Those transformations can be found in \cite{WessBagger}\cite{book_sugra}.

\subsection{Gauge interactions of chiral superfields}\label{sec:gaugexchiral}
Chiral superfields transform under gauge transformations as $\Phi\rightarrow e^{-2ig\Lambda}\Phi$. Terms like $\Phi^{\dagger}e^{-2gV}\Phi$ are then gauge invariant, as in supersymmetry. The gauge invariant kinetic term in supergravity can then be obtained through the couplings $\Phi^{\dagger}e^{-2gV}$ and $e^{-2gV}\Phi$. We then introduce (anti-)chiral  superfield using the "projector" presented in Subsection \ref{sec:chiralSF}:
\begin{equation}
\mathcal{X} = \left( \bar{\mathcal{D}}\cdot\bar{\mathcal{D}} - 8\mathcal{R}  \right)\Phi^{\dagger}e^{-2gV} \quad , \quad \mathcal{X}^{\dagger} = \left( \mathcal{D}\cdot\mathcal{D} - 8\mathcal{R}^{\dagger} \right)e^{-2gV}\Phi\ .\nn
\end{equation}
In the same manner as for the strength tensor $\mathcal{W}_{\alpha}$ in the Wess-Zumino gauge (where $V^3=0$) \autoref{eq:WZWa}, the superfields $\mathcal{X}$ and $\mathcal{X}^{\dagger}$ reduce to:
\begin{gather}
 \mathcal{X} = \Xi -2g\left( \bar{\mathcal{D}}\cdot\bar{\mathcal{D}} - 8\mathcal{R} \right)\Phi^{\dagger}V + 2g^2\left( \bar{\mathcal{D}}\cdot\bar{\mathcal{D}} - 8\mathcal{R} \right) \Phi^{\dagger} V^2\ ,\nn\\
 \mathcal{X}^{\dagger} = \Xi^{\dagger} -2g\left( \mathcal{D}\cdot\mathcal{D} - 8\mathcal{R}^{\dagger} \right)V\Phi + 2g^2\left( \mathcal{D}\cdot\mathcal{D} - 8\mathcal{R}^{\dagger} \right) V^2\Phi\ .\nn
\end{gather}
Calculation of all of the components of $\mathcal{X}$ can be found in \cite{WessBagger}\cite{book_sugra}.

\section{Construction of the action in curved superspace}
The superfields are now defined in the context of curved superspace. There is, however, still some aspects that must be analysed for constructing the supergravity action. The chiral density and new $\Theta$-variables allowing to simplify the computations will be introduced. We will then be able to construct a compact form of the supergravity Lagrangian invariant under a combination of Super-Weyl and Kähler transformations. 
\subsection{Chiral density and new hybrid variables $\big(\Theta^{\alpha} , \bar{\Theta}_{\dot{\alpha}}\big)$} \label{sec:actionsugra}
As we have seen in \autoref{eq:chiral} and \autoref{eq:antichiral}, the components of superfields are defined through derivatives with flat indices, whereas the expansion involves variables in the Einstein frame. This point leads to some difficulties in the development. It is then convenient to define a new set of variables $\xi^{\overline{M}} = \big(x^{\tilde{\mu}}, \Theta^{\alpha} , \bar{\Theta}_{\dot{\alpha}}  \big)$ such as components of chiral superfields are defined as their derivatives. We have then for the $\Phi$-expansion:
\begin{eqnarray}
\Phi(x,\Theta ) &=& \Phi(x,\theta )\Big| + \Theta\cdot\big( \mathcal{D}\Phi(x,\theta ) \big)\Big| - \frac14 \Theta\cdot\Theta\big( \mathcal{D}\cdot\mathcal{D}\Phi(x,\theta ) \big)\Big| \nonumber \\
&=& \phi(x) + \sqrt{2}\Theta^{\alpha}\chi_{\alpha}(x) - \Theta\cdot\Theta F(x)\ .\nonumber 
\end{eqnarray}
Taking for example the $\Theta$-expansion of $\Xi$ and $\mathcal{X}$ we get:
\begin{eqnarray}
\Xi &=& \Xi\big| + \sqrt{2}\Theta\cdot (\mathcal{D}\Xi) \big| - \frac14 \Theta\cdot\Theta (\mathcal{D}\cdot\mathcal{D}\Xi)\big| = 4F^{\dagger} + \frac43 M\phi^{\dagger}  \label{eq:XiComponents} \\
&& + \Theta\cdot\Big\{ -4i\sqrt{2}(\sigma^{\mu}\hat{D}_{\mu}\bar{\chi})+ \frac{2\sqrt{2}}{3}b_{\mu}(\sigma^{\mu}\bar{\chi})+\phi^{\dagger}\big( \frac83 (\sigma^{\mu\nu} \psi_{\mu\nu} ) -\frac{4i}{3}(\sigma^{\mu}\bar{\psi})M + \frac{4i}{3} \psi^{\mu}b_{\mu}\big)\Big\} \nonumber \\
&& - \Theta\cdot\Theta \Big\{ -4e_{\mu}{}^{\tilde{\mu}}\mathcal{D}_{\tilde{\mu}}\hat{D}^{\mu}\phi^{\dagger} + \frac{8i}{3}b^{\mu}\hat{D}^{\mu}\bar{\chi} - \frac{2\sqrt{2}}{3}\bar{\chi}\sigma^{\mu\nu}\bar{\psi}_{\mu\nu} - \frac83 M^*F^{\dagger} \nonumber\\
&& + \sqrt{2}i\bar{\chi}\cdot\bar{\psi}_{\mu}b^{\mu} + \frac{2\sqrt{2}}{3}\bar{\chi}\bar{\sigma}^{\nu\mu}\bar{\psi}_{\mu}b_{\nu} + \phi^{\dagger}\Big(  -\frac23 e_{\mu}{}^{\tilde{\mu}}e_{\nu}{}^{\tilde{\nu}}R_{\tilde{\mu}\tilde{\nu}}{}^{\mu\nu}\big| - \frac89 MM^* + \frac49 b_{\mu}b^{\mu} \nonumber \\
&& + \frac{4i}{3}e_{\mu}{}^{\tilde{\mu}}\mathcal{D}_{\tilde{\mu}}b^{\mu} + \frac23 \bar{\psi}_{\mu}\cdot \bar{\psi}^{\mu}M + \frac23 \psi_{\nu}\sigma^{\nu}\bar{\psi}_{\mu}b^{\mu} + \frac{4i}{3}\bar{\psi}^{\mu}\bar{\sigma}^{\nu}\psi_{\mu\nu} \nonumber \\
&& +\frac16\epsilon^{\mu\nu\rho\sigma}(\psi_{\mu}\sigma_{\sigma}\bar{\psi}_{\nu\rho} + \bar{\psi}_{\mu}\bar{\sigma}_{\sigma}\psi_{\nu\rho})\Big\} \ ,\nn
\end{eqnarray}
\begin{eqnarray}
\mathcal{X} &=& \mathcal{X}\big| + \sqrt{2}\Theta\cdot (\mathcal{D}\mathcal{X}) \big| - \frac14 \Theta\cdot\Theta (\mathcal{D}\cdot\mathcal{D}\mathcal{X})\big|  \nonumber \\
&=& \Xi + \Theta\cdot \Big\{ -4\sqrt{2}g(\sigma^{\mu}\bar{\chi})v_{\mu} + 8ig\phi^{\dagger}\lambda + 4ig\phi^{\dagger}v_{\nu}(\sigma^{\mu}\bar{\sigma}^{\nu}\psi_{\mu})\Big\} \nonumber \\
&& -\Theta\cdot \Theta \Big\{ 8igv_{\mu}\hat{D}^{\mu}\phi^{\dagger} - 4\sqrt{2}ig\bar{\chi}\cdot\bar{\lambda} + 4ig\phi^{\dagger}\hat{D}_{\mu}v^{\mu} + \frac83 g\phi^{\dagger}v_{\mu}b^{\mu} - 4gD\phi^{\dagger} + 4g^2\phi^{\dagger}v_{\mu}v^{\mu} \Big\} \ .\nonumber 
\end{eqnarray}
We have seen in Subsection \ref{sec:chiralSF} that the variation of the auxiliary field $F$ is not a total derivative. However, by defining the chiral density $\Delta$, which plays the rôle of an invariant measure in curved superspace, the action can be written in a fully covariant form:
\begin{eqnarray}
\mathcal{S}=\int d^4x\,d^2\Theta \Delta f(\Phi) + \mathrm{h.c.} \label{eq:S}
\end{eqnarray}
with $f(\Phi)$ a holomorphic function of chiral superfields. By imposing that the action transforms under a supergravity transformation \autoref{eq:sugratrso} as a total derivative, the chiral density must then transform as:
\begin{eqnarray}
\delta\Delta = - \big( - \big)^{| \bar{M} |}\partial_{\bar{M}}\big( \eta^{\bar{M}}\Delta \big)\label{eq:deltadensity}
\end{eqnarray}
with $\eta^{\bar{M}}$ the supergravity transformation parameters written with the new $\Theta-$variables:
\begin{gather}
\eta^{\bar{M}}(x,\Theta) = \eta^{\bar{M}}_{(0)}(x) + \Theta^{\alpha}\eta^{\bar{M}}_{(1)\alpha}(x) + \Theta\cdot\Theta \eta^{\bar{M}}_{(2)}(x)\ . \nn
\end{gather}
From the invariance of \autoref{eq:S} under supergravity transformations, the superfield $\Delta\Phi$ must then transforms as a chiral density, \textit{i.e.}, $\delta(\Delta\Phi)=-(-)^{\bar{M}}\partial_{\bar{M}}\left( \eta^{\bar{M}}\Delta\Phi \right)$ which leads to the transformation of the chiral superfield $\Phi$:
\begin{eqnarray}
\delta \Phi = - \eta^{\bar{M}}\partial_{\bar{M}}\Phi \ .\nonumber
\end{eqnarray}
Developing this relation reads:
\begin{gather}
\delta\Phi = \big( -\eta^{\tilde{\mu}}_{(0)}\partial_{\tilde{\mu}}\phi - \sqrt{2}\eta^{\alpha}_{(0)}\chi_{\alpha} \big) + \Theta^{\alpha}\big( \sqrt{2}\eta_{\alpha(0)}F - \eta^{\mu}_{(0)}\partial_{\tilde{\mu}}\chi_{\alpha} - \frac{1}{\sqrt{2}}\eta^{\tilde{\mu}}_{(1)\alpha}\partial_{\tilde{\mu}}\phi - \eta^{\beta}_{(1)\alpha}\chi_{\beta} \big) \nn \\
 - \Theta\cdot\Theta \big( -\eta^{\tilde{\mu}}_{(0)}\partial_{\tilde{\mu}}F + \eta^{\tilde{\mu}}_{(2)}\partial_{\tilde{\mu}}\phi - \frac{1}{\sqrt{2}}\eta^{\tilde{\mu}}_{(1)}{}^{\alpha}\partial_{\tilde{\mu}}\chi_{\alpha} + \sqrt{2}\eta^{\alpha}_{(2)}\chi_{\alpha} - \eta^{\alpha}_{(1)\alpha}F \big)\ .\nn
\end{gather}
By comparing this result with (\ref{eq:chiraltrso1}, \ref{eq:chiraltrso2}, \ref{eq:chiraltrso3}), the form of $\eta^{\bar{M}}$ can be found:
\begin{eqnarray}
\eta^{\tilde{\mu}} &=& 2i\Theta\sigma^{\tilde{\mu}}\bar{\epsilon} + \Theta\cdot\Theta \bar{\psi}_{\tilde{\nu}}\bar{\sigma}^{\tilde{\mu}}\sigma_{\tilde{\nu}}\bar{\epsilon} \ ,\nonumber \\
\eta^{\alpha} &=& \epsilon^{\alpha} - i\Theta\sigma^{\tilde{\mu}}\bar{\epsilon}\psi_{\tilde{\mu}}{}^{\alpha} + \Theta\cdot\Theta \Big( \frac13 M^*\epsilon^{\alpha} - i\omega_{\tilde{\mu}}{}^{\alpha\beta}\big( \sigma^{\tilde{\mu}}\bar{\epsilon} \big)_{\beta} + \frac16 b_{\mu}\big( \bar{\epsilon}\bar{\sigma}^{\mu} \big)^{\alpha} - \frac12 \psi_{\tilde{\nu}}{}^{\alpha}\bar{\psi}_{\tilde{\mu}}\bar{\sigma}^{\tilde{\nu}}\sigma^{\tilde{\mu}}\bar{\epsilon} \Big)\ . \nonumber
\end{eqnarray}
Knowing the form of $\eta^{\bar{M}}(x,\Theta)$ and using \autoref{eq:deltadensity}, the variation of the components of the chiral density can also be calculated. \medskip

The link with the vielbein can now be made. By selecting a chiral density $\mathcal{E}$ such as its lowest component reduces to $\mathcal{E}\big| = \mathrm{det}\big( e_{\tilde{\mu}}{}^{\mu} \big)\equiv e$, we then introduce a relation between the chiral density and the vierbein. Step-by-step, it is possible to identify the components of $\mathcal{E}$, following \autoref{eq:deltadensity} and the transformation of the supergravity multiplet under supergravity transformations (\ref{eq:sugramltp_trsfo1}, \ref{eq:sugramltp_trsfo2}):
\begin{eqnarray}
\mathcal{E} &=& e\Big( 1 + i\Theta\sigma^{\tilde{\mu}}\bar{\psi}_{\tilde{\mu}}-\Theta\cdot\Theta \big( M^* + \bar{\psi}_{\tilde{\mu}}\bar{\sigma}^{\tilde{\mu}\tilde{\nu}}\bar{\psi}_{\tilde{\nu}} \big) \Big)\ . \label{eq:chiraldensity}
\end{eqnarray}

\subsection{Compact form of the supergravity action}
We will now turn to the construction of the invariant action in supergravity. We suppose a compact real Lie algebra $\mathfrak{g}$ and the vector superfields $V=V^aT_a$ associated to the adjoint representation of $\mathfrak{g}$. The chiral superfields $\Phi^i$ are in a representation $\mathfrak{R}$ of $\mathfrak{g}$ whilst the anti-chiral superfields $\Phi^{\dagger}_{i^*}$ are in the complex conjugate representation $\overline{\mathfrak{R}}$. The (anti-) chiral superfields contain the matter fields whereas the vector superfields $V$ introduce the gauge interaction in the theory.\medskip

All the necessary tools have been introduced to construct a compact form of the $N=1$ supergravity action in four dimensions. First of all, we recall the action of an $N=1$ supersymmetry theory (derived in \cite{WessBagger}\cite{book_sugra}). Three functions are needed to fully specify the dynamics of the model: the superpotential $W(\Phi)$ a holomorphic function leading to the Yukawa interactions, the Kähler potential $K(\Phi,\Phi^{\dagger})$, a real function defining the kinetic terms and the gauge kinetic function $h_{ab}$, a holomorphic function ($a$ and $b$ are indices in the adjoint representation of $G$). Recall that $N=1$ supersymmetry action can be written as:
\begin{eqnarray}
\mathcal{S}_{SUSY} &=& \Bigg(\frac12 \int d^4x d^2\theta \left(-\frac14 \bar{D}\cdot\bar{D}K(\Phi,\Phi^{\dagger}e^{-2gV})\right) \nonumber \\
&& +  \int d^4x d^2\theta W(\Phi) +\frac{1}{16g^2}\int d^4xd^2\theta h_{ab} \mathcal{W}^{\alpha a}\mathcal{W}_{\alpha}^b + \mathrm{h.c.} \Bigg) \label{eq:susyaction}
\end{eqnarray}
with $D_{\alpha}$ the covariant derivative of supersymmetry. \medskip

For extending this action in the supergravity context we must:
\begin{itemize}
\item[1)] substitute the operator $D\cdot D$ by $\mathcal{D}\cdot \mathcal{D} - 8\mathcal{R}^{\dagger}$,
\item[2)] change the Grassmann variable $\theta$ to the new variables $\Theta$,
\item[3)] add the chiral density $\mathcal{E}$ to have invariance under supergravity transformations.
\end{itemize}
The $N=1$ supergravity action in four dimensions can then be written as:
\begin{eqnarray}
\mathcal{S} &=& \frac38 \int d^4x\, d^2\Theta\mathcal{E}\Big( \bar{\mathcal{D}}\cdot\bar{\mathcal{D}} - 8\mathcal{R} \Big)e^{-\frac13 K(\Phi,\Phi^{\dagger}e^{-2gV})} \nonumber \\
&& + \int d^4x\, d^2\Theta \mathcal{E}W(\Phi) + \frac{1}{16g^2}\int d^4x\, d^2\Theta \mathcal{E}h_{ab}(\Phi)\mathcal{W}^{\alpha a}\mathcal{W}_{\alpha}{}^b + \mathrm{h.c.} \ . \label{eq:sugraaction}
\end{eqnarray}
It can be observed that the term involving the Kähler potential is not obtained from the process mentioned above. This form is essential to get a correctly normalised action. Moreover, we see that reintroducing the Planck mass in the exponential and taking the low energy limit ($m_p\rightarrow \infty$):
\begin{gather}
\frac38 m_p^2 e^{-\frac{1}{3m_p^2}K(\Phi,\Phi^{\dagger}e^{-2gV})} = \frac{3}{8}m_p^2-\frac18 K(\Phi,\Phi^{\dagger}e^{-2gV}) + \mathcal{O}\left( 1/m_p \right)\nn 
\end{gather}
leads to the correct kinetic terms in \autoref{eq:susyaction} with the pure gravity part of the action (given by \autoref{eq:PURE_SUGRA}).

\subsection{Super-Weyl rescaling and Kähler transformation}
In Subsection \ref{subsec:weyl}, we have introduced Super-Weyl transformations and obtained the transformation laws of the fundamental objects of supergravity. We will see that Super-Weyl transformations coupled to Kähler transformations are symmetries of the supergravity action \autoref{eq:sugraaction}. \medskip

The transformation of the chiral density $\mathcal{E}$ under super-Weyl transformations can be complex to compute. It is then more convenient to analyse the variation components-by-components, \textit{i.e.}, identifying $\delta_{\Sigma}e$ and $\delta_{\Sigma}\bar{\psi}_{\tilde{\mu}}$ with $\Sigma$ the transformation parameter (a chiral superfield). Introducing the spinorial superfield:
\begin{gather}
S^{\alpha} = \Theta^{\alpha} \big( 2\Sigma^{\dagger} - \Sigma \big)\big| + \Theta\cdot\Theta \mathcal{D}^{\alpha}\Sigma\big|\nn
\end{gather}
and regrouping the variations of the various fields in $\mathcal{E}$ we can rewrite the transformation $\delta_{\Sigma}\mathcal{E}$ as:
\begin{eqnarray}
\delta_{\Sigma}\mathcal{E} &=& 6\Sigma\mathcal{E} + \partial_{\Theta^{\alpha}}\big( S^{\alpha}\mathcal{E} \big) \ . 
\end{eqnarray}
General Super-Weyl transformations acting on chiral and vector superfields are defined as:
\begin{gather}
\delta_{\Sigma}\Phi = w\Sigma\Phi \quad , \quad \delta_{\Sigma}V = w' \big( \Sigma + \Sigma^{\dagger} \big)V \ ,\nn
\end{gather}
where $w$ and $w'$ are the Weyl weights of respectively the chiral and the vector superfield. In the $\Theta$-formalism, the variation of the chiral superfield modifies as:
\begin{eqnarray}
\delta_{\Sigma}\Phi &=& w\Sigma\Phi - S^{\alpha}\partial_{\Theta^\alpha}\Phi\ .\nn
\end{eqnarray}
By assuming a vanishing weight for the chiral superfield, \textit{i.e.}, $\delta_{\Sigma}\Phi = -S^{\alpha}\partial_{\Theta^{\alpha}}\Phi$, as well as for the vector superfield, the superpotential $W(\Phi)$ and the gauge kinetic function $h_{ab}$ have also vanishing weight. Thus:
\begin{gather}
\delta_{\Sigma}\big( \mathcal{E}W \big) = 6\Sigma\mathcal{E}W + \partial_{\Theta^{\alpha}}\big( S^{\alpha}\mathcal{E}W \big)\quad , \quad \delta_{\Sigma} \mathcal{W}_{\alpha} = -3\Sigma\mathcal{W}_{\alpha} - S^{\beta}\partial_{\Theta^{\beta}}\mathcal{W}_{\alpha}\ ,\nn \\
\delta_{\Sigma}\Big( \mathcal{E}\left( \bar{\mathcal{D}}\cdot\bar{\mathcal{D}}-8\mathcal{R} \right)e^{-\frac13 K(\Phi,\Phi^{\dagger}e^{-2gV})} \Big) =\mathcal{E}\left( \bar{\mathcal{D}}\cdot\bar{\mathcal{D}} - 8\mathcal{R} \right)\left( 2\left( \Sigma + \Sigma^{\dagger}\right)e^{-\frac13K(\Phi,\Phi^{\dagger}e^{-2gV})} \right) \nn\\
+ \partial_{\Theta^{\alpha}}\left( S^{\alpha}\mathcal{E}\left( \bar{\mathcal{D}}\cdot\bar{\mathcal{D}} - 8\mathcal{R} \right)e^{-\frac13K(\Phi,\Phi^{\dagger}e^{-2gV})} \right) \nn
\end{gather}
which leads (considering now a finite transformation) to the following variation of the supergravity Lagrangian
\begin{eqnarray}
\mathcal{L} &\rightarrow& \frac38 \int d^2\Theta \mathcal{E} \big( \bar{\mathcal{D}}\cdot\bar{\mathcal{D}} -8\mathcal{R}\big)e^{2\big( \Sigma + \Sigma^{\dagger} \big)}e^{-\frac13 K(\Phi, \Phi^{\dagger}e^{-2gV})}+ \int d^2\Theta\mathcal{E}e^{6\Sigma}W(\Phi) \nonumber \\
&& + \frac{1}{16g^2}\int d^2\Theta \mathcal{E}h(\Phi)_{ab}\mathcal{W}^{\alpha a}\mathcal{W}_{\alpha}^{b} + \mathrm{h.c.}\ .\label{eq:action:SuperWeyl}
\end{eqnarray}
For further use, it is also important to give the transformations of the various fields of the theory under Weyl transformations:
\begin{eqnarray}
e_{\mu}{}^{\tilde{\mu}} \rightarrow e^{(\Sigma + \Sigma^{\dagger})|}e_{\mu}{}^{\tilde{\mu}} \quad , \quad e&\rightarrow& e^{2(\Sigma + \Sigma^{\dagger})|}e \quad , \quad \chi^i \rightarrow e^{(\Sigma - 2 \Sigma^{\dagger})|}\chi^i \ ,\label{eq:rescale} \\
\lambda_a\rightarrow e^{-3\Sigma|}\lambda_a \quad , \quad \bar{\psi}_{\tilde{\mu}} & \rightarrow & e^{(2\Sigma - \Sigma^{\dagger})|}\big( \bar{\psi}_{\tilde{\mu}} - i\bar{\sigma}_{\tilde{\mu}}\mathcal{D}\Sigma | \big) \quad , \quad v_\mu^a\rightarrow e^{-(\Sigma + \Sigma^{\dagger})|}v_{\mu}^a\ , \nonumber \\
D^a&\rightarrow& e^{-2(\Sigma + \Sigma^{\dagger})|}D^a \ .\nonumber 
\end{eqnarray}
Note that this transformation induces also a gravitino-shift which will be important in the sequel. \medskip

Two points can be mentioned:
\begin{itemize}
\item[1)] the action is not invariant under Super-Weyl (superconformal) transformations,
\item[2)] the gauge term is invariant under Super-Weyl (superconformal) transformations.
\end{itemize}  
We now consider another type of transformation called Kähler transformation. This transformation involves a modification of the Kähler potential as
\begin{eqnarray}
K(\Phi,\Phi^{\dagger}) \rightarrow K(\Phi,\Phi^{\dagger}) + F(\Phi) + F^*(\Phi^{\dagger})\ ,\nn
\end{eqnarray}
where $F(\Phi)$ is a gauge-singlet holomorphic function of the chiral superfields $\Phi$. Imposing that the superpotential transforms as:
\begin{eqnarray}
W \rightarrow e^{-F} W\ , \nonumber
\end{eqnarray}
the covariant derivative of the superpotential $\mathcal{D}_iW=W_i + K_i W$ transforms as:
\begin{gather}
\mathcal{D}_iW \rightarrow e^{-F} \mathcal{D}_iW\ .\nn
\end{gather}
Applying a Kähler transformation to the supergravity Lagrangian leads to:
\begin{eqnarray}
\mathcal{L}\rightarrow\mathcal{L}'&=& \frac38 \int d^2\Theta \mathcal{E} \big( \bar{\mathcal{D}}\cdot\bar{\mathcal{D}} -8\mathcal{R}\big)e^{-\frac13 K}e^{-\frac13 \big( F + F^{\dagger} \big)}+ \int d^2\Theta\mathcal{E}e^{-F}W(\Phi) \nonumber \\
&& + \frac{1}{16g^2}\int d^2\Theta \mathcal{E}h(\Phi)_{ab}\mathcal{W}^{\alpha a}\mathcal{W}_{\alpha}^{b} + \mathrm{h.c.} \ .\label{eq:KtrsfoAction}
\end{eqnarray}
Comparing the transformations \autoref{eq:KtrsfoAction} and \autoref{eq:action:SuperWeyl}, the Kähler transformation can then be seen as a superconformal transformation. Combining the two transformations with $F=6\Sigma$ produce then a symmetry of the supergravity action.

\section{Computation of the supergravity action}
All the tools are now present for expanding in components the Lagrangian of supergravity (\autoref{eq:sugraaction}). The pure supergravity action will be given, and the gauge \& matter coupled to supergravity will be expanded. The second part is simple but lengthy. The explicit calculation can be found in \cite{WessBagger}\cite{book_sugra}. \newline
We will also see that this expansion generates a Brans-Dicke normalisation \cite{BransDicke} in the kinetic terms. A Super-Weyl rescaling must then be applied in order to recover correctly normalised couplings. 
\subsection{The pure supergravity action}
The first term in \autoref{eq:sugraaction} contains the pure supergravity Lagrangian (when the Kähler potential vanish $K=0$):
\begin{eqnarray}
\mathcal{L}_{\mathrm{Pure\, SUGRA}} = -3 \int d^2\Theta \, \mathcal{E}\mathcal{R} + \mathrm{h.c.}\ .\label{eq:PURE_SUGRA}
\end{eqnarray}
Using the various derivatives of the chiral superfield $\mathcal{R}$ \autoref{eq:chiraldensity} and the chiral density $\mathcal{E}$ \autoref{eq:chiraldensity}, the computation is straightforward:
\begin{eqnarray}
\mathcal{L}_{\mathrm{Pure\, SUGRA}} &=& \frac12 ee_{\mu}{}^{\tilde{\mu}}e_{\nu}{}^{\tilde{\nu}}R_{\tilde{\mu}\tilde{\nu}}{}^{\mu\nu}\Big| + \frac14 e\epsilon^{\mu\nu\rho\sigma}\Big( \psi_{\mu}\sigma_{\sigma}\bar{\psi}_{\nu\rho} - \bar{\psi}_{\mu}\bar{\sigma}_{\sigma}\psi_{\nu\rho} \Big) \nn\\
&&- \frac13 e \big( |M|^2 + b_\mu b^\mu \big) \ .\label{eq:lagPureSUGRA}
\end{eqnarray}
The first term corresponds to the Einstein-Hilbert Lagrangian for the helicity-2 graviton whilst the second is the Rarita-Schwinger Lagrangian describing the spin-$3/2$ gravitino. The two last terms involving $M$ and $b_{\mu}$ show that those fields does not propagate, \textit{i.e.}, they are the two auxiliary fields of the supergravity multiplets.

\subsection{Matter / Gauge coupling to Supergravity}\label{subsec:mattergauge}
To calculate the matter-gauge interactions couplings, we define the kinetic energy $\Omega$:
\begin{equation}
\Omega (\Phi, \Phi^{\dagger}) = -3e^{-\frac13 K(\Phi,\Phi^{\dagger})} = W_I(\Phi^{\dagger})W^I(\Phi)\ ,\label{eq:Omega}
\end{equation}
where $W^{I}$ and $W_I$ are (anti-)holomorphic functions depending on chiral (anti-chiral) superfields. Analogously as in supersymmetric theory (see \cite{WessBagger}, \cite{book_sugra} and \cite{book_susy}) we have:
\begin{eqnarray}
W^I(\Phi) &=& W^{I} + \sqrt{2}\Theta\cdot \big( W^{I}_i\chi^i \big) - \Theta\cdot\Theta \big( W^{I}_iF^i + \frac12 W_{ij}^{I} \chi^i\cdot\chi^j \big)\ , \label{eq:Wdev} \\
W_I(\Phi^{\dagger}) &=& W_{I} + \sqrt{2}\bar{\Theta}\cdot \big( W_{I}^{i*}\bar{\chi}_{i*} \big) - \bar{\Theta}\cdot\bar{\Theta} \big( W_{I}^{i*}F_{i*}^{\dagger} + \frac12 W^{i^*j^*}_{I}  {}{\bar{\chi}_{i*}\cdot\bar{\chi}_{j*}} \big) \label{eq:Wbardev}
\end{eqnarray}
(we introduce now $W^{I}\equiv W^{I}(\phi)$ and similarly for $W_{I}$). The development in components of the various terms is relatively easy but time-consuming. We will only give the main steps of the calculation. \medskip 

Considering the Kähler part of the supergravity Lagrangian, using the definition of $\Omega(\Phi, \Phi^{\dagger})$ \autoref{eq:Omega}, it follows:
\begin{equation}
\mathcal{L}_{K} = \int d^2\Theta\, \frac38 \mathcal{E}\big( \bar{\mathcal{D}}\cdot\bar{\mathcal{D}}-8\mathcal{R} \big)e^{-\frac13 K(\Phi,\Phi^{\dagger}e^{-2gV})} = -\int d^2\Theta\, \frac18 \mathcal{E}\big( \bar{\mathcal{D}}\cdot\bar{\mathcal{D}}-8\mathcal{R} \big)\Omega(\Phi,\Phi^{\dagger}) \ .\label{eq:lagkahler}
\end{equation}

We first observe that since $\bar{\mathcal{D}}\cdot\bar{\mathcal{D}} - 8\mathcal{R}$ acts as a projector from anti-chiral to chiral superfields (see Subsection \ref{sec:chiralSF}):
\begin{eqnarray}
\big( \bar{\mathcal{D}}\cdot\bar{\mathcal{D}}-8\mathcal{R} \big)\Omega(\Phi,\Phi^{\dagger}) &=& \Big( \big( \bar{\mathcal{D}}\cdot\bar{\mathcal{D}}-8\mathcal{R} \big)W_{I}(\Phi^{\dagger}e^{-2gV}) \Big)W^I(\Phi) \nonumber \\
&\equiv & \mathcal{X}_{I} W^I(\Phi)\ . \nonumber 
\end{eqnarray}
Using the Wess-Zumino gauge \autoref{eq:WZgauge} (with $V^3=0$), the expression of $\mathcal{X}_{I}$ can be decomposed in three terms:
\begin{eqnarray}
\mathcal{X}_{I} &=& \left( \bar{\mathcal{D}}\cdot\bar{\mathcal{D}} - 8\mathcal{R} \right)W_{I}(\Phi^{\dagger}e^{-2gV}) \nn\\
&=& \left( \bar{\mathcal{D}}\cdot\bar{\mathcal{D}} - 8\mathcal{R} \right)\Big(  W_{I}(\Phi^{\dagger}) + \frac{\partial W_{I}(\Phi^{\dagger})}{\partial\Phi^{\dagger}_{i^*}}\left( -2g(\Phi^{\dagger}V)_{i^*} + 2g^2(\Phi^{\dagger}V^2)_{i^*} \right) \nn\\
&& + 2g^2\frac{\partial^2 W_{I}(\Phi^{\dagger})}{\partial\Phi^{\dagger}_{i^*}\partial\Phi^{\dagger}_{j^*}}(\Phi^{\dagger}V)_{i^*}(\Phi^{\dagger}V)_{j^*}\Big)\nn \\
&=& \Xi_{I} + \mathcal{X}_{I}^{(g)}+ \mathcal{X}_{I}^{(g^2)}\ ,	\nn
\end{eqnarray}
with $\Xi_{I}$ defined as in \autoref{eq:XiComponents} by substituting the components of the superfield $\Phi^{\dagger}$ by the components of $W_{I}(\Phi^{\dagger})$. The product of this term with the chiral density $\mathcal{E}$ \autoref{eq:chiraldensity} leads then to $\mathcal{L}_{K}$.\medskip

The computation of the superpotential and the gauge parts of the supergravity Lagrangian
\begin{eqnarray}
\mathcal{L}_{W+gauge} &=& \int d^2\Theta \mathcal{E}W(\Phi) + \frac{1}{16g^2}\int d^2\Theta \mathcal{E}h_{ab}(\Phi)\mathcal{W}^{\alpha a}\mathcal{W}^{b}_{\alpha} + \text{h.c.} \nn
\end{eqnarray}
is easier after decomposing the gauge kinetic function $h_{ab}(\Phi)$ in the same manner as $W_{I}(\Phi)$ \autoref{eq:Wdev} and using $\mathcal{W}_{\alpha}$ in \autoref{eq:Walphacompo1}.\medskip

The final Lagrangian takes the form:
\begin{gather}
\mathcal{L} = \mathcal{L}_{kin} + \mathcal{L}_{int} + \mathcal{L}_{aux}\nn  
\end{gather}
with $\mathcal{L}_{kin}$ containing the kinetic terms, $\mathcal{L}_{int}$ the interaction terms and $\mathcal{L}_{aux}$ the terms evolving the auxiliary fields $(F^i,D^a,M,b_{\mu})$. Note that this last term can be eliminated through the Euler-Lagrange relations leading to new couplings:
\begin{equation}
\frac{\partial\mathcal{L}_{aux}}{\partial N} = \frac{\partial\mathcal{L}_{aux}}{\partial N^*} = \frac{\partial\mathcal{L}_{aux}}{\partial F^i} = \frac{\partial\mathcal{L}_{aux}}{\partial F^{\dagger}_{i^*}} = \frac{\partial\mathcal{L}_{aux}}{\partial b_{\mu}} = \frac{\partial\mathcal{L}_{aux}}{\partial D^a} = 0 \nonumber
\end{equation}
with $N=M+3(\mathrm{ln}\Omega)^{i^*}F^{\dagger}_{i^*}$. \medskip

However, it turns out that after eliminated the auxiliary fields, the various kinetic terms are not correctly normalised. For instance, we obtain for gravity:
\begin{eqnarray}
\mathcal{L}_{K} = -\frac16 e\Omega e_{\mu}{}^{\tilde{\mu}}e_{\nu}{}^{\tilde{\nu}}R_{\tilde{\mu}\tilde{\nu}}{}^{\mu\nu}\big| \label{eq:LK}
\end{eqnarray} 
which corresponds to a non-proper normalisation (called Brans-Dicke normalisation). To obtain a correctly normalised action, we can perform a Weyl rescaling (see \autoref{eq:rescale}):
\begin{eqnarray}
e_{\mu}{}^{\tilde{\mu}} \rightarrow e^{-\lambda} e_{\mu}{}^{\tilde{\mu}} \quad \quad (e\rightarrow e^{4\lambda} e) \ .\label{eq:rescalee}
\end{eqnarray}
Choosing $e^{2\lambda} = - \frac{3}{\Omega}$, we get for the gravity kinetic term:
\begin{equation}
-\frac16 e\Omega e_{\mu}{}^{\tilde{\mu}}e_{\nu}{}^{\tilde{\nu}}R_{\tilde{\mu}\tilde{\nu}}{}^{\mu\nu}\Big| \rightarrow  -\frac16 e^{2\lambda}e\Omega e_{\mu}{}^{\tilde{\mu}}e_{\nu}{}^{\tilde{\nu}}R_{\tilde{\mu}\tilde{\nu}}{}^{\mu\nu}\Big| + \cdots = \frac12 e e_{\mu}{}^{\tilde{\mu}}e_{\nu}{}^{\tilde{\nu}}R_{\tilde{\mu}\tilde{\nu}}{}^{\mu\nu}\Big| + \cdots \nonumber
\end{equation}
where the dots indicate other terms coming from the transformation of the spin-connection $\omega_{\tilde{\mu}MN}$. Similarly, the fermions kinetic terms read after the expansion, 
\begin{gather}
\mathcal{L}_{kin.\ fermions} = -\frac{1}{12}e\Omega\epsilon^{\mu\nu\rho\sigma}\left( \psi_{\mu}\sigma_{\sigma}\bar{\psi}_{\nu\rho} - \bar{\psi}_{\mu}\bar{\sigma}_{\sigma}\psi_{\nu\rho} \right) + \frac{\sqrt{2}}{3}e\left( \Omega_i\chi^i\sigma^{\mu\nu}\psi_{\mu\nu} + \Omega^{i^*}\bar{\chi}_{i^*}\bar{\sigma}^{\mu\nu}\bar{\psi}_{\mu\nu} \right) \nn\\
-\frac{i}{2}e\Omega^{i^*}{}_{i}\left( \chi^i\sigma^{\mu}\tilde{\mathcal{D}}_{\mu}\bar{\chi}_{i^*} - \tilde{\mathcal{D}}_{\mu}\chi^i\sigma^{\mu}\bar{\chi}_{i^*}\right)\ ,\nn
\end{gather}
After the Weyl rescaling
\begin{gather}
\psi_{\tilde{\mu}}\rightarrow e^{\frac12\lambda}\psi_{\tilde{\mu}} \quad , \quad \bar{\psi}_{\tilde{\mu}}\rightarrow e^{\frac12\lambda}\bar{\psi}_{\tilde{\mu}} \quad , \quad \chi^i\rightarrow e^{-\frac12 \lambda}\chi^i \quad , \quad \bar{\chi}_{i^*}\rightarrow e^{-\frac12 \lambda}\bar{\chi}_{i^*}\ , \nn
\end{gather}
the fermions kinetic part are correctly normalised, but the gravitino-fermion mixing term remains present. This rescaling must then be followed by the gravitino-shift:
\begin{equation}
\psi_{\tilde{\mu}} \rightarrow \psi_{\tilde{\mu}} + \frac{\sqrt{2}i}{2}\Omega^{-1}\Omega^{i^*}\sigma_{\tilde{\mu}}\bar{\chi}_{i^*}\ ,\label{eq:shift32}
\end{equation}
which suppress this mixing term. Those two transformations can be embedded in a single Super-Weyl transformations \autoref{eq:rescale} with the transformation parameters\footnote{The transformation can be fully embedded in a single Super-Weyl transformation since the Weyl rescaling and the gravitino-shift commute.}:
\begin{gather}
\Sigma= \frac12\lambda + \sqrt{2}\Theta\cdot \lambda_i\chi^i - \Theta\cdot\Theta\mathcal{F} \quad , \quad \Sigma^{\dagger}= \frac12\lambda + \sqrt{2}\bar{\Theta}\cdot \bar{\lambda}^{i^*}\bar{\chi}_{i^*} - \bar{\Theta}\cdot\bar{\Theta}\mathcal{F}^{\dagger}\ . \label{eq:SigmaSUGRA}
\end{gather}
This transformation is the key point of the complexity of the calculus.\medskip

Applying the Weyl transformations associated with the gravitino-shift to the supergravity action \autoref{eq:sugraaction} allow to obtain a correctly normalised action. This is a very tedious computation, mostly due to the presence of the gravitino-shift \autoref{eq:shift32}. \medskip

The properly normalised Lagrangian of the $N=1$ supergravity in four dimensions follows:
\begin{gather}
\mathcal{L}_{SUGRA} = \mathcal{L}_{kin.} + \mathcal{L}_{int.}\ . \label{eq:SUGRAlagrangian}
\end{gather}
The kinetic part reads:
\begin{gather}
\mathcal{L}_{kin.} = \frac12 e_{\mu}{}^{\tilde{\mu}}e_{\nu}{}^{\tilde{\nu}}R_{\tilde{\mu}\tilde{\nu}}{}^{\mu\nu}\Big| + \frac14 e\epsilon^{\mu\nu\rho\sigma}\big[ \psi_{\mu}\sigma_{\sigma}\bar{\psi}_{\nu\rho} - \bar{\psi}_{\mu}\bar{\sigma}_{\sigma}\psi_{\nu\rho} \big] -\frac14e h_{ab}{}^{R}F^a_{\mu\nu}F^{\mu\nu b}+\frac14e h_{ab}{}^{I}F^a_{\mu\nu}{}^*F^{\mu\nu b} \nn\\
-\frac{i}{2}eh_{ab}^{R}\big( \lambda^a\sigma^{\mu}\widecheck{\mathcal{D}}_{\mu}\bar{\lambda}^b - \widecheck{\mathcal{D}}_{\mu}\lambda^a\sigma^{\mu}\bar{\lambda}^b \big) + \frac12 e h_{ab}^{I}\widecheck{\mathcal{D}}_{\mu}\left( \lambda^a\sigma^{\mu}\bar{\lambda}^b \right) - \frac{i}{2}eK^{i^*}{}_{i}\Big[ \chi^i\sigma^{\mu}\widecheck{\mathcal{D}}_{\mu}\bar{\chi}_{i^*} - \widecheck{\mathcal{D}}_{\mu}\chi^i\sigma^{\mu}\bar{\chi}_{i^*} \Big] \nn\\
+ eK^{i^*}{}_{i}\tilde{\mathcal{D}}_{\mu}\phi^i\tilde{\mathcal{D}}^{\mu}\phi^\dagger_{i^*}-\frac{\sqrt{2}}{2}eK^{i^*}{}_{i}\left( \chi^i\sigma^{\mu}\bar{\sigma}^{\nu}\psi_{\mu}\tilde{\mathcal{D}}_{\nu}\phi^{\dagger}_{i^*} + \bar{\chi}_{i^*}\bar{\sigma}^{\mu}\sigma^{\nu}\bar{\psi}_{\mu}\tilde{\mathcal{D}}_{\nu}\phi^i \right)\nn\\
+\frac{i}{4}eh^R_{ab}\left( \lambda^a\sigma_{\mu}\bar{\sigma}^{\nu\rho}\bar{\psi}^{\mu}+\bar{\lambda}^a\bar{\sigma}_{\mu}\sigma_{\nu\rho}\psi^{\mu} \right)\left( F^{\nu\rho b } + \hat{F}^{\nu\rho b}\right) - \frac{\sqrt{2}}{4}e\left( h_{ab i}\chi^i\sigma^{\mu\nu}\lambda^b + h^{\star}_{ab}{}^{i^*}\bar{\chi}_{i^*}\bar{\sigma}^{\mu\nu}\bar{\lambda}^b \right)F^a_{\mu\nu} \nn
\end{gather}
containing a kinetic term for the graviton $e_{\tilde{\mu}}{}^{\mu}$, the gravitino $\psi_{\mu}$, the gauge bosons $v_{\mu}^a$, the gauginos $\lambda^a$, the scalar and fermionic fields $\phi^i$ and $\chi^i$.\medskip

The interactions part is given by:
\begin{gather}
\mathcal{L}_{int.} = \frac{e}{4}f\left( K_i\left(T_a\phi\right)^i + K^{i^*}\left( \phi^{\dagger}T_a\right)_{i^*} \right)\left( \bar{\lambda}^a\bar{\sigma}^{\mu}\psi_{\mu}+ \bar{\psi}_{\mu}\bar{\sigma}^{\mu}\lambda^a \right)\label{eq:lagrangianint}\\
+ee^{\frac{K}{2}}\Bigg\{ \frac{\sqrt{2}}{2}i\left( \mathcal{D}_{i}W\bar{\psi}_{\mu}\bar{\sigma}^{\mu}\chi^i + \bar{\mathcal{D}}^{i^*}W^*\psi_{\mu}\sigma^{\mu}\bar{\chi}_{i^*}\right) \nn\\
-W\bar{\psi}_{\mu}\bar{\sigma}^{\mu\nu}\bar{\psi}_{\nu} - W^*\psi_{\mu}\sigma^{\mu\nu}\psi_{\nu} - \frac12\mathcal{D}_{i}\mathcal{D}_{j}W\chi^{i}\cdot\chi^{j} - \frac12\bar{\mathcal{D}}^{i^*}\bar{\mathcal{D}}^{j^*}W^*\bar{\chi}_{i^*}\cdot\bar{\chi}_{j^*} \nn\\
+ \frac14 K^{i}{}_{i^*}\left( h_{ab i}\bar{\mathcal{D}}^{i^*}W^* \lambda^a\cdot\lambda^b + h^{\star}_{ab}{}^{i^*} \mathcal{D}_{i}W\bar{\lambda}^a\cdot\bar{\lambda}^b\right)\Big\}\nn\\
+\sqrt{2}iegK^{i^*}{}_{i}\left( \left(\phi^{\dagger}T_a\right)_{i^*}\lambda^a\cdot\chi^i - \left(T_a\phi\right)^{i}\bar{\lambda}^a\cdot\bar{\chi}_{i^*}\right)\nn\\
-\frac{\sqrt{2}}{8}ieh^{Rab}\left( K^{i^*}\left( \phi^{\dagger}T_a\right)_{i^*} + K_{i}\left(T_a\phi\right)^{i} \right)\left( h_{bdj}\chi^j\cdot\lambda^d - h^{\star}_{bd}{}^{j^*}\bar{\chi}_{j^*}\cdot\bar{\lambda}^{d} \right) - V + \mathcal{L}_{4-fermions}\ .\nn
\end{gather}
The first line corresponds to the gaugino-matter-gravitino interaction, and the second line is the goldstino-gravitino mixing term. The third, fourth and fifth lines are the mass terms of the fermionic sector. The last line contains the gauge-matter interaction, the scalar potential $V$ and $\mathcal{L}_{\text{4-fermions}}$ the four-fermions interactions. This lagrangian is invariant under gauge, supergravity and Kähler transformations (with the $U(1)$ transformation in the fermionic sector) through the covariant derivatives:
\begin{eqnarray}
\tilde{\mathcal{D}}_{\mu}\phi^i &=& e_{\mu}^{\tilde{\mu}}\left( \partial_{\tilde{\mu}}\phi^i + igv^a_{\tilde{\mu}}\left( \phi T_a \right)^i \right)\ ,\nn\\
\widecheck{\mathcal{D}}_{\mu}\chi^{i\alpha} &=& e_{\mu}{}^{\tilde{\mu}}\left( \partial_{\tilde{\mu}}\chi^{i\alpha} + \chi^{i\beta}\omega_{\tilde{\mu}\beta}{}^{\alpha} + igv_{\tilde{\mu}}^a\left( T_a\chi^{\alpha} \right)^{i} + \Gamma_{j}{}^{i}{}_{k}\tilde{\mathcal{D}}_{\tilde{\mu}}\phi^j - \frac14 \left( K_{j}\tilde{\mathcal{D}}_{\tilde{\mu}}\phi^j - K^{j^*}\tilde{\mathcal{D}}_{\tilde{\mu}}\phi^{\dagger}_{i^*} \right)\chi^{\alpha i} \right)\ ,\nn\\
\widecheck{\mathcal{D}}\lambda^{\alpha a} &=& e_{\mu}{}^{\tilde{\mu}} \left( \partial_{\tilde{\mu}}\lambda^{\alpha a} + \lambda^{\beta a}\omega_{\tilde{\mu}\beta}{}^{\alpha} - gf_{bc}{}^{a}v_{\tilde{\mu}}^{b}\lambda^{\alpha c} + \frac14 \left( K_{i}\tilde{\mathcal{D}}_{\tilde{\mu}}\phi^i - K_{i^*}\tilde{\mathcal{D}}_{\tilde{\mu}}\phi^{\dagger}_{i^*} \right)\lambda^{\alpha a} \right)\ ,\nn \\
\widecheck{\mathcal{D}}_{\mu}\psi_{\tilde{\nu}}{}^{\alpha} &=& e_{\tilde{\mu}}{}^{\mu} \left( \partial_{\mu}\psi_{\tilde{\nu}}{}^{\alpha} + \psi_{\tilde{\nu}}{}^{\beta}\omega_{\tilde{\mu}\beta}{}^{\alpha} + \frac14 \left( K_j \tilde{\mathcal{D}}_{\tilde{\mu}}\phi^j - K^{j^*}\tilde{\mathcal{D}}_{\tilde{\mu}}\phi^{\dagger}_{i^*} \right) \psi_{\tilde{\nu}}^{\alpha} \right)\ ,\nn
\end{eqnarray}
with the Christoffel symbol $\Gamma_{i}{}^{k}{}_{j}=K^{k}{}_{k^*}K_{i}{}^{k^*}{}_{j}$.\medskip

The scalar potential present in the Lagrangian \autoref{eq:lagrangianint} can be decomposed in a $F$-term, a $D$-term and a $M$-term:
\begin{eqnarray}
V &=& V_{F} + V_D + V_M\nonumber \\
&=& e^{K/m_p^2}\Big( \mathcal{D}^{i}W (K^{-1} )_{i}{}^{j^*}\mathcal{D}_{j^*}\bar{W} -3|W|^2 \Big) \nonumber \\
&& + \frac{g^2}{8}h^{Rab} \Big( K^{i*}(\phi^{\dagger}T_{a} )_{i^*} + K_i(T_a\phi)^i\Big)\Big( K^{j^*}(\phi^{\dagger}T_b)_{j^*} + K_j(T_b\phi)^j \Big) \label{eq:VSUGRA}
\end{eqnarray}
with $\mathcal{D}^iW = W^i + K^iW$. Note that it is common to denote $V_F+V_M$ as $V_F$. One important remark can be made on the structure of this potential. Due to the presence of a negative term in the potential \autoref{eq:VSUGRA}, the potential is then no longer positive. This will play an important rôle in the context of cosmology, more precisely with respect to the vanishing cosmological constant.

\subsection{Supersymmetry as a low energy supergravity theory}\label{subsec:lowenergylimit}
In the preceding sections, the action of the $N=1$ supergravity in four dimensions has been constructed. We will see that taking the low energy limit, $N=1$ supersymmetry can be obtained.\medskip

We assume a gauge group $G$ and a superfields content containing chiral superfields $\Phi^i$ in a representation $\mathfrak{R}$ of $G$ and vector superfields $V^a$. The chiral superfields $\Phi^i=(\phi^i,\chi^i,F^i)$ define the matter content with $\phi^i$ the scalar fields, $\chi^i$ the fermions and $F^i$ the auxiliary fields whilst the vector supermultiplets $V^a=(v_{\mu}^a,\lambda^a,D^a)$ contain the gauge bosons $v_{\mu}^a$, the gauginos $\lambda^a$ and the auxiliary fields $D^a$.\medskip

The Kähler potential $K$ and the gauge kinetic function $h_{ab}$ are taken in a canonical form, \textit{i.e.}, 
\begin{gather}
K(\Phi,\Phi^{\dagger}) = \Phi^i\Phi^{\dagger}_i ,\quad   h_{ab} = \delta_{ab}\ . \label{eq:Khsusy}
\end{gather}
The superpotential $W$ contains all couplings between the superfields that are invariant under the considered gauge group $G$. Note that $\Phi^{\dagger}_i$ are expressed not using star indices since we are in the canonical case. We consider for the following a renormalisable theory where the superpotential is a cubic polynomial:
\begin{gather}
W(\Phi) = \frac13 \lambda_{ijk} \Phi^i\Phi^j\Phi^k + \frac12 \beta_{ij}\Phi^i\Phi^j + \alpha_i\Phi^i\ .\nn
\end{gather}

By dimensional analysis, the reduced Planck-mass $\displaystyle m_p^r=m_p/(\sqrt{8\pi})$ can be reintroduced in the supergravity Lagrangian \autoref{eq:SUGRAlagrangian}. We now obtain in the low energy limit of \autoref{eq:SUGRAlagrangian}
\begin{gather}
\displaystyle \lim_{m_p\rightarrow \infty} \Big( \mathcal{L}_{SUGRA} -\mathcal{L}_{Pure\ SUGRA} \Big) = \mathcal{L}_{SUSY}^{kin.} +\mathcal{L}_{SUSY}^{int.} \label{eq:sugralimitsusy}
\end{gather}
with
\begin{gather}
\label{eq:lag_susy}
\mathcal{L}_{SUSY}^{kin.}  =  -\frac{1}{4}F_a^{\mu\nu}F^a_{\mu\nu}-\frac{i}{2} \Big( \lambda_a \sigma^{\mu}\mathcal{D}_{\mu} \bar{\lambda}^a - \mathcal{D}_{\mu}\lambda_a\sigma^{\mu}\bar{\lambda}^a \Big) + \mathcal{D}_{\mu}\phi^i\mathcal{D}^{\mu}\phi^{\dagger}_i \nn\\
- \frac{i}{2}\Big[ \chi^i\sigma^{\mu}\mathcal{D}_{\mu}\bar{\chi}_i - \mathcal{D}_{\mu}\chi^i\sigma^{\mu}\bar{\chi}_i \Big] \ , \nonumber
\end{gather}
corresponding to the kinetic terms of, respectively, gauge bosons, gauginos, scalar fields $\phi^i$ and fermions $\chi^i$, and 
\begin{gather}
\label{eq:lag_susy}
\mathcal{L}_{SUSY}^{int.}  =  \sqrt{2}ig\Big[ \left( \phi^{\dagger}T_a \right)_i\lambda^a\cdot\chi^i - \left( T_a\phi \right)^i\bar{\lambda}^a\cdot\bar{\chi}_i \Big] -\frac12\Big( \partial_i\partial_j W(\phi )\chi^i\cdot\chi^j + \mathrm{h.c.} \Big)  - V\ , \nonumber \\
V = \partial_i W(\phi ) \partial^i W^\star (\phi^\dagger ) +\frac{1}{2}g^2(\phi^\dagger T^a\phi ) ( \phi^\dagger T_a \phi ) \ , \label{eq:potSUSY}
\end{gather}
(with the index $a$ is raised using the Killing metric $\kappa^{ab}=\delta^{ab}$, \textit{i.e.}, $T^a=\delta^{ab}T_b$) containing the gaugino-matter interactions, Yukawa interactions and mass terms for the fermions. We then recover the Lagrangian of a $N=1$ supersymmetric theory with the covariant derivatives containing now only gauge couplings.

\section{Supergravity breaking}
In the previous section, the $N=1$ supergravity action has been constructed. This construction is based on an invariance under supergravity transformations \autoref{eq:sugratrso}. Nonetheless, the implication of such symmetry in particle physics implies the existence of new particles (in particular in the low energy limit, see Subsection \ref{subsec:lowenergylimit}),  which should have already been detected in the actual experiments. Indeed, particles from the same supermultiplets must be degenerated in mass, which leads to various states such as massless fermion (partner of the photon) or scalar fields with low masses (partners of the leptons of the Standard Model). Since, to date, no signatures of those particles have been highlighting in the various collider experiments, then supergravity has to be spontaneously broken.\medskip

Historically, the first supersymmetry breaking mechanisms have been proposed in supersymmetry. Those mechanisms, namely the \textit{Fayet-Iliopoulos} \cite{Fayet} and the \textit{O'Raifeartaigh} mechanisms \cite{ORaifeartaigh} suppose that, respectively, the auxiliary fields $F^i$ and $D^a$ take a non-zero \textit{v.e.v.}, \textit{i.e.}, $\big<F^i\big>\neq 0$ and $\big<D^a\big>\neq 0$. Unfortunately, those mechanisms generate two problems: 
\begin{itemize}
\item supersymmetric particles can be lighter than particles from the Standard Model,
\item a massless fermion (the goldstino) naturally appears as a signature of supersymmetry breaking. 
\end{itemize}
However, it is possible to construct supersymmetry breaking models compatible with the experimental measurements in the context of supergravity.\medskip

In order to bypass this issue in supergravity, it is usual to introduce a hidden sector $Z=(\zeta,\chi_{\zeta},F_{\zeta})$ where supergravity will be broken following the \textit{Fayet-Iliopoulos} or the \textit{O'Raifeartaigh} mechanisms. The massless goldstino can then be absorbed by the gravitino, defining a mass term for the gravitino. However, the effects of supergravity breaking in the hidden sector must be communicated from the hidden to the matter sector. This communication can be done through various types of interactions, defining the type of supergravity breaking mechanism. Several mechanisms will be presented, focusing on the \textit{Gravity Mediated Supersymmetry Breaking}.\medskip

For the continuation, we reintroduce in the theory the mass parameters by dimensional analysis (with $[W]=M^3$, $[K]=M^2$ and $[h_{ab}]=M^0$).

\subsection{Hidden Sector \& Gravitino mass}\label{subsec:breakingmodel}
The supergravity breaking mechanism is quite similar to the Higgs mechanism of the Standard Model. We consider a hidden sector $Z^A=(\zeta^A,\chi_{\zeta}^A,F_{\zeta}^A)$, singlet under a gauge group $G$\footnote{But can be charged under an other gauge group $G'$, not coupling to the matter sector.} ($G=SU(3)_C\times SU(2)_L\times U(1)_Y$ or GUT gauge group), to be distinguished to the matter sector $\Phi^i=(\phi^i,\chi^i,F^i)$. Assuming that at least one field $\zeta^A$ acquire a non-zero vacuum expectation value, a goldstino state $\psi_G$ appears as a signature of supergravity breaking. A gravitino-shift then allows to eliminate the goldstino, leading to a mass term for the gravitino $\psi_{\mu}$ 
\begin{eqnarray}
m_{3/2} = \frac{1}{m_p^2}\Big< e^{K/(2m_p^2)} W \Big> \ .\nonumber 
\end{eqnarray}
This mechanism is called the Super-Brout-Englert-Higgs mechanism. \medskip

The interactions between the hidden and the matter sector can be of different natures:
\begin{itemize}
\item \textit{Gauge Mediated Supersymmetry Breaking} or GMSB \cite{GMSB}: the breaking is mediated by means of messenger fields $\mathbf{S}$ and $\overline{\mathbf{S}}$ (corresponding, for example, to the representation $\underset{\sim}{5}$ and $\overline{\underset{\sim}{5}}$ in $SU(5)$ models). A superfield $Z$ from the hidden sector gets a non-zero \textit{v.e.v.} $\big< Z\big> = \big< \zeta \big> -\theta\cdot\theta \big< F_{\zeta} \big>$ and a hidden/messenger interaction in the superpotential $W=\lambda \mathbf{S}\overline{\mathbf{S}}Z$ generates at loop-level quantum corrections that break supersymmetry. This mediation will be used in Section \ref{llp} but will not be studied in details.
\item \textit{Anomaly Mediated Supersymmetry Breaking} or AMSB \cite{AMSB}: The breaking is mediated via quantum loop effects related to anomalous calculations. This mechanism will be not useful for this manuscript. 
\item \textit{Gravity Mediated Supersymmetry Breaking} \cite{ChamseddineGMSB}\cite{BFSGMSB}: the breaking is mediated from the hidden to the matter sector through gravitational effects. 
\end{itemize}
The last symmetry breaking mechanism is the one mainly studied in this manuscript.\medskip

The superpotential $W$ and the Kähler potential $K$ are central in this context. Defining a general superfield $\mathcal{A}^{I}=\{ Z^A , \Phi^i \}$, the Kähler potential $K(\mathcal{A},\mathcal{A}^{\dagger})$ and the superpotential $W(\mathcal{A^I})\equiv W$, the scalar potential of supergravity $V$ can be written as:
\begin{gather}
V=e^{K/m_p^2}\left( \mathcal{D}_{I}W\left( K^{-1} \right)^{I}{}_{J^*}\mathcal{D}^{J^*}\overline{W} -\frac{3}{m_p^2}\left|W\right|^2\right)\ .\label{eq:potentialmp}
\end{gather}
By taking a non-zero \textit{v.e.v} in the hidden sector $\left< \zeta^A \right> \sim \mathcal{O}(m_p)$, the perturbative expansion of the potential \autoref{eq:potentialmp} induces supersymmetry breaking terms in the Lagrangian and so dynamically generates supersymmetry breaking. Since the interactions between the matter and hidden sectors are defined at the Planck scale $m_p$, the superpotential $W(\mathcal{A})$ and the Kähler potential $K(\mathcal{A}^I,\mathcal{A}^{\dagger}_{I^*})$ must be chosen correctly so that $V$ does not involve divergent terms for the matter sector in the low energy limit, \textit{i.e.}, when $m_p\rightarrow \infty$.\medskip

One possible solution for $W(\mathcal{A}^I)$ and $K(\mathcal{A}^I,\mathcal{A}^{\dagger}_{I^*})$ leading to this situation was firstly obtained by Soni \& Weldon \cite{soni_analysis_1983}\footnote{Other solutions can be obtained in the context of \textit{No-Scale} supergravity model, see \cite{noscale1} \cite{noscale2} \cite{noscale3}}.

\subsection{Soni-Weldon solutions}\label{sub:SW}
In 1983, Sanjeev K. Soni and H. Arthur Weldon submitted an article \cite{soni_analysis_1983} on an analysis of the low energy limit of the supergravity scalar potential \autoref{eq:potentialmp} in the context of \textit{Gravity Mediated Supersymmetry Breaking}. Imposing that the theory leads to phenomenologically acceptable solutions when the Planck mass goes to infinity, they have identified a general form of the superpotential $W(\mathcal{A})$ and the Kähler potential $K(\mathcal{A},\mathcal{A}^{\dagger})$.\medskip

For this study, they made the following assumptions:
\begin{itemize}
\item One field from the hidden sector gets a non-zero \textit{v.e.v} $\left< \zeta \right> \sim \mathcal{O}(m_p)$. To facilitate the computation, it is helpful to define the field
\begin{equation}
\zeta^i \equiv m_p z^i\ , \nn
\end{equation}
where $\big< z^i \big> = \mathcal{O}(1)$,
\item The superpotential and Kähler potential can be written as:
\begin{eqnarray}
W(\Phi, Z) &=& \displaystyle\sum_{n=0}^{N}m_p^n W_n (Z,\Phi)\ ,  \label{eq:WSW}\\
K(\Phi,\Phi^{\dagger},Z,Z^{\dagger}) &=& \displaystyle\sum_{n=0}^{M}m_p^nK_n(\Phi, \Phi^{\dagger}, Z,Z^{\dagger})\ , \label{eq:KSW}
\end{eqnarray}
where $W_n (Z,\Phi) $ are holomorphic functions and $K_n(\Phi, \Phi^{\dagger}, Z,Z^{\dagger})$ are real functions. 
\item All interactions of fields from the visible sector $\Phi$ must appear as $m_p^{-n}$ in the potential with $(n\geq 0)$, \textit{i.e.}, 
\begin{equation}
\lim_{m_p\rightarrow \infty} V_{\Phi} \text{ is non-divergent}\nn
\end{equation}
where $V_{\Phi}$ is the part of the potential containing the interactions of the matter sector.
\end{itemize}
By explicitly calculating the scalar potential \autoref{eq:potentialmp} using those assumptions, they have restricted the form of $W(\mathcal{A})$ and $K(\mathcal{A},\mathcal{A}^{\dagger})$ as follow:
\begin{eqnarray}
W(\Phi, z) &=& m_p^2 W_2 (z) + m_p W_1 (z) + W_0(z,\Phi) \ ,\label{SWsolution1}\\
K(\Phi,\Phi^{\dagger},z,z^{\dagger}) &=& m_p^2K_2(z,z^{\dagger}) + m_p K_1(z,z^{\dagger}) + K_0(\Phi, \Phi^{\dagger}, z,z^{\dagger})\ , \label{SWsolution2}
\end{eqnarray}
where $ W_2 (z)$, $ W_1 (z)$ are arbitrary holomorphic functions of the hidden sector, $K_2(z,z^{\dagger})$ \& $K_1(z,z^{\dagger})$ are arbitrary real functions of the hidden sector, $W_0(z,\Phi)$ is an arbitrary holomorphic function of both sectors and $K_0(\Phi, \Phi^{\dagger}, z,z^{\dagger})$ is a real function of both sectors too. Remark that the functions containing the matter and hidden sector only appear in the lowest power of Planck mass (when $n=0$ in the expansion). We can also notice that those results were reconsidered in a recent paper \cite{tant:tel-01546150}\cite{moultaka_low_2018} and new solutions were obtained. A thoroughgoing study of those new solutions will be performed in the next chapter. \medskip

Using the superpotential and the Kähler potential obtained from the Soni-Weldon solutions, the low-energy potential can then be computed. Those solutions will generate the so-called soft-breaking terms that explicitly break supersymmetry in a \textit{soft} way, \textit{i.e.}, the loop-corrections induced by those terms lead to logarithmic divergences.

\subsection{General soft-breaking terms}\label{subsec:soft}
As an example, we present the calculation of such supersymmetric soft breaking terms following the Giudice and Masiero approach \cite{GIUDICEMASIERO} for a general renormalisable theory \cite{BRIGNOLE}. The Kähler potential and the superpotential are then taken to be:
\begin{eqnarray}
K(\Phi,\Phi^{\dagger},z,z^{\dagger}) &=& \hat{K}(z,z^{\dagger})  + \Phi^{\dagger}_{a^*}\Lambda^{a^*}{}_{a}(z,z^{\dagger})\Phi^a + \Big( \frac12 Z_{ab}(z,z^{\dagger})\Phi^a\Phi^b + \mathrm{h.c.} \Big) \ ,\nn\\
W(\Phi,z) &=& \hat{W}(z) + \frac12 m_{ab}(z)\Phi^a\Phi^b + \frac16 \lambda_{abc}(z)\Phi^a\Phi^b\Phi^c \ ,\label{eq:Wgeneral}
\end{eqnarray}
where the orders of magnitude of $\hat{K}$ and  $\hat{W}$ are respectively $m_p^2$ and $m_p^2M$ with $M$ an energy scale much lower than the Planck mass $M \ll m_p$. In this case, the gravitino mass is 
\begin{gather}
m_{3/2}= \frac{1}{m_p^2}\left< \hat{W} \right>e^{\left<\hat{K}\right>/(2m_p^2)} = Me^{\left<\hat{K}\right>/(2m_p^2)} \ .\nn
\end{gather}
\medskip

For the computation of the scalar potential \autoref{eq:potentialmp}, we introduce the metric of the Kähler manifold:
\begin{gather}
K^{I^*}{}_{J} = \begin{pmatrix}
K^{a^*}{}_{a} & K^{a^*}{}_{i} \\
K^{i^*}{}_{a} & K^{i^*}{}_{i}
\end{pmatrix} 
=  \begin{pmatrix}
A & B \\
C & D
\end{pmatrix} \ ,
\end{gather}
with
\begin{eqnarray}
A^{a^*}{}_{a} &=& \Lambda^{a^*}{}_{a}\ ,\nn\\
B^{a^*}{}_{i} &=& \partial_i\Lambda^{a^*}{}_{a}\phi^a + \partial_i \overline{Z}^{a^*b^*}\phi^{\dagger}_{b^*}\ ,\nn\\
C^{i^*}{}_{a} &=& \phi^{\dagger}_{a^*}\partial^{i^*}\Lambda^{a^*}{}_{a} + \partial^{i^*}Z_{ab}\phi^b\ ,\nn\\
D^{i^*}{}_{i} &=& \hat{K}^{i^*}{}_{i} + \phi^{\dagger}_{a^*}\partial_i\partial^{i^*}\Lambda^{a^*}{}_{a}\phi^a + \frac12\left( \partial_i\partial^{i^*}Z_{ab}\phi^a\phi^b + \text{h.c.}\right)\ .\nn
\end{eqnarray}
The inverse metric must also be calculated. In accordance with the previous notations, we have:
\begin{gather}
K^{J}{}_{I^*} = \begin{pmatrix}
\left( A-BD^{-1}C \right)^{-1} & -\left( A-BD^{-1}C \right)^{-1}BD^{-1} \\
-D^{-1}C\left( A-BD^{-1}C \right)^{-1} & D^{-1}C\left( A-BD^{-1}C \right)^{-1}BD^{-1} + D^{-1}
\end{pmatrix}=\begin{pmatrix}
K^{a}{}_{a^*} & K^{a}{}_{i^*} \\
K^{i}{}_{a^*} & K^{i}{}_{i^*}
\end{pmatrix} .
\end{gather} 
Calculating the inverse metric perturbatively in the Planck mass, we obtain
\begin{eqnarray}
K^{i}{}_{i^*} &=& \hat{K}^{i}{}_{i^*} + \hat{K}^{i}{}_{j^*}\left( \left( \phi^{\dagger}_{a^*}\partial^{j^*}\Lambda^{a^*}{}_{a} + \partial^{j^*}Z_{ab}\phi^b \right) \left(\Lambda^{-1}\right)^{a}{}_{b^*}\left( \partial_j\Lambda^{b^*}{}_{c}\phi^c + \partial_j\overline{Z}^{b^*c^*}\phi^{\dagger}_{c^*} \right) \right.\nn\\
&& \left. -\phi^{\dagger}_{a^*}\partial_j\partial^{j^*}\Lambda^{a^*}{}_{a}\phi^a - \frac12\left( \partial_j\partial^{j^*}Z_{ab}\phi^a\phi^b + \text{h.c.}\right)\right)\hat{K}^{j}{}_{i^*}\ ,\nn\\
K^{a}{}_{a^*} &=& \left( \Lambda^{-1} \right)^a{}_{b^*}\left( \partial_i\Lambda^{b^*}{}_{b}\phi^b + \partial_i\overline{Z}^{b^*c^*}\phi^{\dagger}_{c^*} \right)\hat{K}^i{}_{i^*}\left( \phi^{\dagger}_{d^*}\partial^{i^*}\Lambda^{d^*}{}_{d}+ \partial^{i^*}Z_{de}\phi^e \right)\left(\Lambda^{-1}\right)^d{}_{a^*}\nn\\
&& + \left( \Lambda^{-1} \right)^a{}_{a^*}\ ,\nn\\
K^{a}{}_{i^*} &=& -\left(\Lambda^{-1}\right)^a{}_{a^*}\left( \partial_i\Lambda^{a^*}{}_{b}\phi^b + \partial_i\overline{Z}^{a^*b^*}\phi^{\dagger}_{b^*} \right)\hat{K}^i{}_{i^*}\ ,\nn\\
K^{i}{}_{a^*} &=& - \hat{K}^i{}_{i^*}\left( \phi^{\dagger}_{b^*}\partial^{i^*}\Lambda^{b^*}{}_{a} + \partial^{i^*}Z_{ab}\phi^b \right)\left(\Lambda^{-1}\right)^a{}_{a^*}\ . \nn
\end{eqnarray}
After lengthy calculations, the potential can be rewritten in the form:
\begin{equation}
V = m_p^2 \Lambda + V_{SUSY} + V_{SOFT}  \ ,\nn
\end{equation}
where
\begin{equation}
\Lambda = \frac{1}{m_p^2}\big< F_{i^*}^{\dagger} \big>\big< \hat{K}^{i^*}{}_{i} \big>\big< F^i\big> - 3m_{3/2}^2\quad \text{with} \quad F^i=\left< e^{\hat{K}/(2m_p^2)}\hat{K}^I{}_{I^*}\mathcal{D}^{I^*}W^* \right>\nn
\end{equation}
is the cosmological constant. We can see that the minimum of the potential is not positively defined, contrary as in supersymmetry. The cosmological constant $\Lambda$ can then be fine-tuned, imposing $\Lambda \approx 0$. The second term $V_{SUSY}$ is the classical supersymmetry potential,
\begin{equation}
V_{SUSY} = \partial_a W_m(\Lambda^{-1})^a{}_{a^*}\partial_{a^*}W_m^*\ ,\nn
\end{equation}
with 
\begin{equation}
W_m = \frac16 \hat{\lambda}_{abc}\phi^a\phi^b\phi^c + \frac12 \big( \hat{m}_{ab} + m_{3/2} \big< Z_{ab} \big> - F^{\dagger}_{i^*}\big< \partial^{i^*}Z_{ab} \big> \big)\phi^a\phi^b \ ,\nn
\end{equation}
where $\hat{\lambda}_{abc}=\left< \lambda_{abc} \right>e^{\left<\hat{K}\right>/(2m_p^2)}$ and $\hat{m}_{ab}=\left<m_{ab} \right>e^{\left<\hat{K}\right>/(2m_p^2)}$.\medskip

The Z-term in the Kähler potential has then dynamically introduced in the effective superpotential a bilinear coupling which is known as the Giudice Masiero mechanism \cite{GIUDICEMASIERO}. This mechanism is a natural solution to the $\mu$-problem (see Subsection \ref{subsec:mssm}). We will see in Subsection \ref{sec:N2MSSMDes} an other possible solution using singlet superfields.\medskip
 
The last part of the potential contains the so-called \textit{soft-breaking terms} which spontaneously break supersymmetry:
\begin{equation}
V_{SOFT} = \phi^{\dagger}_{a^*}\big( m_{3/2}^2S^{a^*}{}_{a} + \Lambda\big< \Lambda^a{}_{a^*} \big> \big)\phi^a + \left( \frac16 A_{abc}\phi^a\phi^b\phi^c + \frac12 B_{ab}\phi^a\phi^b +\mathrm{h.c.}\right)\label{eq:vsoftcalcul}
\end{equation}
with:
\begin{eqnarray}
A_{abc} &=& e^{\frac{\big< \hat{K} \big>}{2m_p^2}} \big< F^i \big>\Big\{ \frac{1}{m_p^2}\big< \partial_i \hat{K} \big>\big< \lambda_{abc} \big> + \big< \partial_i\lambda_{abc} \big>  \label{eq:softpar1}  \\
&& - \Big( \big< (\Lambda^{-1})^d{}_{a^*}\partial_i\Lambda^{a^*}{}_{a}\lambda_{dbc} \big> + ( a \leftrightarrow b ) + ( a \leftrightarrow c ) \Big)\Big\}\ ,\nn \\
B_{ab} &=& e^{\frac{\big< \hat{K} \big>}{2m_p^2}} \big< F^i \big> \Big\{ \frac{1}{m_p^2}\big< \partial_i \hat{K} \big>\big< m_{ab} \big> + \big< \partial_i m_{ab} \big> - \Big( \big< (\Lambda^{-1})^c{}_{a^*}\partial_i\Lambda^{a^*}{}_{b}m_{ac} \big> + ( a \leftrightarrow b ) \Big)\Big\} \nn \\
&& -m_{3/2}^{\dagger} e^{\frac{\big< \hat{K} \big>}{2m_p^2}} \big<m_{ab}\big> + (2m_{3/2}^2 + \Lambda)\big< Z_{ab} \big> - m_{3/2}^{\dagger}\big< F^{\dagger}_{i^*} \big>\big< \partial^{i^*}Z_{ab} \big> \label{eq:softpar2} \\
&& + m_{3/2} \big< F^i \big>\Big\{ \big<  \partial_i Z_{ab} \big> - \big( \big< (\Lambda^{-1})^c{}_{a^*}\partial_i\Lambda^{a^*}{}_{b}Z_{ac} \big> + \big( A \leftrightarrow b \big) \big) \big) \Big\} \nonumber \\
&& - \big< F^iF^{\dagger}_{i^*} \big>\Big\{ \big< \partial_i\partial^{i^*}Z_{ab} \big> - \Big( \big< (\Lambda^{-1})^c{}_{a^*}\partial_i\Lambda^{a^*}{}_{b}\partial^{i^*}Z_{ac} \big> + (a \leftrightarrow b ) \Big) \Big\} \ ,\nonumber \\
S^{a^*}{}_{a} &=& \big< \Lambda^{a^*}{}_{a} \big> + \frac{1}{m_p^2}\big< F^{\dagger}_{i^*} \big>\Big\{ \big< \partial^{i^*}\Lambda^{a^*}{}_{b} \big>\big< (\Lambda^{-1})^b{}_{b^*} \big>\big<  \partial_i \Lambda^{b^*}{}_ {a}\big> - \big< \partial_i\partial^{i^*}\Lambda^{a^*}{}_{a} \big> \Big\} \big< F^i \big> \ .\label{eq:softpar3}
\end{eqnarray}
Those terms are then dynamically generated through supergravity breaking in the hidden sector $\{Z\}$.\medskip

At this stage, we can point out some remarks:
\begin{itemize}
\item Putting the Giudice Masiero mechanism aside, the introduction of interaction terms in the superpotential will automatically generate associated soft breaking terms. It can also be shown that defining a linear term $\xi_a\Phi^a$ in the superpotential \autoref{eq:Wgeneral} will generate a linear soft breaking term $\xi^S_a\phi^a$ in $V_{SOFT}$. 
\item The soft-mass terms $S^{a^*}{}_{a}$ in $V_{SOFT}$ are different for each scalar fields $\phi^a$. It is a consequence of the non-universal term in the Kähler potential $\Phi^{\dagger}_{a^*}\Lambda^{a^*}{}_{a}\Phi^a$. A simpler model can be identified by taking the matrix $\Lambda^{a^*}{}_{a} = \delta^{a^*}{}_{a}$, which leads to a universal mass term for scalar supersymmetric particles.
\item Those terms are introduced at high energy. Renormalisation group equations (RGEs) must then be used to express these parameters at low energy.
\item Other breaking mediation mechanisms (such as GMSB or AMSB) will also lead to the same form of supersymmetric soft-breaking terms \autoref{eq:vsoftcalcul}.
\item Those terms are called \textit{soft}, \text{i.e.}, their contributions to loop-corrections will only give logarithmic divergences. We will see in the next section that new solutions for \textit{Gravity Mediated Supersymmetry Breaking} scenarii can be obtained. Those solutions also generate hard breaking terms, leading to quadratic quantum corrections scaled by a new energy scale. They will be analysed in Chapter \ref{NSW}. 
\end{itemize} 
In phenomenology and experimental particle physics, it is standard to introduce the soft-breaking terms as \textit{ad-hoc} couplings in the Lagrangian, \textit{i.e.}, introducing $V_{SOFT}$ \autoref{eq:vsoftcalcul} in the Lagrangian without taking into account any mechanism of supersymmetry breaking. We will consider this method in the analysis of the N2MSSM Chapter \ref{chap:N2MSSM}.

\subsection{Supersymmetry breaking}\label{subsec:susybreak}
We now suppose one of the mechanisms mentioned in Subsection \ref{subsec:breakingmodel}. Taking the low energy limit of the supergravity action \autoref{eq:SUGRAlagrangian} with the Kähler potential and the gauge kinetic function as \autoref{eq:Khsusy} and the superpotential \autoref{eq:Wgeneral}, we obtain:
\begin{gather}
\lim_{m_p\rightarrow \infty}\left(\mathcal{L}_{SUGRA} - \mathcal{L}_{Pure\ SUGRA}\right) = \mathcal{L}^{kin}_{SUSY} + \mathcal{L}^{int}_{SUSY} - V_{SOFT} \ .\nn
\end{gather}
The terms $\mathcal{L}^{kin}_{SUSY}$ and $\mathcal{L}^{int}_{SUSY}$ are similar to \autoref{eq:sugralimitsusy}. The new term $V_{SOFT}$ corresponds to the supersymmetry soft-breaking terms. Supposing that $V_{SOFT}$ is generated by \textit{Gravity Mediated Supersymmetry Breaking}, we have:
\begin{gather}
V_{SOFT} = m_{3/2}^2\phi^{\dagger}_{a}\phi^a + \left(\frac16 A_{abc}\phi^a\phi^b\phi^c + \frac12 B_{ab}\phi^a\phi^b +\mathrm{h.c.}\right) \nn
\end{gather}
with (assuming a vanishing cosmological constant $\Lambda=0$):
\begin{eqnarray}
A_{abc} &=& e^{\frac{\big< \hat{K} \big>}{2m_p^2}} \big< F^i \big>\Big\{ \frac{1}{m_p^2}\big< \partial_i \hat{K} \big>\big< \lambda_{abc} \big> + \big< \partial_i\lambda_{abc} \big>  \Big\}\ ,\nn \\
B_{ab} &=& e^{\frac{\big< \hat{K} \big>}{2m_p^2}} \big< F^i \big> \Big\{ \frac{1}{m_p^2}\big< \partial_i \hat{K} \big>\big< m_{ab} \big> + \big< \partial_i m_{ab} \big> \Big\}  -m_{3/2}^{\dagger} e^{\frac{\big< \hat{K} \big>}{2m_p^2}} \big<m_{ab}\big>  \nn 
\end{eqnarray}
The soft breaking mass terms are now fully controlled by the gravitino mass 
\begin{gather}
m_{3/2} = \frac{\left< W \right>}{m_p^2}e^{\left<K\right>/(2m_p^2)}\ .\nn
\end{gather}
We also see that taking $\lambda_{abc} =0$ ($m_{ab} = 0$) directly impose $A_{abc}=0$ ($B_{ab}=0$), \textit{i.e.}, each coupling in the superpotential $W$ is associated to a soft-breaking term in $V_{SOFT}$.

\section{Construction of supersymmetric models}

We now construct some supersymmetric models obtained as a low energy limit of supergravity, which embed the Standard Model. Those models include the corresponding soft-supersymmetric breaking terms generated through an analogue mechanism to Subsection \ref{subsec:soft}. In this section, we will present two models of supersymmetry. The first model is the simplest supersymmetry extension of the Standard Model called the MSSM (for \textit{Minimal Supersymmetric Standard Model}). The second is an extension of the MSSM with two extra singlet superfields $\hat{S}^p$ $(p=1,\ 2)$ called the N2MSSM. \medskip

\subsection{MSSM: Minimal Supersymmetric Standard Model}\label{subsec:mssm}
Since we want to recover the Standard Model at low energy:
\begin{itemize}
\item[(1)] the gauge group $G$ is taken as the one of the Standard Model\footnote{GUT theory can also be reconstructed by considering gauge group which allows gauge unification such as $SU(5)$ or $SO(10)$.}, \textit{i.e.}, 
\begin{gather}
G_{SM} = SU(3)_C\times SU(2)_L\times U(1)_Y \ .\nn
\end{gather}
\item[(2)] We then add three vector superfields $V^a$ defined in the adjoint representation of $G_{SM}$ (see \autoref{tab:VSM}).
\begin{table}[H]
\begin{tabular}{ |c|c|c|c|c| } 
 \hline
 simple factor of $G_{SM}$ & vector superfields & particles & superpartners & representation \\
 \hline
 \hline\xrowht{20pt}
 $SU(3)_C$ & $V_3$ & gluons $g$ & gluinos $\tilde{g}$ & $(\underset{\sim}{8},\underset{\sim}{1},0)$ \\ 
 \hline\xrowht{20pt}
 $SU(2)_L$ & $V_2$ & W-bosons $W$ & Winos $\tilde{W}$ & $(\underset{\sim}{1},\underset{\sim}{3},0)$ \\ 
 \hline\xrowht{20pt}
 $U(1)_Y$ & $V_1$ & B-boson $B$ & Bino $\tilde{B}$ & $(\underset{\sim}{1},\underset{\sim}{1},0)$ \\ 
 \hline
\end{tabular}
\caption{Vector superfields associated to the gauge group $SU(3)_C\times SU(2)_L\times U(1)_Y$}
\label{tab:VSM}
\end{table}

The notation $(\underset{\sim}{d}{}_C,\underset{\sim}{d}{}_L,Y)$ corresponds to representations of $G_{SM}$. Each vector superfield $V_a$ contains the gauge bosons of the gauge group  with their corresponding superpartners, the gauginos. \medskip

In order to include particles of the Standard Model, the chiral superfields in \autoref{tab:PSM} are defined through their representations under the group gauge $G_{SM}$.
\begin{table}[H]
\begin{tabular}{ |c|c|c|c|c|c| } 
 \hline
 Chiral superfields $\Phi$ & name & particle & superpartner & representation \\
 \hline
 \hline\xrowht{20pt}
 $\hat{Q}^i$ & \multirow{3}{4cm}{quarks \& squarks} & $q^i_L=\begin{pmatrix} u^i_L \\ d^i_L \end{pmatrix}$ &  $\tilde{q}^i_L=\begin{pmatrix} \tilde{u}^i_L \\ \tilde{d}^i_L \end{pmatrix}$ & $(\underset{\sim}{3},\underset{\sim}{2},\frac{1}{6})$ \\ 
 $\hat{U}^i$ &  & $u_L^i{}^c$ & $\tilde{u}_R^i{}^{\dagger}$ & $(\overline{\underset{\sim}{3}},\underset{\sim}{1},-\frac{2}{3})$  \\ 
 $\hat{D}^i$  & &$d_L^i{}^c$ & $\tilde{d}_R^i{}^{\dagger}$ & $(\overline{\underset{\sim}{3}},\underset{\sim}{1},\frac{1}{3})$ \\ 
 \hline\xrowht{20pt}
 $\hat{L}^i$ & \multirow{3}{4cm}{leptons \& sleptons} & $l^i_L=\begin{pmatrix} \nu^i_L \\ e^i_L \end{pmatrix}$ & $\tilde{l}^i_L=\begin{pmatrix} \tilde{\nu}^i_L \\ \tilde{e}^i_L \end{pmatrix}$ & $(\underset{\sim}{1},\underset{\sim}{2},-\frac{1}{2})$  \\ 
 $\hat{E}^i$ & & $e_L^i{}^c$ & $\tilde{e}_R^i{}^{\dagger}$ & $(\underset{\sim}{1},\underset{\sim}{1},1)$  \\ 
 $\hat{N}^i$ & & $\nu_L^i{}^c$ & $\tilde{\nu}^i_R{}^{\dagger}$ & $(\underset{\sim}{1},\underset{\sim}{1},0)$  \\ 
 \hline\xrowht{20pt}
 $\hat{H}_D$ & \multirow{2}{4cm}{Higgs \& Higgsinos} & $H_D=\begin{pmatrix} H_D^0 \\ H_D^{-} \end{pmatrix}$ & $\tilde{H}_D=\begin{pmatrix} \tilde{H}_D^0 \\ \tilde{H}_D^{-} \end{pmatrix}$ & $(\underset{\sim}{1},\underset{\sim}{2},-\frac{1}{2})$  \\ 
 $\hat{H}_U$ & & $H_U=\begin{pmatrix} H_U^{+} \\ H_U^{0} \end{pmatrix}$ & $\tilde{H}_U=\begin{pmatrix} \tilde{H}_U^{+} \\ \tilde{H}_U^{0} \end{pmatrix}$ & $(\underset{\sim}{1},\underset{\sim}{2},\frac{1}{2})$  \\ 
 \hline
\end{tabular}
\caption{Chiral superfields content associated to the Standard Model}
\label{tab:PSM}
\end{table}
The index $i=1,2,3$ corresponds to the three generations of quarks \& leptons. Remark that two Higgs doublets are introduced to avoid chiral anomaly in triangle diagrams and allow Higgs couplings with both up and down type fermions.
\item[(3)] The Kähler potential $K$ and the gauge kinetic function $h_{ab}$ is taken as \autoref{eq:Khsusy}
\end{itemize}
The field content presented in \autoref{tab:VSM} and \autoref{tab:PSM} is sufficient in order to recover the particle spectrum of the Standard Model. It can be, however, widened by adding new superfields, which will lead to new particles in the spectrum (as we will see in \autoref{sec:N2MSSMDes}). Considering first only the superfields \autoref{tab:VSM} and \autoref{tab:PSM} leads to the simplest supersymmetric model called MSSM for \textit{Minimal Supersymmetric Standard Model} \cite{NILLESMSSM}\cite{FAYETMSSM1}\cite{FAYETMSSM2}\cite{martin_supersymmetry_1998}.\medskip

Allowing all the possible couplings in the superpotential of the MSSM, $W_{MSSM}$ write:
\begin{eqnarray}
W_{MSSM} &=& -\hat{L}\cdot \hat{H}_D\mathbf{Y_E}\hat{E}- \hat{Q}\cdot \hat{H}_D \mathbf{Y_D} \hat{D} +  \hat{Q}\cdot \hat{H}_U \mathbf{Y_U} \hat{U} \label{mssmW}\\
&&+ \mu \hat{H}_U\cdot \hat{H}_D + W_{N} +W_{RPV} \ ,\nn
\end{eqnarray}
where $\mathbf{Y_I}$ are $3\times 3$ matrices containing Yukawa couplings, $W_N$ the Yukawa couplings and the Dirac mass of the neutrino superfield
\begin{gather}
W_N = L\cdot \hat{H}_D \mathbf{Y_N} \hat{N} + m_N \hat{N}\hat{N} \nn
\end{gather}
and the dot product "$\cdot$" the $SU(2)$-product which gives, for example, $\hat{H}_U\cdot\hat{H}_D=\hat{H}_U^{+}\hat{H}_D^{-} - \hat{H}_U^{0}\hat{H}_D^{0}$. The last term, $W_{RPV}$, 
\begin{gather}
W_{RPV} = \lambda_{ijk}\hat{L}^i\cdot \hat{L}^j\hat{E}^k+\lambda'_{ijk}\hat{L}^i\cdot \hat{Q}^j\hat{D}^k + \kappa_i\hat{L}^i\cdot \hat{H}_U + \lambda''_{ijk}\hat{U}^i\hat{D}^j\hat{D}^k \nn 
\end{gather}
contains terms that violate a discrete symmetry called $R$-parity \cite{MSSMRPV}. This symmetry involves a multiplicative quantum number $R=(-1)^{3B+L+2S}$ to all particles defined as $R=+1$ for particles of the Standard Model and $R=-1$ for superpartners. \medskip

The violation of R-parity has an essential impact on phenomenology. Indeed, those new couplings allow in specific configurations ($\lambda''_{112}\neq 0$ and $\lambda'_{i12}\neq 0$) the decay of the proton through the process $p=uud\rightarrow\tilde{s}^\dag u\rightarrow e^+ \pi_0$. Those types of processes are already constrained by experimental measurements \cite{protondecay}. Differently, in a conserved R-parity model ($W_{RPV} = 0$), supersymmetric particles are only created by pairs because the product of the R-number of all particles in a vertex is $+1$. It then leads to a stable SUSY particle called \textit{LSP} (for \textit{Lightest-Supersymmetric-Particle}), which potentially solves the problem of dark matter.\medskip

It is usual for simplifying the study of such supersymmetric models to consider $R$-parity conservation, \textit{i.e.}, $W_{RPV}=0$ and not taking into account the neutrino sector, \textit{i.e.}, $W_N=0$. The first and second family of fermions can also be omitted, taking:
\begin{eqnarray}
\mathbf{Y_U} \approx \begin{pmatrix}
0 & 0 & 0 \\
0 & 0 & 0 \\
0 & 0 & y_t
\end{pmatrix},\quad  \mathbf{Y_D} \approx \begin{pmatrix}
0 & 0 & 0 \\
0 & 0 & 0 \\
0 & 0 & y_b
\end{pmatrix},\quad \mathbf{Y_E} \approx \begin{pmatrix}
0 & 0 & 0 \\
0 & 0 & 0 \\
0 & 0 & y_\tau
\end{pmatrix}\ ,\label{yukawa}
\end{eqnarray}
with $y_t,\ y_b,\ y_\tau$ the Yukawa couplings of the top quark, bottom quark and tau. \medskip
 
The supersymmetric breaking terms $\mathcal{L}_{\cancel{SUSY}}$ in the Lagrangian can be written as:
\begin{gather}
\mathcal{L}_{\cancel{SUSY}} = V_{SOFT}^{\phi-mass} + V_{SOFT}^{W} + V_{SOFT}^{\lambda-mass}\label{eq:VsoftMSSM}
\end{gather}
defining
\begin{gather}
V_{SOFT}^{\lambda-mass} = \frac12 \left( M_1 \tilde{B}\cdot\tilde{B} + M_2 \tilde{W}^a\cdot\tilde{W}_a + M_3 \tilde{g}^a\cdot\tilde{g}_a + \mathrm{h.c.} \right)\ , \label{eq:vsoftgauginos}\\
V_{SOFT}^{\phi-mass} = m_{H_U}^2 \left|H_U\right|^2+ m_{H_D}^2\left|H_D\right|^2 + m_Q^2 |\hat{Q}|^2 +  m_L^2 |\hat{L}|^2 + m_U^2 |\hat{U}|^2 + m_D^2 |\hat{D}|^2\ ,\label{eq:vsoftphi}\\
V_{SOFT}^{W} = \Big( B\mu H_U\cdot H_D + y_UA_U\tilde{q}_L\cdot H_U \tilde{u}_R^\dag -y_D A_D \tilde{q}_L\cdot H_D \tilde{d}^\dag_R \nn\\
-y_EA_E\tilde{l}_L\cdot H_D \tilde{e}^\dag_R + \mathrm{h.c.} \Big) \ ,\label{eq:vsoftw}
\end{gather}
where $V_{SOFT}^{\lambda-mass}$ and $V_{SOFT}^{\phi-mass}$, the mass terms for gauginos and scalar fields ($\tilde{W}^a=\delta^{ab}\tilde{W}_b$ and $\tilde{g}^a=\delta^{ab}\tilde{g}_b$), and $V_{SOFT}^{W}$ supersymmetric breaking terms deduced by associating to each coupling in $W_{MSSM}$ a term in \autoref{eq:VsoftMSSM}. \medskip

Imposing that the two Higgs fields $H_U^0$ and $H_D^0$ acquire non-zero \textit{v.e.vs.} $v_U$ and $v_D$, a mechanism similar to the well-known Brout-Englert-Higgs generates the mass terms for the Z and W-bosons:
\begin{gather}
M_W^2 = \frac14 g_2^2 v^2 \ , \quad M_Z^2=\frac14\left( g_1^2 + g_2^2 \right) v^2 \ , 
\end{gather}
with $v^2 = v_U^2+ v_D^2$ and $g_1$ and $g_2$ the gauge couplings of respectively $U_Y(1)$ and $SU_L(2)$. \medskip

After imposing the minimum of the potential and the calculation of the mass matrices, the remaining particle spectrum is constituted by two scalar Higgs $h_i$ ($i=1,2$), one pseudoscalar $A_0$ and a charged Higgs $h^\pm$ in the Higgs sector. In the sfermions sector, two states are associated to each fermion of the Standard Model, meaning two stops $\tilde{t}_i$, two sbottoms $\tilde{b}_i$ and two staus $\tilde{\tau}_i$ ($i=1,2$). Furthermore, four new states called neutralinos $\tilde{\chi}^0_i$ ($i=1,...,4$) are generated through the mixing of neutral Higgsinos and gauginos. The charged Higgsinos and Wino states also mix themselves, leading to two charginos $\tilde{\chi}^\pm_i$ ($i=1,2$). A more complete presentation of the MSSM can be found in \cite{book_susy}\cite{martin_supersymmetry_1998}. \medskip

Finally, we would like to end this section with an observation. The parameter $\mu$ in the superpotential \autoref{mssmW} contributes to the mass of the Higgs bosons. In order to reconstruct a proper electroweak scale, this parameter should be of the order of $100\ \mathrm{GeV}$, but this value is not natural. Indeed, because this parameter has a dimension of mass, it must of the order of $M_{Planck}\approx 10^{19}\ \mathrm{GeV}$, which leads to the famous \textit{$\mu$-problem}. One possible solution to this issue has been found by Giudice \& Masiero by introducing a new term in the Kähler potential (see \autoref{subsec:soft}). Another solution is to add a new singlet superfield $S=(\underset{\sim}{1},\underset{\sim}{1},0)$ to the particle content. This new model is called the NMSSM (for \textit{Next-to-Minimal Supersymmetric Standard Model}) \cite{Fayetnmssm}\cite{NMSSM} and will be briefly discussed in Subsection \ref{sub:difficultNMSSM}.\medskip

The next section will be devoted to a similar supersymmetric where two singlets superfields $\hat{S}^1$ and $\hat{S}^2$ are appended to the superfield content of the MSSM.

\subsection{Description of the N2MSSM}\label{sec:N2MSSMDes}
\subsubsection{Superpotential and Potential calculation}
The N2MSSM is defined as the \textit{Minimal Supersymmetric Standard Model} (MSSM) with two extra singlet superfields $\hat{S}^p=(\underset{\sim}{1},\underset{\sim}{1},0)$ $(p=1,\ 2)$. Not considering the neutrino sector and the couplings which violate R-parity, the general form of the superpotential can be written as:
\begin{eqnarray}
W_{N2MSSM} &=& \mu\hat{H}_U\cdot \hat{H}_D  + \lambda_i \hat{S}^i \hat{H}_U\cdot \hat{H}_D  + \frac13\kappa_{ijk}\hat{S}^i\hat{S}^j\hat{S}^k + \frac12 \mu'_{ij}\hat{S}^i\hat{S}^j + \xi_{F,i}\hat{S}^i \nn\\
&& + \hat{Q}\cdot \hat{H}_U \mathbf{Y_U} \hat{U} - \hat{Q}\cdot \hat{H}_D \mathbf{Y_D} \hat{D} - \hat{L}\cdot \hat{H}_D \mathbf{Y_E} \hat{E} \ .\label{eq:n2mssm}
\end{eqnarray}
We assume only real parameters to avoid breaking CP-symmetry and only the third family of fermions (see \autoref{yukawa}). The NMSSM is obtained as some limit of the N2MSSM, \textit{i.e.}, taking all the couplings of the second singlet $\hat{S}^2$ to zero.\medskip

We note that that the $\mu$-coupling between the two Higgs superfields can be eliminated. Indeed, by considering the shift $\hat{S}^1 \rightarrow \hat{S}^1 - \frac{\mu}{\lambda_1}$ for the singlet superfield,
\begin{gather}
\mu \hat{H}_U\cdot \hat{H}_D + \lambda_1\hat{S}^1\hat{H}_U\cdot \hat{H}_D \underset{shift}{\rightarrow} \lambda_1\hat{S}^1\hat{H}_U\cdot \hat{H}_D \ ,\nn
\end{gather}
 this term can be reabsorbed in the lambda coupling. Following the same approach for the second singlet, the bilinear term $\mu'_{12}$ can also be assimilated. In this manner, we reduce the parameters space which will be helpful for the phenomenological analysis (see Chapter \ref{chap:N2MSSM}).\medskip

Assuming a $\mathbb{Z}_3$-invariance ($\hat{\Phi}\rightarrow e^{\frac{2\pi i}{3}}\hat{\Phi}$) of the superpotential can also simplify the model. This symmetry is obtained by forbidding  all dimensional parameters in $W_{N2MSSM}$ ($\mu'_{ij} = \xi_{F,i} = 0$). \footnote{We can assume a superconformal invariance of the theory, which impose a $\mathbb{Z}_3$ symmetry of the superpotential (see \cite{book_sugra})} This assumption, however, leads to cosmological inconsistency known as the \textit{domain wall problem} \cite{domainwall}. The non-vanishing \textit{v.e.vs.} of the Higgs and singlet fields break this discrete symmetry. It generates then in the universe regions with the same vacuum energy but with different phases from the transformation $\hat{\Phi}\rightarrow e^{\frac{2\pi i}{3}}\hat{\Phi}$. The separation surface between those regions (the domain wall) contains a lot of energy not detected by the experimental measurements. Some solutions to this problem have been proposed \cite{domwalsol1}\cite{domwalsol2}. Since this phenomenon is out of the scope of this manuscript, we will not consider this possible issue.\medskip

With the superpotential \autoref{eq:n2mssm}, using a \textit{Gravity Mediated Supersymmetry Breaking} mechanism, we obtain:
\begin{gather}
V_{SOFT}^{N2MSSM} = V_{SOFT}^{\phi-mass} + V_{SOFT}^{\lambda-mass} + V_{SOFT}^{W}\ ,\nn
\end{gather}
where $V_{SOFT}^{\lambda-mass}$ describes the mass terms for the gauginos (see \autoref{eq:vsoftgauginos}), $V_{SOFT}^{\phi-mass}$ corresponds to the mass terms of the scalar fields,
\begin{gather}
V_{SOFT}^{\phi-mass} = V_{SOFT}^{\phi-mass}\big|_{MSSM} +  \frac12 m^2_{S^{ij}}S^i S^{\dagger}_j \ , \nn
\end{gather}
(with $V_{SOFT}^{\phi-mass}\big|_{MSSM}$ the MSSM mass content, see \autoref{eq:vsoftphi}), and $V_{SOFT}^{W}$ the soft breaking terms associated to the superpotential $W_{N2MSSM}$ \autoref{eq:n2mssm},
\begin{gather}
V_{SOFT}^{W} = V_{SOFT}^{W}\big|_{MSSM\ }^{(\mu = 0)} + \lambda_i A_{\lambda_i}S^iH_U\cdot H_D + \frac13 \kappa_{ijk} A_{\kappa_{ijk}}S^iS^jS^k + \frac12 m'_{i}\left( S^i \right)^2 + \xi_{S,i}S^i \ ,\nn
\end{gather}
(where $ V_{SOFT}^{W}\big|_{MSSM\ }^{(\mu = 0)}$ are the soft breaking terms associated to the MSSM, see \autoref{eq:vsoftw}) with $\mu=0$. \medskip

From these definitions, it is then possible to compute the full scalar potential by calculating the F and D-term in \autoref{eq:potSUSY}.\medskip
	
We now study, after electroweak symmetry breaking, the fields content. Assuming that neutral fields develop a non-zero \textit{v.e.v.}, we get:
\begin{equation}
H_D=\begin{pmatrix}
\frac{1}{\sqrt{2}}(v_D + h_{D}^0 + iA_{D}^0) \\
H_{D}^- 
\end{pmatrix}\qquad , \qquad 
H_U= \begin{pmatrix}
H_{U}^+ \\
\frac{1}{\sqrt{2}}(v_U + h_{U}^0 + iA_{U}^0) 
\end{pmatrix}\ ,\nonumber
\end{equation}
\begin{equation}
S^i=\frac{1}{\sqrt{2}}(v_i + \mathrm{Re}(S^i) + i\mathrm{Im}(S^i)) \ .\label{eq:re&im}
\end{equation}
 \newline
Since we must work at the minimum of the potential, the minimisation equations
\begin{eqnarray}
\left<\frac{\partial V}{\partial H_U^0}\right> = 0,\quad \left<\frac{\partial V}{\partial H_D^0}\right> = 0,\quad \left<\frac{\partial V}{\partial S^1}\right> = 0,\quad \left<\frac{\partial V}{\partial S^2}\right> = 0 \nn\label{eq:minin2}
\end{eqnarray}
allow us to eliminate the mass parameters $m_{H_U}^2,\ m_{H_D}^2,\ m_{S^1}^2,\ m_{S^2}^2$ in the potential: 
\begin{eqnarray}
m_{H_U}^2 &=& \left(\tan\beta\right)^{-1}\Big( \lambda_1 ( \kappa_1v_1^2 + 2\kappa_{12}v_1v_2 +\kappa_{21}v_2^2 \mu_1v_1 + \xi_{F,1} ) \nn\\
&&+ \lambda_2 ( \kappa_2v_2^2 + 2\kappa_{21}v_1v_2 +\kappa_{12}v_1^2 \mu_2v_2 + \xi_{F,2} )\Big) \nn\\
&&-(\lambda_1^2+\lambda_2^2)v_D^2 -(\lambda_1v_1 + \lambda_2v_2)^2-\frac{g_1^2+g_2^2}{4}(v_U^2-v_D^2)\ ,\label{eq:param_min1}\\
m_{H_D}^2 &=& \tan\beta\Big( \lambda_1 ( \kappa_1v_1^2 + 2\kappa_{12}v_1v_2 +\kappa_{21}v_2^2 \mu_1v_1 + \xi_{F,1} ) \nn\\
&&+ \lambda_2 ( \kappa_2v_2^2 + 2\kappa_{21}v_1v_2 +\kappa_{12}v_1^2 \mu_2v_2 + \xi_{F,2} )\Big) \nn\\
&&-(\lambda_1^2+\lambda_2^2)v_U^2 -(\lambda_1v_1 + \lambda_2v_2)^2-\frac{g_1^2+g_2^2}{4}(v_D^2-v_U^2)\ ,\label{eq:param_min2}\\
m_{S^1}^2 &=& \lambda_1v_1^{-1}\Big(A_{\lambda_1}v_Uv_D-(v_U^2+v_D^2)(\lambda_1v_1+\lambda_2v_2)\Big)-v_1^{-1}(v_2m_{S^1S^2}^2+\kappa_{21}A_{\kappa_{21}}v_2^2 + \xi_{S,1})\nn\\
&&+v_1^{-1}\Big( 2\kappa_{1}v_1+2\kappa_{12}v_2 +\mu_1 \Big)\Big( \lambda_1v_Uv_D-\kappa_{1}v_1^2-2\kappa_{12}v_1v_2 - \kappa_{21}v_2^2-\mu_1v_1-\xi_{F,1} \Big)\nn\\
&&+2v_1^{-1}\Big( \kappa_{12}v_1+\kappa_{21}v_2 \Big)\Big( \lambda_2v_Uv_D-\kappa_{12}v_1^2-2\kappa_{21}v_1v_2 - \kappa_2v_2^2-\mu_2v_2-\xi_{F,2} \Big)\nn\\
&& -{m'}_1^2-v_1\kappa_1 A_{\kappa_1}-2\kappa_{12}A_{\kappa_{12}}v_2\ ,\label{eq:param_min3}\\
m_{S^2}^2 &=& \lambda_2v_2^{-1}\Big(A_{\lambda_2}v_Uv_D-(v_U^2+v_D^2)(\lambda_1v_1+\lambda_2v_2)\Big)-v_2^{-1}(v_1m_{S^1S^2}^2+\kappa_{12}A_{\kappa_{12}}v_1^2 + \xi_{S,2})\nn\\
&&+v_1^{-2}\Big( 2\kappa_{21}v_1+2\kappa_{2}v_2 +\mu_2 \Big)\Big( \lambda_1v_Uv_D-\kappa_{12}v_1^2-2\kappa_{21}v_1v_2 - \kappa_{2}v_2^2-\mu_2v_2-\xi_{F,2} \Big)\nn\\
&&+2v_1^{-1}\Big( \kappa_{12}v_1+\kappa_{21}v_2 \Big)\Big( \lambda_2v_Uv_D-\kappa_{1}v_1^2-2\kappa_{12}v_1v_2 - \kappa_{21}v_2^2-\mu_1v_1-\xi_{F,1} \Big)\nn\\
&& -{m'}_2^2-v_2\kappa_{2} A_{\kappa_{2}})-2\kappa_{21}A_{\kappa_{21}}v_1 \ ,\label{eq:param_min4}
\end{eqnarray} 
defining $\beta=\frac{v_U}{v_D}$, $\kappa_{ij}=\kappa_{iij}$ and $A_{\kappa_{ij}}=A_{\kappa_{iij}}$.
\subsubsection{Higgs sector}

\sloppy After using \autoref{eq:re&im} and (\ref{eq:param_min1} - \ref{eq:param_min4}), the calculation of the mass matrix in the basis $\{h_U^0,h_D^0,S^1_R,S^2_R,A_U^0,A_D^0,S^1_I,S^2_I,H_U^+,(H_D^-)^\dagger\}$ leads to the following block-diagonal structure:
\begin{eqnarray}
\mathbb M^2_{H} &=& 
\begin{pmatrix} 
\mathbb M^2_{h^0} & 0 & 0 \\
0 & \mathbb M^2_{A^0} & 0 \\
0 & 0 & \mathbb M^2_{h^\pm} \\
\end{pmatrix}\ .\nn
\end{eqnarray}
The exact form of $\mathbb{M}^2_{h^0}$ and $\mathbb{M}^2_{A^0}$ can be found in Appendix \ref{app:higgssector}.\medskip 

The scalar sector $\{h_U^0,h_D^0,S^1_R,S^2_R\}$ is now four-dimensional and leads to four different scalar Higgs: $\{h_i^0\}$ $(i=1, ...4)$. In the pseudoscalar sector $\{A_U^0,A_D^0,S^1_I,S^2_I\}$, a Goldstone boson remains in the spectrum. We can then rotate the matrix in the new basis $\{G,A,S^1_I,S^2_I\}$:
\begin{eqnarray}
\begin{pmatrix}
A_U^0\\
A_D^0\\
S^1_I\\
S^2_I
\end{pmatrix} &=& \begin{pmatrix}
\sin\beta & \cos\beta & 0 & 0 \\
-\cos\beta & \sin\beta & 0 & 0 \\
0 & 0 & 1 & 0 \\
0 & 0 & 0 & 1 \\
\end{pmatrix}\begin{pmatrix}
G\\
A\\
S^1_I\\
S^2_I
\end{pmatrix}\ ,\nn
\end{eqnarray} 
where $G$ is the Goldstone boson. For the continuation, we redefine the pseudoscalar mass matrix $\mathbb{M}^2_{A^0}$ as the $3\times 3$ matrix in the $\{A,S^1_I,S^2_I\}$ basis.\newline 
For the charged Higgs sector $\{H_U^+,(H_D^-)^\dagger\}$, by defining $m_A^2\equiv\left(\mathbb{M}^2_{A^0}\right){}_{11}$, we can rewrite $\mathbb{M}^2_{h^\pm}$ in an equivalent form to the MSSM and the NMSSM :
\begin{eqnarray}
\mathbb{M}^2_{h^\pm} &=& \left( \frac12 m_A^2\sin 2\beta+\left( \frac{g_2^2}{2} - \left(\lambda_1^2 + \lambda_2^2\right)\right) v_Uv_D \right) \begin{pmatrix}
\tan^{-1}\beta & 1 \\
1 & \tan\beta 
\end{pmatrix}\ .\nn
\end{eqnarray}
The eigenstates are one Goldstone boson and a charged Higgs with the mass:
\begin{eqnarray}
m_{h^\pm}^2 &=& m_A^2 + m_W^2 - (\lambda_1^2+\lambda_2^2)v^2\ .\nn
\end{eqnarray}
We see that the charged Higgs mass can be smaller than in the MSSM due to the coupling between the Higgs doublets and the singlet fields. We can also note that taking $\lambda_1^2 + \lambda_2^2=\lambda^2$ gives an equivalent form of the NMSSM Higgs sector.
\subsubsection{Neutralinos \& charginos sector}
Compare to the MSSM, the addition of two singlets modify the neutralino sector and turns out to be six dimensional $\psi_0=\{-i\tilde{B},-i\tilde{W}_3,\tilde{h}^0_U,\tilde{h}^0_D,\tilde{S}^1,\tilde{S}^2\}$ with the mass term:
\begin{eqnarray}
\mathcal L = -\frac12 (\psi_0)^T \mathbb M_{\chi^0} \psi_0 + \mathrm{h.c.} \nn
\end{eqnarray}
with
\begin{eqnarray}
\mathbb M_{\chi^0} &=& \begin{pmatrix}
M_1 & 0 & -\frac{g_1v_D}{\sqrt{2}} & \frac{g_1v_U}{\sqrt{2}} & 0 & 0 \\
    & M_2 &  \frac{g_2v_D}{\sqrt{2}} &  \frac{g_2v_U}{\sqrt{2}} & 0 & 0 \\
    &     & 0 & -\frac{\lambda_1v_1+\lambda_2v_2}{\sqrt{2}} & -\frac{\lambda_1v_D}{\sqrt{2}} & -\frac{\lambda_2 v_D}{\sqrt{2}} \\
    &     &   &  0  & -\frac{\lambda_1v_U}{\sqrt{2}} & -\frac{\lambda_2 v_U}{\sqrt{2}} \\
    &     &   &     &  \sqrt{2}\kappa_1v_1 + \sqrt{2}\kappa_{12}v_2 +\mu_1  &  \sqrt{2}\kappa_{12}v_1 + \sqrt{2} \kappa_{21}v_2 \\
    &     &   &     &                                         & \sqrt{2}\kappa_2v_2 + \sqrt{2}\kappa_{21}v_1 + \mu_2    
\end{pmatrix} \ .\nn
\end{eqnarray}
The charginos sector is analogous as in the MSSM. The states are defined by mixing between $\{\tilde{W}^+,\tilde{W}^-\}$ (with $\tilde{W}^- = \frac1{\sqrt{2}}( \tilde{W}^1 + i\tilde{W}^2 )$ and $\tilde{W}^- = \frac1{\sqrt{2}}( \tilde{W}^1 - i\tilde{W}^2 )$) and $\{\tilde{H}_U^+,\tilde{H}_D^-\}$: 
\begin{eqnarray}
\psi^+ = \begin{pmatrix}
-i\tilde{W}^+ \\
\tilde{H}_U^+
\end{pmatrix},\quad \psi^- = \begin{pmatrix}
-i\tilde{W}^- \\
\tilde{H}_D^-
\end{pmatrix}\ , \nn
\end{eqnarray}
the mass term read
\begin{eqnarray}
\mathcal L = -\frac12 \begin{pmatrix}
\psi^+ & \psi^- 
\end{pmatrix}\begin{pmatrix}
0 & X^T \\
X & 0
\end{pmatrix}\begin{pmatrix}
\psi^+ \\
\psi^-
\end{pmatrix} + \mathrm{h.c.}\ ,\nn
\end{eqnarray}
with :
\begin{eqnarray}
X = \begin{pmatrix}
M_2 & g_2v_U \\
g_2v_D & \lambda_1 v_1 + \lambda_2 v_2
\end{pmatrix}\ .\nn
\end{eqnarray}
\subsubsection{Squarks \& Sleptons sector}
In the same manner as in the chargino sector, the new contributions of the second singlet will appear in the squarks and sleptons sector :
\begin{itemize}
\item stop mass matrix in the basis $\{\tilde{t}_L,\tilde{t}_R\}$:
\begin{eqnarray}
\mathbb M^2_{\tilde{t}} = \begin{pmatrix}
m_{U_3}^2+y_t^2v_U^2 - \frac{g_1^2}{3}(v_U^2-v_D^2) & y_t\left(A_tv_U - (\lambda_1v_1 + \lambda_2v_2)v_D)\right) \\
y_t\left(A_tv_U - (\lambda_1v_1 + \lambda_2v_2)v_D)\right) & m_{Q_3}^2+y_t^2v_U^2 + \left(\frac{g_1^2}{12}-\frac{g_2^2}{4}\right)(v_U^2-v_D^2)
\end{pmatrix}\ ,\nn
\end{eqnarray}
\item sbottom mass matrix in the basis $\{\tilde{b}_L,\tilde{b}_R\}$:
\begin{eqnarray}
\mathbb M^2_{\tilde{t}} = \begin{pmatrix}
m_{D_3}^2+y_b^2v_D^2 + \frac{g_1^2}{6}(v_U^2-v_D^2) & y_b\left(A_bv_D - (\lambda_1v_1 + \lambda_2v_2)v_U)\right) \\
y_b\left(A_bv_D - (\lambda_1v_1 + \lambda_2v_2)v_U)\right) & m_{Q_3}^2+y_b^2v_D^2 + \left(\frac{g_1^2}{12}+\frac{g_2^2}{4}\right)(v_U^2-v_D^2)
\end{pmatrix}\ ,\nn
\end{eqnarray}
\item stau mass matrix in the basis $\{\tilde{\tau}_L,\tilde{\tau}_R\}$:
\begin{eqnarray}
\mathbb M^2_{\tilde{t}} = \begin{pmatrix}
m_{E_3}^2+y_\tau^2v_D^2 + \frac{g_1^2}{2}(v_U^2-v_D^2) & y_\tau\left(A_\tau v_D - (\lambda_1v_1 + \lambda_2v_2)v_U)\right) \\
y_\tau\left(A_\tau v_D - (\lambda_1v_1 + \lambda_2v_2)v_U)\right) & m_{L_3}^2+y_\tau^2v_D^2 - \left(\frac{g_1^2-g_2^2}{4}\right)(v_U^2-v_D^2)
\end{pmatrix}\ .\nn
\end{eqnarray}
\end{itemize}

-

\label{sugra}

\chapter{New Solutions in Gravity Mediated Supersymmetry Breaking}
\label{NSW}
In section \ref{sub:SW}, we have mentioned the solutions obtained by Soni and Weldon for \textit{Gravity Mediated Supersymmetry Breaking} \cite{soni_analysis_1983}. The form of the Kähler potential $K(\Phi,\Phi^{\dagger},Z,Z^{\dagger})$ and the superpotential $W(\Phi, Z)$ are constrained in a way that couplings of the matter sector $\Phi$ do not diverge in the limit $m_p\rightarrow\infty$ (see \autoref{SWsolution1} and \autoref{SWsolution2}). Those types of solutions are the cornerstone of all current studies involving gravity mediated solutions.\medskip

In a recent paper \cite{moultaka_low_2018}, G. Moultaka, M. Rausch de Traubenberg and D. Tant have found that Soni \& Weldon's classification of the Kähler potential and the superpotential was incomplete. By making the analytic expansion of the potential in Planck mass $m_p$, they have found two types of solutions (considering a canonical Kähler potential): the already known Soni-Weldon solutions and a new form of solution, which will be introduced in this chapter.\medskip

Those new solutions correspond to a specific form of the superpotential with new type of singlet superfields (noted $S^p$, $p=1,\dots, n$) called \textit{hybrid} fields, which have the properties of both hidden and matter sectors. The interaction between those superfields and the usual matter superfields generate new types of supersymmetry breaking terms. Indeed, the usual Soni \& Weldon's solutions lead only to soft-breaking terms (terms leading to logarithmic divergences), whilst the new solutions (which will be called Non-Soni \& Weldon, or NSW solutions) generate in the Lagrangian hard-breaking terms (generating quadratic divergences) controlled by the gravitino-mass $m_{3/2}$ and suppressed by an energy-scale lower than the Planck scale. \medskip

In this chapter, we briefly present the new solutions and define a specific model assuming a non-canonical Kähler potential (and by the way extends the results of \cite{moultaka_low_2018}) called the S2MSSM. This model involves two hybrid fields $S^p$ ($p=1,2$), which lead to some relationship with the two-singlet extension of the MSSM (called the N2MSSM, see \autoref{chap:N2MSSM}). The full potential is calculated, and some discussions are performed on the possible effects of the new hard-breaking terms on the spectrum. A first study on the effects of those new solutions on the Higgs sector is achieved. The order of magnitude of the loop-corrections on the Higgs boson mass is calculated in order to discuss the reduction of the fine-tuning of the parameters to recover a Higgs boson of about $125\ \mathrm{GeV}$. \medskip

This work is done in collaboration with Gilbert Moultaka (L2C) and Michel Rausch de Traubenberg (IPHC) \cite{NSW_Higgs}.
\section{Non-Soni Weldon solutions}
This section is dedicated to a discussion of the Non-Soni-Weldon solutions. After introducing the form of the superpotential and the new hybrid superfields $S^p$, the main differences between those solutions and the Soni-Weldon's ones are exhibited. The calculation of the Coleman-Weinberg potential is done to emphasise the properties of the hard-breaking terms regarding the one-loop corrections. 
\subsection{Presentation of the general solutions}
We describe (without going into the details) the results of \cite{moultaka_low_2018}. We assume the existence of two sectors: a hidden sector $\{z^i\}$ (with $\zeta^i = z^im_p$) and a visible sector $\{\Phi^a\}$. Assuming further a canonical Kähler potential and a superpotential written as:
\begin{eqnarray}
W(Z,\Phi ) &=& \displaystyle\sum_{n=0}^{M}m_p^n W_n(z,\Phi)\ , \label{eq:Wexpansion} \\
K(\Phi,\Phi^{\dagger}, z , z^{\dagger}) &=& m_p^2 z_i^{\dagger}z^i + \Phi^{\dagger}_a\Phi^a \ , \label{eq:Kcanonical}
\end{eqnarray}
we can calculate the F-term of $N=1$ supergravity and expend $V_F$ in power of $m_p$: 
\begin{eqnarray}
V_F &=& e^{K(\Phi,\Phi^{\dagger},z,z^{\dagger})/m_p^2}\Big\{ \mathcal{D}_{I}W(K^{-1})^I{}_{J^\ast}\mathcal{D}^{J^\ast}\overline{W} -\frac{3}{m_p^2}|W|^2\Big\} \label{eq:SUGRApot} \\
 &=& e^{K(\Phi,\Phi^{\dagger},z,z^{\dagger})/m_p^2}\displaystyle\sum_{c=0}^{2M} V_{M,c}(z,z^{\dagger},\Phi,\Phi^{\dagger})m_p^c \label{eq:potential_mp}
\end{eqnarray}
(the complete form of $V_{M,c}$ can be found in the paper \cite{moultaka_low_2018}). We now impose the constraint on the low-energy coupling of the matter sector. Denoting $V_{\Phi}$ the part of the potential containing the interactions to the sector $\{\Phi^a\}$, we require (in the same manner as in \autoref{sub:SW}) that
\begin{equation}
\lim_{m_p\rightarrow \infty} V_{\Phi} \text{ is non-divergent}\ .\label{eq:constraintNSW}
\end{equation}
An equivalent formulation of this constraint using the potential \autoref{eq:potential_mp} is given by
\begin{eqnarray}
\frac{\delta V_{M,c}(z,z^{\dagger},\Phi,\Phi^{\dagger})}{\delta\Phi^a(x)} =\frac{\delta V_{M,c}(z,z^{\dagger},\Phi,\Phi^{\dagger})}{\delta\Phi_a^{\dagger}(x)} = 0 \ \forall x,\ M,\ a, c\geq 1 \ .\nn
\end{eqnarray}
This relation leads to 2M partial differential equations which strongly constrain the form of $W_n(\Phi,z)$ in \autoref{eq:Wexpansion}. After resolving this system, the authors have found that:
\begin{itemize}
\item there is no solutions satisfying the constraint \ref{eq:constraintNSW} for $M>2$ with $W_{n}(\Phi,z)$ ($n\geq 3$) as non-vanishing functions,
\item in the case $M\leq 2$, there is two distinct solutions:
\begin{itemize}
\item the classical Soni-Weldon solutions (SW):
\begin{gather}
W(z,\Phi) = m_p^2 W_2(z) + m_pW_1(z) + W_0(z,\Phi) \nn
\end{gather}
and already introduced in the first chapter of this manuscript. 
\item a new solution with $W_2 = 0$ and where the matter sector must be divided into two different subsectors $\{\Phi^a\} = \{\tilde{\Phi}^a\}\oplus\{S^p\}$ where $\{\tilde{\Phi}^a\}$ denotes the \textit{standard} matter sector (for example, the matter content of the MSSM). The structure of the superpotential is given by: 
\begin{gather}
W(Z,\Phi)=m_p W_1(z,S) + W_0(z,S,\tilde{\Phi})\ , \label{eq:Wexp}
\end{gather}
where
\begin{gather}
W_1(z,S) = W_{1,0}(z) + \displaystyle\sum_{p\geq 1}^{P} W_{1,p}(z)\displaystyle\sum_{s\geq 1}^{n_p}\mu^\ast_{p_s}S^{p_s} \label{eq:W1}
\end{gather}
and
\begin{gather}
W_0(z,S,\tilde{\Phi}) = \displaystyle\sum_{q\geq 1}^{k_1} W_{0,q}(z)S^q + \Xi(\mathcal{U}_S^{pp_s},\tilde{\Phi}^a, z^i )\ . \label{eq:W0}
\end{gather}
The functions $W_{1,0}(z)$, $W_{1,p}(z)$ and $W_{0,q}(z)$ are arbitrary holomorphic functions. The function $\Xi$ is the only holomorphic function containing the matter sector $\tilde{\Phi}^a$ and a linear combination of a pair of $S$ through the $\mathcal{U}$ fields:
\begin{gather}
\mathcal{U}_S^{pp_s} \equiv \xi_{p_s}(z)S^{p_s} - \xi^{p_s}(z)S^{p_1} \label{eq:Ustruc}
\end{gather}
with the two $\xi$ functions related by $\mu_{p_s}\xi_{p_s}(z) = \mu_{p_1}\xi^{p_s}(z)$ (with $\mu_{p}$ the complex conjugate of $\mu^{\ast}_{p}$ from \ref{eq:W1}). 
\end{itemize}

\end{itemize}
Some remarks are now in order. The $\{S^p\}$ sector is defined as \textit{hybrid} since it is present in $W_1$, \textit{i.e.} the $m_p$-term in the expansion of the superpotential \ref{eq:Wexp} whilst its couplings do not diverge at low-energy. Moreover, the fields $\{S^p\}$ must be singlet under the gauge group of the standard model. The \textit{v.e.vs} $\lag S^p\rag$, as well as the one the matter sector $\{\tilde{\Phi}^a\}$, must be much lower than the Planck mass ($\lag\tilde{\Phi}^a\rag \ll m_p$ and $\lag S^p\rag \ll m_p$), which avoids dangerous terms at low energy. \medskip

The matter sector $\{\tilde{\Phi}^a\}$ is only present in $W_0$ \autoref{eq:Wexp} (more precisely in $\Xi(\mathcal{U}^{pp_s}_S,\tilde{\Phi}^a,z^i)$ \autoref{eq:W0}). It is similar to the classical Soni-Weldon solutions. Note that $\Xi$ contains the usual superpotential of the supersymmetric models (such as the MSSM or the NMSSM) with new interactions with the $\mathcal{U}^{pp_s}_S$ fields. Due to the form of the $\mathcal{U}$-field \ref{eq:Ustruc}, couplings between the matter and the hybrid sector only appear with models containing at least two hybrid fields $S^1$ and $S^2$. \medskip

We can also note that while the superpotential is a function of the complex parameters $\mu^{\ast}_{p_s}$ (see \autoref{eq:W1}), the conjugate parameters $\mu_{p_s}$ are present in the definition of the linear combination $\mathcal{U}_S^{pp_s}$ \ref{eq:Ustruc}. This is a crucial point for prohibiting dangerous terms at low-energy.\medskip

Introducing the mass scale
\begin{gather}
M^2\equiv \big| \lag W_{1,0}(z) + \displaystyle\sum_{p\geq 1}^{P} W_{1,p}(z)\displaystyle\sum_{s\geq 1}^{n_p}\mu^\ast_{p_s}S^{p_s} \rag \big|\ , \label{eq:M2}
\end{gather}
the gravitino mass is (we neglect the contribution from the matter sector):
\begin{gather}
m_{3/2} = \frac{1}{m_p^2}\lag W\rag e^{K/(2m_p^2)} = \frac{M^2}{m_p}e^{\sum\limits_{i}|\lag z^i\rag|^2}\ .\nn
\end{gather}
We then see that in the case of Non-Soni \& Weldon solutions, the gravitino mass is Planck-suppressed which is not the case for classical Soni-Weldon solutions. We have thus a hierarchy $m_{3/2}^{SW}>m_{3/2}^{NSW}$ (since we have the hierarchy $M<m_p$).\medskip

Note that this classification of \textit{Gravity Mediated Supersymmetry Breaking} solutions has been made by assuming a canonical Kähler potential \ref{eq:Kcanonical}. A full classification (with non-canonical Kähler potential) has not been obtained yet. Only some specific solutions have been derived\footnote{Without taking into account the S2MSSM solutions presented in \ref{sec:S2MSSM}, a No-Scale Supergravity solution has also been discovered where the vanishing of the cosmological constant is simply a geometrical property, see \cite{noscale1}\cite{noscale2}\cite{noscale3}\cite{moultaka_low_2018}}. For instance, a Giudice-Masiero-like solution has also been developed (see Appendix \ref{app:GMsolutions} for the full calculation of the potential).\medskip

The last property of such new solutions is that the interactions between the hybrid fields $\{\mathcal{U}^{pp_s}_S\}$ and the hidden sector $\{z^i\}$ generate the classical soft-breaking terms together with new hard-breaking terms. Those new terms have the property to generate quadratic radiative corrections whilst the soft-breaking terms lead to only logarithmic divergences. A priori, quadratic divergences can be seen as dangerous since the cancellation of $\Lambda^2$-corrections in supersymmetry is usually seen as advantageous\footnote{Note however that quantum instability has been already pointed out on the conventional supergravity solutions, see \cite{destabilSUGRA1}\cite{destabilSUGRA2}}. Nonetheless, those dangerous terms are controlled by the gravitino mass $m_{3/2}^{NSW}$, the \textit{v.e.vs} of the S ($\mathcal{U}$) fields and in addition are suppressed by the intermediate energy scale $M\ll m_p$ (see \autoref{eq:M2}). At low-energy, the full potential can then be written as:
\begin{gather}
V = V_{SUSY} + V_{HARD} + V_{SOFT}\nn
\end{gather}
with $V_{SUSY}$ the classical potential of supersymmetry (see \autoref{eq:potSUSY}) and (the fields $\phi^a$ denote the scalar part of the sector $\{\Phi^a\}=\{\tilde{\Phi}^a\}\oplus\{S^p\}$):
\begin{gather}
V_{SOFT} = \Big(C_a\phi^a + \frac{1}{2}B_{ab}\phi^a\phi^b + \frac{1}{6}A_{abc}\phi^a\phi^b\phi^c +\mathrm{h.c.}\Big) + (m^2)^{i}{}_{j}\phi^j\phi^{\dagger}_i\equiv s+ (m^2)^{i}{}_{j}\phi^j\phi^{\dagger}_i \ , \label{eq:vsoftCW}
\end{gather}
\begin{gather}
V_{HARD} = \left(\frac12 D_{ij}{}^k \phi^i\phi^j\phi^\dag_k + \frac16F_{ijk}{}^l\phi^i\phi^j\phi^k\phi^\dag_l + \mathrm{h.c.} \right) + \frac14E_{ij}{}^{kl}\phi^i\phi^j\phi^\dag_k\phi^\dag_l\equiv h \ ,\label{eq:vhardCW}
\end{gather}
the soft and hard-breaking terms. A complete computation of the potential with NSW solutions considering several models are presented in Section \ref{subsec:potential_calculation} and in Appendix \ref{app:GMsolutions}. Since those hard-breaking terms are parametrically suppressed, it is legitimate to ask whether those effects are negligible or not. A first rough calculation has already been done in the article \cite{moultaka_low_2018}. They showed that, assuming several energy scales in the superpotential (as \autoref{eq:scales}), the loop-effects coming from the hard terms can be sizable.  \medskip

A last comment can be added on the quantum stability of the superpotential \ref{eq:Wexp}. The new solutions are valid at tree-level for all (renormalisable or non-renormalisable) operator as long as the superpotential stay in the same form as \ref{eq:Wexp}, \ref{eq:W1} and \ref{eq:W0}. However, the effects of the radiative corrections on the superpotential's structure have not been analysed yet. 
\subsection{Coleman-Weinberg potential with Hard Breaking terms}\label{sec:CW}
In order to analyse the effects of those new hard-breaking terms, we calculate the one-loop correction of the potential. For this purpose, we use the Coleman-Weinberg potential \cite{CWpotential}\cite{DerendingerCW}. We remind the form of the Coleman-Weinberg potential (written in the $\overline{DR}$-scheme):
\begin{eqnarray}
\Delta V_{eff} = \frac1{64\pi^2}\mathrm{STr}\mathcal{M}^4\left[ \ln\left(\mathcal{M}^2/\Lambda^2\right)-\frac32 \right]+ \frac{\Lambda^2}{32\pi^2}\mathrm{STr}\mathcal{M}^2\label{eq:CWpot}
\end{eqnarray}
where $\Lambda$ is the cut-off. We have introduced $\mathrm{STr}\mathcal{M}$, the supertrace of the mass matrices:
\begin{eqnarray}
\mathrm{STr}\mathcal{M}^n &\equiv& \mathrm{Tr}\mathcal{M}_0^n-2\mathrm{Tr}\mathcal{M}_{1/2}^n+3\mathrm{Tr}\mathcal{M}_1^n \ .\label{eq:Str}
\end{eqnarray}
with $\mathcal{M}^2_0$, $\mathcal{M}_{1/2}$ and $\mathcal{M}_1$ the mass matrix of the scalar, fermion and gauge sector, respectively. We can identify in this formula a logarithmic and a quadratic correction. We will mainly be interested in the effects of the hard breaking terms on the last term (proportional to $\Lambda^2$).
We define a general parametrization of the superpotential $W_m$:
\begin{eqnarray}
W_m &=& \alpha_i\phi^i + \frac12 \beta_{ij}\phi^i\phi^j + \frac16\lambda_{ijk}\phi^i\phi^j\phi^k\ , \nn
\end{eqnarray}
whilst the soft breaking terms $V_{SOFT}$ and the hard breaking terms $V_{HARD}$ are taken as \ref{eq:vsoftCW} and \ref{eq:vhardCW}. The full form of the scalar potential is then :
\begin{eqnarray}
V &=& W_{m}{}_iW_m^i +\frac12 D^aD_a + \left(  m^2\right)^i{}_j\phi^i\phi^\dag_j + h + 	s\nn\\
&=& \left( \alpha_i + \beta_{il}\phi^l + \frac12 \lambda_{ilm}\phi^l\phi^m\right)\left( \alpha^i + \beta^{ik}\phi^\dag_k + \frac12\lambda^{ikn}\phi^\dag_k\phi^\dag_n \right)+\frac12 D^aD_a + \left(  m^2\right)^i{}_j\phi^j\phi^\dag_i \nn\\
&&+ h + s\nn\\
&=& \alpha^i\alpha_i + \beta_{il}\beta^{ik}\phi^l\phi^\dag_k + \frac14\lambda_{ilm}\lambda^{ikn}\phi^l\phi^m\phi^\dag_k\phi^\dag_n + \frac12 D^aD_a + \left(  m^2\right)^i{}_j\phi^j\phi^\dag_i + h + s  \nn\\
&& + \left\{ \alpha_i\beta^{ik}\phi^\dag_k + \frac12 \alpha_i\lambda^{ikn}\phi^\dag_k\phi^\dag_n + \frac12\beta_{il}\lambda^{ikn}\phi^l\phi^\dag_k\phi^\dag_n + \mathrm{h.c.} \right\}\nn
\end{eqnarray}
(with $D^a=g\phi^\dag T^a\phi$). \medskip

The mass matrices thus read:
\begin{itemize}
    \item for the scalar sector $\left(\mathcal{M}_0^2\right)$:
\begin{eqnarray}
\mathcal{M}_0^2 &=& \begin{pmatrix}
\frac{\partial^2 V}{\partial \phi^\dag_j \partial\phi^i} & \frac{\partial^2 V}{\partial \phi^\dag_i \partial\phi^\dag_j} \\
\frac{\partial^2 V}{\partial \phi^i \partial\phi^j} & \frac{\partial^2 V}{\partial \phi^\dag_i \partial\phi^j}
\end{pmatrix} \nn
\end{eqnarray}
with :
\begin{eqnarray}
\frac{\partial^2 V}{\partial\phi^i\partial\phi^\dag_j} &=& W_{ik}W^{kj} + D^{aj}{}_iD_a + D^{aj}D_{ai} + (m^2)^j{}_i+h^j{}_i\ , \nn \\
\frac{\partial^2 V}{\partial \phi^i\phi^j} &=& W_{ijk}W^k + D^a{}_iD_{aj} + s_{ij} + h_{ij}\ , \nn \\
\frac{\partial^2 V}{\partial \phi^\dag_i\phi^\dag_j} &=& W^{ijk}W_k + D^{ai}D_{a}{}^j + s^{ij} + h^{ij}\ , \nn
\end{eqnarray}
\end{itemize}
\begin{itemize}
    \item for the fermion sector ($\mathcal{M}_{1/2}^2$) (note that we include the gaugino masses $V_{gaugino} = \frac{1}{2}m_{ab}\lambda^a\lambda^b + \mathrm{h.c.}$):
\begin{eqnarray}
\mathcal{M}^2_{1/2} &=& \begin{pmatrix}
W_{ij}W^{jk} + 2D_b{}^kD^b{}_i & -\sqrt{2}iW_{ij}D_d{}^j + \sqrt{2} i m_{bd}D^b{}_i \\
\sqrt{2}iD^a{}_j W^{jk} - \sqrt{2}i D{}_b^km^{ab} & 2D^a{}_jD_d{}^j + m^{ab}m_{bd}
\end{pmatrix}\ .\nn
\end{eqnarray}
Note that the matrix has the same structure for models without hard-breaking terms.  
\end{itemize}
\begin{itemize}
    \item for the gauge sector ($\mathcal{M}_{1}^2$):
\begin{eqnarray}
\mathcal{M}^2_{1} &=& D^a{}_iD^b{}^i + D^b{}_iD^a{}^i\ . \nn
\end{eqnarray}
The contributions come only from the D-term.
\end{itemize}
The calculus of $\mathrm{STr}\mathcal{M}^2$ is then straightforward :
\begin{eqnarray}
\mathrm{STr}\mathcal{M}^2 &=& 2\Big( \mathrm{Tr}m_0^2 -\mathrm{Tr}m_{1/2}^2 + h^i{}_i\Big)\nn\\
&=& 2\Big( \mathrm{Tr}m_0^2 -\mathrm{Tr}m_{1/2}^2 + \left\{ D_{ik}{}^i\phi^k + \frac12 F_{ikl}{}^i\phi^k\phi^l + \mathrm{h.c.} \right\} + E_{ik}{}^{il}\phi^\dag_l\phi^k\Big)\label{eq:Str2}\ . \label{eq:strm2}
\end{eqnarray}
For the $\mathrm{STr}\mathcal{M}^4$ term we get : 
\begin{eqnarray}
\mathrm{Tr}\mathcal{M}_0^4 &=& 2\left( W^{ik}W_{kj} + D^{ai}{}_jD_a + D^{ai}D_{aj} + (m^2)^i{}_j + h^i{}_j \right)\nn\\
&&\times \left( W^{jl}W_{li} + D^{aj}{}_iD_a + D^{aj}D_{ai} + (m^2)^j{}_i + h^j{}_i \right)\nn\\
&& + 2 \left( W^{ijk}W_k + D^{ai}D_a{}^j+ \overline{s}^{ij} + h^{ij} \right)\left( W_{jin}W^n + D^{b}{}_jD_{bi}+ s_{ij} + h_{ij}\right)\ , \nn \\
3\mathrm{Tr}\mathcal{M}_1^4 &=& 6 D^a{}_iD^{bi}D_a{}^jD_{bj} + 6D^a{}_iD^{bi}D_{aj}D_b{}^j\ , \nn \\
-2\mathrm{Tr}\mathcal{M}_{1/2}^4 &=& -2W_{ij}W^{jk}W_{kl}W^{li}-2m^{ab}m_{bd}m^{dc}m_{ca}-16D_a{}^im^{ab}m_{bd}D^d{}_i\nn\\
&& -16D^a{}_jD_d{}^jD^d{}_iD_a{}^i - 16 D_b{}^kD^b{}_iW_{kl}W^{li} + 8W_{ij}D_d{}^jm^{de}D_e{}^i + 8W^{ij}D^b{}_im_{bd}D^d{}_j\ . \nn
\end{eqnarray}
For the following we suppose :
\begin{eqnarray}
(m^2)^i{}_j = m^2_i\delta^i{}_j ,\qquad m_{ab} = m_{a}\delta_{ab}\ . \nn
\end{eqnarray}
Since the superpotential $W_m$, the soft breaking terms $s$ and the hard breaking terms $h$ are gauge invariant, we have:
\begin{eqnarray}
W_iD_a{}^i = 0,\quad s_iD_a{}^i = 0,\quad h_iD_a{}^i - h^iD_{ai}= 0 \ . \nn
\end{eqnarray}
Introducing the quadratic Casimir operator $C$ and the Dynkin label $T$: 
\begin{eqnarray}
T(R)\delta^{ab} &=& \mathrm{Tr}\left(  T^aT^b \right)\ ,  \nn\\
C(R) &=& C(R)\delta^i{}_j = T^a{}^i{}_kT_a{}^k{}_j\ , \nn\\
C(G)\delta^a{}_c &=& f_{bd}{}^af^{bd}{}_c\ ,  \nn
\end{eqnarray}
we finally obtain :
\begin{eqnarray}
\mathrm{STr}\mathcal{M}^4 &=&  2g^2\Big( T(R) + 2C(R) - 3C(G)  \Big)D^aD_a +2\left(\mathrm{Tr}m_0^4 - \mathrm{Tr}m_{1/2}^4\right) \nn\\
&& + 4 \Big((m_0^2)\delta^j{}_i + h^j{}_i\Big)W^{ik}W_{jk} + 2\left\{ W^{ijk}W_k\left( s_{ij} + h_{ij} \right) + \mathrm{h.c.} \right\}\nn\\
&& -2g^2C(R_i)\left( s_i\phi^i + h_i\phi^i + \mathrm{h.c.} \right) + 2W^{ijk}W_{jin}W_kW^n\nn\\
&& + 4g^2C(R_i) (m^2)^j{}_i\phi^i\phi^\dag_j + 4W^{ik}W_{kj}D^{bj}{}_iD_b -16g^2\sum_a m_a^2C(R_i)\phi^i\phi^\dag_i \nn\\
&& - 8g^2C(R_i)\sum_a m_a\left( W_i\phi^i + \mathrm{h.c.} \right) - 2g^2C(R_i)\left( \phi^iW^jW_{ij} + \mathrm{h.c.} \right) \nn\\
&& + 2\left( \overline{s}^{ij} + h^{ij} \right)\left(  {s}_{ij} + h_{ij} \right) -8g^2C(R_i)W^iW_i \nn \\
&& + 4m_0^2h^i{}_i + 2h^i{}_jh^j{}_i + 4D_{ai}h^i{}_jD^{aj} + 4h^j{}_i\Big( D^{ai}{}_jD_a + D^{ai}D_{aj}\Big)\ .\label{eq:STr4}
\end{eqnarray}
Considering a pure supersymmetric theory (without SUSY breaking terms), the quadratic correction \ref{eq:strm2} vanishes
. We also see that we obtain in the framework of classical Soni-Weldon solutions (with soft-breaking terms $s$):
\begin{gather}
\mathrm{STr}\mathcal{M}^2 = 2\big( \mathrm{Tr}m_0^2 - \mathrm{Tr}m_{1/2}^2 \big)\ .\nn
\end{gather} 
Quadratic corrections are then present in the theory. However, those corrections are field-independent and do not contribute to the mass matrices. It is no longer the case with hard-breaking terms. The presence of $V_{HARD}$ in the NSW solutions generates new contributions in \ref{eq:Str2} which are field-dependent, leading to new corrections in the mass matrices. Considering now the logarithmic corrections through the $\mathrm{STr}\mathcal{M}^4$-contribution (see \autoref{eq:STr4}), we see that the $V_{HARD}$ contributions do not drastically change the structure of the corrections. \medskip

Within the scope of Soni-Weldon solutions (\textit{i.e.} with only soft-breaking terms), the computation of the Coleman-Weinberg potential \ref{eq:CWpot} allows to calculate the one-loop renormalisation group equations (or RGEs). The RGEs are fundamental to realise a phenomenological analysis at low energy. This calculation can be performed by applying the non-renormalisation theorem, \textit{i.e.} that the superpotential is not renormalised (only the fields are renormalised). A full calculation of the RGEs following this approach can be found in \cite{DerendingerCW}. Assuming now hard-breaking terms in the framework of NSW solutions, there is no reason that the non-renormalisation theorem holds. A diagrammatic approach could then be mandatory to compute the complete set of RGEs in NSW models. 
\section{NSW solutions with two hybrid fields: the S2MSSM}\label{sec:S2MSSM}
By analysing the new canonical solutions pointed out by the article \cite{moultaka_low_2018}, we have obtained a non-canonical solution. Assuming at least two hybrid fields $S^p$ in $W_0$ to generate interactions between the matter sector $\{\tilde{\Phi}^a\}$ and the $S^p$ fields, we construct a model called the S2MSSM\footnote{To be compared with the N2MSSM defined in \autoref{chap:N2MSSM} in the case of Soni-Weldon solutions.}. The scalar potential is calculated as well as the minimum of the potential, the minimisation equations of the potential and the mass matrices.
\subsection{Presentation of the S2MSSM}\label{subsec:S2MSSMPres}
The S2MSSM is defined as a model based on the Non-Soni \& Weldon solutions. We assume a hidden sector $\{Z\}$ with $Z=(\zeta, \chi_{\zeta}, F_{\zeta})$ and $\zeta = m_p z$, $n$ hybrid fields $\{S^p\}$ (with $p=1,\dots , n$) and a matter sector $\{\tilde{\Phi}^a\}$ corresponding to the content of the MSSM (introduced in \autoref{tab:PSM}). The superpotential and the Kähler potential is defined as:
\beqa
W(z, S, \widetilde{\Phi}) &=& 
m_{p\ell}\left(W_{1,0}(z) + 
 S^p W_{1, p} (z)\right)
+ \   W_0(z, \widetilde{\Phi},\mathcal{U}^{12}) +  S^p W_{0, p} (z) \ ,  \label{eq:Non-SW}\ \\
K &=& m_p^2z^iz^\dag_i + S_p^\dag S^p + \Phi_{a*}^\dag\Lambda^{a*}{}_a(z,z^{\dagger})\Phi^a \nn\\
&=& m_p^2\hat{K}(z,z^\dag) + \tilde{K}(z,z^\dag,S,S^\dag,\Phi,\Phi^\dag),  \label{eq:kahler} 
\eeqa
where:
\begin{gather}
\mathcal{U}^{12} = \mu_2 S^1 - \mu^1S^2 (\equiv \mathcal{U})\nn
\end{gather}
and with:
\begin{eqnarray}
W_0(z,\tilde{\Phi},S) &=& \lambda(z) \mathcal{U}\hat{H}_U\cdot\hat{H}_D + \kappa(z)\mathcal{U}^3 + y_u(z) \hat{Q}\cdot\hat{H}_U\hat{U} -y_d(z) \hat{Q}\cdot\hat{H}_D\hat{D} \nn\\
&& -y_e(z) \hat{L}\cdot\hat{H}_D\hat{E} \label{eq:Ws2mssm}
\end{eqnarray}
(with $\{\tilde{\Phi}^a\}=\{\hat{H}_U,\hat{H}_D,\hat{Q},\hat{U},\hat{D},\hat{L},\hat{E}\}$, \textit{i.e.}, the field content of the MSSM). Note that the two hybrid fields $S^1$ and $S^2$ couple with the matter sector $\tilde{\Phi}^a$ through the combination $\mathcal{U}^{12}$ (denoted as $\mathcal{U}$). The Kähler potential is supposed to be non-canonical in the matter sector $\{\Phi^a\}$ in order to generate non-universal soft-supersymmetric breaking mass terms. For simplicity, we also assume only one field $z$ from the hidden sector.\medskip

We introduce some notations that will be useful for the following. Firstly, we denote the various energy scales: 
\beqa
\begin{aligned}
 {W}_{1,0}  &=  M_1^2 \omega_1(z) , \\
 {W}_{1, p}  &=  M_{2, p} \omega_{1p}(z) , (\text{no sum on $p$}), \\
 {W}_{0, p}  &=  M_{3, p} ^2 \omega_{0p}(z) , (\text{no sum on $p$}), \\
 {W}_0  &=  M_4^3 \omega_0(z,{\cal U} , \phi) , \\
 \zeta &= m_{p} z , \\
 {\tPhi} &=  M_4 \phi\ .
\end{aligned}
\label{eq:scales}
\eeqa
with $M_1, M_{2, p}, M_{3, p}, M_4 \ll m_{p}$. Following those definitions, we have $\langle z \rangle \sim {\cal O}(1)$ and $\langle \phi \rangle \sim {\cal O}(1)$. Note that in the case where several energy scales are set in the matter sector, \textit{i.e.} $\Phi_{GUT}=M_{GUT}\phi_{GUT}$ and $\Phi_{EW}=M_{EW}\phi_{EW}$ (with $M_{GUT} \gg M_{EW}$), rescaling by $M_{GUT}$ the fields $\Phi_{GUT}$ and $\Phi_{EW}$ leads to $\lag\phi_{GUT}\rag \sim\mathcal{O}(1)$ and $\lag\phi_{EW}\rag \ll 1$. We will show later that several energy scales in the matter sector are mandatory to increase the loop effects. We set the functions $I_p(z)$ (and ${\cal I}_p$) as:
\begin{gather}
I_p(z) = m_pM_{2,p}\omega_{1,p}(z) + M_{3,p}^2\omega_{0,p}(z)\ , \quad {\cal I}_p = I_p (\lag z\rag) \ . \nn
\end{gather}
Assuming non vanishing \textit{v.e.vs} for the matter and hybrid fields ($\lag \phi^a\rag\neq 0$ and $\lag S^p\rag\neq 0$), we write the gravitino mass as follow
\begin{gather}
m_{3/2} = \frac{1}{m_p^2}e^{K/(2m_p^2)}\lag W\rag = e^{|\lag z \rag |^2/2} \frac{1}{m_p^2}\big(M_4^3\lag w \rag + {\cal I}_p \lag S^p \rag \big) \nn
\end{gather}
with $w = \omega_0 + \frac{m_pM_1^2}{M_4^3}\omega_1$. We also define the variables $m'_{3/2}$ and $m''_{3/2}$ as:  
\begin{gather}
m'_{3/2} = e^{|\lag z \rag |^2/2} \frac{M_4^3}{m_p^2}\big(\lag \mathfrak{d}_zw \rag + \frac{1}{M_4^3}\mathfrak{d}_z{\cal I}_p \lag S^p \rag \big)\ , \quad m''_{3/2} = e^{|\lag z \rag |^2/2} \frac{M_4^3}{m_p^2}\big(\lag \mathfrak{d}^2_zw \rag + \frac{1}{M_4^3}\mathfrak{d}_z^2{\cal I}_p \lag S^p \rag \big) \nn
\end{gather}
where $\mathfrak{d}_zw = \partial_z w + z^{\dagger}w$ is the covariant derivative. We denote: 
\begin{gather}
\xi_{3/2} = \frac{m_{3/2}}{m'_{3/2}}\ . \nn
\end{gather}
\subsection{Potential calculation}\label{subsec:potential_calculation}
We now compute the low-energy scalar potential using equation \ref{eq:SUGRApot}. We start with the calculation of the covariant derivatives $\mathcal{D}_zW$, $\mathcal{D}_aW$ and $\mathcal{D}_pW$ with:
\begin{gather}
D_X W = \partial_XW + \frac{1}{m_p^2}K_XW\ .\nn
\end{gather}
Supergravity is broken in the hidden sector when the field $z$ acquire a non-zero \textit{v.e.v} $\lag z\rag\neq 0$. We provide the following substitution for the calculation of the scalar potential:
\begin{gather}
z \rightarrow \lag z \rag\ , \nn \\
S^p \rightarrow S^p + \lag S^p \rag\ , \nn\\
\phi^a \rightarrow \phi^a + \lag\phi^a \rag\ .\nn
\end{gather}
We thus introduce the notations:
\begin{gather}
X\equiv X(\lag z \rag, \phi + \lag \phi \rag, S + \lag S \rag)\ , \quad \lag X \rag \equiv X(\lag z \rag, \lag \phi\rag, \lag S\rag )\ , \nn \\
\Delta X \equiv X - \lag X\rag \ . \nn 
\end{gather} 
The covariant derivatives read then:
\beqa
\frac{1}{m_p}\mathcal{D}_zW &\rightarrow & e^{-\frac{\hat{K}}{2}}m_pm'_{3/2}  + \frac1{m_p}\left(\delta_z{\cal I}_pS^p + M_4^3\delta_z\Delta \omega_0 \right) \nn\\
&& +\frac1{m_p}{\tilde K}_z\left( e^{-\frac{\hat{K}}{2}}m_{3/2} + \frac{1}{m_p^2}\left({\cal I}_pS^p + M_4^3\Delta \omega_0\right)   \right)\ ,   \nn\\
\mathcal{D}_pW &\rightarrow & {\cal I}_p + M_4^3\partial_p\omega_0 +\left( S_p^\dag + \left< S_p^\dag \right> \right)\left(  e^{-\frac{\hat{K}}{2}}m_{3/2} + \frac{1}{m_p^2}\left({\cal I}_pS^p + M_4^3\Delta \omega_0\right)  \right)\ , \nn\\
\frac{1}{M_4}\mathcal{D}_aW &\rightarrow & M_4^2\partial_a \omega_0 + M_4^2\left(\Phi^\dag_{a*}+\left<\Phi^\dag_{a*}\right>\right)\Lambda^{a*}{}_a\left( e^{-\frac{\hat{K}}{2}}m_{3/2} + \frac{1}{m_p^2}\left({\cal I}_pS^p + M_4^3\Delta \omega_0\right)  \right)\ .\nn 
\eeqa
To calculate the scalar potential, we need the inverse of the Kähler metric. Using the method presented in \autoref{subsec:soft}, the perturbative inversion of the metric gives then:
\beqa
\label{eq:K-1}
(K^{-1})^a{}_{a^\ast}&=& (\Lambda^{-1})^a{}_{a^\ast} + \frac{M_4^2}{m_p^2}  (\Lambda^{-1})^a{}_{b^\ast} \tilde \Lambda^{b^\ast}{}_b   (\Lambda^{-1})^b{}_{a^\ast} +o(1/m_p^2)\ , \nn\\
(K^{-1})^p{}_{q}&=& \delta^p{}_q\ ,   \\
(K^{-1})^z{}_z &=& (\hat{K}^{-1})^z{}_z -\frac{M_4^2}{m_p^2} \phi^\dag_{a^\ast} (\hat{K}^{-1})^z{}_z\partial^z\partial_z \Lambda^{a^\ast}{}_a \phi^a\hat{K}^z{}_z\nn\\
&& + \frac{M_4^2}{m_p^2} \phi^\dag_{a^\ast}\hat{K}^z{}_z \partial^z \Lambda^{a^\ast}{}_a (\Lambda^{-1})^a{}_{b^\ast} \partial_z \Lambda^{b^\ast}{}_b \phi^b\hat{K}^z{}_z +o(1/m_p^2)\ , \nn \\
(K^{-1})^z{}_{a^\ast}&=& - \frac 1{m_p} \hat{K}^z{}_z\phi^\dag_{b^\ast} \partial^z \Lambda^{b^\ast}{}_c (\Lambda^{-1})^c{}_{a^\ast} +o(1/m_p)\ , \nn\\
(K^{-1})^a{}_{z}&=& - \frac 1{m_p} (\Lambda^{-1})^a{}_{b^\ast} \partial_z \Lambda^{b^\ast}{}_c \Phi^c\hat{K}^z{}_z +o(1/m_p)\ , \nn\\
(K^{-1})^p{}_{z}&=&(K^{-1})^p{}_{a^\ast}= (K^{-1})^z{}_{p}=(K^{-1})^{a^\ast}{}_{p} =0 \ , \nn
\eeqa
with :
\beqa
\tilde \Lambda^{m^\ast}{}_m \equiv \phi^\dag_{n^\ast} \partial^z \Lambda^{n^\ast}{}_m\hat{K}^z{}_z\partial_z \Lambda^{m^\ast}{}_n \phi^n \ . \nn
\eeqa
We observe a mixing between the hidden sector $\{z^i\}$ and the matter sector $\{\phi^a\}$ coming from the non-canonical of the Kähler potential in the $\phi^a$-sector.  
We can now compute the scalar potential (with a perturbative expansion of the exponential):
\beqa
V&=&e^{\hat{K}}\left( 1+\frac{\tilde{K}}{m_p^2}\right)\Bigg[ \frac{1}{m_p^2}\mathcal{D}_iW (K^{-1})^{i}{}_{j} \mathcal{D}^j\overline{W}+\mathcal{D}_pW\mathcal{D}^p\overline{W}+\mathcal{D}_aW( K^{-1})^a{}_{a*}\mathcal{D}^{a^\ast}\overline{W}\nn\\
&& +\frac{1}{m_p}\Big(\mathcal{D}_iW(K^{-1})^{i}{}_{a^\ast}\mathcal{D}^{a^\ast}\overline{W}+\mathrm{h.c.}\Big)-\frac{3}{m_p^2}\left| W\right|^2\Bigg]\ . \nn
\eeqa
After a lengthy computation, the scalar potential is:
\beqa 
V&=& m_p^2\lb m_{3/2}\lb^2 \big(\frac 1{\lb \xi_{3/2}\lb^2} -3 \big)+ \eK\Big(\sum \limits_p
\lb {\cal I}_p +M_4 ^3\partial_p \omega_0 \lb^2 + M_4^4 \partial_a \omega_0 \partial^{a^\ast}\bar \omega_0 \lag (\Lambda^{-1})^a{}_{a^\ast}\rag\Big)\nn\\
&& +M_4^2\Big(\pp{a} \pb{a^\ast}  \Big) \Big(\lb m_{3/2}\lb^2 {\cal S}^{a^\ast}{}_a +\frac {1}{m_p^2}\eKs
 \big[ \bar m_{3/2} S^p \; ({\cal S}_p)^{a^\ast}{}_a + {\mbox h.c.}\big]\nn\\
&&  \hskip 1.8truecm +\frac 1 {m_p^4} \eK
   S^p S^\dag_q \; ({\cal S}^q{}_p)^{a^\ast}{}_a \Big)\nn\\
 && + \Big(\bS{p} \bSb{p}\Big)\Big( \lb m_{3/2}\lb^2 T+ 
 \frac 1{m_p^2} \eKs \big[\bar m_{3/2}S^r T_r + \mbox{h.c.} \big]\nn\\
&& \hskip 1.8truecm +\frac 1{m_p^4} \eK S^r S^\dag_t T^t{}_s\Big)  +\frac 1{m_p^2} \eK S^p S^\dag_q \Big( \mathfrak{d}_z {\cal I}_p 
\mathfrak{d}^z  \bar{\cal I}^q -3 {\cal I}_p \bar{\cal I}^q\Big)\nn\\ 
&& +\frac 1{m_p^2} \eKs \Bigg\{ \Big(M_4^2 \pb{a^\ast} \pp{a} \lag \Lambda^{a^\ast}{}_a \rag +\bSb{p} \bS{p}\Big)\times\nn\\
&& \hskip 2.3truecm   \bS{q} {\cal I}_q \Big(\bar m_{3/2}+ \frac 1 {m_p^2} \eKs S^\dag_r \bar {\cal I}^r\Big) 
+ \mbox{h.c.} \Bigg\} 
\label{eq:potentialS2MSSM}\\
 &&+ \frac1 {m_p^2}\eK \Big(M_4^2
 \pp{a} \pb{a^\ast}  \lag \Lambda^{a^\ast}{}_a \rag +\bS{p} \bSb{p} \Big)\times\nn\\
&& \hskip 1.8truecm  \Big(\sum \limits_ r \lb {\cal I}_r\lb^2+ M_4^3{\cal I}^r \partial_r \omega_0+M_4^3 \bar{{\cal I}}_r \partial^r \bar \omega^0\Big)\nn\\
 &&+ \eKs\Bigg\{\bar m_{3/2}  M_ 4^3 R^b{}_a \pp{a} \partial_b \omega_0 +
 \frac{M_4^3} {m_p^2} \eKs (R^p)^b{}_a\pp{a} S^\dag_p\partial_b \omega_0 \nn
 \\
 && \hskip 1.8truecm + \big[\bar m_{3/2} + \frac 1{m_p^2} \eKs S^\dag_q \bar {\cal I}^q\big] \bS{p}
\big[{\cal I}_p +M_4^3 \partial_p \omega_0\big]\nn\\
&& \hskip 1.8truecm  +
\frac{M_4^3}{m_p^2}\eKs \bSb{p} \bar {\cal I}^p \Delta \omega_0 + \mbox{h.c.}
 \Bigg\}+ \eKs \bar m_{3/2}S^p \big(\frac1 {\bar \xi_{3/2}} \mathfrak{d}_z {\cal I}_p -3 \ {\cal I}_p + \mbox{h.c.}\big)\nn\\
&&+ \eKs M_4 ^3\Big(\Delta \mathfrak{d}_z \omega_0\big[
\frac {\bar m_{3/2}}{\bar\xi_{3/2}}  + \frac1 {m_p^2} \eKs S^\dag_q \mathfrak{d}^z \bar {\cal I}^q\big]
-3 \Delta  \omega_0\big[\bar m_{3/2} + \frac1 {m_p^2} \eKs S^\dag_q \bar {\cal I}^q\big]+ \mbox{h.c.} \Big)\nn
\eeqa
where we recall that $m'_{3/2} = m_{3/2}/\xi_{3/2}$ and we define
\beqa
{\cal S}^{a^\ast}{}_a&=&\frac 1 {\lb \xi_{3/2}\lb^2}\Big( \lag \partial^z \Lambda^{a^\ast}{}_b (\Lambda^{-1})^{b}{}_{b^\ast} \partial_z \Lambda^{b^\ast}{}_a -
\partial^z \partial_z \Lambda^{a^\ast}{}_a\rag\Big) + \lag \Lambda^{a^\ast}{}_a\rag\big(\frac 1{\lb \xi_{3/2}\lb^2} -2\big)\nn\\
&=& -2 \Lambda^{a^*}{}_a + \frac 1 {\big|\xi_{3/2}\big|^2}\ \tilde {\cal S}^{a^*}{}_a\ , \nn\\
({\cal S}_p)^{a^\ast}{}_a&=&\frac 1 {\bar \xi_{3/2}}\Big( \lag \partial^z \Lambda^{a^\ast}{}_b (\Lambda^{-1})^{b}{}_{b^\ast} \partial_z \Lambda^{b^\ast}{}_a -
\partial^z \partial_z \Lambda^{a^\ast}{}_a\rag\Big)
\mathfrak{d}_z {\cal I}_p +
\lag \Lambda^{a^\ast}{}_a\rag\Big(\frac 1 {\bar \xi_{3/2}}
 \mathfrak{d}_z {\cal I}_p  -2 {\cal I}_p \Big)\ ,\nn\\
( {\cal S}^q{}_p)^{a^\ast}{}_a&=&\Big( \lag \partial^z \Lambda^{a^\ast}{}_b (\Lambda^{-1})^{b}{}_{b^\ast} \partial_z \Lambda^{b^\ast}{}_a-
\partial^z \partial_z \Lambda^{a^\ast}{}_a\rag\Big)
\mathfrak{d}_z{\cal I}_p\mathfrak{d}^z \bar {\cal I}^q+
\lag \Lambda^{a^\ast}{}_a\rag (\mathfrak{d}_z{\cal I}_p\mathfrak{d}^z \bar {\cal I}^q-2 {\cal I}_p\bar {\cal I}^q\big)\ ,\nn\\
T&=&\frac 1 {\lb \xi_{3/2}\lb^2} -2\ ,\nn\\
T_p&=& \frac 1{\bar \xi_{3/2}} \mathfrak{d}_z {\cal I}_p -2   {\cal I}_p\ ,\nn\\
T^p{}_q&=& \mathfrak{d}_z {\cal I}_q  \mathfrak{d}^z \bar {\cal I}^q-
2 {\cal I}_q  \bar {\cal I}^q \ ,\nn\\
R^a{}_b &=& \delta^a_b - \frac 1{\bar \xi_{3/2}} \lag (\Lambda^{-1})^a{}_{b^\ast} \partial_z \Lambda^{b^\ast}{}_b\rag\ ,\nn\\
(R^p)^a{}_b&=& \bar {\cal I}^p \delta^a{}_b -  \mathfrak{d}^z \bar{\cal I}^p \lag (\Lambda^{-1})^a{}_{b^\ast} \partial^z \Lambda^{b^\ast}{}_b\rag\ .\nn
\eeqa

At this stage, we can identify several types of contributions:
\begin{itemize}
\item a non-vanishing cosmological constant $\Lambda=\lag V \rag$:
\beqa
\big\lag V \big \rag
&=& m_p^2\big(\lb m'_{3/2}\lb^2 -3 \lb m_{3/2}\lb^2 \big)+ \eK\Big(\sum \limits_p
\lb {\cal I}_p +M_4 ^3\lag \partial_p \omega_0\rag \lb^2 + M_4^4 \lag \partial_a \omega_0 \partial^{a^\ast}\bar \omega_0 (\Lambda^{-1})^a{}_{a^\ast}\rag\Big)\nn\\
&&+\lb m_{3/2}\lb^2  M_4^2\vp{a} \vpb{a^\ast}   {\cal S}^{a^\ast}{}_a 
  + \lb m_{3/2}\lb^2 \vS{p} \vSb{p}T
 \nn\\
&& +\frac 1{m_p^2} \eKs \Big( M_4^2\vpb{a^\ast} \vp{a} \lag \Lambda^{a^\ast}{}_a \rag +\vSb{p} \vS{p}\Big) \Bigg\{
\Big(\bar m_{3/2} \vS{q} {\cal I}_q 
+ \mbox{h.c.} \Big) \label{eq:minV}\\
&& +\eKs \Big(\sum \limits_ r \lb {\cal I}_r\lb^2 + M_4^3{\cal I}^r \lag\partial_r \omega_0\rag +M_4^3 \bar{{\cal I}}_r \lag\partial^r \bar \omega^0\rag \Big) \Bigg\} \nn
\\
 &&+ \eKs\Big(\bar m_{3/2}  M_ 4^3 R^b{}_a \vp{a} \lag\partial_b \omega_0\rag + \bar m_{3/2}  \vS{p}
\big[{\cal I}_p +M_4^3 \lag\partial_p \omega_0\rag\big] 
+ \mbox{h.c.}
 \Big) \ .
\nn
\eeqa 
We assume:
\begin{gather}
\lag V \rag = 0 \nn
\end{gather}
which may lead to a fine-tuning of the parameters and is usual in supergravity. 
\item the classical SUSY potential (see \autoref{eq:potSUSY}):
\begin{gather}
\eK\Big(\sum \limits_p
\lb {\cal I}_p +M_4 ^3\partial_p \omega_0 \lb^2 + M_4^4 \partial_a \omega_0 \partial^{a^\ast}\bar \omega_0 \lag (\Lambda^{-1})^a{}_{a^\ast}\rag\Big)\ . \nn
\end{gather}
\item soft-breaking terms associated with the $\{\phi^a\}$ and $\{S^p\}$ sectors. For example, from the second and fourth lines of \ref{eq:potentialS2MSSM}, we can identify contributions to the soft-breaking mass terms
\begin{gather}
|m_{3/2}|^2\Big( M_4^2\phi^{\dagger}_{a^\ast}\mathcal{S}^{a^\ast}{}_a\phi^a + |S^p|^2T \Big)\ . \nn
\end{gather}
\item the new hard-breaking terms. For instance, we found in the third and fifth lines the couplings \ref{eq:vhardCW}
\begin{gather}
\frac{1}{m_p^4}\eK S^rS^{\dagger}_{t}\Big( M_4^2\phi^a\phi^{\dagger}_{a^\ast}(\mathcal{S}^t{}_{r})^{a^\ast}{}_a  + |S^p|^2T^{t}{}_{r}\Big) \label{eq:hardphi2S2}
\end{gather} 
which contribute to the last term of \ref{eq:vhardCW}.
\end{itemize} 

Based on the scalar potential \autoref{eq:potentialS2MSSM}, the minimisation equations of the potential can be obtained as well as the element of the mass matrix. They can be found in Appendix \autoref{app:S2MSSM}.
\subsection{Differences between S2MSSM and N2MSSM}\label{subsec:diffS2N2}
The structure of the Non-Soni-Weldon solutions imposes the presence of at least two hybrid fields $S^p$ ($p=1,2$) to generate couplings between the $\mathcal{U}$ field and the matter sector $\{\tilde{\Phi}^a\}$. Since those new fields must be singlet under the gauge group of the matter sector, the field content is equivalent to a two-singlets extension of the MSSM, the N2MSSM (see Section \ref{sec:N2MSSMDes}). The N2MSSM is, however, defined in the context of Soni-Weldon solutions, resulting so to only soft-breaking terms in the potential.\medskip

Nonetheless, the presence/absence of hard SUSY breaking-terms is not the only difference between those two models. For instance, 
 the two couplings $\lambda_1S^1H_U\cdot H_D$ and $\lambda_2S^2H_U\cdot H_D$ in the N2MSSM are completely independent whilst the equivalent couplings in the S2MSSM are correlated from the structure of the superpotential and the form of the $\mathcal{U}$-field.\\

\begin{figure}[H]
    \centering
      \includegraphics[width=0.5\linewidth]{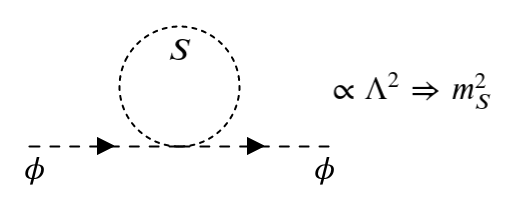}
   \caption{One-loop contribution on the mass of the field $\phi$.}
   \label{eq:phi_loop}
\end{figure}

Of course, the hard-breaking terms will contribute to all the phenomenological analysis of the NSW solutions. The electroweak symmetry breaking conditions will be modified by the presence of such new breaking terms. In addition, new contributions on the mass matrices could also affect the mass spectrum of the model. Indeed, the presence of $\phi\phi^{\dagger}SS^{\dagger}$ term in the potential will contribute at tree-level and loop-level through diagrams in \autoref{eq:phi_loop} (proportional to the singlet mass after renormalisation). The effects of the new S-sector on the Higgs sector is investigated in \autoref{sec:HS_effects}. Moreover, those couplings could also push the sfermion sector to a higher energy scale, leading potentially to Split-SUSY-like models \cite{Split1}\cite{Split2}. We recall that Split-SUSY models are based on the paradigm that the naturalness problem is ignored, assuming the anthropic principle. The supersymmetry breaking scale is assumed to be very high (for example, at the GUT scale $M_{SUSY}\approx M_{GUT} \approx 10^{15}\ \mathrm{GeV}$). As a consequence, the masses of the fermionic superpartners of the standard model are in the order of $M_{GUT}$, which leads to interesting phenomenological properties (such as long-lived gluino) and to highly fine-tuned models. The differences between Split-SUSY models and NSW solutions is essentially due to the presence of hard-breaking terms. Nonetheless, a heavy sfermionic sector coupled with a heavy Higgs state can be a consequence of a really heavy singlet state pushing all the mass spectrum to higher energy (from	 the loop effect \autoref{eq:phi_loop}). Following a see-saw effect, a light singlet state can then appear in the spectrum. Finally, as mentioned in \autoref{sec:CW}, the NSW solutions provide new effects which will modify the structure of the RGEs.

\section{$W_0$ not containing $S^p$ fields: constraints and mass matrix}\label{sec:HS_effects}
The potential \autoref{eq:potentialS2MSSM} leads to several new contributions coming from loop-effects involving $S$ fields. While the terms in the form $(Q_a)^{q}{}_p\phi^a\phi^{\dagger}_aS^pS^{\dagger}_q$ and \sloppy ${Q'}^{r}{}_tS^pS^{\dagger}_pS^tS^{\dagger}_r$ act on the diagonal elements of the mass matrix, the terms coming from the superpotential such as $D^q\mathcal{U}H_U\cdot H_D S^{\dagger}_q$ contribute to the off-diagonal elements and potentially generate push-up effects. Those mechanisms could end up with a reduction of the fine-tuning on the Higgs boson mass. \medskip

In this section, we present a numerical analysis of the effects of the $S$ fields on the $\{\phi^a\}$ sector ($a=1,\dots,m$). Considering several constraints such as the vanishing of the cosmological constant and the minimisation conditions of the potential, we investigate the order of magnitude of the $S$-loop in a simplified model:
\begin{itemize}
\item the \textit{v.e.vs} of the hybrid and matter sector are assumed to vanish $\lag S\rag = \lag \phi \rag = 0$;
\item we also assume a canonical Kähler potential, $\Lambda^{a^\ast}{}_a(z,z^{\dagger}) = \delta^{a^\ast}{}_a$.
\end{itemize}
Following those approximations, the $S$-contribution to the gravitino mass vanish:
\begin{gather}
m_{3/2} = \frac{1}{m_p^2}e^{K/(2m_p^2)}\lag W\rag = e^{|\lag z \rag |^2/2} \frac{M_4^3}{m_p^2}\lag w \rag\nn
\end{gather}
(see \autoref{subsec:S2MSSMPres}). A gravitino mass $m_{3/2}$ lower than $M_1$ and $M_4$ can then be obtained by assuming:
\begin{gather}
m_pM_1^2 \approx M_4^3 \ll m_p^3 \quad \text{and} \quad |\lag w\rag | \ll 1 \label{eq:OoM_w}
\end{gather}
with $|\lag\omega_0\rag|\sim |\lag\omega_1\rag|\sim \mathcal{O}(1)$, and $\omega_0$ and $\omega_1$ have opposite sign phases. This parametrisation may include a little fine-tuning that will not be considered for this first study.
\subsection{Corrections on the Higgs sector}\label{sec:corrHiggsSector}
The mass matrix of the Higgs/$S$ sector is:
\begin{gather}
\mathcal{M}_{H/S}^2 = \left(\begin{array}{c|c}
\mathcal{S} & (\mathcal{SH})^t \\
\hline 
\mathcal{SH} & \mathcal{H} 
\end{array}\right) \label{eq:matH}
\end{gather}
with $\mathcal{S}$ the submatrix in the hybrid sector, $\mathcal{H}$ the submatrix in the Higgs sector and $\mathcal{SH}$ the mixing elements. This matrix is quite similar to a Soni-Weldon solution of \textit{Gravity-Mediated Supersymmetry Breaking} with singlets fields (such as NMSSM or N2MSSM). The presence of the hybrid sector can then already modify the Higgs spectrum at tree-level in the same manner as singlets from the N2MSSM do. In a desirable case, they can contribute to the Higgs boson mass of the Standard Model and possibly reduce a potential fine-tuning on the parameters. The two main methods (described in \ref{sub:difficultNMSSM} for the NMSSM) to push the Higgs boson mass at tree-level are
\begin{itemize}
\item contributing to the diagonal elements;
\item generating a push-up effect \textit{via} a mixing term in the $\mathcal{SH}$ submatrix.
\end{itemize}

In the context of the S2MSSM, the hard-breaking terms contribute at tree-level but at one-loop level too. We can identify in the full potential \autoref{eq:potentialS2MSSM} new supersymmetric breaking terms which contribute at one-loop level to each submatrix in \autoref{eq:matH}. The hard-breaking terms in \ref{eq:hardphi2S2} contribute to the diagonal elements of $\mathcal{S}$ and $\mathcal{H}$ as (using the assumptions mentioned in this section):
\begin{gather}
\frac{1}{m_p^4}e^{|\lag z\rag|^2}\big( M_4^2\phi^a\phi^{\dagger}_a + S^pS^{\dagger}_p \big)\big( 4|\xi_{3/2}|^2 - 2 \big){\cal I}_r\bar{\cal I}^t S^r S^{\dagger}_t \label{eq:hardcorrection1}
\end{gather}
where the two couplings involving $\Delta \omega_0$ (in the two last lines of \ref{eq:potentialS2MSSM}) generate correction in the mixing part $\mathcal{SH}$ through:
\begin{gather}
-\frac{2}{m_p^2}e^{|\lag z\rag|^2}\bar{\cal I}^qS^{\dagger}_q\mathcal{U}H_U\cdot H_D \label{eq:hardcorrection2}
\end{gather}
by closing the $S$-loop. The negative sign can be problematic since it can lower the tree-level contribution of the Higgs boson mass. However, the sign can be reabsorbed in the definition of the $\mathcal{U}$-field and so resolve this issue. We will study the order of magnitude of those corrections. 
\subsection{Cosmological constant and minimisation of the potential}   \label{subsec:s2matrix}
In this section, we present the calculation of the minimum of the potential, the minimisation equations and the mass matrix in the general case of $n$ hybrid singlet fields $\{S^p\}$ ($p=1,\dots, n$). We assume no direct couplings between the hybrid and the matter sector for a first approach, \textit{i.e.} $\partial_p\omega_0=0$. Note that this simplification naturally emerges in the case of one singlet $S$. Indeed, since the coupling between the matter and the hybrid sector is obtain \textit{via} the linear combination of two $S$-fields, there is no direct coupling between $\phi$ and $S$ in the superpotential $W_0$. In this configuration, the minimum of the potential of \autoref{eq:minV} can be written as:
\begin{eqnarray}
\label{eq:V1}
{\left \langle V \right \rangle}&=&
e^{\big|\langle z \rangle\big|^2}\Big( \sum\limits_p|{\cal I}_p|^2 +
M_4^4\sum \limits_a \Big|\langle \partial_a \omega_0\rangle\Big|^2\Big)+ m_p^2
( |m'_{3/2}|^2 -3|m_{3/2}|^2)
\end{eqnarray}
where supersymmetry breaking is signaled by $\langle F^I F^\dag_I \rangle \neq 0$ with
\begin{eqnarray}
\label{eq:FF}
\langle F^I F^\dag_I \rangle =e^{\big|\langle z \rangle\big|^2}\Big( \sum\limits_p|{\cal I}_p|^2+    
M_4^4\sum \limits_a \Big| \langle \partial_a \;\omega_0\rangle \Big|^2\Big)+ 
m_p^2 \, |m_{3/2}'|^2 \ .
\end{eqnarray}
For simplification, we introduce the notation $|{\cal I}|^2=\sum\limits_p|{\cal I}_p|^2$. Assuming the vanishing of the cosmological constant allows us to express the parameter $|{\cal I}|^2$:
\begin{gather}
|{\cal I}|^2=  e^{-\big|\langle z \rangle\big|^2}m_p^2 ( 3|m_{3/2}|^2- |m'_{3/2}|^2)-M_4^4\sum \limits_a \Big|\langle \partial_a \omega_0\rangle\Big|^2\ . \label{eq:ConsVmin}
\end{gather}
Through this relation, it is possible to suppress the $|{\cal I}|^2$-dependency in the mass matrix. Note also that due to the negative sign on the last term, and since $|{\cal I}|^2$ must be positive, we have
\begin{gather}
|m'_{3/2}|  < \sqrt{3}|m_{3/2}| \Rightarrow |\lag\mathfrak{d}_z w\rag| < \sqrt{3} |\lag w\rag| \ll 1\ .
\end{gather}
Since $|{\cal I}|^2$ controls the loop contributions (see \autoref{eq:hardcorrection1} and \autoref{eq:hardcorrection2}), a first hint to obtain non-negligible corrections is to suppress the negative terms in \autoref{eq:ConsVmin}. Since the partial derivative $\partial_a\omega_0$ will be eliminated using minimisation equations, a possible solution is to assume low value of $m'_{3/2}$ (\textit{i.e.}, high values of $\xi_{3/2}$). Note however that taking too high value of $\xi_{3/2}$ may correspond to a fine-tuning. The first derivatives of the potential can also be easily obtained:
\begin{eqnarray}
\Big<\partial_z V\Big>&=&
e^{\big|\langle z \rangle\big|^2}\Big(
\mathfrak{d}_z {\cal I}_p \bar {\cal I}^p   +
M_4^4    \langle \partial_a \mathfrak{d}_z \;\omega_0\rangle  \langle   \partial^a \;\bar \omega_0\rangle \Big)\nn\\
&&+
m_p^2
\Big( m''_{3/2} \bar m'_{3/2}  -2 m'_{3/2} \bar m_{3/2}\Big) \label{eq:dzV}\\
\Big<\partial_ p V \Big>&=&
e^{\frac12\big|\langle z \rangle\big|^2} \Big(\mathfrak{d}_z {\cal I}_p \bar m'_{3/2} -2 {\cal I}_p \bar m_{3/2}\Big)\ , \quad \forall p =1,\ \dots, \ n  \label{eq:dpV}
\\
\Big< \partial_a V \big>&=& 
e^{ \big|\langle z \rangle\big|^2}  
M_ 4^4\langle\partial_a\partial_b \omega_0\rangle  
\langle \partial^b \bar \omega_0\rangle  \nn\\
&& +  
e^{\frac12 \big|\langle z \rangle\big|^2}      M_4^3
\Big(\langle\mathfrak{d}_z\partial_a \omega_0
\rangle \bar m'_{3/2}-2 \langle\partial_a \omega_0\rangle \bar m_{3/2}\Big)\ , \quad \forall a=1,\ \dots,\ m\label{eq:daV}
\end{eqnarray}
The derivative along the $S$ field \ref{eq:dpV} leads to a relation between ${\cal I}_p$ and the covariant derivative $\mathfrak{d}_z{\cal I}_p$:
\begin{gather}
\lag\partial_p V\rag = 0 \qquad \Rightarrow \qquad \mathfrak{d}_z{\cal I}_p=2\bar{\xi}_{3/2}{\cal I}_p \quad \forall p=1,\dots,\ n \ .\label{eq:dIp}
\end{gather}

Assuming one field $\phi$ in the matter sector for simplicity, the two others constraints (\ref{eq:dzV} and \ref{eq:dpV}) can be rewritten in the following form:

\begin{eqnarray}
M_4^2&=&   \frac{ Re \lag \partial_\phi^2\omega_0\partial^\phi\bar{\omega}_0 \rag}{ 
Re \lag 2\partial_\phi\omega_0\bar{w} - \mathfrak{d}_z\partial_\phi\omega_0\mathfrak{d}_z\bar{w} \rag } m_{p}^2\label{eq:M4_1}
\\
&=&     \frac{ Im \lag \partial_\phi^2\omega_0\partial^\phi\bar{\omega}_0 \rag}{ 
Im \lag 2\partial_\phi\omega_0\bar{w} - \mathfrak{d}_z\partial_\phi\omega_0\mathfrak{d}_z\bar{w} \rag } m_{p}^2 \ , \quad \forall a=1,\ \dots,\ m  \nn
\end{eqnarray}
for the $\lag\partial_aV\rag$ derivative and 
\begin{eqnarray}
M_4^2&=& \frac{Re\lag \partial^\phi\bar{\omega}_0\big( 2\bar{w}\partial_\phi\omega_0 - \mathfrak{d}_z\partial_\phi\omega_0 \mathfrak{d}_z\bar{w} \big) \rag
}{Re \left\langle6 |w|^2 \overline w -4 |\mathfrak{d}_z w|^2\overline w +
(\overline{\mathfrak{d}}_z \overline w)^2 \mathfrak{d}^2_z w
\right\rangle} m_{p}^2\label{eq:M4_2}\\
&=&  \frac{Im\lag \partial^\phi\bar{\omega}_0\big( 2\bar{w}\partial_\phi\omega_0 - \mathfrak{d}_z\partial_\phi\omega_0 \mathfrak{d}_z\bar{w} \big) \rag
}{Im \left\langle6 |w|^2 \overline w -4 |\mathfrak{d}_z w|^2\overline w +
(\overline{\mathfrak{d}}_z \overline w)^2 \mathfrak{d}^2_z w
\right\rangle} m_{p}^2\nn
\end{eqnarray}
for $\lag\partial_zV\rag$.
\subsection{Mass matrix}\label{subsec:massmatrix}
To pursue our analysis, we first check the mass matrix structure. The mass matrix is calculated in the $\{\phi,S,z\}$ basis. Denoting $\mathcal{M}_1$ the real $m\times m$ symmetric submatrix in the basis $\{\phi^ a\}$ (with their respecting eigenvalues $\lambda_1,\dots, \lambda_m$) and $\mathcal{M}_2$ the real $(n+1)\times (n+1)$ symmetric submatrix in the basis $\{S^p,z\}$ (with their eigenvalues $\mu_1,\dots, \mu_{n+1)}$), the mass matrix $\mathcal{M}$ has then the following structure:
\beqa
\mathcal{M}= \bpm\phantom{\varepsilon} \mathcal{M}_1& \varepsilon P^t\\
       \varepsilon P & \varepsilon \mathcal{M}_2 \epm \nn
\eeqa
with $\varepsilon\sim 0$ and $P$ a real $(n+1)\times m$ matrix. It can be shown that the eigenvalues of the matrix $\mathcal{M}$ are at the first order of $\varepsilon$: $\lambda_1,\dots ,\lambda_m$ and $\varepsilon \mu_1,\dots ,\varepsilon \mu_{n+1}$, meaning that the matrix $\mathcal{M}_1$ and $\mathcal{M}_2$ can be diagonalised separately. The complete proof of this statement can be found in Appendix \ref{eq:screenmatmass}. In order to estimate the order of magnitude of the S-loop, we consider the submatrix $\{S^p,z\}$. \medskip

Implementing the constraints \ref{eq:dIp} in the general mass matrix (see Appendix \ref{app:S2MSSM}) in the case of the simplified model:
\begin{eqnarray}
{\left\langle\frac{\partial^2 V}{\partial S^p\partial
S_q^\dagger}\right\rangle} &=& \delta_p{}^q |\xi _{3/2}|^2 |m'_{3/2}|^2 + 
\frac{e^{|\left\langle z \right\rangle |^2}}{m_{p}^2} \left(4 |\xi_{3/2}|^2-1\right) \mathcal{I}_p\overline{\mathcal{I}}^q, \nn\\
m_{p}^{-1}{\left\langle\frac{\partial^2 V}{\partial S^p\partial z^\dagger}\right\rangle}&=&
-\frac{e^{|\left\langle z \right\rangle|^2/2}}{m_{p}}\left(\overline{m}'_{3/2}-2\overline{\xi}_{3/2}\overline{m}''_{3/2}\right)\mathcal{I}_p \label{eq:d2VdSpdzdag}\\
m_p^{-2} {\left\langle \frac{\partial^2 V}{\partial z\partial z^\dag} \right \rangle}&=&
\frac 1 {m_p^2}e^{\big|\langle z \rangle\big|^2}\Bigg[
\big|{\cal I}\big|^2\Big(1+
4|\xi_{3/2}|^2\Big)+
M_4^4 \sum \limits_a \Big(\Big|\langle \partial_a \mathfrak{d}_z \omega_0\rangle \Big|^2+
\Big|\langle \partial_a \omega_0\rangle \Big|^2\Big)
\Bigg] \nn\\
&& +
\big|m''_{3/2}\big|^2-2 \big| m_{3/2}\big|^2\nn
\end{eqnarray}
Note that we do not compute the elements ${\left\langle \frac{\partial^2 V}{\partial S^p\partial z} \right \rangle}$ and ${\left\langle \frac{\partial^2 V}{\partial z^2} \right \rangle}$ since they can be fine-tuned to zero using the parameters $m'''_{3/2}$ and $\mathfrak{d}_z^2{\cal I}_p$ and so simplify the mass matrix structure. The others elements ${\left\langle \frac{\partial^2 V}{\partial \phi^a\partial \phi^b} \right \rangle}$, ${\left\langle \frac{\partial^2 V}{\partial z\partial \phi^a} \right \rangle}$ and ${\left\langle \frac{\partial^2 V}{\partial S^p\partial \phi^a} \right \rangle}$ are not relevant due to the screening property of the mass matrix and ${\left\langle \frac{\partial^2 V}{\partial S^p\partial S^q} \right \rangle}=0$ in the special case $\partial_p\omega_0=0$. Furthermore, all the elements in the form ${\left\langle \frac{\partial^2 V}{\partial X^I\partial X^J} \right \rangle}$ and ${\left\langle \frac{\partial^2 V}{\partial X^{\dag}_I\partial X^{\dag}_J} \right \rangle}$ automatically vanish if the hidden sector is invariant under a $U(1)$ symmetry.\nl
If we want a high effect from the S-loop, the off-diagonal term must be important to boost the mixing between the $z$ and $S^p$ fields. Since $\bar{m}'_{3/2}$ is already constrained by the cosmological constant, we can assume:
\begin{gather}
|\lag\mathfrak{d}_z^2w\rag | \sim  \mathcal{O}(1) \ .\label{eq:OoM-d2w}
\end{gather}
Due to this structure, the mass matrix has some interesting properties which we now demonstrate. We first rewrite the mass matrix in a more general form by highlighting the ${\cal I}_p$ dependencies:	
 \begin{eqnarray}
{\left\langle \frac{\partial^2 V}{\partial S^p\partial S_p^\dag} \right \rangle} =
a + b |\mathcal{I}_p|^2 , \quad
{\left\langle\frac{\partial^2 V}{\partial S^p\partial
S_q^\dagger}\right\rangle}=
b \mathcal{I}_p\overline{\mathcal{I}}^q\ , \\
m_{p}^{-1}{\left\langle\frac{\partial^2 V}{\partial S^p\partial z^\dagger}\right\rangle}=
c \mathcal{I}_p \label{eq:d2VdSpdzdag-1} \ , \quad
m_{p}^{-2}{\left\langle \frac{\partial^2 V}{\partial z \partial z^\dag} \right \rangle}&=& d \ .
\end{eqnarray}
The $(n+1)\times (n+1)$ squared mass matrix can then be written in the following form:
\beqa
\label{eq:Mass2}
\mathcal{M}^2  \equiv a \mathbb{I} + \mathbb{A}\ , \quad \text{with}\quad \mathbb{A} = \bpm [B]  &  [C]  \\
                  [C^\dag]  &  [D] \epm , 
\eeqa
where $\mathbb{A}$ is a block matrix and we introduce $[B], [C], [D]$, $n\times n$, $1\times n$ and $1\times 1$ matrices respectively:
\beqa
\begin{aligned}
\left[B\right]_{p}{}^q&= b \mathcal{I}_p\overline{\mathcal{I}}^q\ , \\
\left[C\right]_p &= c \mathcal{I}_p,  \ [C^\dag]^q = c^* \overline{\mathcal{I}}^q \label{eq:defabcd}
\left[D\right]_{1}\!\!{}^1 &= d-a \ .
\end{aligned}
\eeqa
The squared mass matrix $\mathbb{A}^2$ is
\begin{eqnarray}
 \mathbb{A}^2  &=& \bpm [B].[B] + [C].[C^\dag]  &  [B].[C] + [D] [C]  \\
                  [C^\dag].[B] + [C^\dag] [D]  &  [C^\dag].[C] + [D]^2 \epm \nn\\
                &=& \bpm (b  |\mathcal{I}|^2 +\frac{|c|^2}{b})[B] &  (b |\mathcal{I}|^2 + d -a) [C]  \\
                  (b |\mathcal{I}|^2 + d -a) [C^\dag]  &  (\frac{|c|^2}{d-a} + d-a) [D] \epm  \ ,\nn
\end{eqnarray}
showing that $\mathbb{A}$ has the same block matrix structure of $\mathbb{A}$ up to multiplicative factors that are function of $|\mathcal{I}|^2=\sum_p|{\cal I}_p|^2$. We can then conclude that $\mathbb{A}^s$ (with $s \in \mathbb{N}$) is of the same form:
\begin{equation}
\mathbb{A}^s=  \bpm f_s(|\mathcal{I}|^2) [B] &  g_s(|\mathcal{I}|^2) [C]  \\
                  g_s(|\mathcal{I}|^2) [C^\dag]  &  h_s(|\mathcal{I}|^2) [D] \epm   \nn
\end{equation}
where $f_s, g_s, h_s$ denote some polynomial functions. One thus has
\begin{equation}
\mbox{Tr}[\mathbb{A}^s] = b f_s(|\mathcal{I}|^2) |\mathcal{I}|^2 + (d-a) h_s(|\mathcal{I}|^2)\ ,\nn
\end{equation}
i.e., a function of only $|\mathcal{I}|^2$. Using the constraint on the cosmological constant \ref{eq:ConsVmin}, we can then suppress the $|{\cal I}|^2$-dependency in the matrix. Since the eigenvalues can be written in terms of the invariants of the matrix (which are obtained from $\mbox{Tr}[\mathbb{A}^s]$), the mass states are thus not function of the intermediate energy scale $M_{2,p}$ and $M_{3,p}$. \medskip
We can also show that $\mathcal{M}^2$ can be expressed in a simpler form:
\beqa
{\mathcal{ M}'}^2 = \begin{pmatrix} a \mathbb I_{n-1}&0&0\\
  0&a + b |{\cal I}|^2&c |{\cal I}|\\
  0 &\bar c |{\cal I}|&d\\
  \end{pmatrix}
  \label{eq:matrixprimee}
\eeqa
with the parameters $a,\ b,\ c$ and $d$ already defined in \autoref{eq:defabcd}. Since $a=|\xi_{3/2}|^2|m'_{3/2}|^2 = |m_{3/2}|^2$, the mass matrix leads to $(n-1)$ fields with a mass equal to the gravitino mass. The two last states mix themselves through a submatrix equivalent to the case of only one singlet and one field from the hidden sector. The complete calculation to obtain the matrix structure \ref{eq:matrixprimee} can be found in Appendix \ref{app:eigen}, as well as the determination of the complete eigenvector basis using the Gram-Schmidt process.

\subsection{Simple case: only one $S$-field}
We now consider only one S field and one $\phi$ field. The two relations \ref{eq:M4_1} and \ref{eq:M4_2} imply a dependency on the covariant derivative $\mathfrak{d}_z\partial_\phi\omega_0$. From the simplification $\lag \mathfrak{d}_z\partial_a\omega_0\rag = z^{\dagger} \partial_a\omega_0$ (which corresponds to a decoupled contribution to the hidden sector and the matter sector in the superpotential $W_0$, or more generically $\partial_a\partial_z\omega_0 = 0$)\footnote{This is typical where the parameters in the superpotential \autoref{eq:Ws2mssm} are no longer function of the hidden sector (for example, $\lambda(z)=\lambda$).}, we obtain some information on the order of magnitude of some parameters. From \autoref{eq:M4_1}, it follows from the inequality $M_4 \ll m_p$ (we assume real parameters):
\begin{gather}
|\lag \partial_\phi^2\omega_0\rag | \ll |\lag 2w -z^{\dagger}\mathfrak{d}_z\bar{w}\rag| \ll 1 \ .\nn
\end{gather}
Moreover, using the relation \ref{eq:OoM-d2w}, we get from the equality \ref{eq:M4_2} and the equality between \ref{eq:M4_1} and \ref{eq:M4_2}:
\begin{gather}
|\lag \partial_\phi\omega_0 \rag| \ll \lag \frac{|\mathfrak{d}_zw|}{\sqrt{|2w - z\mathfrak{d}_zw}} \rag \ll 1 \ , \label{eq:magndphiw}\\
|\lag \partial_\phi^2\omega_0 \rag| \approx \frac{|\lag 2w - z\mathfrak{d}_zw\rag|^2}{|\lag (\mathfrak{d}_zw)^2\mathfrak{d}_z^2w \rag|}|\lag \partial_\phi\omega_0\rag|^2 \sim \mathcal{O}(|\lag \partial_\phi\omega_0 \rag|^2)\ .
\end{gather} 

In the case where we relax the condition $\partial_a\partial_z\omega_0= 0$, the relations are more involved. We can solve the system of equations (\ref{eq:M4_1},\ref{eq:M4_2}) as function of $\partial_\phi\omega_0$ and $\mathfrak{d}_z\partial_\phi\omega_0$, which gives:
\beqa
\left<\mathfrak{d}_z \partial_\phi \omega_0 \right> &=&
 \left(2 \left<w\right> - \frac{m_{p}^2}{M_4^2} \left<\partial_\phi^2
\omega_0\right>\right)\left<\frac{\partial_\phi \omega_0}{\mathfrak{d}_z w}\right>,  \label{eq:deltadomega0}\\
\left<\partial_\phi \omega_0 \right>_\pm &=&\pm \frac{M_4^2}{m_{p}^2} \sqrt{\left<\frac{ 6 w^3 +
(\mathfrak{d}_z w)^2 ( \mathfrak{d}_z^2 w -4 w )}{\partial_\phi^2 \omega_0}\right>} \label{eq:domega0}
\eeqa
(the sign in the notation $\lag\partial_\phi\omega_0\rag_\pm$ will be omitted when the sign of the solution is irrelevant for the discussion). The expression of $\lag\partial_\phi^2\omega_0\rag$ can also be expressed as function of $\partial_z\partial_a\omega_0$:
\beqa
\left<\partial_\phi^2 \omega_0 \right>_{\mp}=
\frac{1}{2 c_1}\left(c_2^2 - 2 c_1 c_3 \mp |c_2|\sqrt{c_2^2 - 4 c_1 c_3}\right)
\label{eq:d2omega0pm}
\eeqa
where we introduce the coefficients $c_1$, $c_2$ and $c_3$: 
\beqa
c_1 &=& \left < 6 w^3 +(\mathfrak{d}_z w)^2 (\mathfrak{d}_z^2 w -4 w) \right > = \frac{\left <w^3\right >}{\xi_{3/2}^2}\left < 6  \xi_{3/2}^2 + \frac{\mathfrak{d}_z^2 w}{w} -4 \right >, \label{eq:c1}\\ 
c_2 &=& \left < (\mathfrak{d}_z w) \ \partial_z\partial_\phi\omega_0
\right > = \frac{\left <w\right >}{\xi_{3/2}} \left <\partial_z\partial_\phi\omega_0 \right >, \label{eq:c2}\\ 
 c_3 &=&  \left < z \mathfrak{d}_z w -2 w \right > \frac{M_4^2}{m_{p}^2}= \frac{\left <w\right >}{\xi_{3/2}} \left(\left < z \right >-2 \xi_{3/2} \right) \frac{M_4^2}{m_{p}^2}\ . \label{eq:c3}
 \eeqa
Note that the solution does not depend of the sign in \autoref{eq:domega0}. This configuration (with one matter field $\phi$) will be the main studied in this chapter. The two configurations (with and without the assumption $\partial_a\partial_z\omega_0=0$) will be analysed in the sequel.  \medskip

A remark can be made on some limits of $\lag \partial_\phi\omega_0 \rag_{\pm}$. We have already mentioned that high values of $\xi_{3/2}$ may be interesting to push-up the radiative corrections through the parameter $|{\cal I}|^2$. Taking the limit $\xi_{3/2}$ to infinity gives
\begin{gather}
\displaystyle\lim_{\xi_{3/2}\rightarrow \infty}\lag \partial_\phi\omega_0 \rag_{\pm} = 3\frac{m_p^2}{M_4^4}e^{-|\lag z\rag|^2}|m_{3/2}|^2\ . \nn
\end{gather}
From \autoref{eq:ConsVmin}, we see that the first and last term compensate each other leading to
\begin{gather}
\displaystyle\lim_{\xi_{3/2}\rightarrow \infty} |{\cal I}|^2 = 0\ . \nn
\end{gather}
Taking high value of $\xi_{3/2}$ may then not automatically leads to high loop corrections. The full numerical computation of the one-loop Higgs boson mass seems mandatory to understand the evolution regarding the $\xi_{3/2}$ parameter (see Subsection \ref{eq:NumComput}).\medskip

Assuming one singlet $S$, the mass matrix of the $z-S$ sector is a simple $2\times 2$ matrix. We can then easily obtain the mixing angle of $z-S$ with the two eigenvalues denoted $m^2_{S'}$ ($S'$ being the lighter state) and $m^2_{\zeta '}$ ($\zeta '$ being the heavier state). Introducing
\begin{eqnarray}
a&=& { e^{|\langle z \rangle|^2}\frac{M_4^4}     
{m_{p}^2} \Big(  \Big<\big|\partial_\phi \omega_0\big|^2-\big|\mathfrak{d}_z \partial_\phi\omega_0\big|^2 \Big> \Big)+ 2 \big| m_\frac 32'|^2 -3 \big| m_\frac 32|^2   - \big| m_\frac32 ''|^2}\ , \nn\\
\Delta&=&\sqrt{a^2+4 e^{ |\langle z \rangle|^2} \Bigg(\left|
2 \overline m_{3/2}'' \overline \xi_{3/2}  - \overline m_{3/2}'\right|^2 \frac{\big|{\cal I}\big|^2}{m^2_{p}}\Bigg)}.\label{eq:delta}
\end{eqnarray}
the mixing angle and the eigenvalues are then:
\begin{gather}
\cos\theta = \sqrt{\frac{\Delta-a}{2\Delta}}\ ,\sin\theta = \sqrt{\frac{\Delta+a}{2\Delta}}\ ,\label{eq:mixingangle}
\end{gather}
and the eigenvalues 
\beqa
m^2_{\zeta'/S'}&=&\frac12\Bigg( e^{|\langle z \rangle|^2}\frac{M_4^4}     
       {m_{p}^2} \Big(  \Big<\big|\partial_\phi \omega_0\big|^2(1-8 |\xi_\frac32|^2)+\big|\mathfrak{d}_z \partial_\phi\omega_0\big|^2 \Big> \Big)+ \big| m_\frac 32 \big|^2(24 |\xi_\frac32|^2-9) + \big| m_\frac32 ''|^2 \nn\\
&&\pm  \Bigg\{\Bigg( e^{|\langle z \rangle|^2} \frac{M_4^4}{m_{p}^2} \Big(  \Big<\big|\partial_\phi \omega_0\big|^2-\big|\mathfrak{d}_z \partial_\phi\omega_0\big|^2 \Big> \Big)+ \big| m_\frac 32'|^2(2-3 \big|\xi_\frac32\big|^2) - \big| m_\frac32 ''|^2
\Bigg)^2\label{eq:eigenvalues}\\
&&\hskip 2.truecm +4 e^{ |\langle z \rangle|^2} \Bigg(\left|
2 \overline m_{3/2}'' \overline \xi_{3/2}  - \overline m_{3/2}'\right|^2 \frac{\big|{\cal I}\big|^2}{m^2_{p}}\Bigg)\Bigg\}^{1/2}\Bigg)\ .                \nn 
\eeqa
The eigenstates reduce then to
\beqa
\begin{aligned}
S'&=& \cos \theta\; S -\e m _{p} \sin \theta\; z \ ,\\
z'&=& \sin \theta\; S + \frac \e{m_{p}} \cos \theta\;z
\end{aligned}\ \quad \text{with}\quad \e = \frac{\Big| \Big<\frac {\partial^2 V}{\partial S \partial z^\dag}\Big>\Big|}{\Big<\frac {\partial^2 V }{\partial S \partial z^\dag}\Big>} \ . \label{eq:eigenstates}
\eeqa
We have derived all the interesting quantities for our investigation.  
\section{Order of magnitude of the effects from the hybrid sector}\label{sec:HS_effects}
In the previous section, we have investigated the form of the constraints coming from the nullity of the cosmological constant and the minimisation of the potential. The mass matrix has also been calculated in the general case of $n$ singlets. We now discuss the loop effects coming from the hybrid sector.\newline

Following the results \ref{eq:eigenvalues} and \ref{eq:eigenstates}, the state $S$ is described by the two mass eigenstates $S'$ and $\zeta '$ as
\begin{gather}
S = \zeta '\sin\theta + S'\cos\theta\ .\nn
\end{gather}
Since all the one-loop contributions (see \autoref{eq:hardcorrection1} and \autoref{eq:hardcorrection2}) can be rewritten in the general form $S S^\dag O$ with $O$ a specific operator, then the one-loop ($S$) contribution is: 
\begin{equation}
\delta O \sim  \frac 1{16 \pi^2}(m_{\zeta'}^2 \sin^2 \theta +  m_{S'}^2 \cos^2 \theta) \label{eq:deltaO}
\end{equation}
Assuming that the heaviest sector is the main contribution to the correction, we can first analyse the contribution $m_{\zeta'}^2 \sin^2 \theta$. Those studies will draw attention to the relevant regimes which allow a Higgs boson mass near $125\ \mathrm{GeV}$. After this discussion, a more complete numerical computation of the possible loop-effect on the Higgs boson mass is done.  
\subsection{Set of constraints on the model}\label{subsec:set_cons}
Before making a numerical analysis, we must consider all the constraints that prevent unphysical points. Our parameters-space is an eight-dimensional space: $\{m_{3/2},\lag z\rag,\lag \mathfrak{d}_z^2w\rag,\xi_{3/2},M_4,\lag \partial_z\partial_\phi \omega_0\rag,sgn(\lag\partial_\phi\omega_0\rag) \rag, s\}$ (with $s$ the sign present in the solution of $\lag \partial_\phi^2\omega_0\rag$ \autoref{eq:d2omega0pm}). We assume all real parameters. Note also that we have freedom on the sign of the parameters $\{m_{3/2},\lag z\rag,\mathfrak{d}_z^2w,\xi_{3/2},\lag \partial_z\partial_\phi\omega_0\rag\}$. This parameters-space is already constrained by the previous results in \ref{subsec:s2matrix}.\medskip

The vanishing of the cosmological constant express the parameter $|{\cal I}|^2$ as a function of the other parameters through the relation \ref{eq:ConsVmin}. So we must take care of the sign of the parameter $|{\cal I}|^2$. The choice of parameters must then fulfilled the following inequality
\begin{gather}
|{\cal I}|^2\geq 0\label{eq:consI2}
\end{gather}
which can be rewritten in the following form (using the relation \ref{eq:domega0})
\beqa
 |{\cal I}|^2 &=& M_4^4 \langle w \rangle^2 \left(\frac{ (3 \xi_{3/2}^2 -1)}{\xi_{3/2}^2}\frac{M_4^2}{m_{p}^2}- \frac{\left( 2\langle w \rangle (3 \xi_{3/2}^2 -2) + \langle \mathfrak{d}_z^2 w \rangle \right)}{\xi_{3/2}^2 \left<\partial_\phi^2 \omega_0 \right>}\frac{M_4^4}{m_{p}^4} \right)
 \geq 0\ . \label{eq:poscond}
\eeqa
Moreover, the calculation of the mass spectrum (\textit{i.e.}, $m_{\zeta '}^2$ and $m_{S '}^2$) could leads to negative squared masses. To recover a correct spectrum, we must impose:
\begin{gather}
m_{S'}^2 \geq 0 \ , \quad m_{\zeta '}^2\geq 0\ .\nn
\end{gather}
The two last constraints come from $\lag\partial_\phi^2\omega_0\rag$ solution's in \autoref{eq:d2omega0pm}. Writing the equality of \autoref{eq:deltadomega0} and \autoref{eq:domega0} (by expressing $\partial_\phi\omega_0$ as function of the other parameters), we get:
\beqa
\sqrt{\frac{c_1}{\left<\partial_\phi^2 \omega_0 \right>}} = \mp \frac{c_2}{c_3 + \left<\partial_\phi^2 \omega_0 \right>}
\label{eq:implquad}
\eeqa
where the sign in the right term comes from the sign in \ref{eq:domega0}. From those relations, we get:
\beqa
&&\frac{c_1}{\left<\partial_\phi^2 \omega_0 \right>_{\mp}} \ge 0, \label{eq:poscondc1}\\
&&\frac{c_2}{c_3 + \left<\partial_\phi^2 \omega_0 \right>_{\mp}} \le 0, \ge 0 \ \text{for} \ \left<\partial_\phi \omega_0 \right> \ge 0, \le 0 \ , \label{eq:poscondc2}
\eeqa
meaning that not all couples of solutions $(\lag\partial_\phi\omega_0\rag_\pm,\lag\partial_\phi^2\omega_0\rag_\mp)$ are possible. By using the solution \ref{eq:d2omega0pm} in those inequalities, we identify the region which are valid regarding the equation \ref{eq:implquad}. We obtain:
\begin{align}
 i) \ &c_1 c_3 <0 :\nonumber \\
\sloppy &c_1 c_2 <0  \Leftrightarrow \ \rm{the \ only \ 
solution \ is} \left<\partial_\phi^2 \omega_0 \right>_- (\rm{resp.} \ \left<\partial_\phi^2 \omega_0 \right>_+) \ \rm{if} \ \left<\partial_\phi \omega_0 \right> < 0 \ \nn\\
& \hskip 2.0truecm (\rm{resp.} \ \left<\partial_\phi \omega_0 \right> >0) \nonumber \\
\sloppy &c_1 c_2 > 0 \Leftrightarrow \ \rm{the \ only \ 
solution \ is} \left<\partial_\phi^2 \omega_0 \right>_- (\rm{resp.} \ \left<\partial_\phi^2 \omega_0 \right>_+) \ \rm{if} \ \left<\partial_\phi \omega_0 \right> > 0 \ \nonumber \\
& \hskip 2.0truecm (\rm{resp.} \ \left<\partial_\phi \omega_0 \right> <0) \nonumber \\
& \label{eq:cases}\\
ii) \ &c_1 c_3 >0 :\nonumber \\
&c_1 c_2 <0  \land c_2^2 - 4 c_1 c_3 > 0\Leftrightarrow \ \rm{the \ two \ solutions} \left<\partial_\phi^2 \omega_0 \right>_\mp \ \rm{are \ valid \ when} \ \left<\partial_\phi \omega_0 \right> > 0 \nonumber \\
&c_1 c_2 > 0 \land c_2^2 - 4 c_1 c_3 > 0\Leftrightarrow \ \rm{the \ two \ solutions} \left<\partial_\phi^2 \omega_0 \right>_\mp \ \rm{are \ valid \ when} \ \left<\partial_\phi \omega_0 \right> < 0 \nonumber
\end{align}
Note that in the case where $\partial_z\partial_\phi\omega_0=0$, the coefficient $c_2$ drop out of the equations. There is then only one solution for $\lag\partial_\phi^2\omega_0\rag$ and the only constrain is \ref{eq:poscondc1} which can be rewritten as $c_1c_3 <0$. \medskip
From the definitions of the coefficients $c_1$, $c_2$ and $c_3$ in \ref{eq:c2}, assuming $\lag\mathfrak{d}_z^2 w\rag > 4\lag w\rag$ (which is in accordance with the assumptions \ref{eq:OoM-d2w} and \ref{eq:OoM_w}), the sign of $c_1$ is then equal to the sign of $w$, and so $m_{3/2}$. Thus:
\begin{gather}
\text{sgn}(c_1c_3) = \text{sgn}(\xi_{3/2}(\lag z \rag -2\xi_{3/2} )) \ \text{and}\ \text{sgn}(c_1c_2)=\text{sgn}(\xi_{3/2}\lag \partial_z\partial_\phi\omega_0 \rag) \ .
\end{gather}
We have already a restriction on the possible values of $\xi_{3/2}$ which comes from the fine-tuning of the cosmological constant ($|\xi_{3/2}|>\frac{1}{\sqrt{3}}$ in \ref{eq:ConsVmin}). Since we have by definition $\lag z\rag = \mathcal{O}(1)$, we are mostly in the case $\text{sgn}(c_1c_3)<0$ (independent of $\text{sgn}(\xi_{3/2})$ and $\text{sgn}(\lag z \rag)$) as long as the value of $\xi_{3/2}$ is outside the domain $[\frac{1}{\sqrt{3}}, \frac{|\langle z \rangle|}{2}] \cup [-\frac{|\langle z \rangle|}{2},-\frac{1}{\sqrt{3}}]$, which is generally small or closed.\medskip

Note also that the relation \ref{eq:poscondc1} can be rewritten as
\beq
\frac{\left( 2\langle w \rangle (3 \xi_{3/2}^2 -2) + \langle \mathfrak{d}_z^2 w \rangle \right)}{ \left<\partial_\phi^2 \omega_0 \right>} \ge 0\ . \label{eq:poscondc11}
\eeq
Since this positive term is subtracted in \autoref{eq:poscond}, if its contribution is not subleading (\textit{i.e.}, $\left|\left<\partial_\phi^2 \omega_0 \right>\right|$ is relatively small), the configurations could become highly constrained.\medskip

We can now compute the order of magnitude of the correction \autoref{eq:deltaO}. 

\subsection{A first approach: $\partial_a\partial_z\omega_0=0$}
We assume the simplifying assumption $\lag\partial_\phi\partial_z\omega_0\rag = 0$. Starting from $m_{\zeta}^2$ and $\sin\theta$ in \autoref{eq:eigenvalues} and \autoref{eq:mixingangle}, we can substitute $|{\cal I}|^2$ and $\partial_\phi\omega_0$ using the vanishing of the cosmological constant \ref{eq:ConsVmin} and the minimisation equation along the field $z$ \ref{eq:M4_2}. After lengthy computation and Taylor expansion with the gravitino mass $m_{3/2}$ we obtain (we assume real \textit{v.e.vs} for the fields):
\begin{eqnarray}
\sin^2 \theta &=&  4 \, \frac{e^{-\frac{\langle z \rangle^2}{2}} \xi_{3/2} }{(\langle z \rangle -2 \xi_{3/2})\langle {\mathfrak{d}}_z^2 w \rangle} 
\frac{m_{3/2} m_{p}^2}{M_4^3}
+ {\cal O}(m_{3/2}^2) \label{eq:sinexpand},\\
m_{\zeta'}^2 &=&e^{\langle z \rangle^2} \langle {\mathfrak{d}}_z^2 w \rangle^2 \frac{M_4^6}{m_{p}^4} + {\cal O}(m_{3/2}) \ .\label{eq:mzexpand}
\end{eqnarray}
which lead to
\begin{equation}
m_{\zeta'}^2 \sin^2 \theta =  4 \, \frac{e^{\frac{\langle z \rangle^2}{2}} \xi_{3/2} \langle {\mathfrak{d}}_z^2 w \rangle}{(\langle z \rangle -2 \xi_{3/2})} 
\frac{m_{3/2} M_4^3}{m_{p}^2}
+ {\cal O}(m_{3/2}^2)   \label{eq:crucialexpansion}
\end{equation}
(we denote here $m_{3/2}\equiv |m_{3/2}| >0$ and $\xi_{3/2} \equiv |\xi_{3/2}| > 1/\sqrt3$). This relation requires for consistency the inequality
\begin{gather}
\frac{\lag \mathfrak{d}_z^2w \rag}{(\lag  z\rag -2\xi_{3/2})} > 0 \nn
\end{gather}
which is not in contradiction with the constraint $|{\cal I}|^2 >0$. \medskip

Supposing a energy scale in the matter sector of $M_{EW}\approx 10^2\ \text{GeV}$ we see that the correction is relatively small ($m_{\zeta'}^2 \sin^2 \theta\ \propto\ 10^{-32}m_{3/2}$ with $m_p=10^{18}\ \mathrm{GeV}$). To consider large loop-effects, we must then assume models with several energy scales in the matter sector, such as GUT theories with $M_4=M_{GUT}\approx 10^{15}\ \text{GeV}$. In this manner, the value of  $m_{\zeta '}^2$ mass increase and gives $m_{\zeta'}^2 \sin^2 \theta\ \propto\ 10^{7}m_{3/2}$. The superpotential can then be written as:
\begin{gather}
\lag W_0(z,S,\phi)\rag = M_4^3\lag\omega_0(z,S,\phi)\rag = M_4^3 \lag\omega_0(z,S,\phi_{GUT},\phi_{EW})\rag \nn
\end{gather}
with $\lag\phi_{GUT}\rag \approx \mathcal{O}(1)$ and  $\lag\phi_{EW}\rag \ll 1$ (since those fields are rescaled with a GUT scale). Notice that, even we assume only one singlet $S$ (and so no $\mathcal{U}$-field in $\omega_0$), we will estimate the order of magnitude of the one-loop effect \ref{eq:hardcorrection2} (as well as \ref{eq:hardcorrection1}) in Section \ref{eq:NumComput}.

\subsection{A first approach: Analysing several regimes}\label{eq:regimes}
If we relax the assumption on $\lag\partial_z\partial\phi \omega_0\rag$, the relations are more complex. One finds after using the relations \ref{eq:ConsVmin}, \ref{eq:deltadomega0} and \ref{eq:domega0}:
\beqa
\sin^2 \theta &=&  4 \, \frac{e^{-\langle z \rangle^2}}{\langle \partial_\phi^2 \omega_0 \rangle^2} \frac{m_{3/2}^2}{M_4^2}\frac{\Upsilon}{1  + \displaystyle\frac{\langle {\mathfrak{d}}_z^2 w \rangle}{\langle\partial_\phi^2 \omega_0\rangle} \frac{M_4^2}{m_{p}^2}} + {\cal O}(m_{3/2}^3)
\eeqa
where
\beq
\Upsilon=\frac{3 \xi_{3/2}^2 -1 -\displaystyle \frac{\langle {\mathfrak{d}}_z^2 w \rangle}{\langle\partial_\phi^2 \omega_0\rangle} \frac{M_4^2}{m_{p}^2}}{1  + \displaystyle\frac{\langle {\mathfrak{d}}_z^2 w \rangle}{\langle\partial_\phi^2 \omega_0\rangle} \frac{M_4^2}{m_{p}^2}} . \label{eq:Upsilon}
\eeq
and
\beqa
m_{\zeta'}^2 &=&  e^{\langle z \rangle^2} \langle {\mathfrak{d}}_z^2 w \rangle \frac{M_4^4}{m_{p}^2} \, \left(\langle \partial_\phi^2 \omega_0 \rangle  + \langle {\mathfrak{d}}_z^2 w \rangle \frac{M_4^2}{m_{p}^2}\right) + {\cal O}(m_{3/2})\ .
\eeqa
Thus:
\beqa
m_{\zeta'}^2 \sin^2 \theta &=& 4 \,  \frac{\langle {\mathfrak{d}}_z^2 w \rangle}{\langle\partial_\phi^2 \omega_0\rangle} \frac{m_{3/2}^2 M_4^2}{m_{p}^2} \Upsilon + {\cal O}(m_{3/2}^3) \label{eq:crucialexpansion1}
\eeqa
Contrary to the simplifying case (where $m_{\zeta'}^2 \sin^2 \theta\sim m_{3/2}M_4^3/m_p^2$), we see that the loop-effect turns to be $m_{\zeta'}^2 \sin^2 \theta\sim m_{3/2}^2M_4^2/m_p^2$. A $M_4$ prefactor has been replaced by $m_{3/2}$, which seems to reduce the effect in the loop. The parameter $\Upsilon$ \autoref{eq:Upsilon} can however push-up the correction in the case of high values of $\xi_{3/2}$.\medskip

Note that is we tend smoothly to $\lag \partial_z\partial_\phi\omega_0\rag = 0$, the two solutions $\lag\partial_\phi^2\omega_0\rag$ are degenerated and gives back the result \autoref{eq:crucialexpansion} with:
\begin{equation}
 \left<\partial_\phi^2 \omega_0 \right>_{\mp}=-c_3 =
 \frac{\left <w\right >}{\xi_{3/2}} \left(2 \xi_{3/2}-\left < z \right > \right) \frac{M_4^2}{m_{p}^2}. \label{eq:d2omega0pmlim}
\end{equation}

 The study of the leading behavior of the correction in the limit $\lag \partial_z\partial_\phi \omega_0 \rag\rightarrow 0$ is however not so simple. Indeed, since we need $\lag w\rag \ll 1$  and considering the expression \ref{eq:c1}, \ref{eq:c2} and \ref{eq:c3}, we must consider the ratio
\begin{gather}
\rho=\frac{\lag \partial_z\partial_\phi \omega_0 \rag}{\lag w\rag}\ .\nn 
\end{gather}
Three regimes are then studied,  $\rho^{-1} \ll 1$, $\rho \ll 1$ and $\rho ={\cal O}(1)$.
\noindent
\subsubsection{\underline{Regime $\mathbf{\rho^{-1} \ll 1}$:}}

\noindent
After expanding $\left<\partial_\phi^2 \omega_0 \right>_-$
and $\left<\partial_\phi^2 \omega_0 \right>_+$ around $\rho^{-1}=0$ in powers of $\left <w\right >$ (with $\left <w\right > \ll \min[1,\left <\partial_z\partial_\phi\omega_0 \right >]$), one finds
\beqa
\left.\left<\partial_\phi^2 \omega_0 \right>_-\right|_{(\rho^{-1} \ll 1)}&=&
\langle w \rangle^2 \frac{(2 \xi_{3/2} -\langle z \rangle)^2}{\xi_{3/2}^2}
\frac{\langle \mathfrak{d}_z^2 w \rangle}{\langle \partial_z\partial_\phi\omega_0 \rangle^2} \frac{M_4^4}{m_{p}^4} + {\cal O}(\langle w \rangle^3) ,
\label{eq:rhoinvll1m}\\
\left.\left<\partial_\phi^2 \omega_0 \right>_+\right|_{(\rho^{-1} \ll 1)}&=& \frac{\langle \partial_z\partial_\phi\omega_0 \rangle^2}{\langle \mathfrak{d}_z^2 w \rangle} + 
2 \langle w \rangle \left( \left(2 - 3 \xi_{3/2}^2\right)\frac{\langle \partial_z\partial_\phi\omega_0 \rangle^2}{\langle \mathfrak{d}_z^2 w \rangle^2} + \frac{\left(2 \xi_{3/2} -\langle z \rangle \right)}{\xi_{3/2}}
\frac{M_4^2}{m_{p}^2} \right)\nn\\
&& + {\cal O}(\langle w \rangle^2).
\label{eq:rhoinvll1p}
\eeqa
The solutions are totally decoupled. Obviously, these expressions are not valid for $\left <\partial_z\partial_\phi\omega_0 \right > \to 0$ (which is clearly the case in the leading term of $\left<\partial_\phi^2 \omega_0 \right>_-$, while the singularity of $\left<\partial_\phi^2 \omega_0 \right>_+$ appears in higher orders not shown here).\medskip

We now tend to analyse the order of magnitude of \ref{eq:crucialexpansion1}, assuming the two solutions \autoref{eq:rhoinvll1m} and \autoref{eq:rhoinvll1m}.\medskip

We first consider the solution \autoref{eq:rhoinvll1m} since it seems to be the only configuration that leads to an enhancement of $m_{\zeta'}^2 \sin^2 \theta$. Indeed, the leading term of $\lag \partial_\phi^2\omega_0\rag_{-}$ being proportional to $\lag w^2\rag$, the $m_{3/2}^2$ factor in $m_{\zeta '}^2\sin^2\theta$ is cancels out and can possibly increase the magnitude of the loop. Nevertheless, the first order contribution in \ref{eq:rhoinvll1m} cancels the $\frac{M_4^4}{m_{p}^4}$ suppression in \ref{eq:poscond}, which make the subtracted positive
term important and potentially leads to $|{\cal I}|^2<0$ for most values of parameters. One can recast the two equations  (\ref{eq:poscond}) and (\ref{eq:poscondc11})
respectively in a simple form
\beqa
\begin{aligned}[l]
&& A \langle w \rangle^2 - (B \langle w \rangle + C) \ge 0, \label{eq:poscond2}\\
&& B \langle w \rangle + C \ge 0, \label{eq:poscondc12}
\end{aligned}\quad \text{with :}\quad
\begin{aligned}[l]
&A=M_4^2 \frac{(3 \xi_{3/2}^2 -1) (2 \xi_{3/2} - \langle z \rangle)^2}{\langle \partial_z\partial_\phi\omega_0 \rangle^2},& \\
&B=2 m_{p}^2  \frac{\xi_{3/2}^2 (3 \xi_{3/2}^2 -2)}{\langle {\mathfrak{d}}_z^2 w \rangle},& \\
&C=m_{p}^2 \xi_{3/2}^2  .&
\end{aligned}
\eeqa
The coefficient $B$ can be positive or negative while $C$ and $A$ are defined positive (since $|\xi_{3/2}| > \frac{1}{\sqrt{3}}$). Recalling that $|\langle w \rangle|\ll 1$ we identify two possible ranges of values for $\lag w\rag$\footnote{The two other solutions, namely $\langle w \rangle > \frac{B + \sqrt{B^2 + 4 A C}}{2 A}$ when
$B>0$ and $\langle w \rangle < \frac{B - \sqrt{B^2 + 4 A C}}{2 A}$ when $B<0$ lead to high values of $\langle w \rangle \gg 1$ and so are not considered here.}:

\beqa
\Big( B > 0 \; \text{and}\; -\frac{C}{B} < &\langle w \rangle& < \frac{B - \sqrt{B^2 + 4 A C}}{2 A} \Big)\\
&\text{or}&   \nonumber \\ 
 \Big(B < 0 \; \text{and}\;  \frac{B + \sqrt{B^2 + 4 A C}}{2 A}  < &\langle w \rangle& < -\frac{C}{B}\Big)
\eeqa
which corresponds to a width $\langle \Delta w \rangle$ of the allowed interval of
\beqa
\langle \Delta w \rangle = \sgn(B) \frac{M_4^2}{8 m_{p}^2} \frac{(3 \xi_{3/2}^2 -1) (2 \xi_{3/2} - \langle z \rangle)^2 \langle {\mathfrak{d}}_z^2 w \rangle^3}{\xi_{3/2}^2 (3 \xi_{3/2}^2 -2)^3 \langle \partial_z\partial_\phi\omega_0 \rangle^2} + {\cal O}(\frac{M_4^4}{ m_{p}^4})\ .
\eeqa
The range of values for $\lag w \rag$ seems then extremely small for relatively high values of $\xi_{3/2}$, \textit{i.e.}, the solution is highly fine-tuned. A possible solution to increase the range $\lag \Delta w \rag$ can be to choose $\xi_{3/2}^2 \approx 2/3$. However, this value drives the intervals outside the condition $|\langle w \rangle|\ll 1$. We thus conclude that the regime given by Eq.~(\ref{eq:rhoinvll1m}) is actually not viable.\medskip

Considering now the solution \autoref{eq:rhoinvll1p}, we see that the $\frac{M_4^4}{m_{p}^4}$ term in \autoref{eq:poscond} is not reduced by $\lag \partial_\phi^2\omega_0\rag$. The positivity of $|{\cal I}|^2$ is generally always satisfied, leaving to the only constraint \autoref{eq:poscondc1}, which is relatively easy to achieve. However, due to the non-dependency of the leading term of \autoref{eq:rhoinvll1p} in $\lag w \rag$, the effect still remains suppressed by the $m_{3/2}^2$ factor in \autoref{eq:crucialexpansion1}. The regime $\rho^{-1} \ll 1$ seems then not appropriate to generate huge loop effects.\medskip

\noindent
\subsubsection{\underline{Regime $\mathbf{\rho \ll 1}$ and $\mathbf{\rho \sim \mathcal{O}(1)}$:}}

\noindent
We now consider the two other regimes, namely $\rho \ll 1$ and $\rho \sim \mathcal{O}(1)$. The expansion of $\left<\partial_\phi^2 \omega_0 \right>_-$
and $\left<\partial_\phi^2 \omega_0 \right>_+$ in the two mentioned limits gives:
\beqa
\left.\left<\partial_\phi^2 \omega_0 \right>_\mp\right|_{\rho \ll 1}&=& \frac{\left <w\right >}{\xi_{3/2}} \left(2 \xi_{3/2}-\left < z \right > \right) \frac{M_4^2}{m_{p}^2} 
\mp \langle \partial_z\partial_\phi\omega_0 \rangle \sqrt{\frac{\left <w\right > \left(2 \xi_{3/2} -\langle z \rangle \right)}{\xi_{3/2}\left(\langle \mathfrak{d}_z^2 w \rangle  + (6 \xi_{3/2}^2 - 4) \left<w\right >\right)}}  \frac{M_4}{m_{p}}\nonumber \\
&&+ \frac12 \frac{\langle \partial_z\partial_\phi\omega_0 \rangle^2}{\langle \mathfrak{d}_z^2 w \rangle  + (6 \xi_{3/2}^2 - 4)\left <w\right >} + {\cal O}(\langle \partial_z\partial_\phi\omega_0 \rangle^3), \label{eq:rholl1mp}
\eeqa
and
\beqa
\left.\left<\partial_\phi^2 \omega_0 \right>_\mp\right|_{\rho \sim \mathcal{O}(1)}&=& \frac{\left <w\right >}{\xi_{3/2}} \left(2 \xi_{3/2}-\left < z \right > \right) \frac{M_4^2}{m_{p}^2} 
\mp \langle \partial_z\partial_\phi\omega_0 \rangle \sqrt{\frac{\left <w\right > \left(2 \xi_{3/2} -\langle z \rangle \right)}{\xi_{3/2} \langle \mathfrak{d}_z^2 w \rangle  }}  \frac{M_4}{m_{p}}\nonumber \\
&&+ \frac12 \frac{\langle \partial_z\partial_\phi\omega_0 \rangle^2}{\langle \mathfrak{d}_z^2 w \rangle} + {\cal O}(\langle w \rangle^3, (\rho-1)^3),
\label{eq:rhoO1mp}
\eeqa
The two leading orders reproduce the result of \autoref{eq:d2omega0pmlim} in the case $\lag \partial_z\partial_\phi\omega_0\rag=0$, meaning that the loop effects can be sizable. Note also that the expansion \autoref{eq:rholl1mp} is not
valid in the limit $\left <w\right > \to 0$.\footnote{In fact, divergences in the limit $\left <w\right > \to 0$ appears in the terms of order $\langle \partial_z\partial_\phi\omega_0 \rangle^3$ and higher.} A difference between $\mathbf{\rho^{-1} \ll 1}$ and the solutions $\mathbf{\rho \ll 1}$ and $\mathbf{\rho \sim \mathcal{O}(1)}$ can be noted. While \autoref{eq:rhoinvll1m} and  \autoref{eq:rhoinvll1p}) are valid for $\langle w \rangle \ll 1$, the relations \ref{eq:rholl1mp} and \ref{eq:rhoO1mp}) impose the condition $|\left<\partial_z\partial_\phi\omega_0 \right >| \ll \frac{M_4}{m_{p}}$. This is due to the decreasing powers of $\frac{M_4}{m_{p}}$ in the terms with increasing powers of $\left<\partial_z\partial_\phi\omega_0 \right >$ as can be seen from \autoref{eq:rholl1mp} and \autoref{eq:rhoO1mp}.\medskip

\begin{figure}[H]
    \centering
      \includegraphics[width=1.\linewidth]{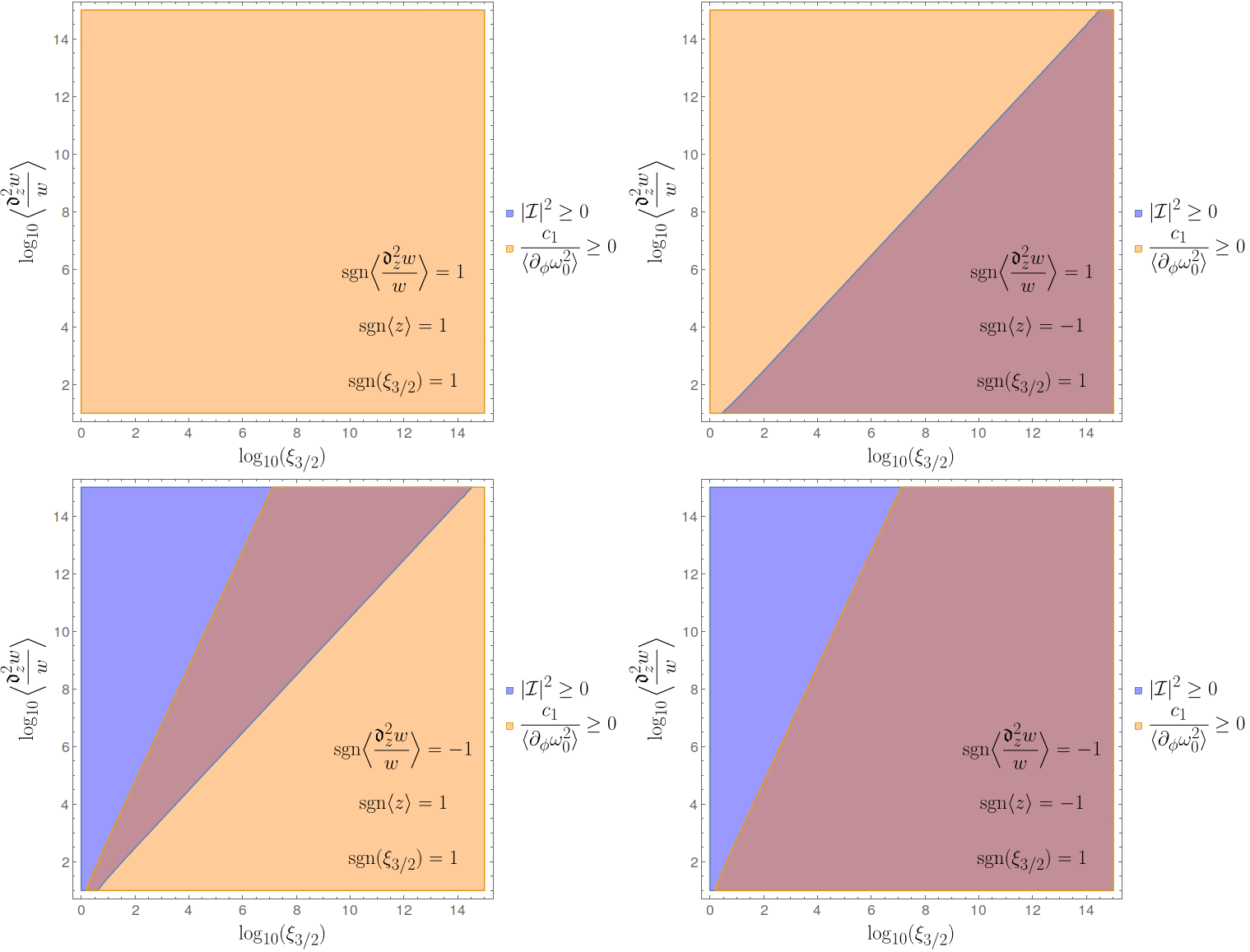}
   \caption{Allowed region by the constraints \autoref{eq:poscondprime} and \autoref{eq:othercons} as function of $\log_{10}(\xi_{3/2})$ and $\log_{10}(\frac{\langle \mathfrak{d}_z^2 w \rangle}{\langle w \rangle})$. Upper-left: the inequality $|{\cal I}|^2 \geq 0$ is not satisfied in all the plane. The red color is the superposition of the two regimes.}
   \label{fig:cons}
\end{figure}

In both cases ($\mathbf{\rho \ll 1}$ and $\mathbf{\rho \sim \mathcal{O}(1)}$), the negative term in \autoref{eq:poscond} is now of order $\frac{M_4^2}{m_{p}^2}$ and now function of $\langle w \rangle$ as:
\beqa
 |{\cal I}|^2 &=&  \frac{M_4^6}{m_{p}^2} \langle w \rangle^2 \left(\frac{ (3 \xi_{3/2}^2 -1)}{\xi_{3/2}^2}- \frac{1}{\xi_{3/2} (2 \xi_{3/2}- \langle z \rangle)}\left( 2 (3 \xi_{3/2}^2 -2) + \frac{\langle \mathfrak{d}_z^2 w \rangle}{\langle w \rangle} \right) \right)
 \geq 0\ . \label{eq:poscondprime}
\eeqa
We also have
\beq
\frac{1}{\xi_{3/2} (2 \xi_{3/2}- \langle z \rangle)}\left( 2 (3 \xi_{3/2}^2 -2) + \frac{\langle \mathfrak{d}_z^2 w \rangle}{\langle w \rangle} \right)
 \geq 0\ .\label{eq:othercons}
\eeq
The last condition is essentially always satisfied if $\lag \mathfrak{d}_z^2 w / w \rag > 4$ and the fact that  $\sgn(c_1 c_3)<0$ (see the discussion in \ref{subsec:set_cons}). However, $|{\cal I}|^2$ is not necessarily positive. By developing the relation \ref{eq:poscondprime}, we note that some the parameters-space zone are forbidden. For example, taking $\xi_{3/2}$, $\lag z\rag$ and $\frac{\langle \mathfrak{d}_z^2 w \rangle}{\langle w \rangle}$ positive automatically lead to $|{\cal I}|^2 <0$. Numerical investigations reveal four different regions described by $\text{sgn}(\frac{\langle \mathfrak{d}_z^2 w \rangle}{\langle w \rangle})=\pm 1$ and $\text{sgn}(z\xi_{3/2})=\pm 1$ (\textit{i.e.}, all the regions with the sign of $\xi_{3/2}$ and $\lag z\rag$ swapped are equivalent). The different regions are shown in \autoref{fig:cons}.

The regimes $\rho \sim \mathcal{O}(1)$ and $\rho \ll 1$ seem to be more interesting for high corrections from the $m_{\zeta '}^2\sin^2\theta$ contribution. However, since the prefactor in \autoref{eq:hardcorrection1} and \autoref{eq:hardcorrection2} can be sizable, the regime $\rho\gg 1$ will also be considered in our numerical analysis. Note that for simplicity, we denote the contribution $\phi^{\dagger}\phi S^{\dagger}S$ as \textit{purely} hard and the $\mathcal{U}S^{\dagger}H_U\cdot H_D$  as the \textit{superpotential} contribution.

\subsection{Numerical computation of the order of magnitude of the corrections}\label{eq:NumComput}
For our numerical study, we assume that the \textit{v.e.v} of the field from the hidden sector is $|\lag z\rag| = 1$ as well as the parameter $|\lag\mathfrak{d}_z^2 w \rag| = 1$ (see \autoref{eq:OoM-d2w}). We also set $M_4=10^{15}\ \mathrm{GeV}$ and $m_p=2\times 10^{18}\ \mathrm{GeV}$. The analysis is done for several values of the parameter $\rho$:
\begin{gather}
\{\rho\} = \{0, 10^{-2},-10^{-2},1,-1,10^2,-10^2\}\nn
\end{gather} 
and by varying the gravitino mass in the range $[10^{-3}\ \mathrm{GeV},10^{3}\ \mathrm{GeV} ]$. All the possibles configurations for the signs of the parameters are investigated. All solutions regarding the choice of $\lag \partial_\phi\omega_0\rag_{\pm}$ and  $\lag \partial_\phi^2\omega_0\rag_{\pm}$ are also taken into account. The complete contribution of the two terms \autoref{eq:hardcorrection1} and \autoref{eq:hardcorrection2} on the Higgs boson mass is written as:
\begin{gather}
m_h^{(1L)} = \sqrt{m_h^2{}^{(TL)} + \frac{1}{16\pi^2m_p^2}e^{|\lag z\rag|^2}\Big( m_{\zeta '}^2\sin^2\theta +  m_{S '}^2\cos^2\theta  \Big)\Big(\sqrt{|{\cal I}|^2} + \frac{1}{m_p^2}\big(4|\xi_{3/2}|^2 - 2\big)|{\cal I}|^2\Big)} \label{eq:mh1L}
\end{gather}
with $m_h^{(TL)}$, the Higgs boson mass containing the tree-level and one-loop contributions coming from the soft-breaking terms, assumed to be $115\ \mathrm{GeV}$. In a first analysis, even though the correction \autoref{eq:hardcorrection2} corresponds to the off-diagonal matrix element, we assume that this term directly contributes to the Higgs boson mass without first diagonalising the matrix. This simplification is done considering the known effect for singlet-extension of the MSSM (see for example \autoref{eq:mlimitNMSSM}). A more proper method will be to diagonalise the corrected mass matrix and compute the eigenvalues of the matrix. \medskip

We first make a remark on the sign of the parameters. Analysing all the previous equations, we see that switching the sign of all the parameters 
\begin{gather}
\{p_i\}=\{\xi_{3/2},\lag z\rag,\lag\mathfrak{d}_z^2 w \rag,m_{3/2},\rho\} \rightarrow \{-p_i\}=\{-\xi_{3/2},-\lag z\rag,-\lag\mathfrak{d}_z^2 w \rag,-m_{3/2},-\rho\}\nn
\end{gather}
leads to equivalent results. This reduces the number of configurations to investigate.\medskip

We consider in a first step the case when $\rho=0$. In this case, we recall that the two solutions $\lag \partial_\phi^2\omega_0\rag_\pm$ are degenerated. Also, the constraint \autoref{eq:poscondc2} is not appropriate since $c_2=0$. However, many configurations lead to no possible solutions in the range of values defined above. All the space is generally forbidden by the other constraints in Section \ref{subsec:set_cons}. The positivity of the lighter state $m_{S '}^2$ seems to be the most restrictive condition. Only a few configurations are possible which can be separated into two different cases.

\begin{figure}[H]
\centering
\begin{tabular}{cc}
\subcaptionbox{Evolution of $\log_{10}\big(|{\cal I}|^2\big)$\label{1}}{\includegraphics[width=.47\linewidth]{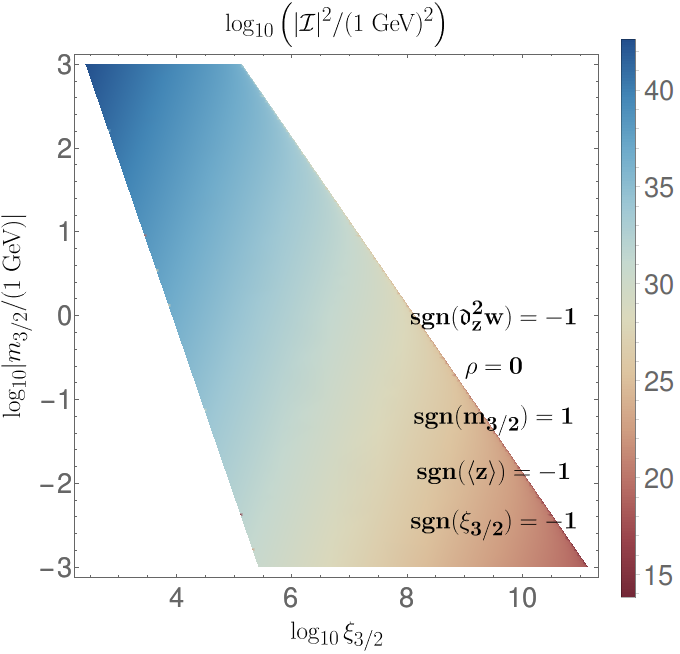}} &
\subcaptionbox{Evolution of $\log_{10}\big(m_{S '}^2\big)$\label{2}}{\includegraphics[width=.5\linewidth]{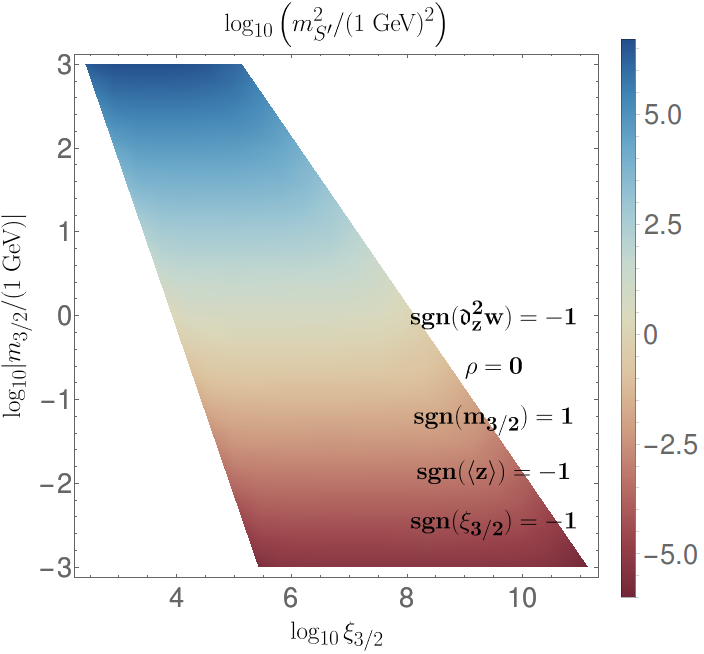}} \\
\subcaptionbox{Evolution of $\big(m_{\zeta '}^2-m_{\zeta_0 '}^2\big)/m_{\zeta_0 '}^2$\label{3} with $m_{\zeta_0 '}^2=10^{17.23017}\text{GeV}^2$}{\includegraphics[width=.5\linewidth]{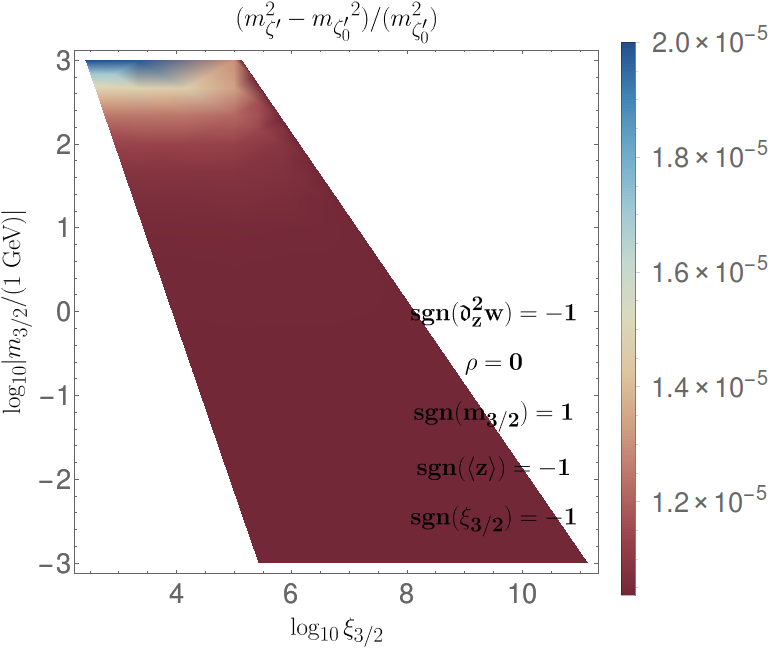}} &
\subcaptionbox{Evolution of $\log_{10}\big(\sin^2\theta\big)$\label{4}}{\includegraphics[width=.47\linewidth]{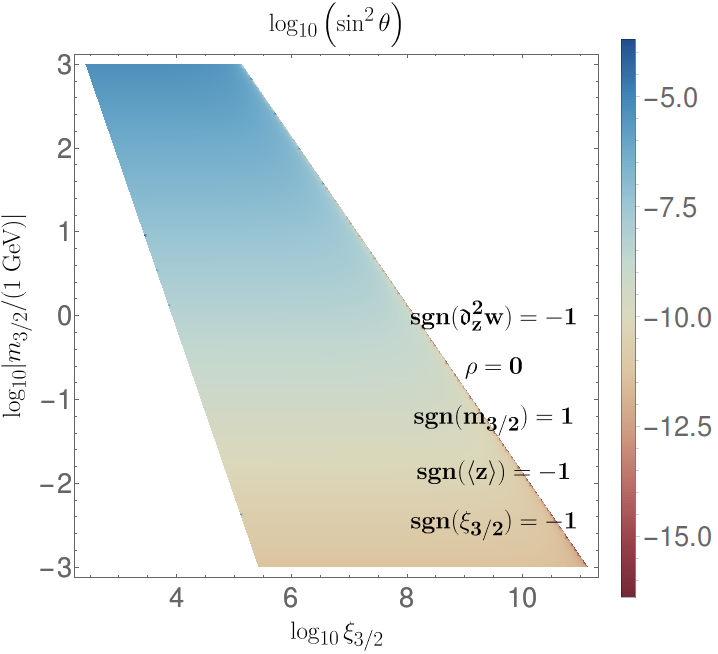}}\\
\end{tabular}
\caption{Evolution of several parameters as function of $\log_{10}\big(|m_{3/2}/(1\ \mathrm{GeV})|\big)$ and $\log_{10}\big(|\xi_{3/2}|\big)$ for $\mathrm{sgn}(\lag\mathfrak{d}_z^2 w \rag)=\mathrm{sgn}(\xi_{3/2})=\lag z \rag = -1$, $\rho=0$ and $\mathrm{sgn}(m_{3/2})=1$.}
\label{fig:rho0}
\end{figure}

\begin{figure}[htb!]
    \centering
\begin{tabular}{cc}
\subcaptionbox{\label{1mh1}}{\includegraphics[width=.5\linewidth]{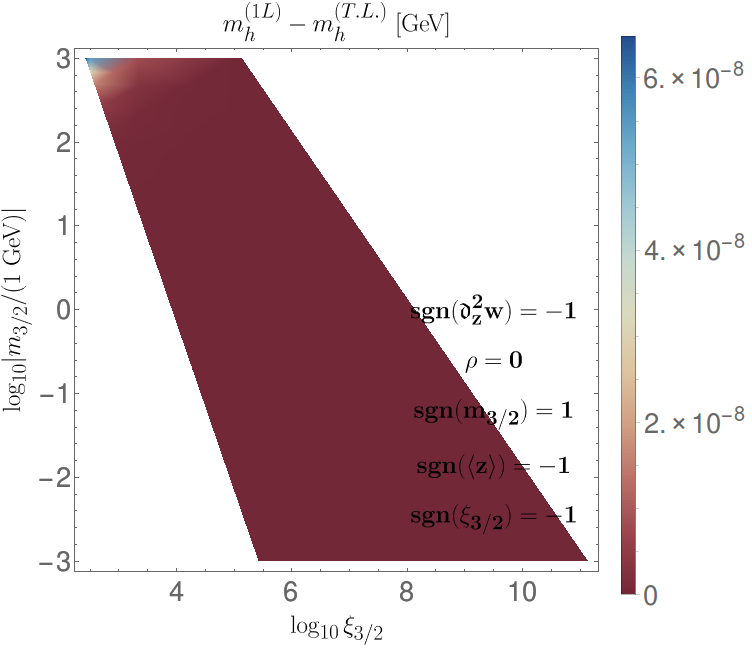}} &
\subcaptionbox{\label{2mh1}}{\includegraphics[width=.465\linewidth]{plot6.png}} 
\end{tabular}
\caption{Evolution of $m_h^{(1L)}$ as function of $\log_{10}\big(|m_{3/2}/(1\ \mathrm{GeV})|\big)$ and $\log_{10}\big(|\xi_{3/2}|\big)$ for two configurations.}
\label{fig:rho0mh1L}
\end{figure}

The configuration $\mathrm{sgn}(\lag z \rag)=\mathrm{sgn}(\lag\mathfrak{d}_z^2 w \rag)=\mathrm{sgn}(\xi_{3/2})=-1$ and $\mathrm{sgn}(m_{3/2})=1$ (see \autoref{fig:rho0} and \autoref{fig:rho0mh1L}) is quite representative to the first part of the solutions. The light state \autoref{2} has a mass between the TeV scale and the MeV while the heavier one \autoref{3} remain at $m_{\zeta '}^2 \sim 10^{17}\ \mathrm{GeV}^2$. The computation of the mixing angle \autoref{4} shows us that the light state $S'$ is mainly singlet. Finally, the magnitude of $|{\cal I}|^2$ in \autoref{1} allows to determinine the order of magnitude of each contribution in \autoref{eq:mh1L}. Taking the highest possible value of $|{\cal I}|^2\approx 10^{40}\ \mathrm{GeV}^2$, the factor $\sqrt{|{\cal I}|^2}/m_p^2$ coming from the contribution \autoref{eq:hardcorrection2} is approximately $10^{-16}$ while the second contribution gives $|{\cal I}\xi_{3/2}|^2/m_p^4\approx 10^{-32}$. Taking the highest value of $\xi_{3/2}$ to boost the second contribution, we obtain respectively for the two factors $10^{-26}$ and $10^{-22}$. Even though the purely hard contribution is suppressed in $m_p^4$, high values of $\xi_{3/2}$ can then push this loop effect and exceed the \textit{superpotential} contribution. However, as we can see in all the figures, increasing the values of $\xi_{3/2}$ must be coordinated with a reduction of the gravitino mass to pass the constraints. This leads to lower the mass $m_{S '}^2$ and the parameters $\sin^2\theta$ and $|{\cal I}|^2$ which reduce the loop effects. Moreover, taking too high values of $\xi_{3/2}$ is non-natural and generate a fine-tuning. The one-loop Higgs Boson mass is shown in Figure \ref{1mh1}. As we can see from this evolution and the previous discussion, the effects are mostly nonexistent. Finally, note that the restriction for $\log_{10}|\xi_{3/2}|<6$ is coming from the negativity of the squared mass $m_{S '}^2$ where the constraint for high $\xi_{3/2}$ values corresponds to both $|{\cal I}|^2 < 0$ and $m_{S '}^2<0$.\medskip

\begin{figure}[h!]
\centering
\begin{tabular}{cc}
\subcaptionbox{Evolution of $\log_{10}\big(|{\cal I}|^2\big)$\label{11}}{\includegraphics[width=.45\linewidth]{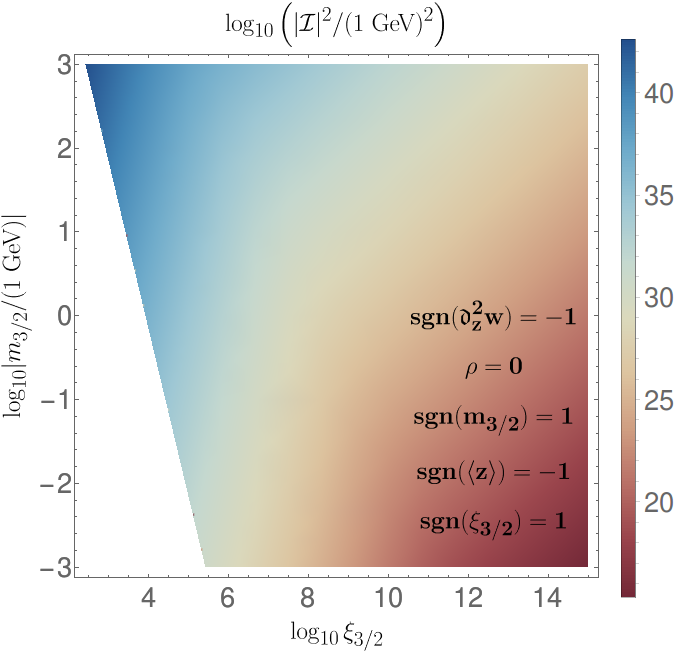}} &
\subcaptionbox{Evolution of $\log_{10}\big(m_{S '}^2\big)$\label{21}}{\includegraphics[width=.45\linewidth]{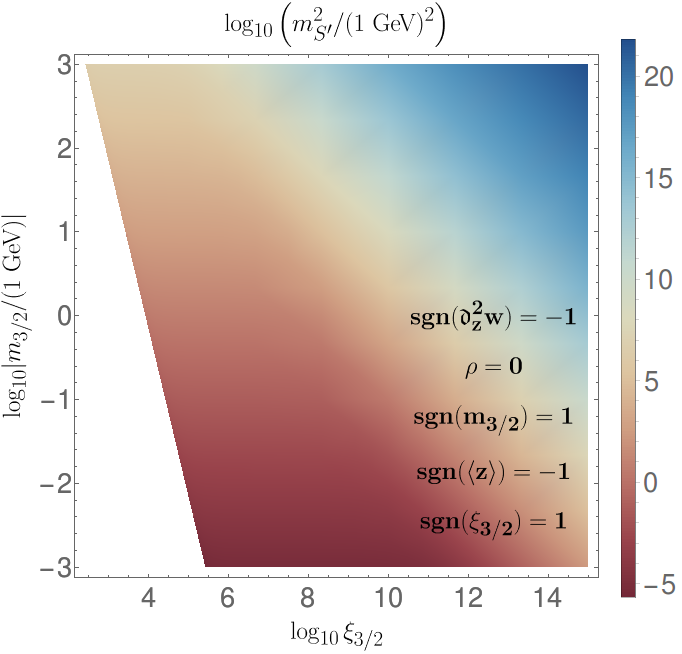}} \\
\subcaptionbox{Evolution of $\log_{10}\big(m_{\zeta '}^2\big)$\label{31}}{\includegraphics[width=.45\linewidth]{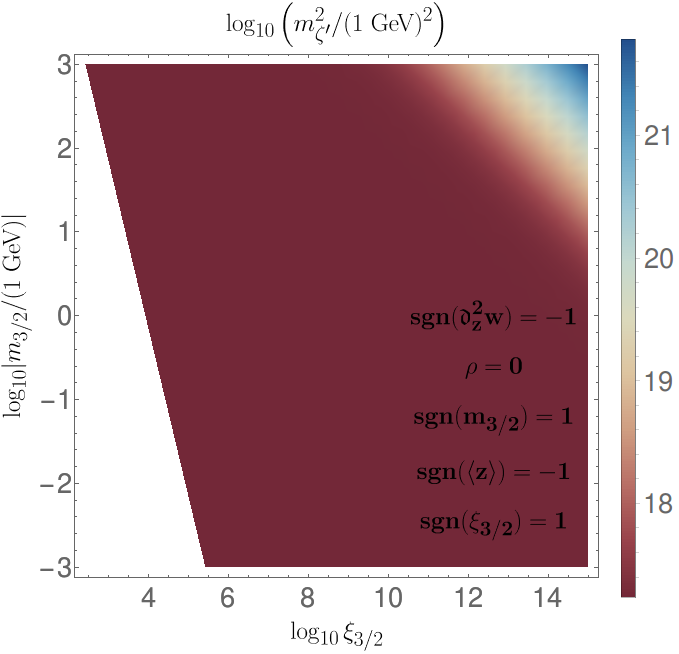}} &
\subcaptionbox{Evolution of $\log_{10}\big(\sin^2\theta\big)$\label{41}}{\includegraphics[width=.45\linewidth]{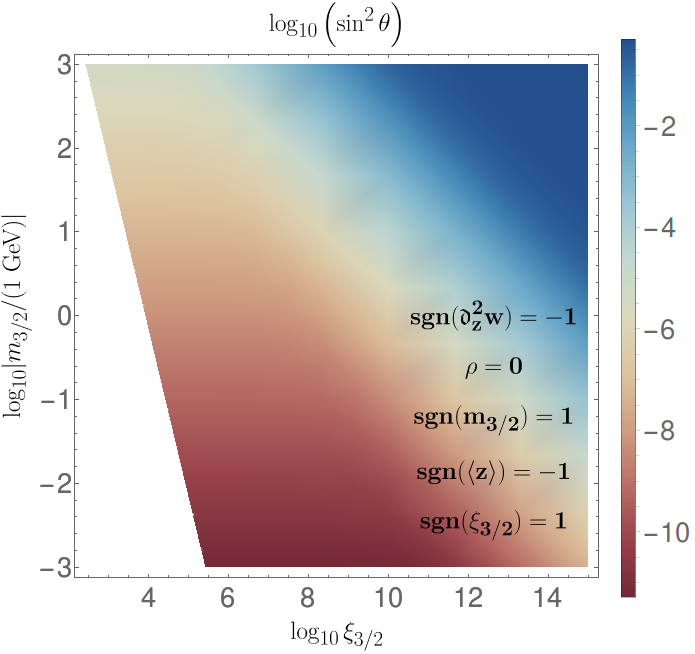}}\\
\end{tabular}
\caption{Evolution of several parameters as function of $\log_{10}\big(|m_{3/2}/(1\ \mathrm{GeV})|\big)$ and $\log_{10}\big(|\xi_{3/2}|\big)$ for $\mathrm{sgn}(\lag\mathfrak{d}_z^2w\rag)=-1$, $\rho=0$, $\mathrm{sgn}(m_{3/2})=1$, $\mathrm{sgn}(\lag z \rag)=-1$ and $\mathrm{sgn}(\xi_{3/2})=1$.}
\label{fig:1}
\end{figure}

A second type of solution has been found. A representative configuration is $\mathrm{sgn}(\lag \mathfrak{d}_z^2 w \rag)=\mathrm{sgn}(\lag z \rag)=-1$ and $\mathrm{sgn}(\xi_{3/2})=\mathrm{sgn}( m_{3/2} )=1$ (see \autoref{2mh1} and \autoref{fig:1}). High values of $\xi_{3/2}$ are not forbidden for fixed $m_{3/2}$ compared to the other solutions. The region with low $\xi_{3/2}$ values is constrained by the negativity of the lightest state's squared mass. The evolution of the various parameters regarding $m_{3/2}$ and $\xi_{3/2}$ are also quite different. While the mass of the heaviest state remains unchanged in the $(\log_{10}(\xi_{3/2}),\log_{10}(m_{3/2}/(1\ \mathrm{GeV}))$-plane in the previous solution, high $\xi_{3/2}$ values lead to quasi non-degenerate states $m_{S '}^2 \approx m_{\zeta '}^2$. The mixing angle is also pushed to $\sin^2\theta\approx \frac12$ which maximise the mixing between the two states. The full one-loop Higgs boson mass is shown in \autoref{2mh1}. Due to the possibility of taking high $\xi_{3/2}$ values, the correction coming from the pure hard term increases intensively, leading to a really high Higgs mass, meaning that possible solutions exist for obtaining $m_h\approx 125\ \mathrm{GeV}$. Nonetheless, assuming $m_{3/2}=1\ \mathrm{TeV}$ as the maximum value for the gravitino mass, those effects only appear for $\xi_{3/2}>10^{14}$ which is non-natural. \medskip

\begin{figure}[htb!]
    \centering
      \includegraphics[width=.6\linewidth]{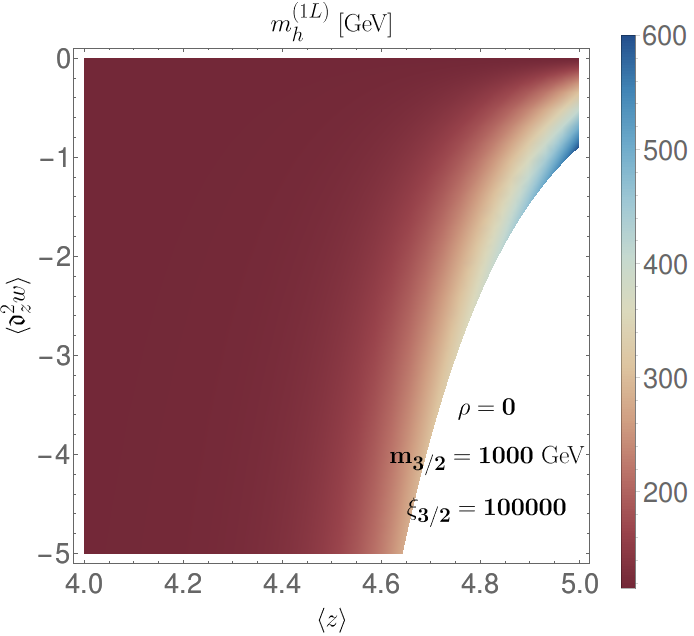}
\caption{Evolution of $m_h^{(1L)}$ as function of $\lag\mathfrak{d}_z^2 w  \rag$ and $\lag z\rag$ for $\rho=0$, $\xi_{3/2}=10^5$ and $m_{3/2}=10^3\ \mathrm{GeV}$. The white region is forbidden due to the restriction $m_{S '}^2> 0$.}
\label{fig:mh1varz}
\end{figure}

We can study the variation of the full correction depending on the possible values of $\lag z \rag$ and $\lag \mathfrak{d}_z^2 w \rag$. We take the gravitino mass as $m_{3/2}=1\ \mathrm{TeV}$ to increase the loop contributions and $\xi_{3/2} = 10^5$ to pass all the constraints (following results from \autoref{1mh1}). The variation of $m_h^{(1L)}$ considering the same configuration as \autoref{fig:rho0} for the inputs is shown in \autoref{fig:mh1varz}. We notice that the loop corrections still remain low for values of $|\lag z \rag|$ close to $1$. However, the mass increase exponentially when $\lag z \rag > 4$, leading to possible configurations with lower value of $\xi_{3/2}$ as in the case of \autoref{2mh1}. This is simply due to the presence of the \textit{v.e.vs} from the hidden sector in the exponential in \autoref{eq:hardcorrection1} and \autoref{eq:hardcorrection2}, and more generally in the scalar potential. This value of $\lag z\rag$ can be seen as non-natural, but may be legitimate for more complex models (see Subsection \ref{subsec:morecomplex}).\medskip

We now analyse the general case $\rho\neq 0$. The two solutions $\lag \partial_\phi^2\omega_0\rag_{\pm}$ correspond in this case to two different results. The choice of $\lag \partial_\phi\omega_0 \rag_{+}$ or $\lag \partial_\phi\omega_0 \rag_{-}$ leads to the same numerical results (since this parameter always appears as $\lag\partial_\phi\omega_0\rag^2$), but to different constraints following \autoref{eq:poscondc2}. Note that the degeneracy of the solutions $\lag \partial_\phi^2\omega_0\rag_{\pm}$ has been obtained for $\rho=0$ but can also be obtained with the more general assumption $c_2\rightarrow 0$ (with $c_2$ in \autoref{eq:c2}) which is also true for $\xi_{3/2}\rightarrow \infty$, \textit{i.e.}, the two solutions $\lag \partial_\phi^2\omega_0\rag_{\pm}$ are equivalent for high $\xi_{3/2}$ values. \medskip

\begin{figure}[H]
\centering
\begin{tabular}{cc}
\subcaptionbox{\label{6}}{\includegraphics[width=.45\linewidth]{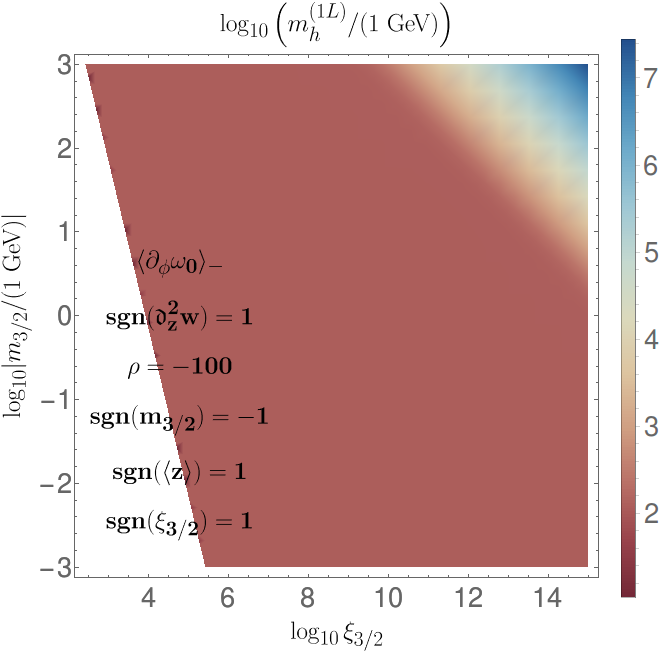}} & 
\subcaptionbox{\label{7}}{\includegraphics[width=.45\linewidth]{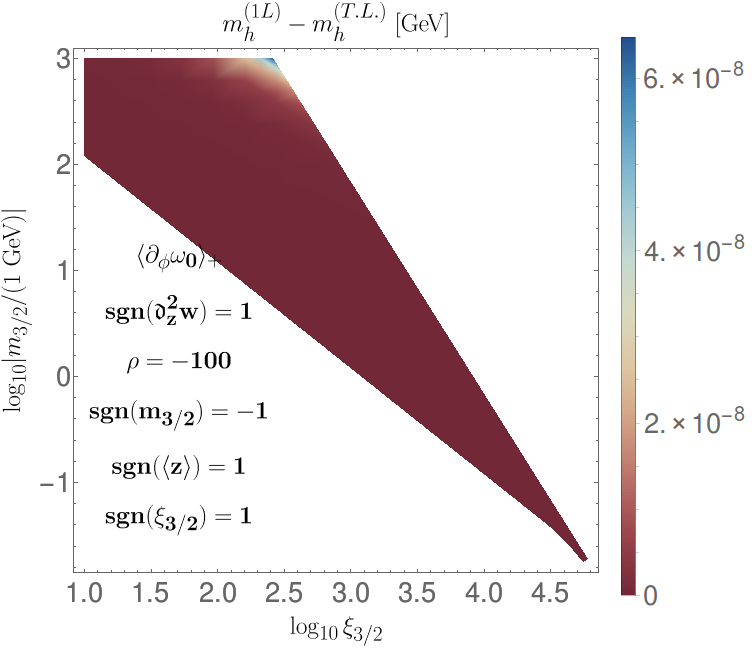}} \\ 
\subcaptionbox{\label{8}}{\includegraphics[width=.45\linewidth]{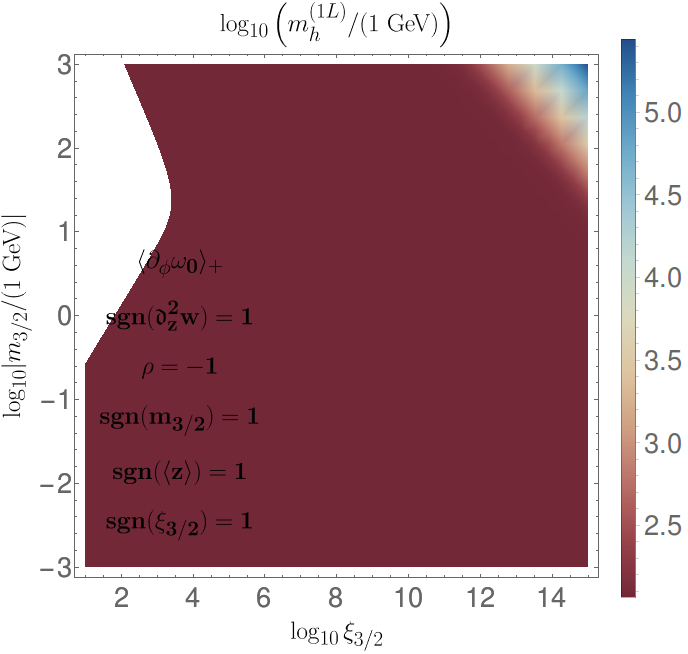}} & 
\subcaptionbox{\label{9}}{\includegraphics[width=.45\linewidth]{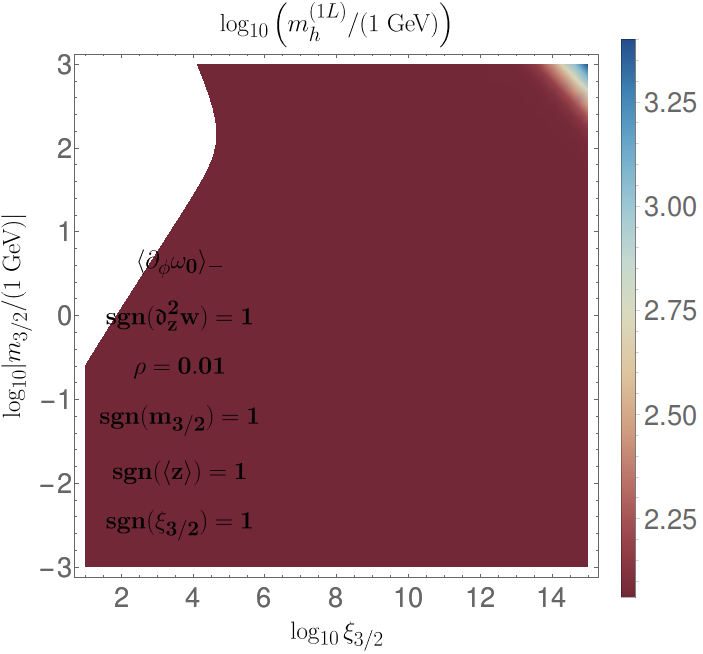}}\\ 
\end{tabular}
\caption{Evolution of the one-loop corrected Higgs boson mass as function of $\log_{10}\big(|m_{3/2}/(1\ \mathrm{GeV})|\big)$ and $\log_{10}\big(|\xi_{3/2}|\big)$ for several configurations with $\rho\neq 0$ and using the solution $\lag \partial_\phi^2\omega_0\rag_{+}$.}
\label{fig:99}
\end{figure}

\begin{figure}[h]
\centering
\begin{tabular}{cc}
\subcaptionbox{Evolution of $\log_{10}\big(|{\cal I}|^2\big)$}{\includegraphics[width=.45\linewidth]{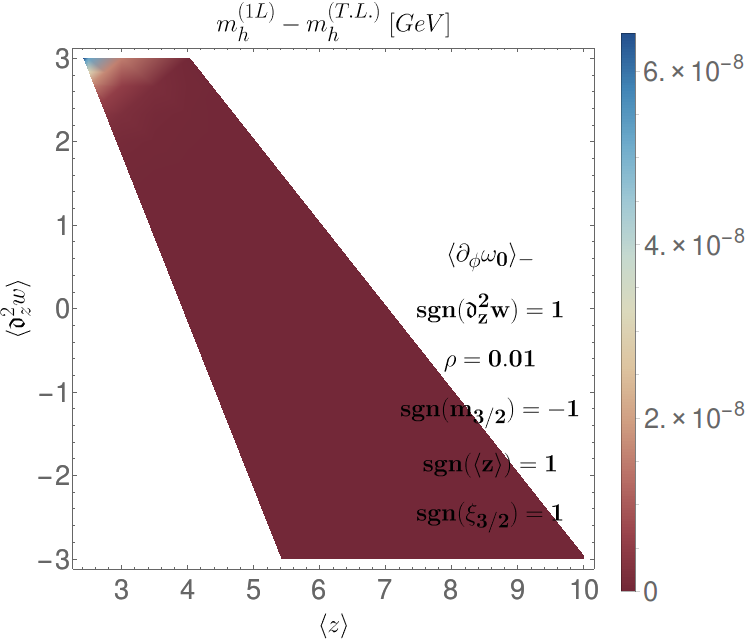}} &
\subcaptionbox{Evolution of $\log_{10}\big(m_{S '}^2\big)$}{\includegraphics[width=.45\linewidth]{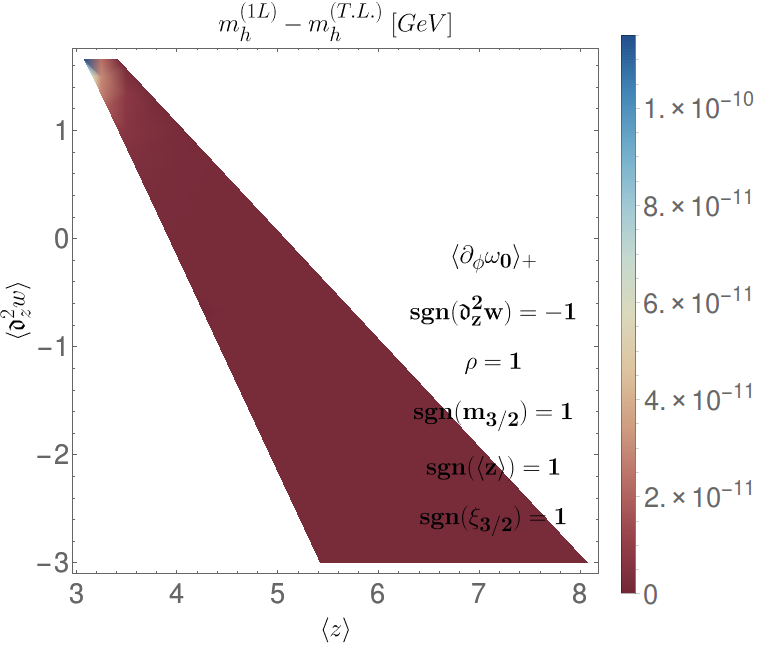}} \\
\end{tabular}
\caption{Evolution of the one-loop corrected Higgs boson mass as function of $\log_{10}\big(|m_{3/2}/(1\ \mathrm{GeV})|\big)$ and $\log_{10}\big(|\xi_{3/2}|\big)$ using $\lag \partial_\phi^2\omega_0\rag_{-}$ with $\rho\neq 0$.}
\label{fig:101}
\end{figure}

For all the values of $\rho=\{\pm 10^2,\pm 1,\pm 10^{-2}\}$, we have found the same types of evolution than for the case $\rho=0$. A lot of configurations lead to inconsistency with the constraints in Subsection \ref{subsec:set_cons}. The distributions for several configurations representative to the general results can be found in \autoref{fig:99} for the solution $\lag \partial_\phi^2\omega_0\rag_{+}$ and in \autoref{fig:101} for $\lag \partial_\phi^2\omega_0\rag_{-}$.\medskip

First considering the $\lag \partial_\phi^2\omega_0\rag_{+}$ solutions. As already said in our first qualitative analysis (see \autoref{eq:regimes}), the solutions for $\rho=\pm 1$ and $\pm 10^{-2}$ must be quite equivalent to the case $\rho=0$. The numerical results are quite in accordance with those statements. Only some differences appear at low value of $\log_{10}\xi_{3/2}$. Possible configurations exists only for high values of $\xi_{3/2}$. Those effects results from the same mechanisms as \autoref{fig:1} and so are non-natural. \nl Even though our qualitative analysis for $\rho \geq 1$ with $\lag \partial_\phi^2\omega_0\rag_{+}$ have pointed out possible difficulties for high corrections, we can numerically find possible configurations for high $\xi_{3/2}$ values (see \autoref{6} or \autoref{7}). However, those configurations are either highly fine-tuned or do not push the Higgs boson mass to $125\ \mathrm{GeV}$. 

%

We consider now the solution $\lag \partial_\phi^2\omega_0\rag_{-}$. Our analytic investigation in Subsection \autoref{eq:regimes} showed a high fine-tuning on the gravitino mass for high values of the parameter $\rho$. The same result has been obtained in the numerical investigation. Indeed, no configurations for $\rho=\pm 10^2$ have passed the imposed restrictions. In the case of $\rho = \{\pm 1, \pm 10^{-2}\}$, for a specific choice of $\mathrm{sgn}(\rho)$ and $\lag \partial_\phi\omega_0\rag_{\pm}$, only a quarter of the configurations gives possible results. When a solution with $\lag \partial_\phi\omega_0\rag_{+}$ pass the constraints, the equivalent configuration with $\lag \partial_\phi\omega_0\rag_{-}$ is completely forbidden (and \textit{vice versa}). Possible solutions are presented in \autoref{fig:101}. In the same manner as before, no interesting zone of the parameters-space has been discovered to increase the Higgs boson mass. The evolution of $|{\cal I}|^2$, $\sin^2\theta$ and $m_{\zeta '}^2$ are equivalent to the case $\rho=0$ with \autoref{fig:rho0}. The only difference comes from the lightest state $m_{S '}^2$ which remains higher than the $\mathrm{GeV}$ scale. \medskip

\begin{figure}[H]
    \centering
      \includegraphics[width=.6\linewidth]{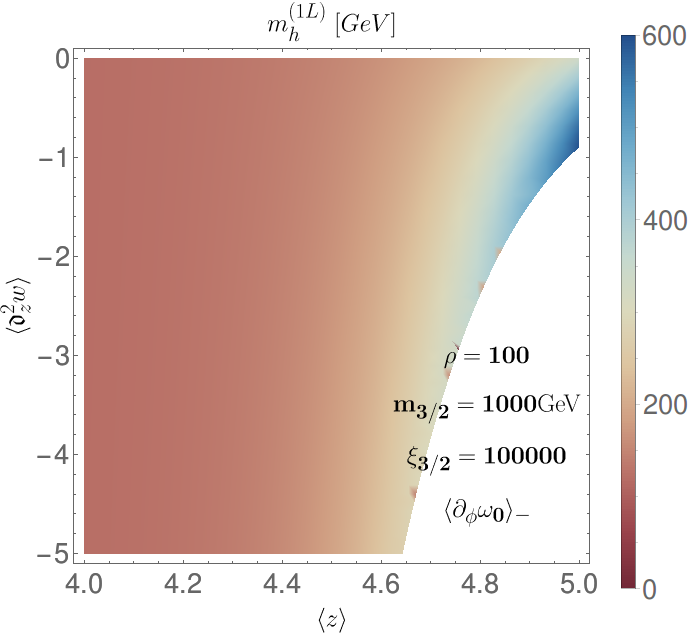}
\caption{Evolution of $m_h^{(1L)}$ as function of $\lag\mathfrak{d}_z^2 w \rag$ and $\lag z\rag$ for $\rho=100$, $\xi_{3/2}=10^5$ and $m_{3/2}=10^3\ \mathrm{GeV}$. The white region is forbidden due to the restriction $m_{S '}^2> 0$. This solution is obtained with $\lag\partial_\phi^2\omega_0\rag_{+}$ and $\lag\partial_\phi\omega_0\rag_{-}$.}
\label{fig:mh1varz2}
\end{figure}

Considering now possible variations on the parameters $\lag z \rag$ and $\lag\mathfrak{d}_z^2 w \rag$, we also remark possible benchmarks which conduct to $m_h^{(1L)}\approx 125\ \mathrm{GeV}$ for $\lag z\rag \approx 4$ (see \autoref{fig:mh1varz2}). These solutions will be discussed in the next section \ref{subsec:morecomplex}.

\section{Towards more complex models}\label{subsec:morecomplex}
As seen in the previous section, two possible configurations have been found for a natural a Higgs boson mass close to $125\ \mathrm{GeV}$: using high values of $\xi_{3/2}$ to increase the \textit{purely hard} contribution or assuming $\lag z \rag\approx 4$. However, those configurations are non-natural and lead to a fine-tuning problem. Nonetheless, the analysis presented in this manuscript is based on a simplified version of the S2MSSM: 
\begin{itemize}
\item the Kähler potential is assumed to be canonical, \textit{i.e.}, $\Lambda^{a^\ast}{}_a(z,z^\dag) = \delta^{a^\ast}{}_a$;
\item the superpotential $W_0$ is not function of the hybrid fields, \textit{i.e.}, $\partial_p\omega_0=0$;
\item the \textit{v.e.vs} from the visible sector and the hybrid are assumed to be $\lag S^p\rag = \lag \Phi^a\rag= 0$;
\item the hidden sector contains only one field $z$. 
\end{itemize}
Analysing more complex models may lead to new effects which could boost or reduce the one-loop corrections.\medskip

Firstly, numerical investigation from Subsection \ref{eq:NumComput} has shown zones of the parameters space with $\lag z\rag \approx 4-5$ leading to a one-loop Higgs boson mass of $m_h^{(1L)}\approx 125\ \mathrm{GeV}$. This value for $\lag z \rag$ may be considered as non-natural for our model since it must be $\lag z \rag \sim \mathcal{O}(1)$ by definition. However, since the high correction is obtained through the exponential $e^{|\lag z \rag|^2}$, then assuming $\lag z\rag \approx 4$ seems equivalent to several fields from the hidden sector $z^i$ with $\lag z^i \rag\sim \mathcal{O}(1)$ contributing as $\mathrm{exp}\big(\displaystyle\sum|\lag z^i\rag|^2\big)$. Obviously, the full calculation must be done considering several fields from the hidden sector to consider those effects properly. Nonetheless, it can be a hint to recover a Higgs boson mass near $125\ \mathrm{GeV}$ naturally. A simple model for several hidden sector fields could be to consider a simple sum in the superpotential, such as:
\begin{gather}
W(\{z\},\{S^p\},\{\tilde{\Phi}^a\}) = \displaystyle\sum_i W^{(i)}(z^i,\{S^p\},\{\tilde{\Phi}^a\})\ , \nn\\
K( \{z\},\{S^p\},\{\tilde{\Phi}^a\}) = \sum\limits_{i}z^iz^\dag_i + S^pS^\dag_p + M_4^2\phi^a\phi^\dag_a \nn
\end{gather} 
where the term $W^{(i)}$ is only function of the field $z^i$. In this context, the \textit{pure hard} and the \textit{superpotential} contributions extend to:
\begin{gather}
\frac{1}{m_p^4}e^{\sum\limits_i |\lag z^i\rag|^2}\big( M_4^2\phi \phi^{\dag} + S^pS^\dag_p\big)\big( 4\sum\limits_{i}|\xi_{3/2}{}_i|^2 -2 \big){\cal I}_r\bar{\cal I}^tS^rS^\dag_t\ ,\nn
\end{gather}
and
\begin{gather}
\frac{2}{m_p^2}e^{\sum\limits_i |\lag z^i\rag|^2}\bar{\cal I}^qS^{\dagger}_q\mathcal{U}H_U\cdot H_D\ .\nn
\end{gather}\medskip

Note also that we have not considered loop contributions involving the hidden field $z$. Such contributions are present in the potential where we assume a perturbation on the field from the hidden sector:
\begin{gather}
z \rightarrow \lag z \rag + z \ . \nn
\end{gather}
Those corrections must be computed for a complete analysis. \medskip

Going beyond the assumptions $\partial_p\omega_0 = 0$ and $\lag S^p\rag=0$ will potentially modify the mass matrix structure \autoref{eq:matrixprimee}. Indeed, this structure is obtained with the help of the proportionality relation between $\mathfrak{d}_z{\cal I}_p$ and ${\cal I}_p$ in \autoref{eq:dIp}. Relaxing one of those two assumptions suppressed this proportionality, making the analysis more complex. Moreover, due to the complexity of the potential, the minimisation of the potential must be solved for $\lag \phi^a\rag$ and $\lag S^p\rag$. \medskip

For this analysis, we have assumed a Higgs boson mass of $m_h^{(T.L.)}=115\ \text{GeV}$. This value contains the tree-level contributions and the usual loop corrections coming from the soft-breaking terms in the MSSM. Nonetheless, as already mentioned in Subsection \ref{sec:corrHiggsSector}, the presence of the two singlets can also push the tree-level Higgs boson mass following several mechanisms. In this manner, a complete calculation of the Higgs boson mass containing loop-corrections form soft-breaking terms can be higher than $115\ \text{GeV}$ (see for instance the \autoref{fig:l_mh2}, for the N2MSSM). In this way, the loop effects coming from the specific form of the new solutions may not need to be relatively high.\medskip

Finally, a complete analysis must be done by calculating and diagonalising the full one-loop corrected mass matrix in the basis $\{z^i,\tilde{\Phi}^a,S^p\}$.
\label{nsw}

\chapter{A two-singlets-extension of the MSSM: the N2MSSM}\label{chap:N2MSSM}
In the first chapter of this manuscript, we have shown that supersymmetry can be constructed as a low energy limit of supergravity (see Subsection \ref{subsec:susybreak}). By defining an appropriate superfield content (which are representations of the gauge group $G=SU(3)_C\times SU(2)_L\times U(1)_Y$) and a superpotential $W$, we have given two models that extend the Standard Model. The first model was the simplest supersymmetric extension of the Standard Model, \textit{i.e.}, the MSSM. This model is still, today, highly studied by the physicist community and signatures to this model are searched by experimentalists. As mentioned in Subsection \ref{subsec:mssm}, there is, however, some formal issues with this model (for example, the $\mu$-problem). A possible solution to those problems is to consider a singlet-extension of the MSSM.\medskip  

The new solutions in \textit{Gravity-Mediated Supersymmetry Breaking} (introduced in \autoref{NSW}) has pointed out models containing at least two singlet fields $S^1$ and $S^2$. It is then natural to define a model based on the classical Soni-Weldon solutions called the N2MSSM (already defined in \autoref{sec:N2MSSMDes}).\medskip

This chapter is devoted to a preliminary analysis of the N2MSSM. We will examine the differences between the NMSSM and the N2MSSM and investigate the reduction the fine-tuning of the Z-boson mass. We will motivate the study of the N2MSSM and its benefits in comparison to the NMSSM. The construction of a spectrum generator is described using the \textsc{Mathematica} package \textsc{Sarah}, which calculates two-loop \textit{Renormalisation Group Equations} (RGEs), mass matrices, one-loop corrected masses (two-loop for the Higgs sector) and some physical observables. The numerical integration of the RGEs (and other calculations) will be performed with the \textsc{Fortran} spectrum generator \textsc{SPheno}.\medskip

The parameters-space of the $\mathcal{Z}_3$-invariant N2MSSM is a high-dimensional space. It can then be complicated to proceed to an efficient scan in order to recover an acceptable phenomenological spectrum. For this purpose, a \textit{Monte-Carlo Markov-Chain} (MCMC) algorithm has been implemented to facilitate this scan. It is then simpler to recover points in the parameters-space with an acceptable Higgs boson mass. Other physical observables can also be reconstructed (using \textsc{SPheno} or third-party programs), such as the anomalous magnetic momentum of the muon $a_\mu = \frac12 (g_\mu - 2)$, the relic density of dark matter (using the program \textsc{MicrOmegas}) or other observables linked to the Higgs boson with \textsc{HiggsBounds}. \medskip

This work is done in collaboration with Eric Conte (IPHC) and Cyril Hugonie (LUPM) \cite{N2MSSM_Pheno}.
\section{Why the N2MSSM?}
There are a plethora of supersymmetric models. They can differ by the superfields content, the gauge group used or by other imposed symmetries leading to different particles spectrum and phenomenology. The N2MSSM (see \autoref{sec:N2MSSMDes}) is a simple extension of the MSSM (since it is obtained by adding two singlets superfield $S^{p}$ ($p=1,2$) in the MSSM content)\medskip

This definition can be seen as a purely arbitrary extension of the NMSSM. There is, however, some formal and phenomenological reasons to consider such a model. We will present three arguments for introducing such a model: comparing Soni-Weldon and Non-Soni-Weldon solutions, solving difficulties of the NMSSM and generating new interesting signatures at the LHC.
\subsection{Comparison between Soni-Weldon \& Non-Soni-Weldon solutions}\label{sub:comparisonNSW}
In the previous chapter (see Chapter \ref{nsw}), we have described new \textit{Gravity Mediated Supersymmetry Breaking} solutions. One of those new solutions allowed us to construct a model of supergravity called the S2MSSM. In this model, two hybrid fields $S^p$ ($p=1,2$) are introduced in such a way that they generate new hard breaking terms which could solve some phenomenological issues. \newline However, to fully understand the phenomenological differences between Soni-Weldon (SW) and Non-Soni-Weldon (NSW) solutions, it is mandatory to first study an equivalent model of the S2MSSM in the context of the classical Soni-Weldon solutions. This will help point out the effects of the hard breaking terms in the spectrum at tree \& loop levels. Therefore, the N2MSSM naturally emerge from this investigation.
%
\subsection{Solutions to difficulties of the NMSSM}\label{sub:difficultNMSSM}
\subsubsection{Reducing the fine-tuning on the Higgs boson mass}
\begin{figure}[!htb]
    \centering
    \begin{minipage}{.5\textwidth}
        \centering
        \includegraphics[width=1.07\linewidth]{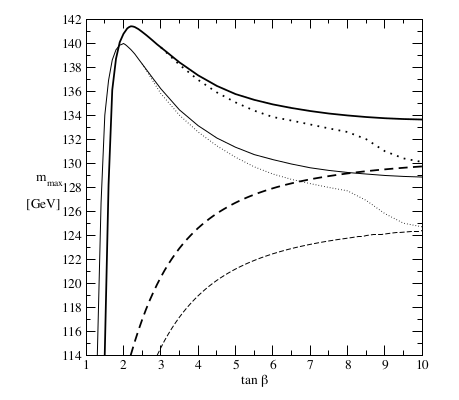}
    \end{minipage}%
    \begin{minipage}{0.5\textwidth}
        \centering
        \includegraphics[width=1.14\linewidth]{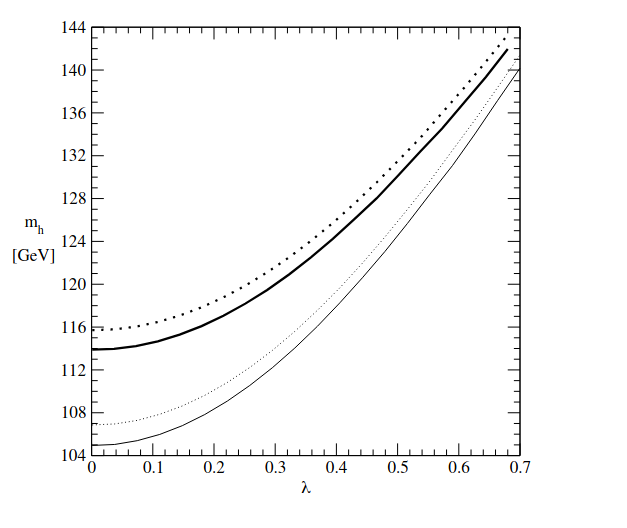}
    \end{minipage}
    \caption{(Taken from \cite{hugonie_lightesthiggs}.) In the left: Upper bound on the lightest Higgs mass (with radiative corrections) in the NMSSM for $m_t=178\ \text{GeV}$ and $m_t=171.4\ \text{GeV}$ (respectively for thick full and dotted lines) as function of $\tan \beta$. The dashed lines correspond to the MSSM case. The dotted and dashed lines are obtained with $M_A=1\ TeV$, whereas the full lines are obtained with an arbitrary $M_A$. The $\lambda$-couplings is taken as $\lambda\sim 0.7$. In the right: Upper bound of the lightest Higgs mass in the NMSSM as function of $\lambda$. The thick full (dotted) line is obtained for $\tan\beta=2.2$ and $m_t=178\ \text{GeV}$ with (without) radiative corrections where the thin full (dotted) line is obtained for $\tan\beta=2$ and $m_t=171.4\ \text{GeV}$ with (without) radiative corrections.}
    \label{fig:mhmaxNMSSM}
\end{figure}

In the context of the MSSM, the lightest Higgs boson has a tree-level mass: 
\begin{gather}
m_{h^0}^2 = M_A^2 + M_Z^2 - \sqrt{(M_A^2 + M_Z^2)^2 - 4 M_A^2M_Z^2\cos^2 2\beta} \nn 
\end{gather} 
(with $M_A$ the mass of the pseudoscalar state and $\tan\beta=v_U/v_D$) which leads to the following upper bound:
\begin{gather}
m_{h^0}^2 < M_Z^2 \cos^2 2\beta < (91 \ \text{GeV})^2\ .\label{eq:limitmhMSSM}
\end{gather}
This upper-limit seems inconsistent with the Higgs boson measured mass achieved at the LHC. Taking into account the approximation of the large radiative corrections coming from the top quark and the stop one-loop on the $\mathcal{M}_S^2{}_{(2,2)}$ (see \autoref{eq:ms211_n2mssm}) \cite{NMSSM}\cite{DJOUADI_2008}:
\begin{gather}
\Delta_{rad.} = \frac{6m_t^4}{4\pi^2v^2}\Big( \ln\Big( \frac{m_T^2}{m_t^2} \Big) + \frac{A_t^2}{m_T^2}\Big( 1 - \frac{A_t^2}{12m_T^2} \Big)\Big)\ , \nn
\end{gather}
(with the top mass $m_t$, $m_T^2$ the geometrical mean of stop masses and $v^2=v_U^2+v_D^2$) the tension can be released by introducing a fine-tuning on the stop parameters (usually called \textit{"little fine-tuning problem"}). The limit on the lightest Higgs can be then pushed-up to approximately $130\ \text{GeV}$ (see \ref{fig:mhmaxNMSSM}).\medskip

Adding a singlet field $S$ in the context of the NMSSM allows to suppress this issue. We can rotate the mass matrix by an angle $\beta=tan^{-1}\left( v_U/v_D \right)$ giving a new matrix $\mathcal{M}{'}_S^2$ in the basis $\{H'_{SM},H',S_r\}$
\begin{gather}
\mathcal{M'}_S^2 = R(\beta)\mathcal{M}_S^2 R^T(\beta) \ , \quad R(\beta) = \begin{pmatrix}
\cos\beta & \sin\beta & 0 \\
 -\sin\beta & \cos\beta & 0 \\
0 & 0 & 1 \\
\end{pmatrix}\nn
\end{gather}
This basis has the advantage that only one Higgs boson gets a non-zero \textit{v.e.v.} ($\langle H'_{SM}\rangle \neq 0$ and $\langle H'\rangle = 0$). Moreover, the mass matrix is nearly diagonal for typical values of parameters \cite{thesis_student_ellwanger}: the state $H'_{SM}$ having SM-like couplings to SM particles, $H'$ being the heavy MSSM scalar state and $S_r$ remaining the pure singlet state. The first element in the diagonal of this matrix gives
\begin{gather}
\mathcal{M'}_S^2{}_{(1,1)} = M_Z^2\cos^2 2\beta + \frac12\lambda^2v^2\sin^2 2\beta \label{eq:mlimitNMSSM}
\end{gather}
corresponding to the high limit of the mass of the lightest Higgs boson (with $\lambda$ the coupling $\lambda \hat{S}\hat{H}_U\cdot \hat{H}_D$ and $v^2 = v_U^2 + v_D^2$). We recognise the first term of \ref{eq:limitmhMSSM} of the MSSM with a new contribution coming from the singlet field and the $\lambda$-coupling \ref{eq:mlimitNMSSM} which lift-up the high-limit. Following those results, there are two different methods to generate a Higgs boson with a mass of $m_h=125\ \text{GeV}$ in the context of the NMSSM:
\begin{itemize}
\item[1)] Boost the element \autoref{eq:mlimitNMSSM} with high $\lambda$ and low $\tan\beta$ value. From \autoref{fig:mhmaxNMSSM}, a limit on $\tan\beta$ can be set to $\tan\beta \lesssim 6$. A high $\lambda$-coupling would be preferable but since $\lambda\gtrsim 1$ leads to Landau singularities, it is more convenient to take $\lambda \lesssim 0.7$.\medskip
\item[2)] Another possibility arises if the effective mass of the singlet state $S_r$ (the $\mathcal{M'}_S^2{}_{(3,3)}$ matrix element) is lower than the lightest Higgs state. In this configuration, the $\mathcal{M'}_S^2{}_{(1,3)}$ element can generate a push-up effect on the lightest Higgs mass \textit{via} diagonalisation. It can be shown that 
\begin{gather}
\mathcal{M'}_S^2{}_{(1,3)} = \sqrt{2}\lambda v\mu_{eff}\Big( 1 - \sin 2\beta \Big( \frac{A_{\lambda}}{2\mu_{eff}} + \frac{\kappa}{\lambda} \Big) \Big)\label{eq:ms13}
\end{gather}
(with $\mu_{eff} = \lambda v_S$). In order to neglect the negative term in \autoref{eq:ms13}, a moderate/high $\tan\beta$ can be taken. The restriction on the $\lambda$-coupling is more complex. Due to the $H'_{SM}-S_r$ mixing, couplings between the singlet state and particles from the Standard Model appear. The $\lambda$-value is then directly constrained by the LEP limit on light scalar state below $114 \ \text{GeV}$ \cite{higgs_limit_114}. This coupling is then usually taken as $\lambda\approx\mathcal{O}(0.1)$ \cite{nmssm_bblambda0.1}\cite{nmssm_pushup}.
\end{itemize}
The mechanism 2) can be more advantageous because it also decreases the lightest scalar mass and helps to obtain a dark matter relic density $\Omega_{DM}h^2$ (with $h$ the Planck constant) coherent with the latest experimental measurements. It can, however, be difficult to identify the Higgs boson and singlet as candidates on dark matter and inflaton (since the singlet parameters will already be settled for a coherent Higgs boson mass and a coherent dark matter density relic).\medskip 

Assuming now the N2MSSM, the mass matrix of the scalar sector is completely similar to the NMSSM case. The various effects presented previously still remain with the two singlets $S^1$ and $S^2$. We can then suppose three different scenarios:
\begin{itemize}
\item the two singlets contribute to the $\mathcal{M'}_{S}^2{}_{(1,1)}$ (the new contributions of \autoref{eq:mlimitNMSSM} generalise easily to $\frac12(\lambda_1^2+\lambda_2^2)v^2\sin^2 2\beta$)
\item if the masses of the singlets are lower than the lightest Higgs state, the two singlets could generate a push-up effect, leading to two light scalar states.
\item a combination of both mechanisms: one singlet acting like a push-up generator and one contributing to the $\lambda_i^2$-term.
\end{itemize}
Those effects could lead then to a reduction on the fine-tuning of the parameters to recover a phenomenologically acceptable Higgs boson mass.
\subsubsection{Passing through $\Omega_{DM}$ measurements with good Higgs or inflaton candidate}
As explained in the previous paragraph, singlet scalars with low masses can emerge from the analysis of some scenarios of the NMSSM. Due to the singlet nature of those states, the couplings to the Standard Model particles can be relatively low, which define good candidates for dark matter \cite{nmssmDM}\cite{nmssmDM2}. Each singlet can then be tuned to reconstruct an acceptable Higgs boson mass and a coherent dark-matter density relic $\Omega_{DM}h^2$ simultaneously.\medskip

Dark matter is not the only cosmological problem that arises in particle physics. Cosmic inflation is an epoch in the early universe where the expansion of the universe increases exponentially. This phenomenon can be understood by assuming a hypothetical field called the \textit{inflaton} \cite{vennin}. Some inflation-scenario as already been proposed in the NMSSM where fields in the Higgs sector can be a good candidate for the inflaton \cite{NMSSMinflation1} \cite{NMSSMinflation2} \cite{NMSSMinflation3}. Assuming two singlets $S^1$ and $S^2$, we can then suppose configurations where candidates for both inflaton and dark-matter (or possibly with coherent Higgs boson mass) can be produced.
\subsection{Searches at colliders}
Supersymmetry signals have been hunted for many decades on the various colliders experiments. Many analyses focus on the search for new scalar particles corresponding to new CP-even or CP-odd Higgs states. Due to the matrix structure of the MSSM and the NMSSM, many searches have been done for hunting signatures of heavy scalar particles.\medskip

As seen before, adding the new field $S^2$ generates new terms in all the matrix elements. It also produces a richer particle spectrum than the other classical supersymmetric models. The phenomenology analysis of the N2MSSM can then be handled in two aspects: searching for new topologies and applying known-signal research to the model  (the action of $S^2$ can help pass the actual limit). Moreover, the action of the second singlet can allow the presence of very light scalar particles in the spectrum, \textit{i.e.}, with a mass below $10\ \text{GeV}$ (using the push-up effect mentioned previously). Such typologies start to be investigated to cover the mass range fully. \cite{lightscalarLHC1}\cite{lightscalarLHC2}\cite{lightscalarLHC3}. 

\section{Spectrum generator for the N2MSSM}\label{sec:sarah_spheno}
Supersymmetry theory is broken at a high energy scale. So, a full spectrum generator is mandatory to study the phenomenological aspects of a model. The construction of such a program can be split into two parts:
\begin{itemize}
\item calculate analytically the one (or two) loop RGEs, all the mass matrices, the minimisation equations of the potential and potentially the self-energy contributions; 
\item implemented those results in a program and numerically solved the RGEs in order to reconstruct the entire mass spectrum of the model 
\end{itemize}
Those implementations can be tedious depending on which RGEs (one or two-loop) we want to use, the radiative corrections we want to implement in the program, and the various possible observables we want to investigate. \medskip

To facilitate those steps, the particle physics community provides several generic spectrum generators, depending on the model studied or the calculated observables. Some generators are specific to a model (\textit{e.g.} \textsc{NumSpec} implemented in \textsc{NmssmTools} for the NMSSM \cite{nmspec} or \textsc{SuSpect} for the MSSM \cite{suspect}) when others are more generic and leave the possibility to implement your own supersymmetric (or non-supersymmetric) models. For our investigation, we will follow the second approach by using the \textsc{Mathematica} package \textsc{Sarah} \cite{Staub_sarah}\cite{staub2012sarah}\cite{sarah_2015} coupled with the \textsc{Fortran} program \textsc{SPheno} \cite{Porod_spheno}. 
\subsection{Analytic calculus with \textsc{Sarah} package}
The program \textsc{Sarah} is a package \textsc{Mathematica} for studying of supersymmetric (and non-supersymmetric) models. The purpose of the implemented methods in \textsc{Sarah} is to go beyond the MSSM by defining any specific model from the definition of the gauge-group and the particles content. Coupled with the \textsc{Fortran} program \textsc{SPheno} (see \autoref{sub:spheno}), the \textsc{Mathematica} package can be seen as a "spectrum-generator generator" as we will see later.\\

After defining several properties of the model:
\begin{itemize}
\item the gauge group $G$ and the possible global symmetries as, \textit{i.e.}, R-parity or Peccei-Quinn symmetry \cite{PQsym1}\cite{PQsym2},
\item the fields content $\{\phi^i,\chi^j,A_{\mu}^k\}$ in their respective gauge group representations with their potentially non-zero \textit{v.e.vs.} ($\langle \phi^i\rangle \neq 0$), 
\item the superpotential $W(\Phi)$ invariant under the defined symmetries,
\end{itemize}
the program can to calculate several useful objects for a complete spectrum generator (see \autoref{fig:sarah_structure}). After checking possible gauge anomalies and charge conservation issues, the complete form of the tree-level Lagrangian is calculated (automatically in the case of SUSY models). Note that soft-breaking terms are automatically generated from the superpotential (for each parameter from $W(\Phi)$, a soft-breaking term is automatically defined). The fields are then rotated in their mass eigenstates after introducing the non-zero \textit{v.e.vs}. The ghost interactions are also derived.\medskip

After implementing the model, \textsc{Sarah} is able to derive the renormalisation group equations (or RGEs). The RGEs are a group of differential equations that allow to run the parameters of the models at any energy scale. Assuming a parameter $\rho$, the $\beta$-function of the parameter can be written as:
\begin{gather}
\frac{d\rho}{dt} \equiv \beta_{\rho} = \frac{1}{16\pi^2}\beta_{\rho}^{(1)} + \frac{1}{(16\pi^2)^2}\beta_{\rho}^{(2)} \nn
\end{gather}
with $t=\ln E$ and $\beta_{\rho}^{(1)}$ ($\beta_{\rho}^{(2)}$), the one-(two-)loop contribution to the $\beta$-function. Taking for example the gauge coupling $g_I$, the one and two-loops contributions to the $\beta$-function are \cite{2loopRGE}:
\begin{gather}
\beta_{g_I}^{(1)} = g_I^3\big[\tau_I(\mathcal{R}_I)-3C_I(\mathfrak{g}_I)\big]\ , \nn\\
\beta_{g_I}^{(2)} = g_I^3\big[2g_I^2\tau_I(\mathcal{R}_I)C_I(\mathfrak{g}_I)-6g_I^2C_I(\mathfrak{g}_I)^2 + 4\displaystyle\sum_J g_J^2\tau_I(\mathcal{R}_J)C_J(\mathfrak{g}_J) \big] -g_I^3\lambda^{ijk}\lambda_{ijk}\frac{C_I(k)}{d_I} \nn
\end{gather}
with the representation of the matter fields $\phi^I$ denoted as $\mathcal{R}_I$ and $\lambda_{ijk}$ the general trilinear coupling of the superpotential. The Dynkin label $\tau_I(\mathcal{R}_I)$ and the Casimir operator $C_I(\mathfrak{g}_I)$ are defined as 
\begin{gather}
T^a(\mathcal{R}_I)T^b(\mathcal{R_I})\delta_{ab} = C_I(\mathcal{R}_I)\ , \nn\\
\mathrm{Tr}\big[T^a(\mathcal{R}_I),T^b(\mathcal{R_I})\big] = \tau_I(\mathcal{R}_I)\delta^{ab}\ , \nn
\end{gather}
with $T^a(\mathcal{R}_I), a =1, \dots,\mathrm{dim}\mathfrak{g}_I$ the generators or $\mathfrak{g}_I$ in the representation $\mathcal{R}_I$. Note that the parameters from the superpotential are only renormalised by the wave function due to the non-renormalisation theorem \cite{nonrenormtheo}. Assuming a linear term $\xi^i$, a bilinear term $\mu^{ij}$ and trilinear term $\lambda^{ijk}$ on the superpotential. The RGEs can then be written as
\begin{gather}
\frac{d\lambda^{ijk}}{dt}=\lambda^{ijk}\big[\frac{1}{16\pi^2}\gamma_l^{(1)k}+\frac{1}{(16\pi^2)^2}\gamma_l^{(2)k}\big] + (k\leftrightarrow i) + (k\leftrightarrow j), \nn\\
\frac{d\mu^{ij}}{dt}=\lambda^{ik}\big[\frac{1}{16\pi^2}\gamma_k^{(1)j}+\frac{1}{(16\pi^2)^2}\gamma_k^{(2)j}\big] + (i\leftrightarrow j)\ , \nn\\
\frac{d\xi^{i}}{dt}=\xi^{j}\big[\frac{1}{16\pi^2}\gamma_j^{(1)i}+\frac{1}{(16\pi^2)^2}\gamma_j^{(2)i}\big]\ , \nn
\end{gather}
with $\gamma$ denoted as the anomalous dimension of the fields. The renormalisation group equations (RGEs) are obtained following general expressions \cite{rge1}\cite{rge2}\cite{rge3}. It is so possible to generate the 1 and 2-loop RGEs for any model.
\medskip

\begin{figure}[h!]
    \centering
      \includegraphics[width=.8\linewidth]{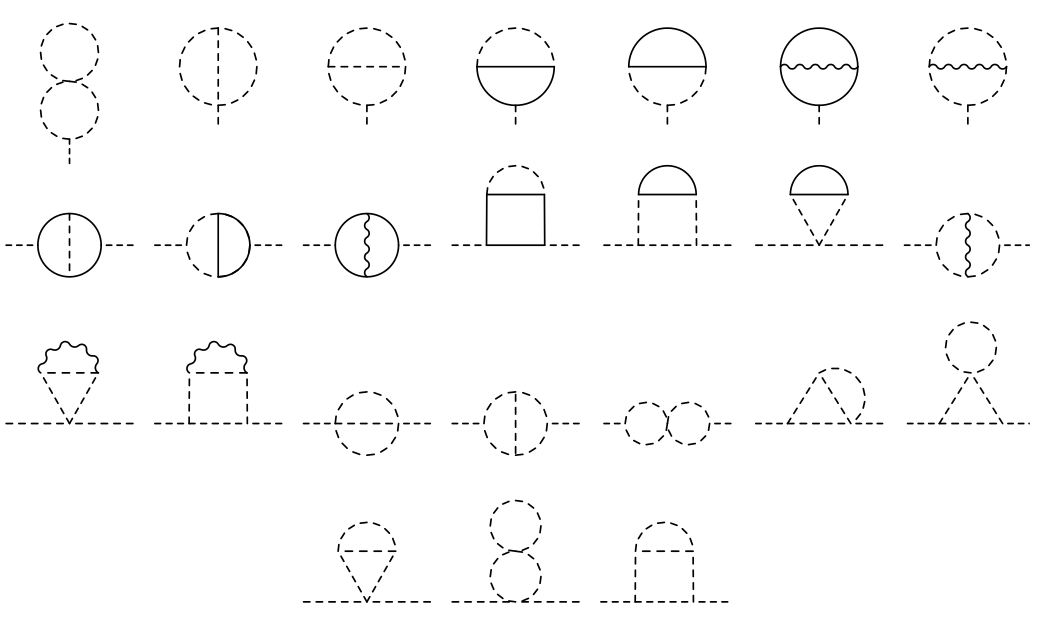}
    \caption{Two-loops radiative corrections to tadpoles and self-energy real scalar \textsc{Sarah} package (taken from \cite{sarah_2015}).}
    \label{fig:sarah_2loop}
\end{figure}

At this stage, all the mass matrices are calculated, as well as the tree-level masses and the tadpole equations (leading to the minimisation equations of the potential). The complete list of the vertices can also be computed, which allow \textsc{Sarah} to produce input files for many third party programs such as the UFO model \cite{ufo}, \textsc{CalcHep} file  \cite{calchep} (allowing to use the \textsc{micrOMEGAs} \cite{micromegas_lastv} program for our model) or \textsc{HiggsBounds} input files \cite{higgsbounds}.
It is also possible to calculate the 1-loop corrections to the self-energy and the tadpoles. Those calculations are done in the $\overline{DR}$-scheme using 't-Hooft gauge. The corrections are analytically implemented in \textsc{SARAH} using the Passarino-Veltman integrals (see \cite{staub2012sarah} for more information). In some cases, it is also possible to compute the two-loop corrections to the self-energies of real scalar. Two equivalent approaches can be used: an effective potential approach (based on the generic results \cite{effectivepot_approach}) and a diagrammatic approach (from \cite{goodsell2015twoloop}). The two approaches lead to equivalent results. In our case, we will consider the diagrammatic approach since the method is described as numerically more robust. The two-loops contributions included in \textsc{Sarah} are shown in \autoref{fig:sarah_2loop}. Those diagrams do not vanish in the gaugeless limit (massless gauge boson) and are taking in limit the $p^2=0$. Those corrections calculated, it is so possible to determine the corrected mass matrices, their corresponding eigenvalues and the minimisation equations. Taking for example a mass matrix $m_{\phi}^2$ of real scalar fields $\phi^i$, the two-loops corrected mass matrix is obtained from 
\begin{gather}
m_{\phi}^2{}^{(2L)}(p_i^2) = m_{\phi}^2{}^{(T)} - \text{Re}(\Pi_{\phi}^{(1L)}(p_i^2)) - \text{Re}(\Pi_{\phi}^{(2L)}(0)) \nn
\end{gather}
with $m_{\phi}^2{}^{(T)}$, the mass matrix at tree-level and $\Pi_{\phi}(p^2)$ the self-energy contributions. The two-loop mass states are then obtained by an iterative procedure on 
\begin{gather}
\text{Det}[p_i^2 - m_{\phi}^2{}^{(2L)}(p_i^2)]=0\ .
\end{gather}
with the on-shell condition $p_i^2 = m_{\phi^i}^2$.\medskip

{For our case, we can inplement the N2MSSM in the SARAH framework by:
\begin{itemize}
\item defining the gauge group as $G_{SM} = SU(3)_C\times SU(2)_L\times U(1)_Y$
\item adding to the MSSM superfields content the two new singlet superfields $\hat{S}^p$ ($p=1,2$) with their non-zero \textit{v.e.vs} on the scalar part $\langle S^p\rangle \neq 0$,
\item implementing the superpotential of the N2MSSM\footnote{Even thought our analysis is based on the $\mathcal{Z}_3$-invariant N2MSSM, we have implemented the general superpotential of the N2MSSM.}
\begin{eqnarray}
W_{N2MSSM} &=& \lambda_i \hat{S}^i \hat{H}_U\cdot \hat{H}_D  + \frac13\kappa_{ijk}\hat{S}^i\hat{S}^j\hat{S}^k + \frac12 \mu'_{ij}\hat{S}^i\hat{S}^j + \xi_{F,i}\hat{S}^i \nn\\
&& + y_U\hat{Q}\cdot \hat{H}_U \hat{U} - y_D\hat{Q}\cdot \hat{H}_D  \hat{D} - y_E\hat{L}\cdot \hat{H}_D  \hat{E} 
\end{eqnarray}
firstly defining in \ref{sec:N2MSSMDes}.
\end{itemize}
The mass matrices obtained by the \textsc{Sarah} method has been compared by the ones presented in \ref{sec:N2MSSMDes} and Appendix \ref{app:higgssector} and validated.\medskip
\begin{figure}[h!]
    \centering
      \includegraphics[width=.8\linewidth]{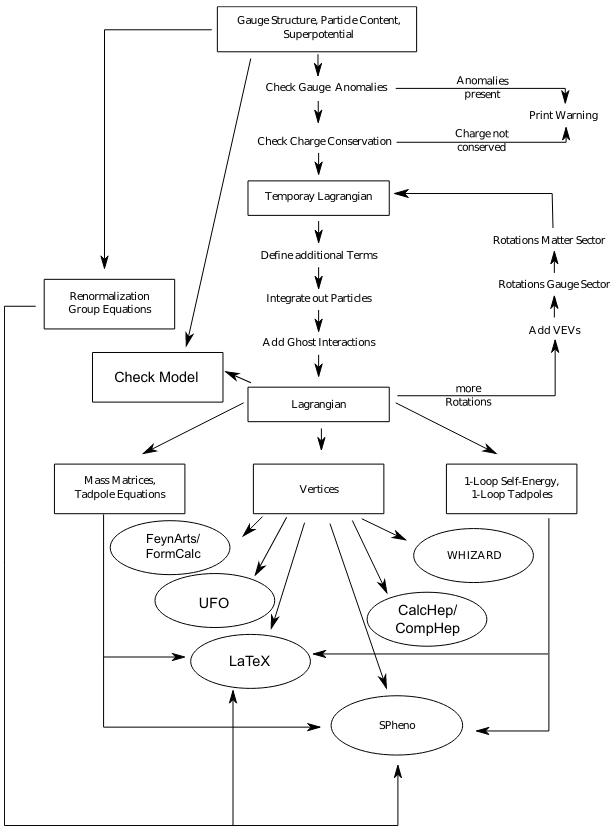}
    \caption{Structure of the \textsc{Sarah} package (taken from \cite{staub2012sarah}).}
    \label{fig:sarah_structure}
\end{figure}
\subsection{From \textsc{Sarah} to \textsc{SPheno}}
At this stage, we have all the information to create our own spectrum generator for the N2MSSM. As mentioned previously, the package \textsc{Sarah} allows to generate a full spectrum generator automatically using the \textsc{SPheno} source code. We first need to impose the minimisation of the potential from the tadpole equations:
\begin{gather}
\Big\langle \frac{\partial V}{\partial h_U^0}\Big\rangle = 0 \ , \Big\langle \frac{\partial V}{\partial h_D^0}\Big\rangle = 0 \ , \Big\langle \frac{\partial V}{\partial S^2_r}\Big\rangle = 0 \ , \Big\langle \frac{\partial V}{\partial S^1_r}\Big\rangle = 0 \ . \label{eq:miniN2}
\end{gather}
(see \ref{eq:param_min1}, \ref{eq:param_min2}, \ref{eq:param_min3} and \ref{eq:param_min4}). In the MSSM and the NMSSM, we usually eliminate the soft-breaking mass terms of the scalar sector through the minimisation relations, \textit{i.e.}, $m_{H_U}^2$, $m_{H_S}^2$ and $m_{S}^2$ (for the NMSSM). Similarly, we will eliminate the four soft-breaking mass terms \sloppy $\{m_{H_U}^2,m_{H_D}^2,m_{S^1}^2,m_{S^2}^2\}$ using \ref{eq:miniN2}. Note that there is still one soft-mass breaking term in our model which can not be eliminated, \textit{i.e.}, the singlet-mixing term $\frac12 m_{S^{12}}^2\left( S^1(S^2)^{\dagger}+ S^2(S^1)^{\dagger} \right)$. \medskip

The various input for running the parameters at high energy must also be defined. We suppose a Gravity-Mediated Supersymmetry Breaking Mechanism with universality (for the soft scalar masses) at GUT scale $M_{GUT}$. The universality is obtained by taking a canonical Kähler potential, \textit{i.e.}, in the form 
\begin{gather}
K(\phi,\phi^{\dagger}, z , z^{\dagger}) = \phi^a\phi^{\dagger}_a + m_p^2 z^iz^{\dagger}_i \nn
\end{gather} 
with $\phi^a$ the fields from the matter sector and $z^i$ fields from the hidden sector (see \autoref{sec:sugra} and \autoref{NSW} for more information). This will leave us with four generic inputs at the GUT scale: the universal scalar soft-breaking mass term $m_0$ (note that the parameters eliminated from the minimisation equations are not impacted by this parameter), the universal gaugino mass $m_{1/2}$, the universal trilinear coupling $A_0$ and the parameter $\tan\beta$. We suppose a universality in trilinear couplings only in the sfermion sector (\textit{i.e.}, $A_t=A_b=A_{\tau}=A_0$) where the couplings in the Higgs and singlet sector ($A_{\lambda_i}$ and $A_{\kappa_{ijk}}$) are specified as input. The GUT scale $M_{GUT}$ is defined by matching the $U_Y(1)$ and $SU_L(2)$ couplings $g_1=g_2$. As far as the parameters from the superpotential are concerned, they are all defined at SUSY scale (defined as the geometrical mean of the two stop masses: $M_{SUSY}=\sqrt{m_{\tilde{t}_1}m_{\tilde{t}_2}}$). We also assume a first guess on the SUSY scale defined as 
\begin{gather}
\left(M_{SUSY}^2\right)^{\text{guess}} = m_0^2 + 4m_{1/2}^2\ . \label{eq:msusy_firstguess}
\end{gather}
for the first iteration with \textsc{SPheno} (see next section).
\subsection{Numerical solutions with \textsc{SPheno}}\label{sub:spheno}
\textsc{SPheno} is a \textsc{Fortran} program that numerically solves the renormalisation group equations and reconstructs the full mass spectrum. Some methods are also implemented to calculate various low energy observables. In his first version, this program was specific to the reconstruction of the MSSM spectrum. It is now possible to adapt the generic methods of \textsc{SPheno} for more generic models with the help of \textsc{Sarah}.  
\subsubsection{Running the two-loops RGEs}
There exist several methods to solve differential equations numerically. The program \textsc{SPheno} uses a method called RK4 (for Runge-Kutta at the fourth-order). Starting from the initialisation point $y_0$, the next iteration $y_1$ is obtained by a first approximation. This calculation allows to estimate a second approximation of $y_1$, and so on until the fourth approximation (in the case of RK4). The complete numerical solution up to $y_n$ can then be obtained. The calculation of the mass spectrum is done as follow (see \autoref{fig:spheno_schema} for a schematic representation): 
\begin{itemize}
\item[1)] The coupling constants $g_i$ and the Yukawa couplings are firstly computed (at tree-level) at the electroweak scale ($M_{EW}\equiv m_Z$).
\item[2)] Using the renormalisation group equations at one loop, the parameters are run to first guess the SUSY scale \autoref{eq:msusy_firstguess}, where we introduce the parameters from the superpotential. The integration then continues to the GUT scale, where we introduce the universal soft breaking terms $\{m_0,m_{1/2},A_0\}$, $\tan\beta$ and the trilinear singlets couplings $A_{\kappa_{ijk}}$ and $A_{\lambda_i}$.
\item[3)] We now run to the electroweak scale to impose the minimum of the potential by identifying the soft-mass parameters $\{m_{H_U}^2,m_{H_D}^2,m_{S^1}^2,m_{S^2}^2\}$. All the parameters are now defined. The tree-level masses are obtained as eigenvalues of the mass matrices. 
\item[4)] One-loop radiative corrections from Standard Model and supersymmetry to $g_i$ and the Yukawa couplings $y_i$ are calculated. The corrections to the coupling constant are obtained through the one-loop corrections of the Z-boson and W-boson masses $m_Z$ and $m_W$, on $\alpha^{\overline{DR}}(m_Z)$ and $\alpha_S^{\overline{DR}}(m_Z)$ and on the Weinberg angle $\sin^2\Theta^{\overline{DR}}_W$. The corrections to the Yukawa couplings are calculated from the loop corrections to the fermion masses (more precision on those calculations in \cite{sarah_2015}\cite{mssmprecision_usebySPheno}).

\item[5)] The same running as 3) can now be performed with the two-loop RGEs following an equivalent pattern. Note that we now use the dynamic SUSY scale $M_{SUSY}=\sqrt{m_{\tilde{t}_1}m_{\tilde{t}_2}}$. The potential is minimised at the electroweak scale, and the two-loops corrected masses are computed using \textsc{Sarah}'s two-loops self-energy diagrams. 
\item[6)] The convergence of the program is evaluated through the reconstructed masses. Assuming that we are at iteration $i$, the variation on the masses between two iterations is calculated: 
\begin{gather}
\frac{|m_i - m_{i-1}|}{m_i} < \delta \nn
\end{gather}
with $\delta$ automatically set as $10^{-4}$ in the program. While the inequality is not satisfied, the program restart at the step 4) and continue the loop. 
\end{itemize}
During all the process, some checks are performed on the presence of negative squared masses in the scalar-pseudoscalar sector and the possible divergences of the RGEs (from Landau pole) due to the  input parameters. 
\begin{figure}[h!]
    \centering
      \includegraphics[width=.8\linewidth]{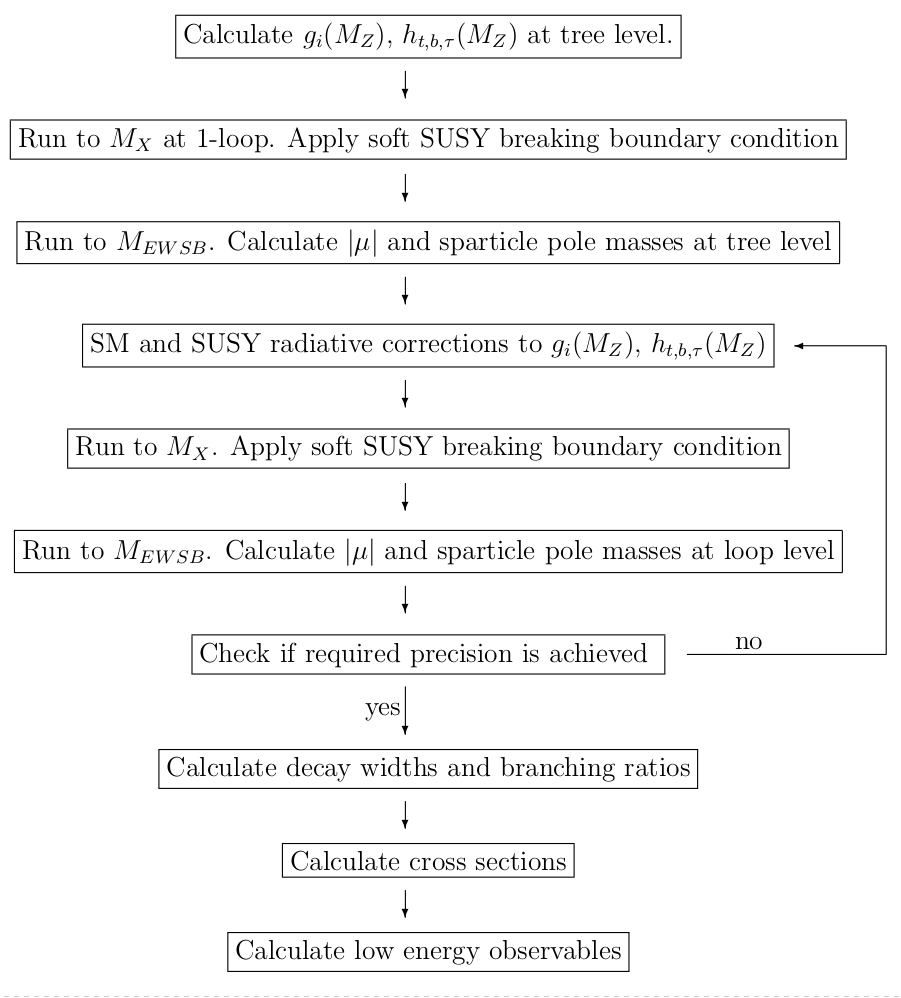}
    \caption{Algorithm implemented in the program \textsc{SPheno} for the spectrum and generation and the calculation of the various observables (taken from \cite{Porod_spheno}). The notation $h_X$ corresponds to the Yukawa couplings $y_X$. Note that the $\mu$-parameter is present since the first of \textsc{SPheno} was specific for the MSSM. }
    \label{fig:spheno_schema}
\end{figure}
\subsubsection{Low energy observables}
After reaching the precision $\delta$ on the reconstructed masses, the program \textsc{SPheno} calculate various observables. The decay width of all two-body decays are computed, as well as some specific three-body decays. A plethora of (lepton and quark) flavour violation observables are also obtained from the \textsc{Sarah} extension \textsc{FlavourKit} \cite{fkit1}\cite{fkit2}. Supersymmetric contributions to the anomalous magnetic moment of leptons $a_{\ell}$ are also calculated (based on \cite{gm2_calculus}), considering contributions from neutralinos and charginos exchange (see diagrams in \autoref{fig:gm2-SUSY}). \medskip
\begin{figure}[h!]
    \centering
      \includegraphics[width=.7\linewidth]{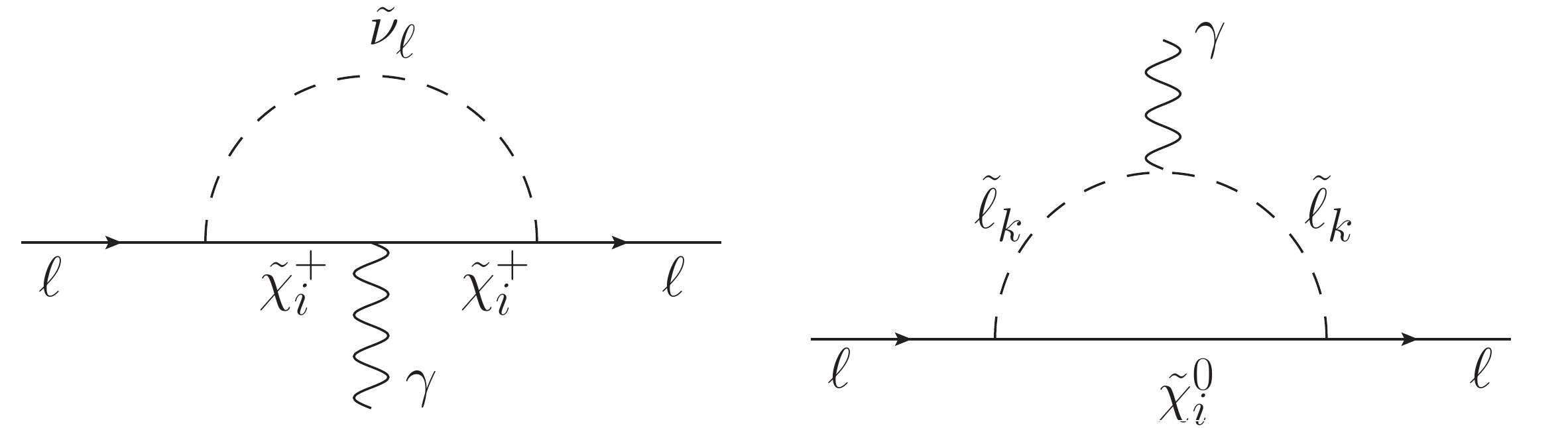}
    \caption{SUSY contributions to the anomalous magnetic moment of leptons $a_{\ell}$}
    \label{fig:gm2-SUSY}
\end{figure}

We will go into details on an observable, which will be helpful in the following. From \cite{fine_tuning1}\cite{fine_tuning2}, a measurement of the fine-tuning of the elecroweak scale has been defined by:
\begin{eqnarray}
\Delta_{FT}=\max\lvert\Delta_{i}\rvert\quad\mathrm{with}\quad \Delta_{i}=\frac{\partial\ln m_Z^2 }{\partial\ln \mathfrak{q}_i}=\frac{\mathfrak{q}_i}{m_Z^2}\frac{\partial m_Z^2}{\partial \mathfrak{q}_i}\label{eq:DeltaFT}
\end{eqnarray}
The variable $\Delta_{FT}$ is defined as the "fine-tuning of the electroweak" (or simply "fine-tuning") relatively to the parameters $\{\mathfrak{q}_i\}$. This parameter gives an estimation of how the parameters $\{q_i\}$ must be fine-tuned at high energy to get a correct electroweak scale.   The program \textsc{Sarah} allows to choose the list of parameters $\{\mathfrak{q}_i\}$ for the calculation of $\Delta_{FT}$. In our case, we have implemented:
\begin{gather}
\{\mathfrak{q}_i\} = \{m_0, m_{1/2}, m_{H_U}^2, m_{H_D}^2, m_{S^1}^2, m_{S^2}^2,A_0, \lambda_i, A_{\lambda_i}, \kappa_{ijk}, A_{\kappa_{ijk}},  y_U, \tan\beta \} \label{eq:par_FT}
\end{gather} 
The computation is done using the two-loop RGEs by slightly varying the parameters from \ref{eq:par_FT}. The one-loop correction to the Z-boson mass $m_Z$ is considered for this calculation. This observable will be extensively used for comparing the N2MSM with the NMSSM. 
\subsection{Other observables from third party programs}\label{sub:prog}
As mentioned previously, it is also possible to generate input files for third-party programs.  This can be used to constrain the parameters space following other observables that are not directly implemented in the program. It is the case for two programs that will be useful for a future complete analysis: \textsc{HiggsBounds} and \textsc{micrOMEGAs}. \medskip

The program \textsc{micrOMEGAs} \cite{micromegas_lastv} is dedicated to the computation of dark matter (or DM) properties. By solving the Boltzmann equation for the N2MSSM, we can deduce the temperature of the chemical decoupling of the dark matter with the primordial plasma (or freeze-out) and the density relic of dark matter of the universe $\Omega_{DM}h^2$. Since dark matter signals are also searched in the direct (and indirect) detection experiments, the direct (and indirect) cross-section of dark matter with the baryonic matter are also calculated. \medskip

The \textsc{Fortran} program \textsc{HiggsBounds} \cite{higgsbounds} checks the validity of the parameter point against the experimental measurement on the Higgs sector. Using the various output from the \textsc{SPheno} spectrum generator of the N2MSSM, \textsc{HiggsBounds} determines if the parameter point has been excluded at $95\%$ C.l. by computing the parameter
\begin{gather}
k_0 = \max k_i = \max \frac{Q_{model}(X_i)}{Q_{obs}(X_i)},\ \quad \big( Q(X_i)=\sigma(H_i)BR(X_i) \big)\nn
\end{gather}
with $X_i$ some process of the Higgs $H_i$ and $\sigma(H_i)$ the production cross-section of the Higgs $H_i$. If the parameter $k_0>1$, the parameter point is excluded at $95\%$ C.L..
\section{Parameters-Space scan with a \textit{Markov-Chain Monte-Carlo} (MCMC) algorithm}
We wave now at our disposal all the information to restrict the parameter space of the N2MSSM. For this purpose, simple methods can be used: defining a random scan in a specific zone of the parameter space or defining a grid scan where the parameters are iterated on a specific range. Those methods can, however, be not adapted for our spectrum generator. The parameter space of the N2MSSM is a high-dimensional space (17 parameters for our case, see \autoref{tab:1stpt}). A complete grid scan needs then a high computing time. Moreover, a considerable part of the parameter space leads to unphysical results such as divergences of the renormalisation group equations or negative scalar squared masses. Random scans can then be not efficient. \medskip
Another solution is to develop an algorithm that efficiently scans the parameter space regarding the issues mentioned in the previous paragraph. This algorithm is called \textit{Markov-Chain Monte-Carlo} algorithm (or MCMC). It helps to explore zones of the parameter space, which lead to phenomenologically acceptable parameter points (correct SM-Higgs mass and sfermions masses in accordance with the actual limits). 
\subsection{\textit{Markov-Chain Monte-Carlo} (MCMC) algorithm}\label{sub:MCMC}
Markov-chain are stochastic processes with the Markov property, \textit{i.e.}, the probability of obtaining a specific future state is only determined by the present state and not by all the past states. Those processes can be combined with Monte-Carlo methods based on randomness. \medskip
The MCMC algorithm is described as follow:
\begin{itemize}
\item[1)] we start the implementation by defining a first point corresponding to a list of all the input parameters $\{\mathfrak{q}_i\}$. This point is called the centre point $\{\mathfrak{q}_i\}_{CENTRE}$. The mass spectrum, as well as the other observables mentioned in \ref{sec:sarah_spheno}, are calculated with the parameters $\{\mathfrak{q}_i\}$.
\item[2)] At this stage, we introduce a new function called the penalty function $\mathcal{X}_{CENTRE}\in[1,\infty [$, which embeds the point $\{\mathfrak{q}_i\}_{CENTRE}$'s phenomenological properties. This function is defined so that a high value of $\mathcal{X}_{CENTRE}$ corresponds to a non-phenomenologically acceptable point (the exact definition of $\mathcal{X}$ for our study is given in \ref{subsec:MCMC4N2MSSM}).The algorithm's goal is then to minimise this function to efficiently explore the parameter space.
\item[3)] The next point is obtained from a stochastic draw:
\begin{gather}
\mathfrak{p}_i{}_{NEXT} = \mathfrak{p}_i{}_{CENTRE}\big(1 + \sigma_i\mathcal{N}(0,1,x)\big) \label{eq:next_pt}
\end{gather}
with $\mathcal{N}(0,1,x)$ a symmetric normalised and centred gaussian and $\sigma_i$ an input parameter for each parameter $\mathfrak{p}_i$. The value of $x$ is drawn randomly. A high value of $\sigma_i$ allows a big variation on the parameter $\mathfrak{p}_i$ for the next iteration.  
\item[4)] After generating all the spectrum for the new point $\{\mathfrak{p}_i\}_{NEXT}$, the penalty function associated with this point can be computed $\mathcal{X}_{NEXT}$. The two points $\{\mathfrak{p}_i\}_{NEXT}$ and $\{\mathfrak{p}_i\}_{CENTRE}$ can now be compared with the help of the two penalty functions. For this purpose, the probability $p(\mathcal{X}_{NEXT},\mathcal{X}_{CENTRE},\sigma_{\mathcal{X}})$ (see \autoref{fig:probaMCMC}) to define the next point as the new centre point $\{\mathfrak{p}_i\}_{CENTRE}=\{\mathfrak{p}_i\}_{NEXT}$ is obtained from
\begin{gather}
p(\mathcal{X}_{NEXT},\mathcal{X}_{CENTRE},\sigma_{\mathcal{X}}) = \frac{1}{2}\left(1+\mathrm{erf}\left( \sqrt{2}\frac{\mathcal{X}_{CENTRE}-\mathcal{X}_{NEXT}}{\sigma_{\mathcal{X}}\min\left(\mathcal{X}_{CENTRE},\mathcal{X}_{NEXT}\right)} \right)\right) \label{eq:pMCMC}
\end{gather}
with $erf(x)$ the error function. If $p(\mathcal{X}_{NEXT},\mathcal{X}_{CENTRE},\sigma_{\mathcal{X}})<f(x)$ (with $f(x)\in[0,1]$ a uniform probability distribution), we set $\{\mathfrak{p}_i\}_{CENTRE}=\{\mathfrak{p}_i\}_{NEXT}$. This case occurs when typically $\mathcal{X}_{NEXT}<\mathcal{X}_{CENTRE}$, \textit{i.e.}, the point $\{\mathfrak{p}_i\}_{NEXT}$ is a better phenomenological solution than $\{\mathfrak{p}_i\}_{CENTRE}$. If it is not the case, the centre point remains the same. The algorithm continues by restarting to the step 3) with the $\{\mathfrak{p}_i\}_{CENTRE}$ defined from the stochastic test. 
\end{itemize}
After several iterations, the penalty function must decrease and converge to the interesting parameter space zone. Note that the parameter $\sigma_{\mathcal{X}}$ in the function \autoref{eq:pMCMC} modifies the function's slope where $\{\mathcal{X}\}_{NEXT}\approx \{\mathcal{X}\}_{CENTRE}$. This parameter must be chosen before the computation.
\begin{figure}[h!]
    \centering
    \includegraphics[scale=0.6]{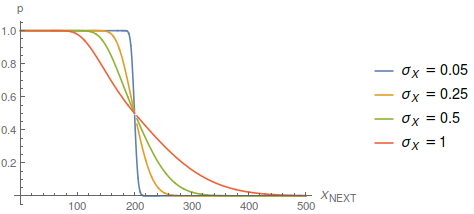}
    \caption{Evolution of $p(\mathcal{X}_{NEXT},\mathcal{X}_{CENTRE},\sigma_{\mathcal{X}})$ with $\mathcal{X}_{CENTRE}=200$ for several values of $\sigma_{\mathcal{X}}$.}
\label{fig:probaMCMC}
\end{figure}
\subsection{Penalty function $\mathcal{X}$ for the MCMC}\label{subsec:MCMC4N2MSSM}
All the phenomenological validity of the points $\{\mathfrak{p}_i\}$ has been embedded in the penalty function $\mathcal{X}$. We will go into the details on the numerical definition of this function. First, we need to define two different types of points in the parameter space: the points that lead to numerical (from Landau pole or problems with matrix diagonalisation) or theoretical (negative squared mass in the spectrum of scalar particles) issues and the points that do not satisfy the actual phenomenological constraints. We define $\mathcal{X}$ as follow:
\subsubsection{Constraints on the theoretical and numerical validity of $\{\mathfrak{p}_i\}$}
\begin{itemize}
\item $\mathcal{X}=10^{80}$ if divergences on the numerical integration of the RGEs or \textit{NaN} appear during the calculation.
\item $\mathcal{X}=10^{70}$ if problems appear on the radiative corrections at the electroweak scale (see step 4 in \ref{sub:spheno}). The penalty function is set at the same value if problems in the computation and diagonalisation of the mass matrices arise.
\item $\mathcal{X}=(|\min\{m_{\phi^i}^2\}|+1.0)\times 10^{60}$ (with $m_{\phi^i}^2$ the squared masses of the scalar field $i$) if negative squared masses are present on the scalar sector during the calculation. 
\end{itemize}
If the point passes all those restrictions, \textsc{SPheno} produces the N2MSSM spectrum. We can now choose the observables that must fit the experimental measurements (or limits). Obviously, the produce spectrum must satisfy all the actual constraints to be defined as a valid point.Nevertheless, not all the given constraints will be applied for this preliminary analysis due to the time computation. For the experimental constraints, we have the following definitions.
\subsubsection{Constraints on the phenomenological validity of $\{\mathfrak{p}_i\}$}
\begin{itemize}
\item We firstly want to reproduce an appropriate SM-like Higgs boson mass. Assuming an experimental measurement of $m_h^{exp}=125\ \mathrm{GeV}$ and an uncertainty $\Delta m_h$, we define
\begin{gather}
\mathcal{X} = \Bigg\{\max\Bigg(0,\frac{m_h^{SPheno} - \big(m_h^{exp} + \Delta m_h\big)}{m_h^{exp} + \Delta m_h}\Bigg) \nn\\
- \min\Bigg(0,\frac{m_h^{SPheno} - \big(m_h^{exp} - \Delta m_h\big)}{m_h^{exp} - \Delta m_h}\Bigg)\Bigg\}\times 10^{50}.\nn
\end{gather}
The actual experiment uncertainty is $\Delta m_h^{exp}= 0.17\ \mathrm{GeV}$ \cite{pdg} while the usual theoretical uncertainty is assumed to be $\Delta m_h^{theo}= 3\ \mathrm{GeV}$ (this value has been obtained by comparing spectrum from various programs, see \cite{uncert_mh}). However, for this preliminary analysis, we will assume (in general) an uncertainty of $\Delta m_h= 7\ \mathrm{GeV}$ due to a problem of time computation. A future (more complete) analysis will be extended with $\Delta m_h= 3\ \mathrm{GeV}$. 
\item We assume a squark sector higher than the $\mathrm{TeV}$ scale \cite{pdg}. We have 
\begin{gather}
\mathcal{X} =\max\Big((m_{\tilde{q}}^2)^{limit}-\min\{m_{\tilde{q}^i}^2\},0\Big)\times 10^{40} \label{eq:minsquark}
\end{gather}
with $(m_{\tilde{q}}^2)^{limit} = 1\ \mathrm{TeV}$. 
\item We have also implemented a constraint on the gluino mass $M_3$:
\begin{gather}
\mathcal{X}=\max\Big(M_3^{limit}-M_3^{SPheno},0\Big)\times 10^{30}\label{eq:PenFunM3}
\end{gather}
with $M_3^{limit}=1.6\ \mathrm{TeV}$. 
\item The output from the program \textsc{HiggsBounds} can also be used as input for the algorithm. In the Subsection \ref{sub:prog}, we have mentioned the parameter $k_0$ which determined the validity of the point regarding the actual limit on the Higgs measurements. We define:
\begin{gather}
\mathcal{X} =\max\Big(k_0 - 1,0\Big)\times 10^{20}\ .\nn
\end{gather}
\item In the same manner, using the output of \textsc{micrOMEGAs}, we can force a correct relic density of dark matter: 
\begin{gather}
\mathcal{X} = \Bigg\{\max\Bigg(0,\frac{\Omega_{DM} - \big(\Omega_{DM}^{exp} + \Delta \Omega_{DM}\big)}{\Omega_{DM}^{exp} + \Delta \Omega_{DM}}\Bigg) \nn\\
- \min\Bigg(0,\frac{\Omega_{DM}^{SPheno} - \big(\Omega_{DM}^{exp} - \Delta \Omega_{DM}\big)}{\Omega_{DM}^{exp} - \Delta \Omega_{DM}}\Bigg)\Bigg\}\times 10^{10}\ .
\end{gather}
It is also possible to only consider the upper bound of the experimental measurement, assuming that other sources to $\Omega_{DM}h^2$ occur. We can also consider in the same manner the measurements on $\sigma_{direct}$ and  $\sigma_{indirect}$ from the various experiments. 
\end{itemize}
A last observable is added to the definition of the penalty function. We have already defined the calculation of the fine-tuning of the electroweak scale $\Delta_{FT}$ in Subsection \ref{sub:spheno}. When all the above constraints are satisfied, we define the penalty function as
\begin{gather}
\mathcal{X} = \Delta_{FT}\ . \nn 
\end{gather} 
From this definition, the fine-tuning $\Delta_{FT}$ is naturally minimised. After several iterations, the algorithm must then find the minimum of the penalty function (and so the fine-tuning $\Delta_{FT}$). A convergence test is performed to check if we are at the minimum
\begin{gather}
\frac{|\mathcal{X}_{CENTRE}-\mathcal{X}_{NEXT}|}{\mathcal{X}_{CENTRE}} < \Delta_{\mathcal{X}}\label{eq:Xconvergence}
\end{gather}   
with $\Delta_{\mathcal{X}}$ the demand precision as input. The maximum number of iterations is also defined to converge to the minimum.
\section{Scan for N2MSSM and NMSSM}
We now have all the tools to study the N2MSSM. This section focuses on one of the three aspects mentioned before: the differences between the NMSSM and the N2MSSM. Indeed, the second singlet $S^2$ leads to new contributions that could help resolve some tension in the NMSSM. This analysis can be done in various directions. The following analysis only considers two points: 
\begin{itemize}
\item the reduction of the fine-tuning of the electroweak scale in the N2MSSM;
\item reducing the tension on the reconstruction of a phenomenological acceptable Higgs boson. 
\end{itemize}
Another important analysis could also be done on the dark matter properties (relic density $\Omega_{DM}h^2$ and direct (indirect) detection rates $\sigma_{direct}$ ($\sigma_{indirect}$)). 
This study is not included in this manuscript (mainly due to time processing).
\subsection{Spectrum generator for the NMSSM}
Since the NMSSM is a well-known model, a plethora of programs already exists for the complete calculation of the NMSSM spectrum (\textsc{NMSSMTools} to only cite on). In order to avoid possible differences between the two programs, an equivalent spectrum generator for the NMSSM is obtained with the program \textsc{Sarah}.\medskip

We define the NMSSM following the notations of the N2MSSM\footnote{The general NMSSM has been implemented for the generation. However, the $\mathcal{Z}_3$-invariant model will be used in the following, \textit{i.e.}, taking $\xi_F=0$ and $\mu'=0$.}:
\begin{gather}
W_{NMSSM}(\Phi) = \lambda \hat{S}\hat{H}_U\cdot \hat{H}_D + \frac13\kappa \hat{S}^3 + \frac12\mu' \hat{S}^2 + \xi_F\hat{S} + W_{MSSM}(\Phi)\Big|_{\mu = 0}\ .\nn 
\end{gather}
The soft-breaking terms are equivalent to those of the N2MSSM not considering the $S^2$-couplings. A comparison between \textsc{Sarah}/\textsc{SPheno} and \textsc{NMSSMTools} has been made under various benchmark points from the LHC NMSSM Subgroup\footnote{\url{https://twiki.cern.ch/twiki/bin/view/LHCPhysics/NMSSMBenchmarkPoints}}, leading to only a few per cent differences on the mass spectrum.\medskip

Note that the parameters-space of the NMSSM is a smaller dimensional space than for the N2MSSM (9 parameters for our case). 
\subsection{Method for the MCMC-scan}\label{subsec:methodMCMC}
The method of the parameter-scan is reported here, as well as the various inputs for the program. \\
The first point has been chosen by analysing the already known benchmark of the NMSSM and assuming a light singlet sector. This configuration allows a push-up effect on the Higgs sector (see \autoref{sub:difficultNMSSM} for more information about the various ways to generate SM-like Higgs boson in the NMSSM) and is more attractive by giving a possible light scalar candidate for dark matter. The SM-like Higgs is then supposed to be the second lightest Higgs in the spectrum. The first point $\{\mathfrak{p}_i\}_{BEGIN}$ is presented in the \autoref{tab:1stpt}.
\begin{table}[h!]
\begin{center}
\begin{tabular}{ |c||c|c|c|c|c|c|c| } 
 \hline
 Parameter &  $\tan\beta$ & $A_0\ [\text{GeV}]$ & $\lambda_i$ & $\kappa_{ijk}$ & $\mu_{eff}{}_{i}\ [\text{GeV}]$ & $A_{\lambda}{}_i\ [\text{GeV}]$ & $A_{\kappa}{}_{ijk}\ [\text{GeV}]$\\
\hline
\hline 
 Input & 30 & -1500 & 0.1 & 0.1 & 100 & -200 & -80 \\
 \hline
 Variation $\sigma_i$ & \multicolumn{7}{|c|}{0.1} \\
 \hline
\end{tabular}
\end{center}
\caption{First point $\{\mathfrak{p}_i\}_{BEGIN}$ for the algorithm. The parameter $m_0$ and $m_{1/2}$ are defined separately to produce a grid scan. The index $i$ is introduced for the N2MSSM-parameters. For the NMSSM, we set $x_i=x$ where $x$ is a parameter.}
\label{tab:1stpt}
\end{table}

We select $m_0$ and $m_{1/2}$ such that $m_0^{min}(m_{1/2}^{min})<m_0(m_{1/2})<m_0^{max}(m_{1/2}^{max})$ with a number of points $N_{m_0}(N_{m_{1/2}})$ in the scan. This permit a grid-scan on the $(m_0, m_{1/2})$-plane while the other parameters are obtained through the MCMC algorithm. For our scan, we set:
\begin{itemize}
\item $m_0^{min}=0\ \text{GeV}$, $m_0^{max}=1000\ \text{GeV}$ and $N_{m_0}=10$; 
\item $m_{1/2}^{min}=750\ \text{GeV}$, $m_{1/2}^{max}= 1250\ \text{GeV}$ and $N_{m_{1/2}}=10$.
\end{itemize}
Those parameters set a $10\times 10$ grid on the $(m_0, m_{1/2})$-plane. While $m_0$ and $m_{1/2}$ are fixed for all the iterations, the other parameters run according to the MCMC algorithm. The allowed variations $\sigma_i$ on the parameters $\{\mathfrak{p}_i\}$ are set to be 0.1 (see \autoref{eq:next_pt}). The maximum number of iterations is set at $N_{pts}=1000$ for each cell defined by the grid-scan. The iterations end if the penalty function (\textit{i.e.}, the fine-tuning of the electroweak scale $\Delta_{FT}$) converges at a minimum defined by \autoref{eq:Xconvergence} with $\Delta_{\mathcal{X}}= 10^{-2}$.\medskip

Concerning the penalty function (see Subsection \ref{subsec:MCMC4N2MSSM}), the allowed uncertainty on the Higgs mass is set to be $\Delta m_h=7\ \text{GeV}$, which compromises a short computing time and a relevant mass reconstruction. In order to check the validity of the algorithm regarding the theoretical uncertainty of $\Delta m_h=3\ \text{GeV}$, some convergence tests are performed in \ref{sub:convergence}. We also suppose a possible deviation of the penalty function of $\sigma_{\mathcal{X}}=0.01$, meaning that a point with a lower penalty function will be generally defined as the centre point (see \autoref{fig:probaMCMC}). For our analysis, we do not consider constraints on dark matter properties (from \textsc{micrOMEGAs}) and Higgs measurements (with \textsc{HiggsBounds}). \medskip

In order to optimize the computation of the grid-scan, the calculation is done as follow:
\begin{itemize}
\item Starting at a high value of $m_0$ and $m_{1/2}$ (\textit{i.e.}, for the mentioned inputs, $m_0=950\ \text{GeV}$ and $m_{1/2}=1225\ \text{GeV} $), the iterations are done for the first cells. When the convergence occurs, we go to the next cells by changing the value of $m_{0}$, leaving $m_{1/2}$ unchanged ($m_0=850\ \text{GeV}$ and $m_{1/2}=1225\ \text{GeV} $ for our case). The parameters which have minimised the penalty function are used as input parameters for the next cells. Cell after cell, all the points with $m_{1/2}$ are then generated.
\item Fixing now the values of $m_0$, all the cells are computed by taking as input the parameters of the previous cells. 
\end{itemize}  
\subsection{Convergence of the MCMC}\label{sub:convergence}
Before introducing the full grid scan, we would like to check the convergence of the MCMC calculation. As mentioned previously, the full scan can be very time consuming depending on the required precision. Many factors that can affect the computation time such as:
\begin{itemize}
\item the choice of the first point $\{\mathfrak{p}_i\}_{BEGIN}$ (since starting with a point with Landau pole or tachyonic states need in general more iterations to converge compared to points which already generate a consistent spectrum);
\item the admitted variation on the parameters $\sigma_i$ (from \autoref{eq:next_pt});
\item the allowed deviation on the penalty function $\sigma_{\mathcal{X}}$ and the precision of convergence $\Delta_{\mathcal{X}}$;
\item the constraints implemented on the penalty function ($\Delta m_h$, $M_3^{limit}$ or $m_{\tilde{q}}^{limit}$).
\end{itemize}
The preliminary study in this manuscript supposes an uncertainty on the Higgs boson mass of $\Delta m_{h} =7\ \mathrm{GeV}$ with a precision on the convergence of $\mathcal{X}_{CEN}$ of $\Delta_{\mathcal{X}}=10^{-2}$. Those inputs do not permit a high precision on the phenomenological validity of the various points. It can then be essential to test the convergence of the program,  assuming more realistic input parameters. Then we test the program assuming $\Delta m_{h} =3\ \mathrm{GeV}$ with $\Delta_{\mathcal{X}}=10^{-3}$.\medskip

The evolution of the penalty function for three different points in the N2MSSM is given in Figures \ref{fig:N1}, \ref{fig:N2} and \ref{fig:N3}, following the new input parameters. We first notice four plateaus in the evolution of $\log\mathcal{X}_{CEN}$. Those levels represent various regions in the penalty function definition. The first iterations correspond to $\log\mathcal{X}_{CEN}\approx {60}$ , meaning negative squared scalar masses are present in the spectrum. After some iterations, the tachyonic states disappear while the second lightest Higgs is pushed to $m_h^{exp} \pm \Delta m_h$ (\textit{i.e.}, a penalty function of $\log\mathcal{X}_{CEN}\approx 50$). In the evolution \autoref{fig:N3}, we also see the region where $\min\{m^2_{\tilde{q}}\} < (1\ TeV)^2$. The two other points (\ref{fig:N1} and \ref{fig:N2}) seems to directly converge in the last region where the fine-tuning $\Delta_{FT}$ is minimised. This can be explained by the high value of the $m_0$ input ($950\ \text{GeV}$ and $550\ \text{GeV}$ to be compared with $250\ \text{GeV}$) (the effects from the running of the RGEs and the loop corrections must be important to pass the constraints \autoref{eq:minsquark}).
\begin{figure}[H]
    \centering
    \includegraphics[scale=0.7,origin=c]{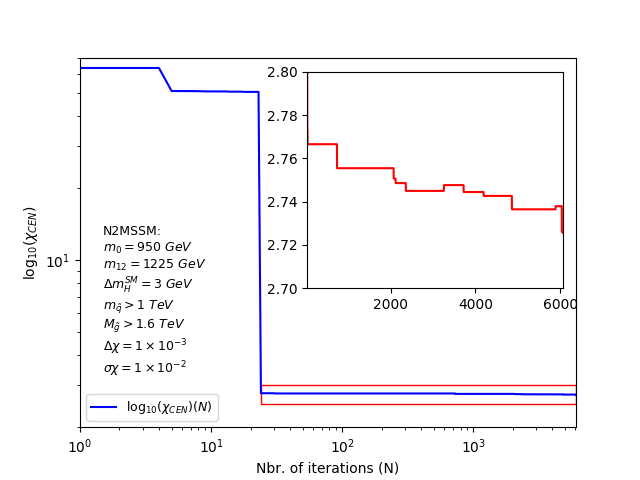}
    \caption{Convergence of the penalty function $\mathcal{X}_{CEN}$ in the N2MSSM with $\Delta m_h=3\ \text{GeV}$, $\Delta_{\mathcal{X}}=10^{-3}$ and $\sigma_{\mathcal{X}}=10^{-2}$ for the point $m_0=950\ \text{GeV}$ and $m_{1/2}=1225\ \text{GeV}$. The second evolution corresponds to the zoomed-in red rectangle.}
  \label{fig:N1}
\end{figure}
Two observations are in order. It can be seen in Figures \ref{fig:N1} and \ref{fig:N2} that the penalty function $\log\mathcal{X}_{CEN}$ does not just decrease after each iteration. Indeed, the penalty function can increase too. This is a specific signature of the MCMC algorithm since the probability of defining the new point as the central (even if $\mathcal{X}_{NEXT} > \mathcal{X}_{CEN}$) is never zero. This property helps to go over walls with high $\mathcal{X}$ values.\newline
The second point is that the convergence time can vary to five between the different points ($2500$ iterations for \ref{fig:N3} and $12 500$ for \ref{fig:N2}). The point with the lowest $m_0$-value (see \autoref{fig:N3}) take the most iterations to reconstruct a coherent spectrum. \medskip

Following those checks, we see that assuming more constraining (and more realistic) input parameters lead to a convergence of the program after more than $1000$ iterations. It is then clear that assuming a limit number of iterations of $N_{pts} = 1000$ do not permit a good convergence in all cells. We will assume $\Delta m_h = 7 \ \mathrm{GeV}$ and $\Delta_{\mathcal{X}} = 10^{-2}$ in the following preliminary study.
\begin{figure}[!ht]
    \centering
    \includegraphics[scale=0.7,origin=c]{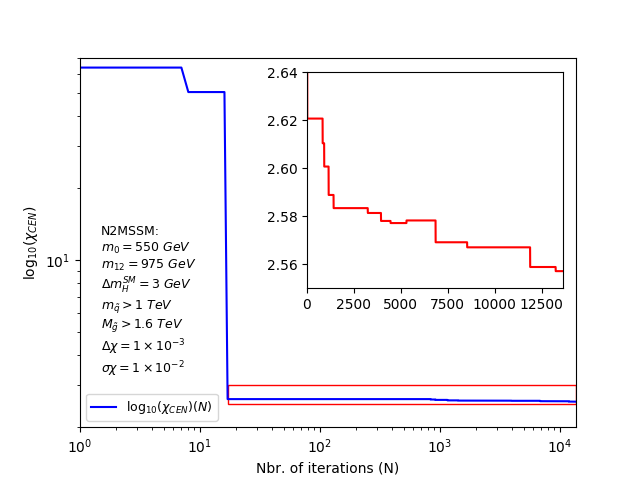}
    \caption{Convergence of the penalty function $\mathcal{X}_{CEN}$ in the N2MSSM with $\Delta m_h=3\ \text{GeV}$, $\Delta_{\mathcal{X}}=10^{-3}$ and $\sigma_{\mathcal{X}}=10^{-2}$ for the point $m_0=550\ \text{GeV}$ and $m_{1/2}=975\ \text{GeV}$. The second evolution corresponds to the zoomed-in red rectangle.}
  \label{fig:N2}
\end{figure}
\begin{figure}[H]
    \centering
    \includegraphics[scale=0.7,origin=c]{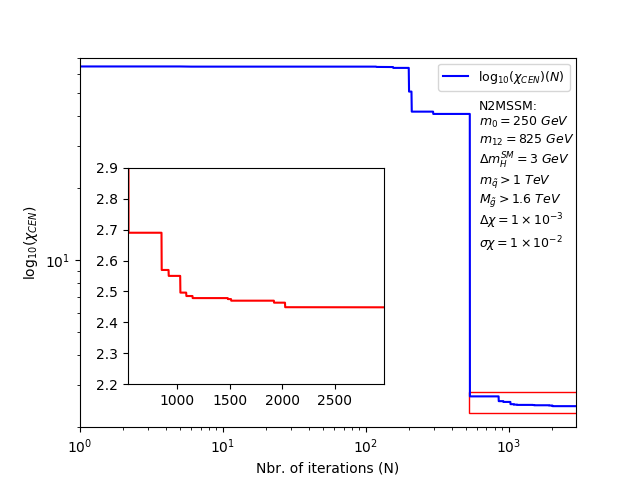}
    \caption{Convergence of the penalty function $\mathcal{X}_{CEN}$ in the N2MSSM with $\Delta m_h=3\ \text{GeV}$, $\Delta_{\mathcal{X}}=10^{-3}$ and $\sigma_{\mathcal{X}}=10^{-2}$ for the point $m_0=250\ \text{GeV}$ and $m_{1/2}=825\ \text{GeV}$. The second evolution corresponds to the zoomed-in red rectangle.}
  \label{fig:N3}
\end{figure}

\subsection{Reproductiblity of the results}\label{subsec:reprod}
We add some remarks on the reproducibility of the results. Since the MCMC algorithm is based on stochastic processes, the results follow a probability distribution. The choice of the parameter $\Delta_{\mathcal{X}}$ defines the precision of the convergence to the \textit{real} minimum of the penalty function. Choosing $\Delta_{\mathcal{X}} = 10^{-2}$ may lead to a non-negligible variation of the final penalty function $\mathcal{X}$. For this purpose, we can produce several scans in order to obtain distributions of $\Delta_{FT}$ in the NMSSM and in the N2MSSM. To deals with the computation time, we generated hundred times the minimised fine-tuning for four different cells in the $(m_0,m_{1/2})$-plane: $(950\ \mathrm{GeV},1225\ \mathrm{GeV})$, $(950\ \mathrm{GeV},775\ \mathrm{GeV})$, $(550\ \mathrm{GeV},1225\ \mathrm{GeV})$ and $(550\ \mathrm{GeV},775\ \mathrm{GeV})$. Some of the distributions are shown in \autoref{fig:FT_distrib1} (for the NMSSM) and \autoref{fig:FT2_distrib1} (for the N2MSSM). \medskip 

We note that some distributions are antisymmetric, since a tail appears for high values of $\Delta_{FT}$ (the other distributions, not presented here, also have the same structure). This is a signature of the convergence of the penalty function in a local minimum. Obviously, choosing a smaller $\Delta_{\mathcal{X}}$ value may lead to a smaller standard deviation $\sigma$ and possibly suppress the tail in the distributions. To model those distributions, we can assume two different functions:
\begin{itemize}
\item a Crystal Ball function (denoted as $f_{CB}$):
\begin{gather}
f_{CB}(x,\alpha,\bar{x},\sigma, n)=\begin{cases}
 e^{-\frac{(x-\bar{x})^2}{2 \sigma ^2}} & \frac{\bar{x}-x}{\sigma }>-\alpha  \\
 n^n e^{-\frac{\left| \alpha \right| ^2}{2}} \left(\frac{1}{\left| \alpha \right| }\right)^n \left(-\left| \alpha \right|
   +\frac{n}{\left| \alpha \right| }+\frac{x-\bar{x}}{\sigma }\right)^{-n} & \frac{\bar{x}-x}{\sigma }\leq -\alpha 
\end{cases}\ , 
\label{eq:CB}
\end{gather}
\item a gaussian with exponential distribution (denoted as $f_{ET}$):
\begin{gather}
f_{ET}(x,k,\bar{x},\sigma)=\begin{cases}
 e^{-\frac{(x-\bar{x})^2}{2 \sigma ^2}} & \frac{\bar{x}-x}{\sigma }\geq -k \\
 e^{\frac{k^2}{2}+\frac{k (\bar{x}-x)}{\sigma }} & \frac{\bar{x}-x}{\sigma }<-k
\end{cases}\ .
\label{eq:ET}
\end{gather}
\end{itemize}
The fits are superimposed on the distributions (see \autoref{fig:FT_distrib1} and \autoref{fig:FT2_distrib1}). From those fits, we extract the parameters and compute a $\chi^2$ test (see \autoref{tab:fitdistrib} for the results).
\begin{figure}[H]
    \centering
    \includegraphics[scale=0.6]{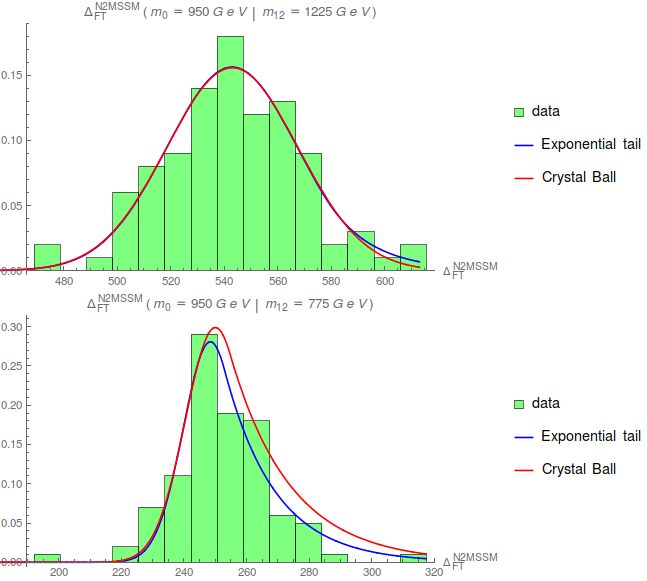}
    \caption{Distribution of $\Delta_{FT}^{N2MSSM}$ for the cells $(950\ \mathrm{GeV},1225\ \mathrm{GeV})$ and $(950\ \mathrm{GeV},775\ \mathrm{GeV})$. The histogram corresponds to the data where the red and blue lines are fits with a Crystal Ball and an Exponential Tail distribution, respectively.}
\label{fig:FT2_distrib1}
\end{figure}
\begin{figure}[H]
    \centering
    \includegraphics[scale=0.6]{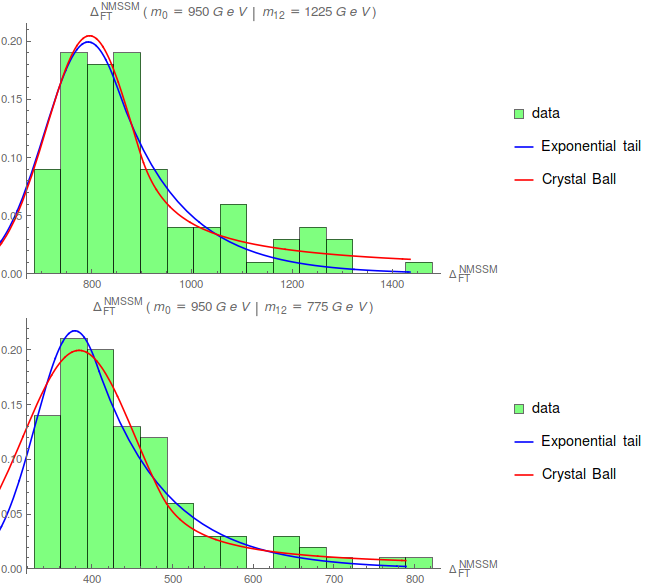}
    \caption{Distribution of $\Delta_{FT}^{NMSSM}$ for the cells $(950\ \mathrm{GeV},1225\ \mathrm{GeV})$ and $(950\ \mathrm{GeV},775\ \mathrm{GeV})$. The histogram corresponds to the data where the red and blue lines are fits with a Crystal Ball and an Exponential Tail distribution, respectively.}
\label{fig:FT_distrib1}
\end{figure}
The results of the various fits are shown in Table \autoref{tab:fitdistrib}.
\begin{table}[H]
\begin{center}
\begin{tabular}{ |c|c|c||c|c|c|c| } 
 \hline
  & & Cells & $(950,1225)$ & $(950,775)$ & $(550,1225)$ & $(550,775)$ \\
\hline
\hline 
 \multirow{9}{*}{NMSSM} & \multirow{5}{*}{$f_{CB}$} & $\alpha$ & 1.17 & 1.46 & 1.42 & 0.32 \\
 & & $\bar{x}$ & 796.18 & 384.09 & 832.47  & 366.04  \\
 & & $\sigma$ & 86.66 & 69.40 & 94.39 & 36.38 \\
 & & $n$ & 1.00 & 1.18 & 1.00 & 6.83   \\
 & & $\chi_{CB}^2/ndf$ & $1.03\times 10^{-2}$ & $5.39\times 10^{-3}$ & $9.72\times 10^{-3}$ & $1.89\times 10^{-2}$ \\
 \cline{2-7}
 & \multirow{4}{*}{$f_{ET}$} & $k$ & 0.69 & 0.56 & 0.86 & 0.29 \\
 & & $\bar{x}$ & 793.53 & 378.87 & 829.79 & 366.80 \\
 & & $\sigma$ & 87.65 & 48.98 & 92.73 & 37.34  \\
 & & $\chi_{ET}^2/ndf$ & $3.17\times 10^{-2}$ & $1.07\times 10^{-2}$ & $5.09\times 10^{-2}$ & $4.00\times 10^{-2}$ \\
 \hline
 \multirow{9}{*}{N2MSSM} & \multirow{6}{*}{$f_{CB}$} & $\alpha$ & 10.45 & 0.50 & 0.94 & 25046.90 \\
 & & $\bar{x}$ & 543.10 & 250.05 & 549.40  & 256.00  \\
 & & $\sigma$ & 24.56 & 9.60 & 19.77 & 14.96 \\
 & & $n$ & 10.45 & 133.109 & 50.60 & 1929.71   \\
 & & $\chi_{CB}^2/ndf$ & $1.64\times 10^{-2}$ & $1.49\times 10^{2}$ & $6.14\times 10^{-2}$ & $2.59\times 10^{28}$ \\
 & & $\chi_{CB}^2/ndf_{corr.}$ & $1.64\times 10^{-2}$ & $1.15\times 10^{-2}$ & $4.37\times 10^{-3}$ & $3.07\times 10^{-1}$ \\
 \cline{2-7}
 & \multirow{5}{*}{$f_{ET}$} & $k$ & 1.46 & 0.53 & 0.92 & 0.87 \\
 & & $\bar{x}$ & 543.02 & 250.87 & 549.38 & 255.13 \\
 & & $\sigma$ & 24.37 & 12.46 & 19.78 & 13.68  \\
 & & $\chi_{ET}^2/ndf$ & $1.27\times 10^{-2}$ & $1.76\times 10^{3}$ & $9.42\times 10^{-2}$ & $1.75$ \\
 & & $\chi_{ET}^2/ndf_{corr.}$ & $1.27\times 10^{-2}$ & $2.02\times 10^{-2}$ &  $4.34\times 10^{-3}$ & $2.60\times 10^{-3}$ \\
\hline
\end{tabular}
\end{center}
\caption{Results of the fits on the distributions $\Delta_{FT}^{NMSSM}$ and $\Delta_{FT}^{N2MSSM}$.}
\label{tab:fitdistrib}
\end{table}
We note that some $\chi^2/ndf$-values can be relatively high. This effect is due to extremum values in the distributions of $\Delta_{FT}^{N2MSSM}$ (for example, the first bin in \autoref{fig:FT2_distrib1} with $m_{1/2}=775\ \text{GeV}$). In order to obtain reasonable values, we also compute a corrected $\chi^2/ndf$ by not taking into account problematic bins. This issue may not be present considering more statistics. 

\subsection{Presentation of the main scans}
We keep on with the main scan of the N2MSSM and NMSSM. By assuming the input parameters as in \ref{subsec:methodMCMC}, we generate a $10\times 10$ grid scan in the $(m_0,m_{1/2})$-plane which minimise the fine-tuning of the Z-boson mass $\Delta_{FT}$. The results for the N2MSSM are shown in \autoref{fig:scanN2} whilst the grid of the NMSSM can be found in \autoref{fig:scanN}.\medskip

Several remarks are relevant. Firstly, most of the cells in the $m_{0}=50\ \text{GeV}$ line for the NMSSM grid scan have not converged (which is not the case for the N2MSSM). The penalty function for those cells have reached a plateau of $\log\mathcal{X}_{CEN}\approx 50$ without going any further, \textit{i.e.}, the Higgs boson mass has not been properly reconstructed. The only two cells that have converged ($m_{1/2}=925\ \mathrm{GeV}$ and $m_{1/2}=975\ \mathrm{GeV}$) have minimised their penalty functions with a more higher fine-tuning than the neighbour cells. The minimisation of the penalty function could have been done in a local minimum and not in a global minimum. We will then not take into account the $m_0=50\ \mathrm{GeV}$ line in the following study.  
\begin{figure}[H]
    \centering
    \includegraphics[scale=0.85,origin=c]{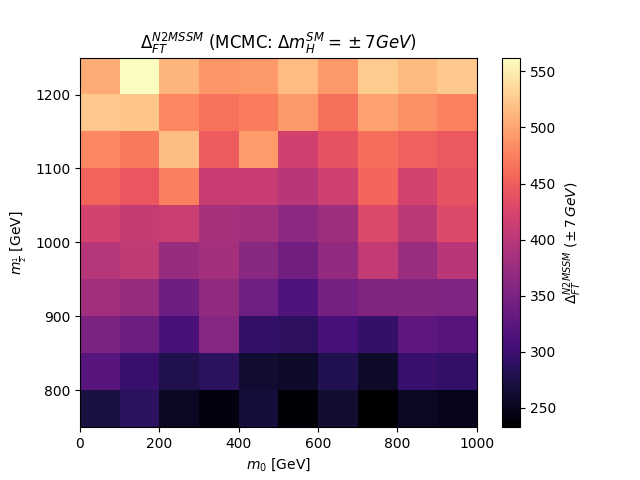}
    \caption{$10\times 10$ grid scan in the $(m_0,m_{1/2})$-plane for the N2MSSM. After reconstructing a valuable squark sector, gluino and (SM) Higgs boson mass, the fine-tuning $\Delta_{FT}$ is minimised. The inputs parameters are defined in \ref{subsec:methodMCMC}.}
    \label{fig:scanN2}
\end{figure}
\begin{figure}[H]
    \centering
    \includegraphics[scale=0.85,origin=c]{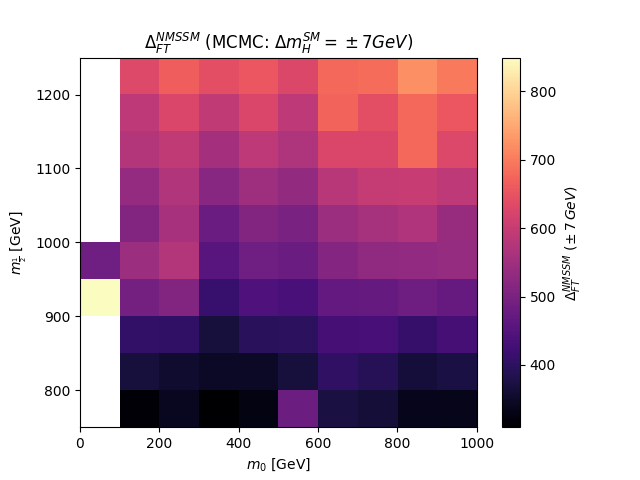}
    \caption{$10\times 10$ grid scan in the $(m_0,m_{1/2})$-plane for the NMSSM. After reconstructing a valuable squark sector, gluino and (SM) Higgs boson mass, the fine-tuning $\Delta_{FT}$ is minimised. The inputs parameters are defined in \ref{subsec:methodMCMC}.}
    \label{fig:scanN}
\end{figure}
Secondly, we can see in the scans of the two models that the value of the fine-tuning $\Delta_{FT}$ is sensitive to the value of the universal soft-breaking mass gaugino term $m_{1/2}$, whilst the value of $m_{0}$ seems not to impact $\Delta_{FT}$. These effects can be explained by the definition of the fine-tuning $\Delta_{FT}$ (sse \autoref{eq:DeltaFT}) and the RGEs structures. Indeed, the fine-tuning $\Delta_{FT}$ is defined as the maximum value of all the fine-tuning of the Z-boson mass $\{\Delta_i\}$ in relation to the parameter $\mathfrak{p}_i$. It turns out that the maximum value for all the scans corresponds to the fine-tuning according to the soft-gaugino mass $\Delta_{m_{1/2}}$, \textit{i.e.}, $\Delta_{FT}=\Delta_{m_{1/2}}$. The high sensibility of the gluino soft mass $M_3$ in the running of the soft-scalar masses is directly related to the Z-boson mass through radiative corrections. \medskip

To compare the results from the two models, we have to reconstruct the difference of the fine-tuning $\Delta_{FT}^{NMSSM} - \Delta_{FT}^{N2MSSM}$ (see \autoref{fig:scanDelta}). We can remark that the inequality $\Delta_{FT}^{NMSSM}>\Delta_{FT}^{N2MSSM}$ stands for all the grid scan. It is, however, not obvious to directly conclude to the reduction of the fine-tuning in the case of the N2MSSM. Indeed, the MCMC algorithm being based on stochastic processes and the convergence being only defined numerically, several iterations on the same cell could lead to some differences on the final spectrum, and so on the final fine-tuning (as already seen in Subsection \ref{subsec:reprod}). We must then check if those differences on the fine-tuning are statistically signicative.
\begin{figure}[H]
    \centering
    \includegraphics[scale=0.85,origin=c]{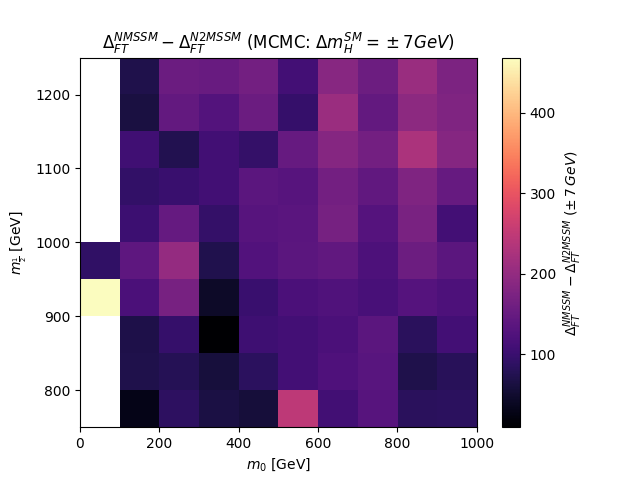}
    \caption{Difference of the fine-tuning $\Delta_{FT}^{NMSSM}-\Delta_{FT}^{N2MSSM}$ in the NMSSM and the N2MSSM in the $10\times 10$ grid scan in the $(m_0,m_{1/2})$-plane.}
    \label{fig:scanDelta}
\end{figure}
For this purpose, we need to get estimations on the uncertainties of the difference of fine-tuning $\Delta_{FT}^{NMSSM} - \Delta_{FT}^{N2MSSM}$ after convergence. The best way to get those estimations is through the distributions of $\Delta_{FT}$ in \autoref{subsec:reprod}. Since we have only four points in the $(m_0-m_{1/2})$-plane, the full plane is obtained by linear regression on all the grid. The standard deviation is get from the distributions of $\Delta_{FT}^{NMSSM} - \Delta_{FT}^{N2MSSM}$. The distributions for ($m_0=950\ \text{GeV}$, $m_{1/2}=1225\ \text{GeV}$) and ($m_0=950\ \text{GeV}$, $m_{1/2}=775\ \text{GeV}$) are represented in \autoref{fig:DFT_distrib1}.
\begin{figure}[H]
    \centering
    \includegraphics[scale=0.6]{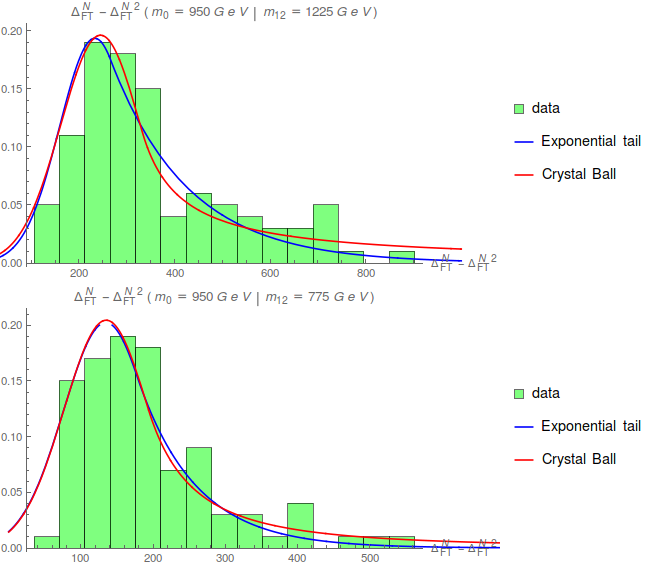}
    \caption{Distribution of the difference $\Delta_{FT}^{NMSSM}-\Delta_{FT}^{N2MSSM}$ for the cells $(950\ \mathrm{GeV},1225\ \mathrm{GeV})$ and $(950\ \mathrm{GeV},775\ \mathrm{GeV})$. The histogram corresponds to the data where the red and blue lines are fits with a Crystal Ball and an Exponential Tail distribution, respectively.}
\label{fig:DFT_distrib1}
\end{figure}

A $\chi^2$-test is performed for each fit (results of the fits are set in \autoref{tab:chi2_test}). Following those results, the two functions correctly represent the data, with the exponential tail $f_{ET}$ which seems to better represent the distributions for all cells.\newline
\begin{table}[H]
\begin{center}
\begin{tabular}{ |c|c||c|c|c|c| } 
 \hline
 & Cells & $(950,1225)$ & $(950,775)$ & $(550,1225)$ & $(550,775)$ \\
\hline
\hline 
 \multirow{5}{*}{$f_{CB}$} & $\alpha$ & $1.05$ & $0.97$ & $1.41$ & $0.60$ \\
 & $\bar{x}$ & $245.11$ & $135.08$ & $268.01$ & $107.15$ \\
 & $\sigma$ & $83.17$ & $58.60$ & $100.90$ & $43.90$ \\
 & $n$ & $1.00$ & $2.06$ & $1.00$ & $2.21$ \\
 & $\chi_{CB}^2/ndf$ & $6.57\times 10^{-3}$ & $1.66\times 10^{-1}$ & $4.44\times 10^{-3}$ & $1.10\times 10^{-1}$ \\
 \hline
 \hline
  \multirow{4}{*}{$f_{ET}$} & $k$ & $0.455$ & $0.729$ & $0.91$ & $0.378$ \\
 & $\bar{x}$ & $233.2$ & $134.5$ & $266.8$ & $103.5$ \\
 & $\sigma$ & $73.67$ & $58.78$ & $100.94$ & $42.78$ \\
 & $\chi_{ET}^2/ndf$ & $1.19\times 10^{-2}$ & $7.89\times 10^{-1}$ & $9.34\times 10^{-3}$ & $1.48\times 10^{-1}$ \\
 \hline
\end{tabular}
\end{center}
\caption{Results for the different fits with the functions \autoref{eq:CB} and \autoref{eq:ET}.}
\label{tab:chi2_test}
\end{table}
\subsection{Reduction of the fine-tuning $\Delta_{FT}$ in the N2MSSM}
After getting the standard deviations for the four cells, we can generate a complete grid of $\sigma$-value on the $(m_0,m_{1/2})$-plane by assuming a linear regression between those cells. Checking the standard deviation
\begin{gather}
 X_{\sigma} = \frac{\big(\Delta_{FT}^{NMSSM} - \Delta_{FT}^{N2MSSM}\big)}{\sigma_{FT}^{N-N2}} \nn
\end{gather}
(where $\sigma_{FT}^{N-N2}$ is the value of standard deviation after the linear regressions) gives an estimation of the p-value associated with the hypothesis: 
\begin{center}
\textit{The values of $\Delta_{FT}^{NMSSM}$ and $\Delta_{FT}^{N2MSSM}$ are not statistically compatible.}
\end{center}
The grid of $X_{\sigma}$ in the $(m_0,m_{1/2})$-plane assuming a Crystall Ball fit $f_{CB}$ and an exponential tail fit $f_{ET}$ can be found in \ref{fig:ratioCB} and \ref{fig:ratioET}.\medskip

As well as the grid scan of the NMSSM and the N2MSSM, the value of $m_{0}$ does not interfere with the value of $X_{\sigma}$. While the $m_0 > 1000\ \mathrm{GeV}$ plane allows a standard deviation higher than $3\sigma$, the other part of the plane corresponds to a standard deviation lower than $3\sigma$. Some cells at high values of the soft-scalar masses barely exceed the $5\sigma$ limit. Depending on the threshold we assume for the calculated p-value, the results are then statistically significant for high value of $m_0$ only. If we suppose a $5\sigma$ result, only few cells agree with the previously mentioned hypothesis.\medskip

Those results do not permit to conclude the reduction of the fine-tuning of the Z-boson mass in the N2MSSM. Nevertheless, the $3\sigma$ deviation in parts of the scan shows that a more precise analysis is needed to get a statistically significant deviation (higher than $5\sigma$) between the N2MSSM and the NMSSM. Allowing more computation time with a more strict convergence parameter ($\Delta_{\mathcal{X}}<10^{-3}$) could be a possible solution to this issue.
\begin{figure}[!ht]
    \centering
    \includegraphics[scale=0.7,origin=c]{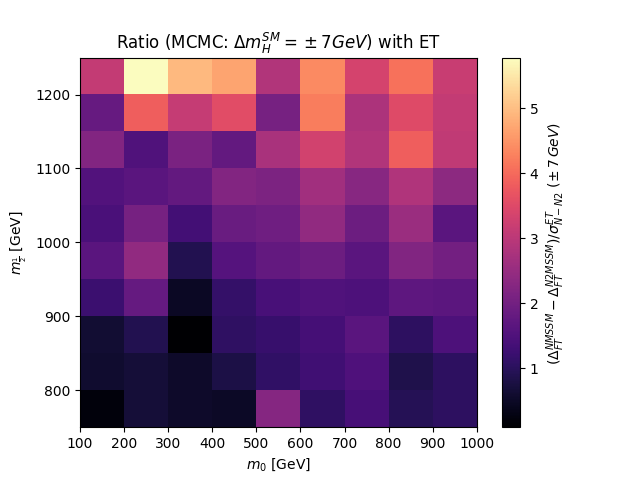}
    \caption{$X_{\sigma}$-grid with an Exponential tail $f_{ET}$ fit.}
\label{fig:ratioET}
\end{figure}
\begin{figure}[H]
    \centering
    \includegraphics[scale=0.7,origin=c]{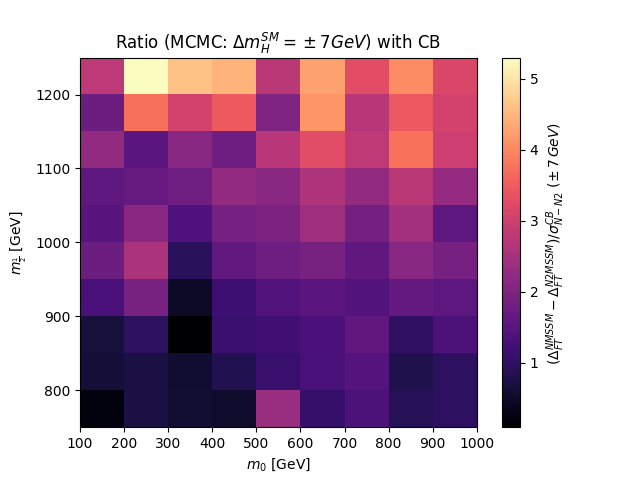}
    \caption{$X_{\sigma}$-grid with an Exponential tail $f_{CB}$ fit.}
\label{fig:ratioCB}
\end{figure}

\subsection{Release of constraints on parameters in the N2MSSM}
In the previous section, we have discussed the reduction of the fine-tuning of the electroweak scale in the N2MSSM. A grid-scan in the $(m_0,m_{1/2})$-plane has been made, assuming some phenomenological constraints. This section analyse the variety of spectrums obtained through the MCMC algorithm. \medskip

Let us first consider the reconstruction of the SM Higgs boson mass. The parameters which are the most important for the eigenvalues of the scalar Higgs masses are $\tan\beta$ and $\lambda_i$. We will then focus on the evolution of the second lightest Higgs mass regarding those parameters. The mass of the second lightest Higgs mass as function of the $\lambda$-parameter (for the NMSSM) and $\sqrt{\lambda_1^2 + \lambda_2^2}$ (for the N2MSSM) are shown in \autoref{fig:l_mh2}. We note that the reconstructed Higgs masses for the NMSSM remain close to the upper limit value whilst points of the N2MSSM can be found in the acceptable Higgs mass range. Even though the Higgs boson mass has been restricted to $125\pm7\ \mathrm{GeV}$, we can already find spectrums with a Higgs boson mass $m_{h2}\approx 125\ \mathrm{GeV}$. Moreover, the value of the $\lambda$-parameter of the NMSSM stay in a small range $0.1 < \lambda < 0.2$ where the effective parameter for the N2MSSM take a larger range of values ($0.1 < \sqrt{\lambda_1^2 +\lambda_2^2} < 0.35$). Note that those results do not mean that a Higgs boson mass near $125\ \mathrm{GeV}$ in the NMSSM is impossible since it has already been shown that such spectrum can be generated (see for example \cite{Ellwanger_2012}). A more precise analysis (by imposing $\Delta m_h=3\ \mathrm{GeV}$) will allow to properly compare the two models with phenomenologically acceptable Higgs boson mass.
\begin{figure}[!htb]
    \centering
    \begin{minipage}[t]{.45\textwidth}
        \centering
        \includegraphics[scale=0.5]{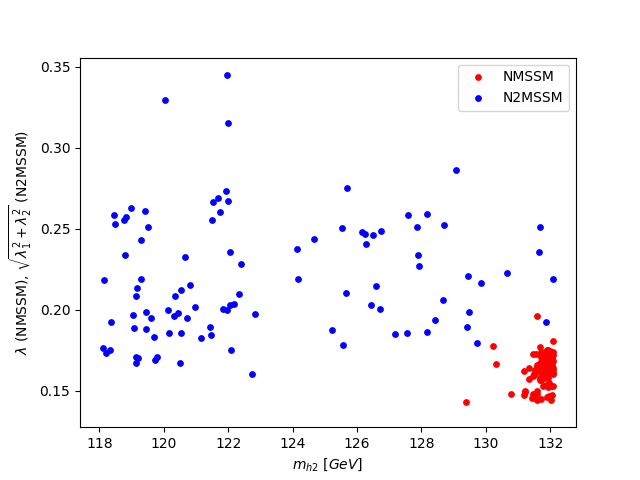}
        \caption{Results of the grid scan: $\lambda$ (for the NMSSM, in red) and $\sqrt{\lambda_1^2 + \lambda_2^2}$ (for the N2MSSM, in blue) as function of the second lightest Higgs boson mass.}
        \label{fig:l_mh2}
    \end{minipage}\quad%
    \begin{minipage}[t]{0.45\textwidth}
        \centering
        \includegraphics[scale=0.5]{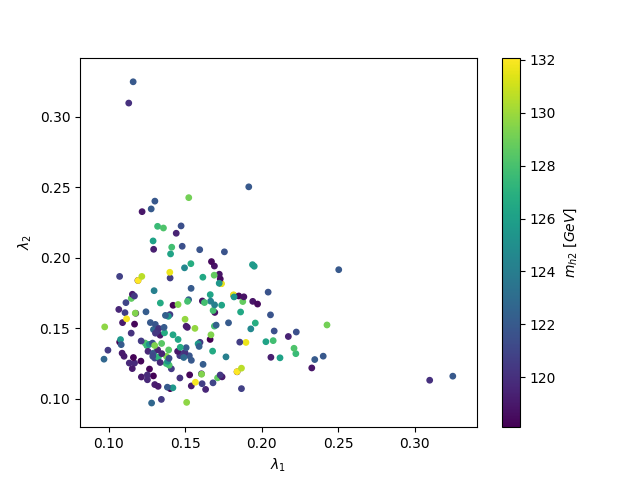}
        \caption{Results of the grid scan: distribution of the $\lambda_1$ and $\lambda_2$ values as function of the second lightest Higgs boson mass.}
        \label{fig:l1l2}
    \end{minipage}
\end{figure}
The distribution of $\lambda_1$ and $\lambda_2$ for the N2MSSM is also given in \autoref{fig:l1l2} (the values have been symmetrised in the $(\lambda_1,\lambda_2)$-plane). Most values have been spread in the range $[0.1, 0.2]$ with top values for $\lambda_i\approx 0.3$. The deviation parameters $\sigma_{\lambda i}$ (presented in \autoref{tab:1stpt}) could be enlarged in order to allow a higher $\lambda_i$-parameters value (with $\lambda_i\sim 0.7$ which is one of the configurations presented in \ref{sub:difficultNMSSM}).\medskip

A small remark on gluino mass $M_3$ can also be made. The evolution of the universal gaugino mass $m_{1/2}$ as function of the reconstructed gluino mass $M_3$ for the NMSSM and the N2MSSM is given in \autoref{fig:M3}. The evolution is linear which directly comes from the structure of the RGEs of the gluino mass. We note that the results are equivalent for the two models, which is coherent since the new superfields are singlets under the gauge group of the Standard Model. Considering the constraint imposed on the gluino mass in the penalty function (\autoref{eq:PenFunM3}), we can then directly constrain the universal gaugino mass term. Indeed, assuming a gluino mass $M_3 > 1.6\ \mathrm{TeV}$ gives an upper bound of $m_{1/2} > 700\ \mathrm{GeV}$\footnote{The first grid scans for this preliminary analysis was performed with $m_{1/2}\in [500\ \mathrm{GeV},1000\ \mathrm{GeV}]$. The range of values for $m_{1/2}$ has been changed to  $m_{1/2}\in [750\ \mathrm{GeV},1250\ \mathrm{GeV}]$ to avoid this issue.}.

\begin{figure}[!htb]
    \centering
    \begin{minipage}[t]{.45\textwidth}
        \centering
        \includegraphics[scale=0.5]{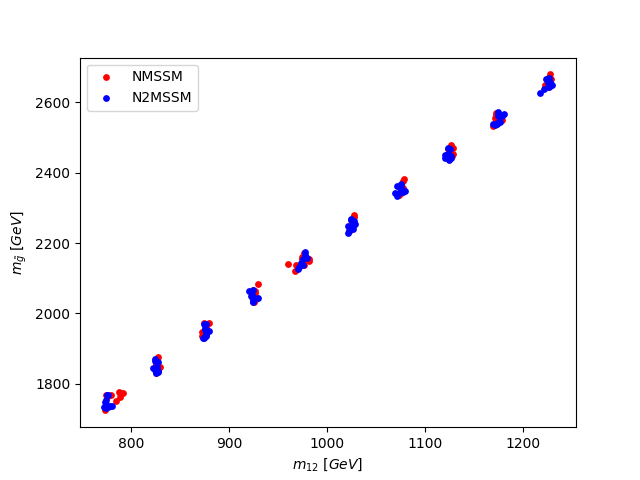}
        \caption{Results of the grid scan: universal gaugino mass $m_{1/2}$ (in red for the NMSSM and in blue for the N2MSSM) as function of the gluino mass.}
        \label{fig:M3}
    \end{minipage}\quad%
    \begin{minipage}[t]{0.45\textwidth}
        \centering
        \includegraphics[scale=0.5]{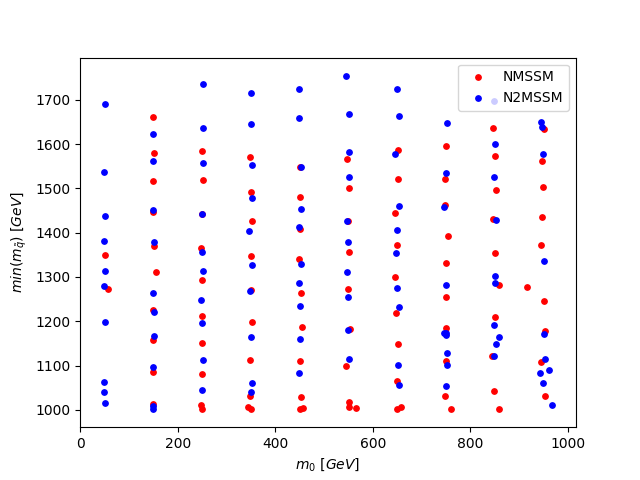}
        \caption{Results of the grid scan: universal scalar mass $m_{0}$ (in red for the NMSSM and in blue for the N2MSSM) as function of the lightest squark mass $\min (m_{\tilde{q}})$.}
        \label{fig:msq}
    \end{minipage}
\end{figure}

\begin{figure}[H]
    \centering
    \includegraphics[scale=0.5]{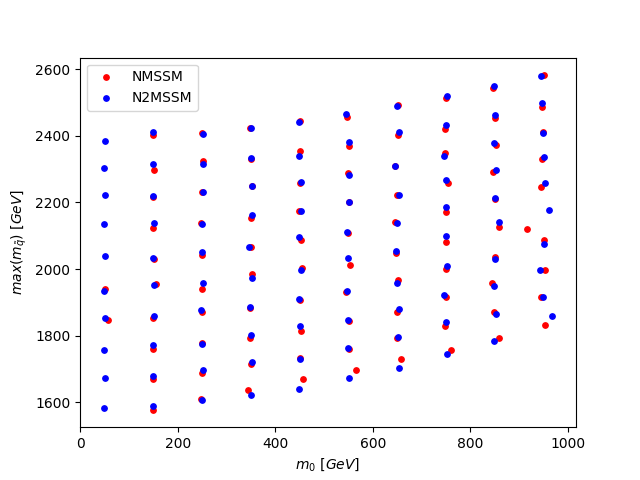}
    \caption{Results of the grid scan: universal scalar mass $m_{0}$ (in red for the NMSSM and in blue for the N2MSSM) as function of the heaviest squark mass $\max (m_{\tilde{q}})$.}
    \label{fig:maxsq}
\end{figure}

During the minimisation of the penalty function, we have also imposed an upper bound of the squark masses of $1\ \mathrm{TeV}$. In order to check the new effects of the N2MSSM, the lightest squark mass as function of $m_0$ can be found in \autoref{fig:msq}. We can note that even though the singlets can act on the squark sector (through the Higgs sector), it seems to not highly affect the calculated masses. The loops corrections to the squark mass matrices are also very similar in the NMSSM and the N2MSSM (between $40$ and $50\ \mathrm{GeV}$).\medskip

We now conclude with a last remark. We can also check the calculus of observables that have not been restricted in the penalty function. In the Section \ref{sub:spheno}, we have mentioned that the \textsc{SPheno} program computes the anomalous magnetic moment of the leptons $a_{\ell}=\frac{1}{2}(g-2)_{\ell}$. Considering the latest measurement of the anomal magnetic moment of the muon \cite{gm2_calculus}\cite{gm2_exp}, it can be interesting to analyse the distribution of the supersymmetric contribution to $a_{\mu}$ in the NMSSM and in the N2MSSM. The measured deviation of $4.2\sigma$ corresponds to $\Delta a_{\mu} \equiv a_{\mu}^{exp.} - a_{\mu}^{SM} = (251\pm 59)\times 10^{-11}$ (where $a_{\mu}^{exp.}$ is the measure obtained by the Muon $g-2$ Collaboration and $a_{\mu}^{SM}$ the SM contributions including QED, hadronic and electroweak processes). The distributions of $\Delta a_{\mu}^{SUSY}$ in the NMSSM and the N2MSSM obtained by the MCMC-algorithm are displayed in \ref{fig:amu_N2} and \ref{fig:amu_N}.
\begin{figure}[H]
    \centering
    \begin{minipage}[t]{.45\textwidth}
        \centering
        \includegraphics[scale=0.5]{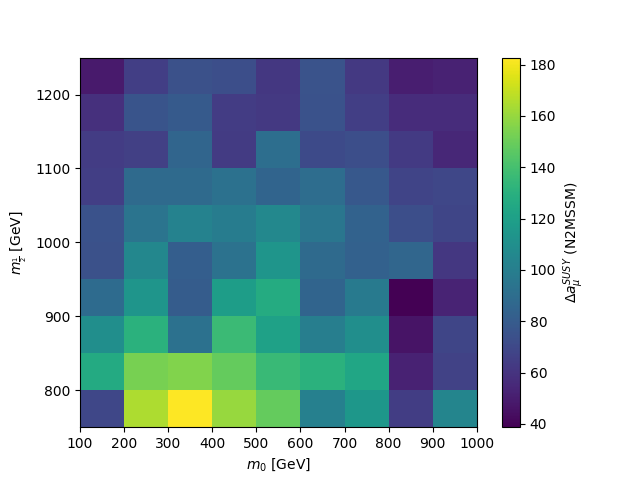}
        \caption{Supersymmetric contribution to the anomalous magnetic moment of the muon $\Delta a_{mu}^{SUSY}$ ($\times 10^{11}$) in the N2MSSM.}
        \label{fig:amu_N2}
    \end{minipage}\quad%
    \begin{minipage}[t]{0.45\textwidth}
        \centering
        \includegraphics[scale=0.5]{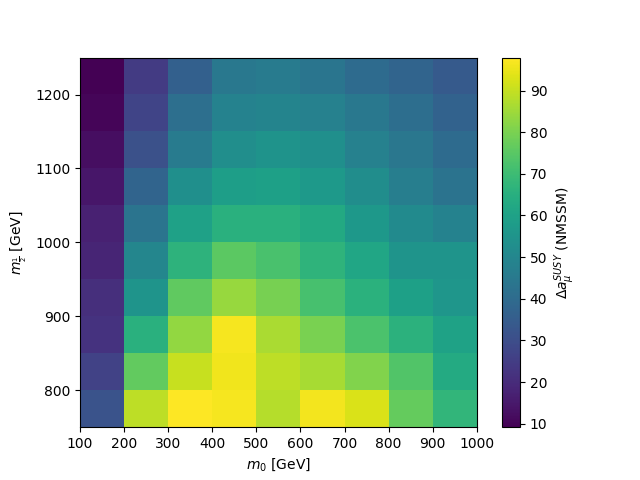}
        \caption{Supersymmetric contribution to the anomalous magnetic moment of the muon $\Delta a_{mu}^{SUSY}$ ($\times 10^{11}$)  in the NMSSM.}
        \label{fig:amu_N}
    \end{minipage}
\end{figure}
Assuming the constraints mentioned in Subsection \ref{subsec:MCMC4N2MSSM}, we see that the N2MSSM framework allows a higher supersymmetric contribution on the anomalous magnetic moment $a_{\mu}$ (potentially due to the new neutralino state). However, no values from those scans are coherent with the experimental measurement $\Delta a_{\mu} = (251\pm 59)\times 10^{-11}$ \cite{gm2_exp}. Implementing this constraint on the penalty function could help to recover a valid SUSY contribution.


\section{Toward a more complete analysis of the N2MSSM}
All of those results represent a preliminary analysis on the N2MSSM. The spectrum generator obtained by the programs \textsc{SARAH} and \textsc{SPheno} allows to calculate all the mass spectrum with some low-energy observables. One of the issues from the presented scans is the computing time. Scanning high-dimensional space parameters (for the NMSSM and the N2MSSM) is a complex process and the MCMC algorithm helps to find a phenomenologically acceptable zone of the parameters-space. A choice must be made between assuming strong constraints and demanding a short computing time. The strategy used for this preliminary study is the second approach by assuming a high uncertainty on the Higgs boson mass of $\Delta m_{h} = \pm 7\  \mathrm{GeV}$ (where the theoretical constraint is $\Delta m_h = \pm 3\ \mathrm{GeV}$) and a convergence parameter of the penalty function of $\Delta_{\mathcal{X}}=10^{-2}$. We have seen from our study that a convergence of the algorithm assuming $\Delta m_h=\pm 3\ \mathrm{GeV}$ and $\Delta_{\mathcal{X}} = 10^{-3}$ is possible (as illustrated in \ref{sub:convergence}) but requires more iterations. A possible solution to those issues is grid computing. Some preliminary tests on grid have already been done but have been interrupted due to access issues (and COVID-19 crisis). \medskip

The reduction of the fine-tuning between the N2MSSM and the NMSSM is not statistically significant. However, it shows some cells with a deviation higher than $3\sigma$, meaning that assuming a more constrained scan with more statistics could result in more interesting results ($X_\sigma > 5$) with high $m_{1/2}$ values.\medskip

We have also shown several observables calculated by \textsc{SPheno} or other third party programs. The implementation of the outputs from \textsc{MicrOmegas} and \textsc{HiggsBounds} in the penalty function has been explained in Subsection \ref{subsec:MCMC4N2MSSM}. The constraints on the dark matter and Higgs properties have, however, not been taken into account for this study. A more precise analysis will use those programs to restrict the parameters space. The implementation of the $a_{\mu}$ measurement in the program can also be done. \medskip

Finally, we have mentioned in Subsection \ref{sub:difficultNMSSM} several mechanisms in the N2MSSM for constructing a Higgs boson mass near $125\ \mathrm{GeV}$. A full study of the various scenarios could be interesting in order to fully highlight the new contributions coming from the second singlet $S^2$.




\label{n2mssm}

\chapter{Long Lived Stop in a simplified MSSM-like SUSY model}
To this day, the particle physics experiments continue to hunt some deviations of the Standard Model, which could be identified as signatures of new physics such as supersymmetry and supergravity\footnote{Many other models are also searched, \textit{e.g.}, GUT theory, composite models or extra-dimensional theory.}.
\begin{figure}
    \centering
      \includegraphics[width=.9\linewidth]{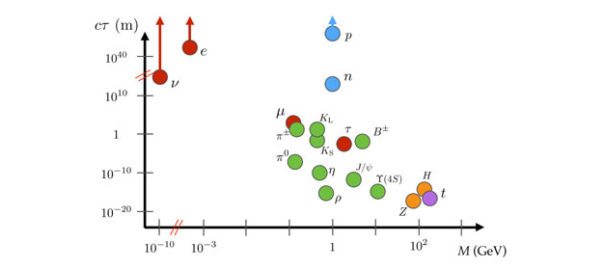}
    \caption{Average flight distance of particles of the Standard Model in the particle frame \cite{llpSM}.}
    \label{fig:llpSM}
\end{figure}
The first searches of supersymmetry at the LHC were based on several signatures, such as the calculation of invariant masses or the search of excess of $MET$ (\textit{Missing Transverse Energy}) (based on the presence of LSP neutralinos which generate large $MET$ events):
\begin{gather}
\vec{MET} = -\sum \vec{p}_T{}_{visible}\label{eq:MET}
\end{gather}
(with the transverse momentum $\vec{p}_T$ of the visible particles in the detector). Some other sophisticated transverse observables have also been used, such as $\alpha_{T}$ , $M_{T2}$ or $M_{T2W}$ (respectively introduced in \cite{alphaT}, \cite{MT2} and \cite{MT2W}). Those new models bring additional signals that could be identified through the Standard Model background (events from the Standard Model with an equivalent final state to the searched one). Possible signatures coming from those models are particles with a long lifetime and, therefore, long flight distance. Those types of particles are called long-lived particles and can be interesting to study since long-lived particles from the Standard Model have relatively low-masses and are well-known (see \autoref{fig:llpSM}).The non-detection of new signals steered the experimentalists to hunt more exotic topologies like long-lived particles.\medskip

\section{Long-lived particles}
For this analysis, we define the long-lived particles as follow:
\begin{center}
\textit{their lifetimes allow them to decay in a 4D volume defined by the tracker volume (between $4\ \text{cm}$ and $100\ \text{cm}$) and the collision period (a collision every $25\ \text{ns}$).}
\end{center} 
They can be charged and coloured. In the case where the particles are not electrically charged and not coloured, it can be challenging to reconstruct the tracks of the particles. Depending on the nature (and the possible decays) of the long-lived particle, the experimental signature can be completely different (see Figure \ref{fig:displaced}). For example, a heavy-stable charged particle (or HSCP) can cross the entire tracker of the detector and create energy deposits in the electromagnetic calorimeter. In contrast, a long-lived neutral particle that decays in the tracker will generate displaced vertices. Note also that long-lived particles are not a specific signature of supersymmetry. There exist several models (which do not include supersymmetry) that generate long-lived scenarios such as \textit{hidden valley} \cite{hiddenvalley} or \textit{Heavy neutrinos} models \cite{neutrinos}.
\begin{figure}[h]
    \centering
      \includegraphics[width=.8\linewidth]{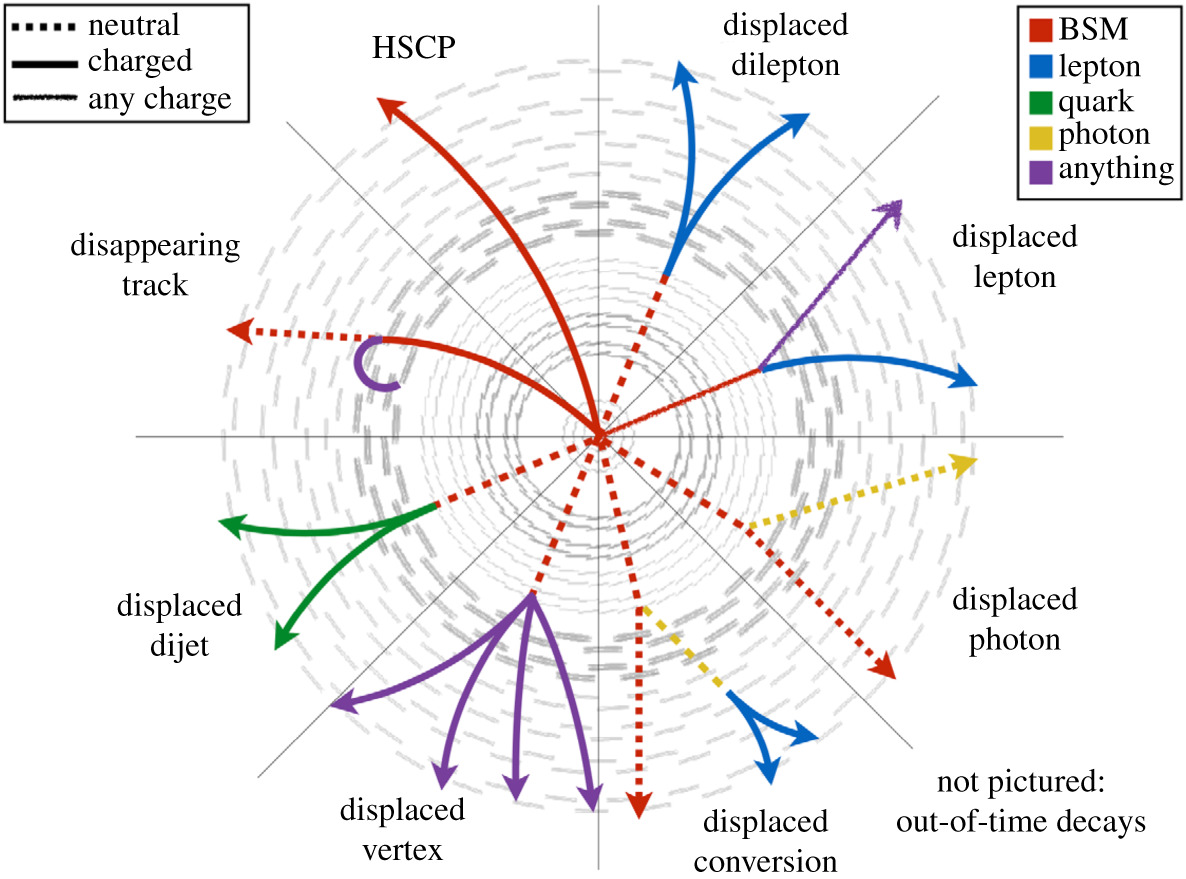}
    \caption{Schematic presentation of the various signatures coming from the presence of long-lived particles \cite{llp_signatures}.}
    \label{fig:displaced}
\end{figure}

There exist several strategies in BSM (\textit{Beyond the Standard Model}) analysis. The two main methods are: 
\begin{itemize}
\item a \textit{top-down} approach, \textit{i.e.}, analysing the phenomenological properties of a specific model (e.g. the N2MSSM) 
\item a \textit{bottom-up} approach, \textit{i.e.}, constructing a simplified model leading to specific experimental signatures (e.g. displaced-top), which can lead to constraints on existing models (like supersymmetric models, GUT theory, composite models or extra-dimension models). 
\end{itemize}

For this manuscript, we follow a \textit{bottom-up}-like approach and focus on a displaced-top signature associated with a production of an LSP as a source of MET in the CMS and ATLAS detector. Displaced-tops correspond to top quarks, which decay after the detector's first slices. This analysis was done in collaboration with Strasbourg's CMS Team, specialising in top quark physics at the LHC. The possibility of such a signal in the context of the MSSM (see Subsection \ref{subsec:mssm}) will be studied using two different processes:
\begin{itemize}
\item the decay $\tilde{t}_i\rightarrow \chi^0_1 t$ in the context of \textit{Gravity Mediated Supersymmetry Breaking} with an NLSP (\textit{Next-Lightest Supersymmetric Particle}) stop and an LSP neutralino,
\item the decay $\tilde{t}_i\rightarrow \psi_\mu t$ using \textit{Gauge Mediated Supersymmetry Breaking} (GMSB) mechanism \cite{GMSB} with an NLSP stop and an LSP gravitino
\end{itemize} 
(the same signal in the context of RPV-MSSM and Split-SUSY was also considered, see \cite{llpCMS}). The last process is the main study. Results of this analysis will be given by defining benchmarks allowing long-lived stop $\tilde{t}$. Some words for further studies will also be given. \medskip

We reintroduce the velocity of light c, the Planck mass $m_p$ and the reduced Planck constant $\hbar$ in the relations during this chapter.
\section{Topology of the searched signal}
Long-lived particles already exist in the Standard Model. However, plenty of such processes emerges from BSM models (\textit{Beyond the Standard Model}). The main mechanisms in order to obtain a long-lived particle in supersymmetric models are:
\begin{itemize}
\item a low coupling coming from a Planck suppressed couplings,
\item a heavy mediator in the process leading to highly off-shell decay with models like Split-SUSY \cite{Split1}\cite{Split2} (soft-breaking mass terms are taking to high-energy, decoupling the sfermions to the other sectors, see Subsection \ref{subsec:diffS2N2}). 
\item a small phase-space region coming from a quasi-symmetry (for example, the isospin symmetry between proton and neutron leading to the small decay width of the neutron or the R-parity violation in supersymmetric models \cite{MSSMRPV}).  
\end{itemize}
In the framework of the LHC experiments, the long-lived particles can travel the collision area and decay after the first part of the detector. The particle coming from the decay is then defined as \textit{non-prompt}. During this analysis, a configuration is searched, such as the particle decay in the tracker volume (see \autoref{fig:CMS}) corresponding to a flight distance between $4\ \text{cm}$ and $1\ \text{m}$. The particles coming from the decay are then reconstructed from the various part of the detectors\footnote{This statement is true only if the long-lived particle decays in the tracker volume.}. The expected amount of background events is low. Nonetheless, the reconstruction efficiency is also weak since the detector has not been specifically designed for such signals.\medskip

The tracker part of the CMS and ATLAS detectors combined with the various calorimeters enables to reconstruct the tracks and the nature of charged and neutral particles (except for invisible particles such as neutrinos). 
\begin{figure}[h]
    \centering
      \includegraphics[width=.7\linewidth]{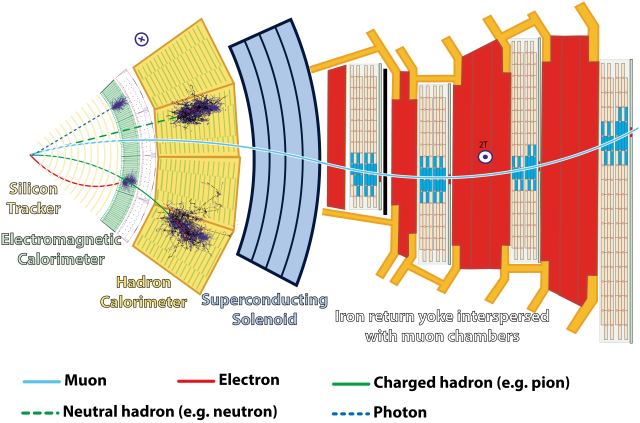}
    \caption{Slice of the CMS detector at the LHC \cite{sliceCMS}. The silicon tracker with the magnetic field of $3.8\ T$ generated by the supraconducting solenoid allows the reconstruction of charged particles' trajectory. The energy of electrons and photons are measured by the electromagnetic calorimeter, whilst the hadronic calorimeter measures the energy of the hadrons. The muon chamber is the last part of the detector. The trajectories of various particles in the detector are shown. The structure of the ATLAS detector is quite similar.}
    \label{fig:CMS}
\end{figure}
The two processes studied in this section correspond to the same topology in an experimental context. The neutralino and the gravitino are the LSP in their respective models and are so $MET$ (or \textit{Missing Transverse Energy}) sources for the analysis of collisions. The top quarks will mainly decay as $t\rightarrow W^+ b$, which generate jets coming from the hadronisation of the b-quark identified through \textit{b-tagging} algorithms (algorithms that allows the identification of jets coming from the hadronisation of b quark). Moreover, the W-boson can decay in a leptonic ($W^+\rightarrow l^+\nu_l$) or fully hadronic channel. Those decays will be studied in the context of the MSSM with \textit{Gravity Mediated Supersymmetry Breaking} and \textit{Gauge Mediated Supersymmetry Breaking}.

\section{Definition of the simplified models}
{Models of supersymmetry are defined through a lot of free parameters. Phenomenological analysis of such models is then complex to compute since it corresponds to an analysis of a high-dimensional parameters space. It can be then useful to build simplified models that are easy to study and constrain, and also to reinterpret the results in the framework of more complete but sophisticated models.\medskip

For this analysis, we construct a simplified supersymmetric model based on the MSSM. Since we only consider top quarks, the superpotential that we consider is:
\begin{eqnarray}
W & = & \hat{Q}\cdot \hat{H}_U \mathbf{Y_U} \hat{U}\ .\label{eq:Wmodel}
\end{eqnarray}
However, this analysis will consider two different supersymmetry breaking mechanisms: the \textit{Gravity Mediated Supersymmetry Breaking} mechanism widely presented in this manuscript and the \textit{Gauge Mediated Supersymmetry Breaking} mechanism briefly presented in Subsection \ref{subsec:breakingmodel}. 

\subsection*{$\bullet$ MSSM with Gravity mediated model:}
We consider for this model an NLSP stop decaying into the lightest neutralino and a top quark ($\tilde{t}_i\rightarrow\ \tilde{\chi}^0_1 t$). We impose that the lightest neutralino is the LSP. The neutralino $\tilde{\chi}^0_1$ is a mixing state of Bino $\tilde{B}$, Wino $\tilde{W}^3$ and Higgsinos $\tilde{H}^0_U,\tilde{H}^0_D$. For this analysis, we will consider a simplified model where the lightest neutralino $\tilde{\chi}^0_1$ corresponds to a pure state. Supposing no mixing between $\tilde{t}_L$ and $\tilde{t}_R$, we end up with several cases considering the different natures of the neutralino and the stop:
\begin{eqnarray}
\tilde{t}_L\rightarrow  t \tilde{B} \quad , \quad \tilde{t}_R \rightarrow t \tilde{H}^0_U \quad , \quad \tilde{t}_L \rightarrow t \tilde{W}^3 \quad , \quad \tilde{t}_R\rightarrow t \tilde{B} \quad , \quad \tilde{t}_L \rightarrow t \tilde{H}_U^0 \ .\nonumber
\end{eqnarray}
The supersymmetry breaking mechanism used for this model is the gravity mediated. In this framework, the gravitino mass is (we reintroduce the Planck mass, contrary to what was done in Chapter \ref{sec:sugra}):
\begin{equation}
m_{3/2} = Me^{\left<K\right>/(2m_p)}\ ,\nonumber
\end{equation}
where $M\ll m_p$ (which is usually at the TeV scale) and will not interfere in this analysis. \medskip

We then have four free parameters for this analysis: the type of squark and neutralino, the squark mass $m_{\tilde{t}_i}$ and the mass of the lightest neutralino $m_{\chi^0_1}$.

\subsection*{$\bullet$ MSSM with Gauge mediated model:}
We also consider the same supersymmetric model with \textit{Gauge Mediated Supersymmetry Breaking}. A couple of messenger superfields $\mathbf{S}=(S,\chi_S,F_S)$ and $\mathbf{S}^{\dagger}=(S^{\dagger},\bar{\chi}_S,F_S^{\dagger})$ in non-trivial complex conjugate representations of the considered gauge group ($SU(5)$ in the simplest model) are added to the superpotential \autoref{eq:Wmodel} and coupled to the gauge-singlet superfield $Z=(\zeta, \chi_\zeta, F_\zeta)$ from the hidden sector:
\begin{eqnarray}
W_{mess} = \lambda  \mathbf{S}\bar{\mathbf{S}} Z\ .\nonumber
\end{eqnarray} 
Supposing that the gauge-singlet superfield gets a non-zero \textit{v.e.v}:
\begin{equation}
\big< Z \big> = \big< \zeta \big> - \theta\cdot\theta \big< F_\zeta \big>\ ,\nonumber
\end{equation}
soft supersymmetry breaking terms are produced from loop corrections. Gauginos mass terms are generated at the one-loop level (see \autoref{fig:gmsb} with $\lambda=\{ \tilde{B}, \tilde{W}^{\mu} , \tilde{g}^a \}$):
\begin{equation}
V_{SOFT}^{gaug.} = \frac12\big( m_{\tilde{B}} \tilde{B}\cdot\tilde{B} + m_{\tilde{g}} \tilde{g}^a\cdot\tilde{g}_a + m_{\tilde{W}}\tilde{W}^{\mu}\cdot\tilde{W}_{\mu} + \text{h.c.}\big) \nonumber
\end{equation}
where scalar masses are only present at two-loop (see for example \autoref{fig:gmsb}).
\begin{figure}[H]
    \centering
    \begin{subfigure}[t]{.5\textwidth}
      \centering
      \includegraphics[width=.8\linewidth]{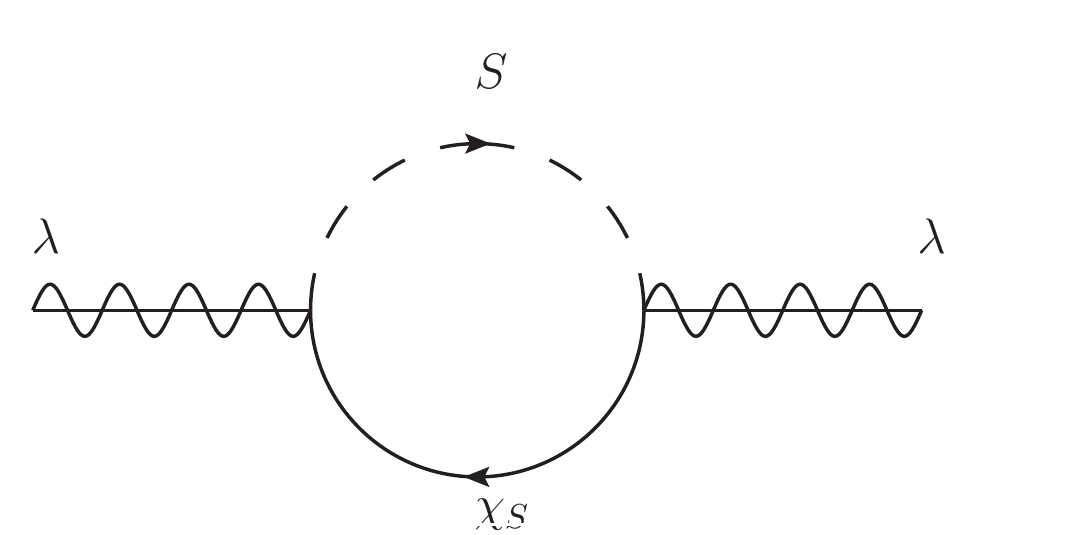}
      \caption{One-loop correction to gauginos masses}
   \end{subfigure}%
   \begin{subfigure}[t]{.5\textwidth}
      \centering
      \includegraphics[width=.75\linewidth]{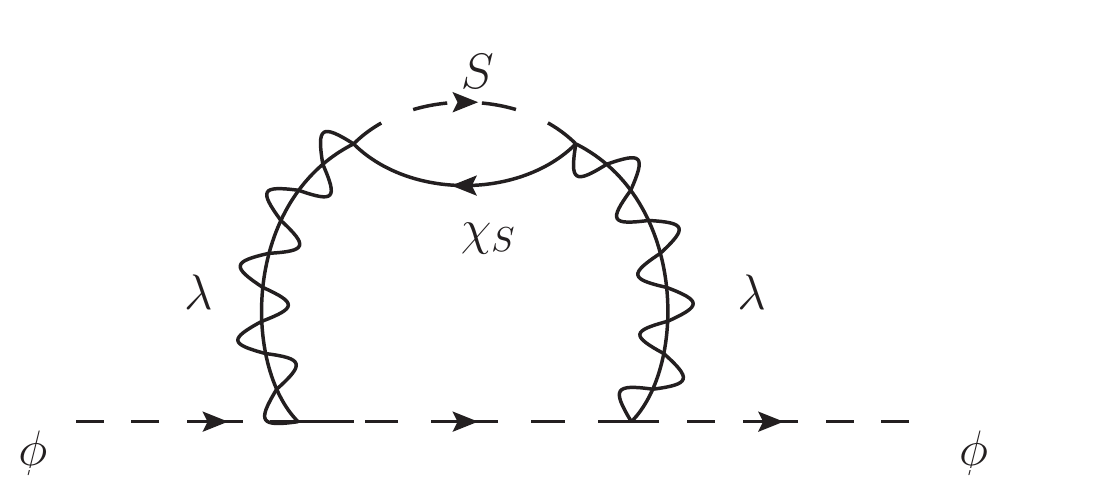}
      \caption{Two-loop correction to scalar mass}
   \end{subfigure}
   \caption{Loop corrections from GMSB model}
\label{fig:gmsb}
\end{figure}
Taking the MSSM in the context of supergravity, the gravitino mass is:
\begin{eqnarray}
m_{3/2}\sim \frac{M_{SUSY}^2}{m_p}\ .\nonumber
\end{eqnarray}
A SUSY scale $M_{SUSY}$ of $1\  \mathrm{TeV}$ generates then a gravitino mass of $1\times 10^{-13}\, \mathrm{GeV}$. Therefore, the gravitino is the LSP. We then consider a (NLSP) stop squark decaying into a gravitino and a top quark ($\tilde{t}_i \rightarrow t \psi_{\mu}$). \medskip

In this model, we have 2 independent free parameters: the squark stop mass $m_{\tilde{t}}$ and the gravitino mass $m_{3/2}$. We will see later that the mixing angle $\theta$ between $\tilde{t}_L$ and $\tilde{t}_R$ do not much modify the final results. 

\section{Stop production at LHC with $\sqrt{s}=14\ \text{TeV}$}
\subsection{Cross section of stop pair production}
The Run 3 of LHC, which will start in 2022, will be able to produce proton-proton collisions with a center-energy mass of $\sqrt{s}=14\ \text{TeV}$. Due to the partonic environment, the significant contribution to the cross-section $\sigma(pp\rightarrow \tilde{t}_1\bar{\tilde{t}}_1)$ comes from two processes: quark-antiquark annihilation \autoref{fig:qqstop} and gluon-gluon fusion \autoref{fig:ggstop}. Other supersymmetric processes which could contribute to the cross section are neglected.   
\begin{figure}[H]
    \centering
      \includegraphics[width=.4\linewidth]{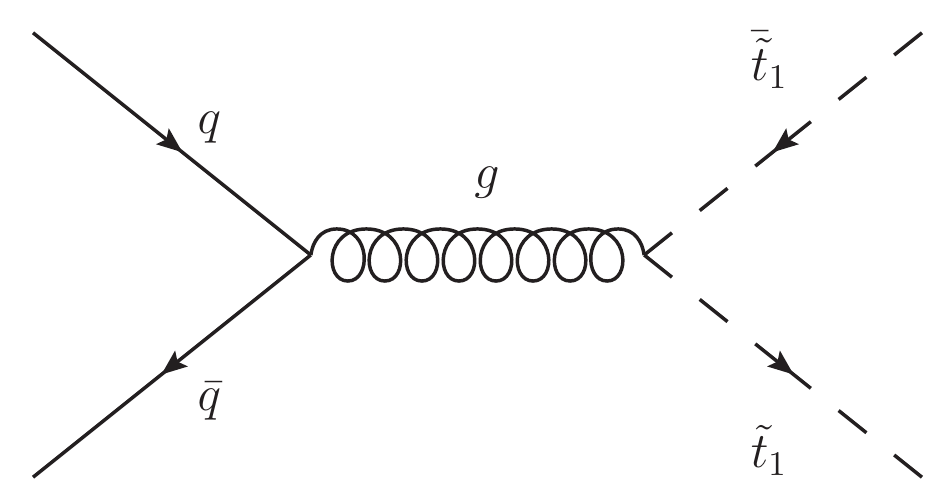}
    \caption{Production of a stop $\tilde{t}_1$ pair from quark-antiquark annihilation (LO QCD).}
    \label{fig:qqstop}
\end{figure}
\begin{figure}[H]
    \centering
      \includegraphics[width=0.7\linewidth]{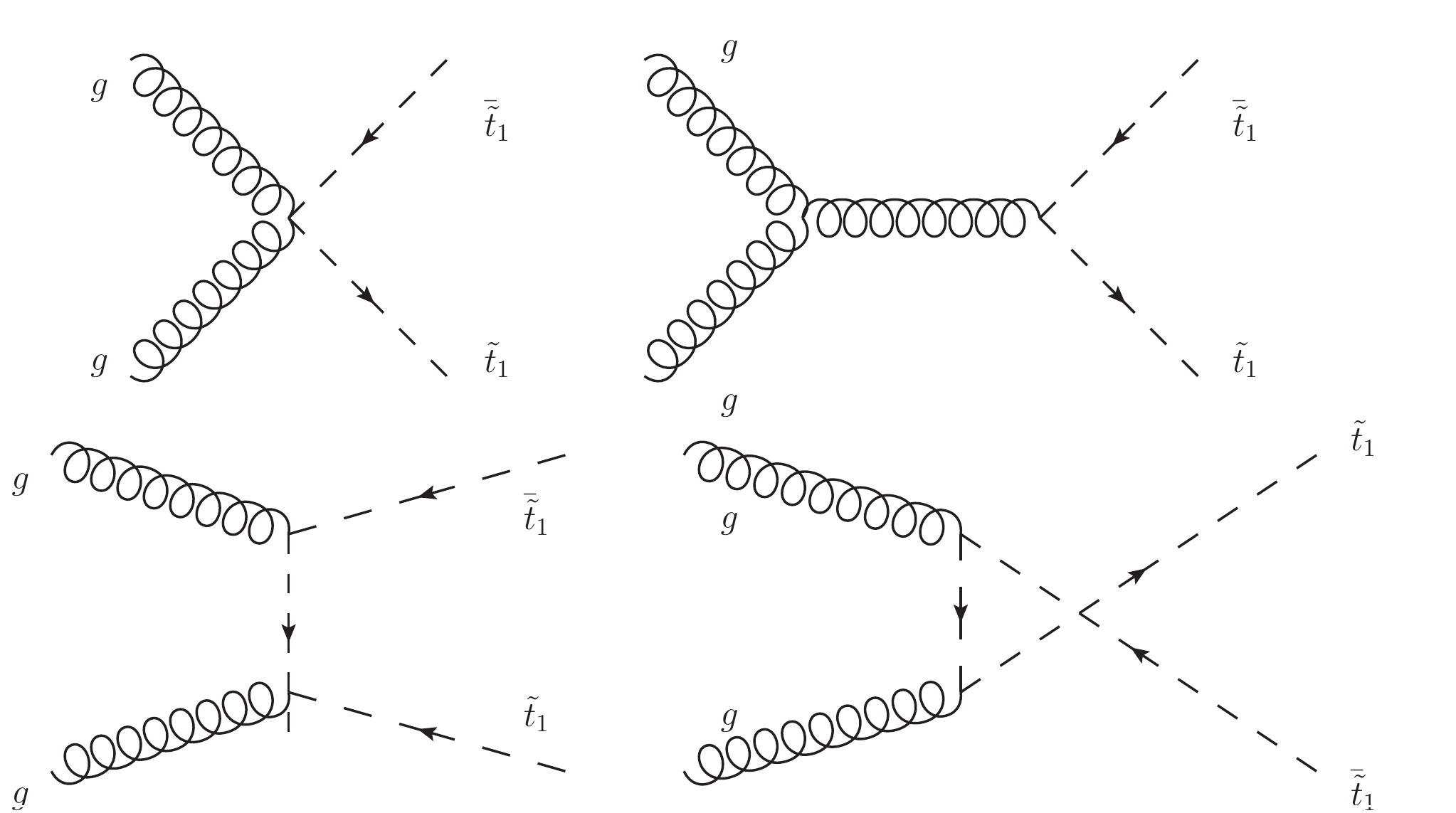}
    \caption{Production of a stop $\tilde{t}_1$ pair through gluon-gluon fusion  (LO QCD).}
    \label{fig:ggstop}
\end{figure}
Since the two models presented previously are based on the same supersymmetric model, the production cross-section are similar. Moreover, the only free parameter which leads to the stop pair production is the stop mass $m_{\tilde{t}_1}$.\medskip

The evolution of $\sigma (pp\rightarrow \tilde{t} \bar{\tilde{t}})$ (with $\tilde{t}$ can be $\tilde{t}_L$ or $\tilde{t}_R$) with $m_{\tilde{t}_1}$ can be obtained using various programs. 
The UFO model (for \textit{Universal FeynRules Output}) \cite{ufo} containing all Feynman rules is made using the \textsc{FeynRules} package (version 2.3) \cite{feynrules} of \textsc{Mathematica}\footnote{MSSM with goldstino/gravitino couplings: \url{https://feynrules.irmp.ucl.ac.be/wiki/goldstino}}. The hard process $pp\rightarrow \tilde{t}_1\bar{\tilde{t}}_1$ and the cross-section calculation are performed using the Monte-Carlo generator \textsc{MadGraph\_aMC@NLO} \cite{madgraph} at LO (\textit{Leading Order}) QCD (see \autoref{fig:stop_xsec}). Note, however, that high order calculations are already available \cite{stop_prod_NLO}\cite{stop_prod_NNLL} (with a K-factor $K_{NLO+NNLL}=\sigma_{NLO+NNLL}/\sigma_{LO}\approx 1.30$). We recall that the cross section of a process $pp\rightarrow X$ can be calculated (using the factorisation theorem) as:
\begin{gather}
\sigma (pp\rightarrow X)=\displaystyle\sum_{a,b\in protons}\int_0^1dx_a\int_0^1dx_b f_{a/p}(x_a,\mu_F^2)f_{b/p}(x_b,\mu_F^2)\hat{\sigma}(ab\rightarrow X, \hat{s},\mu_R^2,\mu_F^2) \label{eq:XS}
\end{gather}
with the "Feynman x" $x=\frac{p_L(partons)}{p_L(hadrons)}$ (with $p_L$ the longitudinal momentum), the factorisation scale $\mu_F$, the renormalisation scale $\mu_R$, $\hat{s}=x_ax_bs$ and $f$ the \textit{parton density function} (PDF). The following generation is obtained by assuming $\mu_R=\mu_F$ and with the PDF (\textit{Parton Density Function}) NNPDF30\_lo\_as\_0130 ($lhaid=263000$) of \textsc{LHAPDF6} \cite{lhapdf}. The factorisation scale $\mu_F$ (and so the renormalisation scale $\mu_R$) is dynamically set for each event as the transverse mass of the $2\rightarrow 2$ system resulting from a $k_t$-clustering algorithm \cite{MG5_dynamical_scheme}\cite{kt_clustering}. The uncertainties related to the scales $\mu_F$ and $\mu_R$ (see \autoref{fig:stop_xsec}) are obtained by varying those scales with factors $\frac12$ and $2$. 

          \begin{figure}[H]
      \centering
		    \includegraphics[scale=0.9]{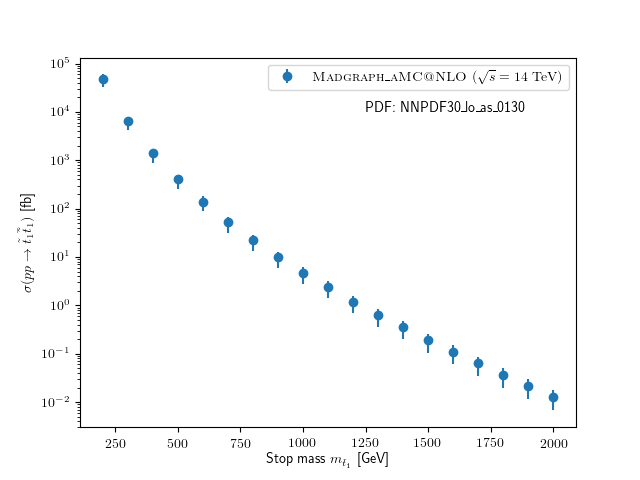}
		    \caption{Evolution of $\sigma (pp\rightarrow \tilde{t}_1\bar{\tilde{t}}_1)$ as function of $m_{\tilde{t}_1}$ using the PDF NNPDF30\_lo\_as\_0130. The uncertainties are obtained by varying the factorisation and renormalisation scales $\mu_F$ and $\mu_R$.}
		    \label{fig:stop_xsec}
          \end{figure}



Imposing at least a hundred of events for our analysis, and supposing an integrated luminosity of $\mathcal{L}=300\ \text{fb}^{-1}$ for the Run 3 of the LHC, a limit of the stop mass can be set to approximately $m_{\tilde{t}}<1.4\ \text{TeV}$. A low limit on the stop mass can also be set to $1\ \text{TeV}$, considering the general limits on the squark mass \footnote{Note that this limit is in general model dependent an can be lowered to fully cover the modern analysis.}.\medskip

\subsection{Kinematic of the stop squark}
The goal of this analysis is to calculate the average flight distance $c\tau$ of the stop squark (in the particle frame) from the calculation of the decay width $\Gamma$:
\begin{eqnarray}
\Gamma(A\rightarrow BC) &=& \frac{\sqrt{\lambda (m_{A}^2,m_B^2,m_C^2)}}{16\pi m_{A}^3}|\Bar{\mathcal{M}}|^2\nonumber
\end{eqnarray}
with the Källén function $\lambda (x,y,z)=(x - y - z)^2 - 4yz$ and the matrix element $\mathcal{M}(A\rightarrow BC)$. The relationship between those two observables is:
\begin{equation}
c\tau = \frac{\hbar c}{\Gamma}\ .\nonumber
\end{equation}
This flight distance must be corrected from relativistic effects using the Lorentz factor $\gamma$ and $\beta=v/c$ (with $v$ the velocity of the particle) to get the flight distance in the laboratory frame $c\tau_{corr.}$:
\begin{equation}
c\tau_{corr.} = \beta\gamma c\tau ,\qquad \text{with}\,\gamma =\frac{1}{\sqrt{1-\beta^2}}\ .\label{eq:ctau}
\end{equation}
Those corrective factors can be obtained using the events generation of \sloppy\textsc{Madgraph\_aMC@NLO} and the framework \textsc{MadAnalysis 5}\cite{MA5} in expert mode. The distributions of the $\gamma\beta$-factor for various stop masses are collected as well as the mean values $\big< \gamma\beta \big>$ (see \autoref{fig:beta_distrib}).
\begin{figure}[H]
    \centering
    \includegraphics[scale=0.9,origin=c]{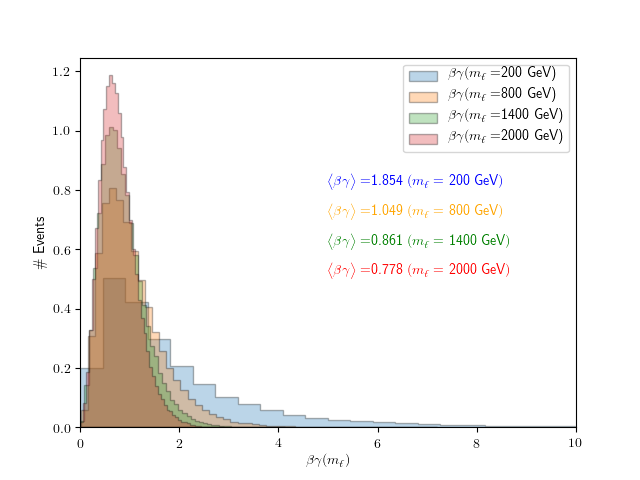}
    \caption{$\gamma\beta$-distribution of the stop squark with $m_{\tilde{t}_1}(\text{GeV})\in \big[ 200, 800, 1400, 2000\big]$.}
    \label{fig:beta_distrib}
\end{figure}
From those results, we can derive the mean-value of the $\beta\gamma$-factor as function of $m_{\tilde{t}_1}$ from fitting a $4^{th}$-order polynomial \autoref{fig:stop_betagamma} on the mean values $\lag \beta\gamma\rag$.
\begin{figure}[H]
    \centering
      \includegraphics[width=.9\linewidth]{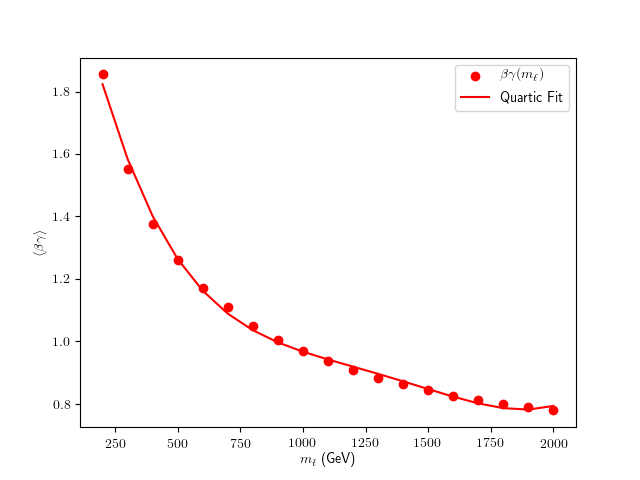}
    \caption{Evolution of $\big< \beta \big>$ as function of $m_{\tilde{t}_1}$}
    \label{fig:stop_betagamma}
\end{figure}
After fitting, the polynomial reads:
\begin{eqnarray}
\big< \beta\gamma \big> (m_{\tilde{t}_1}) &=& 2.51823 - 0.0043249 m_{\tilde{t}_1} + 4.70873\times 10^{-6} m_{\tilde{t}_1}^2 - 2.38266\times 10^{-9} m_{\tilde{t}_1}^3 \nonumber \\
&& + 4.46899\times 10^{-13} m_{\tilde{t}_1}^4\ . \label{eq:betagamma}
\end{eqnarray}
The associated $\chi^2$-value to this fit is $\chi^2/ndf = 1.89746\times 10^{-4}$. The relativistic effects will then not highly modify the flight distance $c\tau$ of the stop squark. The two models mentioned previously can now be investigated.

\section{Decay of NLSP stop to LSP neutralino: $\tilde{t}\rightarrow \chi_j^0 t $}\label{sec:NeutLSP}
We firstly consider the decay of the NLSP stop squark to LSP neutralino and top quark ($\tilde{t}\rightarrow \chi^0_j t$) in the MSSM with \textit{Gravity Mediated Supersymmetry Breaking}.\medskip

In order to determine the $(\tilde{t}_i,t,\chi^0_1)$ vertex, we remind that the superpotential $W(\Phi)$ take the form \autoref{eq:Wmodel}. The interaction terms generating the vertex $(\tilde{t}_i,t,\chi^0_1)$ are (see \autoref{eq:potSUSY}):
\begin{eqnarray}
\mathcal{L}_{int} &=& \underbrace{\Big( -\frac{1}{2}W_{ij}\psi^i\cdot\psi^j}_{(i)} +\underbrace{i\sqrt{2}g\Bar{\lambda}^a\cdot\Bar{\psi}T_a\phi \Big)}_{(ii)}+ \mathrm{h.c.} \nonumber
\end{eqnarray}
where (i) generates higgsinos-matter couplings and (ii) leads to gauginos-matter interactions. \medskip

In four-component notation, we denote the up-quark and the gauginos as:
\begin{gather}
u^{(D)}=\begin{pmatrix}
u_L \\
u_R
\end{pmatrix} \equiv u\ , \ \lambda^{(M)}=\begin{pmatrix}
\lambda \\
\bar{\lambda}
\end{pmatrix}\ . \nonumber
\end{gather}
The various states of the lightest neutralino are:
\begin{gather}
\chi^0_1{}^{(M)}=\{-i\tilde{B}^{(M)}, -i\tilde{W}^3{}^{(M)}, \tilde{H}_D^0{}^{(M)}, \tilde{H}_U^0{}^{(M)}\}\ .\nonumber
\end{gather}
with the elements denoted $\chi^0_1{}^{(M)}{}_j$.
\subsubsection*{(i) Yukawa interactions}
Only the stop coupling in the superpotential \autoref{eq:Wmodel} is relevant for the analysed process. 
Recalling the notations for the superfields:
\begin{eqnarray}
\hat{Q} = \begin{pmatrix}\begin{pmatrix}
\tilde{u}_L \\
\tilde{d}_L
\end{pmatrix},\begin{pmatrix}
u_L \\
d_L
\end{pmatrix},\begin{pmatrix}
F_{u_L} \\
F_{d_L}
\end{pmatrix}\end{pmatrix}\, &,& \, \hat{H}_U =  \begin{pmatrix}\begin{pmatrix}
H_U^+ \\
H_U^0
\end{pmatrix},\begin{pmatrix}
\tilde{H}_U^+ \\
\tilde{H}_U^0
\end{pmatrix},\begin{pmatrix}
F_{H_U^+} \\
F_{H_U^0}
\end{pmatrix}\end{pmatrix},\nonumber \\
U &=& \begin{pmatrix}
\tilde{u}_R^\dagger, u_R^\dagger, F_U
\end{pmatrix},\nonumber
\end{eqnarray}
the higgsinos-stop interactions are then:
\begin{eqnarray}
\mathcal{L}_{\tilde{H}\tilde{t}t} &=&  -\frac{1}{2}W_{ij}\psi^i\cdot\psi^j +\mathrm{h.c.} \nonumber \\
&=& -Y_u \left( \tilde{H}_U^0\tilde{u}_R^\dagger u_L+\bar{\tilde{H}}_U^0\tilde{u}_L^\dagger u_R\right) +\mathrm{h.c.} \nonumber
\end{eqnarray}
\subsubsection*{(ii) Gauginos/Squark/Quark interactions}
Using the gauge group $SU_c(3)\times SU_L(2)\times U_Y(1)$, the gauge structure of the superfields $\hat{Q},\ \hat{H}_u\, \hat{U}$ are:
\begin{equation}
\hat{Q} = \left(\underset{\sim}{3},\underset{\sim}{2},\frac{1}{6}\right)\quad , \quad \hat{U} = \left(\underset{\sim}{3},\underset{\sim}{1},-\frac{2}{3}\right)\quad , \quad \hat{H}_U = \left(\underset{\sim}{1},\underset{\sim}{2},\frac{1}{2}\right),\nonumber
\end{equation}
The neutralinos $\chi_j^0$ being a mixing state of higgsinos, bino $\tilde{B}$ and wino $\tilde{W}^3$ the interactions coming from $U_Y(1)$ and $SU_L(2)$ can only be considered: 
\begin{itemize}
    \item $U(1)_Y$ Hypercharge:
    \begin{eqnarray}
    ig_1\sqrt{2}\Bar{\tilde{B}}\Big(\frac{1}{6}\Bar{u}_L\tilde{u}_L-\frac{2}{3}u_R\tilde{u}_R^\dagger \Big) + \mathrel{h.c.}\nonumber 
    \end{eqnarray}
    \item $SU(2)_L$ Weak isospin:
    \begin{eqnarray}
    i\frac{g_2}{\sqrt{2}}\Bar{\tilde{W^3}}\Bar{u}_L\tilde{u}_L +\mathrm{h.c.}\nonumber
    \end{eqnarray}
\end{itemize}
The full $(\tilde{t}_i,t,\chi_j^0)$-coupling can then be written in the Dirac-formalism with the projector $P_R$ and $P_L$:
\begin{eqnarray}
\mathcal{L}_{int} &=& \tilde{u}_i^\dagger \bar{\chi}^0_1{}^{(M)}{}_j\Bigg[\Bigg(\frac{-e}{\sqrt{2}s_Wc_W}\delta^i{}_1\Big( \frac{1}{3}\delta^j{}_1 s_W + \delta^j{}_2 c_W \Big)-Y_u \delta^j{}_4 \delta^i{}_2\Bigg)P_L \label{eq:lint} \\
&&\qquad+\; \Bigg(\frac{2e\sqrt{2}}{3c_W}\delta^j{}_1 \delta^i{}_2-Y_u \delta^j{}_4 \delta^i{}_1\Bigg)P_R\Bigg] u + \mathrm{h.c.}\nonumber
\end{eqnarray}

For the following, the vertex term is denoted as $V_{\tilde{t}\chi t}$. Using $\mathcal{L}_{int}$, it follows:
\begin{eqnarray}
V_{\tilde{t}\chi t} &=& V_LP_L + V_RP_R\nonumber \\
&=& \frac{1}{2}V_L(1-\gamma^5)+\frac{1}{2}V_R(1+\gamma^5)\nonumber \\
&=&\frac{1}{2}(V_R+V_L)+\frac{1}{2}\gamma^5(V_R-V_L)\nonumber \\
&\equiv& V_{\tilde{t}\chi t}^{+}\nonumber
\end{eqnarray}
with $V_L$ and $V_R$ easily defined from \autoref{eq:lint}. We also introduce:
\begin{equation}
V_{\tilde{t}\chi t}^{-}=\frac{1}{2}(V_R+V_L)-\frac{1}{2}\gamma^5(V_R-V_L)\ .\nonumber
\end{equation}

The mean-value of the matrix element is calculated using trace techniques:
\begin{eqnarray}
i\mathcal{M} &=& \bar{u}_t(p_1,s_1)iV_{\tilde{t}\chi t}^{+}v_\chi(p_2,s_2)\nonumber \\
\Rightarrow\sum_{s_1,s_2}|\mathcal{M}|^2&=&\mathrm{Tr}[(\slashed{p}_1+m_t)V_{\tilde{t}\chi t}^{+}(\slashed{p}_2-m_\chi)V_{\tilde{t}\chi t}^{-}] \nonumber \\
&=& \underbrace{\mathrm{Tr}[\slashed{p}_1V_{\tilde{t}\chi t}^{+}\slashed{p}_2V_{\tilde{t}\chi t}^{-}]}_\text{\circled{A}}-\underbrace{\mathrm{Tr}[m_tm_\chi V_{\tilde{t}\chi t}^{+}V_{\tilde{t}\chi t}^{-}]}_\text{\circled{B}}\nonumber
\end{eqnarray}
where:
\begin{eqnarray}
    \mathrm{\circled{A} } \mathrm{Tr}[ \slashed{p}_1V_{\tilde{t}\chi t}^+\slashed{p}_2V_{\tilde{t}\chi t}^-]&=& \frac{1}{4}\mathrm{Tr}\Big[\slashed{p}_1\left\{ \left(V_R + V_L\right) + \gamma^5\left( V_R-V_L\right) \right\}\slashed{p}_2\left\{ \left(V_R + V_L\right) - \gamma^5\left( V_R-V_L\right) \right\} \Big]\nonumber \\
    &=& 2p_1\cdot p_2(|V_L|^2 +|V_R|^2) \nonumber \\
    &=& (m_{\tilde{t}_i}^2-m_t^2-m_\chi^2)(|V_L|^2 +|V_R|^2)\nonumber 
\end{eqnarray}
and
\begin{eqnarray}
\mathrm{\circled{B} } \mathrm{Tr}[m_tm_\chi V_{\tilde{t}\chi t}^+V_{\tilde{t}\chi t}^-] &=& -\frac{1}{4}m_tm_\chi\mathrm{Tr}\left[ \left\{ (V_R+V_L)+\gamma^5(V_R-V_L) \right\}\left\{ (V_R+V_L)-\gamma^5(V_R-V_L) \right\} \right]\nonumber \\
&=& -4m_tm_\chi V_RV_L \ .\nonumber 
\end{eqnarray}
Summing on colour indices, the mean-value of matrix element can be written as:
\begin{eqnarray}
|\bar{\mathcal{M}}|^2&=&\frac{1}{N_c}\sum_{s_1,s_2}\sum_{colours}|\mathcal{M}|^2 \nonumber \\
&=& (m_{\tilde{t}_i}^2-m_t^2-m_\chi^2)(|V_R|^2 + |V_L|^2) -4m_tm_\chi V_LV_R \ .\nonumber
\end{eqnarray}
The decay width is thus:
\begin{eqnarray}
\Gamma(\tilde{t}_i\rightarrow t\chi_j^0) &=& \frac{\sqrt{\lambda (m_{\tilde{t}_i}^2,m_t^2,m_\chi^2)}}{16\pi m_{\tilde{t}_i}^3}\left( (m_{\tilde{t}_i}^2-m_t^2-m_\chi^2)(|V_R|^2 + |V_L|^2) -4m_tm_\chi V_LV_R\right)\ .\nonumber
\end{eqnarray} 
All the possible decays are presented in \autoref{table:VLVR} and \autoref{table:Decay}:
\begin{table}[H]
\begin{center}
\begin{tabular}{ ||c||c|c|c| } 
\hline\xrowht[()]{20pt}
$\mathbf{\Gamma(\tilde{t}_i\rightarrow \chi^0_1 t)}$ & $\mathbf{\tilde{t}_i=\tilde{t}_L}$ & $\mathbf{\tilde{t}_i=\tilde{t}_R}$\\
\hline
\hline\xrowht[()]{20pt}
 $\mathbf{\chi_1^0=\tilde{B}}$  & $V_L=\frac{-e}{3\sqrt{2}c_W},\ V_R=0$ & $V_L=0,\ V_R=\frac{2e\sqrt{2}}{3c_W}$ \\ 
\hline\xrowht[()]{20pt}
 $\mathbf{\chi_1^0=\tilde{W}^3}$  & $V_L=\frac{-e}{3\sqrt{2}s_W},\ V_R=0$ & $V_L=0,\ V_R=0$ \\ 
\hline\xrowht[()]{20pt}
 $\mathbf{\chi_1^0=\tilde{H}_d^0}$  & $V_L=0,\ V_R=0$ & $V_L=0,\ V_R=0$ \\ 
\hline\xrowht[()]{20pt}
 $\mathbf{\chi_1^0=\tilde{H}_u^0}$  & $V_L=0,\ V_R=Y_U$ & $V_L=Y_U,\ V_R=0$ \\ 
\hline
\end{tabular}
\caption{Values for $V_L$ and $V_R$ parameters depending on the process.}
\label{table:VLVR}
\end{center}
\end{table}
\begin{table}[hbt!]
\begin{center}
\begin{tabular}{ ||c||c| } 
\hline\xrowht[()]{20pt}
$\mathbf{\Gamma(\tilde{t}_i\rightarrow \chi^0_1 t)}$ & $\mathbf{\Gamma(\tilde{t}_i\rightarrow \chi^0_1 t)}$\\
\hline
\hline\xrowht[()]{20pt}
$\mathbf{\tilde{t}_L\rightarrow \tilde{B} t}$  & $\frac{\alpha}{72c_W^2 m_{\tilde{t}_i}^3}\sqrt{\lambda (m_{\tilde{t}_i}^2,m_t^2,m_\chi^2)}(m_{\tilde{t}_i}^2-m_t^2-m_\chi^2)$ \\ 
\hline\xrowht[()]{20pt}
$\mathbf{\tilde{t}_L\rightarrow \tilde{W}^3 t}$  & $\frac{\alpha}{8c_W^2 m_{\tilde{t}_i}^3}\sqrt{\lambda (m_{\tilde{t}_i}^2,m_t^2,m_\chi^2)}(m_{\tilde{t}_i}^2-m_t^2-m_\chi^2)$\\ 
\hline\xrowht[()]{20pt}
$\mathbf{\tilde{t}_L\rightarrow \tilde{H}_u^0 t}$  & $\frac{\sqrt{\lambda (m_{\tilde{t}_i}^2,m_t^2,m_\chi^2)}}{16\pi m_{\tilde{t}_i}^3}(m_{\tilde{t}_i}^2-m_t^2-m_\chi^2)Y_U^2$\\ 
\hline\xrowht[()]{20pt}
$\mathbf{\tilde{t}_R\rightarrow \tilde{B} t}$  & $\frac{2 \alpha}{9c_W^2}\frac{\sqrt{\lambda (m_{\tilde{t}_i}^2,m_t^2,m_\chi^2)}}{m_{\tilde{t}_i}^3}(m_{\tilde{t}_i}^2-m_t^2-m_\chi^2)$ \\ 
\hline\xrowht[()]{20pt}
$\mathbf{\tilde{t}_R\rightarrow \tilde{H}_u^0 t}$  & $\frac{\sqrt{\lambda (m_{\tilde{t}_i}^2,m_t^2,m_\chi^2)}}{16\pi m_{\tilde{t}_i}^3}(m_{\tilde{t}_i}^2-m_t^2-m_\chi^2)Y_U^2$ \\ 
\hline
\end{tabular}
\caption{Decay width of the various processes.}
\label{table:Decay}
\end{center}
\end{table}

Note that some decay channels are forbidden due to the gauge invariance. From \autoref{table:Decay}, the corrected flight distance can then be computed using \autoref{eq:ctau} and \autoref{eq:betagamma} (see \autoref{fig:ctauNeut} and in Appendix \ref{app:meanflight}).

\begin{figure}
    \centering
      \includegraphics[width=0.8\linewidth]{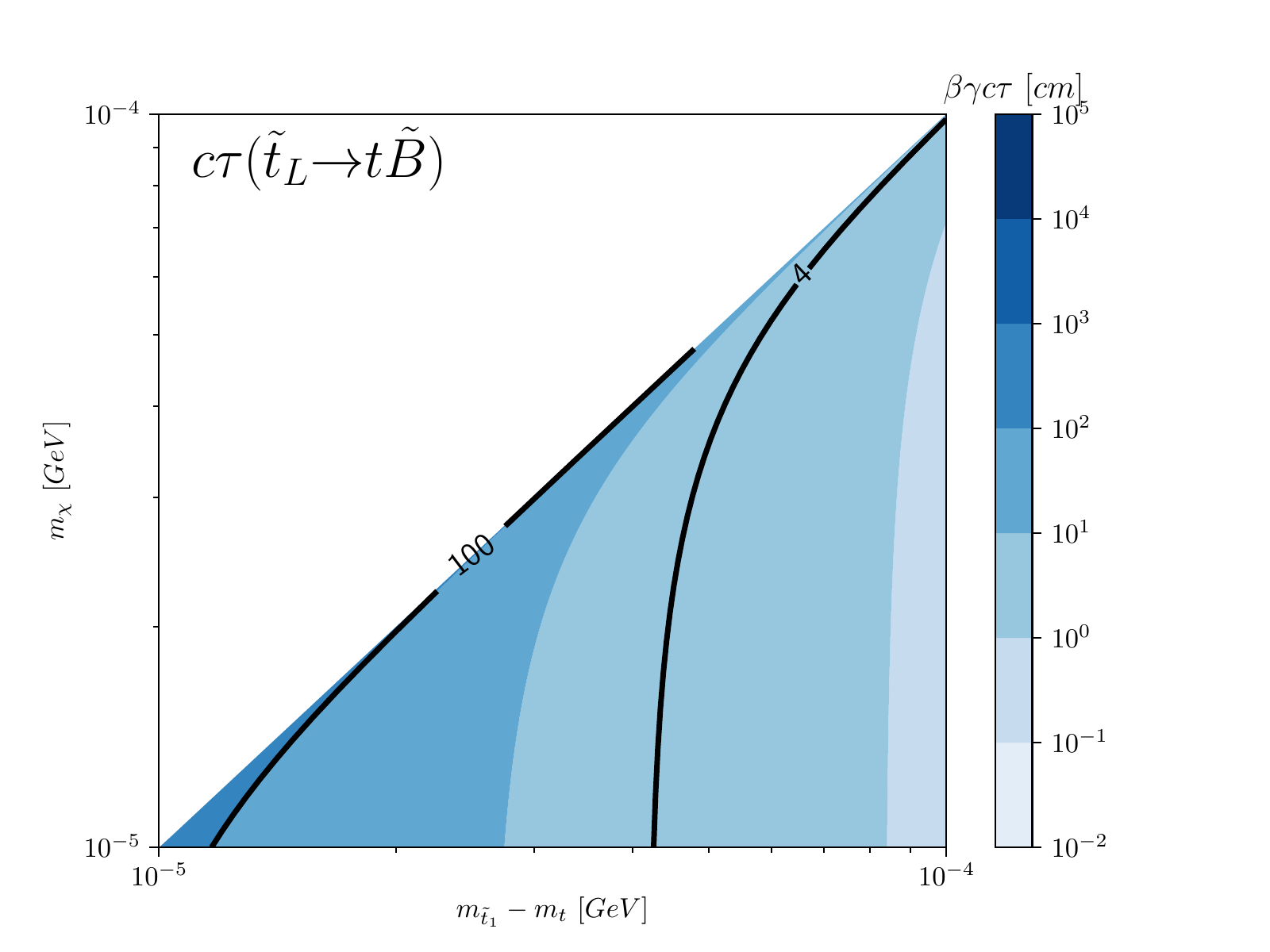}
      \centering
      \includegraphics[width=0.8\linewidth]{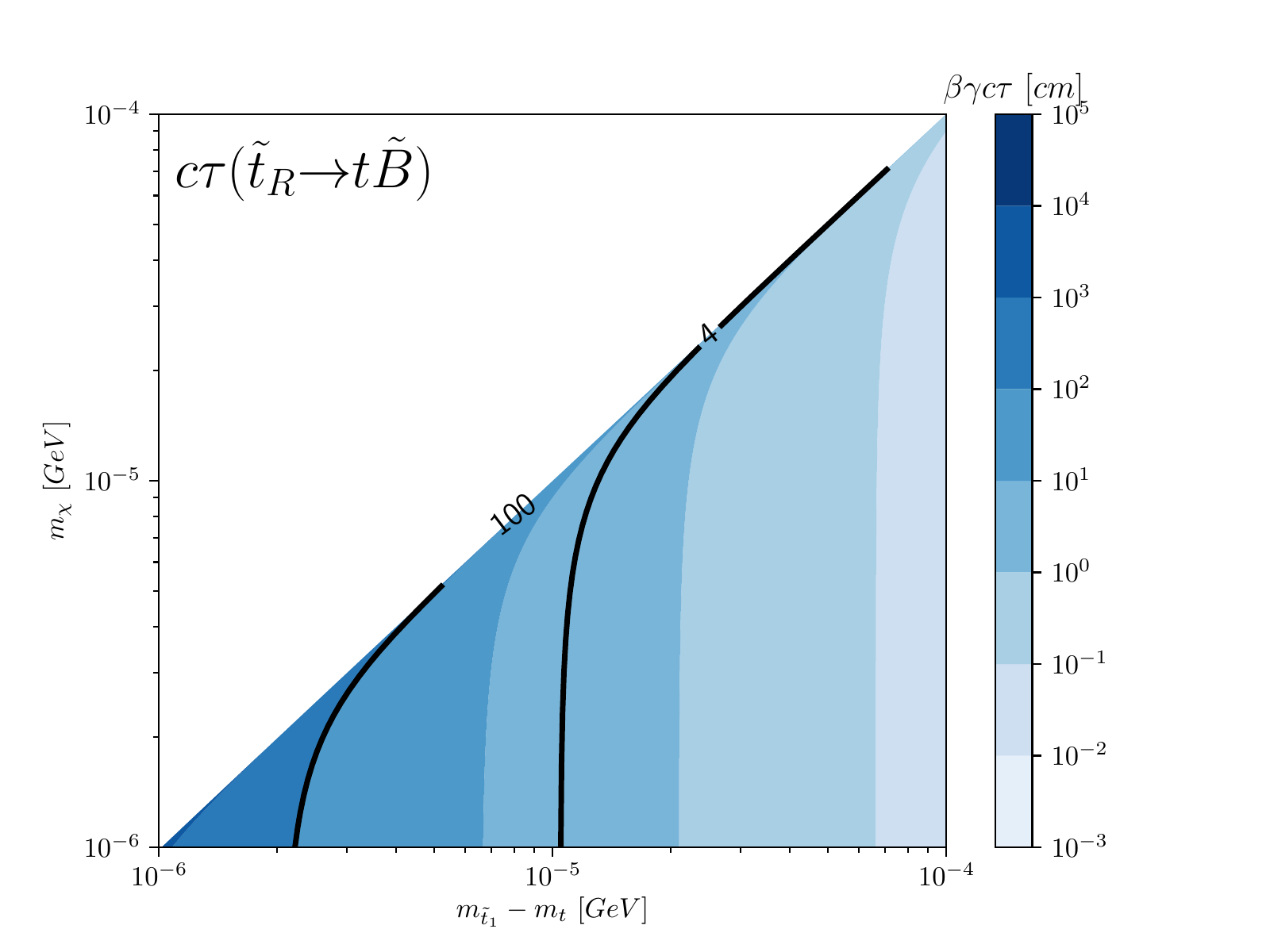}
      \label{fig:neutralino-flightdistance2} 
   \caption{Mean flight distance of the stop squark $\tilde{t}$ in the laboratory frame as function of the neutralino mass $m_{\chi^0_1}$ (with $\chi_1^0=\tilde{B}$) and the difference between stop and top mass $m_{\tilde{t}}-m_t$ for the various processes of \autoref{table:Decay}. The tracker volume corresponds to the region between the two black lines (between $4\ \text{cm}$ and $100\ \text{cm}$).}
\label{fig:ctauNeut}
\end{figure}

In order to agree with actual research on supersymmetric particles, we must take into account the constraints on squarks and neutralinos masses \cite{pdg}\footnote{We assume that those constraints can be applied to long-lived particles.}. We assume a low limit of the squark mass of $10^3\ \text{GeV}$, while the actual limit on the lowest mass of the neutralino is $m_{\tilde{\chi}_j^0}> 46\ \text{GeV}$. Looking at the results \autoref{fig:ctauNeut} and Appendix \autoref{app:meanflight}, it is evident that a displaced-top signal in the CMS tracker is only possible with a major fine-tuning on the neutralino mass $m_{\tilde{\chi}}$ and the stop mass $m_{\tilde{t}_i}$. A vast area in the $(m_{\tilde{\chi}},m_{\tilde{t}_i}-m_t)$ with $m_{\tilde{\chi}}\approx m_{\tilde{t}_i}-m_t\approx 10^{-5}-10^{-6}$ seems the only way to generate such processes.

\section{Decay of NLSP stop with LSP Gravitino $\psi_{\mu}$: $\tilde{t}\rightarrow \psi_{\mu} t $}\label{sec:gravitinoLSP}
Reintroducing the scale $M_p=m_p/\sqrt{8\pi}$ in the supergravity Lagrangian, the $(\psi_\mu , \phi , \chi)$-interaction terms are (see \autoref{eq:SUGRAlagrangian}): 
\begin{eqnarray}
\mathcal{L}_{int} &=& -\frac{1}{\sqrt{2}M_p}\left( \chi^i\sigma^{\mu}\bar{\sigma}^{\nu}\psi_{\mu}\tilde{\mathcal{D}}_{\nu}\phi^{\dagger}_{i} + \bar{\chi}_{i}\bar{\sigma}^{\mu}\sigma^{\nu}\bar{\psi}_{\mu}\tilde{\mathcal{D}}_{\nu}\phi^i \right) \nonumber
\end{eqnarray}
which can be rewritten in the Dirac formalism:
\begin{eqnarray}
\mathcal{L}_{int} &=& -\frac{1}{\sqrt{2}M_p}\big( \tilde{\mathcal{D}}_\nu \phi^i\bar{\chi}_L{}_i\gamma^{\mu}\gamma^{\nu}\Psi_{\mu}^{(M)} + \tilde{\mathcal{D}}_{\nu}\phi^{\dagger}_{i}\bar{\Psi}_{\nu}^{(M)}\gamma^{\nu}\gamma^{\mu}\chi_R^i \big)\ . \nonumber
\end{eqnarray}
with $\bar{\chi}_L{}_i$ and $\chi_R^i$ obtained with the chiral projector acting on the Dirac spinor $\chi^{(D)}$ and
\begin{gather}
\Psi_{\mu}^{(M)} = \begin{pmatrix}
\psi_{\mu} \\
\bar{\psi}_{\mu}
\end{pmatrix}\ . \nn
\end{gather} 
In a first part, we assume the process $\tilde{t}_L \rightarrow t_L \psi_{\mu}$.
Introducing $p_{\tilde{t}}^{\mu}$, the impulsion of the stop squark $\tilde{t}$, the vertex term corresponds to:
\begin{equation}
i\mathcal{M} = \frac{i}{\sqrt{2}M_p}\bar{u}p_{\tilde{t}}^{\mu}\big(1+\gamma_5\big)\Psi_{\mu}\label{eq:iM32} \ .
\end{equation}
For the following, we recall some calculus rules for the gravitino field $\psi_{\mu}$. The $\lambda$-helicity sum on the gravitino field gives  \cite{Pmunu32}
\begin{eqnarray}
P_{\mu\nu}(p) &=& \displaystyle\sum_{\lambda} \Psi_{\nu}\bar{\Psi}_{\mu} \\
&=& -\big( \slashed{p} - m_{3/2} \big)\Bigg\{ \Bigg( \eta_{\mu\nu} - \frac{p_{\mu}p_{\nu}}{m_{3/2}^2} \Bigg) -\frac13\Bigg( \eta_{\mu\sigma} - \frac{p_{\mu}p_{\sigma}}{m_{3/2}^2} \Bigg)\Bigg( \eta_{\lambda\nu} - \frac{p_{\lambda}p_{\nu}}{m_{3/2}^2} \Bigg)\gamma^{\sigma}\gamma^{\lambda} \Bigg\}  \nonumber
\end{eqnarray} 
where $P_{\mu\nu}$ obey the relations:
\begin{eqnarray}
\gamma^{\mu}P_{\mu\nu}(p) &=& P_{\mu\nu}(p)\gamma^{\nu} = 0\ , \nonumber \\
p^{\mu}P_{\mu\nu}(p) &=& P_{\mu\nu}(p)p^{\nu} = 0\ , \nonumber \\
\big( \slashed{p} - m_{3/2} \big)P_{\mu\nu}(p) &=& P_{\mu\nu}(p)\big( \slashed{p} - m_{3/2} \big) = 0\ . \nonumber 
\end{eqnarray}
From \autoref{eq:iM32}, the mean-value of the matrix-element reads:
\begin{equation}
|\bar{\mathcal{M}}|^2 = \frac{1}{2M_p^2}p_{\tilde{t}}^{\nu}p_{\tilde{t}}^{\mu}\text{Tr}\Big( \big( \slashed{p}_{t} + m_{t} \big)\big( 1 + \gamma_5 \big)P_{\nu\mu} \big(1 - \gamma_5\big) \Big)\nonumber
\end{equation}
with $p_t^{\mu}$ and $m_t$ the impulsion and the mass of the top quark. After computation using trace techniques, the decay width $\Gamma (\tilde{t}_L \rightarrow \psi_{\mu} t)$ follows:
\begin{eqnarray}
\Gamma (\tilde{t}_L\rightarrow \psi_{\mu}t ) & \approx & \frac{1}{48\pi M_p^2m_{3/2}^2m_{\tilde{t}}^3}\Big((-m_{3/2}^2+m_t^2+m_{\tilde{t}}^2)^2-4m_t^2m_{\tilde{t}}^2\Big)^{3/2}\nonumber \\
&& \times (m_{\tilde{t}}^2-m_{3/2}^2-m_t^2)
\end{eqnarray}
where $M_p=m_{p}/\sqrt{8\pi}$. Note also that if we consider a general mixing angle $\theta$ defined as:
\begin{equation}
\begin{pmatrix}
\tilde{u}_1 \\
\tilde{u}_2 
\end{pmatrix} = \begin{pmatrix}
c_{\theta} & -s_{\theta} \\
s_{\theta} & c_{\theta}
\end{pmatrix}\begin{pmatrix}
\tilde{u}_L \\
\tilde{u}_R 
\end{pmatrix}\ , \nonumber 
\end{equation}
the width reads then
\begin{eqnarray}
\Gamma (\tilde{t}_i\rightarrow \psi_{\mu}t ) &=& \frac{1}{48\pi M_p^2m_{3/2}^2m_{\tilde{t}}^3}\Big((-m_{3/2}^2+m_t^2+m_{\tilde{t}}^2)^2-4m_t^2m_{\tilde{t}}^2\Big)^{3/2}\nonumber \\
&& \times (m_{\tilde{t}}^2-m_{3/2}^2-m_t^2 +2\sin 2\theta m_tm_{3/2}) \ .
\end{eqnarray}
From the structure of $\Gamma$ above,  the mixing angle seems not to impact the decay width highly. The mixing angle can then be taken as $\theta=0$ for simplification. From the function $\big< \beta\gamma \big> (m_{\tilde{t}_1})$ in \autoref{eq:betagamma}, the corrected flight distance of the stop squark $\beta\gamma c\tau_{corr.}$ can directly be computed.
\begin{figure}[H]
    \centering
      \includegraphics[width=.9\linewidth]{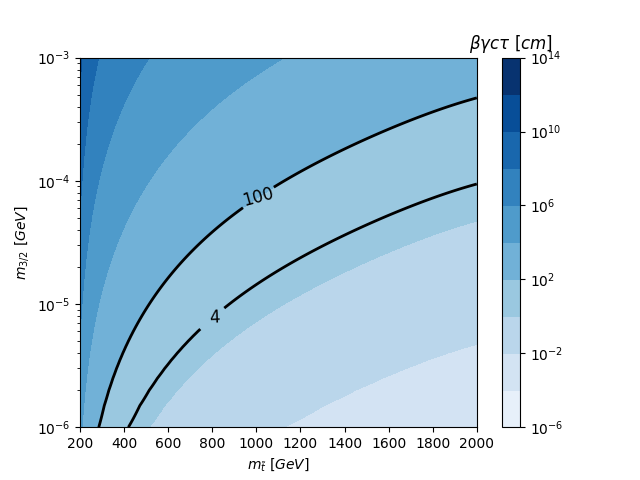}
    \caption{Mean flight distance of the stop in the laboratory frame as a function of stop and gravitino mass.}
    \label{fig:stop_flightdistance}
\end{figure}
Imposing a decaying stop in the tracker volume, \text{i.e.}, $\beta\gamma c\tau \in \left[4\ \text{cm},\ 100\ \text{cm}  \right]$, the \autoref{fig:stop_flightdistance} shows us a large data range in $(m_{3/2},m_{\tilde{t}_1})$-plane. Moreover, those results are coherent with the actual stop mass limit $(m_{\tilde{t}}\gtrapprox 1\ \text{TeV})$. A gravitino mass higher than $m_{3/2} \gtrapprox 10^{-5}\ \text{GeV}$ seems then appropriate. 
\subsection{Benchmarks definition and observables distributions}\label{subsec:bench}
To pursue the analysis, we define several benchmarks that represent all the parameters space's kinematics. The different topologies of signals can be defined with:
\begin{itemize}
\item a stop mass of $m_{\tilde{t}}=1\ \mathrm{TeV}$ leading to a high cross-section and $m_{\tilde{t}}=1.4\ \mathrm{TeV}$ corresponding to a few amounts of events (see \autoref{fig:stop_xsec}),
\item a flight distance of $10,\ 30,\ 50$ and $70\ \mathrm{cm}$ representing the complete geometry of the tracker
\end{itemize}
which define eight different benchmarks represented in Table \ref{table:bchmrk} from the results of \autoref{fig:stop_flightdistance}.
\begin{table}[H]
\begin{center}
\begin{tabular}{ ||c||c|c| } 
\hline\xrowht[()]{20pt}
 & $m_{\tilde{t}}=1\ \mathrm{TeV}$ & $m_{\tilde{t}}=1.4\ \mathrm{TeV}$\\
\hline
\hline\xrowht[()]{20pt}
$\beta\gamma c\tau = 10\ \mathrm{cm}$  & $m_{3/2}=2.28\times 10^{-5}\ \mathrm{GeV}$ & $m_{3/2}=5.73\times 10^{-5}\ \mathrm{GeV}$  \\ 
\hline\xrowht[()]{20pt}
$\beta\gamma c\tau = 30\ \mathrm{cm}$  & $m_{3/2}=3.94\times 10^{-5}\ \mathrm{GeV}$ & $m_{3/2}=9.93\times 10^{-5}\ \mathrm{GeV}$  \\ 
\hline\xrowht[()]{20pt}
$\beta\gamma c\tau = 50\ \mathrm{cm}$  & $m_{3/2}=5.09\times 10^{-5}\ \mathrm{GeV}$ & $m_{3/2}=1.28\times 10^{-4}\ \mathrm{GeV}$  \\ 
\hline\xrowht[()]{20pt}
$\beta\gamma c\tau = 70\ \mathrm{cm}$  & $m_{3/2}=6.02\times 10^{-5}\ \mathrm{GeV}$ & $m_{3/2}=1.52\times 10^{-4}\ \mathrm{GeV}$  \\ 
\hline
\end{tabular}
\caption{Benchmarks definition.}
\label{table:bchmrk}
\end{center}
\end{table}
We can now compute the generation of events for each benchmarks. 
\subsection{Event generation}
Each benchmark can be used to simulate events of collisions at the LHC and so study the distributions of several observables based on the kinematic of the particles. The partonic generation is done with the program \textsc{MadGraph\_aMC@NLO}. We generated 10 000 events for each benchmark. In order to simulate the hadronic environment of proton-proton collisions and hadronise the coloured particles, we use the program \textsc{Pythia} \cite{pythia} following the MCTunes \textsc{CUETP8M2T4} \cite{CMS-PAS-TOP-16-021}. \medskip

When a coloured supersymmetric particle has a sufficient lifetime (bigger than the hadronisation time), it can interact with coloured particles from the Standard Model and create new bound-states called R-hadrons \cite{FARRAR_Rhadron}. We can define different types of R-hadrons such as R-mesons $(\tilde{g}q\bar{q},\ \tilde{q}\bar{q})$, R-baryons $(\tilde{g}qqq,\ \tilde{q}qq)$ or gluino ball $(\tilde{g}g)$. \medskip

From the partonic background of the LHC experiments, those states can be generated and possibly impact the searched signal in many ways. In our case, the decay of the long-lived stop squark $\tilde{t}\rightarrow t\psi_{\mu}$ can be modified by the formation of R-mesons or R-baryons.\newline
On the other hand, the formation of such bound-states can appear \textit{as} underlying processes and pollute the final state. This may have an impact on the reconstruction efficiency. R-hadrons can also interact with the matter of the detector which could impact the kinematic of the signal. Two different models generally describe those interactions:
\begin{itemize}
\item a "cloud model" \cite{Rhadron_model11}\cite{Rhadron_model1} where the R-hadron is surrounded by a cloud of coloured light particles. Those particles interact then with the material of the detector during scattering. 
\item a model of complete charge suppression \cite{Rhadron_model2} where they assume neutral R-hadrons when entering the detector's muon-system.
\end{itemize}}
Moreover, strong interaction between light quarks and the material of the detector can increase the energy loss and allow to modify the R-hadron electric charge (a charged R-hadron can switch of electric charge or become neutral).\medskip

Those aspects are simulated within the \textsc{Pythia} program.\footnote{A deeper analysis has shown that we are mainly in the context of the model of spectator quarks, \textit{i.e.}, quarks that are not part of the hard process do not impact the decay.}\medskip

Finally, all mesons and hadrons are merged into jets using \textsc{FastJet} with anti-$k_t$ algorithm \cite{fastjet}.
\subsection{Observable distributions}\label{subsec:obs}
The \textsc{MadAnalysis5} framework is used (in \textit{expert} mode) to display distributions of several observables such as transverse momentum $p_T$ (transverse to the beam axis) and pseudorapidity $\eta=-\ln\tan(\theta/2)$ (with $\theta$ the angle between the momentum and the beam) of top quarks, leptons and leading jets. Global observable such as Total Transverse Energy (usually denoted $E_T$ or $TET$), Missing Transverse Energy (already defined in \autoref{eq:MET}), Missing Transverse Hadronic Energy (called $MHT$ and equivalent to the $MET$ focusing only on the hadronic particles) and Total Hadronic Energy ($H_T$ or $THT$):
\begin{gather}
E_T = \displaystyle\sum_{visible\ particles} ||\vec{p}_T|| , \qquad H_T = \displaystyle\sum_{hadronic\ particles} ||\vec{p}_T||  \nn
\end{gather}
are also reconstructed. Finally, some other specific observables are calculated: 
\begin{itemize}
\item  an observable called $\alpha_T$ \cite{alphaT} defined for dijet events:
\begin{gather}
\alpha_T = \frac{p_{T_2}}{m_{jj}} \nn
\end{gather}
with $p_{T_2}$ the transverse momentum of the second hardest jet and $m_{jj}$ the invariant mass of the two hardest jets. For background events (where the jets are generally produced back-to-back), we have $\alpha_T \lessapprox 1/2$. High values of $\alpha_T$ are then possible hints to new physics events.  
\item the transverse and longitudinal impact parameters $d_0$ and $d_z$ of particles defined as (the position of the vertex production if defined as $(x,y,z)$): 
\begin{gather}
d_0 = \frac{1}{p_T}\big( xp_y - yp_x \big), \qquad d_z = z - \frac{p_z}{p_T^2}\big( xp_x + yp_y \big)\ . \nn 
\end{gather}
(we follow the definition of the CMS collaboration\footnote{The definition of $d_z$ comes from internal codes of the CMS collaboration.} \cite{TDR1}). Note that we do not take into account the impact of the magnetic field. 
\end{itemize}
Some distributions can be found in Figures [\ref{fig:do_t}, \ref{fig:met}, \ref{fig:ptj}, \ref{fig:dr1}], the others can be seen in Appendix \autoref{chap:analysis_distrib}. \medskip
\begin{figure}[H]
    \centering
      \includegraphics[width=.9\linewidth]{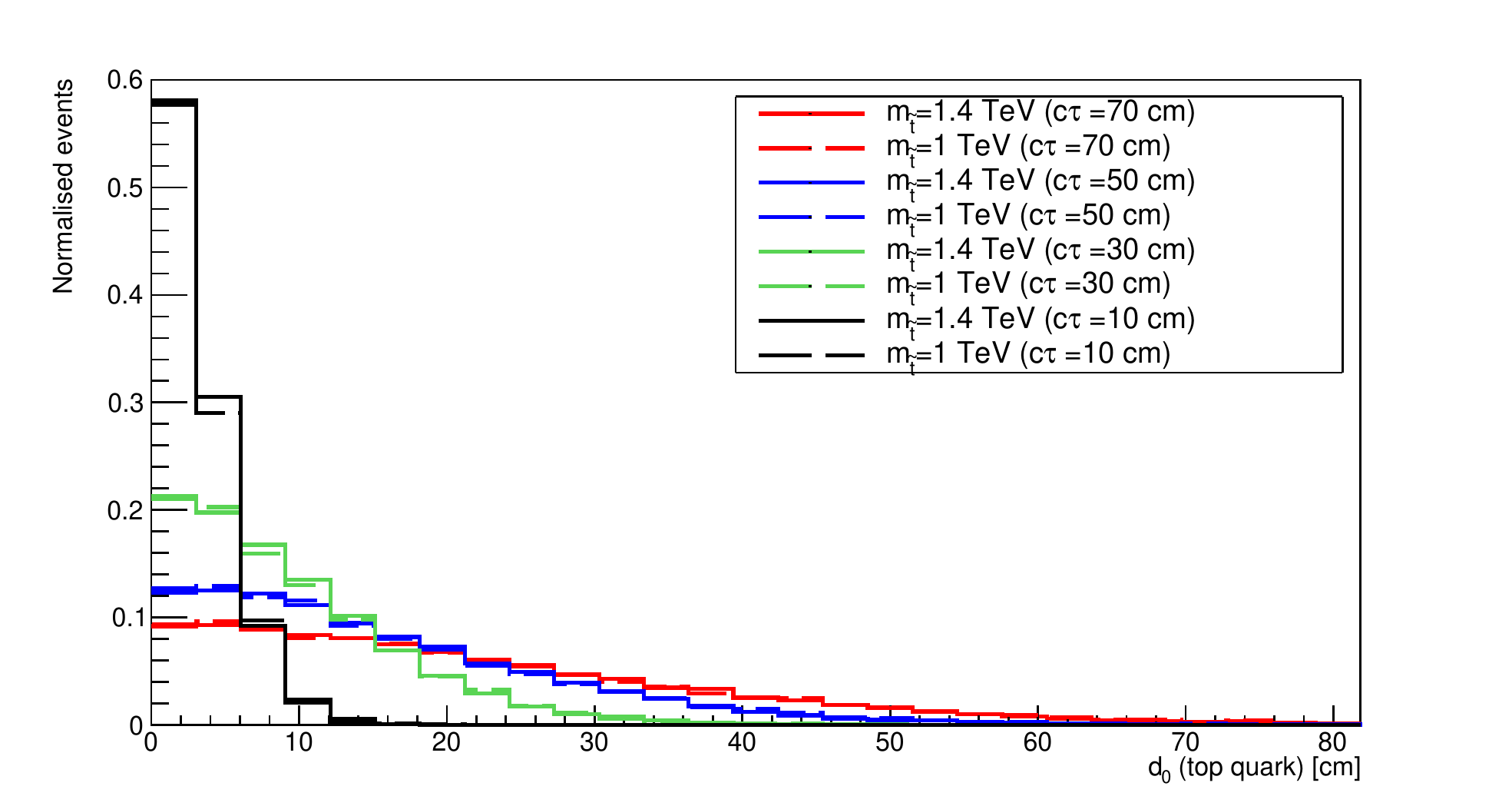}
    \caption{Distributions of $d_0$ of the top quark for $m_{\tilde{t}}=1\ \text{TeV}$ and $m_{\tilde{t}}=1.4\ \text{TeV}$ (the choice of $m_{3/2}$ do not impact the kinematic).}
    \label{fig:do_t}
\end{figure}
We note from \autoref{fig:do_t} that the stop mass does not modify the transverse impact parameter $d_0$ (the same observation can be done for all the distributions of $d_0$ and $d_z$ in Appendix \autoref{chap:analysis_distrib}, Sections \ref{app:d0} and \ref{app:dz}). We obtain for the four values of $c\tau$ four different distributions of $d_0$ and $d_z$ of the top quark, the b-quark and leptons coming from the decay of the top quark. We then note that there are displaced tracks in our signal.  \medskip 

Since the gravitino mass is low, the different values of $m_{3/2}$ do not impact the kinematic of the process. We can then only analyse the distributions regarding the two values of $m_{\tilde{t}}$. Due to the structure of the detectors, a cut on the pseudorapidity is usually done at $| \eta | < 2.4$. Our investigation has pointed out that the majority of the particles pass these constraints (see Section \ref{app:eta}). The study on the global transverse observables (see for example \autoref{fig:met} and Section \ref{app:transv}) and the $p_T$ of the leading jet (see \autoref{fig:ptj} and Section \ref{app:pt}) show the highly energetic nature of the events. We also note that for a fixed value of $m_{\tilde{t}}$, the distributions are similar for different gravitino masses $m_{3/2}$. This is due to the smallness of $m_{3/2}$. The leading jets have high transverse momentum $p_T$, since they come from the decay of a heavy stop.
\begin{figure}[H]
    \centering
      \includegraphics[width=.9\linewidth]{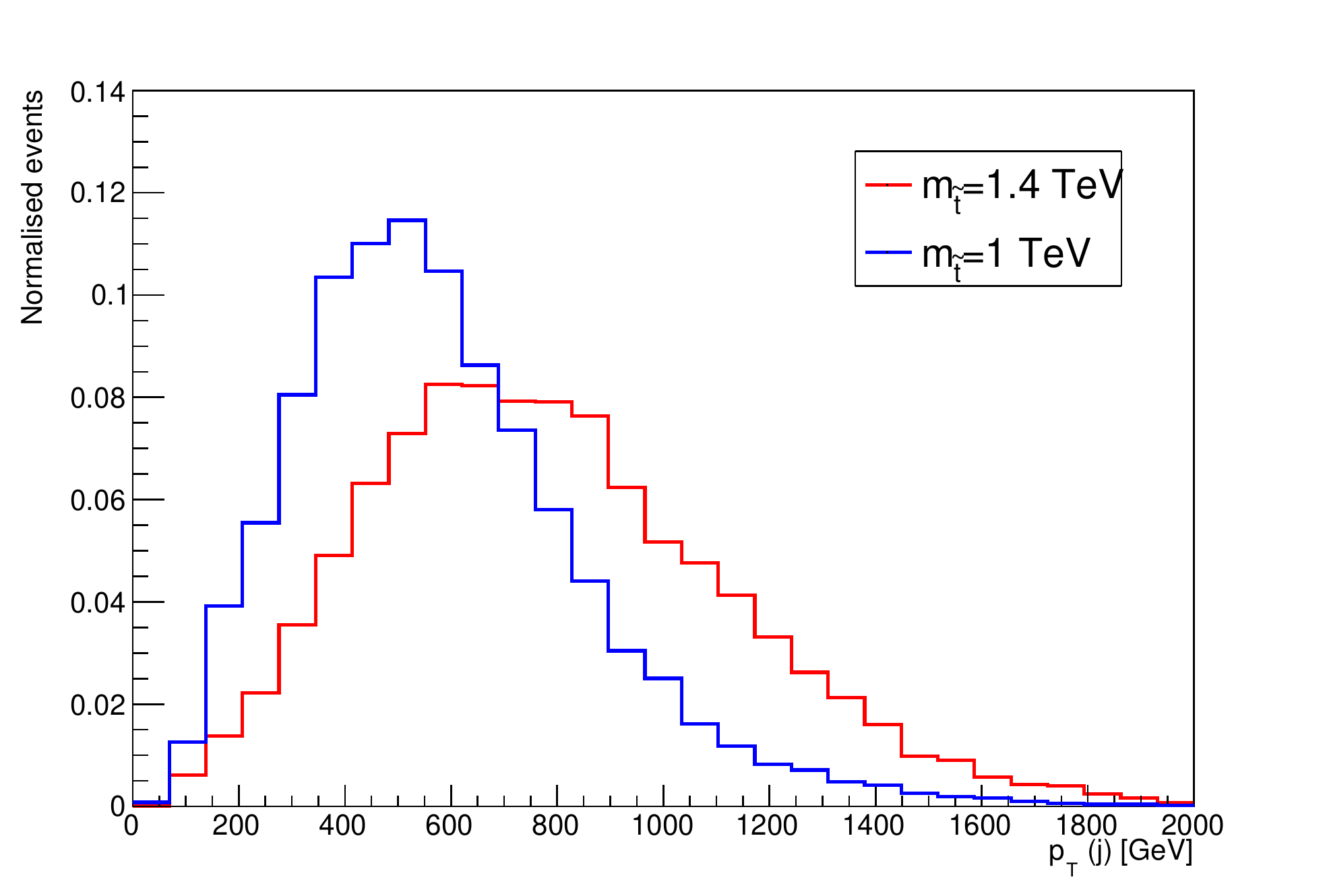}
    \caption{Distributions of the transverse momentum of the leading jet $p_T(j)$ (b-jet included) for $m_{\tilde{t}}=1\ \text{TeV}$ and $m_{\tilde{t}}=1.4\ \text{TeV}$ (the choice of $m_{3/2}$ do not impact the kinematic).}
    \label{fig:ptj}
\end{figure}
\begin{figure}[H]
    \centering
      \includegraphics[width=.9\linewidth]{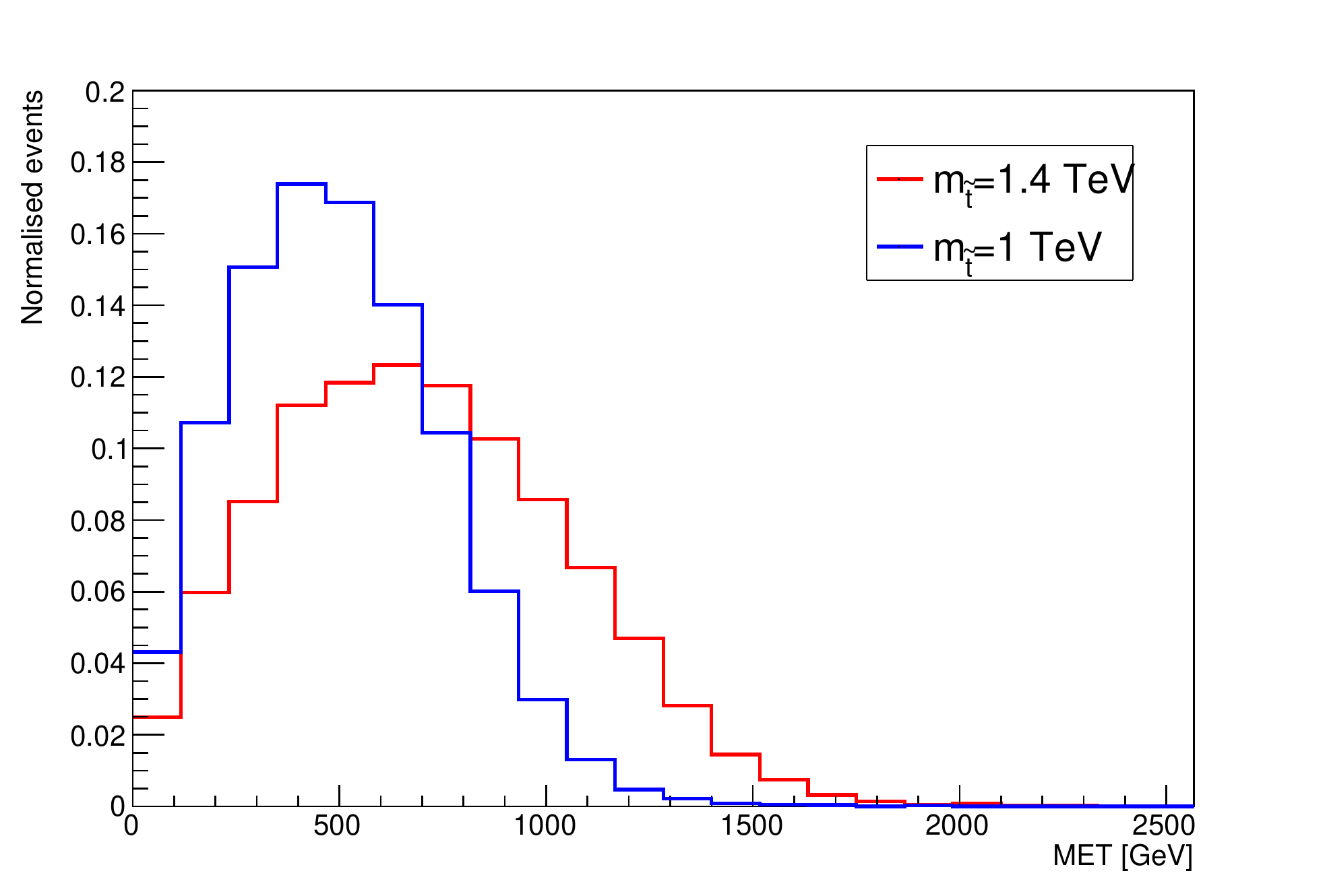}
    \caption{Distributions of $MET$ for $m_{\tilde{t}}=1\ \text{TeV}$ and $m_{\tilde{t}}=1.4\ \text{TeV}$ (the choice of $m_{3/2}$ do not impact the kinematic).}
    \label{fig:met}
\end{figure}
Since the stop mass is in the order of the TEV scale, the decay products can be highly boosted. It leads then to b-quarks coming from the decay close to light quarks coming from the decay of the W boson. This effect may be a problem for the reconstruction. Indeed, if the quarks are too close to each other, all the particles from the decay can be defined into a single heavy jet, which modifies the assumed signature. In order to properly study such events, a specific algorithm for the analysis of sub-structure of jets must be used, which complicates the study. For this purpose, we reconstruct the variable $\Delta R$ defined as
\begin{gather}
\Delta R = \sqrt{(\Delta\eta)^2 + (\Delta \phi)^2}\nn
\end{gather}
(with $\eta$ the pseudorapidity and $\phi$ the angle in the transverse plane) which is the distance between two quarks. We have analysed the distributions of the distance between b-quarks and the closest (second closest) light quark coming from the top quark (denoted as $\Delta R_1$ ($\Delta R_2$)). The distribution of $\Delta R_1$ is shown in \autoref{fig:dr1}.
\begin{figure}[H]
    \centering
      \includegraphics[width=.9\linewidth]{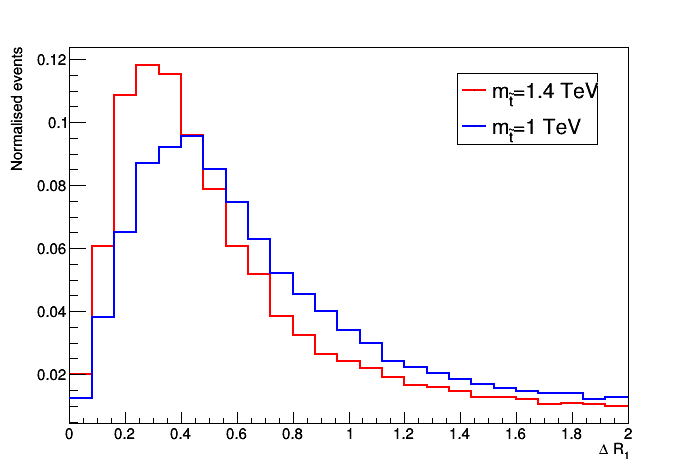}
    \caption{Distributions of $\Delta R_1$ ($\Delta R$ between b-quark and the closest quark) for $m_{\tilde{t}}=1\ \text{TeV}$ and $m_{\tilde{t}}=1.4\ \text{TeV}$ (the choice of $m_{3/2}$ do not impact the kinematic).}
    \label{fig:dr1}
\end{figure}
In general, the quarks which are close to each others with $\Delta R \lessapprox 0.4$ are combined into a single jet by the jet reconstruction algorithms. We note from the distributions that there is a non-negligible amount of quarks from the top decay, which are relatively close with each other and have $\Delta R < 0.4$.\medskip

All the important quantities are displayed in the Tables \ref{table:meanvalue}, \ref{table:eventninvol1} and \ref{table:eventninvol2}. The mean values of several observables are shown in Table \ref{table:meanvalue}. The rate of events with top quarks decaying in the tracker volume (between $4\ \text{cm}$ and $100\ \text{cm}$) are also presented in Tables \ref{table:eventninvol1} and \ref{table:eventninvol2}. We note that for short flight-distance (with $c\tau = 10\ \text{cm}$), around 30\% of the events will not contain signatures of displaced top quarks since top quarks decay before entering the tracker volume. The highest amount of events with two quark tops decaying in the tracker is obtained for $c\tau =50\ \text{cm}$, which corresponds to the middle of the tracker. Note that in the case where the top quarks decay after the tracker volume (with $c\tau > 100\ \text{cm}$), then there will pollute the $MET$ distributions (or the $THT$ distributions if there is interaction with the hadronic calorimeter). 
\begin{table}[H]
\begin{center}
\begin{tabular}{ ||c||c|c| } 
\hline\xrowht[()]{20pt}
 & $m_{\tilde{t}}=1\ \text{TeV}$ & $m_{\tilde{t}}=1.4\ \text{TeV}$ \\
\hline
\hline\xrowht[()]{20pt}
$\lag MET \rag$ [GeV]  & 503.6 & 702.3 \\ 
\hline\xrowht[()]{20pt}
$\lag TET \rag$ [GeV]  & 1265.5  & 1585.8 \\ 
\hline\xrowht[()]{20pt}
$\lag THT \rag$ [GeV]  & 1117.5 & 1412.8 \\ 
\hline\xrowht[()]{20pt}
$\lag MHT\rag $ [GeV]  & 489.7 & 679.0 \\ 
\hline\xrowht[()]{20pt}
$\lag p_T (j_1)\rag $ [GeV]  & 581.0 & 785.8 \\ 
\hline
\end{tabular}
\caption{Mean values of some observables for $m_{\tilde{t}}=1\ \text{TeV}$ and $m_{\tilde{t}}=1.4\ \text{TeV}$.}
\label{table:meanvalue}
\end{center}
\end{table}
\vspace{-2em}
\begin{table}[H]
\begin{center}
\begin{tabular}{ ||c||c|c|c|c| } 
\hline\xrowht[()]{20pt}
\multirow{2}{*}{\begin{tabular}{@{}c@{}}\% events in \\ the tracker volume 
\end{tabular} }
& \multicolumn{4}{|c|}{$m_{\tilde{t}}=1\ \text{TeV}$} \\\cline{2-5}
& $c\tau = 10\ \text{cm}$ & $c\tau = 30\ \text{cm}$ & $c\tau = 50\ \text{cm}$ & $c\tau = 70\ \text{cm}$ \\
\hline
\hline\xrowht[()]{20pt}
\begin{tabular}{@{}c@{}}2 top decay in \\ the tracker volume 
\end{tabular} & 58.8 & 92.7 & 96.7 & 95.1 \\ 
\hline\xrowht[()]{20pt}
\begin{tabular}{@{}c@{}}1 top decay in \\ the tracker volume
\end{tabular} & 11.1 & 4.9 & 2.3 & 2.1 \\ 
\hline
\end{tabular}
\caption{Number of events with top quarks decaying in the tracker volume for $m_{\tilde{t}}=1\ \text{TeV}$. There is three different cases: the two tops coming from the stop squarks decay in the tracker volume, only one top quark decay in the volume or not any.}
\label{table:eventninvol1}
\end{center}
\end{table}
\vspace{-2em}
\begin{table}[H]
\begin{center}
\begin{tabular}{ ||c||c|c|c|c| } 
\hline\xrowht[()]{20pt}
\multirow{2}{*}{\begin{tabular}{@{}c@{}}\% events in \\ the tracker volume 
\end{tabular} }
&  \multicolumn{4}{|c|}{$m_{\tilde{t}}=1.4\ \text{TeV}$} \\\cline{2-5}
 & $c\tau = 10\ \text{cm}$ & $c\tau = 30\ \text{cm}$ & $c\tau = 50\ \text{cm}$ & $c\tau = 70\ \text{cm}$ \\
\hline
\hline\xrowht[()]{20pt}
\begin{tabular}{@{}c@{}}2 top decay in \\ the tracker volume 
\end{tabular}  & 60.6 & 93.1 & 97.1 & 96.7\\ 
\hline\xrowht[()]{20pt}
\begin{tabular}{@{}c@{}}1 top decay in \\ the tracker volume
\end{tabular}  & 9.4 & 4.0 & 2.0 & 1.6\\ 
\hline
\end{tabular}
\caption{Number of events with top quarks decaying in the tracker volume for $m_{\tilde{t}}=1.4\ \text{TeV}$. There is three different cases: the two tops coming from the stop squarks decay in the tracker volume, only one top quark decay in the volume or not any.}
\label{table:eventninvol2}
\end{center}
\end{table}
Now that we have analysed the kinematic of the process following the different benchmarks, we can conclude this analysis with a discussion. 
\section{Discussion}
The observables distributions are, however, not sufficient to conclude on an analysis. There are still several topics that must be investigated, such as the trigger system or the associated background. This work must be performed by the experimentalists to complete this analysis. 
\subsection{Trigger for the process $\tilde{t}\rightarrow t\psi_{\mu}$}
Due to the high frequency of collisions at the LHC (particle beams cross every $25\ \text{ns}$ at the LHC, which corresponds to a collision frequency of $40\ \text{MHz}$), the CMS and ATLAS experiments use various trigger levels to select events with physical interests among the numerous collisions events \cite{CMS_Trigger}\cite{ATLAS_Trigger}. The CMS collaboration uses a two-level trigger-system: the Level-1 Trigger (L1), which use information from the calorimeters, and the muon chamber and the High Level Trigger (HLT) using all the detector.). The first layer allows reducing the data flow from $40\ \text{MHz}$ to $100\ \text{kHz}$ whilst the HLT restrict the flow to approximately $1\ \text{kHz}$ containing events with physics and calibration interests.\newline
In the case of the signal $\tilde{t}\rightarrow t\psi_{\mu}$, the observables which can be used are:
\begin{itemize}
\item the missing transverse energy, or $MET$ (defined in \autoref{eq:MET}), coming from the non-interaction of the LSP gravitino,
\item the scalar sum of the transverse momentum of jets since a high $H_T$ can signify a heavy coloured particle in the process such as squark stop, 
\item in the case where $\beta\gamma c\tau$ is relatively small, the identification of light leptons coming from the decay of a W-boson ($W^- \rightarrow \ell^- \bar{v}_\ell$ or $W^+ \rightarrow \ell^+ v_\ell$)
\item in the case of a high flight distance, information from the muon system can be useful to detect such signals (those types of trigger are commonly used in the case of, \textit{e.g.}, HSCP searches). They also work on the rate of energy loss via ionisation (or $dE/dx$) \cite{hscp}. 
\end{itemize}
We saw in the previous section the presence of high $MET$ and $H_T$ events in our analysis. Actual search on HSCP scenario in the CMS collaboration use as trigger a missing transverse energy $MET>170\ \text{GeV}$ or a reconstructed muon with $p_T>50\ \text{GeV}$ (see \cite{hscp}). In the case of short-lived stop (see \cite{stoprodCMS}), one or two isolated leptons with $p_T> 23\ \text{GeV}$ (leading lepton) and $p_T> 17\ \text{GeV}$ (subleading lepton) are required. For our analysis, the trigger on $MET$ seems relevant, as well as a search of high $THT$ events. These triggers may be used if they are implemented for the Run 3 of the LHC. 
\subsection{Associated background}
The Standard Model background associated with the process $\tilde{t}\rightarrow t\psi_{\mu}$ must also be produced for a complete analysis. This part is, in general, the most time-consuming since a lot of events must be generated. The main background for low flight distance must be composed of production of top-pair quarks $t\bar{t}$, Drell-Yan processes with an associated $MET$ coming from an incorrect reconstruction, $tW/\bar{t}W$ processes with fake b-tagged jet or $tt+X$ (with, for example, $X=Z$ with $Z$ decaying in neutrinos). See for example the analysis \cite{stoprodCMS}. Cuts on various observables such as $\alpha_T$ with $\alpha_T>0.5$ (see Section \ref{app:distribother}) may be a solution to suppress these backgrounds. In the specific case of HSCP particles, the totality of the background is identified as experimental and algorithmic issues.\medskip

There is also specific background associated with displaced vertices in the tracker which can be divided into two categories. Firstly, cosmic rays can interact with the atmosphere and induce displaced muon signals in the muon system or displaced jets in the calorimeters. One possible solution to avoiding such backgrounds is to impose high $p_T$ muon and jets tracks since particles from cosmic-ray quickly lose their momentum. They can also be estimated from simulation or from dedicated data-taking runs. Secondly, nuclear interactions with parts of the detector can also provoke such signals. In order to suppress such backgrounds, a precise density map of the detector material is done and potential LLP signature coming from a high densities area is vetoed (see \cite{llp_eric} for more information). 
\section{LSP gravitino with Non-Soni-Weldon Solutions?}
In this chapter, we have analysed the decay of a stop squark $\tilde{t}$ into a top quark $t$ and a gravitino $\psi_{\mu}$ in the context of \textit{Gauge Mediated Supersymmetry Breaking} mechanism. In such configuration, the gravitino mass is in the order of 
\begin{equation}
m_{3/2} = \frac{\langle W \rangle }{m_p^2}e^{|\langle z\rangle |^2/2} \sim \frac{M_{SUSY}^2}{m_p}\label{eq:m32GMSB}
\end{equation}
which naturally leads to a LSP-gravitino. Thus, the process $\tilde{t}\rightarrow t\psi_{\mu}$ experimentally leads to displaced top tracks with source of $MET$.\\
In the context of the \textit{Gravity Mediated Supersymmetry Breaking} (mainly presented in chapter \ref{sugra}) and following the classical Soni-Weldon solutions, the gravitino mass is in the order of magnitude of an intermediate energy scale $M$, leading to gravitino mass in the order of the TeV scale. The gravitino is then naturally heavier in the context of \textit{Gravity Mediated Supersymmetry Breaking} with Soni-Weldon solutions and the presented analysis can not take place in this context.\medskip

However, following the new solutions presented in \ref{nsw}, the gravitino mass takes a similar form of the \textit{Gauge Mediated Supersymmetry Breaking} case (see \autoref{eq:m32GMSB}) with $M_{SUSY}$ replaced by a new intermediate energy scale much lower than $m_p$
\begin{equation}
m_{3/2}^{(NSW)} \sim \frac{M^2}{m_p}\nonumber
\end{equation}
(NSW meaning Non-Soni-Weldon solutions). Depending on the energy scale $M$, $m_{3/2}$ can then be relatively small and so compared to the GMSB case. The presented analysis may then be applied to \textit{Gravity Mediated Supersymmetry Breaking} mechanism. \medskip

However, since those new solutions have been developed recently, deeper study must be done to clarify if such scenario is possible.

\label{llp}

\chapter{General conclusion \& Perspectives}


%
The work carried out during this thesis deals with the framework of supersymmetry and supergravity. Supersymmetry is a field theory involving a symmetry between fermions and bosons, while supergravity is a local theory of supersymmetry that naturally includes gravitational interactions and leads to global supersymmetry at low energy.
In order to match experimental measures and theoretical predictions, supersymmetry has to be spontaneously broken. In the context of supergravity, supersymmetry breaking can be mediated from a hidden field sector to the usual fields (visible sector) through, among others, gravitational interactions. These mechanisms are called \textit{Gravity Mediated Supersymmetry Breaking} mechanisms and have been studied by Soni \& Weldon, who classified the possible forms of the two fundamental functions in supergravity: the superpotential and the Kähler potential. 
\medskip

Recently, new mechanisms of \textit{Gravity Mediated Supersymmetry Breaking} have been identified, in particular leading to new solutions (defined as Non-Soni-Weldon solutions) with new characteristics. These new solutions introduce a new field sector $\{S^p\}$, called hybrid, which have properties of both hidden and visible sector and must be gauge singlets. In order to generate non-trivial couplings between the hybrid fields and the visible sector, the structure of the superpotential imposes the presence of at least two hybrid fields $\{S^1,S^2\}$. Along the lines of the new solutions, two models involving two gauge singlets have been identified: a simple extension of the MSSM, called the N2MSSM, and a model inspired by the Non-Soni-Weldon solutions, the S2MSSM. Studying these two models will help understand the differences between Non-Soni-Weldon (NSW) and Soni-Weldon (SW) solutions.\medskip

More into detail, the general form of the $N=1$, $D=4$ supergravity Lagrangian has been reproduced. The mechanisms of supersymmetry breaking classified by Soni \& Weldon (which generate soft-breaking terms in the Lagrangian) has also been investigated. The Non-Soni-Weldon solutions, which lead to hard-breaking terms (parametrically suppressed) in the Lagrangian have been explicitly computed. The form of the hard-breaking terms have been classified and a specific model, the S2MSSM has been properly defined. The complete calculation of the scalar potential, the minimisation equations of the potential and the mass matrix has been done for this model.\medskip

A natural question arises: does the S2MSSM naturally generate a Higgs boson mass near $125\ \text{GeV}$ and explain the non-detection of supersymmetric particles at the LHC? In order to understand these important aspects, we have considered, in a first step toward a general study, a simple model considering only one $S$ hybrid field and one field $z$ from the hidden sector. The order of magnitude of the one-loop radiative corrections of the S-field on scalar masses, and more specifically on the Higgs boson mass has been calculated. Our study has pointed out some possible configurations leading to $m_h\approx 125\ \text{GeV}$. Such configurations are, however, fine-tuned. Nonetheless, we have identified more complex models, including several fields from the hidden sector, which may solve this naturality issue. An analysis considering several S-fields must also be performed for a complete study. \medskip

Simultaneously, the analogue of the S2MSSM in the framework of the standard solutions of Soni \& Weldon, the N2MSSM, has been studied. This manuscript mainly focused on the differences between the NMSSM and the N2MSSM. Spectrum generators have been generated by the programs \textsc{SARAH} and \textsc{SPheno}, and an MCMC algorithm (Markov-Chain Monte-Carlo) has been implemented to scan the parameters space easily. The Higgs boson mass has been constrained with the range $[118\ \text{GeV}, 132\ \text{GeV}]$. The squark sector is supposed to be heavier than $1\ \text{TeV}$ and a limit on the gluino mass is imposed such that $m_{\tilde{g}}>1.6\ \text{TeV}$. In this preliminary analysis, a reduction of the fine-tuning on the electroweak scale $\Delta_{FT}$ in the N2MSSM was expected. However, due to the stochastic nature of the MCMC algorithm, statistical variations on the fine-tuning exist. We concluded that the difference of fine-tuning between the NMSSM and the N2MSSM is not statistically significant. More accurate scans of the parameters space, including a reduction of the allowed range on the reconstructed Higgs boson mass can be obtained with the help of grid-computing resources. Other experimental limits should also be taken into account, especially the dark matter relic density $\Omega_{DM} h^2$. Moreover, the excluded regions of the NMSSM parameters space should also be investigated in the context of the N2MSSM.\\

We have now all the tools for a complete analysis of the new solutions. Since the S2MSSM and the N2MSSM are closely related, it is important to understand the phenomenological differences between these two models in particular and between the classical solutions of Soni and Weldon and the new solutions in general. The N2MSSM spectrum generator can be adjusted for the S2MSSM for a complete phenomenological analysis of NSW solutions. The preliminary analysis on the order of magnitude of the one-loop Higgs boson mass has revealed the regions of interest in the parameters-space of the S2MSSM to naturally obtain $m_h\approx 125\ \text{GeV}$. A deeper analysis will provide the phenomenological interest of these new solutions in the context of particle physics and cosmology. This work is beyond this thesis and will be done further. \medskip




In parallel to these studies, an investigation of an experimental signature of a long-lived particle has been performed. Two possible scenarios have been analysed, both containing displaced top tracks signal in the tracker volume (with a flight distance between $4$ and $100\ \text{cm}$) associated with an excess of MET coming from the lightest supersymmetric particle (LSP). The two channels are: a long-lived stop decaying into a top and a LSP neutralino $\tilde{t}\rightarrow \chi^0_i t$ and a long-lived stop with \textit{Gauge Mediated Supersymmetry Breaking} (GMSB) mechanism decaying to a LSP gravitino and a top quark ($\tilde{t}\rightarrow t \psi_{\mu}$). The analytic approach pointed out the feasibility of the second signature, whilst the first involving pure states of neutralinos required highly fine-tuned masses. Considering only the second model, we have defined several benchmarks by considering the cross section values at the LHC, the displacement of the top quark, and relevant final states distributions. From the latter, a discussion was engaged on how to trigger these events and build a selection of the signal. Experimentalists may use both models and benchmarks to investigate the sensibility of the CMS and ATLAS detector to the displaced quark-top signatures. \medskip

\begin{appendices}
\chapter{Conventions and Notations}\label{app:conventions}
We introduce the conventions and notations used in this manuscript. \medskip

Vector indices are written using letters from the middle of the Greek alphabet $(\mu,\ \nu,\ \rho,\ \dots)$ whilst the beginning of the alphabet is devoted to describe spinor indices ($(\alpha,\ \beta,\ \dots)$ for left-handed spinors and $(\dot{\alpha},\ \dot{\beta},\ \dots)$ for right-handed spinors).\medskip

Latin alphabet is used for general superspace coordinates ($M,N,Q,P,\dots$). The Lorentz indices in flat space are defined as untilded $M=(\mu,\ \alpha,\ \dot{\alpha})$ whereas Einstein indices in curved space are taken as tilded, $\tilde{M}=(\tilde{\mu},\ \tilde{\alpha},\ \tilde{\dot{\alpha}})$.\medskip

The Minkowsky metric is given by:
\begin{gather}
\eta_{\mu\nu} = \text{diag}(1,\ -1,\  -1,\  -1)\ .\nn
\end{gather}
We use the Van der Waerden notation for spinors. Left-handed (right-handed) Weyl spinors are written as $\psi_{\alpha}$ ($\bar{\psi}^{\dot{\alpha}}$). The indices are raised and lowered using the following conventions:
\begin{gather}
\psi_{\alpha}=\epsilon_{\alpha\beta}\psi^{\beta} \quad , \quad \psi^{\alpha} = \epsilon^{\alpha\beta}\psi_{\beta} \quad , \quad \bar{\psi}_{\dot{\alpha}} = \epsilon_{\dot{\alpha}\dot{\beta}}\bar{\psi}^{\dot{\beta}}\quad , \quad \bar{\psi}^{\dot{\alpha}} = \epsilon_{\dot{\alpha}\dot{\beta}}\bar{\psi}^{\dot{\beta}}\nn
\end{gather}
with
\begin{gather}
\epsilon_{12} = \epsilon_{\dot{1}\dot{2}}=1\quad , \quad \epsilon^{12} = \epsilon^{\dot{1}\dot{2}}=-1\ .\nn
\end{gather}
Scalar products of spinors are also defined by:
\begin{gather}
\psi\cdot\lambda = \psi^{\alpha}\lambda_{\alpha} \ , \quad \bar{\psi}\cdot\bar{\lambda} = \bar{\psi}_{\dot{\alpha}}\bar{\lambda}^{\bar{\alpha}}\ .\nn
\end{gather}  
We define the Dirac matrices as:
\begin{gather}
\gamma^{\mu} = \begin{pmatrix}
0 & \sigma^{\mu} \\
\bar{\sigma}^{\mu} & 0 
\end{pmatrix}\nn
\end{gather}
with
\begin{gather}
\sigma^{\mu}{}_{\alpha\dot{\alpha}} = (1, \sigma^i) \quad , \quad \bar{\sigma}^{\mu}{}^{\dot{\alpha}\alpha}=(1,-\sigma^i) \ .\nn
\end{gather}
The $\sigma^i$ matrices  ($i=1,2,3$) are the usual Pauli matrices:
\begin{gather}
\sigma^1 = \begin{pmatrix}
0 & 1 \\
1 & 0
\end{pmatrix} \quad , \quad 
\sigma^2 = \begin{pmatrix}
0 & -i \\
i & 0
\end{pmatrix} \quad , \quad 
\sigma^3 = \begin{pmatrix}
1 & 0 \\
0 & -1
\end{pmatrix} \ ,\nn
\end{gather}
The Lorentz generator in the spin representation are 
\begin{gather}
(\sigma^{\mu\nu})_{\alpha}{}^{\beta} = \frac14 \left( \sigma^{\mu}{}_{\alpha\dot{\alpha}}\bar{\sigma}^{\nu}{}^{\dot{\alpha}\beta} - \sigma^{\nu}{}_{\alpha\dot{\alpha}}\bar{\sigma}^{\mu}{}^{\dot{\alpha}\beta}  \right) \ , \nn\\
(\bar{\sigma}^{\mu\nu})^{\dot{\alpha}}{}_{\dot{\beta}} = \frac14 \left( \bar{\sigma}^{\mu}{}^{\dot{\alpha}\alpha}\sigma^{\nu}{}_{\alpha\dot{\beta}} - \bar{\sigma}^{\nu}{}^{\dot{\alpha}\alpha}\sigma^{\mu}{}_{\alpha\dot{\beta}}  \right) \ . \nn
\end{gather}
Pauli matrices $\sigma^{\mu}_{\alpha\dot{\alpha}}$ and  $\bar{\sigma}^{\mu\dot{\alpha}\alpha}$ can be used to convert vector indices into spinor indices (and \textit{vice versa}):
\begin{gather}
v_{\alpha\dot{\alpha}}=\sigma^{\mu}_{\alpha\dot{\alpha}}v_{\mu}\ , \quad v_{\mu} = \frac12 \bar{\sigma}^{\mu\dot{\alpha}\alpha}v_{\alpha\dot{\alpha}} \ .\nn
\end{gather}
We denote chiral superfields as $\Phi^i$ and anti-chiral superfields as $\Phi^\dag_{i^\ast}$. The Kähler potential allows to deduce several geometrical quantities associated to the Kähler manifold:
\begin{itemize}
\item the Kähler metric
\begin{gather}
K^{i^\ast}{}_i = \frac{\partial^2 K}{\partial\phi^i\partial\phi^\dag_{i^\ast}}\ , \quad (K^{-1})^{i}{}_{i^\ast}\equiv K^i{}_{i^\ast}\ , \nn
\end{gather}
\item the Christoffel symbols
\begin{gather}
\Gamma_{i}{}^{k}{}_{j} = K^k{}_{i^\ast}\frac{\partial^3 K }{\partial\phi^i\partial\phi^j\partial\phi^\dag_{j^\ast}} \ , \quad \Gamma^{i^\ast}{}_{k^\ast}{}^{j^\ast} = K^i{}_{k^\ast}\frac{\partial^3 K }{\partial\phi^\dag_{i^\ast}\partial^\dag_{j^\ast}\partial\phi^i}\ ,\nn 
\end{gather}
\item the curvature tensor
\begin{gather}
R_i{}^{i^\ast}{}_{j}{}^{j^\ast} = \frac{\partial^4 K}{\partial\phi^i\partial\phi^j\partial\phi^\dag_{i^\ast}\partial\phi^\dag_{j^\ast}} - K^{k^\ast}{}_{k}\Gamma_i{}^k{}_j\Gamma^{i^\ast}{}_{k^\ast}{}^{j^\ast} \ .\nn 
\end{gather}
\end{itemize}
Note that we use the notation $X_i=\partial_iX=\frac{\partial X}{\partial \phi^i}$.

\chapter{Scalar Higgs sector of the N2MSSM}\label{app:higgssector}
\begingroup
\allowdisplaybreaks

In this appendix, we give the the complete form of the scalar and pseudoscalar mass matrix in the N2MSSM (defined in the Subsection \ref{sec:N2MSSMDes}). We assume the superpotential to be
\begin{eqnarray}
W_{N2MSSM} &=& \lambda_i \hat{S}^i \hat{H}_U\cdot \hat{H}_D  + \frac13\kappa_{ijk}\hat{S}^i\hat{S}^j\hat{S}^k + \xi_{F,i}\hat{S}^i \nn\\
&& + y_U\hat{Q}\cdot \hat{H}_U \hat{U} - y_D\hat{Q}\cdot \hat{H}_D  \hat{D} - y_E\hat{L}\cdot \hat{H}_D  \hat{E} \ . \nn
\end{eqnarray}
The tree level Higgs mass matrices are obtained by expanding the scalar potential around the real neutral vevs $v_U$, $v_D$, $v_1$ and $v_2$ as in \autoref{eq:re&im}.  The complete mass matrix in the basis $\{h_U^0,h_D^0,\text{Re}(S^1),\text{Re}(S^2),A_U^0,A_D^0,\text{Im}(S^1),\text{Im}(S^2),H_U^+,(H_D^-)^\dagger\}$ can then be written in the following block-diagonal structure:
\begin{eqnarray}
\mathbb {\cal M}^2_{H} &=& 
\begin{pmatrix} 
\mathbb {\cal M}_{S}^2 & 0 & 0 \\
0 & \mathbb {\cal M}_{P}^2 & 0 \\
0 & 0 & \mathbb {\cal M}^2_{h^\pm} \\
\end{pmatrix}\ .\nn
\end{eqnarray}

The elements of the $4 \times 4$ CP-even mass matrix ${\cal M}_S^2$ in the basis $(h_D^0, h_{U}^0, \text{Re}(S^1), \text{Re}(S^2))$ after the elimination of $m_{H_U}^2$, $m_{H_D}^2$, $m_{S^1}^2$ and $m_{S^2}^2$ read:

\bea
{\cal M}_{S,11}^2 &=& m_{H_U}^2 + (\l_1^2 + \l_2^2)v_D^2 + (\l_1 v_1 + \l_2 v_2)^2 + \frac{g^2}{2}(3v_U^2 - v_D^2) \label{eq:ms211_n2mssm} \nn\\
&=& g^2 v_U^2 + \Big( \left(\l_1 \k_1 + \l_2 \k_{12}\right) v_1^2 + 2 \left(\l_1 \k_{12} + \l_2 \k_{21}\right)v_1 v_2 + \left(\l_1 \k_{21} + \l_2 \k_2\right) v_2^2 \nn \\
&& +\ \l_1A_{\l1} v_1 + \l_2 A_{\l2} v_2 + \l_1 \xi_{F1} + \l_2 \xi_{F2} \Big)/\tan\b \nn \\
{\cal M}_{S,22}^2 &=& m_{H_D}^2 + (\l_1^2 + \l_2^2)v_U^2 + (\l_1 v_1 + \l_2 v_2)^2 + \frac{g^2}{2}(3v_D^2 - v_U^2) \nn \\
& = & g^2 v_D^2 + \Big( \left(\l_1 \k_1 + \l_2 \k_{12}\right) v_1^2 + 2 \left(\l_1 \k_{12} + \l_2 \k_{21}\right)v_1 v_2 + \left(\l_1 \k_{21} + \l_2 \k_2\right) v_2^2  \nn \\
&& +\ \l_1A_{\l1} v_1 + \l_2 A_{\l2} v_2 + \l_1 \xi_{F1} + \l_2 \xi_{F2} \Big) \tan\b \nn \\
{\cal M}_{S,33}^2 &=& m_{S^1}^2 + \l_1^2 \left( v_U^2 + v_D^2 \right) - 2 \left(\l_1 \k_1 + \l_2 \k_{12}\right) v_U v_D + 6 \left(\k_1^2 + \k_{12}^2\right) v_1^2 \nn \\
&& +\ 12 \left(\k_1 \k_{12} + \k_{12} \k_{21} \right) v_1 v_2 + 2 \left(\k_1 \k_{21} + \k_2 \k_{12} + 2 \k_{12}^2 + 2 \k_{21}^2\right) v_2^2 \nn \\
&& +\ 2 \k_1 A_{\k1} v_1 + 2 \k_{12} A_{k12} v_2 + 2 \k_1 \xi_{F1} + 2 \k_{12} \xi_{F2} \nn \\
&=& 4 \left(\k_1^2 + \k_{12}^2\right) v_1^2 + 6 \left(\k_1 \k_{12} + \k_{12} \k_{21} \right) v_1 v_2 + \k_1 A_{\k1} v_1 \nn \\
&& +\ \Big( 2 \left(\l_1 \k_{12} + \l_2 \k_{21}\right) v_U v_D v_2 - \l_1 \l_2 \left(v_U^2 + v_D^2\right) v_2 + \l_1 A_{\l1} v_U v_D \nn \\
&& -\ 2 \left(\k_{12} \k_{21} + \k_2 \k_{21}\right) v_2^3 - \k_{21} A_{\k21} v_2^2 - 2 \left(\k_{12} \xi_{F1} + \k_{21} \xi_{F2}\right) v_2 - \xi_{S1} \Big)/v_1 \nn \\
{\cal M}_{S,44}^2 &=& m_{S^2}^2 + \l_2^2 \left( v_U^2 + v_D^2 \right) - 2 \left(\l_1 \k_{21} + \l_2 \k_2\right) v_U v_D + 6 \left(\k_{21}^2 + \k_2^2\right) v_2^2 \nn \\
&& +\ 12 \left(\k_2 \k_{21} + \k_{12} \k_{21} \right) v_1 v_2 + 2 \left(\k_1 \k_{21} + \k_2 \k_{12} + 2 \k_{12}^2 + 2 \k_{21}^2\right) v_1^2 \nn \\
&& +\ 2 \k_{21} A_{\k21} v_1 + 2 \k_2 A_{k2} v_2 + 2 \k_{21} \xi_{F1} + 2 \k_2 \xi_{F2} \nn \\
&=& 4 \left(\k_2^2 + \k_{21}^2\right) v_2^2 + 6 \left(\k_2 \k_{21} + \k_{12} \k_{21} \right) v_1 v_2 + \k_2 A_{\k2} v_2 \nn \\
&& +\ \Big( 2 \left(\l_1 \k_{12} + \l_2 \k_{21}\right) v_U v_D v_1 - \l_1 \l_2 \left(v_U^2 + v_D^2\right) v_1 + \l_2 A_{\l2} v_U v_D \nn \\
&& -\ 2 \left(\k_{12} \k_{21} + \k_1 \k_{12}\right) v_1^3 - \k_{12} A_{\k12} v_1^2 - 2 \left(\k_{12} \xi_{F1} + \k_{21} \xi_{F2}\right) v_1 - \xi_{S2} \Big)/v_2 \nn \\
{\cal M}_{S,12}^2 &=& \left(2 \l_1^2 + 2 \l_2^2 - g^2\right) v_U v_D - \left(\l_1 \k_1 + \l_2 \k_{12}\right) v_1^2 - 2 \left(\l_1 \k_{12} + \l_2 \k_{21}\right) v_1 v_2  \nn \\
&& -\ \left(\l_1 \k_{21} + \l_2 \k_2\right) v_2^2 - \l_1A_{\l1} v_1 - \l_2 A_{\l2} v_2 - \l_1 \xi_{F1} - \l_2 \xi_{F2} \nn\\ 
{\cal M}_{S,13}^2 &=& 2 \l_1^2 v_1 v_U + 2 \l_1 \l_2 v_2 v_U - 2 \left(\l_1 \k_1 + \l_2 \k_{12}\right) v_1 v_D - 2 \left(\l_1 \k_{12} + \l_2 \k_{21}\right) v_2 v_D - \l_1 A_{\l1} v_D \nn \\
{\cal M}_{S,14}^2 &=& 2 \l_2^2 v_2 v_U + 2 \l_1 \l_2 v_1 v_U - 2 \left(\l_2 \k_2 + \l_1 \k_{21}\right) v_2 v_D - 2 \left(\l_2 \k_{21} + \l_1 \k_{12}\right) v_1 v_D - \l_2 A_{\l2} v_D \nn \\
{\cal M}_{S,23}^2 &=& 2 \l_1^2 v_1 v_D + 2 \l_1 \l_2 v_2 v_D - 2 \left(\l_1 \k_1 + \l_2 \k_{12}\right) v_1 v_U - 2 \left(\l_1 \k_{12} + \l_2 \k_{21}\right) v_2 v_U - \l_1 A_{\l1} v_U \nn \\
{\cal M}_{S,24}^2 &=& 2 \l_2^2 v_2 v_D + 2 \l_1 \l_2 v_1 v_D - 2 \left(\l_2 \k_2 + \l_1 \k_{21}\right) v_2 v_U - 2 \left(\l_2 \k_{21} + \l_1 \k_{12}\right) v_1 v_U - \l_2 A_{\l2} v_U \nn \\
{\cal M}_{S,34}^2 &=& 6 \left(\k_1 \k_{12} + \k_{12} \k_{21}\right) v_1^2 + 4 \left(\k_1 \k_{21} + \k_2 \k_{12} + 2 \k_{12}^2 + 2 \k_{21}^2\right) v_1 v_2 \nn \\
&& +\ 6 \left(\k_2 \k_{21} + \k_{12} \k_{21}\right) v_2^2 + \l_1 \l_2 \left(v_U^2 + v_D^2\right) - 2 \left(\l_1 \k_{12} + \l_2 \k_{21}\right) v_U v_D \nn \\
&& +\ 2 \k_{12} A_{k12} v_1 + 2 \k_{21} A_{k21} v_2 + 2 \k_{12} \xi_{F1} + 2 \k_{21} \xi_{F2}\nn
\eea

where we have $\tan\beta=v_U /v_D$ and $\kappa_{ij} = \kappa_{iij}=\kappa_{iji}=\kappa_{jii}$.\nl

\noi The elements of the $4 \times 4$ CP-odd mass matrix ${\cal M}_{P}^2$ in the basis $(A_{D}^0, A_{U}^0, \text{Im}(S^{1}), \text{Im}(S^{2})$ after the elimination of $m_{H_U}^2$, $m_{H_D}^2$, $m_{S^1}^2$ and $m_{S^2}^2$ read:

\bea
{\cal M}_{P,11}^2 &=& m_{H_U}^2 + (\l_1^2 + \l_2^2)v_D^2 + (\l_1 v_1 + \l_2 v_2)^2 + \frac{g^2}{2}(v_U^2 - v_D^2) \nn \\
&=& \Big( \left(\l_1 \k_1 + \l_2 \k_{12}\right) v_1^2 + 2 \left(\l_1 \k_{12} + \l_2 \k_{21}\right)v_1 v_2 + \left(\l_1 \k_{21} + \l_2 \k_2\right) v_2^2 \nn \\
&& +\ \l_1A_{\l1} v_1 + \l_2 A_{\l2} v_2 + \l_1 \xi_{F1} + \l_2 \xi_{F2} \Big)/\tan\b \nn \\
{\cal M}_{P,22}^2 &=& m_{H_D}^2 + (\l_1^2 + \l_2^2)v_U^2 + (\l_1 v_1 + \l_2 v_2)^2 + \frac{g^2}{2}(v_D^2 - v_U^2) \nn \\
&=& \Big( \left(\l_1 \k_1 + \l_2 \k_{12}\right) v_1^2 + 2 \left(\l_1 \k_{12} + \l_2 \k_{21}\right)v_1 v_2 + \left(\l_1 \k_{21} + \l_2 \k_2\right) v_2^2  \nn \\
&& +\ \l_1A_{\l1} v_1 + \l_2 A_{\l2} v_2 + \l_1 \xi_{F1} + \l_2 \xi_{F2} \Big) \tan\b \nn \\
{\cal M}_{P,33}^2 &=& m_{S^1}^2 + \l_1^2 \left( v_U^2 + v_D^2 \right) + 2 \left(\l_1 \k_1 + \l_2 \k_{12}\right) v_U v_D + 2 \left(\k_1^2 + \k_{12}^2\right) v_1^2 \nn \\
&& +\ 4 \left(\k_1 \k_{12} + \k_{12} \k_{21} \right) v_1 v_2 - 2 \left(\k_1 \k_{21} + \k_2 \k_{12} - 2 \k_{12}^2 - 2 \k_{21}^2\right) v_2^2 \nn \\
&& -\ 2 \k_1 A_{\k1} v_1 - 2 \k_{12} A_{k12} v_2 - 2 \k_1 \xi_{F1} - 2 \k_{12} \xi_{F2} \nn \\
&=& - 4 \left(\k_1 \k_{21} + \k_2 \k_{12}\right) v_2^2 - 2 \left(\k_1 \k_{12} + \k_{12} \k_{21}\right) v_1 v_2 - 3 \k_1 A_{\k1} v_1 \nn \\
&& -\ 4 \k_{12} A_{\k12} v_2 - 4 \k_1 \xi_{F1} - 4 \k_{12} \xi_{F2} + 4 \left(\l_1 \k_1 + \l_2 \k_{12}\right) v_U v_D \nn \\
&& + \Big( \l_1 A_{\l1} v_U v_D - \l_1 \l_2 \left(v_U^2 + v_D^2\right) v_2 + 2 \left(\l_1 \k_{12} + \l_2 \k_{21}\right) v_U v_D v_2 - \xi_{S1} \nn \\
&& -\ 2 \left(\k_{12} \k_{21} + \k_2 \k_{21}\right) v_2^3 - 2 \left(\k_{12} \xi_{F1} + \k_{21} \xi_{F2}\right) v_2 - \k_{21} A_{\k21} v_2^2 \Big)/v_1 \nn \\
{\cal M}_{P,44}^2 &=& m_{S^2}^2 + \l_2^2 \left( v_U^2 + v_D^2 \right) + 2 \left(\l_1 \k_{21} + \l_2 \k_2\right) v_U v_D + 2 \left(\k_{21}^2 + \k_2^2\right) v_2^2 \nn \\
&& +\ 4 \left(\k_2 \k_{21} + \k_{12} \k_{21} \right) v_1 v_2 - 2 \left(\k_1 \k_{21} + \k_2 \k_{12} - 2 \k_{12}^2 - 2 \k_{21}^2\right) v_2^2 \nn \\
&& -\ 2 \k_{21} A_{k21} v_1 - 2 \k_2 A_{\k2} v_2 - 2 \k_{21} \xi_{F1} - 2 \k_2 \xi_{F2} \nn \\
&=& - 4 \left(\k_1 \k_{21} + \k_2 \k_{12}\right) v_1^2 - 2 \left(\k_2 \k_{21} + \k_{12} \k_{21}\right) v_1 v_2 - 3 \k_2 A_{\k2} v_2 \nn \\
&& -\ 4 \k_{21} A_{\k21} v_1 - 4 \k_{21} \xi_{F1} - 4 \k_2 \xi_{F2} + 4 \left(\l_2 \k_2 + \l_1 \k_{21}\right) v_U v_D \nn \\
&& + \Big( \l_2 A_{\l2} v_U v_D - \l_1 \l_2 \left(v_U^2 + v_D^2\right) v_1 + 2 \left(\l_1 \k_{12} + \l_2 \k_{21}\right) v_U v_D v_1 - \xi_{S2} \nn \\
&& -\ 2 \left(\k_{12} \k_{21} + \k_1 \k_{12}\right) v_1^3 - 2 \left(\k_{12} \xi_{F1} + \k_{21} \xi_{F2}\right) v_1 - \k_{12} A_{\k12} v_1^2 \Big)/v_2 \nn \\
{\cal M}_{P,12}^2 &=& \left(\l_1 \k_1 + \l_2 \k_{12}\right) v_1^2 + 2 \left(\l_1 \k_{12} + \l_2 \k_{21}\right)v_1 v_2 + \left(\l_1 \k_{21} + \l_2 \k_2\right) v_2^2 \nn \\
&& +\ \l_1A_{\l1} v_1 + \l_2 A_{\l2} v_2 + \l_1 \xi_{F1} + \l_2 \xi_{F2} \nn \\ 
{\cal M}_{P,13}^2 &=& - 2 \left(\l_1 \k_1 + \l_2 \k_{12}\right) v_1 v_D - 2 \left(\l_1 \k_{12} + \l_2 \k_{21}\right) v_2 v_D + \l_1 A_{\l1} v_D \nn \\
{\cal M}_{P,14}^2 &=& - 2 \left(\l_2 \k_2 + \l_1 \k_{21}\right) v_2 v_D - 2 \left(\l_2 \k_{21} + \l_1 \k_{12}\right) v_1 v_D + \l_2 A_{\l2} v_D \nn \\
{\cal M}_{P,23}^2 &=& - 2 \left(\l_1 \k_1 + \l_2 \k_{12}\right) v_1 v_U - 2 \left(\l_1 \k_{12} + \l_2 \k_{21}\right) v_2 v_U + \l_1 A_{\l1} v_U \nn \\
{\cal M}_{P,24}^2 &=& - 2 \left(\l_2 \k_2 + \l_1 \k_{21}\right) v_2 v_U - 2 \left(\l_2 \k_{21} + \l_1 \k_{12}\right) v_1 v_U + \l_2 A_{\l2} v_U \nn \\
{\cal M}_{P,34}^2 &=& \l_1 \l_2 \left(v_U^2 + v_D^2\right) + 2 \left(\l_1 \k_{12} + \l_2 \k_{21}\right) v_U v_D + 2 \left(\k_1 \k_{12} + \k_{12} \k_{21}\right) v_1^2 \nn \\
&& +\ 4 \left(\k_1 \k_{21} + \k_2 \k_{12}\right) v_1 v_2 + 2 \left(\k_2 \k_{21} + \k_{12} \k_{21}\right) v_2^2 \nn \\
&& -\ 2 \k_{12} A_{k12} v_1 - 2 \k_{21} A_{k21} v_2 - 2 \k_{12} \xi_{F1} - 2 \k_{21} \xi_{F2}\nn
\eea

${\cal M}_{P}^2$ contains a massless Goldstone mode ${G}$. We can rotate this mass matrix into the basis ($G, A, \text{Im}(S^{1}), \text{Im}(S^{2})$), where $A = \cos\b\, A_{U}^0+ \sin\b\, A_{D}^0$:
\beq
\left(\ba{c} A_{U}^0 \\  A_{D}^0 \\ \text{Im}(S^{1}) \\ \text{Im}(S^{2}) \ea\right) = 
\left(\ba{cccc}
\sin\b & \cos\b & 0 & 0 \\ 
-\cos\b & \sin\b & 0 & 0 \\
0 & 0 & 1 & 0 \\
0 & 0 & 0 & 1 \\
\ea\right)
\left(\ba{c} {G} \\ {A} \\  \text{Im}(S^{1}) \\ \text{Im}(S^{2}) \ea\right)\nn
\eeq

\noi By not considering the Goldstone boson $G$, the remaining $3 \times 3$ mass matrix ${\cal M'}_{P}^2$ in the basis ($A, \text{Im}(S^{1}), \text{Im}(S^{2})$) reads

\bea
{\cal M'}_{P,11}^2 & = & 2 \Big( \left(\l_1 \k_1 + \l_2 \k_{12}\right) v_1^2 + 2 \left(\l_1 \k_{12} + \l_2 \k_{21}\right)v_1 v_2 + \left(\l_1 \k_{21} + \l_2 \k_2\right) v_2^2  \nn \\
&& +\ \l_1A_{\l1} v_1 + \l_2 A_{\l2} v_2 + \l_1 \xi_{F1} + \l_2 \xi_{F2} \Big) / \sin2\b \nn \\
{\cal M'}_{P,22}^2 & = & - 4 \left(\k_1 \k_{21} + \k_2 \k_{12}\right) v_2^2 - 2 \left(\k_1 \k_{12} + \k_{12} \k_{21}\right) v_1 v_2 - 3 \k_1 A_{\k1} v_1 \nn \\
&& -\ 4 \k_{12} A_{\k12} v_2 - 4 \k_1 \xi_{F1} - 4 \k_{12} \xi_{F2} + 4 \left(\l_1 \k_1 + \l_2 \k_{12}\right) v_U v_D \nn \\
&& + \Big( \l_1 A_{\l1} v_U v_D - \l_1 \l_2 \left(v_U^2 + v_D^2\right) v_2 + 2 \left(\l_1 \k_{12} + \l_2 \k_{21}\right) v_U v_D v_2 - \xi_{S1} \nn \\
&& -\ 2 \left(\k_{12} \k_{21} + \k_2 \k_{21}\right) v_2^3 - 2 \left(\k_{12} \xi_{F1} + \k_{21} \xi_{F2}\right) v_2 - \k_{21} A_{\k21} v_2^2 \Big)/v_1 \nn \\
{\cal M'}_{P,33}^2 & = &  - 4 \left(\k_1 \k_{21} + \k_2 \k_{12}\right) v_1^2 - 2 \left(\k_2 \k_{21} + \k_{12} \k_{21}\right) v_1 v_2 - 3 \k_2 A_{\k2} v_2 \nn \\
&& -\ 4 \k_{21} A_{\k21} v_1 - 4 \k_{21} \xi_{F1} - 4 \k_2 \xi_{F2} + 4 \left(\l_2 \k_2 + \l_1 \k_{21}\right) v_U v_D \nn \\
&& + \Big( \l_2 A_{\l2} v_U v_D - \l_1 \l_2 \left(v_U^2 + v_D^2\right) v_1 + 2 \left(\l_1 \k_{12} + \l_2 \k_{21}\right) v_U v_D v_1 - \xi_{S2} \nn \\
&& -\ 2 \left(\k_{12} \k_{21} + \k_1 \k_{12}\right) v_1^3 - 2 \left(\k_{12} \xi_{F1} + \k_{21} \xi_{F2}\right) v_1 - \k_{12} A_{\k12} v_1^2 \Big)/v_2 \nn \\
{\cal M'}_{P,12}^2 & = & \Big( \l_1 A_{\l1} - 2 (\l_1 \k_1 + \l_2 \k_{12}) v_1 - 2 (\l_1 \k_{12} + \l_2 \k_{21}) v_2 \Big) v \nn \\
{\cal M'}_{P,13}^2 & = & \Big( \l_2 A_{\l2} - 2 (\l_1 \k_{12} + \l_2 \k_{21}) v_1 - 2 (\l_1 \k_{21} + \l_2 \k_2) v_2 \Big) v \nn \\
{\cal M'}_{P,23}^2 & = & \l_1 \l_2 \left(v_U^2 + v_D^2\right) +2 \left(\l_1 \k_{12} + \l_2 \k_{21}\right) v_U v_D + 2 \left(\k_1 \k_{12} + \k_{12} \k_{21}\right) v_1^2 \nn \\
&& +\ 4 \left(\k_1 \k_{21} + \k_2 \k_{12}\right) v_1 v_2 + 2 \left(\k_2 \k_{21} + \k_{12} \k_{21}\right) v_2^2 - 2 \k_{12} A_{\k12} v_1 \nn \\
&& -\ 2 \k_{21} A_{\k21} v_2 - 2 \k_{12} \xi_{F1} - 2 \k_{21} \xi_{F2}\nn
\eea

\noi Defining the heavy doublet mass $m_A^2 \equiv {\cal M'}_{P,11}^2$, we can write:
\bea
{\cal M}_{S,11}^2 &=& M_Z^2 \sin^2\b + m_A^2 \cos^2\b \ , \nn \\
{\cal M}_{S,22}^2 &=& M_Z^2 \cos^2\b + m_A^2 \sin^2\b \ , \nn \\
{\cal M}_{S,12}^2 &=& \left(2 \l_1^2 v^2 + 2 \l_2^2 v^2 - M_Z^2 -m_A^2\right)\cos\b\sin\b \ .\nn
\eea
The charged Higgs boson mass matrix ${\cal M}^2_{h^{\pm}}$ has already been presented in the Subsection \ref{sec:N2MSSMDes}. 

\endgroup

\chapter{Potential for Non-Soni-Weldon solutions: Giudice-Masiero-like \& Renormalisable theory case}\label{app:GMsolutions}
In the Subsection \ref{subsec:potential_calculation}, the full calculation of the S2MSSM scalar potential has been done. We present in this appendix an equivalent calculation starting from a more general model inspired by Giudice \& Masiero solutions. \nn The results are then applied to a renormalisable theory and a S2MSSM-like model. \medskip

Consider a Non-Soni-Weldon solution with  fields content:
$$Z^M=(z^i,S^1,S^2,\Phi^a) \ .$$
The fields in the hidden sector $z^i$  are taken dimensionless. The fields in the observable 
sector $\Phi^a$ will be specified later on, and we consider only two fields in the $S-$sector.
\section{The Giudice-Masiero case}
We display the calculation of the scalar potential in a Giudice-Masiero like case \cite{GIUDICEMASIERO}. We assume that the K\"ahler potential and the superpotential take the form
\beqa
W(z,S,\Phi)&=& m_p\big[\hat W_0(z) + S^p \mu_p^\ast \hat W_1(z)\big] + S^p \tilde W_p(z) + \tilde W_k(z) g^k({\cal U}, \Phi)\nn\\
&\equiv& m_p \hat W(z,S) + \tilde W(z,S,\Phi) \ , \nn\\ 
K(z,z^\dag,S,S^\dag,\Phi,\Phi^\dag) &=& m_p^2  \hat K(z,z^\dag) + S^\dag_p S^p + \Phi^\dag_{a^\ast} \Lambda^{a^\ast}{}_a(z)   \Phi^a   + \big[Z_\ell(z,z^\dag) h^\ell(\Phi) + \text{h.c.}\big] \nn\\
&\equiv &m_p^2 \hat K(z,z^\dag) + \tilde K(z,z^\dag,S,S^\dag,\Phi,\Phi^\dag)\ ,  \nn
\eeqa
where the choice for the functions  $g^k({\cal U}, \Phi)$ parameterize the supersymmetric model (MSSM or its extensions). The inverse of the K\"ahler metric is obtained perturbatively:
\beqa
(K^{-1})^a{}_{a^\ast}&=&
(\Lambda^{-1})^a{}_{a^\ast} + \frac1 {m_p^2}  (\Lambda^{-1})^a{}_{b^\ast} \Big(\partial_i \Lambda^{b^\ast}{}_b \Phi^b + {\partial_i Z_\ell^\ast \partial^{b^\ast}  h^{\ast \ell}}\Big)
{(\hat K^{-1})^i{}_{i^\ast}}\nn\\
&& \hskip 0.5truecm \times \Big(\Phi^\dag_{c^\ast} \partial^{i^\ast} \Lambda^{c^\ast}{}_c  + {\partial^{i^\ast} Z_\ell \partial_c h^\ell}\Big)(\Lambda^{-1})^c{}_{a^\ast} +o(1/m_p^2)\ , \nn\\
(K^{-1})^p{}_{q}&=& \delta^p{}_q \ ,  \nn\\
(K^{-1})^i{}_{i^\ast} &=& {(\hat K^{-1})^i{}_{i^\ast}}\nn\\
&&+\frac 1{m_p^2} {(\hat K^{-1})^i{}_{j^\ast}} \Bigg[ \Big(\Phi^\dag_{a^\ast} \partial^{j^\ast} \Lambda^{a^\ast}{}_a
+ {\partial^{j^\ast} Z_\ell \partial_a h^\ell}\Big) (\Lambda^{-1})^a{}_{b^\ast} \Big(\partial_j \Lambda^{b^\ast}{}_b \Phi^b + {\partial_j Z^\ast_{\ell'} \partial^{b^\ast} h^{\ast \ell'}}\Big)
 \nn\\
&&-\partial_j\partial^{j^\ast} \Lambda^{a^\ast}{}_a \Phi^\dag_{a^\ast} \Phi^a -{( \partial_j \partial^{j^\ast} Z_\ell h^\ell + \text{h.c.})} \Bigg] {(\hat K^{-1})^j{}_{i^\ast}}+o(1/m_p^2) \ , \nn\\
(K^{-1})^a{}_{i^\ast}&=& - \frac 1{m_p} (\Lambda^{-1})^a{}_{b^\ast}\Big[ \partial_i \Lambda^{b^\ast}{}_c \Phi^c + {\partial_i Z_\ell^\ast \partial^{b^\ast} h^{\ast \ell} }\Big]
{(\hat K^{-1})^i{}_{i^\ast}}+o(1/m_p)\ , \nn\\
(K^{-1})^p{}_{i}&=&(K^{-1})^p{}_{a^\ast}= (K^{-1})^i{}_{p}=(K^{-1})^{a^\ast}{}_{p} =0 \ . \nn
\eeqa
Supergravity is broken in the hidden sector when some of the $z-$fields develop a vacuum expectation value. The $S-$field can also develop a v.e.v. We proceed now to the following substitution: 
\beqa
z &\to& \big<z\big> \ , \nn\\
S &\to&S + \big<S\big> \ .\nn
\eeqa
We then obtain
\beqa
\hat W &\to& \big<\hat W\big> + S^p\mu_p^\ast\big<\hat W_1\big> \equiv M^2 + S^p a_p^\ast M\ ,  \nn\\
\tilde W&\to&\big<S^p  \tilde W_p \big> + S^p   \big<\tilde W_p\big> + \big<W_{k}\big> g^{k}({\cal U} + \big<{\cal U}\big>,\Phi)\ .\nn
\eeqa
To simplify the notations, $g^{k}({\cal U} + \big<{\cal U}\big>,\Phi)$ is noted throughout $g^{k}({\cal U},\Phi)$. Furthermore the gravitino mass is given by
\beqa
m_\frac 32 = e^{\frac 12 \big<z^\dag z \big>}\frac {M^2} {m_p}\ . \nn
\eeqa
For further use, we define the $\rho$-functions:
\beqa
\hat \rho_i&=&  \partial_i \Big(\hat K + \ln \frac{\hat W}{m_p^2}\Big)\ ,  \nn\\
\hat \rho_{0i} &=& \partial_i \Big(\hat K + \ln \frac{\hat W_0}{m_p^2}\Big) \ , \nn\\
\hat \rho_{1i} &=& \partial_i \Big(\hat K + \ln \frac{  \hat W_1}{m_p}\Big)\ , \nn\\
\tilde \rho_i&=& \partial_i \Big(\hat K + \ln \frac{\tilde W}{m_p^3}\Big) \ , \nn\\
\tilde \rho_{pi} &=& \partial_i \Big(\hat K + \ln \frac{\tilde W_p}{m_p^3}\Big)\ ,  \nn\\
\tilde \rho_{ki} &=& \partial_i \Big(\hat K + \ln \frac{\tilde W_k}{m_p^3}\Big)\ . \nn
\eeqa
The covariant derivatives take the form:
\beqa
{\cal D}_i W &=&\phantom{+} \Big(M^2\big<\hat \rho_i\big> + S^pa_p^\ast M \big<\hat \rho_{1i} \big> \Big) +
\frac 1 {m_p} \Big( \big<S^p\tilde W_p \tilde \rho_{pi}\big> + S^p \big<\tilde W_p \tilde \rho_{pi}\big> + \big<\tilde W_k \tilde \rho_{k}\big> g^k \Big)\nn\\
&&+ \frac 1 {m_p^2} \Big(M^2 + M a_p^\ast S^p\Big) \Big(\Phi^\dag_{a^\ast} \partial_i \Lambda^{a^\ast}{}_a \Phi^a + \big[\partial_i Z_\ell h^\ell + \partial_i Z^\ast_\ell {h^\ast}^\ell\big]\Big)      + o(1/m_p^2)\ , \nn\\
{\cal D}_a W &=& \big<\tilde W_k\big> \partial_a g^k + \frac 1 {m_p}\Big(M^2 + S^p a_p^\ast M\Big)\Big( \Phi^\dag_{a^\ast} \Lambda^{a^\ast}{}_a  + Z_\ell \partial_a h^\ell \Big) + o(1/m_p)
\ ,\nn\\
{\cal D}_p W &=& \phantom{+} m_pa_p^\ast M + \big<\tilde W_p\big> + \big<\tilde W_k\big> \partial_p g^k \nn\\
&&+
\frac{S_p^\dag + \big<S^\dag_p\big>}{m_p}\Big(M^2 + S^p a_p^\ast M + \frac 1{m_p} \big[\big<S^p\tilde W_p\big> + S^p  \big<\tilde W_p\big> + \big<\tilde W_k\big> g^k\big]\Big)\ .\nn
\eeqa
The scalar potential can now be computed:
\begin{gather}
V = e^{\hat K}\Bigg( 1 + \frac{\tilde{K}}{m_p^2}\Bigg) \Bigg[ \mathcal{D}_IW(K^{-1})^{I}{}_{J^\ast}\mathcal{D}^{J^\ast}\overline{W} - \frac{3}{m_p^2}| W |^2\Bigg]\ .\nn
\end{gather}
To regroup the various terms in the potential, we introduce the following functions:
\begin{enumerate}
\item the $A-$ and the $C-$functions:
\beqa
A_k&=&\big< \tilde \rho_{ki^\ast}\hat K^{i^\ast}{}_i \hat \rho^{\ast i}\big>\ , \nn\\
A_{1k}&=&\big< \tilde \rho_{ki^\ast} \hat K^{i^\ast}{}_i \hat \rho_1^{\ast i}\big>\ , \nn\\
C_p&=&\big< \tilde \rho_{pi^\ast} \hat K^{i^\ast}{}_i \hat \rho^{\ast i} \widehat{\widetilde {W_p}}\big>\ , \nn\\
C_{1p}&=&\big< \tilde \rho_{pi^\ast} \hat K^{i^\ast}{}_i \hat \rho_1^{\ast i} \widehat{\widetilde {W_p}}\big>\ .\nn
\eeqa
\item the $\mu-$ and $B-$functions:
\beqa
\mu_\ell&=& m_\frac 32\Big(\big<Z_\ell\big> -\big<\rho_{i^\ast}\partial^{i^\ast} Z_\ell\big>\Big)\ ,  \nn  \\
\mu_{1\ell}&=& m_\frac 32\Big(\big<Z_\ell\big>-\big<\rho_{1 i^\ast}\partial^{i^\ast} Z_\ell\big>\Big)\ ,\nn
\eeqa
\beqa
B_\ell&=&m_\frac 32\Big(\big<\hat \rho_{i^\ast} \partial^{i^\ast} Z_\ell\big>  + \big<\hat \rho^{\ast i} \partial_i Z_\ell\big>-\big<\hat \rho_{i^\ast} \hat \rho^{\ast i} \partial_i \partial^{i^\ast} Z_\ell\big>\Big)\ , \nn\\
B_{1\ell}&=&m_\frac 32\Big(\big<\hat \rho_{1i^\ast} \partial^{i^\ast} Z_\ell\big>  + \big<\hat \rho^{\ast i} \partial_i Z_\ell\big>-\big<\hat \rho_{1i^\ast} \hat \rho^{\ast i} \partial_i \partial^{i^\ast} Z_\ell\big>\Big)
 \ ,  \nn \\
B^\ast_{1\ell}&=&m_\frac 32\Big(\big<\hat \rho_{i^\ast} \partial^{i^\ast} Z_\ell\big>  + \big<\hat \rho_1^{\ast i} \partial_i Z_\ell\big>-\big<\hat \rho_{i^\ast} \hat \rho_1^{\ast i} \partial_i \partial^{i^\ast} Z_\ell\big>\Big)\ , \nn\\
B_{11,\ell}&=&m_\frac 32\Big(\big<\hat \rho_{1i^\ast} \partial^{i^\ast} Z_\ell\big>  + \big<\hat \rho_1^{\ast i} \partial_i Z_\ell\big>-\big<\hat \rho_{1i^\ast} \hat \rho_1^{\ast i} \partial_i \partial^{i^\ast} Z_\ell\big>\Big)\ .\nn
\eeqa
\item the supersymmetric superpotential 
\beqa
W_m = W_m' + W_{GM}  \nn
\eeqa
with :
\beqa
W_m' = \sum_k \hat g^k + \big<\widehat{\widetilde{W}}_p\big> S^p \ , \nn 
\eeqa
\beqa
W_{GM} = \mu_\ell h^\ell + \frac 1 M \mu_{1\ell} a^\ast_p S^p h^\ell\ . \nn
\eeqa
\item The $R-$ and $S-$ functions:
\beqa
S^{a^\ast}{}_a &=& \Big<\Lambda^{a^\ast}{}_a(\hat \rho_{i^\ast }\hat K^{i^\ast}{}_i\hat \rho^{\ast i} -2) \ +\hat \rho_{i^\ast}\Big(\partial^{i^\ast} \Lambda^{a^\ast}{}_b (\Lambda^{-1})^b{}_{b^\ast} \partial_j\Lambda^{b^\ast}{}_a -
\partial_j\partial^{i^\ast} \Lambda^{a^\ast}{}_a\Big)\hat \rho^{\ast j}\Big>\ , \nn\\
S_1^{a^\ast}{}_a&=& \Big<\Lambda^{a^\ast}{}_a(\hat \rho_{1i^\ast }\hat K^{i^\ast}{}_i\hat \rho^{\ast i} -2) \ +\hat \rho_{1i^\ast}\Big(\partial^{i^\ast} \Lambda^{a^\ast}{}_b (\Lambda^{-1})^b{}_{b^\ast} \partial_j\Lambda^{b^\ast}{}_a-
\partial_j\partial^{i^\ast} \Lambda^{a^\ast}{}_a\Big)\hat \rho^{\ast j}\Big>\ , \nn\\
S_{1,1}^{a^\ast}{}_a&=&\Big<\Lambda^{a^\ast}{}_a(\hat \rho_{1i^\ast }\hat K^{i^\ast}{}_i\hat \rho_1^{\ast i} -2) \ +\hat \rho_{1i^\ast}\Big(\partial^{i^\ast} \Lambda^{a^\ast}{}_b (\Lambda^{-1})^b{}_{b^\ast} \partial_j\Lambda^{b^\ast}{}_a-
\partial_j\partial^{i^\ast} \Lambda^{a^\ast}{}_a\Big)\hat \rho_1^{\ast j}\Big>\ , \nn\\
R^a{}_{b}&=&\delta^a{}_b-\Big< (\Lambda^{-1})^a{}_{b^\ast} \partial_i \Lambda^{b^\ast}{}_b\hat \rho^{\ast i}\Big> \ , \nn\\
R_1^a{}_{b}&=&\delta^a{}_b-\Big< (\Lambda^{-1})^a{}_{b^\ast} \partial_i \Lambda^{b^\ast}{}_b\hat \rho_1^{\ast i}\Big> \nn\ .
\eeqa
\end{enumerate}
We then finally obtain :
{\footnotesize
\beqa
\label{eq:pot3}
V&=&e^{\big<z^\dag z\big>}\Bigg[(M M^\dag)^2 \Big\{\big< \hat \rho_{i^\ast} \hat K^{i^\ast}{}_i \hat \rho^{\ast i}\big> -3\Big\}+ m_p^2 |M|^2 a_p^\ast a^p+ \Big(m_p M^\dag a^p \big<\tilde W_p\big>+ \text{h.c.}\Big)
\Bigg]\label{eq:Vapp}\\
&&+ \partial_p W_m' \partial^p W'{}^*_m +
\partial_a W_m \partial^{a^\ast} W_m^\ast(\Lambda^{-1})^a{}_{a^\ast} +
 |m_\frac32|^2\Phi^\dag_{a^\ast} S^{a^\ast}{}_a \Phi^a\nn\\
&&
 +|m_\frac 32|^2\Big\{\frac 1{M} a^\ast_p S_1^{a^\ast}{}_a S^p\Phi^\dag_{a^\ast}  \Phi^a + \text{h.c.}\Big\}+ \frac{|m_\frac32|^2} {|M|^2}S_{1,1}^a{}_{a^\ast} a^p a^\ast_q  S^\dag_p S^q \Phi^\dag_{a^\ast}\Phi^a\nn\\
\nn\\
 && + |m_\frac 32|^2(S^\dag_p + \big<S^\dag_p\big>)(S^p + \big<S^p\big>)\nn\\
 && \times \Bigg[\Big( \big<\hat \rho_{i^\ast} \hat K^{i^\ast}{}_i\hat \rho^{\ast i}\big> -2 \Big) + \Bigg\{\frac 1 M \big(\big<\hat \rho_{1i^\ast} \hat K^{i^\ast}{}_i\hat \rho^{\ast i}\big> -2 \big)a^\ast_r S^r + \text{h.c.}\Bigg\} + \frac 1 {|M|^2} \big(\big<\hat \rho_{1i^\ast} \hat K^{i^\ast}{}_i\hat \rho_1^{\ast i}\big> -2 \big)a^\ast_r a^t S^\dag_t S^r \Bigg]\nn\\
 &&+e^{\big<z^\dag z\big>}|M|^2\Big(\big<\hat \rho_{1i^\ast} \hat K^{i^\ast}{}_i\hat \rho_1^{\ast i}\big> -3 \Big)a_q^\ast a^p S^\dag_p S^q \nn\\
  && + \Bigg\{ e^{\big<z^\dag z\big>}M (M^\dag)^2\Bigg(\Big( \big< \hat \rho_{1i^\ast} \hat K^{i^\ast}{}_i\hat \rho^{\ast i}\big> -3\Big) a_p^\ast S^p +
\big(1 + \frac 1{M^\dag} a^q S^\dag_q\big)a^\ast_p\big(S^p + \big<S^p\big>\Big)\Bigg) +\text{h.c.}\Bigg\}\nn\\
 \nn\\
 && + |m_\frac32|^2 \Bigg[\Big(\big<\hat \rho_{i^\ast} \hat K^{i^\ast}{}_i \hat \rho^{\ast i}\big> -3 \Big) + \bigg\{\frac 1M \Big(\big<\hat \rho_{1i^\ast} \hat K^{i^\ast}{}_i \hat \rho^{\ast i}\big> -3 \Big)
a^\ast_ p S^p + \text{h.c.}\bigg\} \nn\\
&&+ \frac 1{|M|^2} \Big(\big<\hat \rho_{1i^\ast} \hat K^{i^\ast}{}_i \hat \rho_1^{\ast i}\big> -3 \Big) a_p^\ast a^q S^\dag_q S^p\Bigg]
\Big\{Z_\ell h^\ell + {\rm h.c.}  \Big\}\nn
\\
&&+ \Bigg\{m^\dag_\frac 32\Bigg[\sum_k\big(A_k-3\big)\hat g^k+ \big(C_p-3\big<\widehat{\widetilde{W}}_p\big>\big)\big( S^p+ \big<S^p\big>\big) +
\big( B_\ell h^\ell + \frac 1 M B_{1\ell} a^\ast_p S^p h^\ell\big)
\nn\\
&&
+R^a{}_b  \Phi^b  \partial_a W_m+\Big(S^p + \big<S^p\big>\Big)\partial_p W'_m \Bigg] +\text{h.c.}
 \Bigg\}\nn\\
&& + \Bigg\{\frac{m_\frac 32^\dag}{M^\dag}\Bigg[\bigg\{\sum_k(A_{1k}-3) \hat g^k + (C_{1p}-3\big<\widehat{\widetilde{W}}_p\big>) \big[S^p + \big<S^p\big>\big] 
+ \Big(B^\ast_{1\ell} h^\ell + \frac1 M B_{11\ell} a^\ast_p S^p h^\ell\Big) \bigg\}a^q S^\dag_q
\nn\\
&&+R_1{}^a{}_b a^p S^\dag_p \Phi^b  \partial_a W_m
+ a^p\Big(S^\dag_p + \big<S^\dag_p\big>\Big)\Big(\big<S^q \widehat{\widetilde{W}}_p\big> + W_m\Big)+\Big(S^p + \big<S^p\big>\Big) a^q S^\dag_q\partial_p W'_m \Bigg] +\text{h.c.}\Bigg\}\nn\\
& & +  \left[ (S^p + \langle S^p \rangle)(S^{\dagger}_p + \langle S^{\dagger}_p \rangle ) + \Phi^{\dagger}_{a^*} \langle \Lambda^{a^*}{}_a \rangle \Phi^a +\Big\{ Z_\ell h^\ell + {\rm h.c.} \Big\} \right]  \nn \\
& & \times  \left( |M|^2 a^*_q a^q e^{\langle z^{\dagger} z \rangle} + \left\{\frac{m^{\dagger}_{3/2}}{M^{\dagger}} a^q \langle \bar{\hat{\tilde{W}}}_q \rangle + \frac{|m_{3/2}|^2}{M} \left( 1 + \frac{S^{\dagger}_r a^r}{M^{\dagger}} \right) a^*_p (S^p + \langle S^p \rangle ) + \text{h.c.}  \right\} \right)  \nn 
\eeqa
}
\newline

We interpret all terms line-by-line:
\begin{itemize}
\item[Line 2:] the unbroken supersymmetry part:
{\footnotesize
$$ \partial_p W'_m \partial^p W'_m +
\partial_a W_m \partial^{a^\ast} W_m^\ast(\Lambda^{-1})^a{}_{a^\ast}\ .$$
}
\item[Line 2-3:] Mass term (soft) for $\Phi$ and their corresponding  (hard) $S \Phi \Phi^\dag$ and $S S^\dag \Phi \Phi^\dag-$terms:
{\footnotesize
$$|m_\frac32|^2\Phi^\dag_{a^\ast} S^{a^\ast}{}_a \Phi^a
 +|m_\frac 32|^2\Big\{\frac 1{M} a^\ast_p S_1^{a^\ast}{}_a S^p\Phi^\dag_{a^\ast}  \Phi^a + \text{h.c.}\Big\}+ \frac{|m_\frac32|^2} {|M|^2}S_{1,1}^a{}_{a^\ast} a^p a^\ast_q  S^\dag_p S^q \Phi^\dag_{a^\ast}\Phi^a\ .$$
}
 \item[Line 4-7:] Mass term (soft) for $S$ and their corresponding (hard) $S  S S ^\dag$ and $S S^\dag S S ^\dag-$terms:
{\footnotesize
 \beqa
 && |m_\frac 32|^2(S^\dag_p + \big<S^\dag_p\big>)(S^p + \big<S^p\big>)\nn\\
 && \times \Bigg[\Big( \big<\hat \rho_{i^\ast} \hat K^{i^\ast}{}_i\hat \rho^{\ast i}\big> -2 \Big) + \Bigg\{\frac 1 M \big(\big<\hat \rho_{1i^\ast} \hat K^{i^\ast}{}_i\hat \rho^{\ast i}\big> -2 \big)a^\ast_r S^r + \text{h.c.}\Bigg\} \nn\\
 && + \frac 1 {|M|^2} \big(\big<\hat \rho_{1i^\ast} \hat K^{i^\ast}{}_i\hat \rho_1^{\ast i}\big> -2 \big)a^\ast_r a^t S^\dag_t S^r \Bigg]\nn\\
 &&+e^{\big<z^\dag z\big>}|M|^2\Big(\big<\hat \rho_{1i^\ast} \hat K^{i^\ast}{}_i\hat \rho_1^{\ast i}\big> -3 \Big)a_q^\ast a^p S^\dag_p S^q \nn\\
  && + \Bigg\{ e^{\big<z^\dag z\big>}M (M^\dag)^2\Bigg(\Big( \big< \hat \rho_{1i^\ast} \hat K^{i^\ast}{}_i\hat \rho^{\ast i}\big> -3\Big) a_p^\ast S^p +
\big(1 + \frac 1{M^\dag} a^q S^\dag_q\big)a^\ast_p\big(S^p + \big<S^p\big>\Big)\Bigg) +\text{h.c.}\Bigg\}\ .\nn\\
 \nn
 \eeqa
}
 \item[Line 8-9:] Terms in $h^\ell$ coming from the expansion of the exponential (soft and hard):
{\footnotesize
 \beqa
 && |m_\frac32|^2 \Bigg[\Big(\big<\hat \rho_{i^\ast} \hat K^{i^\ast}{}_i \hat \rho^{\ast i}\big> -3 \Big) + \bigg\{\frac 1M \Big(\big<\hat \rho_{1i^\ast} \hat K^{i^\ast}{}_i \hat \rho^{\ast i}\big> -3 \Big)
a^\ast_ p S^p + \text{h.c.}\bigg\} \nn\\
&&+ \frac 1{|M|^2} \Big(\big<\hat \rho_{1i^\ast} \hat K^{i^\ast}{}_i \hat \rho_1^{\ast i}\big> -3 \Big) a_p^\ast a^q S^\dag_q S^p\Bigg]
\Big\{Z_\ell h^\ell + {\rm h.c.}  \Big\}\ .\nn
\eeqa
}
\item[Line 10-11:] Soft $A,B,C-$terms and $\Phi \partial W, S\partial W$ terms:
{\footnotesize
\beqa
&&+ \Bigg\{m^\dag_\frac 32\Bigg[\sum_k\big(A_k-3\big)\hat g^k+ \big(C_p-3\big<\widehat{\widetilde{W}}_p\big>\big)\big( S^p+ \big<S^p\big>\big) +
\big( B_\ell h^\ell + \frac 1 M B_{1\ell} a^\ast_p S^p h^\ell\big)
\nn\\
&&
+R^a{}_b  \Phi^b  \partial_a W_m+\Big(S^p + \big<S^p\big>\Big)\partial_p W'_m \Bigg] +\text{h.c.}
 \Bigg\}\ .\nn
 \eeqa
}
 \item[Line 12-13:] Equivalent hard terms to Line 10-11:
{\footnotesize
 \beqa
&&  \Bigg\{\frac{m_\frac 32^\dag}{M^\dag}\Bigg[\bigg\{\sum_k(A_{1k}-3) \hat g^k + (C_{1p}-3\big<\widehat{\widetilde{W}}_p\big>) \big[S^p + \big<S^p\big>\big] 
+ \Big(B^\ast_{1\ell} h^\ell + \frac1 M B_{11\ell} a^\ast_p S^p h^\ell\Big) \bigg\}a^q S^\dag_q
\nn\\
&&+R_1{}^a{}_b a^p S^\dag_p \Phi^b  \partial_a W_m
+ a^p\Big(S^\dag_p + \big<S^\dag_p\big>\Big)\Big(\big<S^q \widehat{\widetilde{W}}_p\big> + W_m\Big)+\Big(S^p + \big<S^p\big>\Big) a^q S^\dag_q\partial_p W'_m \Bigg] +\text{h.c.}\Bigg\}\ .\nn
\eeqa
}
\end{itemize}
We also found several contributions to the cosmological constant, see for instance the first line of \autoref{eq:Vapp}:
{\footnotesize
$$e^{\big<z^\dag z\big>}\Bigg[(M M^\dag)^2 \Big\{\big< \hat \rho_{i^\ast} \hat K^{i^\ast}{}_i \hat \rho^{\ast i}\big> -3\Big\}+ m_p^2 |M|^2 a_p^\ast a^p+ \Big(m_p M^\dag a^p \big<\tilde W_p\big>+ \text{h.c.}\Big)
\Bigg]\ .$$
}

\section{Case of a renormalisable theory}
We now consider the most general renormalisable theory, for which we have the substitution:
\beqa
\tilde W_k(z)g^k({\cal U},\Phi) &\to&\frac12 m_{ab}(z) \Phi^a \Phi^b + \frac12  \sigma_{ab}(z) {\cal U} \Phi^a \Phi^b + \frac16 y_{abc} \Phi^a \Phi^b \Phi^c + \frac16 \lambda(z) {\cal U}^3 + \frac12 \mu(z) {\cal U}^2\ ,\nn\\
Z_\ell(z,z^\dag) h^\ell(\Phi) &\to& \frac12 Z_{ab}(z,z^\dag) \Phi^a \Phi^b \ .\nn
\eeqa
We thus consider,
\beqa
W(z,\mathcal{U},\Phi)&=& m_p\big[\hat W_0(z) + S^p \mu^\ast_p \hat W_1(z)\big] + S^p \xi_p(z) + \frac12 m_{ab}(z) \Phi^a \Phi^b \nn\\
&&+ \frac12  \sigma_{ab}(z) {\cal U} \Phi^a \Phi^b + \frac16 y_{abc} \Phi^a \Phi^b \Phi^c + \frac16 \lambda(z) {\cal U}^3 + \frac12 \mu(z) {\cal U}^2 \ , \nn\\
K(z,z^\dag,\Phi,\Phi^\dag, S,S^\dag)&=&m_p^2 K(z,z^\dag) + S_p^\dag S^p + \Lambda^{a^\ast}{}_a(z,z^\dag) \Phi^\dag_{a^\ast} \Phi^a + \frac12\big[Z_{ab}(z,z^\dag) \Phi^a \Phi^b + \text{h.c.}\big] \nn
\eeqa
where we have denoted $\tilde W_p S^p= \xi_p S^p $. The potential is given by
\beqa
V= \Lambda + V_{\text{SUSY}} + V_{\text{soft}} + V_{\text{hard}} \ , \nn
\eeqa
where $\Lambda$ is the cosmological constant. The supersymmetric part of the potential reduces to
\beqa
V_{\text{SUSY}} = \partial_p W'_m \partial^p W'{}^\ast_m
+\partial _a W_m \partial^{a^\ast} W_m^\ast (\Lambda^{-1})^a{}_{a^\ast} \ , \nn
\eeqa
 with the superpotential
\beqa
W_m &=& \frac 12 \hat m_{ab}\Phi^a \Phi^b + \frac 16  \hat y_{abc} \Phi^a \Phi^b \Phi^c + \frac 12 \hat y_{abp} \Phi^a \Phi^b S^p 
+ \frac 12 \hat \sigma_{ab} {\cal U} \Phi^a \Phi^b \nn\\
&&+\hat \xi_p S^p +\frac 12 \hat \mu {\cal U}^2 + \frac16 \hat \lambda {\cal U}^3 \nn
\eeqa
and 
\beqa
W'_m &=& e^{\frac 12 \hat K}\Big(\frac 12 m_{ab}\Phi^a \Phi^b + \frac 16  y_{abc} \Phi^a \Phi^b \Phi^c + \frac 12 \sigma_{ab} {\cal U} \Phi^a \Phi^b \nn\\
&&+\xi_p S^p +\frac 12 \mu {\cal U}^2 + \frac16 \lambda {\cal U}^3 \Big) \ .\nn
\eeqa
We have introduced the notations:
\beqa
\hat m_{ab}&=& e^{\frac 12 \hat K} m_{ab} + m_\frac32 \big[Z_{ab} - \rho_{i^\ast} \partial^{i^\ast} Z_{ab}\big] \ , \nn \\
\hat y_{abc} &=& e^{\frac12 \hat K} y_{abc} \ , \nn\\
\hat y_{abp} &=& \frac {m_\frac32}{M}\big[Z_{ab} - \rho_{1 i^\ast} \partial^{i^\ast} Z_{ab}\big] a^\ast_p \ , \nn\\
\hat \sigma_{ab}&=& e^{\frac12 \hat K}\sigma_{ab} \ , \nn\\
\hat \xi_p&=& e^{\frac12 \hat K}  \xi_p \ , \nn\\
\hat \mu&=& e^{\frac12 \hat K} \mu \ , \nn\\
\hat \lambda&=&  e^{\frac12 \hat K} \lambda \ . \nn
\eeqa
The supersymmetric breaking terms are of two types: soft supersymmetric breaking terms (as usual) and hard supersymmetric breaking terms, controlled by
the chiral superfields $S^1$ and $S^2$. These terms are expressed with respect to the two functions
\beqa
\hat \rho_{i^\ast} &=& \partial_i\Big(\hat K + \ln \frac{\hat W}{m_p^2} \Big)(K^{-1})^i{}_{i^\ast}\ \ ,\ \  \hat W= \hat W_0 + S^p \mu^\ast_p \hat W_1 \ , \nn\\
\hat \rho_1{}_{i^\ast} &=& \partial_i\Big(\hat K + \ln \frac{\hat W_1}{m_p^2}\Big)(K^{-1})^i{}_{i^\ast} \ .  \nn
\eeqa
We also recall that the mass scale are defined by
\beqa
M^2&=&\Big< \hat W_0 + \mu^\ast_pS^p \hat W_1 \Big> \ , \nn\\
a^\ast_p M&=& \mu^\ast_p \big<\hat W_{1} \big> \ ,\nn\\
m_\frac 32 &=& e^{\frac12 \big<\hat K\big>}  \frac {M^2}{m_p}  \ .\nn
\eeqa
From now on to easy the notation the symbol $\big<\cdots\big>$ is omitted when the $z-$fields develop vacuum expectation values.
For instance $\hat K$ is understood for $\big<\hat K\big>$. The soft  supersymmetric breaking terms are
\begin{allowdisplaybreaks}
\beqa
V_{\text {soft}} &=& (m^2)^{a^\ast}{}_a\Phi^\dag_{a^\ast} \Phi^a + (m^2)^p{}_q S^\dag_p S^q \nn \\
&&+\Bigg\{ m^\dag_\frac32 e^{\frac 12 \hat K} \Big[ \mu {\cal U} \partial_p  {\cal U} + \frac 12 \lambda {\cal U}^2 \partial_p  {\cal U}
\Big] \Big[S^p + \lag S^p\rag\Big] +   \frac{m^\dag_\frac32}{M^\dag} e^{\frac 12 \hat K}    \mu  {\cal U}\partial_p {\cal U}   \lag S^p\rag a^q S^\dag_q  + \text{h.c.} \Bigg\}\nn\\
\nn\\
&&+\Bigg\{\frac 16 A_{abc} \Phi^a \Phi^b \Phi^c + \frac 12 B_{ab} \Phi^a \Phi^b + \frac12 B'_{pq}S^pS^q + 
\frac 12 A'_{abp}\Phi^a \Phi^b S^p + \frac 12 A''_{ab} {\cal U} \Phi^a \Phi^b\nn\\
&& +C_p S^p+ \frac12 B''{\cal U}^2 +\frac16 D'{\cal U}^3 + \text{h.c.}\Bigg\} \nn
\eeqa
with:
\beqa
(m^2)^{a^\ast}{}_a&= &|m_\frac32|^2 \Bigg[\Lambda^{a^\ast}{}_a\Bigg\{  \Bigg( \frac{\big<S^p \big>a^\ast_p}{M}  + {\rm h.c.} \Bigg) + (\hat \rho_{i^\ast }\hat K^{i^\ast}{}_i\hat \rho^{\ast i} -2)\Bigg\} \nn\\
 && +\hat \rho_{i^\ast}\Big(\partial^{i^\ast} \Lambda^{a^\ast}{}_b (\Lambda^{-1})^b{}_{b^\ast} \partial_j\Lambda^{b^\ast}{}_a -
\partial_j\partial^{i^\ast} \Lambda^{a^\ast}{}_a\Big)\hat \rho^{\ast j}\Bigg]\nn\\
 &&+ \Bigg(|M|^2a_q^\ast a^qe^{\hat K} + \Bigg\{ \frac{m_{\frac32}^\dag}{M^\dag}a^q\hat\xi_q + {\rm h.c.} \Bigg\} \Bigg)\Lambda^{a^\ast}{}_a\ ,  \nn \\
(m^2)^{p}{}_q&=&|m_\frac32|^2\big[\hat \rho_{i^\ast }\hat K^{i^\ast}{}_i\hat \rho^{\ast i} -2\big] \delta^p{}_q +
e^{\hat K} |M|^2\big[\hat \rho_{1 i^\ast }\hat K^{i^\ast}{}_i\hat \rho_1^{\ast i} -1\big]a^p a^\ast_q \ \nn\\
&&e^{\frac 12 \hat K} \Bigg\{\frac {m_\frac 32^\dag}{M^\dag} 
 \Big[ \Big( \partial_{i} \hat K \xi_{q}+ \partial_{i} \xi_{q}\Big) \hat \rho_{1}^{\ast i}  - \xi_{ q} \Big]a^p   + \text{h.c.}\Bigg\}
 + |M|^2 a_r^\ast a^r e^{\hat K}\delta^p{}_q \nn\\
&& + \frac{|m_\frac32 |^2}{|M|^2}\Bigg( \big< S^\dag_rS^r\big>a_q^\ast a^p \Big( \hat \rho_{1i\ast} \hat K^{i\ast}_i\hat \rho_1^{\ast i} - 2 \Big) + a_q^\ast a^p\big<S_r^\dag S^r\big> + \Big\{ a_r^\ast\big<S^r\big>a^p\big<S_q^\dag\big> + {\rm h.c.} \Big\} \Bigg)\nn\\
&&  + \Bigg\{ \frac{|m_\frac32 |^2}{M}\Bigg( a_r^\ast \big<S^r\big>\delta^p{}_q + a_q^\ast \big<S^p\big> + a_q^\ast \Big( \hat\rho_{1i\ast}\hat K^{i\ast}_i\hat\rho_1^{\ast i} - 2 \Big)\big< S^p \big>  \Bigg) + {\rm h.c.}\Bigg\} \nn\\
&& + e^{\frac {\hat K}2} \Bigg\{ \frac{m_\frac32^\dag}{M^\dag}a^r\xi_r + {\rm h.c. } \Bigg\}\delta^p{}_q\ , \nn\\
A_{abc}&=&e^{\frac 12 \hat K} m^\dag_\frac 32 \Bigg\{
\Big[\partial_i \hat K y_{abc} + \partial_i y_{abc}\Big]\hat \rho^{\ast i}
+\frac 1 {M^\dag} y_{abc} a^p \lag S^\dag_p\rag
\nn\\
&&-
\Big[ (\Lambda^{-1})^d{}_{b^\ast} \partial_i \Lambda^{b^\ast}{}_a \hat \rho^{\ast i}  y_{d bc} + (a \leftrightarrow b) + (a \leftrightarrow c)
\Bigg\}\ , \nn\\
A'_{abp}&=& \frac{|m_\frac 32|^2}{M} \Bigg\{  -\Big[(\Lambda^{-1})^c{}_{b^\ast} \partial_i \Lambda^{b^\ast}{}_a \hat \rho^{\ast i}(Z_{cb} -\hat \rho_{1i^\ast} \partial^{i^\ast} Z_{cb}) +  (a \leftrightarrow b)\Big] \nn\\
&& +\Big[ 3 Z_{ab}- \hat \rho_{1i^\ast} \partial^{i^\ast} Z_{ab} + \rho^{\ast i} \partial_i Z_{ab} -  \hat \rho_{1i^\ast}  \rho^{\ast i} \partial^{i^\ast} \partial_i Z_{ab}\Big]
 \nn\\
&& + \big( \hat \rho_{1i^\ast} \hat K^{i^\ast}{}_i\hat \rho^{\ast i} -3\big) Z_{ab}
+ \frac 1{M^\dag}( 2Z_{ab} - \rho_{1i^\ast} \partial^{i^\ast} Z_{ab})  a^q \lag S^\dag_q\rag\Bigg\}a^\ast_p + e^{\frac 12 \hat K} m^\dag_\frac 32\partial_p {\cal U} \sigma_{ab}\ , \nn\\
A''_{ab}&=&e^{\frac 12 \hat K} m^\dag_\frac 32 \Big[ -\sigma_{ab} + \big( \partial_i \hat K \sigma_{ab} + \partial_i \sigma_{ab} \big) \hat \rho^{\ast i}
-\Big((\Lambda^{-1})^c{}_{b^\ast}\partial_i \Lambda^{b^\ast}{}_a \hat \rho^{\ast i} \sigma_{cb}  +(a \leftrightarrow b)\Big)
\Big] \nn\\
&&+
e^{\frac 12 \hat K} \frac{m^\dag_\frac 32}{M^\dag} a^p <S^\dag_p> \sigma_{ab}\ , 
\nn\\
B_{ab}&=& e^{\frac12 \hat K} m^\dag_\frac 32 \Bigg( \Big[ \partial_i \hat K m_{ab} + \partial_i m_{ab}\Big]\hat \rho^{\ast i}-m_{ab}
-\Big[(\Lambda^{-1})^c{}_{b^\ast} \partial_i \Lambda^{b^\ast}{}_a \hat \rho^{\ast i} m_{cb} + (a \leftrightarrow b)\Big]\nn\\
&& +\frac 1 {M^\dag} m_{ab} a^p \lag S^\dag_p\rag  + \sigma_{ab}  <S^p> \partial_p {\cal U} \Bigg)\nn\\
&&+ |m_\frac 32|^2   \Bigg( 2 Z_{ab}- \hat \rho_{i^\ast} \partial^{i^\ast} Z_{ab} + \rho^{\ast i} \partial_i Z_{ab} -  \hat \rho_{i^\ast}  \rho^{\ast i} \partial^{i^\ast} \partial_i Z_{ab}\nn\\
&&
 +\frac {1} {M^\dag} (Z_{ab} - \rho_{i^\ast} \partial^{i^\ast} Z_{ab} )a^p \lag S^\dag_p\rag \Bigg) \nn \\
&&-|m_\frac 32|^2 \Bigg(\Big[(\Lambda^{-1})^c{}_{b^\ast} \partial_i \Lambda^{b^\ast}{}_a \hat \rho^{\ast i}(Z_{cb} -\hat \rho_{i^\ast} \partial^{i^\ast} Z_{cb}) +  (a \leftrightarrow b)\Big]
 -  \Big[\hat \rho_{i^\ast} \hat K^{i^\ast}{}_{i^\ast} \hat \rho^{\ast i} -3 \Big]Z_{ab}  \Bigg)\  \nn\\
&& + |M|^2 a_q^\ast a^qe^{\hat K}Z_{ab} +\Bigg\{ e^{\frac{\hat K}{2}}\frac{m_\frac32^\dag}{M^\dag}a^q\xi_q + \frac{|m_\frac32|^2}{M}a_p^\ast \big< S^p\big> + {\rm h.c.} \Bigg\}Z_{ab}\ , \nn\\
B'_{pq} & =&  \frac{|m_\frac32|^2}{M}\Bigg\{ a_p^\ast\big<S_q^\dag\big>\big( \hat\rho_{1i\ast}\hat K^{i\ast}{}_i\hat\rho^{\ast i} -1 \big) + \frac{1}{M^\dag}a^ra^\ast_p\big< S^\dag_rS^\dag_q\big> \Bigg\}\ , \nn\\
C_p&=& m^\dag_\frac32 e^{\frac12 \hat K} \Big[ -2 \xi_p+ \big( \partial_i \hat K\xi_p + \partial_i \xi_p\big)\hat \rho^{\ast i}\Big]\nn\\
&&+ \frac {m_\frac32}{M} e^{\frac 12 \hat K}\Bigg(\Big[\big( \partial^{i^\ast} \hat K \xi^{\ast q}+ \partial^{i^\ast} \xi^{\ast q}\big) \hat \rho_{1 i^\ast} - \xi^{\ast q}\Big]  <S^\dag_q>a^\ast_p
\Bigg)+\frac {m_\frac32^\dag}{M^\dag} e^{\frac 12 \hat K}\xi_p \big<S^\dag_q\big> a^q\nn\\
&&+ e^{\hat K}  \Big[
M (M^\dag)^2\big(\hat \rho_{i^\ast} \hat K^{i^\ast}{}_{i} \hat \rho^{\ast i} -2\big) a^\ast_p + |M|^2 a^q a^\ast_p <S^\dag_q> \Big]+ |M|^2a_r^\ast a^r e^{\hat K}\big< S_p^\dag\big>\nn\\
&& + \big< S_p^\dag\big>\Bigg\{ \frac{m_\frac32^\dag}{M^\dag}a^r\xi_r + {\rm h.c.} \Bigg\} + \frac{|m_\frac32|^2}{|M|^2}a^ra_p^\ast \big< S^\dag_r \big> + |m_\frac32 |^2 \Big( \hat\rho_{i \ast}\hat K^{i\ast}{}_i\hat\rho^{\ast i}-2 \Big) \big< S_p^\dag \big> \nn\\
&&  + \frac{|m_\frac32|^2}{M}\Bigg\{ a_q^\ast\big<S^qS_p^\dag\big> +\big< S_r^\dag S^r\big>a_p^\ast \big( \hat\rho_{1i\ast}\hat K^{i\ast}{}_i\hat\rho^{\ast i} - 1 \big) \Bigg\} + \frac{|m_\frac32 |^2}{M^\dag}a^q \big<S_q^\dag S_p^\dag\big>\ , \nn\\
B'' &=&  m_\frac32^\dag e^{\frac12\hat K}\Bigg\{ \big( \partial_i \hat K \mu + \partial_i\mu \big)\hat\rho^{\ast i} + \Big(\frac{1}{M^\dag}a^p\big< S_p^\dag \big> -3\Big)\mu \Bigg\} \ , \nn\\
 D &=&  m_\frac32^\dag e^{\frac12\hat K}\Bigg\{ \big( \partial_i \hat K \lambda + \partial_i\lambda \big)\hat\rho^{\ast i} + \Big(\frac{1}{M^\dag}a^p\big< S_p^\dag \big> -3\Big)\lambda \Bigg\} \ .\nn\\
\eeqa
\end{allowdisplaybreaks}
The hard supersymmetric breaking
terms are
\beqa
V_{\text{hard}}&=& Q_{ap}{}^{a^\ast q} \Phi^\dag_{a^\ast}\Phi^a S^\dag_q S^p + Q'_{pr}{}^{qt} S^p S^r S^\dag_q S^\dag_t +\Big(T_{ap}{}^{b^\ast} \Phi^\dag_{b^\ast} \Phi^a S^p + \text{h.c}\Big)
+\Big(T'_{pq}{}^{r}  S^p S^q S^\dag_r + \text{h.c}\Big)
\nn\\
&&+\Big(\frac 16 D_{abc}{}^p \Phi^a \Phi^b \Phi^c S^\dag_p  +\text{h.c.} \Big)+
\Big(\frac 12 D'_{abp}{}^q\Phi^a \Phi^b S^p S^\dag_q + \text{h.c.}\Big)+\Big(\frac 12 D''_{ab}{}^q\Phi^a \Phi^b {\cal U} S^\dag_q + \text{h.c.}\Big)
\nn\\
&&+\Big(\frac 12 E_{ab}{}^p \Phi^a \Phi^b S^\dag_p + \text{h.c.}\Big)
+\Bigg\{ \frac12 B'^p{\cal U}^2S_p^\dag + \frac16 D'^p{\cal U}^3S_p^\dag + \text{h.c.}\Bigg\}\nn
\\
&&+ \Bigg\{\frac 12e^{\frac12 \hat K}  \frac{m^\dag_\frac 32}{M^\dag} \big(\lambda {\cal U}^2 \partial_p {\cal U}\big)(S^p + <S^p>\big) a^q S^\dag_q
+ e^{\frac12 \hat K}  \frac{m^\dag_\frac 32}{M^\dag} \big(\mu {\cal U} \partial_p {\cal U}\big)S^p  a^q S^\dag_q + \text{h.c.} \Bigg\}\nn
\eeqa
with
\begin{allowdisplaybreaks}
\beqa
Q_{ap}{}^{a^\ast q} & = & \frac{|m_\frac32|^2}{|M|^2}
\Bigg[\Lambda^{a^\ast}{}_a(\hat \rho_{1i^\ast }\hat K^{i^\ast}{}_i\hat \rho_1^{\ast i} ) \ +\hat \rho_{1i^\ast}\Big(\partial^{i^\ast} \Lambda^{a^\ast}{}_b (\Lambda^{-1})^b{}_{b^\ast} \partial_j\Lambda^{b^\ast}{}_a-
\partial_j\partial^{i^\ast} \Lambda^{a^\ast}{}_a\Big)\hat \rho_1^{\ast j}\Bigg] a^\ast_p q^q\ , \nn\\
Q'_{pr}{}^{qt}&=& \frac{|m_\frac32|^2}{|M|^2}\Big(\hat \rho_{1 i^\ast} \hat K^{i^\ast}{}_i\hat \rho_1^{\ast i} \Big)\delta^q{}_p a^\ast_r a^t \ , \nn\\
T_{ap}{}^{a^\ast}&=& \frac{|m_\frac 32|^2}{M}
 \Bigg[\Lambda^{a^\ast}{}_a\Big(\hat \rho_{1i^\ast }\hat K^{i^\ast}{}_i\hat \rho^{\ast i}  -1 + \frac{1}{M^\dag}a^r\big< S_r^\dag \big>\Big) \nn\\
&& +\hat \rho_{1i^\ast}\Big(\partial^{i^\ast} \Lambda^{a^\ast}{}_b (\Lambda^{-1})^b{}_{b^\ast} \partial_j\Lambda^{b^\ast}{}_a-
\partial_j\partial^{i^\ast} \Lambda^{a^\ast}{}_a\Big)\hat \rho^{\ast j}\Bigg]a^\ast_p\ , \nn\\
T'_{rq}{}^p&=&  \frac{|m_\frac32|^2}{M}\Big(\hat \rho_{1 i^\ast} \hat K^{i^\ast}{}_i\hat \rho^{\ast i} -1\Big)\delta^p{}_q a^\ast_r\nn\\
&& + \frac{|m_\frac32|^2}{M^2}\Big\{ a^ta^\ast_q\big<S_t^\dag\big>\delta^p{}_r + 2a_r^\ast a^p\big<S_q^\dag\big> + a_q^\ast a^p\Big( \hat\rho_{1i\ast}\hat K^{i\ast}_{i}\hat\rho_1^{\ast i} -2 \Big) \big<S_r^\dag\big> \Big\}\ , \nn\\
D_{abc}{}^p&=&
e^{\frac 12 \hat K}  \frac {m^\dag_\frac 32}{M^\dag}\Bigg\{
\Big[ \partial_i \hat K y_{abc} + \partial_i y_{abc}\Big]\hat \rho_1^{\ast i}
 \nn\\
&&-  \Big[ (\Lambda^{-1})^d{}_{b^\ast} \partial_i \Lambda^{b^\ast}{}_a \hat \rho_1^{\ast i}  y_{d bc} + (a \leftrightarrow b) + (a \leftrightarrow c)
 +  y_{abc}\Bigg\}a^p \ , \nn\\
D'_{abp}{}^q&=&
 \frac{|m_\frac 32|^2}{|M|^2} \Bigg\{ -\Big[(\Lambda^{-1})^c{}_{b^\ast} \partial_i \Lambda^{b^\ast}{}_a \hat \rho_1^{\ast i}(Z_{cb} -\hat \rho_{i^\ast} \partial^{i^\ast} Z_{cb}) +  (a \leftrightarrow b)\Big]a^\ast_p 
+ 
\nn\\
&& +\Big[{\color{darkgreen} 2} Z_{ab}- 2 \hat \rho_{1i^\ast} \partial^{i^\ast} Z_{ab} + \rho_1^{\ast i} \partial_i Z_{ab} -  \hat \rho_{1i^\ast}  \rho_1^{\ast i} \partial^{i^\ast} \partial_i Z_{ab}\Big]a^\ast_p 
+ \hat \rho_{1i^\ast} \hat K^{i^\ast}{}_i \hat \rho^{\ast i} Z_{ab}a^\ast_p  \Bigg\}a^q\ ,  
\nn\\
&&+ e^{\frac 12 \hat K}\frac{m^\dag_\frac 32}{M^\dag} \sigma_{ab} \partial_p {\cal U}a^q \nn\\
D''_{ab}{}^p &=& e^{\frac12 \hat K} \frac{m^\dag_\frac 32}{M^\dag}\Bigg\{ \big( \partial_i \hat K \sigma_{ab} + \partial_i \sigma_{ab}\big)\hat \rho_1^{\ast i}
-\Big[(\Lambda^{-1})^c{}_{b^\ast} \partial_i \Lambda^{b^\ast}{}_a \hat \rho_1^{\ast i} \sigma_{cd} + (a \leftrightarrow b)\Big]\Bigg\}a^p\ , 
\nn\\
E_{ab}{}^p&=&
 \Bigg\{e^{\frac12 \hat K} \frac{m^\dag_\frac 32}{M^\dag}\bigg( \Big[ \partial_i \hat K m_{ab} + \partial_i m_{ab}\Big]\hat \rho_1^{\ast i}-\Big[(\Lambda^{-1})^c{}_{b^\ast} \partial_i \Lambda^{b^\ast}{}_a \hat \rho_1^{\ast i} m_{cb} + (a \leftrightarrow b)\Big]\bigg)
\nn \\
&&+ \frac{|m_\frac 32|^2}{M^\dag} \bigg( 4 Z_{ab}- 2\hat \rho_{i^\ast} \partial^{i^\ast} Z_{ab}
   + \rho_1^{\ast i} \partial_i Z_{ab} -  \hat \rho_{i^\ast}  \rho_1^{\ast i} \partial^{i^\ast} \partial_i Z_{ab}
   +\big(\hat \rho_{i^\ast} \hat K^{i^\ast}{}_i\hat \rho_1^{\ast i}-3) Z_{ab}
\nn\\
&&-\Big[(\Lambda^{-1})^c{}_{b^\ast} \partial_i \Lambda^{b^\ast}{}_a \hat \rho_1^{\ast i}(Z_{cb} -\hat \rho_{i^\ast} \partial^{i^\ast} Z_{cb}) +  (a \leftrightarrow b)\Big]\bigg)
+\frac{m^\dag_\frac32}{M^\dag} e^{\frac 12 \hat K} \partial_q {\cal U} \sigma_{ab} \big<S^q\big>\nn\\
&& + \frac{|m_\frac32|^2}{|M|^2} a_q^\ast \big< S^p \big>Z_{ab}\Bigg\} a^p
 \ , \nn\\
B'^p&=& \frac{m_\frac32^\dag }{M^\dag}e^{\frac{\hat K}{2}}\Bigg\{ \Big[ \partial_i\hat K\mu + \partial_i\mu \Big]\hat\rho_1^{\ast i} - 2\mu \Bigg\}a^p\ ,  \nn\\
D'^p&=& \frac{m_\frac32^\dag }{M^\dag}e^{\frac{\hat K}{2}}\Bigg\{ \Big[ \partial_i\hat K\lambda + \partial_i\lambda \Big]\hat\rho_1^{\ast i} - 2\lambda \Bigg\}a^p\ . \nn
\eeqa
\end{allowdisplaybreaks}

\section{Application}

In this section we are  considering a NSW-extension closely related to the NMSSM. We denote here $Q,U,D$ the superfields describing the quark/squark sector, $L,E,N$ the superfields for the lepton/slepton sector and $H_U,H_D$ the two Higgs superfields (see \autoref{tab:PSM}). Introduce further two singlet fields $S^1,S^2$ and dimensionless fields in the hidden sector $z^i$. Denote generically $\Phi^a$ the fields in the observable sector
and $\tilde \Phi^I=(\Phi^a,H_U,H_D)$. Recall that
\beqa
{\cal U} = \mu^1 S^2 -\mu^2 S^1 \ . \nn
\eeqa
The K\"ahler potential is taken to be
\beqa
K(z,z^\dag,S,S^\dag,\Phi,\Phi^\dag)= m_p^2 \hat  K(z,z^\dag) + S^\dag_p S^p + \sum_{I} {  \Lambda_I(z,z^\dag)}\tilde \Phi^\dag_I \tilde \Phi^I + \Big[ {Z(z,z^\dag)} H_U\cdot H_D + \text{h.c.} \Big]\nn
\eeqa
and the superpotential
\beqa
W(z,S,\Phi) &=& m_p\big[\hat W_0(z) + S^p \mu_p^\ast \hat W_1(z)\big]+ S^p    \zeta_p(z)\nn\\
&&-{ y_e(z)}L\cdot H_D  E- {  y_d (z)}Q\cdot  H_D D 
      + {  y_u(z)}Q  \cdot  H_U  U + {   \lambda(z)} {\cal U}  H_D \cdot H_U  
      \nn\\
&& + \frac16 {   \kappa(z) }{\cal U}^3  \nn  
\eeqa
Several remarks are in order:
\begin{enumerate}
\item We neglect in  this first analysis the flavour of particles, {\it i.e.}, we only consider one family of quarks and leptons;
\item We do not consider the interactions with neutrinos;
\item We only consider dimensionless coefficients in the K\"ahler potential and in the superpotential $\tilde{W}_k(z)g^k(\mathcal{U},\Phi)$, the dimensionfull terms in ${\cal U}, {\cal U}^2$ or $H_U\cdot H_D$ can be easily taken into consideration if needed. Note that the term $\zeta {\cal U}$ is already present in $\tilde{W}_p(z)S^p=\zeta_p S^p$.
\end{enumerate}

Before using the results previously established, some care is needed. Indeed, after supersymmetry breaking the kinetic part of the scalar  component $\varphi$ and
of the fermion  component $\chi$  of the chiral superfield $\Phi$ are not correctly normalised.
Actually the factor $\Lambda \Phi^\dag \Phi$ in the K\"ahler potential leads to
\beqa
L_{\text{kin scal.}} &=& \big<\Lambda\big> \partial_\mu \varphi \partial^\mu \varphi^\dag \ , \nn\\
L_{\text{ kin ferm.}} &=& -\frac i 2 \big<\Lambda\big>\Big(\chi \sigma^\mu \partial_\mu \bar \chi - \partial_\mu \chi \sigma^\mu \bar \chi\Big) \ . \nn 
\eeqa
Thus  $\big<\Lambda\big>$ must be positive, and in order to have correctly normalised fields, a field redefinition must be performed
\beqa
\varphi \to \frac1 {\sqrt{\big<\Lambda\big> }}\varphi \ , \ \
  \chi \to \frac1 {\sqrt{\big<\Lambda\big> }}\chi \ . \nn
\eeqa
When supergravity is broken in the hidden sector the scalar potential takes the forms
\beqa
V= V_{\text{SUSY}} + V_{\text{soft}} + V_{\text{hard}} \ . \nn
\eeqa

Recall that the gravitino mass is given by
\beqa
m_\frac32 = \frac 1 {m_p} e^{\frac 12 \big<\hat K\big>} \big<\hat W\big> \ , \nn
\eeqa
and the scale $M$ is defined through
\beqa
M^2&=& \big<\hat W_0 + S^p \mu^\ast_p  \hat W_1 \big> \ , \nn\\
a_p^\ast M &=&\mu_p^\ast \big<\hat W_1\big> \ . \nn
\eeqa

We also introduce two basic functions
\beqa
\rho_{\is}&=& \partial_i\Big(\hat K + \ln \frac{\hat W}{m_p^2}\Big) (\hat K^{-1})^i{}_\is\nn\\
\rho_{1,\is}&=& \partial_i\Big(\hat K + \ln \frac{\hat W_1}{m_p}\Big) (\hat K^{-1})^i{}_\is\nn
\eeqa

\bigskip \bigskip
The unbroken supersymmetric part of the potential is 
\beqa
V_\text{SUSY} =  \partial_a W_m \partial^a \bar W_m + \partial_p W'_m \partial^p \bar W'_m  \ , \nn
\eeqa
where the superpotential is given by
\beqa
W_m &=& W'_m + W_{GM}\nn\\
W'_m &=& -{\hat  y_e}L\cdot H_D  E- { \hat  y_d}Q\cdot  H_D D 
      + {\hat   y_u}Q \cdot  H_U  U + {\hat  \lambda}\; {\cal U}  H_D \cdot H_U  
      + \frac16 {\hat  \kappa }\;{\cal U}^3 + {\hat \zeta_p} \;{S^p}\nn\\
W_{GM} &=& {\mu} H_U\cdot H_D+ { y_p} S^p H_U \cdot H_D \label{eq:Wgm}
\eeqa

The terms in \autoref{eq:Wgm} come from the K\"ahler potential, and constitute the usual Giudice-Masiero term supplemented by the corresponding term resulting from
the S-field:
\beqa
\label{eq:mu}
{\mu} &=& m_\frac32 \Big(\frac{\big< Z \big>}{\sqrt{ \Lambda_{H_D}\Lambda_{H_U}}}  -  \frac{\big< \rho_{i^\ast}\partial^{i^\ast} Z\big>}{\sqrt{\Lambda_{H_D}\Lambda_{H_U}}}\Big) \ ,\nn\\
{y_p} &=& \frac{m_\frac32 }{M}\Big(\frac{\big< Z \big>}{\sqrt{\Lambda_{H_D}\Lambda_{H_U}}}  -  \frac{\big< \rho_{1,i^\ast}\partial^{i^\ast} Z\big>}{\sqrt{\Lambda_{H_D}\Lambda_{H_U}	}}\Big)
a_p^\ast\ .\nn
\eeqa
Note that the $y_p$ term is $M-$suppressed. The Yukawa interaction have two origins : the superpotential and the field redefinition due to the $  \Lambda-$term in the K\"ahler potential:
\beqa
\label{eq:yuk}
   \hat y_{e} = e^{\frac 12 \big<\hat K\big>} \frac {\big< y_{e}\big>}{\sqrt{\big< \Lambda_E \Lambda_L \Lambda_{H_D}}\big>} \ , 
   \hat y_{d} = e^{\frac 12 \big<\hat K\big>} \frac {\big< y_{d}\big>}{\sqrt{\big< \Lambda_Q \Lambda_D \Lambda_{H_D}}\big>} \ ,
   \hat y_{u} = e^{\frac 12 \big<\hat K\big>} \frac {\big< y_{u}\big>}{\sqrt{\big< \Lambda_Q \Lambda_U \Lambda_{H_U}}\big>} \ ,\nn
\eeqa
and
\beqa
\hat  \kappa = e^{\frac 12 \big<\hat K\big>} \big< \kappa\big> \ , \ \
 \hat \lambda &=&e^{\frac 12 \big<\hat K\big>} \frac{\big< \lambda\big>}{\sqrt{\big< \Lambda_{H_D} \Lambda_{H_U} \big>}} \ ,\ \ 
\hat \zeta_p = e^{\frac 12 \big<\hat K\big>} {\big< \zeta_p\big>} \ .\nn 
\eeqa

The soft-supersymmetric breaking part is given by 
\beqa
V_{\text{soft}} &=& \phantom{+} |m_\frac 32|^2\sum_{I}  { \varpi_I} \varphi^\dag_{I} \varphi^{I} + \Big(|M|^2{ \varpi' 
a^qa^\ast_p} + (m^2)^q{}_p
\Big) S^p S^\dag_q\nn\\
&&
  + \Bigg[\bigg\{\big|m_\frac 32\big|^2\;{ b} +  \hat \lambda m^\dag_\frac32 \big<{\cal U}\big> \bigg\}\; h_U\cdot h_D + \text{h.c.}\Bigg] +|m_\frac32|^2  { \varpi} \Big( S^p + \big< S^p \big>\Big)\Big( S^\dag_p + \big< S^\dag_p \big>\Big) 
\nn\\
&&+ { \left( |M|^2 a^*_q a^q e^{\langle \hat K\rangle} + \left\{\frac{m^{\dagger}_{3/2}}{M^{\dagger}} a^q
\hat \zeta_q + \text{h.c.}  \right\} \right) }\nn\\
&& { \times \left[ (S^p + \langle S^p \rangle)(S^{\dagger}_p + \langle S^{\dagger}_p \rangle ) +\varphi^{\dagger}_{I} \varphi^I  +\Big\{ \frac{1}{\sqrt{\Lambda_{H_U}\Lambda_{H_D}}}\big<Z\big> h_U\cdot h_D + {\rm h.c.} \Big\}\right] } \nn \\
&&
+   \Bigg\{ m_\frac32^\dag     \bigg(-{ A_e} \tilde \ell\cdot h_D \tilde e
- { A_d} \tilde q \cdot h_D \tilde d + { A_u} \tilde q\cdot h_U \tilde u
+  A' {\cal U} h_{U} \cdot h_{D} + {  A_p} \;S^p h_U\cdot h_D
 \nn\\
&& + C_p S^p + \frac 12   \hat { \kappa} {\cal U}^2\big({\cal U} + \big<{\cal U}\big>\big) +\frac16 { A_{\kappa}} {\cal U}^3\bigg)
+   m_\frac32 C'_p S^p   + C''_p S^p +  \text{h.c.} \Bigg\} \nn
\eeqa
where we have introduced:
\begin{enumerate}
  \item Scalar masses:
    \beqa
  \label{eq:f-scal}   
{\varpi_I}&=& \Big<\rho_\is \hat K^\is{}_i \bar \rho^i -2\Big>+  \left<\rho_\is \Big(\frac{\partial^\is {\Lambda_I} \partial_i{ \Lambda_I}}{ \Lambda_I^2} -\frac{\partial^\is \partial_i {\Lambda_I}}{ \Lambda_I}\Big)\bar \rho^i\right> \ , \nn \\
{\varpi}&=&\Big<\rho_\is \hat K^\is{}_i \bar \rho^i -2\Big>\nn\\
\varpi'&=&  e^{\big<\hat K\big>} \Big<\rho_{1,\is} \hat K^\is{}_i \bar \rho_1^i -1\Big>\label{eq:controlM2}\\
(m^2)^p{}_q&=& e^{\frac12 \big<\hat K\big>} \Bigg\{\frac{m^\dag_\frac32}{M^\dag}\Big[
(\partial_i \hat K \zeta_q +\partial_i \zeta_q)\bar \rho_1^i - \zeta_q\Big]a^p + \text{h.c.}\Bigg\} \label{eq:controlM1}
\eeqa
Note  the terms \autoref{eq:controlM2} and \autoref{eq:controlM1} are controlled by $M$ (second line of $V_{\text{soft}}$).
\item $b-$term:
  \beqa
   \label{eq:f-b}
{b}&=& - \frac{\big< Z \big>}{{\sqrt{ \Lambda_{H_D}\Lambda_{H_U}}}} -\frac{\big<\rho_\is \partial^\is {Z}\big>}{\sqrt{ \Lambda_{H_D}\Lambda_{H_U}}}  +
\frac{\big<\bar \rho^i \partial_i { Z}\big>}{{\sqrt{\Lambda_{H_D}\Lambda_{H_U}}}} - \frac{\big< \rho_\is \bar \rho^i \partial^\is \partial_i { Z}\big>}
{\sqrt{ \Lambda_{H_D}\Lambda_{H_U}}} \nn\\
&&-\Big(\frac {\partial_i  \Lambda_{H_U}}{ \Lambda_{H_U}} + \frac {\partial_i \Lambda_{H_D}}{\Lambda_{H_D}}\Big)\frac 1{\sqrt{ \Lambda_{H_U} \Lambda_{H_D}}}
\bar \rho^i( Z -\rho_\is \partial^\is Z)
+\frac 1{\sqrt{ \Lambda_{H_U} \Lambda_{H_D}}} \rho_\is \hat K^\is{}_i\bar \rho^i Z
\nn\\
&& +\frac 1{M^\dag} \Big(\big< Z\big> -\rho_{\is}\big< \partial^{\is} Z \big>\Big) a^p \big<S^\dag_p\big> \frac 1{\sqrt{ \Lambda_{H_U} \Lambda_{H_D}}} \nn
\eeqa
This term is one of the signature of the Giudice-Masiero mechanism (note that $b=0$ if $Z=0$). It is supplement by a $M-$suppressed term  coming from the vev of S.
\item trilinear $HHS-$ term
 \beqa
   \label{eq:f-b}
{ A_p}&=& \frac1{\sqrt{{ \big< \Lambda_{H_U}\Lambda_{H_D}\big>}}}\frac{m_\frac32}{M}\Bigg[ -\bigg(\frac{\big<\partial_i \Lambda_{H_U} \big>}{ \Lambda_{H_U}}+\frac{\big<\partial_i  \Lambda_{H_D}\big>}{\Lambda_{H_D}}\bigg)
 ( Z -\rho_{1\is} \partial^\is Z)\big<\bar \rho^i\big>\nn\\
&&\hskip 1.truecm - Z -\rho_{1\is} \partial^\is Z + \bar \rho^i \partial_i Z  - \rho_{1 \is} \bar \rho^i \partial_i \partial^\is  Z 
+(\rho_{1\is} \hat K^\is{}_i \bar \rho^i) Z \nn\\
&&\hskip1.truecm
 \frac 1 {M^\dag} (Z - \rho_{1\is} \partial^\is Z) a^q \big<S^\dag_q\big>\Bigg]a^\ast_p \nn
\eeqa
and the trilinear term ${\cal U} HH$
\beqa
   A'= \Bigg( \big(\partial_i \hat K    \hat \lambda+ \partial_i    \hat \lambda\big)\bar\rho^i-
\bigg(\frac{\big<\partial_i   \Lambda_{H_U}\big>}{  \Lambda_{H_U}}+\frac{\big<\partial_i   \Lambda_{H_D}\big>}{ \Lambda_{H_D}}\bigg)\bar \rho^i   \hat \lambda
+{ \frac 1{M^\dag} a^p \big<S^\dag_p\big>}    \hat \lambda\Bigg) . \nn
\eeqa
\item Trilinear terms:
  \beqa
   \label{eq:f-a}
  { A_I}= \Bigg[ \big< \partial_i \hat K\big>  \hat y_I+  \partial_i {\hat  y_I}-  \frac{\big<\partial_i  \Lambda_I \big>}{\big<  \Lambda_I\big>}{\hat  y_I}
-\frac{\big<\partial_i   \Lambda_{I'} \big>}{\big<  \Lambda_{I'}\big>}{\hat  y_I}-
   \frac{\big<\partial_i  \Lambda_{H_I}  \Big>}{\big<  \Lambda_{H_I}\big>}{\hat  y_I} \Bigg]\big<\bar \rho^i\big> +
   \frac 1{ M^\dag} a^p \big<S^\dag_p\big> \hat y_I \nn
  \eeqa
$H_I$ being the Higgs that couples to the scalar $\Phi^I$ as $ y_I \Phi^I\cdot H_{H_I} \Phi^{I'}$. The notation $\partial_i \hat y_I$ is a shorthand notation for $\big<e^{\frac12 \hat K} \partial_i y/\sqrt{\Lambda_I \Lambda_{I'} \Lambda_{H_I}}\big>$.
\item The linear term is given by
\beqa
   C_p&=&-2   \hat \zeta_p + \big(\partial_i \hat K    \hat \zeta_p + \partial_i    \hat \zeta_p\big) \bar \rho^i +\frac1{M^\dag}   \hat \zeta_p a^q \big<S^\dag_q\big>\nn\\
   C_p'&=& { \frac 1{M}}\Big(\big[(\partial^\is \hat K    { \hat {\bar \zeta}}^q + \partial^\is    {\hat {\bar  \zeta}}^q ) \rho_{1 \is} -   {\hat {\bar  \zeta}}^q\big]
{  \big<S^\dag_q\big>\big. a^\ast_p}
\Big)\nn\\
C''_p&=&e^{\hat K} \Bigg[ M (M^\dag)^2 \big(\rho_\is \hat K^\is{}_i \bar \rho^i-2) + |M|^2 a^q \big< S^\dag_q\big> \Bigg] a^\ast_p \ .\nn
\eeqa
\item ${\cal U}^3$ term :
\beqa
A_\kappa&=& \big( \partial_i \hat K \hat\kappa + \partial_i\hat\kappa \big)\hat\rho^{\ast i} + \Big(\frac{1}{M^\dag}a^p\big< S_p^\dag \big> -3\Big)\hat\kappa \nn
\eeqa

\end{enumerate}

The hard supersymmetric breaking terms are given by


\beqa
V_{\text{hard}}&=&|m_\frac32|^2\Bigg[ \sum_I   Q_{Ip}{}^q \varphi^\dag_I \varphi^I S^\dag_q S^p +   Q'_{r}{}^{t} S^\dag_p S^p  S^\dag_t S^r   +
 \Big\{\sum_I   T_{Ip}{} \varphi^\dag_I \varphi^I S^p + \text{h.c.}\Big\}
\nn\\
&&\hskip 1.truecm + \Big\{  T'_{r}S^\dag_p S^p S^r  + \text{h.c.}\Big\}\Bigg]+
\Bigg[\Big\{ |m_\frac32|^2 E^q{}  + \frac {m_\frac32^\dag}{M^\dag}    \hat \lambda \big<{\cal U}\big> a^q \Big\} h_U \cdot h_D  S^\dag_q + \text{h.c.}\Bigg]\nn\\
&&+ { \left\{ \frac{|m_{3/2}|^2}{M} \left( 1 + \frac{S^{\dagger}_r a^r}{M^{\dagger}} \right) a^*_p (S^p + \langle S^p \rangle )  + \text{h.c.} \right\}}\nonumber \\
&&\times {\left[ (S^p + \langle S^p \rangle)( S^{\dagger}_p + \langle S^{\dagger}_p \rangle ) + \varphi^{\dagger}_{I}  \varphi^I+\Big\{ \frac{1}{\sqrt{\Lambda_{H_U}\Lambda_{H_D}}}\big<Z\big> h_U\cdot h_D + {\rm h.c.} \Big\} \right] }
\nn\\
&&+ \Bigg[m_\frac32^\dag   \bigg\{-{ D_e^q} \tilde \ell\cdot h_D \tilde e
- { D_d^q} \tilde q \cdot h_D \tilde d + { D_u^q} \tilde q\cdot h_U \tilde u
+ { D_p{}^q} S^p h_U\cdot h_D +    D'^q {\cal U} h_U \cdot h_D\nn\\
&&
\hskip .5truecm +\frac 16\frac1 {M^\dag}    D_\kappa {\cal U}^3 a^q
+\frac12 \frac 1 {M^\dag}    \hat \kappa {\cal U}^2({\cal U} + \big<{\cal U}\big>) a^q \bigg\}S^\dag_q + 
\text{h.c.} \Bigg]\nn\\
 \nn
\eeqa

The various terms in $V_{\text{hard}}$ are
\begin{enumerate}
\item $\Phi\Phi^\dag S S^\dag-$term
\beqa
  Q_{Ip}{}^q=\frac {a^\ast_p a^q}{|M|^2}\Bigg\{\big[\rho_{1\is} \hat K^\is{}_i \bar \rho_1^i -2\big] + \rho_{1 \is} \bigg[\frac{\partial^\is   \Lambda_I \partial_i \Lambda_I}{ \Lambda_I^2} - \frac{\partial^\is \partial_i  \Lambda_I}{ \Lambda_I}\bigg]
\bar \rho_1^i\ \Bigg\} \ . \nn
\eeqa
\item $\Phi\Phi^\dag S$ + h.c.
\beqa
  T_{Ip}=\frac 1{M}\Bigg\{\big[\rho_{1\is} \hat K^\is{}_i \bar \rho^i -2\big] + \rho_{1 \is} \bigg[\frac{\partial^\is   \Lambda_I \partial_i  \Lambda_I}{ \Lambda_I^2} - \frac{\partial^\is \partial_i  \Lambda_I}{ \Lambda_I}\bigg]
\bar \rho^i\ \Bigg\} a^\ast_p \ . \nn
\eeqa
\item $(SS^\dag)^2-$term
\beqa
  Q'_{r}{}^{t} = \frac{a^\ast_r q^t}{|M|^2}\big[\rho_{1\is}  \hat K^\is{}_i \bar \rho_1^i -2\big]  \ . \nn
\eeqa
\item  $S^2S^\dag-$term + h.c.
\beqa
  T'_{r}{}= \frac{a^\ast_r  }{M}\big[\rho_{1\is}  \hat K^\is{}_i \bar \rho^i -2\big]  \ . \nn
\eeqa
\item The $D_I^q-$terms
\beqa
   \label{eq:f-a}
  { D_I^q}= \frac 1{M^\dag}\Bigg\{\Bigg[ \big< \partial_i \hat K\big> \hat y_I+  \partial_i {\hat  y_I}-  \frac{\big<\partial_i  \Lambda_I \big>}{\big<  \Lambda_I\big>}{\hat  y_I}
-\frac{\big<\partial_i   \Lambda_{I'}  \big>}{\big<  \Lambda_{I'}\big>}{\hat  y_I}-
   \frac{\big<\partial_i  \Lambda_{H_I}  \big>}{\big<  \Lambda_{H_I}\big>}{\hat  y_I} \Bigg]\big<\bar \rho_1^i\big>  + \hat y_I \Bigg\}a^q\nn
  \eeqa
  \item The $D_p{}^q-$term
  \beqa
{D_p{}^q}&=& \frac1{\sqrt{\big< \Lambda_{H_U}\Lambda_{H_D} \big>}}\frac{m_\frac32}{|M|^2}\Bigg[ -\bigg(\frac{\big<\partial_i   \Lambda_{H_U}\big>}{  \Lambda_{H_U}}+\frac{\big<\partial_i   \Lambda_{H_D}\big>}{ \Lambda_{H_D}}\bigg)
 (Z-\rho_\is \partial^\is Z)\bar \rho_1^i\nn\\
&&\hskip1.truecm  -2\rho_{1\is} \partial^\is Z + \bar \rho_1^i \partial_i Z - \rho_{1 \is} \bar \rho_1^i \partial_i \partial^\is Z +
\rho_{1\is} \hat K^\is{}_i \bar \rho^iZ  \Bigg] a^\ast_p a^q  \nn
  \eeqa
  \item The $D'^q-$term
  \beqa
D'^q&=&\frac{1}{M^\dag}\Bigg[\hat \lambda +
\Bigg(\partial_i \hat K    \hat \lambda + \partial_i    \hat \lambda
-\Big(\frac{\partial_i   \Lambda_{H_D}}{  \Lambda_{H_D}} + \frac{\partial_i   \Lambda_{H_U}}{  \Lambda_{H_U}}\Big)   \hat\lambda\Bigg)\bar \rho_1^i \Bigg]a^q\nn
\eeqa
  \item The $E-$term
\beqa
   \label{eq:f-b}
{E^q}&=&\frac{1}{M^\dag}\Bigg\{
\frac{ \rho_{\is} \hat K^\is{}_i \bar \rho_1^i \big< Z\big>}{{\sqrt{  \Lambda_{H_D}\Lambda_{H_U}}}} -2\frac{\big<\rho_\is \partial^\is { Z}\big>}{\sqrt{  \Lambda_{H_D}\Lambda_{H_U}}}  +
\frac{\big<\bar \rho_1^i \partial_i { Z}\big>}{{\sqrt{  \Lambda_{H_D}\Lambda_{H_U}}}} - \frac{\big< \rho_\is \rho_1^i \partial^\is \partial_i { Z}\big>}
{\sqrt{  \Lambda_{H_D}\Lambda_{H_U}}} \nn\\
&&-\frac1{\sqrt{\big< \Lambda_{H_U}\Lambda_{H_D}\big>}}\bigg(\frac{\big<\partial_i   \Lambda_{H_U}\big>}{\big<   \Lambda_{H_U}\big>}+ \frac{\big<\partial_i   \Lambda_{H_D}\big>}{\big<   \Lambda_{H_D}\big>}\bigg)\bar \rho_1^i
\big( Z -\rho_{\is}\partial^\is Z \big)\Bigg\}a^q
\nn . 
\eeqa
   \item The ${\cal U}^3$ term
\beqa
D_\kappa&=&\Big( \partial_i\hat K\hat\lambda + \partial_i\hat\lambda \Big) \hat\bar\rho_1^i-2\hat\lambda\nn
\eeqa
\end{enumerate}

\chapter{Vanishing of the cosmological constant, minimisation equations and mass matrice of the S2MSSM}\label{app:S2MSSM}
In this appendix we give the general form of the mass matrix for the general potential. We recall that the superpotential and the Kähler potential take the form
\beqa
W(z, S, \widetilde{\Phi}) &=& 
m_{p\ell}\left(W_{1,0}(z) + 
 S^p W_{1, p} (z)\right)
+ \   W_0(z, \widetilde{\Phi},\mathcal{U}^{12}) +  S^p W_{0, p} (z) \ ,  \nn\\
K &=& m_p^2zz^\dag + S_p^\dag S^p + \Phi_{a*}^\dag\Lambda^{a*}{}_a(z,z^{\dagger})\Phi^a\ . \nn
\eeqa
with only one field $z$ from the hidden sector. We use the definitions introduced in Section \ref{sec:S2MSSM}. The potential at the minimum is easily obtained from \autoref{eq:potentialS2MSSM}:
{\footnotesize
\beqa
\big\lag V \big \rag
&=& m_p^2\big(\lb m'_{3/2}\lb^2 -3 \lb m_{3/2}\lb^2 \big)+ \eK\Big(\sum \limits_p
\lb {\cal I}_p +M_4 ^3\lag \partial_p \omega_0\rag \lb^2 + M_4^4 \lag \partial_a \omega_0 \partial^{a^\ast}\bar \omega_0 (\Lambda^{-1})^a{}_{a^\ast}\rag\Big)\nn\\
&&+\lb m_{3/2}\lb^2  M_4^2\vp{a} \vpb{a^\ast}   {\cal S}^{a^\ast}{}_a 
  + \lb m_{3/2}\lb^2 \vS{p} \vSb{p}T
 \nn\\
&& +\frac 1{m_p^2} \eKs \Big( M_4^2\vpb{a^\ast} \vp{a} \lag \Lambda^{a^\ast}{}_a \rag +\vSb{p} \vS{p}\Big) \Bigg\{
\Big(\bar m_{3/2} \vS{q} {\cal I}_q 
+ \mbox{h.c.} \Big) \nn\\
&& +\eKs \Big(\sum \limits_ r \lb {\cal I}_r\lb^2 + M_4^3\bar{\cal I}^r \lag\partial_r \omega_0\rag +M_4^3 {{\cal I}}_r \lag\partial^r \bar \omega_0\rag \Big) \Bigg\}\nn
\\
 &&+ \eKs\Big(\bar m_{3/2}  M_ 4^3 R^b{}_a \vp{a} \lag\partial_b \omega_0\rag + \bar m_{3/2}  \vS{p}
\big[{\cal I}_p +M_4^3 \lag\partial_p \omega_0\rag\big] 
+ \mbox{h.c.}
 \Big) 
\nn
\eeqa 
}

The derivative with respect to the matter field takes the form:
{\footnotesize
\beqa
\Big<\partial_a V \Big>&=&
M_4^3 \eKs\Big(\bar m'_{3/2} \lag \partial_a \mathfrak{d}_{z} \omega_0 \rag + \bar m_{3/2} \big[-3 \lag \partial_a \omega_0\rag+ R^b{}_a \lag \partial_b \omega_0\rag + R^b{}_c \lag \phi^c \partial_b\partial_a \omega_ 0 \rag + \lag S^p \partial_a \partial_p \omega_0\rag \Big)\nn\\
&&+ \lb m_{3/2}\lb^2 M_4^2 \lag \phi_{a^\ast}\rag {\cal S}^{a^\ast}{}_a +\eK \Big(M_4^2 \lag \partial_a\partial_p \omega_0 \rag( \bar {\cal I}^p + M_4^3 \lag\partial^p\bar \omega_0)\rag +M_4^4 \lag \partial_a \partial_b\omega_0 \partial^{c^\ast} \bar \omega_0(\Lambda^{-1})^{b}{}_{c^\ast} \rag\Big)\nn\\
&& +\eKs \Big( \frac{M_4^3}{m_p^4} \lag \partial_a \omega_0\rag \lag S^\dag _q\rag \bar{\cal I}^q+\frac{M_4^3}{m_p^2}\lag \partial_a\partial_p\omega_0\rag\bar{\cal I}^p\Big) \Big(
M_4^2\lag \phi^b \phi^\dag_{b^\ast} \Lambda^{b^\ast}{}_b \rag+ \lag S^p S^\dag_p\rag\Big)
\nn\\
&&
\eKs \Bigg\{\frac{M_4^2}{m_p^2}\lag \phi^\dag_{b^\ast} \Lambda^{b^\ast}{}_a \rag\Big((\bar m_{3/2} \vS{q} {\cal I}_q
+ \mbox{h.c.} )+ \eKs \big[\sum \limits_r \lb {\cal I}_r\lb^2 + \big(M_4^3\lag \partial_p\omega_0\rag \bar{\cal I}^p + \mathrm{h.c.}\big)\big]\Big)\nn\\
&& \hskip 1.2truecm+\frac{M_4^3}{m_p^2}\eKs
\lag \partial_a \omega_0\rag \vSb{p} \big(\bar{\cal I}^p+M_4^3\lag \partial^p \bar \omega_0 \rag\Big)\Bigg\}
+\frac{M_4^6}{m_p^2} \eK \lag \partial_a \omega_0 \rag \vpb{b^\ast} \lag \partial^{b^\ast} \bar \omega_0\rag
\nn\\
&&-\frac{M_4^6}{m_p^2}\eK \lag \partial_a \mathfrak{d}_z \omega_0\rag \lag (\Lambda^{-1})^b{}_{a^*} \partial^{\bar z} \Lambda^{b^*}{}_b \rag \vpb{a^*} \lag \partial^{b^*}\bar \omega_0\rag
+ \eKs \frac{M_4^5}{m_p^2}\Big(-2 \bar m_{3/2}\lag \partial_a\omega_0 \Lambda^{b^*}{}_b\rag \nn\\
&& \hskip .5truecm+
\bar m'_{3/2} \lag \partial_a \mathfrak{d}_z\omega_0\rag \big[\lag \partial^{\bar z}\Lambda^{c^*}{}_b (\Lambda^{-1})^b{}_{b^*}\partial_z \Lambda^{b^*}{}_c\rag -\lag \partial_z\partial^{\bar z} \Lambda^{b^*}{}_b\rag + \lag \Lambda^{b^*}{}_b\rag\big]
\Big) \vpb{b^*}\vp{b}\nn\\
&& + \eKs \frac{M_4^3}{m_p^2}\Big(-2\bar m_{3/2}\lag \partial_a \omega_0 \rag+ \bar m'_{3/2} \lag \partial_a \mathfrak{d}_z \omega_0 \rag\Big) \vSb{p} \vS{p}\nn
\eeqa
}
while the derivative with respect to $S$ gives
{\footnotesize
\beqa
\Big<\partial_p V\Big>&=& \eKs \Big( \bar m'_{3/2}(\mathfrak{d}_z {\cal I}_p+M_4^3\lag \partial_p \mathfrak{d}_z \omega_0\rag)
+\bar m_{3/2}\big(-2 [{\cal I}_p + M_4^3 \lag \partial_p \omega_0\rag ]+ M_4^3 R^a{}_b \vp{b} \lag \partial_b \partial_p \omega_0\rag  \nn\\
&& + M_4^3\vS{q}\lag \partial_q\partial_p \omega_0\rag \big)\Big)+ \lb m_{3/2}\lb^2 \vSb{p}T + \eKs \frac{M_4^3}{m_p^2} ({\cal I}_p + M_4^3 \lag \partial_p \omega_0 \rag) \vpb{a^*} \lag \partial^{a^*}\bar \omega_0) \nn\\
&& 
+ \eK \Big(M_4^3\lag \partial_p \partial_q \omega_0\rag (\bar{\cal I}^q + M_4^3 \lag \partial^q \bar \omega_0\rag) +M_4^4 \lag \partial_p \partial_a \omega_0 \partial^{a^\ast}\bar \omega_0 (\Lambda^{-1})^{a}{}_{a^\ast}\rag\Big)\nn\\
&&  +
\frac 1 {m_p^2} \eKs  \Big( M_4^2 \lag \phi^\dag_{a^\ast} \phi^a   \Lambda^{a^\ast}{}_a\rag + \lag S^q S^\dag_q\rag \Big) \times \nn\\
&&\hskip 1.truecm \Big(\frac 1 {m_p^2} \eKs( {\cal I}_p+M_4^3\lag \partial_p \omega_0\rag) \vSb{q} \bar {\cal I}^q
+ \bar m_{3/2} {\cal I}_p
+\eKs M_4^3\lag\partial_p\partial_q\omega_0 \rag \bar{\cal I}^q\Big)
\nn\\
&& + \eK \frac 1{m_p^2}[{\cal I}_p + M_4^3 \lag \partial_p \omega_0 \rag]\vSb{q}(\bar{\cal I}^q+ M_4^3 \lag \partial^q \bar \omega_0\rag)\nn\\
&&- \frac {M_4^3}{m_p^2}  \eK[\mathfrak{d}_z {\cal I}_p + M_4^3 \lag \partial_p \omega_0 \rag] \lag (\Lambda^{-1})^b{}_{a^*} \partial^z \Lambda ^{b^*}{}_b \vpb{a^*} \lag \partial^{b^*}\bar \omega_0\rag] \nn\\
&& + \frac{1}{m_p^2}\eKs S_p^\dag \Big( \big(\bar{m}_{3/2} S^q\bar{\cal I}_q + \mathrm{h.c.}\big) + \eKs \sum_r \big|{\cal I}_r\big|^2 + \eKs \big( M_4^3\lag \partial_q\omega_0\rag \bar{\cal I}^q + \mathrm{h.c.} \big)\Big)\nn\\
&& + \frac 1 {m_p^2}\eKs \Big(-2 \bar m_{3/2} [{\cal I}_p + M_4^3 \lag \partial_p \omega_0\rag]\lag \Lambda^{a^*}{}_a\rag +\bar m'_{3/2}[\mathfrak{d}_z {\cal I}_p+ M_4^3 \lag \mathfrak{d}_z\partial_p \omega_0\rag]\times \nn\\
&& \hskip 2.3truecm  [\lag \partial^z \Lambda^{a^*}{}_b (\Lambda^{-1})^b{}_{b^*} \partial_z \Lambda^{b^*}{}_a\rag - \lag \partial^z \partial_z \Lambda^{a^*}{}_a \rag + \lag \Lambda^{a^*}{}_a \rag]\Big) \vpb{a^*}\vp{a}
\nn\\
&&+ \frac 1 {m_p^2}\eKs \Big(-2 \bar m_{3/2} [{\cal I}_p + M_4^3 \lag \partial_p \omega_0\rag] +\bar m'_{3/2}[\mathfrak{d}_z {\cal I}_p+ M_4^3 \lag \partial_p \omega_0\rag] \Big) \vSb{q}\vS{q}\ .\nn
\eeqa
}
The derivative with respect to the hidden sector is given by
{\footnotesize
\beqa 
\Big< \partial_z V \Big>&=&m_p^2 (m''_{3/2}\bar m'_{3/2}- 2 m'_{3/2}\bar m_{3/2})+
m'_{3/2}\bar m_{3/2}\Big(M_4^2 \vp{a} \vpb{a^\ast} {\cal S}^{a^\ast}{}_a+\vS{p} \vSb{p} T\Big)\nn\\
&&+ \eK \Big(\big({ \mathfrak{d}_z  \cal I}_p + M_4^3 \lag \partial_p \mathfrak{d}_z \omega_0\rag \big)(\bar {\cal I}^p +M_4^3 \lag \partial^p \bar \omega_0\rag)  \nn\\
&& + M_4^4  \lag \partial_a \omega_0(\partial_z\Lambda^{-1})^a{}_{a^\ast}\partial^{a^\ast}\bar \omega_0 \rag + M_4^4 \lag \partial_a \mathfrak{d}_z\omega_0(\Lambda^{-1})^a{}_{a^\ast}\partial^{a^\ast}\bar \omega_0 \rag\Big)\nn\\
&&+\lb m_{3/2}\lb^2\Big(M_4^2 \vp{a}\vpb{a^\ast} \partial_z {\cal S}^{a^\ast}{}_a + \vS{p}\vSb{p} \partial_z T \Big)\nn  \\
&&+\frac {M_4^2}{m_p^2} \eKs \vp{a} \vpb{a^\ast} \lag \partial_z \Lambda^{a^\ast}{}_a\rag \times \nn\\
&& \hskip 1.0truecm \Big((\bar m_{3/2} \vS{p}{\cal I}_p + \mbox{h.c.})
+\eKs \sum \limits_p \big(\lb {\cal I}_p \lb^2+ M_4^3 \bar{\cal I}^q \lag \partial_q \omega_0 \rag + M_4^3 {\cal I}_q \lag \partial^q \bar \omega_0\rag \big)\Big)\nn\\
&&+\frac 1{m_p^2} \eKs\Big(M_4^2 \vp{a} \vpb{a^\ast} \lag \Lambda^{a^\ast}{}_a \rag) + \vS{p} \vSb{p}\Big) \Big( m'_{3/2}\vSb{q} \bar{\cal I}^q +  \bar{m}_{3/2}\vS{q} \mathfrak{d}_z{\cal I}_q \nn\\
&& \hskip 1.5truecm + \eKs \bar{\cal I}^q\big( \mathfrak{d}_z{\cal I}_q + M_4^3\lag \partial_p\mathfrak{d}_z\omega_0\rag \big) + \eKs M_4^3\lag \partial^p \bar{\omega}_0\rag\mathfrak{d}_z{\cal I}_p\Big) \nn\\
&&+\eKs  \bar m_{3/2} \Big(M_4^3R^a{}_b \vp{b} \lag \partial_a \mathfrak{d}_z \omega_0\rag + M_4^3\partial_z R^a{}_b \vp{b} \lag \partial_a  \omega_0\rag +
\vS{p} \Big( \mathfrak{d}_z {\cal I}_p + M_4^3\lag \partial_p \mathfrak{d}_z \omega_0 \rag\big)\Big)\nn\\
&& + \eKs m'_{3/2} \Big(M_4^3 \bar R_{a^\ast}{}^{b^\ast}\vpb{b^\ast} \lag \partial^{a^\ast} \bar \omega_0\rag + \vSb{p}(\bar {\cal I}^p +
M_4^3 \lag \partial^q \bar \omega_0\rag\Big)\nn\\
&&  + \eKs M_4^3m_{3/2} \partial_z \bar R_{a^*}{}^{b^*} \phi^\dag_{b^*} \lag\partial^{a^*} \bar \omega_0\rag \ . \nn
\eeqa
}

\begin{allowdisplaybreaks}
The mass matrix elements take the following form:

$ \bullet: z^\dag-z$ sector
{\footnotesize
\beqa
\frac 1{m_p^2}\langle \partial_z\partial^z V \rangle &=& \big( |m''_{3/2}|^2 -2|m_{3/2}|^2 \big)+ \frac 1{m_p^2}\Big( |m'_{3/2}|^2 + |m_{3/2}|^2 \Big)\Big( M_4^2\langle \phi^{\dagger} S \phi \rangle + |\lag S\rag |^2T \Big) \nn\\
&& + \eK \Bigg\{ \frac{M_4^4}{m_p^2}\lag \mathfrak{d}_z\partial_a\omega_0\rag \Big( \lag \partial^{a^\ast}\bar{\omega}_0\partial^z(\Lambda^{-1})^a{}_{a^\ast}\rag +\lag \mathfrak{d}^z\partial^{a^\ast}\bar{\omega}_0(\Lambda^{-1})^a{}_{a^{\ast}}\rag \Big) \nn\\
&& \hskip 1.5truecm  + \frac{M_4^4}{m_p^2}
\lag \partial_a\omega_0\partial^{a^\ast}\bar{\omega}_0(\Lambda^{-1})^a{}_{a^\ast}\rag \Bigg\} \nn \\
&& + \eK \Bigg\{ \frac{M_4^4}{m_p^2}\lag \partial_a\omega_0\mathfrak{d}^z\partial^{a^\ast}\bar{\omega}_0\partial_z(\Lambda^{-1})^a{}_{a^\ast}\rag +
\frac{M_4^4}{m_p^2}\lag \partial_a\omega_0\partial^{a^\ast}\bar{\omega}_0\partial^z\partial_z(\Lambda^{-1})^a{}_{a^\ast}\rag \Bigg\} \nn\\
&& + \frac 1 {m_p^2}\eKs\Bigg\{ |{\cal I}_p + M_4^3\lag \partial_p\omega_0\rag |^2 + |\mathfrak{d}_z{\cal I}_p + M_4^3\lag \mathfrak{d}_z\partial_p\omega_0\rag |^2 \Bigg\} \nn\\
&& + \frac 1{m_p^2}\Big(m'_{3/2}\bar{m}_{3/2}\big( M_4^2 \langle \phi^{\dagger}\partial^zS\phi\rangle + |\lag S\rag|^2\partial^zT  \big) + \mbox{h.c.}\Big) \nn\\
&& + \frac 1{m_p^2}|m_{3/2}|^2 \Big( M_4^2 \lag \phi^{\dagger}\partial^z\partial_zS\phi\rag + |\lag S\rag|^2\partial^z\partial_zT \Big)\nn\\
&& + \frac{M_4^2}{m_p^4}\eKs \langle \phi^{\dagger} \partial^z\partial_z\Lambda\phi\rangle \Big( \eK |{\cal I}_p|^2 + \big( \bar{m}_{3/2}{\cal I}_p\lag S^p\rag  + M_4^3\eKs\lag \partial_p\omega_0\rag \bar{\cal I}^p + \mbox{h.c.} \big) \Big)\nn\\
&& + \frac{M_4^2}{m_p^4}\eKs \Bigg\{\lag  \phi^{\dagger}\partial_z\Lambda\phi\rag \Big( \mathfrak{d}^z\bar{\cal I}^p{\cal I}_p\eKs + \bar{m}'_{3/2}{\cal I}_p\lag S^p\rag + m_{3/2}\mathfrak{d}^z\bar{\cal I}^p\lag S^{\dagger}_p \rag\nn\\
&&+ M_4^3\eKs\big( \mathfrak{d}^z\bar{\cal I}^p\lag\partial_p\omega_0 \rag + {\cal I}_p\lag \mathfrak{d}^z\partial^p\bar{\omega}_0\rag \big) \Big)  \nn \\
&& + \eKs \langle S^p\rangle\Big( \bar{m}'_{3/2}\big(\mathfrak{d}_z{\cal I}_p + M_4^3\lag\mathfrak{d}_z\partial_p\omega_0\rag\big) + \bar{m}_{3/2} \big( {\cal I}_p + M_4^3\lag \partial_p\omega_0 \rag\big)\Big) \nn\\
&& +  M_4^3 \eKs \bar{m}'_{3/2}\langle \phi^b\rangle \big( \lag\mathfrak{d}_z\partial_a\omega_0\rag R^a{}_b + \lag\partial_a\omega_0 \rag\partial_z R^a{}_b \big) \nn\\
&& + M_4^3\eKs\bar{m}_{3/2}\langle \phi^b\rangle \big( \lag \partial_a\omega_0\rag  R^a{}_b + \lag\mathfrak{d}_z\partial_a\omega_0\rag\partial^z R^a{}_b + \lag \partial_a\omega_0 \rag\partial^z\partial_zR^a{}_b \big) +\mbox{h.c.}\Bigg\}\nn\\
&& + \frac{1}{m_p^4}\eKs \big( M_4^2 \lag \phi^{\dagger}\Lambda\phi \rag + |\lag S\rag|^2 \big)\Bigg\{ \eKs\big( |{\cal I}_p|^2+ |\mathfrak{d}_z{\cal I}_p|^2 \big) \nn\\
&& + \Big(  {\cal I}_p\lag S^p\rag \bar{m}_{3/2} + \mathfrak{d}_z{\cal I}_p\lag S^p\rag \bar{m}'_{3/2} + M_4^3\eKs\big(
\lag \partial_p\omega_0\rag \bar{\cal I}^p + \lag \mathfrak{d}_z\partial_p\omega_0\rag \mathfrak{d}^z\bar{\cal I}^p \big) +\mbox{h.c.}\Big)\Bigg\}\nn
\eeqa
}

$\bullet: z^\dag-\phi$ sector
{\footnotesize
\beqa
\frac 1{ m_p M_4}\langle \partial^z\partial_b V\rangle &=& \eKs \frac{M_4^2} {m_p}\Big( \bar{m}''_{3/2}\lag\mathfrak{d}_z\partial_b\omega_0\rag  -2\bar{m}'_{3/2}\lag\partial_b\omega_0\rag \Big)+ \bar{m}'_{3/2}m_{3/2}\frac{M_4}{m_p}\langle \phi^{\dagger}_{a^\ast}\rangle S^{a^\ast}{}_b \nn\\
&& + \eK\Bigg\{ \frac{M_4^3}{m_p}\lag \partial_a\partial_b\omega_0\rag \Big(\lag  \mathfrak{d}^z\partial^{a^\ast}\bar{\omega}_0(\Lambda^{-1})^a{}_{a^\ast}\rag +
\lag \partial^{a^\ast}\bar{\omega}_0\partial^z (\Lambda^{-1})^a{}_{a^\ast}\rag\Big) \nn\\
&&+ \frac{M_4^2}{m_p}\lag\partial_b\partial_p\omega_0\rag\Big( M_4^ 3\lag\mathfrak{d}^z\partial^p\bar{\omega}_0\rag +
\mathfrak{d}^z\bar{\cal I}^p\Big)\Bigg\} 
 + |m_{3/2}|^2\frac{M_4}{m_p}\langle \phi^{\dagger}_{a^\ast}\rangle \partial^z S^{a^\ast}{}_b   \nn\\
&& + \frac{M_4^2}{m_p}\eKs \bar{m}_{3/2} \partial^zR^a{}_c\Big( \langle \phi^c\partial_b\partial_a\omega_0\rangle + \langle \partial_a\omega_0\rangle \delta^c{}_b \Big) \nn\\
&& + \frac{M_4^2}{m_p}\eKs \bar{m}'_{3/2}\Big( \langle S^p\partial_p\partial_b\omega_0 \rangle + \langle \partial_a\omega_0 \rangle R^a{}_b + \langle \phi^c\partial_b\partial_a\omega_0\rangle R^a{}_c \Big) \nn \\
 && + \eK \frac{M_4^4}{m_p^3}\lag\phi^{\dagger}\partial^z\Lambda\phi\rag\Big(  \frac{1}{m_p^2}\bar{\cal I}^p\lag S_p^\dagger\rag\lag \partial_b\omega_0 \rag
 +\lag \partial_b\partial_p\omega_0\rag \bar{\cal I}^p\Big)\nn\\
 &&+ \eK \frac{M_4^2}{m_p^3}\big( M_4^2\lag\phi^{\dagger}\Lambda\phi\rag + |\lag S\rag|^2 \big)\Big( \frac{1}{m_p^2}\mathfrak{d}^z\bar{\cal I}^p\lag S^{\dagger}_p\partial_b\omega_0\rag +
 \lag \partial_b\partial_p\omega_0\rag\mathfrak{d}^z\bar{\cal I}^p  \Big)\nn\\
&& + \eKs \Bigg\{ \frac{M_4}{m_p^3}\lag (\phi^{\dagger} \partial^z\Lambda)_b\rag\Big( \eKs\big( |{\cal I}_p|^2 + \{M_4^3{\cal I}_p\lag    \partial^p \bar{\omega}_0\rag + \mbox{h.c.}\}\big) + \big(m_{3/2}\langle S_p^{\dagger}\rangle \bar{\cal I}^p + \mbox{h.c.} \big)\Big) \nn\\
 && + \frac{M_4}{m_p^3}\lag (\phi^{\dagger}\Lambda)_b\rag  \eKs \Big( {\cal I}_p\mathfrak{d}^z\bar{\cal I}^p + M_4^3\big( {\cal I}_p\lag \mathfrak{d}^z\partial^p\bar{\omega}_0\rag  + \mathfrak{d}^z
   \bar{\cal I}^p\lag \partial_p\omega_0\rag \big)\Big)   \nn\\
&& + {\cal I}_p\lag S^p\rag \bar{m}'_{3/2} + \mathfrak{d}^z\bar{\cal I}^p\lag S^{\dagger}_p\rag m_{3/2}  + \frac{M_4^2}{m_p^3}\eKs \lag S^{\dagger}_p\partial_b\omega_0\rag\big( \mathfrak{d}^z\bar{\cal I}^p + M_4^3\lag \mathfrak{d}^z\partial^p\bar{\omega}_0 \rag\big)\Bigg\}\nn\\
&&- \frac{M_4^5}{m_p^3}\eKs \Big( \lag \partial_b \mathfrak{d}_z \omega_0\rag \Big[
  \lag (\Lambda^{-1})^b{}_{a^*} \partial^z \Lambda^{b^*}{}_b \rag \vpb{a^*} \lag \partial^{b^*} \mathfrak{d}^z \bar \omega_0\rag \nn\\
&& \hskip 2.3truecm+
   \lag \partial^z[(\Lambda^{-1})^b{}_{a^*} \partial^z \Lambda^{b^*}{}_b] \rag \vpb{a^*} \lag \partial^{b^*} \bar \omega_0\rag \Big]\nn\\
&&+ \frac{M_4^5}{m_p^3}\eKs \Bigg[\lag \partial_b \omega_0\rag\Big(-\lag(\Lambda^{-1})^b{}_{a^*} \partial^z \Lambda^{b^*}{}_b\rag +\delta^{a^*}_{b^*}\Big)\vpb{a^*}\lag\partial^{b^*}\mathfrak{d}^z\bar \omega_0\rag\nn\\
&& \hskip 2.truecm
+\lag \partial_ a\omega_0\rag\lag(\Lambda^{-1})^b{}_{a^*} \partial^z \Lambda^{b^*}{}_b\rag\lag \phi^\dag_{a^*} \partial^{b^*} \bar \omega_0\rag\Bigg]
\nn\\
&&+ \frac{M_4^2}{m_p^3}\eKs\Big( \big[\bar m_{3/2}' \Big(-2\Lambda^{b^\ast}{}_b + \tilde{\cal S}^{b^*}{}_b \Big) -2 \bar m_{3/2}\lag  \partial^z\Lambda^{b^*}{}_b\rag\big]\lag \partial_b \omega_0 \rag
\nn\\
&&\hskip 2.3truecm+\big[\bar m_{3/2}'' \tilde {\cal S}^{b^*}{}_b + m_{3/2}'\partial^z \tilde{\cal S}^{b^*}{}_b\big]\lag \partial_a \mathfrak{d}_z \omega_0\rag\Big) \nn\\
&&+ \frac{M_4^2}{m_p^3}\eKs\Big(- \bar m_{3/2}' \lag \partial_b \omega_0\rag  +\bar m_{3/2}'' \lag \partial_b\mathfrak{d}_z \omega_0\rag \Big) \lag S^\dag_p S^p \rag\nn
\eeqa
}

$\bullet: z^\dag-S$ sector
{\footnotesize
\beqa
\frac 1 {m_p}\langle \partial^z\partial_q V \rangle &=& \frac 1 {m_p}\eKs \bar{m}''_{3/2} \big( M_4^3\lag\mathfrak{d}_z\partial_q\omega_0\rag + \mathfrak{d}_z{\cal I}_q \big)
- \frac 1 {m_p}\bar{m}'_{3/2} \eKs \big( M_4^3\lag \partial_q\omega_0\rag + {\cal I}_q \big) \nn\\
&&   + \frac{M_4^3}{m_p}\eKs \bar{m}'_{3/2}\Big(\lag \phi^b\partial_a\partial_q\omega_0\rag R^a{}_b + M_4^3\lag S^p\rangle \lag\partial_p\partial_q\omega_0 \rag \Big) + \frac{M_4^3}{m_p}\eKs \bar{m}_{3/2}
\lag \phi^b \partial_a\partial_q\omega_0\partial^z R^a{}_b\rag \nn\\
&& + \frac 1{m_p}\eK \Bigg\{ M_4^4\lag\partial_a\partial_q\omega_0\rag\Big(\lag \mathfrak{d}^z\partial^{a^\ast}\bar{\omega}_0\rag ( \lag\Lambda^{-1})^a{}_{a^\ast}\rag + \lag\partial^{a^\ast}\bar{\omega}_0\rag
\lag \partial^z(\Lambda^{-1})^a{}_{a^\ast}\rag \Big)\nn\\
&&\hskip 1.3truecm+ M_4^3\lag \partial_p\partial_q\omega_0\rag \big( \mathfrak{d}^z\bar{\cal I}^p
+ M_4^3\lag \mathfrak{d}^z\partial^p\bar{\omega}_0\rag \big) \Bigg\} 
+\Big(m_{3/2} \bar m'_{3/2} T + \big|m_{3/2}\big|^2 \partial^z T\Big)\vSb{q}\nn\\
&& + \frac{1}{m_p^3}\eKs\big( M_4^2\lag \phi^{\dagger}\Lambda\phi\rag + |\lag S\rag|^2 \big) \Big( {\cal I}_q\bar{m}'_{3/2} \nn\\
&&\hskip 2.3truecm +\eKs\big[ \frac{1}{m_p^2}( {\cal I}_q+m_4^3 \lag\partial_q   \mathfrak{d}_z \omega_0\rag) \mathfrak{d}^z\bar{\cal I}^p \lag S^{\dagger}_p\rangle + M_4^3\lag \partial_p\partial_q\omega_0\rag\mathfrak{d}^z\bar{\cal I}^p  \big]\Big)\nn\\
  && + \frac{M_4^2}{m_p^3}\eKs\lag \phi^{\dagger}\partial^z\Lambda\phi \rag\Big( {\cal I}_q\bar{m}_{3/2} +\eKs \big[ \frac{1}{m_p^2} ({\cal I}_q + M_4^3 \lag \partial_q \bar \omega_0\rag)\bar{\cal I}^p\lag S^{\dagger}_p\rag +
    M_4^3\lag\partial_p\partial_q\omega_0\rag \bar{\cal I}^p \big]\Big) \nn\\
  && + \frac{1}{m_p^3}\eKs \lag S_q^\dag\rag\Big(  \bar m'_{3/2}\vS{p} {\cal I}_p + m_{3/2} \vSb{p}  \mathfrak{d}^z \bar {\cal I}^p \nn\\
&& \hskip 2.3truecm+ \eKs\big[
    {\cal I}_p \mathfrak{d}^z \bar {\cal I}^p+ M_4^3 \mathfrak{d}^z {\cal I}^p\lag \partial_p \omega_0\rag + M_4^3 {\cal I}_p \lag \mathfrak{d}\partial^p \bar \omega_0\rag\big]\Big)
  \nn\\
  &&+ \frac 1{m_p^3}\eK ({\cal I}_p + M_4^2 \lag\partial_p\bar \omega_0\rag)\vSb{p}(\mathfrak{d}^z\bar {\cal I}^p +M_4^3 \lag \partial^p  \mathfrak{d}^z \bar\omega_0\rag)
 \nn\\
&& \hskip 2.3truecm +\frac {M_4^3} {m_p^2}\eK ({\cal I}_q + M_4^2 \lag\partial_q \omega_0\rag)\vpb{a^*} \lag \mathfrak{d}\partial^{a^*} \bar \omega_0\rag\nn\\
  && -\frac{M_4^3}{m_p^3}\Bigg\{\lag (\Lambda^{-1})^b{}_{a^*} \partial^z \Lambda^{b^*}{}_b\rag(\mathfrak{d}_z {\cal I}_q + M_4^3\lag \partial_q\mathfrak{d}_z \omega_0\rag)\vpb{a^*} \lag \partial^{b^*} \mathfrak{d}^z\bar \omega_0\rag \nn\\
  &&
  +  \Big(\lag(\Lambda^{-1})^b{}_{a^*} \partial^z \Lambda^{b^*}{}_b\rag  ({\cal I}_q + M_4^3\lag \partial_q\omega_0\rag)
  \nn\\
&& \hskip 2.5truecm+ \lag\partial^z\Big((\Lambda^{-1})^b{}_{a^*}  \partial^z \Lambda^{b^*}{}_b\Big)\rag ) (\mathfrak{d} {\cal I}_q + M_4^3\lag \partial_q\mathfrak d\omega_0\rag\Big)\vpb{a^*}\lag \partial^{b^*} \bar \omega_0\rag\Big)\nn\\
  && + \frac 1{m_p^3} \Bigg\{  \bar m'_{3/2} ({\cal I}_q + M_4^3 \lag \partial_q \omega_0\rag) {\cal S}^{a^*}{}_a -2\bar m_{3/2} ({\cal I}_q + M_4^3 \lag \partial_q \omega_0\rag) \lag \partial^z\Lambda^{a^*}{}_a\rag \nn\\
  && + \bar m''_{3/2} (\mathfrak{d} {\cal I}_q + M_4^3\lag \partial_q\mathfrak {d}\omega_0\rag) \tilde{\cal S}^{a^*}{}_a \nn\\
&& \hskip 2.3truecm+ \bar m'_{3/2} ({\cal I}_q + M_4^3\lag \partial_q\omega_0\rag\Big) \tilde{\cal S}^{a^*}{}_a +
  \bar m'_{3/2} ({\cal I}_q + M_4^3\lag \partial_q\omega_0\rag) \partial^z\tilde{\cal S}^{a^*}{}_a\Big) \vpb{a^*}\vp{a}\Bigg\}\nn\\
  && + \frac 1{m_p^3} \Bigg\{ -\bar m'_{3/2} ({\cal I}_q + M_4^3 \lag \partial_q \omega_0\rag) +
   \bar m''_{3/2} (\mathfrak{d} {\cal I}_q + M_4^3\lag \partial_q\mathfrak {d}\omega_0\rag) \Bigg\}\lag S^\dag _p S^p\rag \nn
\eeqa
}

$\bullet: S-S^\dag$ sector
{\footnotesize
\beqa
\big<\partial^q \partial_p V \big>&=&
\eK \Big[ M_4^6 \lag \partial_p \partial_r \omega_0 \partial^q \partial^r \bar \omega_0\rag +
  M_4^4 \lag \partial_p \partial_a \omega_0 \partial^q \partial^{a^*} \bar \omega_0 (\Lambda^{-1})^{a^*}{}_a\rag\Big] + \delta^q_p  |m_{3/2}|^2 T  \nn\\
&&+\frac 1 {m_p^2}\eK \Bigg\{ \big( \mathfrak{d}_z{\cal I}_p + M_4^3 \lag \partial_p \mathfrak{d}_z \omega_0\rag\big)
\big( \mathfrak{d}^z\bar {\cal I}^q + M_4^3 \lag \partial^q \mathfrak{d}^z \bar \omega_0\rag\big)
-\big( {\cal I}_p + M_4^3 \lag \partial_p \omega_0\rag\big)
\big( \bar {\cal I}^q + M_4^3 \lag \partial^q \bar \omega_0\rag\big)\nn\\
&&\hskip 1.5truecm + M_4^3 \Big( \lag \phi^a \partial_a \partial_p \omega_0 \rag (\bar {\cal I}^q + M_4^3 \lag \partial^q \bar \omega_0\rag)
+ \lag \phi^\dag_{a^*}\partial^{a^*} \partial^q \bar \omega_0 \rag ( {\cal I}_p + M_4^3\lag \partial_p  \omega_0\rag)\Big)\nn\\
&&\hskip 1.5truecm - M_4^3  \lag \phi^a \partial_b \partial_p \omega_0 \rag \lag (\Lambda^{-1})^b{}_{a^*} \partial_z \Lambda^{a^*}{}_a (\mathfrak{d}^z\bar {\cal I}^q +M_4^3 \lag \partial^q \mathfrak{d}^z\bar \omega_0\rag)\nn\\
&& \hskip 1.5truecm - M_4^3  \lag \phi^\dag_{a^*} \partial^{b^*} \partial^q \bar \omega_0 \rag \lag (\Lambda^{-1})^b{}_{b^*} \partial^z \Lambda^{a^*}{}_b (\mathfrak{d}_z\bar {\cal I}_p +M_4^3 \lag \partial_p \mathfrak{d}_z \omega_0\rag)\Bigg\}\nn\\
&&+ \frac 1 {m_p^4} \eK\Bigg\{
\Big[M_4^2 \lag \phi^\dag \Lambda \phi\rag + \lag S^\dag S\rag\Big]
\Big[{\cal I}_p(\bar {\cal I}^q + M_4^3 \lag \partial^q \bar \omega_0\rag) +\bar {\cal I}^q( {\cal I}_p + M_4^3 \lag \partial_p  \omega_0\rag)\Big]\nn\\
&& \hskip1.6truecm -2\Big[M_4^2 \lag \phi^\dag \Lambda \phi\rag + \lag S^\dag S\rag\Big]
({\cal I}_p+ M_4^3 \lag \partial_p \omega_0\rag)(\bar {\cal I}^q + M_4^3 \lag \partial^q \bar \omega_0\rag) \nn\\
&& \hskip1.8truecm +\Big[M_4^2 \vpb{} \tilde{\cal S}\vp{} + \lag S^\dag S\rag\Big]
(\mathfrak{d}_z{\cal I}_p+ M_4^3 \lag \partial_p \mathfrak{d}_z\omega_0\rag)(\mathfrak{d}^z\bar {\cal I}^q + M_4^3 \lag \partial^q \mathfrak{d}^z\bar \omega_0\rag) \Bigg\}\nn\\
&&+  \frac {\delta^q_p}{ m_p^2}\eKs \Bigg\{ \Big[ \bar m_{3/2} \vS{r} {\cal I}_r + m_{3/2} \vSb{r} \bar {\cal I}^r\Big]
+\eKs\Big[ \sum |{\cal I}_r|^2 + M_4^3 (\bar{\cal I}^r\lag \partial_r \omega_0\rag + {\cal I}_r\lag \partial^r \bar\omega_0\rag)\Big]\Bigg\}\nn\\
&&+\frac 1{m_p^2} \eKs \vS{q} \Bigg\{-\bar m_{3/2}({\cal I}_p + 2 M_4^3 \lag \partial_p \omega_0\rag) + \bar m_{3/2}'( \mathfrak{d}_z {\cal I}_p + M_4^3\lag \partial_p \mathfrak{d}_z \omega_0\rag)\nn\\
&&\hskip 1.5truecm +\eKs \Big[ \frac 1{m_p^2}({\cal I}_p + M_4^3 \lag \partial_p \omega_0\rag)\vSb{r} \bar {\cal I}^r + M_4^3 \bar {\cal I}^r\partial_r\lag \partial_p \omega_0\rag \Big]\Bigg\}\nn\\
&&+\frac 1{m_p^2} \eKs \vSb{p} \Bigg\{- m_{3/2}(\bar {\cal I}^q + 2 M_4^3 \lag \partial^q \bar \omega_0\rag) +  m_{3/2}'( \mathfrak{d}^z \bar {\cal I}^q + M_4^3\lag \partial^q \mathfrak{d}^z \bar \omega_0\rag)\nn\\
&&\hskip 1.5truecm +\eKs \Big[ \frac 1{m_p^2}(\bar {\cal I}^q + M_4^3 \lag \partial^q \bar \omega_0\rag)\vS{r}  {\cal I}_r + M_4^3  {\cal I}_r\lag\partial^r \partial^q \bar \omega_0\rag \Big]\Bigg\}\nn\\
&&+\frac{M_4^3}{m_p^2} \eK\Big[ \lag S^r \partial_r \partial_p \omega_0\rag (\bar {\cal I}^q + M_4^3 \lag \partial^q\bar \omega_0\rag) +
  \lag S^\dag_r \partial^r \partial^q \bar \omega_0\rag ( {\cal I}_p + M_4^3 \lag \partial_p \omega_0\rag)\Big] \nn 
\eeqa
}

$\bullet: \phi^\dag-\phi$ sector
{\footnotesize
\beqa
\frac 1 {M_4^2} \big<\partial^{a^*} \partial _a V \big> &=& M_4^2 \big|m_{3/2}\big|^2 {\cal S}^{a^*}{}_a +
\eK\Big(M_4^2 \lag \partial_a \partial_b \omega_0 \partial^{a^*} \partial^{b^*} \bar \omega_0(\Lambda^{-1})^b{}_{b^*}\rag +
M_4^4 \lag \partial_a \partial_p \omega_0 \partial^{a^*} \partial^p \bar \omega_0\rag \Big)\nn\\
&&+\frac {M_4^4}{m_p^2} \eK \Bigg\{
\lag \partial_a \mathfrak{d}_z \omega_0\partial^{a^*}\mathfrak{d}^z\bar \omega_0\rag
-\lag \partial_a  \omega_0\partial^{a^*}\bar \omega_0\rag -\lag \partial_b \omega_0 \partial^{a^*} \mathfrak{d}^z \bar \omega_0 (\Lambda^{-1})^b{}_{b^*}\partial_z\Lambda^{b^*}{}_a\rag
\nn\\
&&\hskip 1.7truecm 
- \partial_a \mathfrak{d}_z\omega_0 \partial^{b^*}  \bar \omega_0  (\Lambda^{-1})^b{}_{b^*}\partial^z\Lambda^{a^*}{}_b\rag- \lag \phi^\dag_{c^*}\partial_a \mathfrak{d}_z\omega_0\partial^{a^*} \partial^{b^*}  \bar \omega_0 (\Lambda^{-1})^b{}_{b^*}\partial^z\Lambda^{c^*}{}_b \rag\nn\\
&&\hskip 1.7truecm -\lag \phi^c \partial_a\partial_b \omega_0 \partial^{a^*} \mathfrak{d}^z \bar \omega_0 (\Lambda^{-1})^b{}_{b^*}\partial_z\Lambda^{b^*}{}_c\rag +\lag S^p \partial_a \partial_p \omega_0 \partial^{a^*} \bar \omega_0\rag
\nn\\
&& \hskip 1.7truecm 
+\lag S^\dag_p \partial_a  \omega_0 \partial^{a^*}\partial^p \bar \omega_0\rag
+ \lag \phi^b \partial_a \partial_b \omega_0 \partial^{a^*} \bar \omega_0 \rag +
\lag \phi^\dag_{b^*} \partial_a  \omega_0 \partial^{b^*}\partial^{a^*} \bar \omega_0 \rag\Bigg\}\nn\\
&&+ \frac{M_4^4}{m_p^2} \eK \Big(
-2\lag \partial_a \omega_0  \partial^{a^*}\bar \omega_0\rag\big[\lag \phi\Lambda \phi\rag + \lag S^\dag S\rag\big]
+\lag \partial_a \mathfrak{d}_z\omega_0  \partial^{a^*}\mathfrak{d}^z\bar \omega_0\rag\big[\lag \phi \tilde{\cal S} \phi\rag + \lag S^\dag S\rag\big]\Big)\nn\\
&& + \frac{M_4^2}{m_p^2} \eKs\lag \Lambda^{a^*}{}_a\rag
\Bigg\{ m_{3/2} \vS{p} {\cal I}_ p + \bar m_{3/2} \vSb{p} \bar {\cal I}^p\nn\\
&&\hskip 2.9truecm +\eKs \Big[ \sum \big|{\cal I}_p\big| ^2 + M_4^3\bar {\cal I}^p\lag \partial_p \omega_0 \rag+
  M_4^3 {\cal I}_p\lag \partial^p\bar \omega_0 \rag\Big]\Bigg\}\nn\\
&& + \frac{M_4^3}{m_p^2} \eK\Bigg\{ \vpb{b^*}\lag \Lambda^{b^*}{}_a\rag {\cal I}_p \Big[ \frac 1 {m_p^2} \lag\partial^{a^*} \bar \omega_0 \rag \vS{p} 
  + \lag \partial^p \partial^{a^*}\bar \omega_0 \rag\Big]\nn\\
 && \hskip 1.9truecm +
\vp{b}\lag \Lambda^{a^*}{}_b\rag \bar {\cal I}^p \Big[ \frac 1 {m_p^2} \lag\partial_a\ \omega_0 \rag \vSb{p} 
  + \lag \partial_p \partial_{a}\bar \omega_0 \rag\Big] \Bigg\}\nn\\
&& + \frac {M_4^3}{m_p^2} \eKs \Bigg\{ \vpb{b^*}\Big(-2 m_{3/2} \lag \partial^{a^*} \bar \omega_0 \Lambda^{b^*}{}_a\rag +
  m'_{3/2} \lag \partial^{a^*}\mathfrak{d}^z \bar \omega_0\rag \tilde{\cal S}^{b^*}{}_a\Big)\nn\\
&& \hskip 2.1truecm + \vp{b}\Big(-2 \bar m_{3/2} \lag \partial_a \omega_0 \Lambda^{a^*}{}_b\rag +
  \bar m'_{3/2} \lag \partial_a\mathfrak{d}_z \omega_0\rag \tilde{\cal S}^{a*}{}_b\Big)\Bigg\}\nn
\eeqa
}

$\bullet: S^\dag-\phi$ sector
{\footnotesize
 \beqa
  \frac 1 {M_4} \big<\partial^p \partial_a V\big> &=&
  \eK\Big[ M_4^4 \lag \partial_a\partial_q \omega_0 \partial^p\partial^q \bar \omega_0\rag +
    M_4^2 \lag \partial_a\partial_b \omega_0 \partial^p\partial^{b^*}  \bar \omega_0(\Lambda^{-1})^b{}_{b^*}\rag \Big]\nn\\
  &&+\frac{M_4^2}{m_p^2} \eK \Bigg\{
  (\mathfrak{d}^z \bar {\cal I}^p + M_4^3 \lag \partial^p\mathfrak{d}^z \bar \omega_0\rag) \lag \partial_a \mathfrak{d}_z \omega_0\rag
  - ( \bar {\cal I}^p + M_4^3 \lag \partial^p \bar \omega_0\rag) \lag \partial_a \omega_0\rag\nn\\
  &&\hskip1.8truecm - (\mathfrak{d}^z \bar {\cal I}^p + M_4^3 \lag \partial^p\mathfrak{d}^z \bar \omega_0\rag) \lag \partial_b \omega_0\rag \lag
  (\Lambda^{-1})^b{}_{a^*} \partial_z \Lambda^{a^*}{}_a\rag +
  ( \bar {\cal I}^p + M_4^3 \lag \partial^p \bar \omega_0\rag) \lag \phi^b \partial_b\partial_a \omega_0\rag\nn\\
  &&\hskip1.8truecm - (\mathfrak{d}^z \bar {\cal I}^p + M_4^3 \lag \partial^p\mathfrak{d}^z \bar \omega_0\rag) \lag\phi^c \partial_b\partial_a\omega_0\rag  (\Lambda^{-1})^b{}_{a^*} \partial_z \Lambda^{a^*}{}_c\rag \nn\\
  &&\hskip1.8truecm  -M_4^3 \lag\partial_q \mathfrak{d}_z \omega_0 \phi^\dag_{b^*} \partial^{a^*} \partial^p\bar \omega_0\rag
  (\Lambda^{-1})^b{}_{a^*} \partial_z \Lambda^{a^*}{}_b\rag+
  M_4^3 \lag \phi^\dag_{a^*} \partial_a \omega_0 \partial^{a^*} \partial^p\bar \omega_0\rag\Bigg\}\nn\\
  && +\frac{M_4}{m_p^2} \eKs \vpb{a^*}\Big[
    -m_{3/2}( \bar {\cal I}^p + 2M_4^3 \lag \partial^p \bar \omega_0\rag)
    \lag\Lambda^{a^*}{}_{a}\rag
+  m'_{3/2}
 (\mathfrak{d}^z \bar {\cal I}^p + M_4^3 \lag \partial^p\mathfrak{d}^z \bar \omega_0\rag) \tilde{\cal S}^{a^*}{}_a\Big]\nn\\
  && + \frac{M_4}{m_p^2}\eK \lag (\phi^\dag \Lambda_a\rag \Big[   M_4^3 {\cal I}_q \lag \partial^p \partial^q \bar \omega_0\rag +
    \frac 1{m_p^2} ( \bar {\cal I}^p + M_4^3 \lag \partial^p \bar \omega_0\rag)\vS{q}{\cal I}_q\Big]\nn\\
  && + \frac{M_4^2}{m_p^4} \eK\Bigg\{ -\Big( M_4 ^2 \lag \phi^\dag \Lambda \phi\rag + \lag S^\dag S\rag\Big) \Big(
  \bar {\cal I}^p + 2M_4^3 \lag \partial^p \bar \omega_0\rag\Big) \lag \partial_a \omega_0\rag\nn\\
  &&\hskip2.3truecm + \Big( M_4 ^2 \lag \phi^\dag \tilde S\phi\rag + \lag S^\dag S\rag\Big)
  \Big(\mathfrak{d}^z\bar {\cal I}^p + M_4^3 \lag \partial^p \mathfrak{d}^z\bar \omega_0\rag\Big) \lag \partial_a \mathfrak{d}_z
  \omega_0\rag\Bigg\}
  \nn
  \eeqa
 } 
  
  $\bullet: \phi-\phi$ sector
{\footnotesize 
   \beqa
 \frac 1{ M_4^2} \big <\partial_a \partial_b V\big>&=&\eK \Big[
  M_4 \lag \partial_a \partial_b \partial_p \omega_0\rag (\bar {\cal I}^p + M_4^3 \lag \partial^p \bar \omega_0\rag) +
  M_4^2 \lag \partial_a \partial_b \partial_c \omega_0 \partial^{c^*}\bar \omega_0 (\Lambda^{-1})^c{}_{c^*}\rag\Big]\nn\\
&& +
   \eKs M_4\Big[ \bar m'_{3/2} \lag \partial_a \partial_b \mathfrak{d}_z  \omega_0\rag  +   \bar m_{3/2}(-3 \lag \partial_a \partial_b   \omega_0\rag + R^c{}_a \lag\partial_b \partial_c \omega_0 \rag+  R^c{}_b \lag\partial_a \partial_c \omega_0\rag \nn\\
&&\hskip 2.1truecm + R^d{}_c \lag \phi^c \partial_a \partial_b\partial_d \omega_0\rag + \lag S^p \partial_a \partial_b\partial_p \omega_0\rag)\Big]\nn\\
  && +\frac {M_4^3}{m_p^2} \eKs \vpb{b^*}\Big[ -3 \bar m_{3/2}( \lag \partial_a \omega_0 \Lambda^{b^*}{}_b \rag +  \lag \partial_b \omega_0 \Lambda^{b^*}{}_a\rag)
    \nn\\
&& \hskip 3.2truecm + \bar m'_{3/2}( \lag \partial_a\mathfrak{d}_z \omega_0 \rag \tilde {\cal S}^{b^*}{}_b  +  \lag \partial_b \mathfrak{d}_z \omega_0\rag  \tilde {\cal S}^{b^*}{}_a)\Big]\nn\\
  && +\frac{M_4}{m_p^2} \eKs\Big[-2 \bar m_{3/2} \lag \partial_a \partial_b \omega_0\rag \big(M_ 4^2\lag \phi^\dag \Lambda \phi\rag + \lag S^\dag S\rag\big)
   \nn\\
&&\hskip 2.3truecm  + \bar m'_{3/2} \lag \partial_a \partial_b \mathfrak{d}_ z\omega_0\rag\big(M_4^2\lag \phi^\dag \tilde {\cal S} \phi\rag + \lag S^\dag S\rag\big)\Big] \nn\\
    && \frac{M_4}{m_p^2}\eK \Big[ \lag \partial_a\partial_b\partial_q\omega_0 \rag\bar{\cal I}^q + \frac{1}{m_p^2}\lag \partial_a\partial_b\omega_0\rag\lag S^\dag_q\rag\bar{\cal I}^q \Big]\Big( M_4^2\lag\phi^\dag \Lambda \phi\rag + \lag S^\dag S\rag\Big) \nn\\
  &&+ \frac{M_4^3}{m_p^2} \eK\Big[ \lag \phi^\dag_{a^*} \Lambda^{a^*}{}_a\rag \big({\cal I}^p \lag\partial_p \partial_b\omega_0\rag +\frac 1 {m_p^2} \lag \partial_b \omega_0\rag \vSb{p} \bar {\cal I}^p \big)\nn\\
&&\hskip 2.0truecm  +
    \lag \phi^\dag_{a^*} \Lambda^{a^*}{}_b\rag \big({\cal I}^P \lag\partial_p \partial_a\omega_0\rag +\frac 1 {m_p^2} \lag \partial_a \omega_0\rag \vSb{p} \bar {\cal I}^p \big)\Big]\nn\\
  &&+\frac{M_4}{m_p^2} \eK\Big[\lag \partial_a \partial_b \omega_0\rag \vSb{p}(\bar {\cal I}^p + M_4^3 \lag \partial^p \bar \omega_0\rag)\nn\\
&& \hskip 1.3truecm    +M_4^3\big(-\lag \partial_a \partial_b \mathfrak{d}_z \omega_0 (\Lambda^{-1})^c{}_{a^*} \partial^z \Lambda^{b^*}_c \phi^\dag_{b^*} \partial^{a^*} \bar \omega_0\rag+
    \lag\partial_a \partial_b \omega_0 \phi^\dag_{a^*} \partial^{a^*}\bar \omega_0\rag\big)\Big]\nn
  \eeqa
}
  
  $\bullet: S-S$ sector
{\footnotesize
     \beqa
   \big< \partial_p \partial_q V\big>&=&\eK \Big[M_4^3 \lag \partial_p \partial_q \partial_r \omega_0\rag(\bar {\cal I}^r + M_4^3 \lag \partial^r \bar \omega_0\rag) + M_4^4 \lag\partial_a \partial_p \partial_q\omega_0 \partial^{a^*} \bar \omega_0 (\Lambda^{-1})^a{}_{a^*}\rag\Big]\nn\\
&&   + M_4^3\eKs\big[ \bar m'_{3/2} \lag \partial_p\partial_q \mathfrak{d}_z\omega_0\rag
    + \bar m_{3/2}\big\{-   \lag \partial_p\partial_q \omega_0\rag +R^b{}_a \lag \phi^a \partial_b\partial_p\partial_q \omega_0\rag
    +\lag S^r \partial_p \partial_q\partial_r \omega_0\rag\big\}\Big]\nn\\
   &&+\frac 1{m_p^2} \eKs \vSb{q}\Big[ - \bar m_{3/2}({\cal I}_p + 2M_4^3 \lag \partial_p \omega_0\rag) +
     \bar m'_{3/2}(\mathfrak{d}_z {\cal I}_p + M_4^3 \lag \partial_p\mathfrak{d}_z \omega_0\rag)   \Big]\nn\\
    &&+\frac 1{m_p^2} \eKs \vSb{p}\Big[ - \bar m_{3/2}({\cal I}_q + 2M_4^3 \lag \partial_q \omega_0\rag) +
     \bar m'_{3/2}(\mathfrak{d}_z {\cal I}_q + M_4^3 \lag \partial_q\mathfrak{d}_z \omega_0\rag)   \Big]\nn\\
   &&  +\frac 1 {m_p^2}\eK  \Bigg\{ \vSb{q}\bar {\cal I}^r\Big[M_4^3 \lag \partial_r \partial_p \omega_0\rag+\frac 1{m^2_p}\vSb{r}({\cal I}_p+  M_4^3 \lag\partial_p \omega_0\rag)\Big]\nn\\
&& \hskip 1.8truecm +\vSb{p}\bar {\cal I}^r\Big[M_4^3 \lag \partial_r \partial_q \omega_0\rag+\frac 1{m^2_p}\vSb{r}({\cal I}_q + M_4^3 \lag\partial_q \omega_0\rag)\Big]
\Bigg\} \nn \\
&& + \frac {M_4^3} {m_p^2} \eK \Big[\vSb{r}\lag \partial_p \partial_q \omega_0\rag (\bar {\cal I}^r + M_4^3 \lag\partial^r\bar \omega_0\rag)
  \nn\\
&&\hskip 1.8truecm  +M_4^3\vpb{a^*}\Big( \lag\partial_p\partial_q\omega_0\partial^{a^\ast}\bar{\omega}_0\rag -\lag \partial_p \partial_q\mathfrak{d}_z \omega_0 \partial^{b^*} \bar \omega_0 (\Lambda^{-1})^b{}_{b^*}
  \partial^z \Lambda^{a^*}{}_b \rag\Big)\Big] \nn\\
&&+ \frac {M_4^3} {m_p^2} \eK\Big(M_4^2 \lag \phi^\dag \Lambda \phi\rag + \lag S^\dag S\rag\Big) \bar {\cal I}^r
\Big( \lag   \partial_p \partial_q\partial_r \omega_ 0 \rag +  \frac  1{m^2_p} \vSb{r} \lag \partial_p \partial_q \omega_0\rag\Big)\nn\\
&&+\frac {M_4^3} {m_p^2}\eKs\Big[ -2 \bar m_{3/2} \Big(M_4^2 \lag \phi^\dag \Lambda \phi\rag + \lag S^\dag S\rag\Big)  \lag \partial_p\partial_q \omega_0 \rag\nn\\
&& \hskip 2.3truecm + \bar m'_{3/2} \Big(M_4^2 \lag \phi^\dag \tilde{\cal S} \phi\rag + \lag S^\dag S\rag\Big)  \lag \partial_p\partial_q \mathfrak{d}_z\omega_0 \rag
\Big]\nn
   \eeqa
}

    $\bullet: S-\phi$ sector
{\footnotesize    
    \beqa
\frac 1 {M_4} \big< \partial_a\partial_p V \big> &=&
\eK\Big[ M_4^2 \lag \partial_a \partial_p \partial_q \omega_0\rag(\bar {\cal I}^q + M_4^3 \lag \partial^q \bar \omega_0\rag)
  + M_4^3 \lag \partial_a\partial_b \partial_p \omega_0 \partial^{a^*} \bar \omega_0 (\Lambda^{-1})^b{}_{b^*}\rag\Big]
\nn\\
&& + M^2_4 \eKs \Big[ \bar m'_{3/2} \lag \partial_a \partial_p \mathfrak{d}_z \omega_0 \rag +\bar m_{3/2}
  \big\{ -2  \lag \partial_a \partial_p \omega_0 \rag + R^b{}_a \lag \partial_p \partial_b \omega_0\rag
  +R^b{}_c  \lag\phi^c \partial_p \partial_a\partial_b \omega_0\rag \nn\\
&& \hskip 1.truecm + \lag S^q \partial_a \partial_p \partial_q \omega_0\rag \big\} \Big]\nn\\
&&+ \frac{1}{m_p^2}\eKs\Bigg\{M_4 \vpb{a^*} \Big[-\bar m_{3/2}({\cal I}_p + 2\lag\partial_p \omega_0\rag) \lag\Lambda^{a^*}{}_a\rag +
  \bar m'_{3/2}(\mathfrak{d}_z {\cal I}_p + \lag\partial_p \mathfrak{d}_z \omega_0\rag) \tilde {\cal S}^{a^*}{}_a\Big]\nn\\
&& \hskip 2.1truecm +
M_4^2\vSb{p}\Big[ -2 \bar m_{3/2} \lag \partial_a  \omega_0\rag + \bar m'_{3/2} \lag \partial_a \mathfrak{d}_z\omega_0\rag\Big]\Bigg\}\nn\\
&& + \frac  {1}{m_p^2} \eK \Bigg\{M_4\lag \phi^\dag_{a^*} \Lambda^{a^*}{}_a\rag \bar {\cal I}^q\Big[
  \frac {1} {m_p^2}({\cal I}_p + M_4^3 \lag \partial_p \omega_0\rag)\vSb{q} + M_4^3 \lag \partial_p\partial_q\omega_0\rag\Big]\nn\\
&& \hskip 1.9truecm
+M_4^4\vSb{p}\bar {\cal I}^q\Big[\frac 1 {m_p^2}  \lag S^\dag_q \partial_a \omega_0\rag + \lag \partial_q \partial_a \omega_0\rag\Big]\Bigg\}\nn\\
&& + \frac {M_4^2}{m_p^2} \eK \Big[ \lag  \partial_a \partial_p \omega_0  S^\dag_q\rag (\bar {\cal I}^q +M_4^3\lag\partial^q \bar \omega_0\rag) + M_4^3 \lag\partial_a \partial_p \phi^\dag_{a^*} \partial^{a^*}\bar\omega_0\rag\Big]\nn\\
&& \hskip 1.9truecm -M_4^3 \lag \partial_p \partial_a \mathfrak{d}_z \omega_0 \phi^\dag_{b^*} \partial^{a^*}\bar \omega_0(\Lambda^{-1})^b{}_{a^*} \partial^z \Lambda^{b^*}{}_b\rag\Big]\nn\\
&&+ \frac {M_4^2}{m_p^2} \eK \Big(M_4^2\lag \phi^\dag \Lambda \phi\rag + \lag S^\dag S\rag\Big) \bar {\cal I}^q\Big(\frac 1 {m_p^2} \lag S^\dag_ q\partial_p \partial_a \omega_0\rag + \lag\partial_a \partial_p \partial_q \omega_0\rag\Big)\nn\\
&& + \frac {M_4^2}{m_p^2} \eKs\Bigg\{
-2 \bar m_{3/2} \lag\partial_a \partial_p \omega_0\rag \Big(M_4^2\lag \phi^\dag \Lambda \phi\rag + \lag S^\dag S\rag\Big) \nn\\
&& \hskip 2.3truecm +
\bar m'_{3/2} \lag\partial_a \partial_p \mathfrak{d}_z\omega_0\rag \Big(M_4^2\lag \phi^\dag \tilde {\cal S} \phi\rag + \lag S^\dag S\rag\Big)
\Bigg\}\ . \nn
\eeqa
}
\end{allowdisplaybreaks}

Note that the formulas presented in the Subsection \ref{sec:HS_effects} can be directly obtained from those general results by taking :
\begin{gather}
\omega_0 (\Phi, S , z) = \omega_0 (\Phi, z) \quad , \quad \Lambda^{a}{}_{a^\ast}(z,z^{\dag}) = \delta^{a}_{a^\ast} \quad , \quad \lag \phi^a\rag = \lag S^p \rag = 0\; .  \nn
\end{gather}

\chapter{Screening property of the mass matrix}\label{eq:screenmatmass}
We show in this subsection an interesting decoupling property of the mass matrix in the basis $\{z,S^p,\Phi^a\}$. Let $M_1\in {\cal M}_{n_1}(\mathbb R), M_2 \in {\cal M}_{n_2}(\mathbb R)$  where $M_1$ and $M_2$ are symmetric matrices and let $P\in {\cal M}_{n_2,n_1}(\mathbb R)$.
Set $n=n_1+n_2$ and consider a symmetric $n\times n$ matrix of the form
\beqa
M= \bpm\phantom{\varepsilon} M_1& \varepsilon P^t\\
       \varepsilon P & \varepsilon M_2 \epm = \bpm M_1& \sqrt{\eta} P^t\\
       \sqrt{\eta} P & \varepsilon M_2 \epm\label{eq:matrix}
\eeqa
with $\varepsilon \sim 0$.

\begin{proposition}
  Let $\overline{\lambda}_n$, be the eigenvalues of the matrices $M_1$ and $\varepsilon M_2$ defined as
\begin{gather}
\overline{\lambda}_n=\begin{cases}
\tilde{\lambda}_n,\ n=1,\ \cdots,\ n_1\ \mathrm{with}\ \mathrm{det}(\tilde{\lambda}_n\mathds{1}-M_1)=0  \\
\varepsilon\tilde{\lambda}_n,\ n=1+n_1,\ \cdots,\ n_1+n_2\ \mathrm{with}\ \mathrm{det}(\tilde{\lambda}_n\mathds{1}-M_2)=0 \label{eq:vpM1M2}
\end{cases}
\end{gather}
We have
\begin{gather}
\mathrm{det}(\overline{\lambda}_n\mathds{1} - M) = 0 + \mathcal{O}(\varepsilon^2)\ . \nn
\end{gather}
\end{proposition}

\begin{demo}
Taking a general matrix $M$ defined by
\begin{eqnarray}
M = \begin{pmatrix}
A & B \\
C & D \end{pmatrix}
\end{eqnarray}
with $A,\ B,\ C$ and $D$ some matrices, the determinant of $M$ can be obtained using
\begin{eqnarray}
\mathrm{det}(M)&=&(\mathrm{det}A)\mathrm{det}(D-CA^{-1}B)\quad \mathrm{if\ }A\mathrm{\ is\ invertible}\ ,  \label{eq:det1}\\
&=&(\mathrm{det}D)\mathrm{det}(A-BD^{-1}C) \quad \mathrm{if\ }D\mathrm{\ is\ invertible}\ . \label{eq:det2}
\end{eqnarray}
In our case, the relation \autoref{eq:det1} become:
\begin{gather}
\mathrm{det}(\lambda\mathds{1} - M) = \mathrm{det}(\lambda\mathds{1} - M_1)\mathrm{det}(\lambda\mathds{1} - \varepsilon M_2 - \eta P(\lambda\mathds{1} - M_1)^{-1}P^t)\ .\label{eq:1app}
\end{gather}
with $\eta=\varepsilon^2$. Writing for simplicity
\begin{gather}
\begin{aligned}
A &= -\frac{1}{\lambda}M_2\ ,  \\
B &= -\frac{1}{\lambda} P \left( \lambda\mathds{1} - M_1 \right)^{-1}P^t\ , 
\end{aligned}\nn
\end{gather}
the relation \autoref{eq:1app} can be rewritten as
\begin{eqnarray}
\mathrm{det}(\lambda\mathds{1} - M_1)\mathrm{det}(\lambda\mathds{1} - \varepsilon M_2 - \eta P(\lambda\mathds{1} - M_1)^{-1}P^t) &=& \lambda^{n_2}\mathrm{det}(\lambda\mathds{1} - M_1)\mathrm{det}(\mathds{1} + \epsilon A + \eta B)\ . \nn
\end{eqnarray}
Using the identity
\begin{eqnarray}
\mathrm{det}(\mathds{1} + \epsilon A + \eta B) &=& \displaystyle\sum_{k=0}^n \frac{\eta^k}{k!}\frac{d^k}{d\eta^k}\mathrm{det}(\mathds{1} + \epsilon A + \eta B)\Big|_{\eta=0}\nn \\
&=& \mathrm{det}(\mathds{1} + \varepsilon A)\Bigg(1 + \eta \mathrm{Tr}\left( X\right)  \nn\\
&& + \frac12 \eta^2\Big( \mathrm{Tr}\left( X\right)^2 - \mathrm{Tr}\left(  X^2\right) \Big) + \cdots \Bigg)\label{eq:devdeterminant}
\end{eqnarray}
with $X=\frac{B}{\mathds{1} + \varepsilon A}$, we have easily 
\begin{eqnarray}
\mathrm{det}(\lambda\mathds{1} - M ) &=& \mathrm{det}(\lambda\mathds{1}-M_1)\mathrm{det}(\lambda\mathds{1} - \varepsilon M_2) + \mathcal{O}(\varepsilon^2)\ . \nn
\end{eqnarray}
Taking the eigenvalues $\overline{\lambda}_n$ of $M_1$ and $\varepsilon M_2$ (see \autoref{eq:vpM1M2}), we then get:
\begin{eqnarray}
\mathrm{det}(\overline{\lambda}_n\mathds{1} - M ) &=& \mathrm{det}(\overline{\lambda}_n\mathds{1}-M_1)\mathrm{det}(\overline{\lambda}_n\mathds{1} - \varepsilon M_2) + \mathcal{O}(\varepsilon^2) = 0 + \mathcal{O}(\varepsilon^2) \nn
\end{eqnarray}
which completes the proof. 
\end{demo}

\begin{proposition}
  Let $\lambda_n$ be the eigenvalues of $M$:
\begin{gather}
\mathrm{det}(\lambda_n\mathds{1} - M) = 0\ . 
\end{gather} 
The eigenvalues of the matrix $M$ are the eigenvalues of the matrix $M_1$ and $\varepsilon M_2$ up to $\mathcal{O}(\varepsilon^2)$ contribution,
\begin{gather}
\lambda_n = \overline{\lambda}_n + \mathcal{O}(\varepsilon^2),\ n=1,\ \cdots,\ n_1+n_2\ . \nn
\end{gather}
\end{proposition}

\begin{demo}
In the limit $\varepsilon\rightarrow 0$, the eigenvalues $\lambda_n$ of the matrix \autoref{eq:matrix} are
\begin{gather}
\lambda_n(\varepsilon) = \begin{cases}
\tilde{\lambda}_n,\ n=1,\ \cdots,\ n_1 \\
0,\ n=1+n_1,\ \cdots,\ n_1+n_2  
\end{cases}\nn
\end{gather}
Writing the eigenvalues as a power series in $\varepsilon$:
\begin{gather}
\lambda_n(\varepsilon) = \displaystyle\sum_{i=0}^{\infty} a_{n_i}\varepsilon^i\ ,\label{eq:powerlambda}
\end{gather}
we can separate them in two different types: $a_{n_0}\neq 0$ for $n=1,\ \cdots,\ n_1$ (defined as \textit{type-I}) and $a_{n_0}= 0$ for $n=1+n_1\ \cdots,\ n_1+n_2$ (defined as \textit{type-II}). Using the relations \autoref{eq:det1} and \autoref{eq:det2}, we have
\begin{itemize}
\item[(i)] \begin{eqnarray}
\mathrm{det}(\lambda\mathds{1} - M ) = 0 \Leftrightarrow \mathrm{det}(\lambda\mathds{1} - \varepsilon M_2 - \eta P(\lambda\mathds{1} - M_1)^{-1} P^t)=0\label{eqdet1} \label{eq:nottypeI}
\end{eqnarray}
if $\lambda$ is not a \textit{type-I},
\item[(ii)] \begin{eqnarray}
\mathrm{det}(\lambda\mathds{1} - M ) = 0 \Leftrightarrow \mathrm{det}(\lambda\mathds{1} - M_1 - \eta P^t(\lambda\mathds{1} - \varepsilon M_2)^{-1} P)=0\label{eqdet2} \label{eq:nottypeII}
\end{eqnarray}
if $\lambda$ is not a \textit{type-II}. 
\end{itemize}
We first consider the case (i). By analysing the form of the determinant \autoref{eqdet1}, we see that it is equivalent to a polynomial in $\lambda$, $\varepsilon$ and $\eta$ 
\begin{eqnarray}
 \mathrm{det}(\lambda\mathds{1} - \varepsilon M_2 - \eta P(\lambda\mathds{1} - M_1)^{-1} P^t) &=& \displaystyle\sum_{\alpha_1+\alpha_2+\alpha_3\geq n_2}c_{\alpha_1,\alpha_2,\alpha_3}\lambda^{\alpha_1}\varepsilon^{\alpha_2}\eta^{\alpha_3}\nn
\end{eqnarray}
In the case where $\lambda$ are the exact eigenvalues \autoref{eq:powerlambda}, then we can write
\begin{eqnarray}
 \mathrm{det}(\lambda\mathds{1} - \varepsilon M_2 - \eta P(\lambda\mathds{1} - M_1)^{-1} P^t) &=& \displaystyle\sum_{\alpha_1+\alpha_2+\alpha_3\geq n_2}c_{\alpha_1,\alpha_2,\alpha_3}\displaystyle\sum_{i_1,i_2\cdots i_{\alpha_1}}a_{i_1}a_{i_2}\cdots a_{i_{\alpha_1}}\nn\\
&& \times \varepsilon^{\alpha_2 + 2\alpha_3 + \sum_{k=1}^{\alpha_1} i_k} \label{eq:III}\ . 
\end{eqnarray}
Note that the relation \ref{eq:nottypeI} is valid at all order of $\varepsilon$, \textit{i.e.}, all the coefficients of the polynomial \ref{eq:III} must be equal to zero. Following Equation \ref{eq:nottypeI} in the limit $\varepsilon\rightarrow 0$, we see that $a_0=0$ and so $\lambda(\varepsilon)=\sum_{i=1}^{\infty}a_i\varepsilon^i$. The index $i_k$ in \autoref{eq:III} are then all different from zero and so
\begin{gather}
\displaystyle\sum_{k=1}^{\alpha_1} i_k  \geq \alpha_1\ . \nn
\end{gather}
Using the relation $\alpha_1 + \alpha_2 + \alpha_3 \geq n_2$, we have
\begin{gather}
\displaystyle\sum_{k=1}^{\alpha_1} i_k + \alpha_2 + 2\alpha_3 \geq n_2 + \alpha_3\ . \nn
\end{gather}
Since $\alpha_1 + \alpha_2 + \alpha_3 \geq n_2$, the lowest order in $\varepsilon$ is $n_2$ which is only obtained for $\alpha_3=0$ and $i_1=i_2=\cdots=1$, \textit{i.e.}, the term in $\varepsilon^{n_2}$ contains only $a_1$ and no dependency in $\eta$. Using \autoref{eq:devdeterminant} and \autoref{eq:nottypeI}, we deduce that the term in $\varepsilon^{n_2}$ is $\mathrm{det}(\lambda\mathds{1} - \varepsilon M_2)$. Since the relation \ref{eq:nottypeI} holds order by order in $\varepsilon$, we then get 
\begin{gather}
\mathrm{det}(a_1\varepsilon\mathds{1} -\varepsilon M_2) = \varepsilon^{n_2} \mathrm{det}(a_1\mathds{1} - M_2)  =0\ . \label{eq:resultI}
\end{gather} 
In the first order of $\varepsilon$, $\lambda$ gives the eigenvalues of $\varepsilon M_2$.\medskip

We must also check that the higher order in $\varepsilon^{\ell}$ ($\ell > n_2$) are not only function of $a_1$ in such a way that possibles constraints that appear in higher order are still consistent with the result \autoref{eq:resultI}. It is indeed the case. We define $\ell=\displaystyle\sum_{k=1}^{\alpha_1}i_k+\alpha_2 + 2\alpha_3$ with $\alpha_3 \geq 1$,  $i_k=1\ \forall k \in \{1,\ \cdots, \alpha_1\}$ and with $\alpha_1,\ \alpha_2,\ \alpha_3$ satisfying $\alpha_1 + \alpha_2 + \alpha_3 \geq n_2$. We assume that the same order is obtained for $\alpha_1$ fixed and $\alpha_3 - r$, $\alpha_2+r$. We then obtain
\begin{gather}
\displaystyle\sum_{k=1}^{\alpha_1}i'_{k} + \alpha_2 + 2\alpha_3 - r = \ell = \displaystyle\sum_{k=1}^{\alpha_1}i_{k} + \alpha_2 + 2\alpha_3 \Rightarrow \displaystyle\sum_{k=1}^{\alpha_1}i'_{k}  = \displaystyle\sum_{k=1}^{\alpha_1}i_{k} + r,\ \mathrm{i.e.},\exists k,\ \ i'_k \geq 2\ .\nn 
\end{gather}
The factor $a_1$ is then not the only one that appears for higher orders terms, which remain the result \autoref{eq:resultI} true.\medskip

The method is the same for the case (ii). We introduce artificially a parameter $r$ such that:
\begin{gather}
\mathrm{det}(\lambda\mathds{1} - M_1 - \eta P^t(\lambda\mathds{1} - \varepsilon M_2)^{-1}P) \rightarrow \mathrm{det}(\lambda\mathds{1} - r M_1 - \eta P^t(\lambda\mathds{1} - \varepsilon M_2)^{-1}P) \ . \nn
\end{gather}
From the Equation \ref{eq:nottypeII}, we see in the limit $\varepsilon \rightarrow 0$ that $a_0 \neq 0$. The determinant can also be written in a polynomial form
 \begin{eqnarray}
 \mathrm{det}(\lambda\mathds{1} - rM_1 - \eta P^t(\lambda\mathds{1} - \varepsilon M_2)^{-1} P) &=& \displaystyle\sum_{\alpha_1+\alpha_2+\alpha_3\geq n_1}c_{\alpha_1,\alpha_2,\alpha_3}\lambda^{\alpha_1}r^{\alpha_2}\eta^{\alpha_3}\nn\\
 &=& \displaystyle\sum_{\alpha_1+\alpha_2+\alpha_3\geq n_2}c_{\alpha_1,\alpha_2,\alpha_3}\displaystyle\sum_{i_1,i_2\cdots i_{\alpha_1}=0}^{\infty}a_{i_1}a_{i_2}\cdots a_{i_{\alpha_1}}\nn\\
&& \times \varepsilon^{2\alpha_3 + \sum_{k=1}^{\alpha_1} i_k} \label{eq:powerdelta}
\end{eqnarray}	
with
\begin{eqnarray}
\begin{rcases}
\displaystyle\sum_{k=1}^{\alpha_1}i_k = 0 \\
a_{i_k} = 1
\end{rcases}\ \mathrm{for}\ \alpha_1 = 0\ . \nn
\end{eqnarray}
The lowest order in $\varepsilon$ is zero for $\alpha_3=0$ and $i_k=0\, \forall k=1,\ \cdots,\ \alpha_1$ with $\alpha_1 + \alpha_2 = n_1$. It is then independent of $\eta$. Cancel the lowest order of $\varepsilon$ in \autoref{eq:nottypeII} is then equivalent to
\begin{gather}
\mathrm{det}(a_0\mathds{1} -M_1)=0 \label{eq:V}
\end{gather} 
Moreover, since the power of $\varepsilon$ in \autoref{eq:powerdelta} has the form $\displaystyle\sum_{k=1}^{\alpha_1} i_k + 2\alpha_3$, the order 1 in $\varepsilon$ is also given by $\alpha_3=0$ and $\displaystyle\sum_{k=1}^{\alpha_1}=1$ (so $i_k=0\ \forall k$ except for one $k^\ast$ giving $i_{k^\ast}=1$). The order 1 in $\varepsilon$ is then given by 
\begin{gather}
\displaystyle\sum_{\alpha_1 + \alpha_2 = n_1}c_{\alpha_1\alpha_2}a_1a_0^{\alpha_1-1}\varepsilon \label{eq:order1}
\end{gather}
and included in $\mathrm{det}(\lambda\mathds{1} - M_1)$. Since the relation \ref{eq:nottypeII} is valid for all the order in $\varepsilon$, the order one in $\varepsilon$ \autoref{eq:order1} must vanish. We have then:
\begin{gather}
a_1= 0 \ , \quad \text{or}\quad \displaystyle\sum_{\alpha_1 + \alpha_2 = n_1}c_{\alpha_1\alpha_2}a_0^{\alpha_1-1}= 0 \ .\label{eq:VI}
\end{gather}
However, since the relation \autoref{eq:V} is polynomial in order $n_1$ in $a_0$ and the second equation in \autoref{eq:VI} corresponds to a polynomial of order $n_1-1$ in $a_0$, they are then incompatible, leading then to $a_1=0$.
\end{demo}

\chapter{Eigenvalues and eigenvectors of the mass matrix in the $\{S^p,z\}$ basis}\label{app:eigen}
This appendix present the analysis of the structure of the mass matrix $\mathcal{M}$ in the hybrid/hidden subsector mentionned in \ref{subsec:massmatrix} and the computation of their corresponding eigenvectors.\medskip

Due to the structure of the mass matrix $\mathcal{M}$ in the basis $\{S^p,z\}$ it is easy to obtain all eigenvalues and eigenvectors. We recall that the element of the mass matrix take the form
\begin{eqnarray}
{\left\langle\frac{\partial^2 V}{\partial S^p\partial
S_q^\dagger}\right\rangle} &=& \delta_p{}^q |\xi _{3/2}|^2 |m'_{3/2}|^2 + 
\frac{e^{|\left\langle z \right\rangle |^2}}{m_{p}^2} \left(4 |\xi_{3/2}|^2-1\right) \mathcal{I}_p\overline{\mathcal{I}}^q\ , \nn\\
m_{p}^{-1}{\left\langle\frac{\partial^2 V}{\partial S^p\partial z^\dagger}\right\rangle}&=&
-\frac{e^{|\left\langle z \right\rangle|^2/2}}{m_{p}}\left(\overline{m}'_{3/2}-2\overline{\xi}_{3/2}\overline{m}''_{3/2}\right)\mathcal{I}_p \ , \nn\\
m_p^{-2} {\left\langle \frac{\partial^2 V}{\partial z\partial z^\dag} \right \rangle}&=&
\frac 1 {m_p^2}e^{\big|\langle z \rangle\big|^2}\Bigg[
\big|{\cal I}\big|^2\Big(1+
4|\xi_{3/2}|^2\Big)+
M_4^4 \sum \limits_a \Big(\Big|\langle \partial_a \mathfrak{d}_z \omega_0\rangle \Big|^2+
\Big|\langle \partial_a \omega_0\rangle \Big|^2\Big)
\Bigg] \nn\\
&& +
\big|m''_{3/2}\big|^2-2 \big| m_{3/2}\big|^2\ . \nn
\end{eqnarray}
We first rewrite the mass matrix in a more general form by highlighting the ${\cal I}_p$ dependencies:	
 \begin{eqnarray}
{\left\langle \frac{\partial^2 V}{\partial S^p\partial S_p^\dag} \right \rangle} =
a + b |\mathcal{I}_p|^2 , \quad
{\left\langle\frac{\partial^2 V}{\partial S^p\partial
S_q^\dagger}\right\rangle}=
b \mathcal{I}_p\overline{\mathcal{I}}^q\ , \nn\\
m_{p}^{-1}{\left\langle\frac{\partial^2 V}{\partial S^p\partial z^\dagger}\right\rangle}=
c \mathcal{I}_p \label{eq:d2VdSpdzdag-1} \ , \quad
m_{p}^{-2}{\left\langle \frac{\partial^2 V}{\partial z \partial z^\dag} \right \rangle}&=& d \ .\nn
\end{eqnarray}
The $(n+1)\times (n+1)$ squared mass matrix can then be written in the following form:
\beqa
\label{eq:Mass2}
\mathcal{M}^2  \equiv a \mathbb{I} + \mathbb{A}\ , \quad \text{with}\quad \mathbb{A} = \bpm [B]  &  [C]  \\
                  [C^\dag]  &  [D] \epm , \nn
\eeqa
where $\mathbb{A}$ is a block matrix and we introduce $[B], [C], [D]$, $n\times n$, $1\times n$ and $1\times 1$ matrices respectively:
\beqa
\begin{aligned}
\left[B\right]_{p}{}^q&= b \mathcal{I}_p\overline{\mathcal{I}}^q\ , \\
\left[C\right]_p &= c \mathcal{I}_p,  \ [C^\dag]^q = c^* \overline{\mathcal{I}}^q \label{eq:defabcd}
\left[D\right]_{1}\!\!{}^1 &= d-a \ .\nn
\end{aligned}
\eeqa

The matrix $\mathbb A$ and $a \mathbb I$ commute so are simultaneously diagonalisable.
Furthermore, looking to the $n\times n$ submatrix  $\mathbb B$ of the matrix $\mathbb A$  in the $S^p-$space we have
\beqa 
\mathbb B = b I \otimes I^\dag  \nn
\eeqa
where $I$ is a vector and $I^\dag$ a covector of $\mathbb C^n$. Thus the kernel of $\mathbb B$ is $(n-1)$-dimensional since the image of $I^\dag$ is one dimensional.
Due to the specific form of $\mathbb B$ is is obvious that the vectors ($1\le p< q \le n-1$)
\beqa
V_{pq}= \begin{pmatrix} 0 \\ \vdots\\ 0\\\bar {\cal I}^q\\ 0\\ \vdots\\ 0\\-\bar {\cal I}^p \\  0 \\ \vdots \\ 0\end{pmatrix}\nn
  \eeqa
  where $\bar {\cal I}^q$ is in the $p-$th line and $-\bar {\cal I}^p$ in the $q-$th line are in the kernel of $\mathbb B$ so the not normalised and not orthogonal vectors
  \beqa
T_{p p+1} = \bar {\cal I}^{p+1} S^p -\bar {\cal I}^p S^{p+1} \ , \ \ p=1,\cdots,n-1\nn
\eeqa
constitute a basis of the the kernel of $\mathbb B$ and  consequently are eigenvectors of $\mathbb M^2$ with eigenvalue $a$.
Next it is easy to see that the vector
\beqa
\label{eq:V}
V= \begin{pmatrix} {\cal I}_1 \\ \vdots \\ {\cal I}_n \end{pmatrix} 
\eeqa
is a eigenvector of $\mathbb B$ with eigenvalue $b |I|^2$. Thus the vector
\beqa
S= \frac {{\cal I}_p}{|{\cal I}|} S^p \ , \label{eq:Ssum}
\eeqa
is a normalised eigenvector of $\mathbb B$. Note that $S$ and $V_{ab}$ are orthogonal.

To proceed to a change of basis we proceed in two steps. First we go from the canonical basis to the basis $V_{12},V_{21},\cdots, V_{n-1,n}, S$. Denote $P$ the corresponding matrix which takes the
form (in the $S^p-z$ space)
\beqa
P=\begin{pmatrix} \phantom{-}\bar {\cal I}^2 & 0& 0&    \cdots &0&\frac {{\cal I}_1}{|{\cal I}|}&0\\
  -\bar {\cal I}^1&\phantom{-}\bar {\cal I}^3 &0&\cdots&0&\frac {{\cal I}_2}{|{\cal I}|}&0\\
  0&-\bar {\cal I}^2&\bar {\cal I}^4 &\cdots &0&\frac {{\cal I}_3}{|{\cal I}|}&0\\
  \vdots&\vdots&&\ddots&\vdots&\vdots&\vdots\\
  0&0&0&\ddots&\bar {\cal I}^n&\frac {{\cal I}_{n-1}}{|{\cal I}|}&0\\
  0&0&0&0&-\bar {\cal I}^{n-1}& \frac {{\cal I}_{n}}{|{\cal I}|}&0\\
   0&&\cdots&&\cdots&0 &1
  \end{pmatrix} = \begin{pmatrix} Q&0\\0&1\end{pmatrix}\nn
  \eeqa
  where the matrix $Q$ acts only on the $S-$sector and not on the hidden sector $z$.
Since the vectors $V_{12},\cdots, V_{n-1,n}$ are not orthonormal, the matrix $P$  (or $Q$) is not unitary.
Next we introduce a Gram Schmidt matrix $G$  (acting only on the $S-$space) on the form
\beqa
G= \begin{pmatrix} H&0\\0&1 \end{pmatrix}\nn
\eeqa
which only acts on the $V_{12},\cdots,V_{n-1n}$ vectors in such a way that they form an orthonormal basis. Since the vector $V$ is normalised and orthogonal to all $V_{ab}$ the matrix
$G$ only acts on the first $(n-1)-$components, say $V_{12}, \cdots, V_{n,n-1}$ and
enables to obtain an orthormal basis of the kernel of $\mathbb B$. We thus have
\beqa
U= GQ = Q G' \ \ \text{with} \ \ G'=Q^{-1}G Q\nn
\eeqa
where now $U$ is unitary. Thus we have in the new basis
\beqa
\mathbb A' = \bpm U^\dag&0\\0&1\epm \mathbb A \bpm U&0\\0&1\epm =
 \bpm U^\dag&0\\0&1\epm  \bpm \mathbb B& \mathbb C^\dag \\ \mathbb C& \mathbb D \epm \bpm U&0\\0&1\epm =
\bpm U^\dag \mathbb B U & U^\dag \mathbb C \\
                                         C^\dag U&\mathbb D \epm\nn
                                         \eeqa
Of course we have
\beqa
V^\dag \mathbb B V = \text{diag}(0,\cdots,0, c|I|^2) \ .\nn
\eeqa
Furthermore, since
\beqa
Q^\dag C= \begin{pmatrix} 0 \\ \vdots\\0\\ c|{\cal I}| \end{pmatrix}\nn
\eeqa
and since the matrix $G'$ acts not trivially on the first $n-1$ components we obtain
\beqa
\mathcal M'^2 = \begin{pmatrix} a \mathbb {\cal I}_{n-1}&0&0\\
  0&a + b |{\cal I}|^2&c |I|\\
  0 &\bar c |{\cal I}|&d\\
  \end{pmatrix}
  \label{eq:matrixprime}
\eeqa
We now compute the orthornormal vectors of $\mathbb B$ obtained by the Gram-Schmidt process.
The computation of the orthonormal basis is done by induction.
We start with the non-orthonormal basis of the kernel of $\mathbb B$ that we denote
${\cal B} = \{V_{12},\cdots, V_{n-1,n}\}$. We denote ${\cal B}_N=\{W_1,\cdots, W_{n-1}\}$ the orthonormal (not
normalised) vecteurs. Pay attention that the orthonormalisation processe is done with the hermitian scalar
product. For convenience we introduce the following notations:
\beqa
{\cal I}_{12} &=& |{\cal I}_1|^2 + |{\cal I}_2|^2 \ \nn
\eeqa
\begin{enumerate}
\item Step one take
  \beqa
W_1= \Big(\frac{\bar {\cal I}^2}{{\cal I}_{12}}, -\frac{\bar {\cal I}^1}{{\cal I}_{12}},0,\cdots,0\Big)^t \ .\nn
\eeqa
\item Step two take $W_2= W_1 + \lambda V_{23}$ imposing the $W_1^\dag W_2=0$ gives $\lambda$. Rescaling $W_2$ leads to
  \beqa
W_2= \Big(\frac{{\cal I}_1}{{\cal I}_{12}}, \frac{{\cal I}_2}{{\cal I}_{12}},-\frac{1}{\bar {\cal I}^3},\cdots,0\Big)^t \ .\nn
\eeqa
\item Step three. Since $V_{34} \perp W_1$,  take $W_3= W_3 + \lambda V_{34}$ imposing the $W_2^\dag W_3=0$ gives $\lambda$. Rescaling $W_3$ leads to
  \beqa
W_3= \Big(\frac{{\cal I}_1}{{\cal I}_{12}}, \frac{{\cal I}_2}{{\cal I}_{12}}, \frac{{\cal I}_3}{{\cal I}_{12}},-\frac{1}{{\cal I}_{12}\bar{{\cal I}}^4}\displaystyle\sum_{p=1}^{3}|{\cal I}_p|^2,\cdots,0\Big)^t \ .\nn
\eeqa
\item Final steps. The  relation above can be easily extended and proceeding along the same lines we get for $3\le k\le n-1$
\beqa
W_k= \Big(\frac{{\cal I}_1}{{\cal I}_{12}}, \frac{{\cal I}_2}{{\cal I}_{12}}, \frac{{\cal I}_3}{{\cal I}_{12}},
 \frac{{\cal I}_4}{{\cal I}_{12}},\cdots, \frac{{\cal I}_k}{{\cal I}_{12}},-\frac{1}{\bar {\cal I}^{k+1}{\cal I}_{12}}\displaystyle\sum_{p=1}^{k}|{\cal I}_{p}|^2,\cdots,0\Big)^t \ . \nn
\eeqa
\end{enumerate}
Finally the vectors $(W_1,\cdots, W_{n-1})$ are orthonormal. A nice check that these vectors belongs to the kernel of $\mathbb B$
is to check that these vectors are orthogonal to the eigenvectors $V$ (see Eq[\ref{eq:V}]). And it is indeed the case.
The next step is to normalise the vectors $W_k$:
\beqa
\|W_1\|^2 &=& \frac{1}{{\cal I}_{12}} \nn\\
\|W_2\|^2 &=& \frac{\displaystyle\sum_{p=1}^{3}|{\cal I}_p|^2}{{\cal I}_{12}|{\cal I}_3|^2} \nn\\
\|W_k\|^2& = &\frac{1}{({\cal I}_{12})^2}\displaystyle\sum^{k}_{p=1}|{\cal I}_{p}|^2\left( 1+\frac{1}{|{\cal I}_{k+1}|^2}\displaystyle\sum^{k}_{p=1}|{\cal I}_{p}|^2\right),\quad 3\le k\le n-1\nn
\eeqa


\chapter{Mean flight distance of the stop squark $\tilde{t}$}\label{app:meanflight}
In this appendix, we summarize the results on the flight distance of the stop squark in the laboratory frame in the process $\tilde{t}_1\rightarrow t\chi^0_1$. The following figures correspond to the analysis Section \ref{sec:NeutLSP}.
\begin{figure}[h!]
      \centering
      \includegraphics[width=0.68\linewidth]{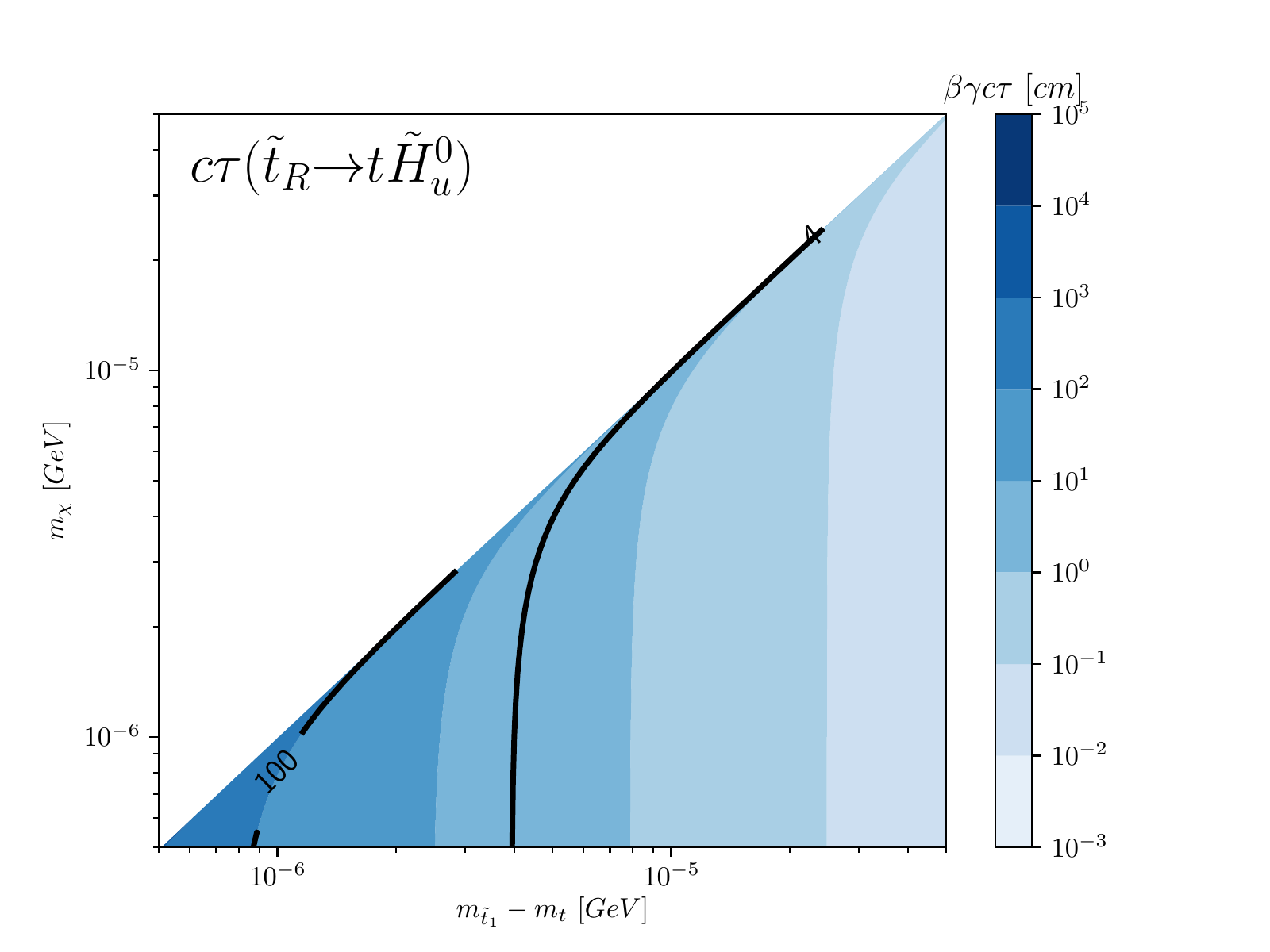}
   \caption{Mean flight distance of the stop squark $\tilde{t}_1$ in the laboratory frame as function of the neutralino mass $m_{\chi^0_j}$ and the difference between stop and top mass $m_{\tilde{t}}-m_t$ for the various processes of \autoref{table:Decay}.}
\end{figure}

\begin{figure}[H]
    \centering
      \includegraphics[width=0.8\linewidth]{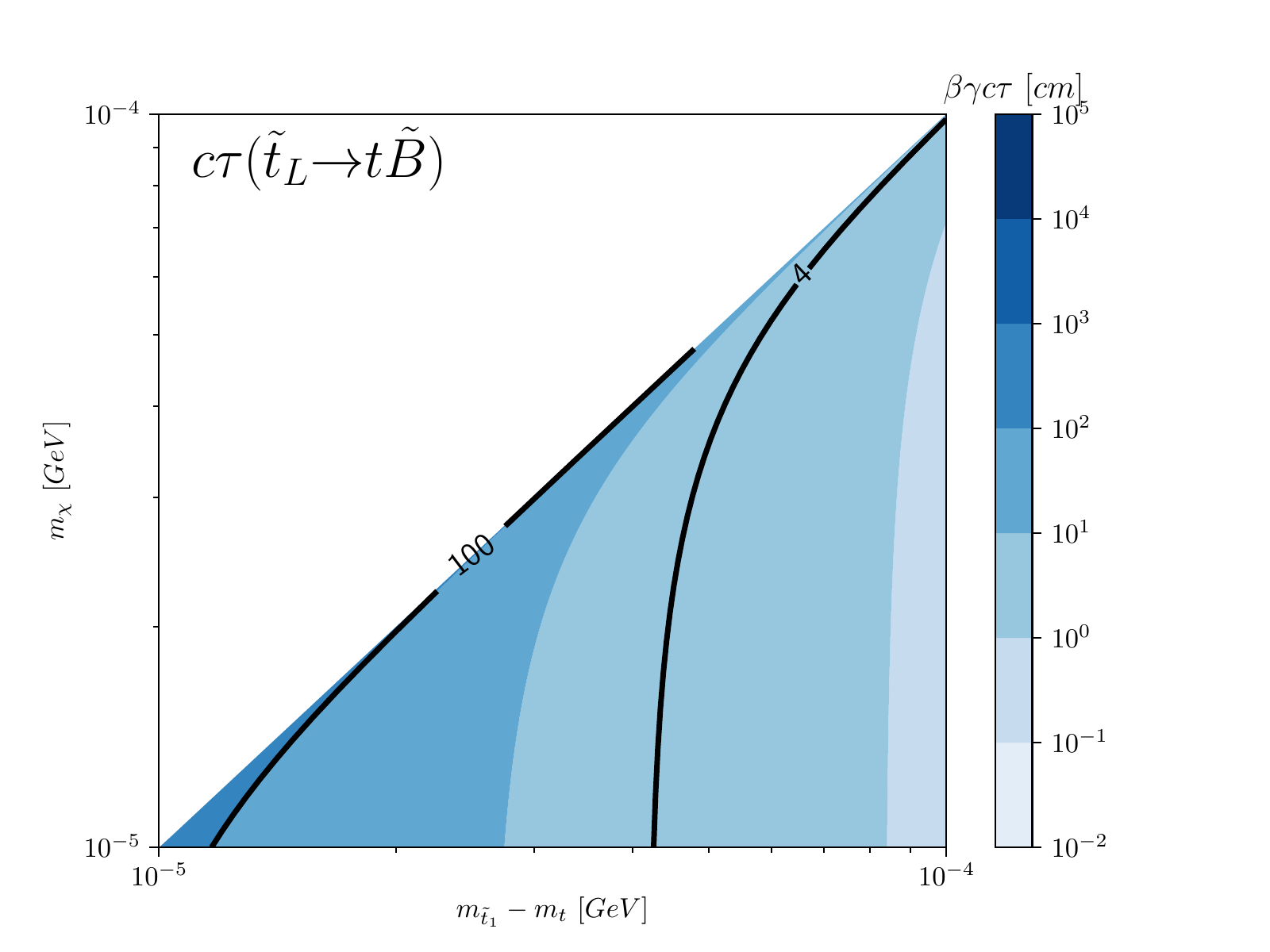}
      \centering
      \includegraphics[width=0.8\linewidth]{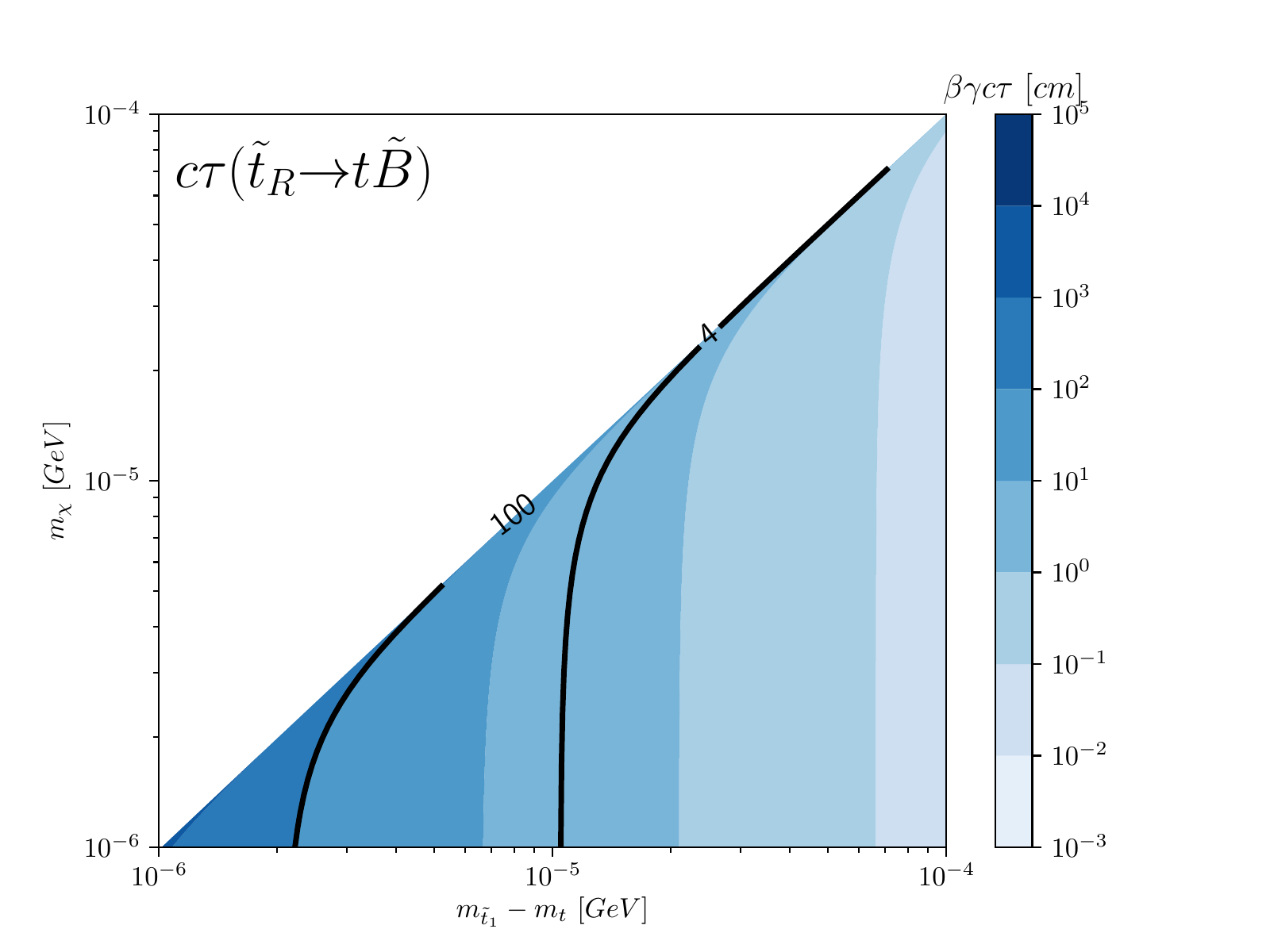}
   \caption{Mean flight distance of the stop squark $\tilde{t}_1$ in the laboratory frame as function of the neutralino mass $m_{\chi^0_j}$ and the difference between stop and top mass $m_{\tilde{t}}-m_t$ for the various processes of \autoref{table:Decay}.}
\end{figure}

\begin{figure}[H]
      \centering
      \includegraphics[width=0.8\linewidth]{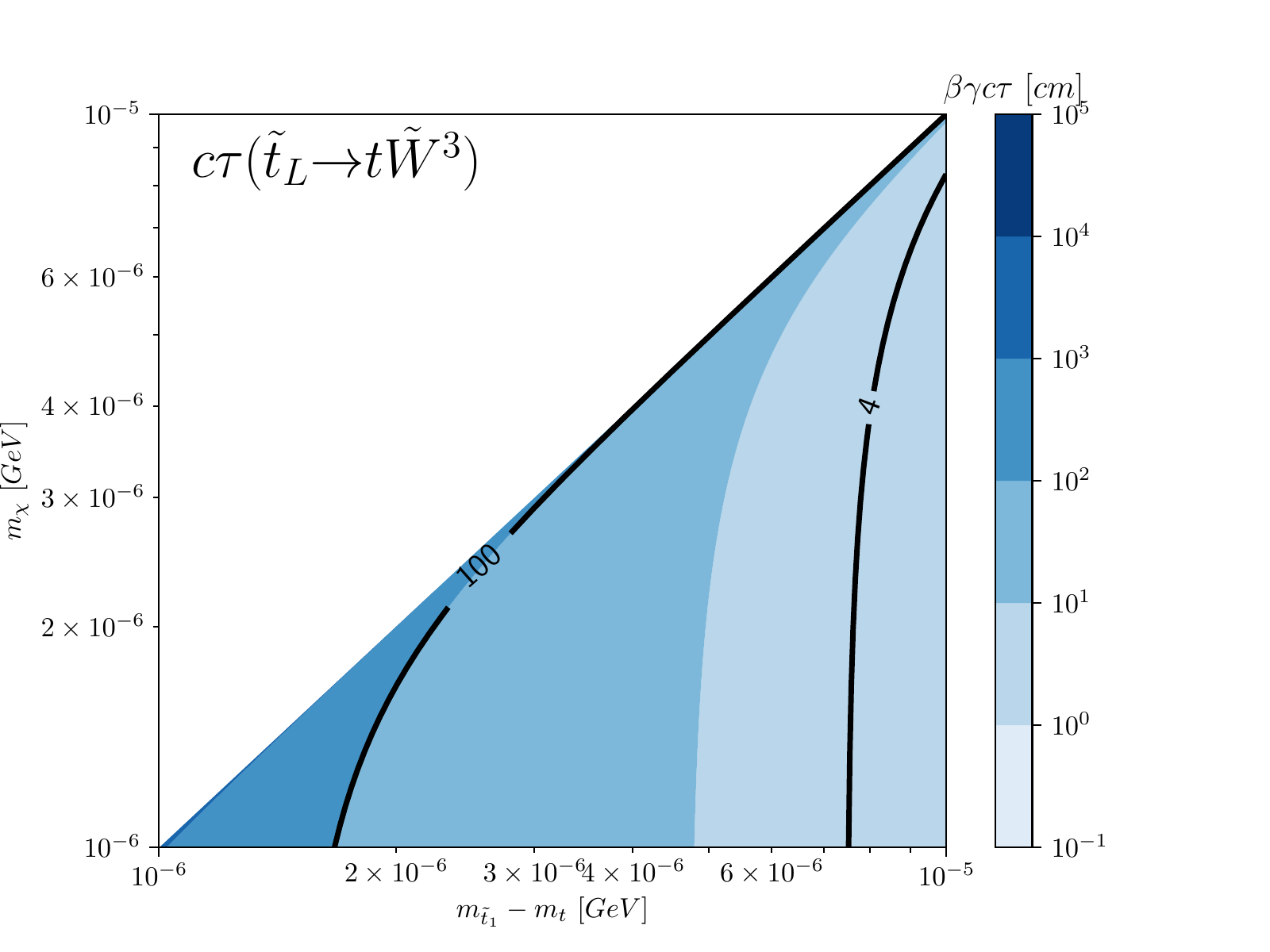}
      \centering
      \includegraphics[width=0.8\linewidth]{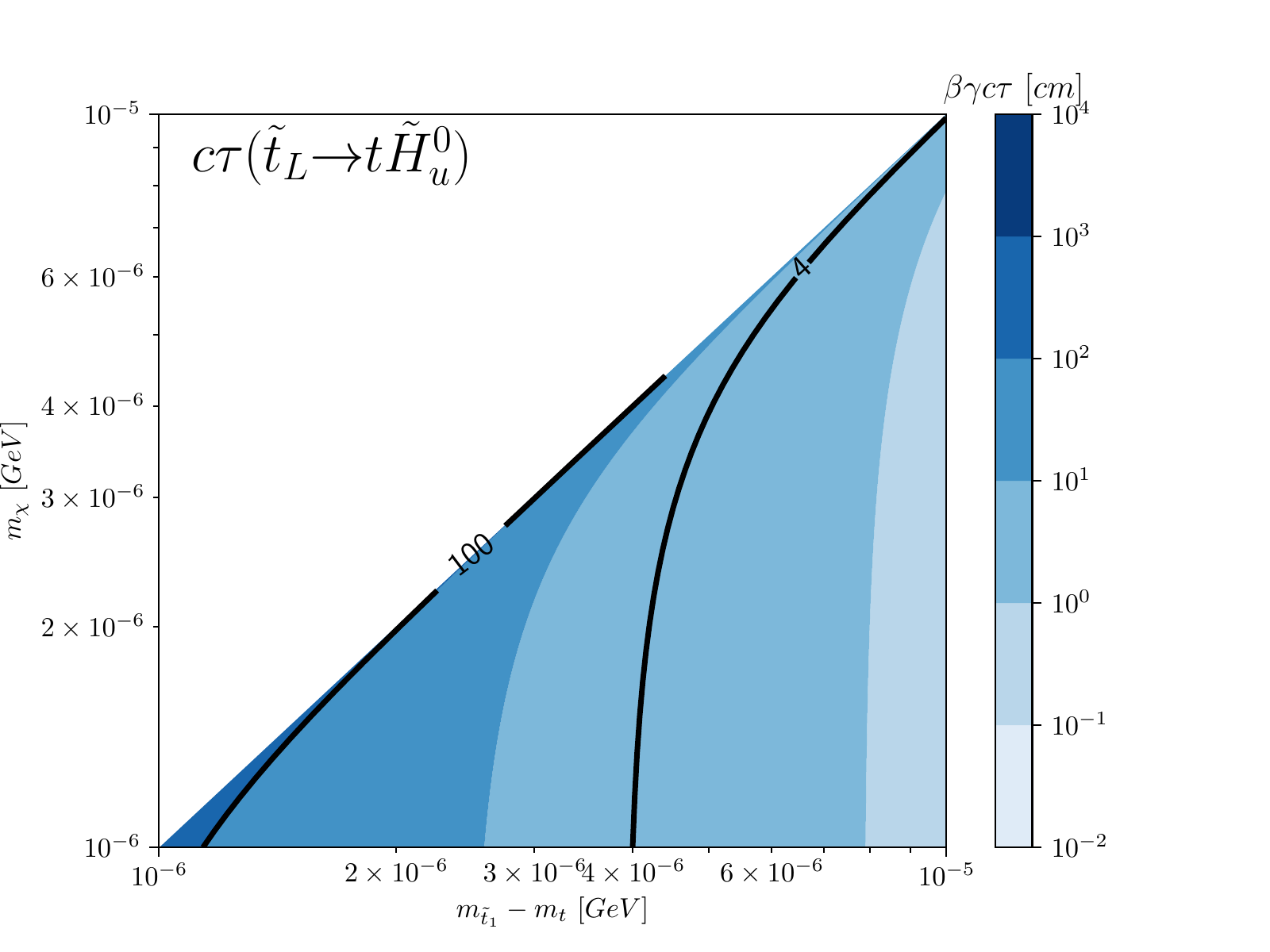}
   \caption{Mean flight distance of the stop squark $\tilde{t}_1$ in the laboratory frame as function of the neutralino mass $m_{\chi^0_j}$ and the difference between stop and top mass $m_{\tilde{t}}-m_t$ for the various processes of \autoref{table:Decay}.}
\end{figure}

\chapter{Distributions of several observables for the analysis $\tilde{t}\rightarrow t \psi_{\mu}$}\label{chap:analysis_distrib}
We present all the distributions produced for the analysis of the long-lived process $\tilde{t}\rightarrow t\psi_{\mu}$ with NLSP stop and LSP gravitino in the MSSM with \textit{Gauge Mediated Supersymmetry Breaking} (see Section \ref{sec:gravitinoLSP}).
\section{Distributions of $d_0$}\label{app:d0}
The distribution of the $d_0$ of the top quark can be found in \autoref{fig:do_t}. The other distributions are shown below. We recall that the impact parameter corresponds to (without taking into account the magnetic field)
\begin{gather}
d_0 = \frac{1}{p_T}( xp_y - yp_x) \nn
\end{gather}
with $(p_x,p_y)$ the components of the transerse impulsion of the particle and $(x,y)$ the position of the displaced vertex. It is a parameter of interest for the study of displaced processes.

\begin{figure}[H]
      \centering
      \includegraphics[width=0.8\linewidth]{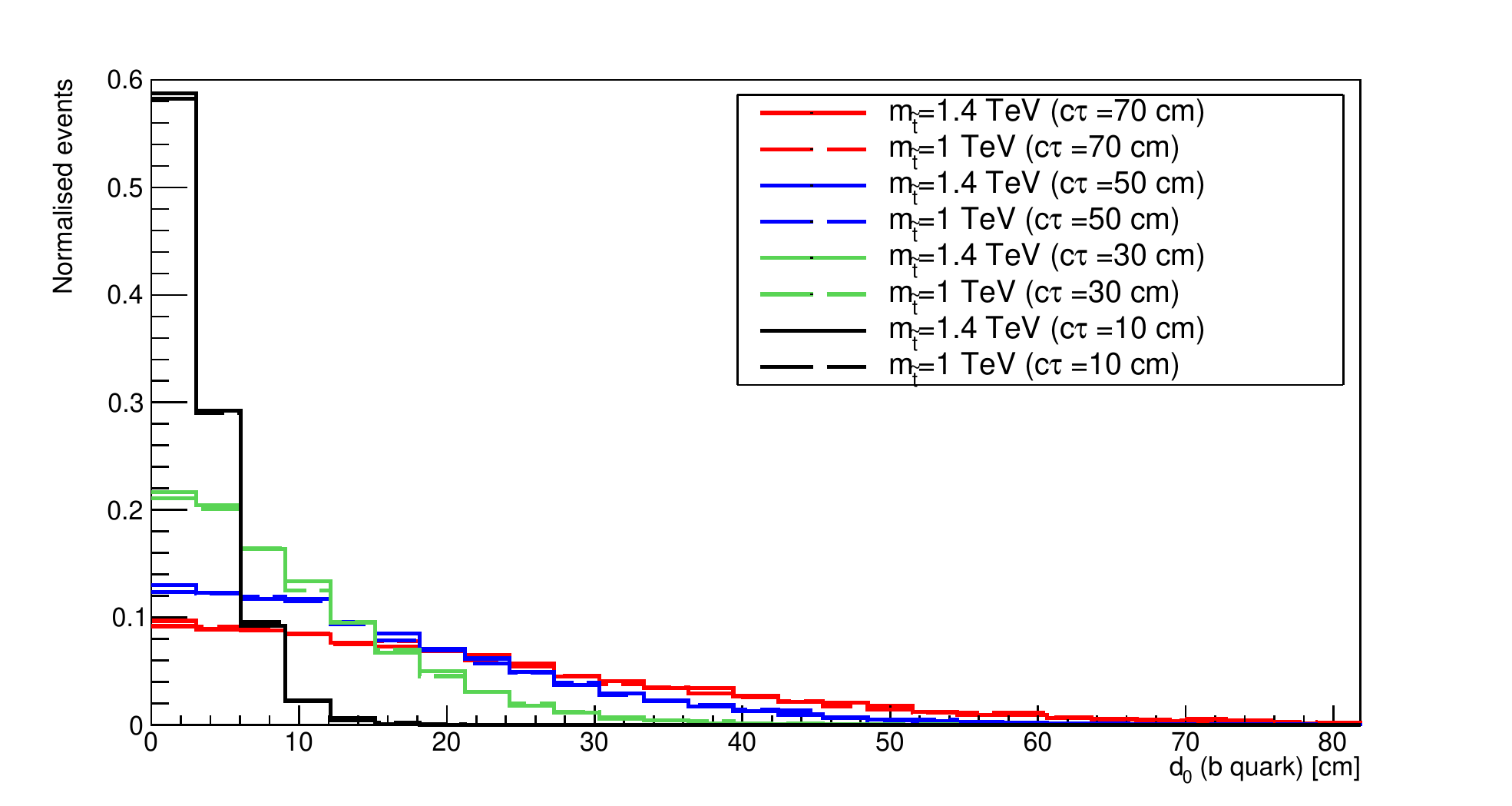}
    \caption{Distributions of $d_0$ of the b-quark coming from the top quark for the eight benchmarks.}
\end{figure}

\begin{figure}[H]
      \centering
      \includegraphics[width=0.8\linewidth]{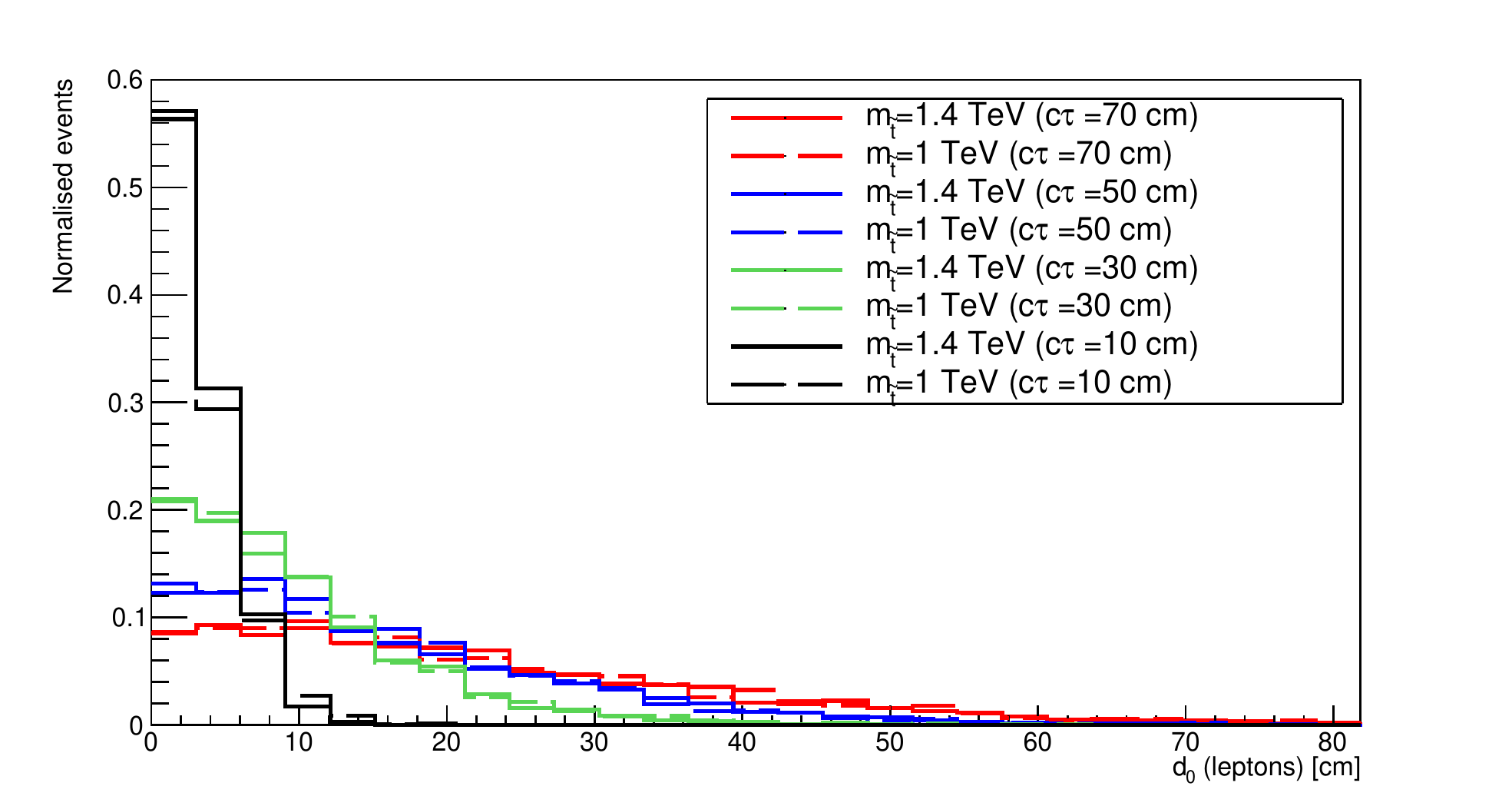}
    \caption{Distributions of $d_0$ of the leptons coming from the decay of top quark for the eight benchmarks.}
\end{figure}

\begin{figure}[H]
      \centering
      \includegraphics[width=0.8\linewidth]{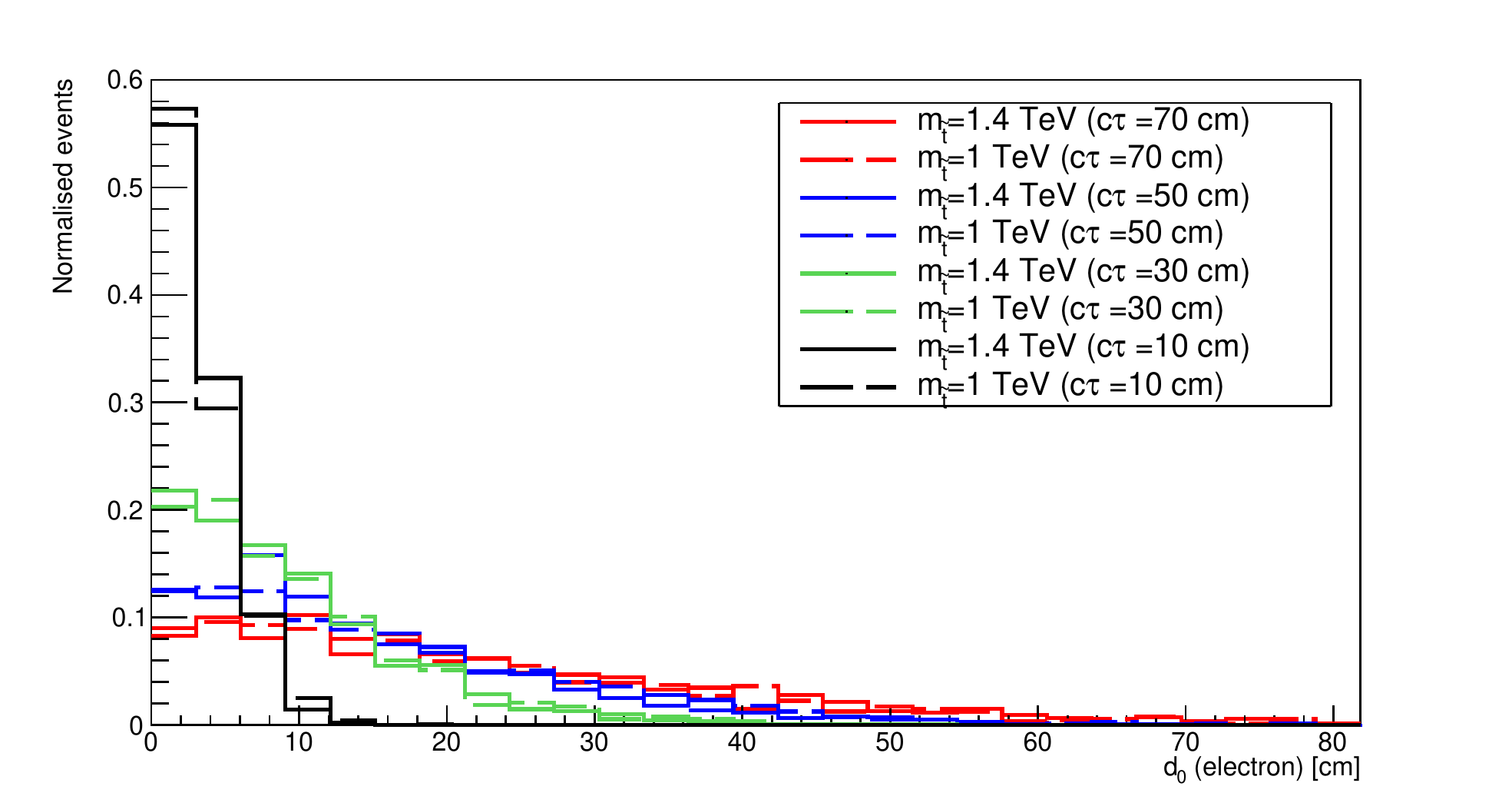}
    \caption{Distributions of $d_0$ of the electrons coming from the decay of top quark for the eight benchmarks.}
\end{figure}

\begin{figure}[H]
      \centering
      \includegraphics[width=0.8\linewidth]{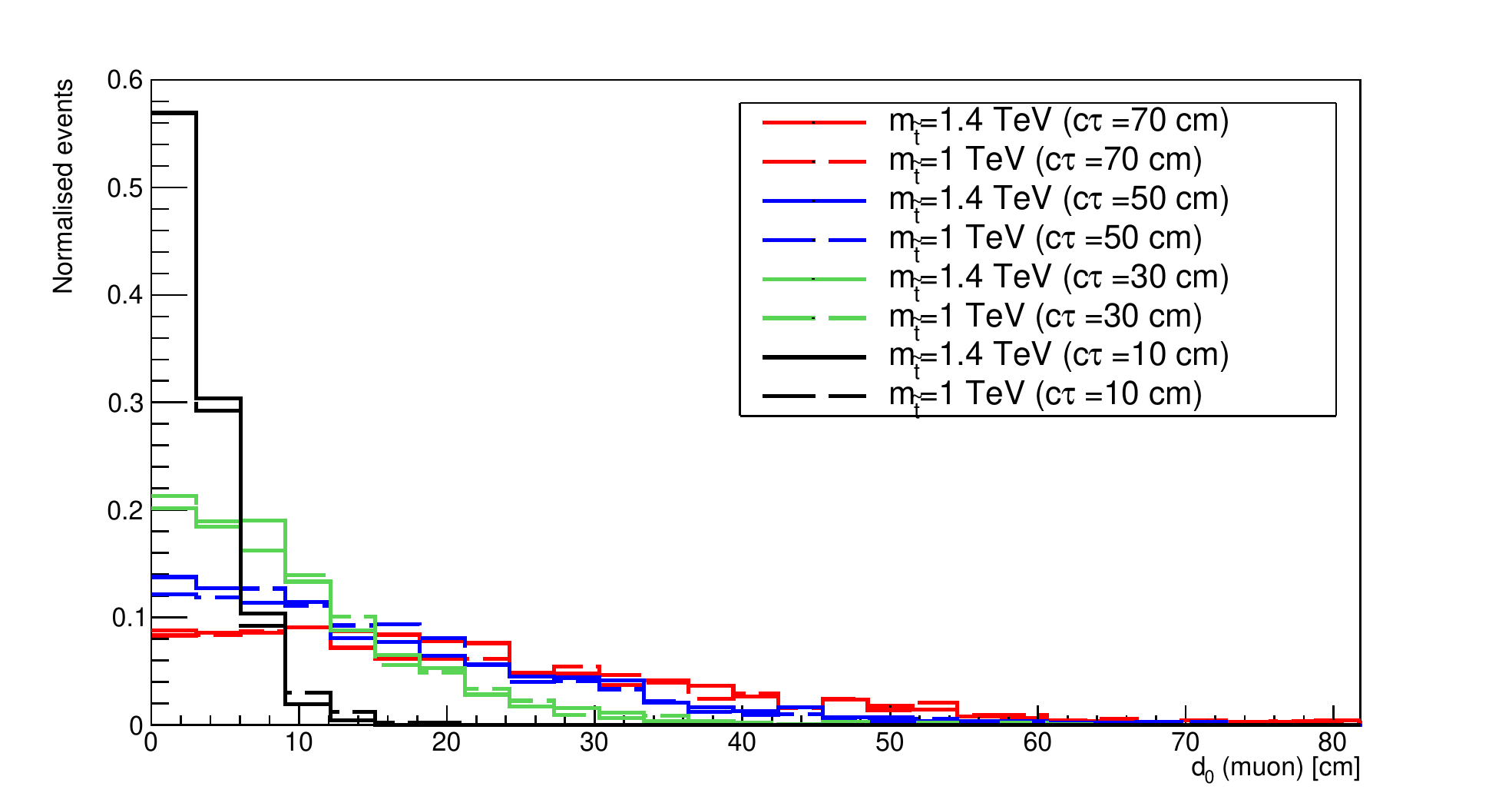}
    \caption{Distributions of $d_0$ of the muons coming from the decay of top quark for the eight benchmarks.}
\end{figure}

\section{Distributions of $d_z$}\label{app:dz}
We recall that the longitudinal impact parameter corresponds to (without taking into account the magnetic field)
\begin{gather}
d_z = z - \frac{p_z}{p_T^2}(xp_x + yp_y) \nn
\end{gather}
with $p_T$ the transerse impulsion of the particle and $(x,y)$ the position of the displaced vertex. 

\begin{figure}[H]
      \centering
      \includegraphics[width=0.8\linewidth]{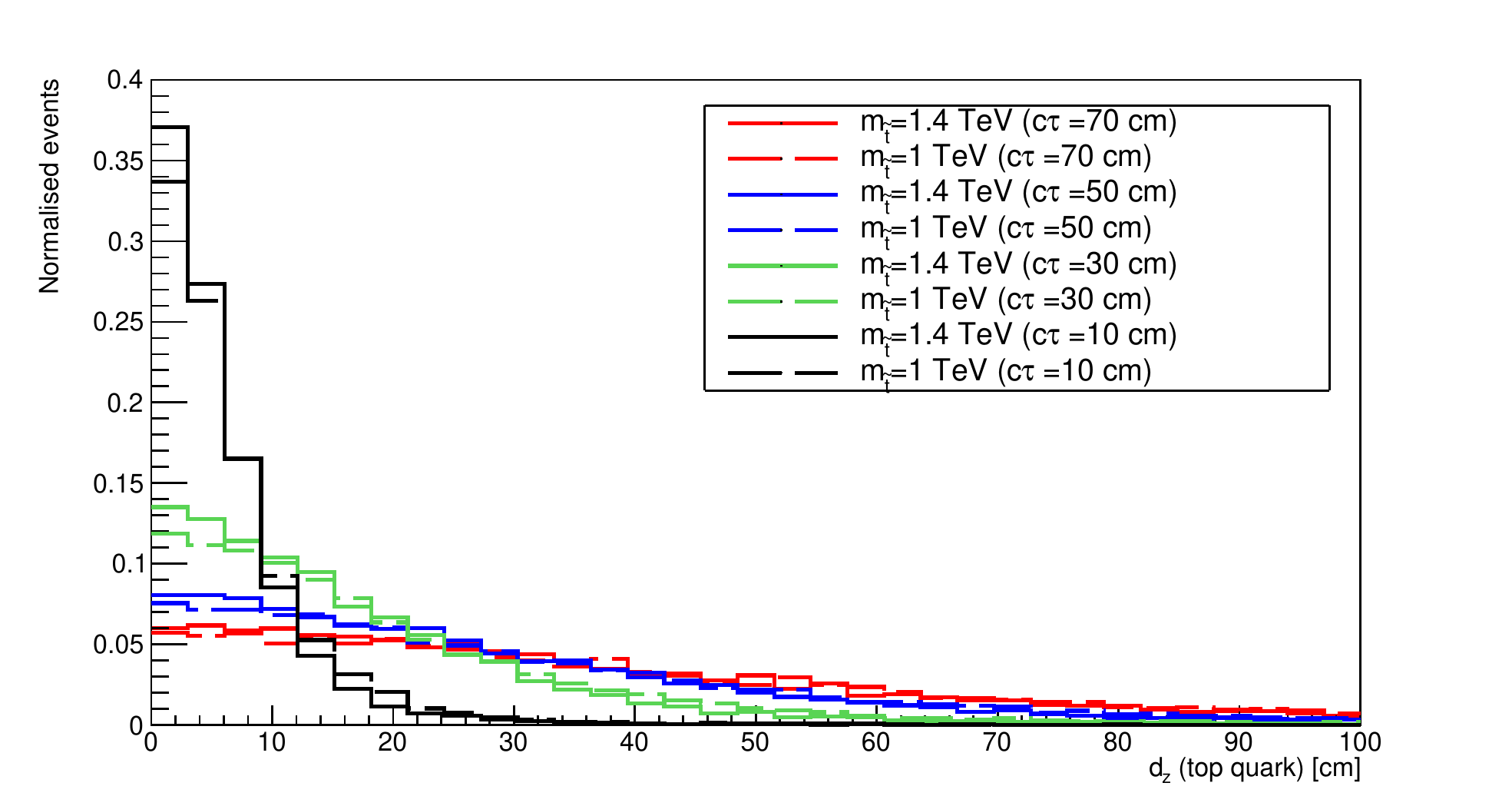}
    \caption{Distributions of $d_z$ of the top quark for the eight benchmarks.}
\end{figure}

\begin{figure}[H]
      \centering
      \includegraphics[width=0.8\linewidth]{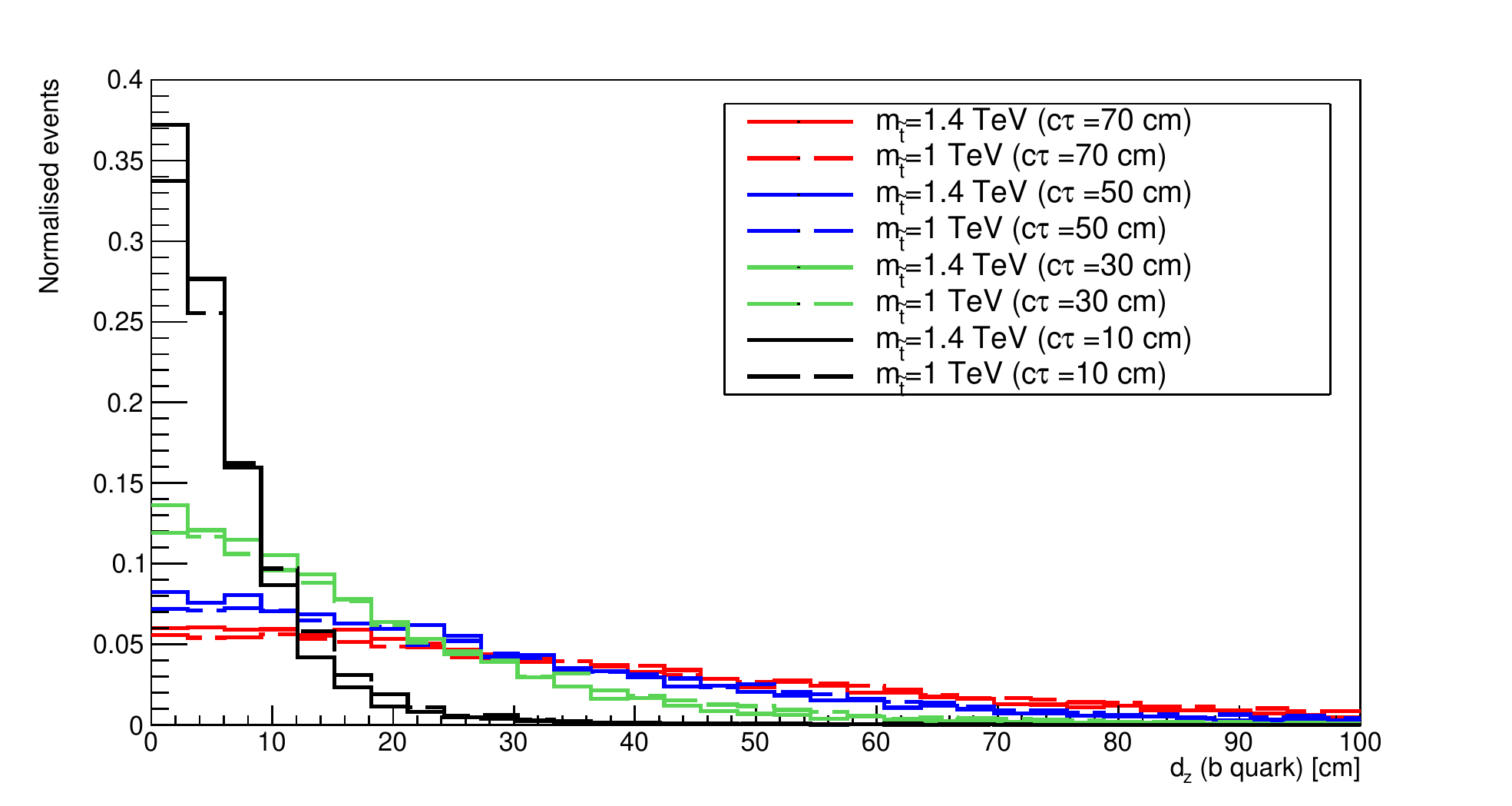}
    \caption{Distributions of $d_z$ of the b-quarks coming from the decay of top quark for the eight benchmarks.}
\end{figure}

\begin{figure}[H]
      \centering
      \includegraphics[width=0.8\linewidth]{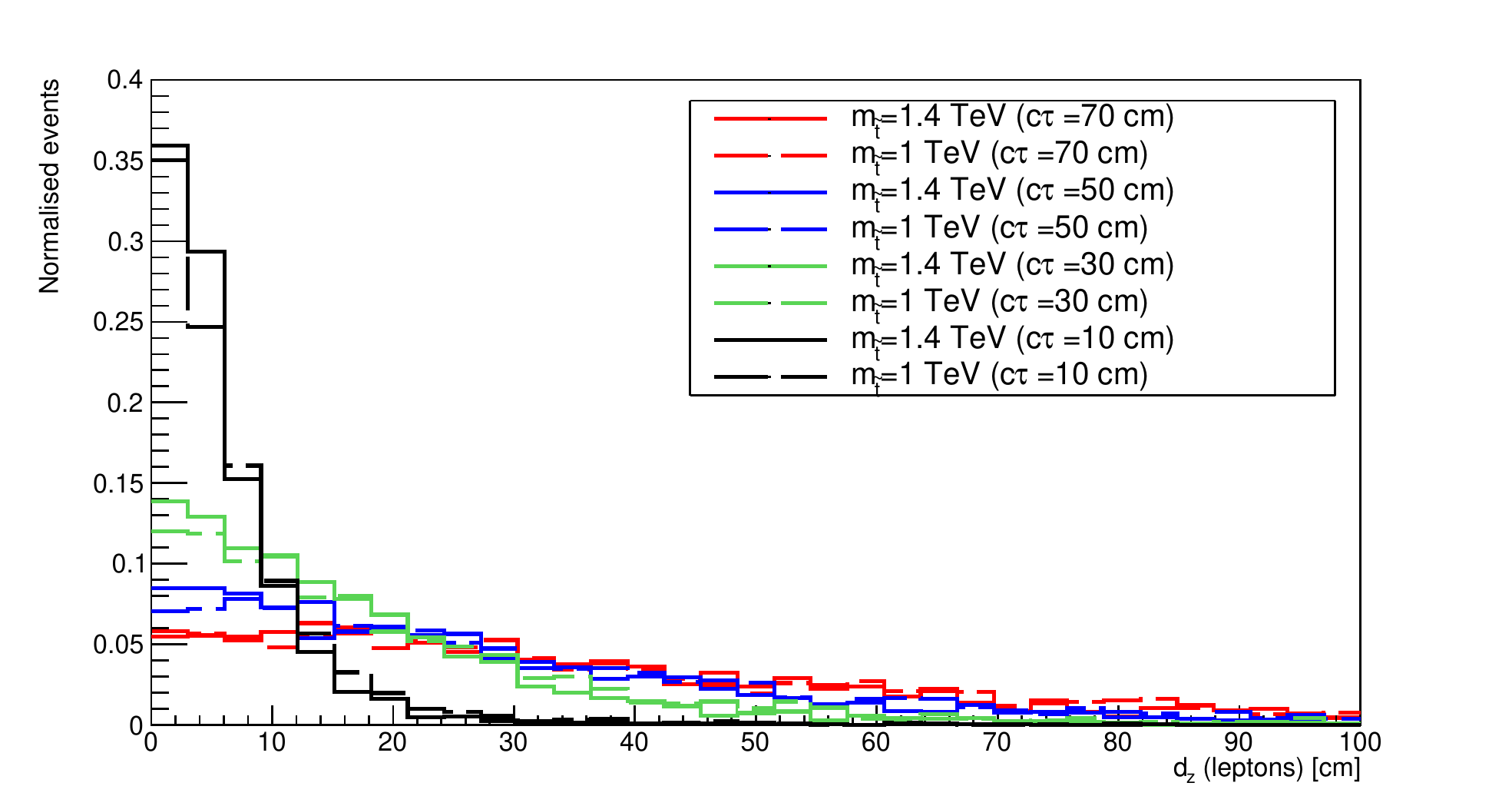}
    \caption{Distributions of $d_z$ of the leptons coming from the decay of top quark for the eight benchmarks.}
\end{figure}

\begin{figure}[H]
      \centering
      \includegraphics[width=0.8\linewidth]{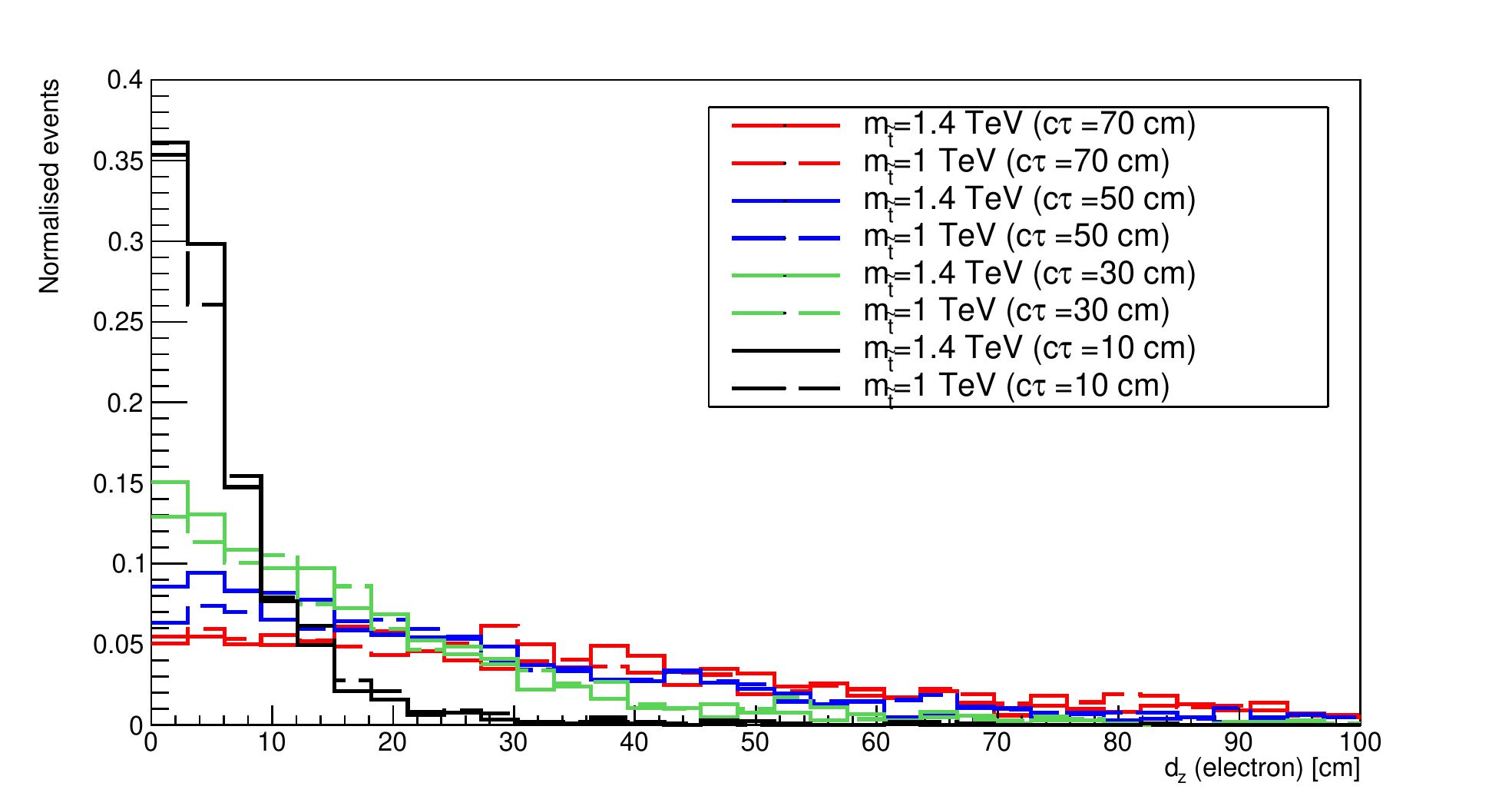}
    \caption{Distributions of $d_z$ of the electrons coming from the decay of top quark for the eight benchmarks.}
\end{figure}

\begin{figure}[H]
      \centering
      \includegraphics[width=0.8\linewidth]{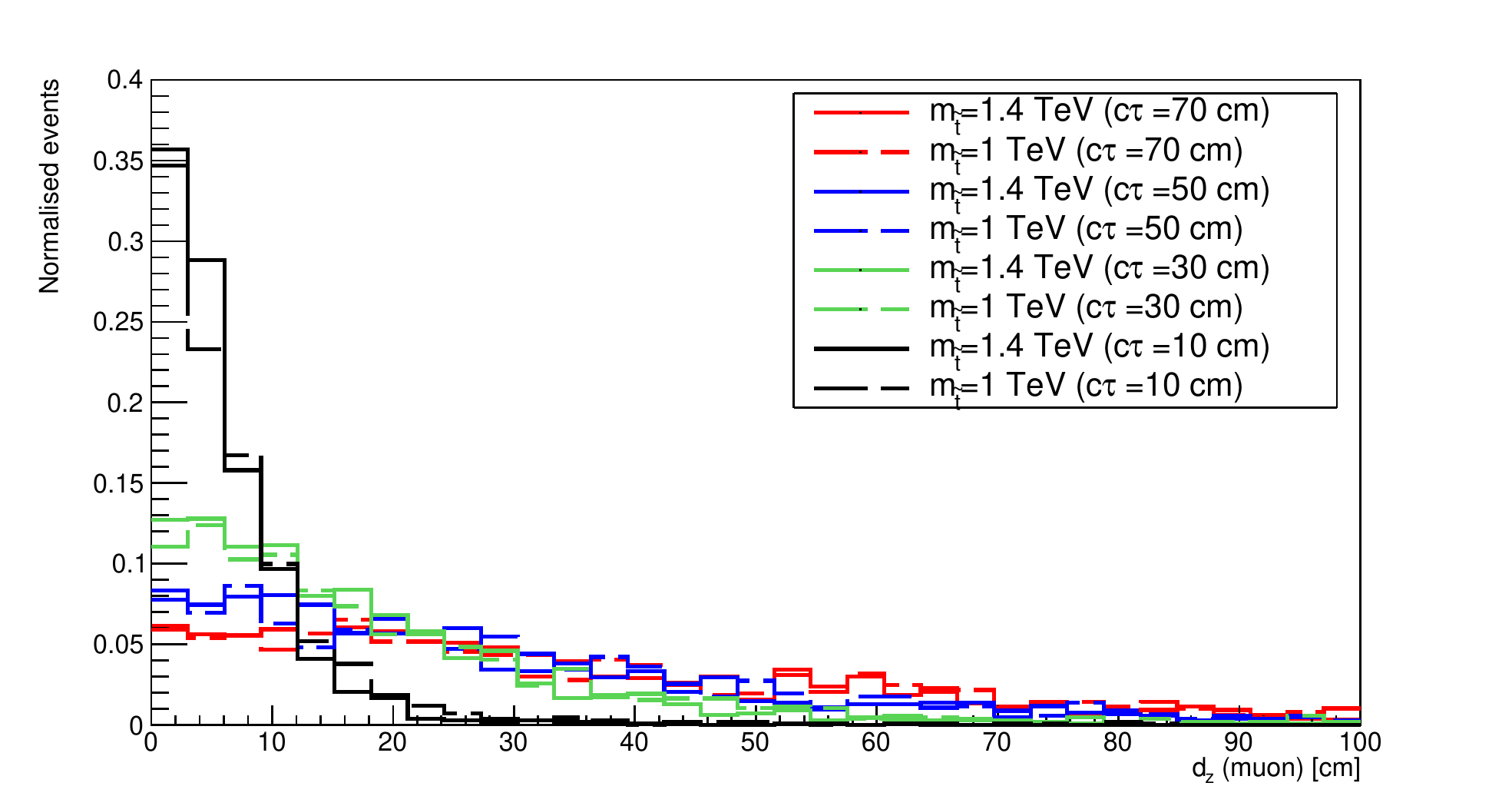}
    \caption{Distributions of $d_z$ of the muons coming from the decay of top quark for the eight benchmarks.}
\end{figure}

\section{Distributions of transverse momentum $p_T$}\label{app:pt}
The distribution of the transverse momentum of the leading jet $p_T(j)$ can be found in \autoref{fig:ptj}. The other distributions are shown below.
\begin{figure}[H]
      \centering
      \includegraphics[width=0.8\linewidth]{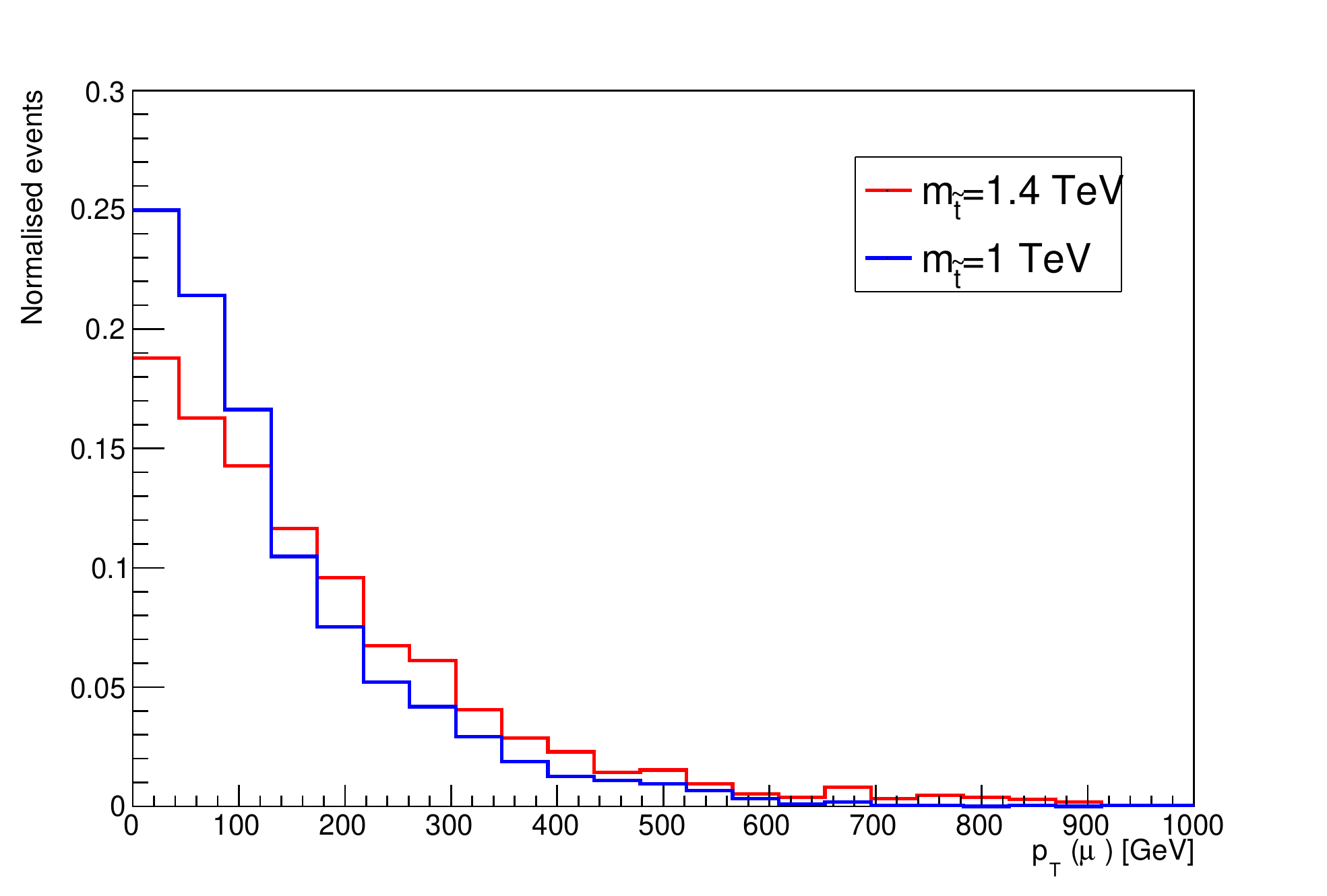}
    \caption{Distributions of the transverse momentum of the muons $p_T(\mu)$ for $m_{\tilde{t}}=1\ \text{TeV}$ and $m_{\tilde{t}}=1.4\ \text{TeV}$ (the choice of $m_{3/2}$ do not impact the kinematic).}
\end{figure}

\begin{figure}[H]
      \centering
      \includegraphics[width=0.8\linewidth]{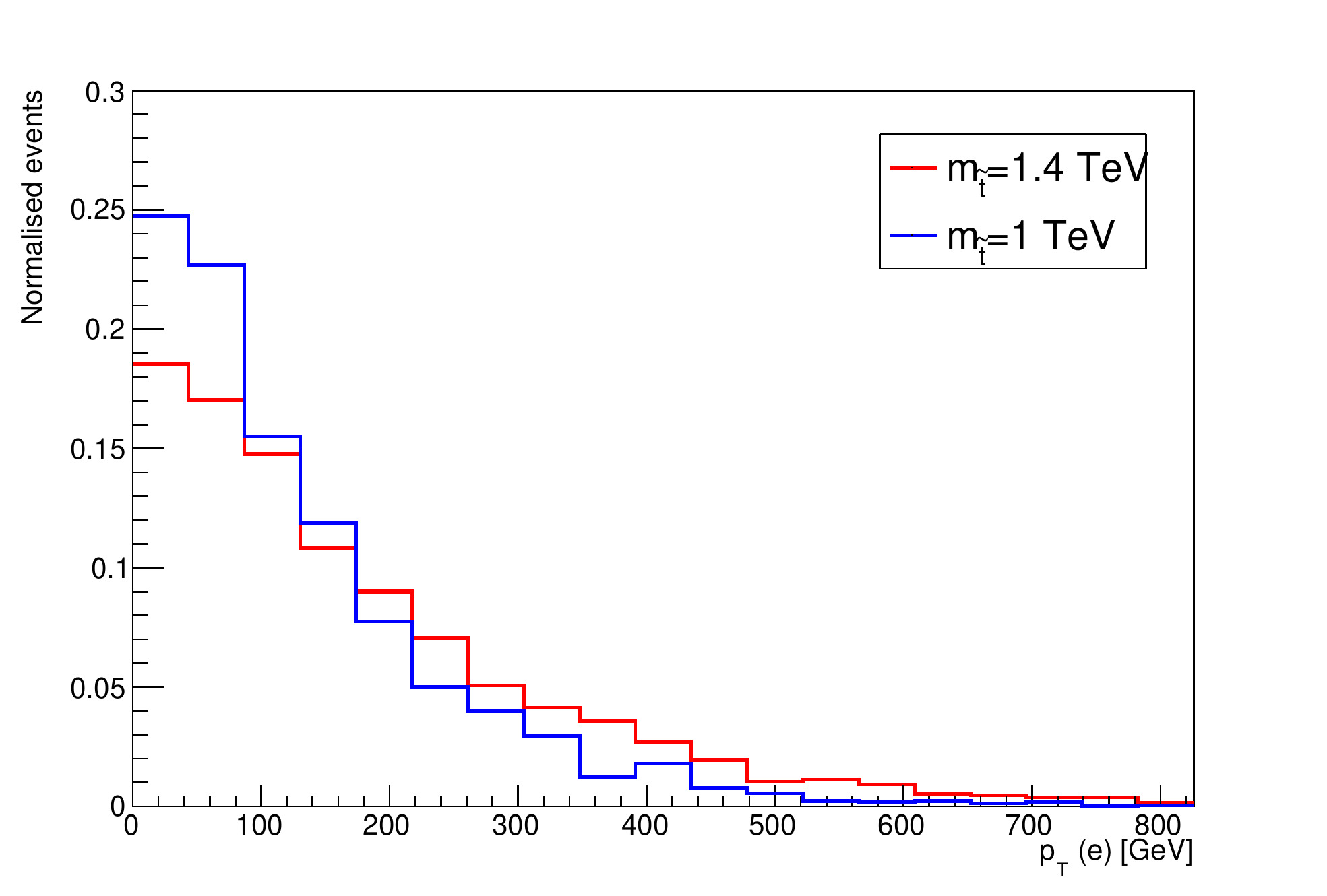}
    \caption{Distributions of the transverse momentum of the electrons $p_T(e)$ for $m_{\tilde{t}}=1\ \text{TeV}$ and $m_{\tilde{t}}=1.4\ \text{TeV}$ (the choice of $m_{3/2}$ do not impact the kinematic).}
\end{figure}

\begin{figure}[H]
      \centering
      \includegraphics[width=0.8\linewidth]{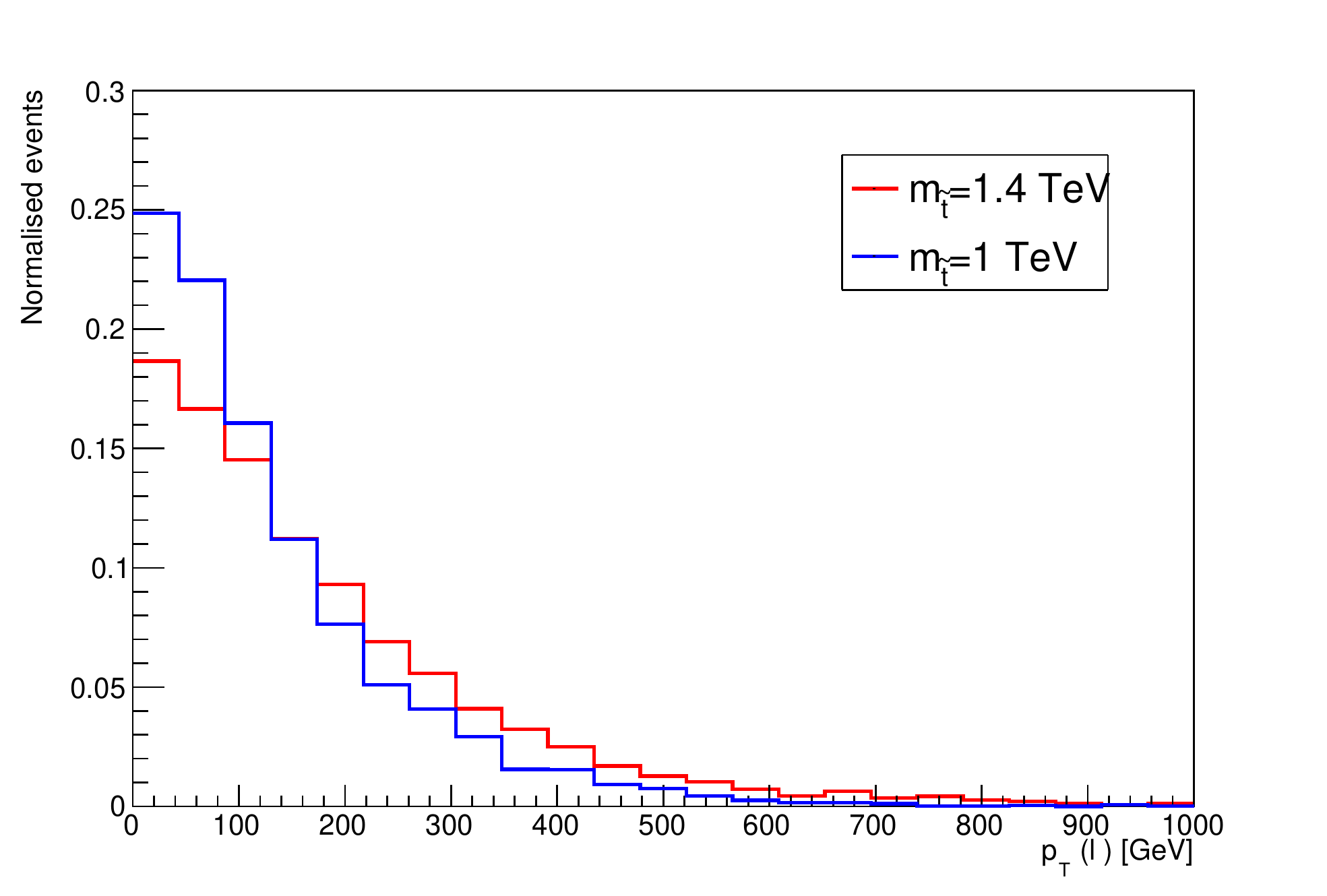}
    \caption{Distributions of the transverse momentum of the leptons $p_T(\ell)$ for $m_{\tilde{t}}=1\ \text{TeV}$ and $m_{\tilde{t}}=1.4\ \text{TeV}$ (the choice of $m_{3/2}$ do not impact the kinematic).}
\end{figure}

\section{Distributions of pseudorapidity $|\eta |$}\label{app:eta}
The pseudorapidity is usually defined as:
\begin{gather}
\eta = -\log\Big(\tan\frac{\theta}{2}\Big) \nn
\end{gather}
with $\theta$ the angle between the impulsion of the particle and the beam axis.
\begin{figure}[H]
      \centering
      \includegraphics[width=0.8\linewidth]{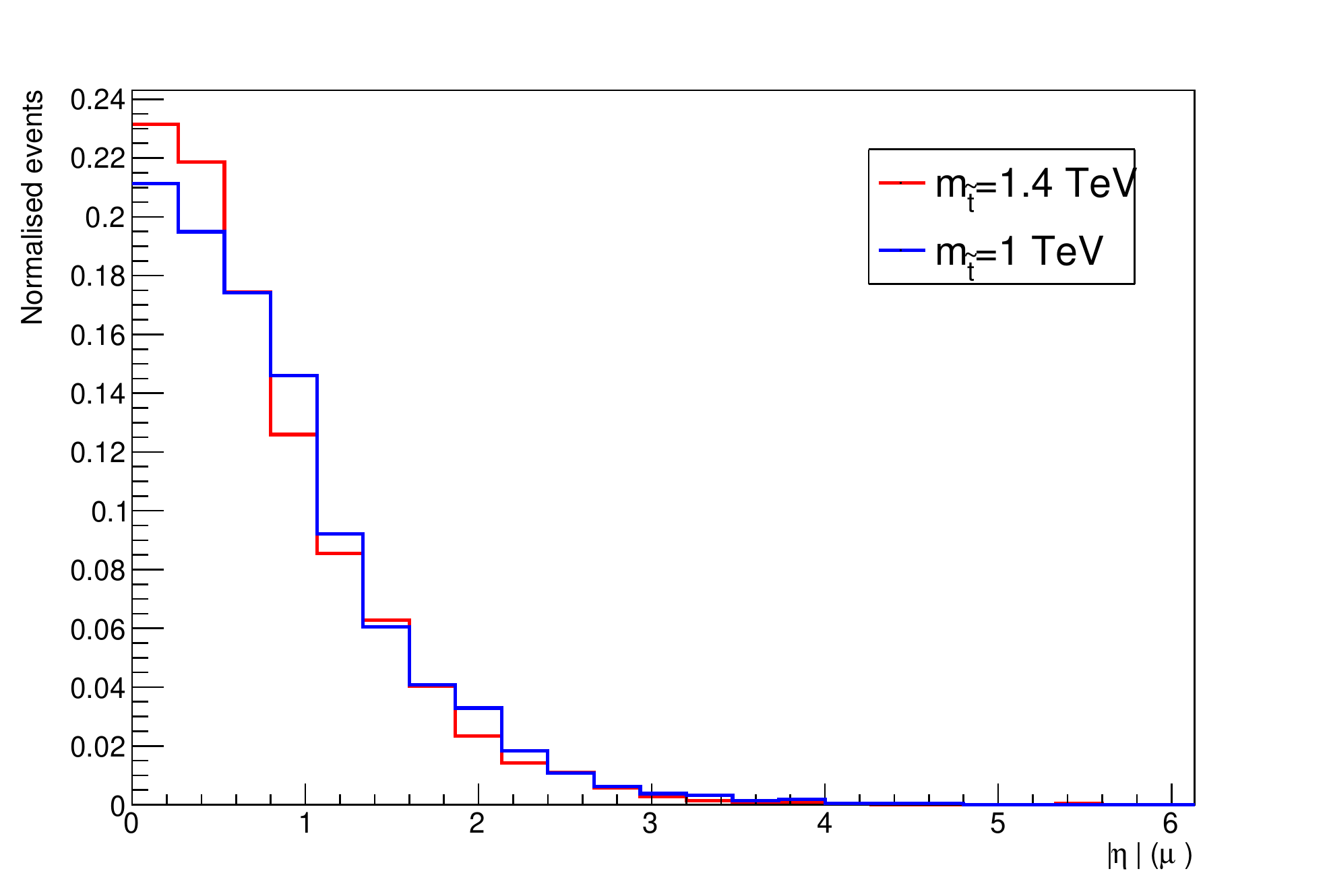}
    \caption{Distributions of the pseudorapidity $|\eta |$ of the muons coming from the decay of top quark for $m_{\tilde{t}}=1\ \text{TeV}$ and $m_{\tilde{t}}=1.4\ \text{TeV}$ (the choice of $m_{3/2}$ do not impact the kinematic).}
\end{figure}

\begin{figure}[H]
      \centering
      \includegraphics[width=0.8\linewidth]{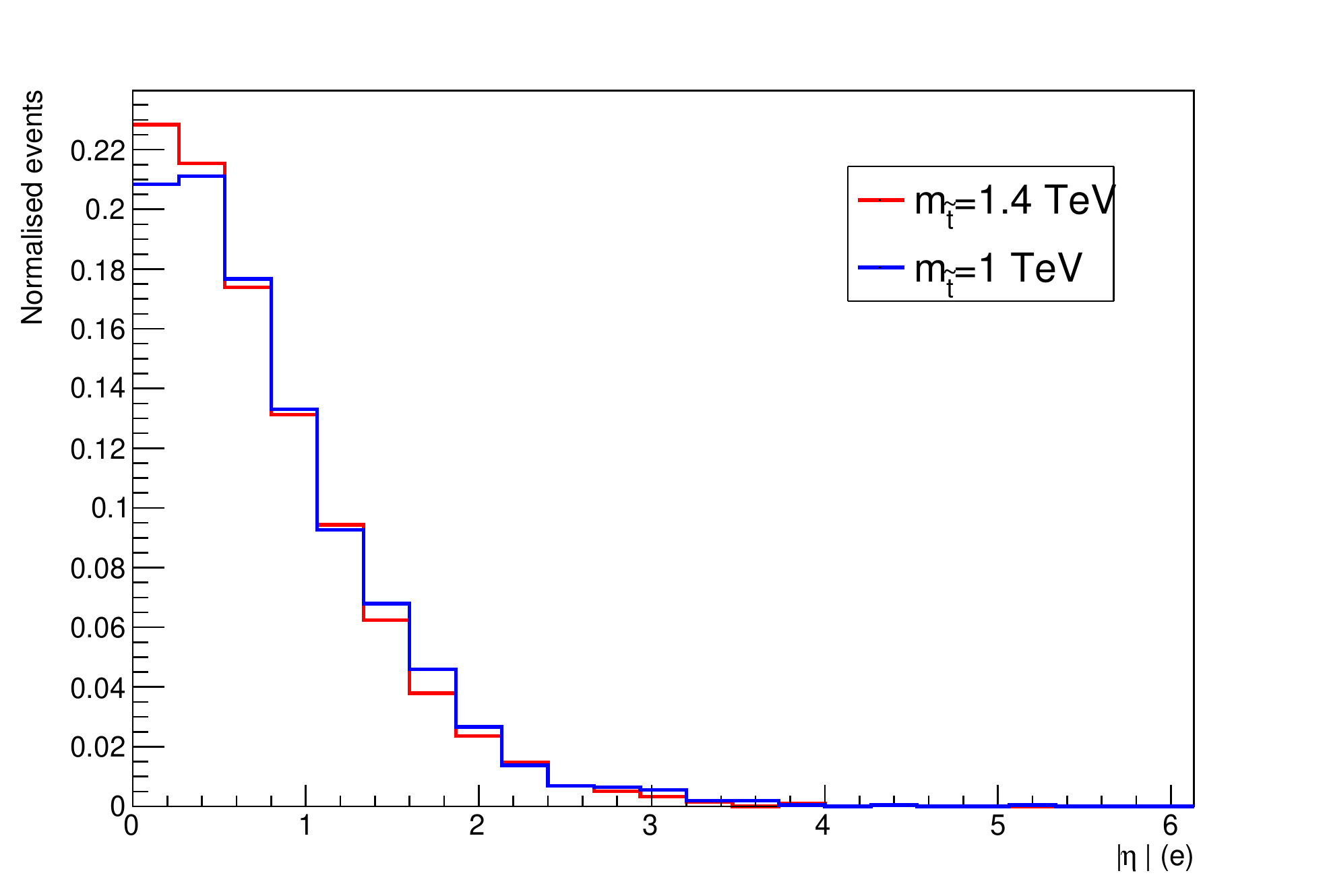}
    \caption{Distributions of the pseudorapidity $|\eta |$ of the electrons coming from the decay of top quark for $m_{\tilde{t}}=1\ \text{TeV}$ and $m_{\tilde{t}}=1.4\ \text{TeV}$ (the choice of $m_{3/2}$ do not impact the kinematic).}
\end{figure}

\begin{figure}[H]
      \centering
      \includegraphics[width=0.8\linewidth]{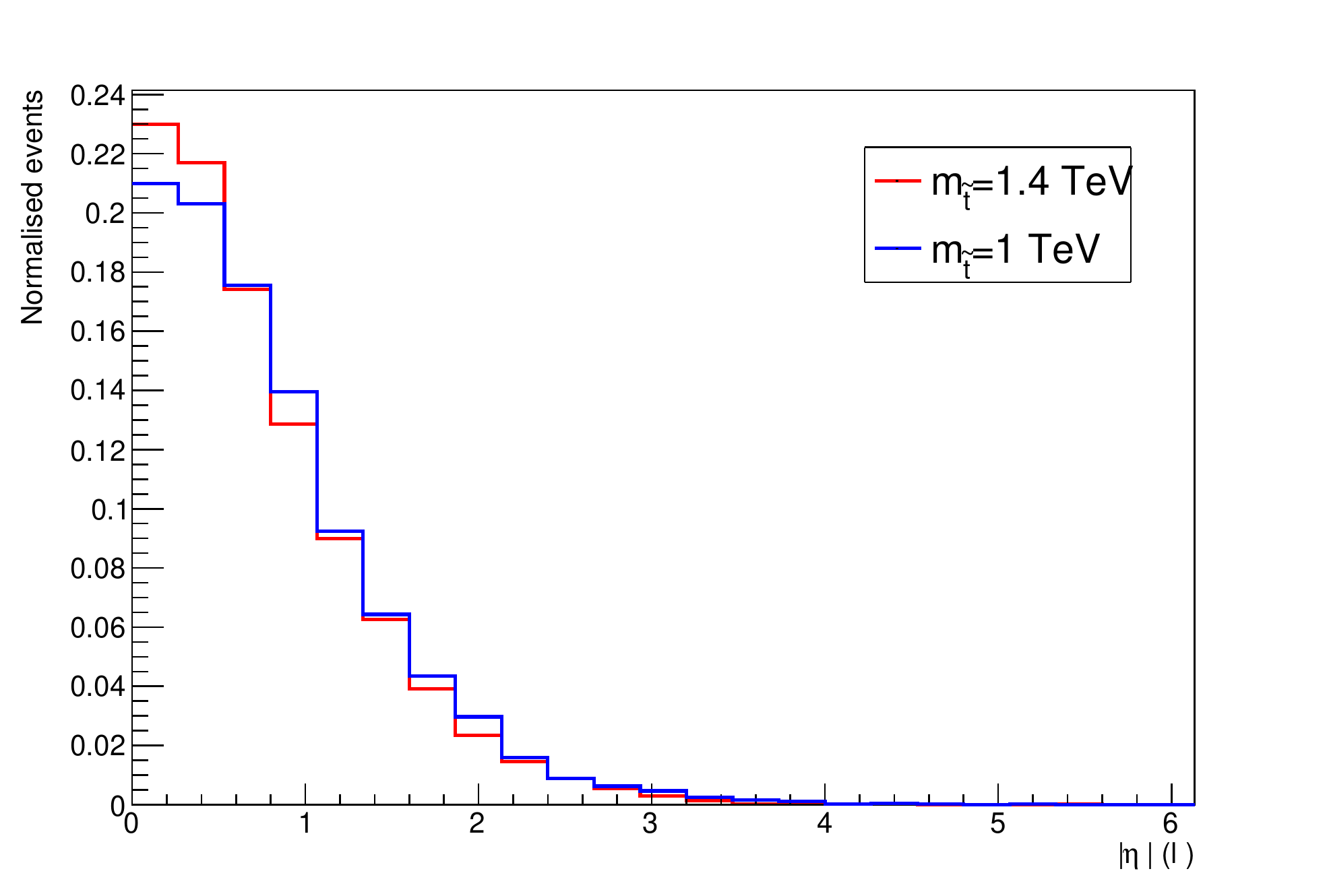}
    \caption{Distributions of the pseudorapidity $|\eta |$ of the leptons coming from the decay of top quark for $m_{\tilde{t}}=1\ \text{TeV}$ and $m_{\tilde{t}}=1.4\ \text{TeV}$ (the choice of $m_{3/2}$ do not impact the kinematic).}
\end{figure}

\begin{figure}[H]
      \centering
      \includegraphics[width=0.8\linewidth]{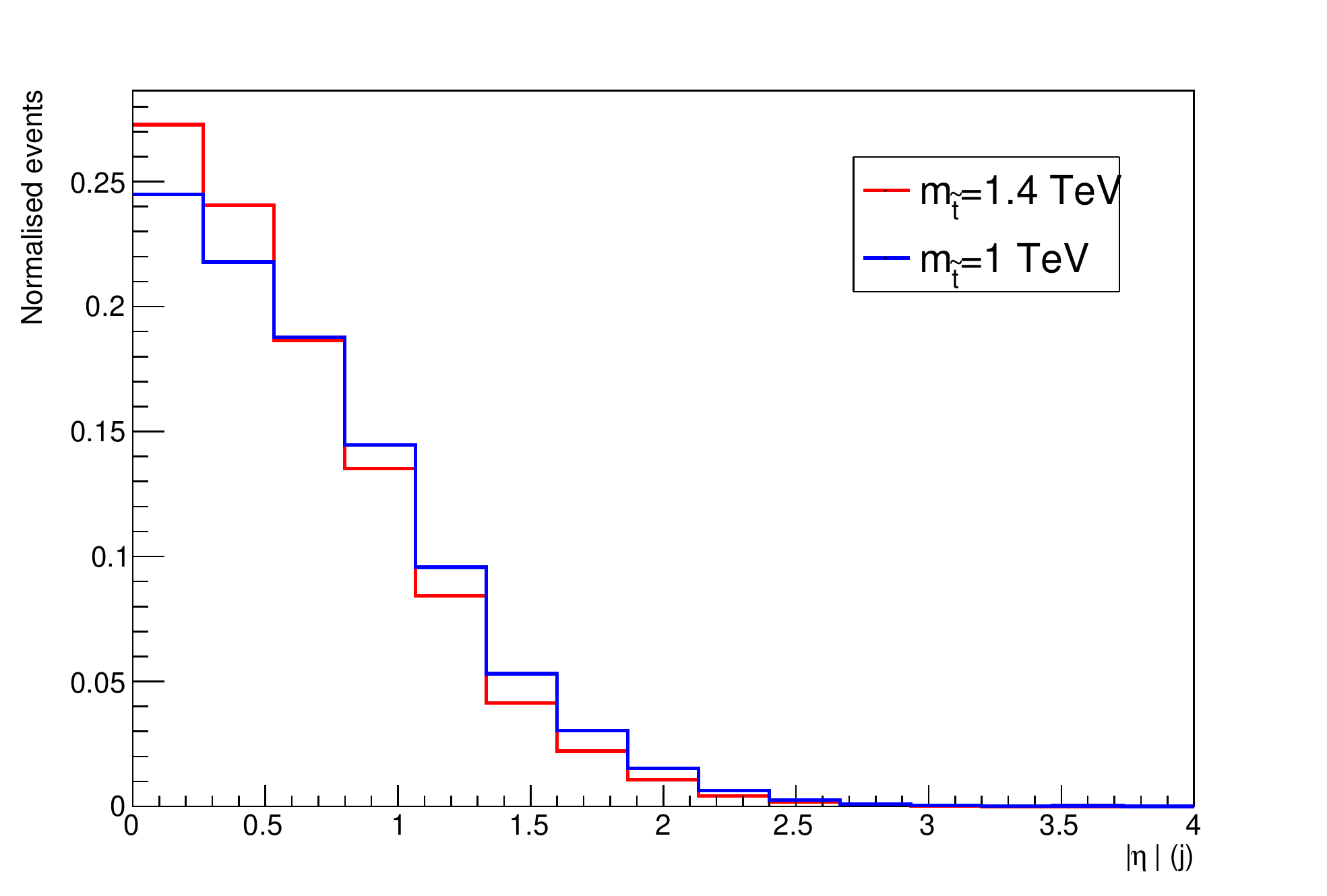}
    \caption{Distributions of the pseudorapidity $|\eta |$ of the leading jet for $m_{\tilde{t}}=1\ \text{TeV}$ and $m_{\tilde{t}}=1.4\ \text{TeV}$ (the choice of $m_{3/2}$ do not impact the kinematic).}
\end{figure}

\section{Distributions of global transverse observables}\label{app:transv}
The distribution of $MET$ can be found in \autoref{fig:met}. The other distributions are shown below.

We recall the various definitions for the total transverse energy $TET$, the total hadronic transverse energy $THT$ and the missing hadronic transverse energy $MHT$: 
\begin{gather}
TET = \sum_{visisble\ particles}||\vec{p}_T|| \quad , \quad THT =  \sum_{hadronic}||\vec{p}_T|| \quad , \quad MHT= \big|\big|\sum_{hadronic}\vec{p}_T\big|\big| \ .\nn  
\end{gather}

\begin{figure}[H]
      \centering
      \includegraphics[width=0.8\linewidth]{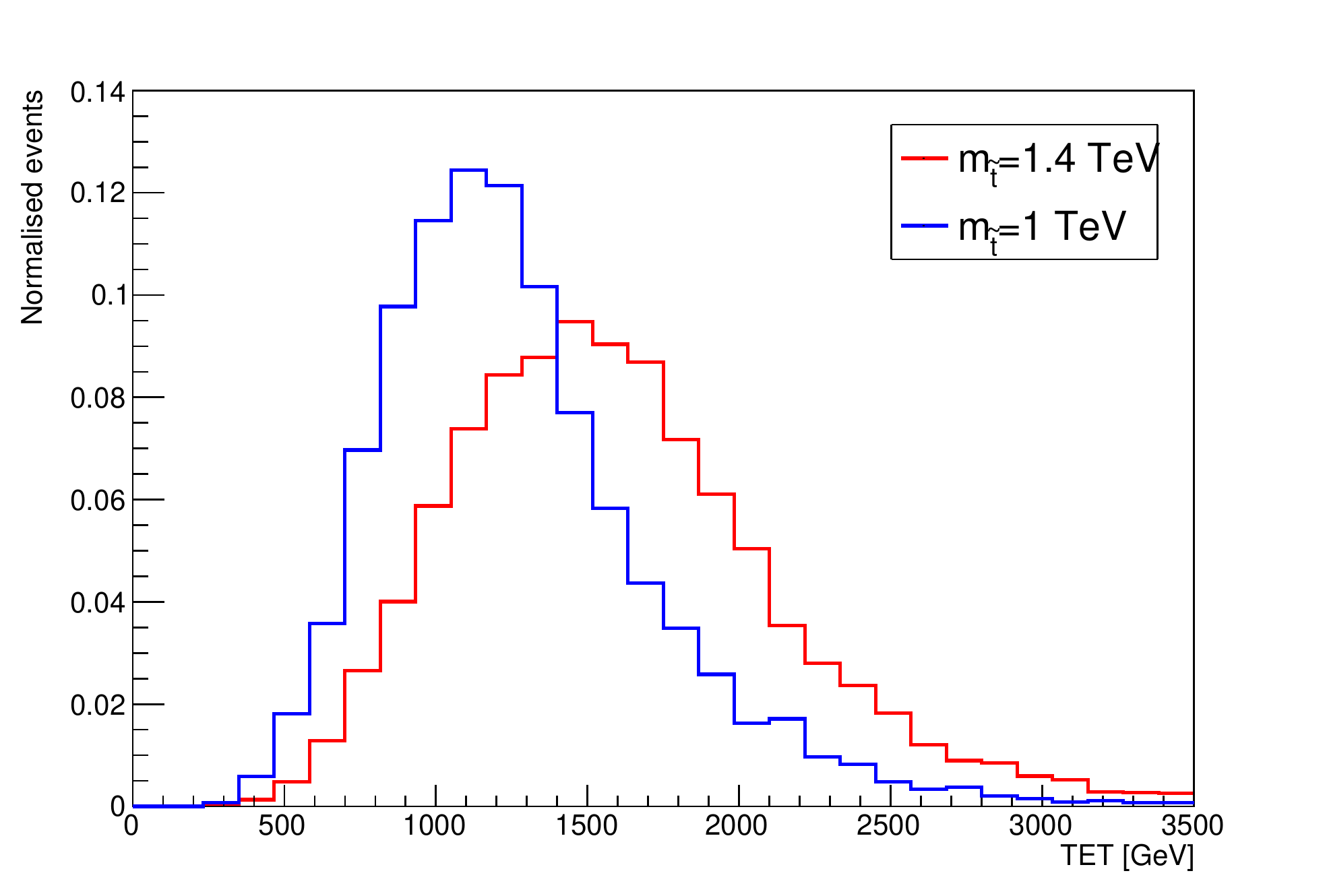}
    \caption{Distributions of total transverse energy $TET$ for $m_{\tilde{t}}=1\ \text{TeV}$ and $m_{\tilde{t}}=1.4\ \text{TeV}$ (the choice of $m_{3/2}$ do not impact the kinematic).}
\end{figure}

\begin{figure}[H]
      \centering
      \includegraphics[width=0.8\linewidth]{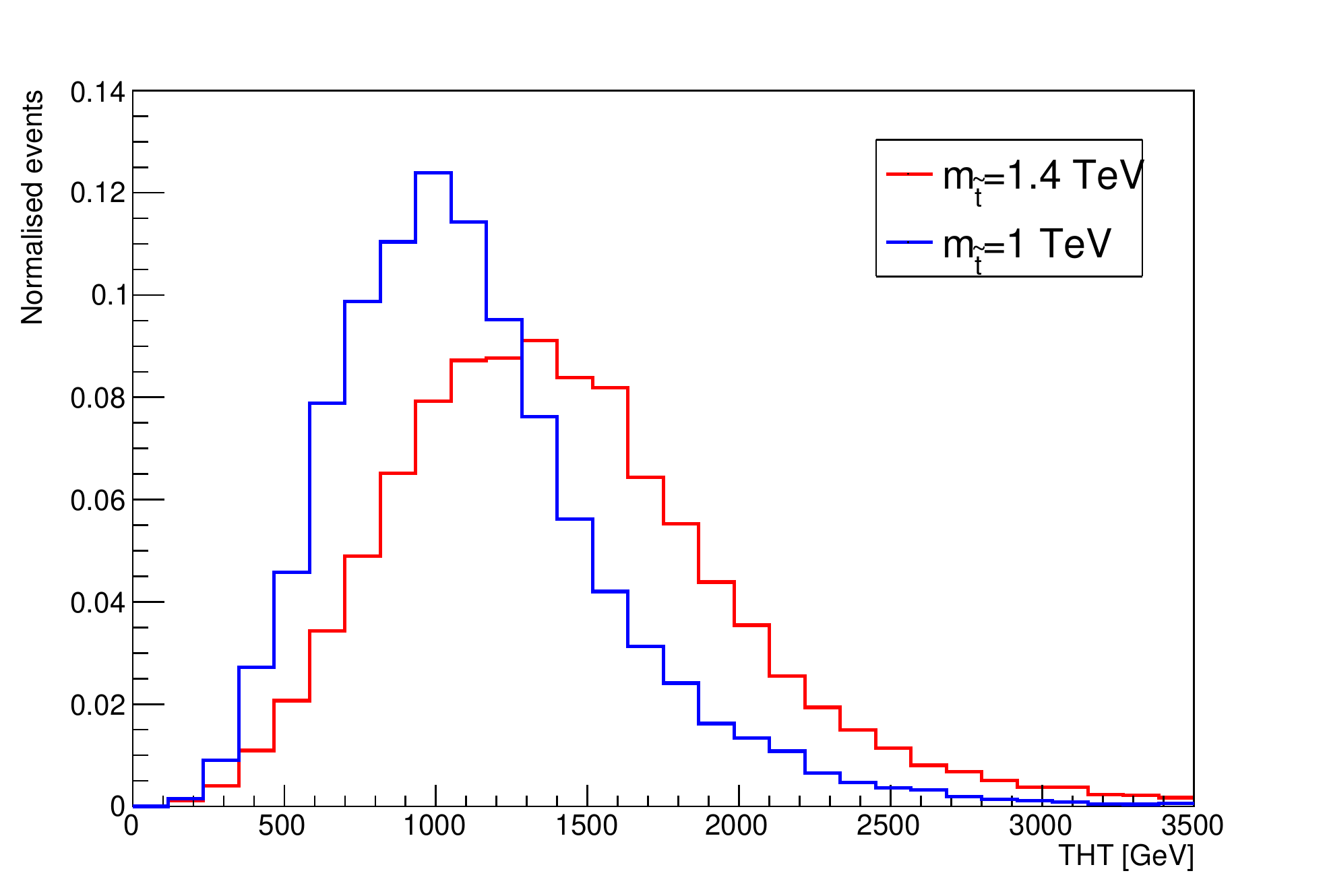}
    \caption{Distributions of the total hadronic transverse energy $THT$ for $m_{\tilde{t}}=1\ \text{TeV}$ and $m_{\tilde{t}}=1.4\ \text{TeV}$ (the choice of $m_{3/2}$ do not impact the kinematic).}
\end{figure}

\begin{figure}[H]
      \centering
      \includegraphics[width=0.8\linewidth]{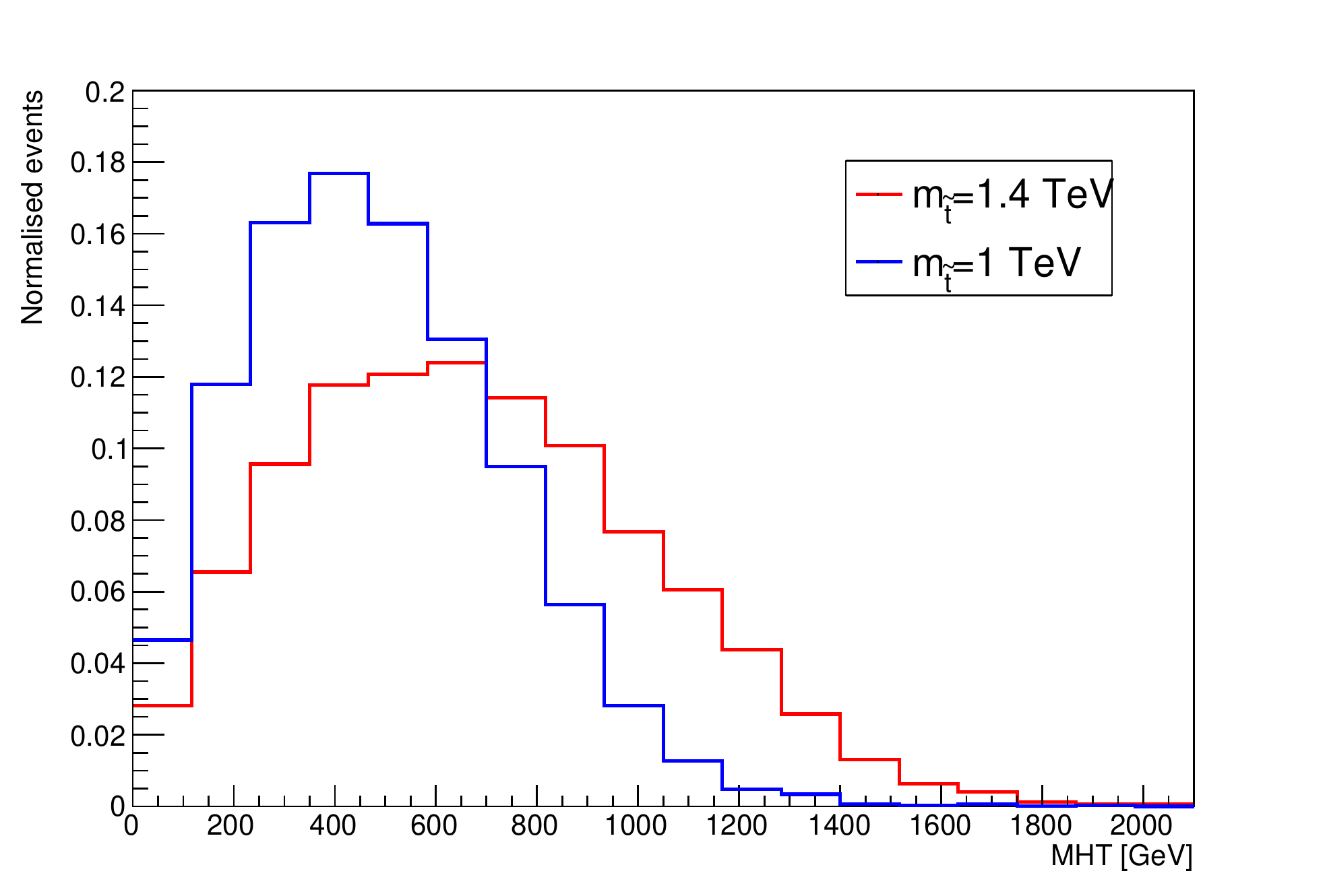}
    \caption{Distributions of $MHT$ for $m_{\tilde{t}}=1\ \text{TeV}$ and $m_{\tilde{t}}=1.4\ \text{TeV}$ (the choice of $m_{3/2}$ do not impact the kinematic).}
\end{figure}

\section{Other observables}\label{app:distribother}
The distribution of $\Delta R_1$ is already shown in \autoref{fig:dr1}. The other distributions ($\Delta R_2$ and $\alpha_T$ already defined in Subsection \ref{subsec:obs}) are shown below.
\begin{figure}[H]
      \centering
      \includegraphics[width=0.8\linewidth]{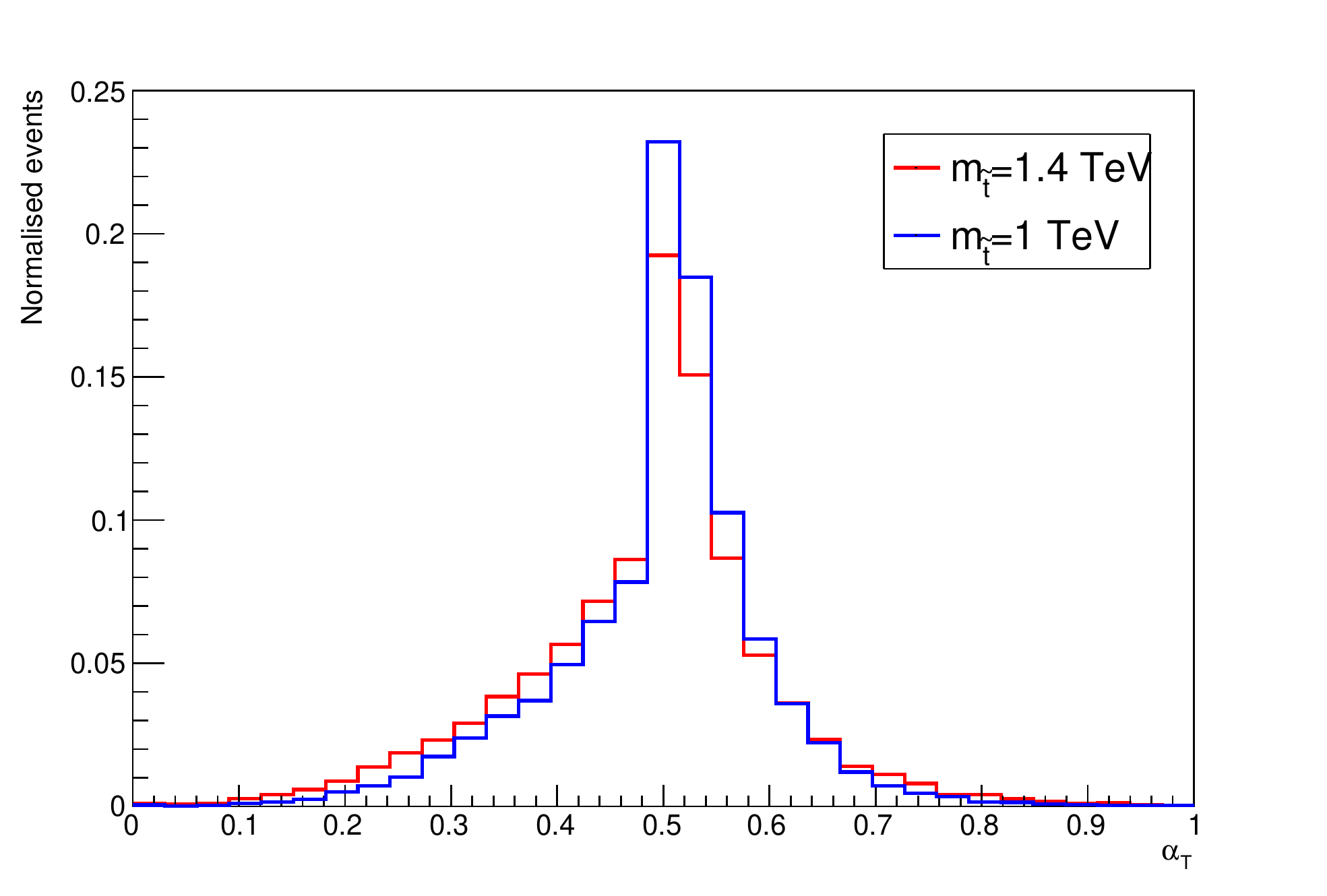}
    \caption{Distributions of the variable $\alpha_T$ for $m_{\tilde{t}}=1\ \text{TeV}$ and $m_{\tilde{t}}=1.4\ \text{TeV}$ (the choice of $m_{3/2}$ do not impact the kinematic).}
\end{figure}

\begin{figure}[H]
      \centering
      \includegraphics[width=0.8\linewidth]{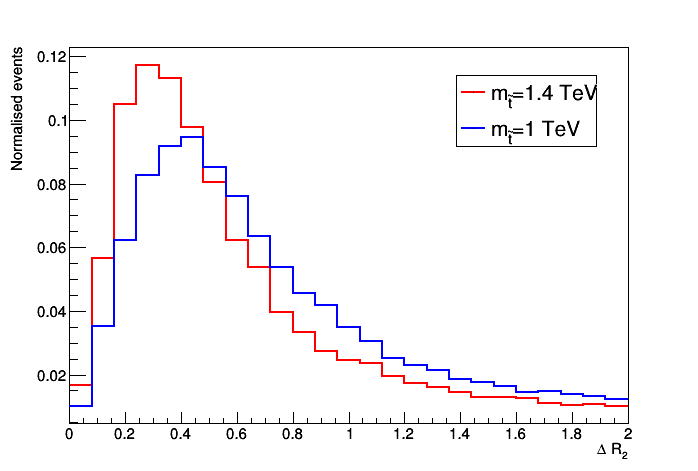}
    \caption{Distributions of the distance between b-quarks and the second closest quarks $\Delta R_2$ for $m_{\tilde{t}}=1\ \text{TeV}$ and $m_{\tilde{t}}=1.4\ \text{TeV}$ (the choice of $m_{3/2}$ do not impact the kinematic).}
\end{figure}

\end{appendices}

\bibliography{bib} 

\begin{thebibliography}{100}

\bibitem{cmsH}
The~CMS Collaboration.
\newblock Observation of a new boson at a mass of 125 gev with the cms
  experiment at the lhc.
\newblock {\em Physics Letters B}, 716(1):30--61, September 2012.
\newblock arXiv: 1207.7235.

\bibitem{atlasH}
The~ATLAS Collaboration.
\newblock Observation of a new particle in the search for the {Standard}
  {Model} {Higgs} boson with the {ATLAS} detector at the {LHC}.
\newblock {\em Physics Letters B}, 716(1):1--29, September 2012.
\newblock arXiv: 1207.7214.

\bibitem{martin_supersymmetry_1998}
Stephen~P. Martin.
\newblock A {Supersymmetry} {Primer}.
\newblock {\em arXiv:hep-ph/9709356}, 18:1--98, July 1998.
\newblock arXiv: hep-ph/9709356.

\bibitem{book_susy}
Benjamin Fuks and Michel Rausch~de Traubenberg.
\newblock {\em Supersymétrie, Exercices avec solutions}.
\newblock Ellipses, 2011.

\bibitem{review}
Robin Ducrocq, Michel Rausch~de Traubenberg, and Mauricio Valenzuela.
\newblock A pedagogical discussion of n = 1 four-dimensional supergravity in
  superspace.
\newblock {\em Modern Physics Letters A}, 0(0):2130015, 0.

\bibitem{WessBagger}
Julius Wess and Jonathan Bagger.
\newblock {\em Supersymmetry and Supergravity: Revised Edition}, volume 103.
\newblock Princeton University Press, rev - revised edition, 1992.

\bibitem{book_sugra}
Michel Rausch~de Traubenberg and Mauricio Valenzuela.
\newblock {\em A Supergravity Primer: From Geometrical Principles to the Final
  Lagrangian}.
\newblock World Scientific, 11 2019.

\bibitem{soni_analysis_1983}
Sanjeev~K. Soni and H.~Arthur Weldon.
\newblock Analysis of the supersymmetry breaking induced by {N} = 1
  supergravity theories.
\newblock {\em Physics Letters B}, 126(3):215--219, June 1983.

\bibitem{moultaka_low_2018}
Gilbert Moultaka, Michel~Rausch de~Traubenberg, and Damien Tant.
\newblock Low {Energy} {Supergravity} {Revisited} ({I}).
\newblock {\em arXiv:1611.10327 [hep-ph, physics:hep-th]}, November 2018.
\newblock arXiv: 1611.10327.

\bibitem{MSSMRPV}
R.~Barbier, C.~Bérat, M.~Besançon, M.~Chemtob, A.~Deandrea, E.~Dudas,
  P.~Fayet, S.~Lavignac, G.~Moreau, E.~Perez, and Y.~Sirois.
\newblock R-parity-violating supersymmetry.
\newblock {\em Physics Reports}, 420(1):1--195, 2005.

\bibitem{domainwall}
Graciela~B. Gelmini, Marcelo Gleiser, and Edward~W. Kolb.
\newblock Cosmology of biased discrete symmetry breaking.
\newblock {\em Phys. Rev. D}, 39:1558--1566, Mar 1989.

\bibitem{domwalsol1}
S.A. Abel.
\newblock Destabilising divergences in the nmssm.
\newblock {\em Nuclear Physics B}, 480(1-2):55–72, Nov 1996.

\bibitem{domwalsol2}
C.~Panagiotakopoulos and K.~Tamvakis.
\newblock Stabilized nmssm without domain walls.
\newblock {\em Physics Letters B}, 446(3-4):224–227, Jan 1999.

\bibitem{Staub_sarah}
Florian Staub.
\newblock Sarah 4: A tool for (not only susy) model builders.
\newblock {\em Computer Physics Communications}, 185(6):1773–1790, Jun 2014.

\bibitem{Porod_spheno}
W.~Porod.
\newblock Spheno, a program for calculating supersymmetric spectra, susy
  particle decays and susy particle production at e+e- colliders.
\newblock {\em Computer Physics Communications}, 153(2):275–315, Jun 2003.

\bibitem{madgraph}
J.~Alwall, R.~Frederix, S.~Frixione, V.~Hirschi, F.~Maltoni, O.~Mattelaer,
  H.-S. Shao, T.~Stelzer, P.~Torrielli, and M.~Zaro.
\newblock The automated computation of tree-level and next-to-leading order
  differential cross sections, and their matching to parton shower simulations.
\newblock {\em Journal of High Energy Physics}, 2014(7), Jul 2014.

\bibitem{spinstat}
W.~Pauli.
\newblock The connection between spin and statistics.
\newblock {\em Phys. Rev.}, 58:716--722, Oct 1940.

\bibitem{Noether}
Emmy Noether.
\newblock Invariant variation problems.
\newblock {\em Transport Theory and Statistical Physics}, 1(3):186–207, Jan
  1971.

\bibitem{ColemanMandula}
Sidney Coleman and Jeffrey Mandula.
\newblock All possible symmetries of the $s$ matrix.
\newblock {\em Phys. Rev.}, 159:1251--1256, Jul 1967.

\bibitem{HLS}
Rudolf Haag, Jan~T. Lopuszanski, and Martin Sohnius.
\newblock {All Possible Generators of Supersymmetries of the s Matrix}.
\newblock {\em Nucl. Phys. B}, 88:257, 1975.

\bibitem{groupmichel}
Rutwig Campoamor-Stursberg and Michel Rausch~de Traubenberg.
\newblock {\em Group Theory in Physics}.
\newblock World Scientific, 2018.

\bibitem{freedman_van_proeyen}
Daniel~Z. Freedman and Antoine Van~Proeyen.
\newblock {\em Supergravity}.
\newblock Cambridge University Press, 2012.

\bibitem{superspace}
Abdus Salam and J.~Strathdee.
\newblock Super-gauge transformations.
\newblock {\em Nuclear Physics B}, 76(3):477--482, 1974.

\bibitem{GRIMM_torsion_constraints}
R.~Grimm, J.~Wess, and B.~Zumino.
\newblock Consistency checks on the superspace formulation of supergravity.
\newblock {\em Physics Letters B}, 73(4):415--417, 1978.

\bibitem{WESS_ZUMINO_torsion_constraints}
J.~Wess and B.~Zumino.
\newblock Superfield lagrangian for supergravity.
\newblock {\em Physics Letters B}, 74(1):51--53, 1978.

\bibitem{Dragon}
N.~Dragon.
\newblock {Torsion and Curvature in Extended Supergravity}.
\newblock {\em Z. Phys. C}, 2:29--32, 1979.

\bibitem{HoweTucker}
P.S. Howe and R.W. Tucker.
\newblock Scale invariance in superspace.
\newblock {\em Physics Letters B}, 80(1):138--140, 1978.

\bibitem{siegelsuperconforme}
Warren Siegel.
\newblock Superconformal invariance of superspace with nonminimal auxiliary
  fields.
\newblock {\em Physics Letters B}, 80(3):224--227, 1979.

\bibitem{new-minimal}
Martin~F. Sohnius and Peter~C. West.
\newblock An alternative minimal off-shell version of n = 1 supergravity.
\newblock {\em Physics Letters B}, 105(5):353--357, 1981.

\bibitem{non-minimal1}
P.~Breitenlohner.
\newblock A geometric interpretation of local supersymmetry.
\newblock {\em Physics Letters B}, 67(1):49--51, 1977.

\bibitem{non-minimal2}
P.~Breitenlohner.
\newblock Some invariant lagrangians for local supersymmetry.
\newblock {\em Nuclear Physics B}, 124(4):500--510, 1977.

\bibitem{WessZumino}
J.~Wess and B.~Zumino.
\newblock The component formalism follows from the superspace formulation of
  supergravity.
\newblock {\em Physics Letters B}, 79(4):394--398, 1978.

\bibitem{BransDicke}
C.~Brans and R.~H. Dicke.
\newblock Mach's principle and a relativistic theory of gravitation.
\newblock {\em Phys. Rev.}, 124:925--935, Nov 1961.

\bibitem{Fayet}
P.~Fayet and J.~Iliopoulos.
\newblock Spontaneously broken supergauge symmetries and goldstone spinors.
\newblock {\em Physics Letters B}, 51(5):461--464, 1974.

\bibitem{ORaifeartaigh}
L.~O'Raifeartaigh.
\newblock {Spontaneous Symmetry Breaking for Chiral Scalar Superfields}.
\newblock {\em Nucl. Phys. B}, 96:331--352, 1975.

\bibitem{GMSB}
Michael Dine and Willy Fischler.
\newblock A phenomenological model of particle physics based on supersymmetry.
\newblock {\em Physics Letters B}, 110(3):227--231, 1982.

\bibitem{AMSB}
Jonathan~A Bagger, Takeo Moroi, and Erich Poppitz.
\newblock Anomaly mediation in supergravity theories.
\newblock {\em Journal of High Energy Physics}, 2000(04):009–009, Apr 2000.

\bibitem{ChamseddineGMSB}
A.~H. Chamseddine, R.~Arnowitt, and Pran Nath.
\newblock Locally supersymmetric grand unification.
\newblock {\em Phys. Rev. Lett.}, 49:970--974, Oct 1982.

\bibitem{BFSGMSB}
R.~Barbieri, S.~Ferrara, and C.A. Savoy.
\newblock Gauge models with spontaneously broken local supersymmetry.
\newblock {\em Physics Letters B}, 119(4):343--347, 1982.

\bibitem{noscale1}
John~R. Ellis, C.~Kounnas, and Dimitri~V. Nanopoulos.
\newblock {No Scale Supersymmetric Guts}.
\newblock {\em Nucl. Phys. B}, 247:373--395, 1984.

\bibitem{noscale2}
A.~B. Lahanas and Dimitri~V. Nanopoulos.
\newblock {The Road to No Scale Supergravity}.
\newblock {\em Phys. Rept.}, 145:1, 1987.

\bibitem{noscale3}
E~Cremmer, Sergio Ferrara, Costas Kounnas, and Dimitri~V Nanopoulos.
\newblock {Naturally vanishing cosmological constant in N = 1 supergravity}.
\newblock {\em Phys. Lett. B}, 133:61--66. 12 p, Jul 1983.

\bibitem{tant:tel-01546150}
Damien Tant.
\newblock {\em {Nouvelles solutions et classification du superpotentiel et du
  potentiel de K{\''a}hler compatibles avec une brisure de la supersym{\'e}trie
  {\`a} basse {\'e}nergie induite par la gravitation}}.
\newblock Theses, {Universit{\'e} de Strasbourg}, December 2016.

\bibitem{GIUDICEMASIERO}
G.F. Giudice and A.~Masiero.
\newblock A natural solution to the $\mu$-problem in supergravity theories.
\newblock {\em Physics Letters B}, 206(3):480--484, 1988.

\bibitem{BRIGNOLE}
A.~Brignole, L.~E. Ib\'{a}\~{n}ez, and C.~Mu\~{n}oz.
\newblock Soft supersymmetry–breaking terms from supergravity and superstring
  models.
\newblock {\em Advanced Series on Directions in High Energy Physics}, page
  125–148, Jul 1998.

\bibitem{NILLESMSSM}
H.P. Nilles.
\newblock Supersymmetry, supergravity and particle physics.
\newblock {\em Physics Reports}, 110(1):1--162, 1984.

\bibitem{FAYETMSSM1}
P.~Fayet.
\newblock Spontaneously broken supersymmetric theories of weak, electromagnetic
  and strong interactions.
\newblock {\em Physics Letters B}, 69(4):489--494, 1977.

\bibitem{FAYETMSSM2}
Pierre Fayet.
\newblock {Supersymmetry and Weak, Electromagnetic and Strong Interactions}.
\newblock {\em Phys. Lett. B}, 64:159, 1976.

\bibitem{protondecay}
Gautam Bhattacharyya and Palash~B. Pal.
\newblock New constraints on r-parity violation from proton stability.
\newblock {\em Physics Letters B}, 439(1):81--84, 1998.

\bibitem{Fayetnmssm}
Pierre Fayet.
\newblock Supergauge invariant extension of the higgs mechanism and a model for
  the electron and its neutrino.
\newblock {\em Nuclear Physics B}, 90:104--124, 1975.

\bibitem{NMSSM}
Ulrich Ellwanger, Cyril Hugonie, and Ana~M. Teixeira.
\newblock The next-to-minimal supersymmetric standard model.
\newblock {\em Physics Reports}, 496(1-2):1--77, November 2010.
\newblock arXiv: 0910.1785.

\bibitem{NSW_Higgs}
Robin Ducrocq, Gilbert Moultaka, and Michel Rausch~de Traubenberg.
\newblock Low energy supergravity revisited (ii).
\newblock In preparation.

\bibitem{destabilSUGRA1}
Vidyut Jain.
\newblock On destabilizing divergences in supergravity models.
\newblock {\em Physics Letters B}, 351(4):481–486, Jun 1995.

\bibitem{destabilSUGRA2}
Jonathan Bagger, Erich Poppitz, and Lisa Randall.
\newblock Destabilizing divergences in supergravity theories at two loops.
\newblock {\em Nuclear Physics B}, 455(1-2):59–82, Nov 1995.

\bibitem{CWpotential}
Sidney Coleman and Erick Weinberg.
\newblock Radiative corrections as the origin of spontaneous symmetry breaking.
\newblock {\em Phys. Rev. D}, 7:1888--1910, Mar 1973.

\bibitem{DerendingerCW}
Jean-Pierre Derendinger.
\newblock {Lecture notes on globally supersymmetric theories in four-dimensions
  and two-dimensions}.
\newblock In {\em {3rd Hellenic School on Elementary Particle Physics}}, pages
  0111--243, 7 1990.

\bibitem{Split1}
G.F. Giudice and A.~Romanino.
\newblock Erratum to: “split supersymmetry” [nucl. phys. b 699 (2004) 65].
\newblock {\em Nuclear Physics B}, 706(1-2):487, Jan 2005.

\bibitem{Split2}
N.~Arkani-Hamed, S.~Dimopoulos, G.F. Giudice, and A.~Romanino.
\newblock Aspects of split supersymmetry.
\newblock {\em Nuclear Physics B}, 709(1-2):3–46, Mar 2005.

\bibitem{N2MSSM_Pheno}
Robin Ducrocq, Cyril Hugonie, and Eric Conte.
\newblock Phenomenological study of the n2mssm.
\newblock In preparation.

\bibitem{DJOUADI_2008}
A~DJOUADI.
\newblock The anatomy of electroweak symmetry breaking tome ii: The higgs
  bosons in the minimal supersymmetric model.
\newblock {\em Physics Reports}, 459(1-6):1–241, Apr 2008.

\bibitem{thesis_student_ellwanger}
Matias Rodriguez~Vazquez.
\newblock {\em {Search for supplementary Higgs Bosons at the LHC}}.
\newblock Theses, {Universit{\'e} Paris Saclay (COmUE)}, September 2017.

\bibitem{higgs_limit_114}
S.~Schael, R.~Barate, R.~Brunelière, I.~De~Bonis, D.~Decamp, C.~Goy,
  S.~Jézéquel, J.-P. Lees, F.~Martin, and et~al.
\newblock Search for neutral mssm higgs bosons at lep.
\newblock {\em The European Physical Journal C}, 47(3), Jul 2006.

\bibitem{nmssm_bblambda0.1}
Marcin Badziak, Marek Olechowski, and Stefan Pokorski.
\newblock Experimental signatures of a light singlet-like scalar in nmssm,
  2014.

\bibitem{nmssm_pushup}
Marcin Badziak, Marek Olechowski, and Stefan Pokorski.
\newblock New regions in the nmssm with a 125 gev higgs.
\newblock {\em Journal of High Energy Physics}, 2013(6), Jun 2013.

\bibitem{hugonie_lightesthiggs}
Ulrich Ellwanger and Cyril Hugonie.
\newblock The upper bound on the lightest higgs mass in the nmssm revisited.
\newblock {\em Modern Physics Letters A}, 22(21):1581–1590, Jul 2007.

\bibitem{nmssmDM}
Ulrich Ellwanger and Cyril Hugonie.
\newblock The higgsino–singlino sector of the nmssm: combined constraints
  from dark matter and the lhc.
\newblock {\em The European Physical Journal C}, 78(9), Sep 2018.

\bibitem{nmssmDM2}
Ulrich Ellwanger and Cyril Hugonie.
\newblock The semi-constrained nmssm satisfying bounds from the lhc, lux and
  planck.
\newblock {\em Journal of High Energy Physics}, 2014(8), Aug 2014.

\bibitem{vennin}
Vincent Vennin.
\newblock {\em {Cosmological Inflation: Theoretical Aspects and Observational
  Constraints}}.
\newblock Theses, {Universit{\'e} Pierre et Marie Curie}, September 2014.

\bibitem{NMSSMinflation1}
Sergio Ferrara, Renata Kallosh, Andrei Linde, Alessio Marrani, and Antoine
  Van~Proeyen.
\newblock Jordan frame supergravity and inflation in the nmssm.
\newblock {\em Physical Review D}, 82(4), Aug 2010.

\bibitem{NMSSMinflation2}
Sergio Ferrara, Renata Kallosh, Andrei Linde, Alessio Marrani, and Antoine
  Van~Proeyen.
\newblock Superconformal symmetry, nmssm, and inflation.
\newblock {\em Physical Review D}, 83(2), Jan 2011.

\bibitem{NMSSMinflation3}
Martin~B. Einhorn and D.~R. Timothy~Jones.
\newblock Inflation with non-minimal gravitational couplings in supergravity.
\newblock {\em Journal of High Energy Physics}, 2010(3), Mar 2010.

\bibitem{lightscalarLHC1}
Yuri Gershtein, Simon Knapen, and Diego Redigolo.
\newblock {Probing naturally light singlets with a displaced vertex trigger}.
\newblock 12 2020.

\bibitem{lightscalarLHC2}
Yuri Gershtein and Simon Knapen.
\newblock Trigger strategy for displaced muon pairs following the cms phase ii
  upgrades.
\newblock {\em Physical Review D}, 101(3), Feb 2020.

\bibitem{lightscalarLHC3}
Jared~A. Evans, Abhijith Gandrakota, Simon Knapen, and Hardik Routray.
\newblock Searching for exotic b meson decays enabled by the cms l1 track
  trigger.
\newblock {\em Physical Review D}, 103(1), Jan 2021.

\bibitem{nmspec}
Ulrich Ellwanger and Cyril Hugonie.
\newblock {NMSPEC: A Fortran code for the sparticle and Higgs masses in the
  NMSSM with GUT scale boundary conditions}.
\newblock {\em {Computer Physics Communications}}, 177(4):399--407, August
  2007.

\bibitem{suspect}
Abdelhak Djouadi, Jean-Loïc Kneur, and Gilbert Moultaka.
\newblock Suspect: A fortran code for the supersymmetric and higgs particle
  spectrum in the mssm.
\newblock {\em Computer Physics Communications}, 176(6):426–455, Mar 2007.

\bibitem{staub2012sarah}
F.~Staub.
\newblock Sarah, 2012.

\bibitem{sarah_2015}
Florian Staub.
\newblock Exploring new models in all detail withsarah.
\newblock {\em Advances in High Energy Physics}, 2015:1–126, 2015.

\bibitem{PQsym1}
R.~D. Peccei and Helen~R. Quinn.
\newblock Constraints imposed by cp conservation in the presence of
  pseudoparticles.
\newblock {\em Phys. Rev. D}, 16:1791--1797, Sep 1977.

\bibitem{PQsym2}
R.~D. Peccei and Helen~R. Quinn.
\newblock Cp conservation in the presence of pseudoparticles.
\newblock {\em Phys. Rev. Lett.}, 38:1440--1443, Jun 1977.

\bibitem{2loopRGE}
Stephen~P. Martin and Michael~T. Vaughn.
\newblock Two-loop renormalization group equations for soft
  supersymmetry-breaking couplings.
\newblock {\em Phys. Rev. D}, 50:2282--2292, Aug 1994.

\bibitem{nonrenormtheo}
Marcus~T. Grisaru, W.~Siegel, and M.~Rocek.
\newblock {Improved Methods for Supergraphs}.
\newblock {\em Nucl. Phys. B}, 159:429, 1979.

\bibitem{rge1}
Stephen~P. Martin and Michael~T. Vaughn.
\newblock Two-loop renormalization group equations for soft
  supersymmetry-breaking couplings.
\newblock {\em Phys. Rev. D}, 50:2282--2292, Aug 1994.

\bibitem{rge2}
Mark~D. Goodsell.
\newblock Two-loop rges with dirac gaugino masses.
\newblock {\em Journal of High Energy Physics}, 2013(1), Jan 2013.

\bibitem{rge3}
Renato~M. Fonseca, Michal Malinský, Werner Porod, and Florian Staub.
\newblock Running soft parameters in susy models with multiple gauge factors.
\newblock {\em Nuclear Physics B}, 854(1):28–53, Jan 2012.

\bibitem{ufo}
Céline Degrande, Claude Duhr, Benjamin Fuks, David Grellscheid, Olivier
  Mattelaer, and Thomas Reiter.
\newblock Ufo – the universal feynrules output.
\newblock {\em Computer Physics Communications}, 183(6):1201–1214, Jun 2012.

\bibitem{calchep}
Alexander Belyaev, Neil~D. Christensen, and Alexander Pukhov.
\newblock Calchep 3.4 for collider physics within and beyond the standard
  model.
\newblock {\em Computer Physics Communications}, 184(7):1729–1769, Jul 2013.

\bibitem{micromegas_lastv}
G.~Bélanger, F.~Boudjema, A.~Goudelis, A.~Pukhov, and B.~Zaldívar.
\newblock micromegas5.0 : Freeze-in.
\newblock {\em Computer Physics Communications}, 231:173–186, Oct 2018.

\bibitem{higgsbounds}
Philip Bechtle, Oliver Brein, Sven Heinemeyer, Oscar Stål, Tim Stefaniak,
  Georg Weiglein, and Karina~E. Williams.
\newblock Higgsbounds-4: improved tests of extended higgs sectors against
  exclusion bounds from lep, the tevatron and the lhc.
\newblock {\em The European Physical Journal C}, 74(3), Mar 2014.

\bibitem{effectivepot_approach}
Stephen~P. Martin.
\newblock Two-loop effective potential for a general renormalizable theory and
  softly broken supersymmetry.
\newblock {\em Physical Review D}, 65(11), May 2002.

\bibitem{goodsell2015twoloop}
M.~Goodsell, K.~Nickel, and F.~Staub.
\newblock Two-loop higgs mass calculation from a diagrammatic approach, 2015.

\bibitem{mssmprecision_usebySPheno}
Damien~M. Pierce, Jonathan~A. Bagger, Konstantin~T. Matchev, and Ren-Jie Zhang.
\newblock Precision corrections in the minimal supersymmetric standard model.
\newblock {\em Nuclear Physics B}, 491(1-2):3–67, Apr 1997.

\bibitem{fkit1}
A.~Vicente.
\newblock Flavorkit: a brief overview.
\newblock {\em Nuclear and Particle Physics Proceedings}, 273-275:1423--1428,
  2016.
\newblock 37th International Conference on High Energy Physics (ICHEP).

\bibitem{fkit2}
Florian Staub.
\newblock {Introduction to SARAH and related tools}.
\newblock Technical report, CERN, Geneva, Sep 2015.
\newblock Comments: 20 pages, Lecture and tutorial given at 'Summer School and
  Workshop on the Standard Model and Beyond', Corfu Summer Institute 2015.

\bibitem{gm2_calculus}
Tarek Ibrahim and Pran Nath.
\newblock Cpviolation and the muon anomaly inn=1supergravity.
\newblock {\em Physical Review D}, 61(9), Apr 2000.

\bibitem{fine_tuning1}
R.~Barbieri and G.F. Giudice.
\newblock Upper bounds on supersymmetric particle masses.
\newblock {\em Nuclear Physics B}, 306(1):63--76, 1988.

\bibitem{fine_tuning2}
R.~Barbieri and G.F. Giudice.
\newblock Upper bounds on supersymmetric particle masses.
\newblock {\em Nuclear Physics B}, 306(1):63--76, 1988.

\bibitem{pdg}
P.A. Zyla et~al.
\newblock {Review of Particle Physics}.
\newblock {\em PTEP}, 2020(8):083C01, 2020.

\bibitem{uncert_mh}
B.C Allanach, A~Djouadi, J.L Kneur, W~Porod, and P~Slavich.
\newblock Precise determination of the neutral higgs boson masses in the mssm.
\newblock {\em Journal of High Energy Physics}, 2004(09):044–044, Sep 2004.

\bibitem{Ellwanger_2012}
Ulrich Ellwanger and Pantelis Mitropoulos.
\newblock Upper bounds on asymmetric dark matter self annihilation cross
  sections.
\newblock {\em Journal of Cosmology and Astroparticle Physics},
  2012(07):024–024, Jul 2012.

\bibitem{gm2_exp}
{Abi, B. and Albahri, T. and Al-Kilani, S. and Allspach, D. and Alonzi, L. and
  Anastasi, A. and Anisenkov, A. and Azfar, F. and Badgley, K. and Baeßler, S.
  and et al.}
\newblock Measurement of the positive muon anomalous magnetic moment to 0.46
  ppm.
\newblock {\em Physical Review Letters}, 126(14), Apr 2021.

\bibitem{llpSM}
Brian Shuve.
\newblock Theory overview of long-lived particles.
\newblock https://www.sciencedirect.com/science/article/pii/0550321377904175,
  2017.

\bibitem{alphaT}
Lisa Randall and David Tucker-Smith.
\newblock Dijet searches for supersymmetry at the large hadron collider.
\newblock {\em Physical Review Letters}, 101(22), Nov 2008.

\bibitem{MT2}
C.G Lester and D.J Summers.
\newblock Measuring masses of semi-invisibly decaying particle pairs produced
  at hadron colliders.
\newblock {\em Physics Letters B}, 463(1):99–103, Sep 1999.

\bibitem{MT2W}
Yang Bai, Hsin-Chia Cheng, Jason Gallicchio, and Jiayin Gu.
\newblock Stop the top background of the stop search.
\newblock {\em Journal of High Energy Physics}, 2012(7), Jul 2012.

\bibitem{hiddenvalley}
Matthew~J. Strassler and Kathryn~M. Zurek.
\newblock Echoes of a hidden valley at hadron colliders.
\newblock {\em Physics Letters B}, 651(5-6):374–379, Aug 2007.

\bibitem{neutrinos}
Marco Drewes.
\newblock Theoretical status of neutrino physics, 2015.

\bibitem{llp_signatures}
Juliette Alimena.
\newblock {Status of searches in the long-lived particle and dark sectors,
  including full run-2 and HL-LHC prospects}.
\newblock Technical report, CERN, Geneva, Jul 2019.

\bibitem{llpCMS}
Jeremy Andrea, Daniel Bloch, Eric Conte, Douja Darej, Robin Ducrocq, and Emery
  Nibigira.
\newblock Searches for displaced top-quarks using reconstructed charged
  particles at the lhc at {14 TeV}.
\newblock In preparation.

\bibitem{sliceCMS}
David Barney.
\newblock Slice showing cms sub-detectors and how particles interact with them.
\newblock https://cds.cern.ch/record/2120661, 2016.

\bibitem{feynrules}
Adam Alloul, Neil~D. Christensen, Céline Degrande, Claude Duhr, and Benjamin
  Fuks.
\newblock Feynrules 2.0 — a complete toolbox for tree-level phenomenology.
\newblock {\em Computer Physics Communications}, 185(8):2250–2300, Aug 2014.

\bibitem{stop_prod_NLO}
W.~Beenakker, M.~Krämer, T.~Plehn, M.~Spira, and P.M. Zerwas.
\newblock Stop production at hadron colliders.
\newblock {\em Nuclear Physics B}, 515(1-2):3–14, Mar 1998.

\bibitem{stop_prod_NNLL}
Alessandro Broggio, Andrea Ferroglia, Matthias Neubert, Leonardo Vernazza, and
  Li~Lin Yang.
\newblock Nnll momentum-space resummation for stop-pair production at the lhc.
\newblock {\em Journal of High Energy Physics}, 2014(3), Mar 2014.

\bibitem{lhapdf}
Andy Buckley, James Ferrando, Stephen Lloyd, Karl Nordström, Ben Page, Martin
  Rüfenacht, Marek Schönherr, and Graeme Watt.
\newblock Lhapdf6: parton density access in the lhc precision era.
\newblock {\em The European Physical Journal C}, 75(3), Mar 2015.

\bibitem{MG5_dynamical_scheme}
Valentin Hirschi and Olivier Mattelaer.
\newblock Automated event generation for loop-induced processes, 2015.

\bibitem{kt_clustering}
Stephen~D. Ellis and Davison~E. Soper.
\newblock Successive combination jet algorithm for hadron collisions.
\newblock {\em Physical Review D}, 48(7):3160–3166, Oct 1993.

\bibitem{MA5}
Eric Conte, Benjamin Fuks, and Guillaume Serret.
\newblock Madanalysis 5, a user-friendly framework for collider phenomenology.
\newblock {\em Computer Physics Communications}, 184(1):222–256, Jan 2013.

\bibitem{Pmunu32}
Takeo Moroi.
\newblock {Effects of the gravitino on the inflationary universe}.
\newblock Other thesis, 3 1995.

\bibitem{pythia}
Torbjörn Sjöstrand, Stefan Ask, Jesper~R. Christiansen, Richard Corke,
  Nishita Desai, Philip Ilten, Stephen Mrenna, Stefan Prestel, Christine~O.
  Rasmussen, and Peter~Z. Skands.
\newblock An introduction to pythia 8.2.
\newblock {\em Computer Physics Communications}, 191:159–177, Jun 2015.

\bibitem{CMS-PAS-TOP-16-021}
{Investigations of the impact of the parton shower tuning in Pythia 8 in the
  modelling of $t\overline{t}$ at sqrt(s)=8 and 13 TeV}.
\newblock Technical report, CERN, Geneva, 2016.

\bibitem{FARRAR_Rhadron}
Glennys~R. Farrar and Pierre Fayet.
\newblock Phenomenology of the production, decay, and detection of new hadronic
  states associated with supersymmetry.
\newblock {\em Physics Letters B}, 76(5):575--579, 1978.

\bibitem{Rhadron_model11}
R.~Mackeprang and A.~Rizzi.
\newblock Interactions of coloured heavy stable particles in matter.
\newblock {\em The European Physical Journal C}, 50(2):353–362, Mar 2007.

\bibitem{Rhadron_model1}
A.~C. Kraan.
\newblock Interactions of heavy stable hadronising particles.
\newblock {\em The European Physical Journal C}, 37(1):91–104, Sep 2004.

\bibitem{Rhadron_model2}
R.~Mackeprang and D.~A. Milstead.
\newblock An updated description of heavy-hadron interactions in geant-4.
\newblock {\em The European Physical Journal C}, 66(3-4):493–501, Feb 2010.

\bibitem{fastjet}
Matteo Cacciari, Gavin~P. Salam, and Gregory Soyez.
\newblock Fastjet user manual.
\newblock {\em The European Physical Journal C}, 72(3), Mar 2012.

\bibitem{TDR1}
G.~L. Bayatian et~al.
\newblock {CMS Physics}: {Technical Design Report Volume 1: Detector
  Performance and Software}.
\newblock 2006.

\bibitem{CMS_Trigger}
V.~Khachatryan, A.~Sirunyan, A.~Tumasyan, W.~Adam, E.~Asilar, T.~Bergauer,
  J.~Brandstetter, E.~Brondolin, M.~Dragicevic, J.~Erö, and et~al.
\newblock The cms trigger system.
\newblock {\em Journal of Instrumentation}, 12(01):P01020–P01020, Jan 2017.

\bibitem{ATLAS_Trigger}
A~Ruiz~Mart{\'{\i}}nez and.
\newblock The run-2 {ATLAS} trigger system.
\newblock {\em Journal of Physics: Conference Series}, 762:012003, oct 2016.

\bibitem{hscp}
V.~Khachatryan, A.~Sirunyan, A.~Tumasyan, W.~Adam, E.~Asilar, T.~Bergauer,
  J.~Brandstetter, E.~Brondolin, M.~Dragicevic, J.~Erö, and et~al.
\newblock {Search for long-lived charged particles in proton-proton collisions
  at $\sqrt{s}=$13 TeV}.
\newblock {\em Physical Review D}, 94(11), Dec 2016.

\bibitem{stoprodCMS}
{Search for direct stop pair production in the dilepton final state at
  $\sqrt{s}=$13 TeV}.
\newblock Technical report, CERN, Geneva, 2017.

\bibitem{llp_eric}
Juliette Alimena, James Beacham, Martino Borsato, Yangyang Cheng, Xabier~Cid
  Vidal, Giovanna Cottin, David Curtin, Albert De~Roeck, Nishita Desai, Jared~A
  Evans, and et~al.
\newblock Searching for long-lived particles beyond the standard model at the
  large hadron collider.
\newblock {\em Journal of Physics G: Nuclear and Particle Physics},
  47(9):090501, Sep 2020.

\end{thebibliography}
\bibliographystyle{unsrt}

\includepdf[pages={1}]{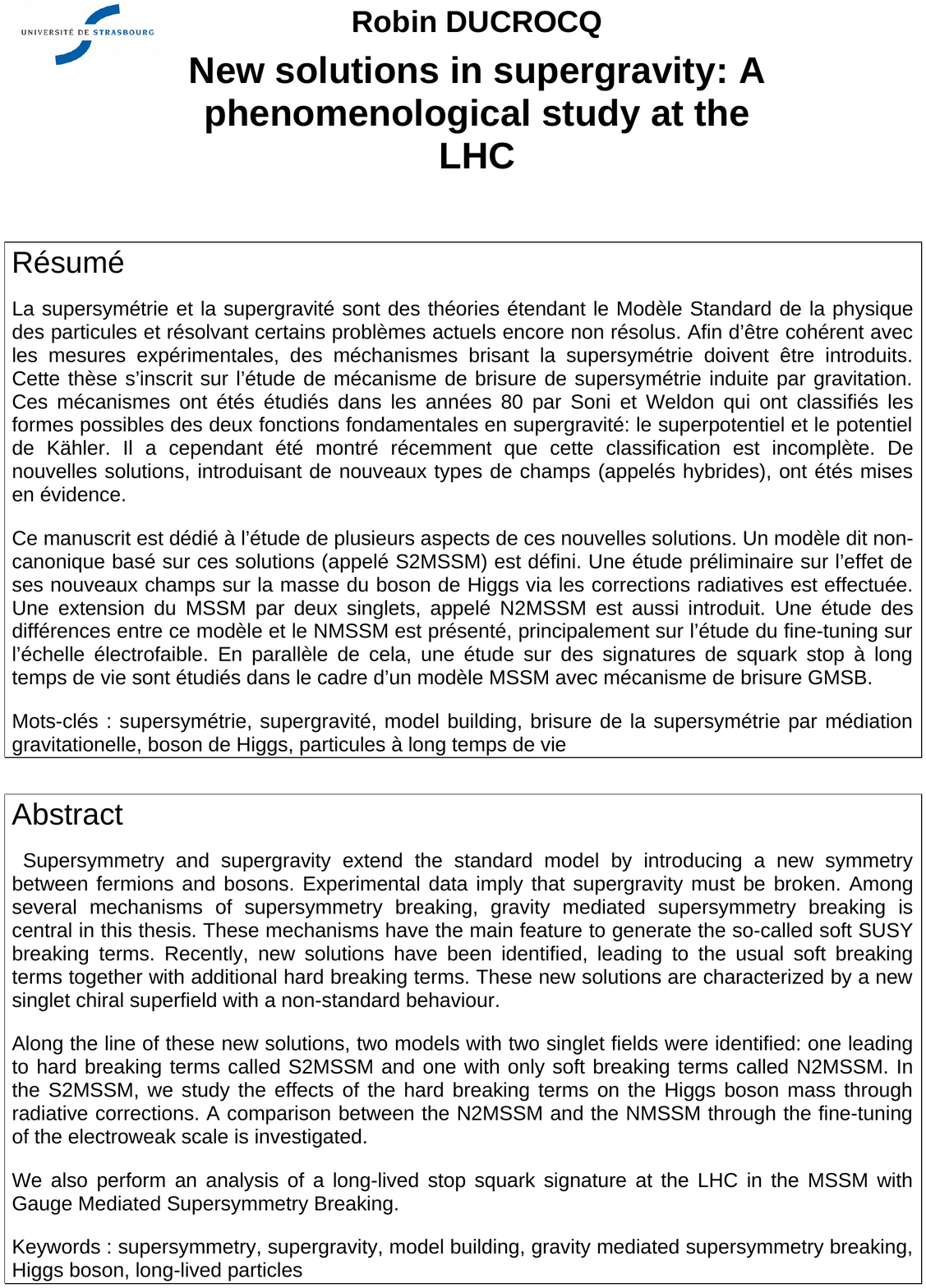}

\end{document}